\documentclass[a4paper,fleqn,usenatbib]{mnras}

\usepackage{newtxtext,newtxmath}

\usepackage[T1]{fontenc}
\usepackage{ae,aecompl}

\usepackage{graphicx}	
\usepackage{amsmath}	
\usepackage{amssymb}	
\usepackage{pdflscape}	
\usepackage{longtable}
\usepackage{hyperref}
\usepackage{xspace}
\usepackage{enumerate}
\usepackage{chemformula}
\usepackage{siunitx}
\usepackage{accents}
\usepackage{tensor}
\usepackage{tikz}
\usepackage{pifont}
\usepackage{longtable}
\newcommand{\kb}{k_\text{B}\xspace}
\newcommand{\kbt}{\kb T\xspace}
\newcommand{\eq}{^{\text{eq}}\xspace}
\newcommand{\st}{^\circ\xspace}
\newcommand{\tot}{_{\text{tot}}\xspace}

\newcommand{\dd}{\mathrm{d}}

\newcommand{\krome}{\textsc{krome}\xspace}

\newcommand{\gaussian}{\textsc{gaussian09}\xspace}

\newcommand{\repothermo}{\url{https://bitbucket.org/JelsB/thermochemistry}\xspace}
\newcommand{\zenodo}{\url{https://zenodo.org/record/3356710}\xspace}

\newcolumntype{h}{>{\setbox0=\hbox\bgroup}c<{\egroup}@{}}
\newcommand{\xquan}[3]{{#1}\st_{#2,#3}\xspace}
\newcommand{\xform}[3]{\Delta_f \xquan{#1}{#2}{#3}\xspace}
\newcommand{\tr}{_{\text{tr}}\xspace}
\newcommand{\rot}{_{\text{rot}}\xspace}
\newcommand{\vib}{_{\text{vib}}\xspace}
\newcommand{\elec}{_{\text{el}}\xspace}
\newcommand{\pd}{\partial\xspace}

\newcommand{\nucspec}{\ch{TiO2}, \ch{MgO}, \ch{SiO} and \ch{Al2O3}\xspace}
\newcommand{\Ti}[1]{{\ifthenelse{\equal{#1}{1}}{\ch{TiO2}\xspace}{\ch{(TiO2){#1}}\xspace}}}
\newcommand{\SiO}[1]{{\ifthenelse{\equal{#1}{1}}{\ch{SiO}\xspace}{\ch{(SiO){#1}}\xspace}}}
\newcommand{\Mg}[1]{{\ifthenelse{\equal{#1}{1}}{\ch{MgO}\xspace}{\ch{(MgO){#1}}\xspace}}}
\newcommand{\Al}[1]{{\ifthenelse{\equal{#1}{1}}{\ch{Al2O3}\xspace}{\ch{(Al2O3){#1}}\xspace}}}

\newcommand{\Ticlusters}{\Ti{1}-clusters\xspace}
\newcommand{\Siclusters}{\SiO{1}-clusters\xspace}
\newcommand{\Mgclusters}{\Mg{1}-clusters\xspace}
\newcommand{\Alclusters}{\Al{1}-clusters\xspace}
\newcommand{\fig}[1]{(Fig.\,#1)\xspace}
\newcommand{\figs}[2]{(Figs.\,#1,\,#2)\xspace}

\def\checkmark{\tikz\fill[scale=0.4](0,.35) -- (.25,0) -- (1,.7) -- (.25,.15) -- cycle;} 
\newcommand{\xmark}{\ding{55}}
\newcommand{\qmark}{{\Large ?}}

\newcommand{\comment}[2]{#2}
\newcommand{\rev}[1]{\textcolor{black}{#1}} 
\newcommand{\old}{\comment} 

\newcommand{\KIDA}{\textsc{KIDA} database \cite{Wakelam2012}}
\newcommand{\UMIST}{\textsc{UMIST} database \cite{McElroy2013} }
\newcommand{\NUCLEATION}{This work}
\newcolumntype{H}{@{}>{\lrbox0}l<{\endlrbox}}

\DeclareMathOperator{\kTroe}{k_\text{Troe}}
\DeclareMathOperator{\EqRatio}{EQR}
\DeclareMathOperator{\kClusterGrowth}{k^+_{N,M}}

\def\ontop#1{\vtop{\null\hbox{#1}}}

\title[Nucleation in AGB winds]{Developing a self-consistent AGB wind model:\\ II. Non-classical, non-equilibrium polymer nucleation in a chemical mixture}
\author[J. Boulangier et al.]{
Jels Boulangier$^{1}$\thanks{E-mail: boulangier.jels@gmail.com},
D. Gobrecht$^{1}$,
L. Decin$^{1,2}$\thanks{E-mail: leen.decin@kuleuven.be},
A. de Koter$^{1,3}$ and
J. Yates$^{4}$
\\
$^{1}$Institute of Astronomy, KU Leuven, Celestijnenlaan 200D, 3001 Leuven, Belgium\\
$^{2}$University of Leeds, School of Chemistry, Leeds LS2 9JT, United Kingdom\\
$^{3}$Anton Pannenkoek Institute for Astronomy, Universiteit van Amsterdam, Science Park 904, NL-1098 XH Amsterdam, The Netherlands\\
$^{4}$Department of Physics and Astronomy, University College London, Gower St., London WC1E 6BT, United Kingdom
}

\date{Accepted 22/08/2019}

\pubyear{2019}

\begin{document}
\label{firstpage}
\pagerange{\pageref{firstpage}--\pageref{lastpage}}
\maketitle

\begin{abstract}
Unravelling the composition and characteristics of gas and dust lost by asymptotic giant branch (AGB) stars is important as these stars play a vital role in the chemical life cycle of galaxies. The general hypothesis of their mass loss mechanism is a combination of stellar pulsations and radiative pressure on dust grains. However, current models simplify dust formation, which starts as a microscopic phase transition called nucleation. Various nucleation theories exist, yet all assume chemical equilibrium, growth restricted by monomers, and commonly use macroscopic properties for a microscopic process. Such simplifications for initial dust formation can have large repercussions on the type, amount, and formation time of dust. By abandoning equilibrium assumptions, discarding growth restrictions, and using quantum mechanical properties, we have constructed and investigated an improved nucleation theory in AGB wind conditions for four dust candidates, \nucspec. This paper reports the viability of these candidates as first dust precursors and reveals implications of simplified nucleation theories. Monomer restricted growth underpredicts large clusters at low temperatures and overpredicts formation times. Assuming the candidates are present, \Al{1} is the favoured precursor due to its rapid growth at the highest considered temperatures. However, when considering an initially atomic chemical mixture, only \Ti{1}-clusters form. Still, we believe \Al{1} to be the prime candidate due to substantial physical evidence in presolar grains, observations of dust around AGB stars at high temperatures, and its ability to form at high temperatures and expect the missing link to be insufficient quantitative data of \ch{Al}-reactions.
\end{abstract}



\renewcommand{\arraystretch}{1.2}

\section{Introduction}\label{sec:intro}
    Low and intermediate mass (initially \SIrange{0.8}{8}{M_{\sun}}) stars evolve through the asymptotic giant branch (AGB) phase at the end of their life time. During this phase, AGB stars lose vast amounts of material to their surroundings via a stellar wind and thereby contribute significantly to the chemical enrichment of the interstellar medium. As low (and intermediate) mass stars dominate the initial mass function, AGB stars are one of the main contributors of this chemical enrichment. The generally accepted hypothesis is that the mechanism triggering the onset of the AGB stellar wind is a combination of stellar pulsations and radiation pressure on newly formed dust grains \citep{Habing2003}. While dynamic models incorporating this scenario can explain observed wind mass loss rates and velocities of carbon-rich winds \citep{Woitke2006}, a substantial fine-tuning is needed for oxygen-rich winds \citep{Woitke2006a} and a model from first principles incorporating all physics and chemistry does not yet exist.\\\\
    Current AGB wind models implement dust growth by accretion of gas onto tiny solid particles, so-called seeds, based on the prescription of \citet{Gail1999}. Such seed particles are either predicted using a nucleation theory \citep[e.g.][]{Gail1988a, Helling2001, Woitke2006}, or are assumed to pre-exist, typically chosen to consist of 1000 monomers or to have a radius of \SI{1}{\nm} \citep[e.g.][]{Ferrarotti2006, Hofner2016, DellAgli2017}. To understand the wind formation mechanism from first principles, it is essential to use a nucleation theory. However, the most complex nucleation theories still assume chemical equilibrium, restrict growth of nucleation clusters to addition of monomers, and apply macroscopic properties of solids to describe clusters of a few molecules. Nonetheless, progress has been made regarding these assumptions, in a range of astrophysical fields where understanding dust formation crucial (e.g. in supernovae, brown dwarf atmospheres, and the interstellar medium). First, the assumption of chemical equilibrium is discarded by e.g. \citet{Sarangi2015, Gobrecht2016, Sluder2018} who treat nucleation as consecutive chemical reactions. From a chosen cluster size, they allow dust growth by coagulation of clusters, controlled by van der Waals forces \citep{Jacobson2013}. The chosen cluster size is typically less than 5 monomer units. As nucleation reaction rate coefficients are rarely known, these coefficients are often estimated and usually neglect the temperature dependence of the reaction. The latter is crucial to infer dust formation rates as a function of the radial distance from the AGB star. Second, the use of bulk solid properties for molecular clusters is abandoned by e.g. \citet{Kohler1997, Goumans2012, Lee2015, Bromley2016, Lee2018} by adopting chemical potential energies from detailed quantum mechanical calculations. When describing the clustering of gas phase molecules it is inaccurate to use extrapolated bulk properties, such as binding energy and surface tension, firstly because cluster binding energies are generally significantly reduced with respect to the bulk one, and secondly because microscopic clusters do not resemble the shape/structure of the solid \citep{Johnston2002, Lamiel-Garcia2017, Gobrecht2017}. E.g., small clusters do not have well-defined surfaces like solids, rendering the use of surface tension meaningless. Third, as far as we know, no astrophysical models exist where the nucleation and the growth are not restricted by specific cluster size additions (e.g. monomers or dimers). Yet polymer and more complex nucleation theories have been developed in non-astrophysics disciplines, e.g. nano and solid-state physics. \citet[and references therein]{Clouet2009} provides a good overview of different complexity levels of nucleation theory from a non-astrophysical perspective.\\\\
    Presolar grains can be identified in meteorites, interplanetary particles, and cosmic dust by isotopic anomalies that cannot be explained by physical or chemical processes within the Solar System. The origin of the grains can be traced by isotopic ratios of atoms in the grains \citep{Nittler1997} and point to other nucleosynthetic environments such as AGB stars or supernovae \citep{McSween2010}. Here we focus on grains with an AGB origin. Since the first discovery of a presolar \Al{1} grain by \citet{Hutcheon1994}, several presolar oxides have been found of which the majority are \Al{1} grains (corundum) and only a few are \ch{MgAl2O4} (spinel) \citep[e.g.][]{Nittler1994, Choi1998, Nittler2008}. Note that \Al{1} grains are often referred to as corundum, which is the thermodynamically most stable solid bulk form, yet \Al{1} exists in a variety of structural forms in presolar grains \citep{Stroud2004, Stroud2007}. Subsequently, \citet{Nittler2008} identified the first \ch{Ti}-oxides in presolar grains, however they did not have any crystallographic data that would allow to determine the structure of the grains or even conclude if they were \Ti{1}-grains. Later, \citet{Bose2010a} claim to have found a \Ti{1}-grain. The occurrence of \ch{Ti}-bearing presolar grains  is low and their rarity is often explained by the low \ch{Ti} abundance in AGB stars. Additionally, presolar silicate grains (containing \ch{Si}-oxides) have been found \citep{Nguyen2009, Bose2010, Bose2012}. A more extended summary of discovered presolar grains can be found in the Presolar Grain Database\footnote{\url{https://presolar.physics.wustl.edu/presolar-grain-database}} \citep{Hynes2009}. Besides physical evidence of presolar grains, there is also observational evidence for different dust precursors in AGB winds. Notably the \SI{13}{\micron} feature, which is found in spectra of half of all AGB stars \citep{Sloan1996,Speck2000,Sloan2003}, is thought to be caused by \Al{1}-grains \citep{Zeidler2013, Takigawa2015, Depew2006}, or \ch{MgAl2O4} \citep{Posch1999}, or by \ch{SiO2} or polymerised silicates \citep{Speck2000}. Since there is no consensus on what causes this feature, there is still a large uncertainty on the composition of dust in AGB winds.\\\\
    We investigated the viability of \nucspec as candidates of oxygen-rich AGB dust precursors with a revised nucleation theory. We have improved on the current nucleation theories by abandoning equilibrium assumptions, discarding growth restrictions, and using quantum mechanical properties of cluster molecules. Firstly, we evolve a nucleation system kinetically, therefore it is time dependent and not in equilibrium. Secondly, the revised theory also allows polymer nucleation, not just interactions via monomers. Thirdly, quantum mechanical properties of molecular clusters are calculated with high accuracy density functional theory. Subsequently, these are used in chemical interactions between the nucleation clusters instead of using extrapolations from bulk material. The abundances and formation times of the largest nucleation clusters are examined in a closed nucleating system (no interaction with other chemical species) and in a large chemical mixture. The former assumes the monomer to be a priori present and is unable to be destroyed into smaller species. The latter allows chemical interactions between all species and starts from a purely atomic composition. To describe the chemical interactions, we used the reduced chemical reaction network of \citet{Boulangier2019} and extended this with additional reactions required to chemically couple to the nucleation candidates.\\\\
    Section~\ref{sec:method} describes the chemical evolution of a closed system and presents the improved nucleation theory. Section~\ref{sec:modelsetup} justifies the chosen nucleation candidates \rev{and explains two different nucleation models. Firstly, a closed nucleating model which only considers one nucleating species without interaction with other chemical species. Secondly, a comprehensive nucleating model which considers all nucleating species simultaneously in a large chemical mixture.} Additionally, it elaborates on the used nucleation networks, the construction thereof, and the details of the used quantum mechanical data. Section~\ref{sec:results} presents the results of the evolution of all nucleation candidates for the different model setups. Section~\ref{sec:impact} focuses on the implications of the model results. Section~\ref{sec:discussion} discusses the limitations of the revised nucleation, the model setups, and compares the results to previous studies. Finally, section~\ref{sec:summary} summarises this work. The appendix consists of detailed description of used calculations (Apps.~\ref{app:minGFE}--\ref{app_sec:GFEoF}) and an overview of all quantum mechanical data sources (App.~\ref{app:quantum_data}). \rev{Additional figures of the model results and the used chemical network are available as appendices~\ref{app:results} and \ref{app:chem_network}.}

\section{Methods}\label{sec:method}
    This section covers the general theory of chemical reactions and how to evolve such a system, i.e. chemical kinetics (Sec.~\ref{sec:chemistry}), and the construction of our improved non-classical, non-equilibrium polymer nucleation theory (Sec.~\ref{sec:KNT}).
    
    \subsection{Chemistry}\label{sec:chemistry}
        The evolution of the composition of a system is dictated by a set of chemical formation and destruction reactions. Mathematically, this is a set of coupled ordinary differential equations where the change in number density of the \textit{i}th species is given by,
        \begin{equation}
        \label{ODEsys}
            \frac{\text{d}n_i}{\text{d}t} = \sum_{j\in F_i} \left( k_j \prod_{r \in R_j} n_r \right)- \sum_{j \in D_i} \left( k_j \prod_{r \in R_j} n_r \right).
        \end{equation}
        Here, the first term, within the summation, represents the rate of formation of the \textit{i}th species by a single reaction $j$ of a set of formation reactions $F_i$. The second term is the analogue for a set of destruction reactions $D_i$. Each reaction $j$ has a set of reactants $R_j$, where $n_r$ is the number density of each reactant. The rate coefficient of this reaction is represented by $k_j$ and has units m$^{3(N-1)}$ s$^{-1}$ where $N$ is the number of reactants involved. To solve the chemical evolution of a system, we use the open source \krome\footnote{\url{http://kromepackage.org/}} package \citep{Grassi2014}, that is developed to model chemistry and microphysics for a wide range of astrophysical applications.\\\\
        In general, the rate coefficient of a two body reaction
        \begin{equation}
            \ch{A + B -> C + D}
        \end{equation}
        is given by
        \begin{equation}\label{eq:coll_rate}
            k = \int_0^\infty \sigma v_r f(v_r) \text{d}v_r,
        \end{equation}
        where $\sigma$ is the total cross section of an  \ch{A}-\ch{B} collision, $v_r$ is the relative speed between \ch{A} and \ch{B}, and $f(v_r)$ is a (relative) speed distribution. The total cross section of a two-particle collision depends on the kinetic energy of both particles and their microphysical interactions. However, the reaction is often reduced to an inelastic collision of two hard spheres due to lack of detailed chemical information. In this case, the total cross section is the geometrical cross section of both spheres, $\sigma = \pi(r_A + r_B)^2$ where $r_A$ and $r_B$ are the radii of both species. The speed distribution can be represented by the Maxwell-Boltzmann relative speed distribution, that considers the motion of particles in an ideal gas,
        \begin{equation}\label{eq:maxwell}
            f(v_r) = \left(\frac{\mu}{2\pi\kb T}\right)^{3/2}4\pi v_r^2 e^{-\frac{\mu v_r^2}{2\kb T}},
        \end{equation}
        where $\mu=\frac{m_A m_B}{m_A + m_B}$ is the reduced mass of the system, $\kb$ is the Boltzmann constant, and $T$ is the temperature of the gas. Note that when the reaction requires an activation energy $E_a$, the integral in equation \eqref{eq:coll_rate} should be evaluated from the equivalent speed $v_a = \sqrt{2E_a/\mu}$, rather than zero. Using the geometrical cross section and the Maxwell-Boltzmann distribution, equation \eqref{eq:coll_rate} results in
        \begin{equation}\label{eq:rate}
            k = \pi(r_A + r_B)^2\sqrt{\frac{8\kb T}{\pi \mu}}\left(1+\frac{E_a}{\kb T} \right)e^{-\frac{E_a}{\kb T}}.
        \end{equation}
        In the limit where $E_a \gg \kb T$ this reduces to
        \begin{equation}
            k = \pi(r_A + r_B)^2\sqrt{\frac{8\kb}{\pi \mu}}\frac{E_a}{\kb} T^{-0.5} e^{-\frac{E_a}{\kb T}},
        \end{equation}
        and has the form of a modified Arrhenius' equation,
        \begin{equation}\label{eq:arrhenius}
            k_{\text{Ar}} = \alpha T^{\beta} e^{-\frac{\gamma}{T}},
        \end{equation}
        where $\alpha, \beta,$ and $\gamma$ are constants. In the limit where there is no activation energy or when $E_a \ll \kb T$, the last two terms in equation \eqref{eq:rate} reduce to $1$ and the rate coefficient is given by
        \begin{equation}\label{eq:easyrate}
             k = \pi(r_A + r_B)^2\sqrt{\frac{8\kb T}{\pi \mu}},
        \end{equation}
        which also has the modified Arrhenius' form. Here, the last factor denotes the average relative speed, often quoted as thermal velocity\footnote{This is, however, not a vector quantity and naming this a velocity is therefore confusing and should be avoided. The correct terminology is average relative speed.}.

    \subsection{Nucleation theory}\label{sec:KNT}
        We assume that the nucleation process is homogeneous and homomolecular. The former states that there are no preferential sites for nucleation to start, and the latter means that nucleation happens by addition of the same molecular type of clusters.  Heteromolecular nucleation is omitted since in this case the number of possible reactions would increase exponentially. Additionally, nucleation occurs in a pure gas-phase condition and as such no preferential nucleation sites exist. This is different compared to nucleation that can occur on solid-state surfaces which can act as a catalyst or where crystal lattice defects can reduce the energy needed for nucleation to start.\\\\
        In general, a nucleation/cluster growth reaction is represented by,
        \begin{equation}
            \ch{C_N + C_M -> C_{N+M}}
        \end{equation}
        where $C_N$ and $C_M$ are clusters\footnote{A cluster $C_N$ of specific size $N$ denotes a molecule that exists of $N$-times molecule $C$, e.g. \SiO{2} is an \SiO{1}-cluster of size 2.} of size $N$ and $M$, respectively. Due to a lack of reaction rate coefficients in the literature, the rate coefficient is determined via equation \eqref{eq:easyrate} by assuming an inelastic collision where the activation energy of the reaction is much smaller than $\kbt$ and is given by
        \begin{equation}\label{eq:k_growth_gen}
            k^+_{N,M} = \pi(r_N + r_M)^2\sqrt{\frac{8\kb T}{\pi \mu_{N,M}}},
        \end{equation}
        where $\mu_{N,M}$ is the reduced mass of the ($N,M$)-system
        , and $r_N$ and $r_M$ are the radii of clusters of size $N$ and $M$, respectively. Assuming that the volume scales linearly with the size of the clusters, the radii can be written as function of the monomer radius\footnote{This assumption reduces the amount of needed information, i.e. just one molecule radius instead of $N$ radii. It does, however, also decrease the accuracy of the description.} $r_1$,
        \begin{equation}\label{eq:k_growth}
            k^+_{N,M} = \pi(N^{1/3} r_1 + M^{1/3}r_1)^2\sqrt{\frac{8\kb T}{\pi \mu_{N,M}}}.
        \end{equation}
        Note that the assumption of a spherical cluster can be generalised to a fractal cluster with a fractal radius $r_{f,N}=N^{1/D_f}r_1$, where $D_f$ is the fractal dimension, which equals 3 for spheres.\\\\
        A cluster destruction process of an $(N+M)$-sized cluster is represented by
        \begin{equation}
            \ch{C_{N+M} -> C_N + C_M},
        \end{equation}
        The rate coefficient can be derived from the principle of detailed balance which states that, at equilibrium, each elementary process is equilibrated by its reverse process. Hereby, we assume that the destruction rate is an intrinsic property of the cluster and does not depend on the embedding system (i.e. no collisional dissociation). We therefore assume that the cluster has enough time to relax to the lowest energy configuration between its formation and spontaneous break-up. This assumption is consistent with the fact that we describe a cluster solely by its size and minimal energy configuration. With the principle of detailed balance, the destruction rate coefficient can be determined via,
        \begin{align}
            n_{N+M}\eq\, k^-_{N,M}  &= n\eq_N\, n\eq_M\, k^+_{N,M}\nonumber\\
             k^-_{N,M} &= \frac{n\eq_N\, n\eq_{M}}{n\eq_{N+M}}\, k^+_{N,M}, \label{eq:kdestruction}
        \end{align}
        where $n\eq_N$ is the equilibrium number density of the $N$-sized cluster and $k^+_{N,M}$ is the growth rate coefficient of the reversed reaction (Eq.\,\ref{eq:k_growth_gen}). For a system at constant pressure and temperature, the equilibrium number distribution is determined by minimising its Gibbs free energy (App.~\ref{app:minGFE}). Consequently, the ratio of the clusters is given by
        \begin{equation}
            \frac{n\eq_N\, n\eq_{M}}{n\eq_{N+M}} = n\tot \exp \left( \frac{G_{N+M} - G_M - G_{N}}{\kbt} \right),
        \end{equation}
        where $G_N$ is the Gibbs free energy of an $N$-sized cluster and $n_{\text{tot}}$ is the total number density of the gas\footnote{Note that this is only valid in the dilute limit, i.e. the number of clusters is small compared to the total number of particles.}. It is more convenient to use the Gibbs free energies at standard pressure $(P\st = 1 \text{ bar}  = \SI{e5}{\Pa} = \SI{1e6}{dyne\per\cm\squared})$. Here, the superscript $\st$ refers to a quantity evaluated at this standard pressure. Using equation \eqref{app:ratioEnd} this ratio is given by
        \begin{equation}\label{eq:equi_ratios}
            \frac{n\eq_N\, n\eq_{M}}{n\eq_{N+M}} = \frac{P\st}{\kbt} \exp \left( \frac{G_{N+M}\st - G_M\st - G_{N}\st}{\kbt} \right).
        \end{equation}
        Substituting this ratio into equation \eqref{eq:kdestruction} yields a cluster destruction rate coefficient
        \begin{equation}\label{eq:k_destruction}
            k^-_{N,M} = k^+_{N,M}\frac{P\st}{\kbt} \exp \left( \frac{G_{N+M}\st - G_M\st - G_{N}\st}{\kbt} \right).
        \end{equation}
        Note that the standard Gibbs free energies are often given in \si{\kJ\per\mole}, in which case the Boltzmann constant $\kb$ in the exponential has to be replaced with the universal gas constant $R$ in \si{\kJ\per\K\per\mole}.

\section{Model setup}\label{sec:modelsetup}
This section explains the two different nucleation descriptions that have been used, a monomer and polymer one (Sec.\,\ref{sec:nucleation_description}). Next, it justifies the choice of nucleation candidates that have been considered, namely \nucspec  (Sec.\,\ref{sec:nucl_candidates}). Additionally, it describes the two different types of chemical nucleation networks, a closed one and a comprehensive one (Secs.~\ref{sec:closed_model} and \ref{sec:comprehensive_model}). \rev{The closed nucleating network assumes the monomer to be a priori present and is unable to be destroyed into smaller species. No assumptions have been made on how the monomer has been formed or its possible existence. The comprehensive nucleating network does not assume the existence of the nucleating monomers and starts from a purely atomic composition. The (possible) formation of the nucleating monomers and other chemical species is determined by a large chemical reaction network.} Finally, this section summarises all the additionally gathered data and performed calculations prior to running the nucleation models (Sec. \ref{sec:data_and_calcs}).

    \subsection{Nucleation description}\label{sec:nucleation_description}
    We consider two different nucleation descriptions, polymer and monomer nucleation. The former is the most general and uses growth and destruction of the corresponding clusters described by equations \eqref{eq:k_growth} and \eqref{eq:k_destruction}, whereas the latter uses those same equation but with $M=1$ reducing it to a monomer. We make this distinction because, to our knowledge, most homomolecular nucleation studies assume monomer nucleation \citep[e.g.][]{Kohler1997,Lee2015,Bromley2016,Lee2018}. However, the monomer assumption is only valid when the number of monomers is much larger than that of any other cluster. There is no quantitative evidence to support this assumption and it turns out to be invalid in our parameter space\footnote{For higher densities this will be even less valid, e.g. brown dwarfs and planetary atmospheres.} (Sec.\,\ref{sec:results}). \citet{Sarangi2015, Gobrecht2016, Sluder2018}, however, do allow polymer nucleation but limit it to small clusters ($N<5$).

    \subsection{Nucleation candidates}\label{sec:nucl_candidates}
\begin{figure}
        \centering
        \includegraphics[width=\columnwidth]{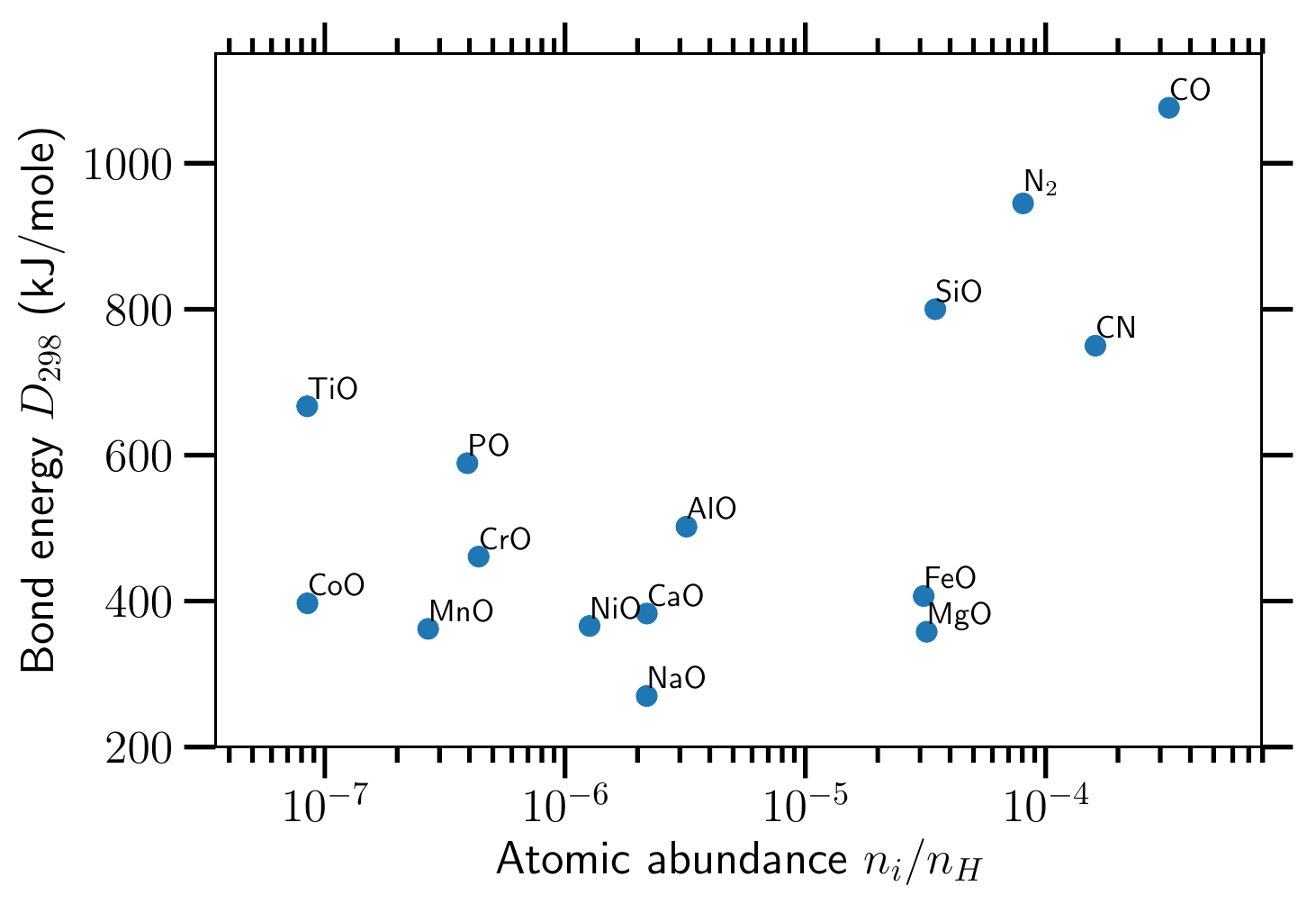}
        \caption{Simple molecules (mainly oxides) with high bond energies at \SI{298}{\K} \citep{Luo2007} and/or a high atomic abundance provide hints at which species play a dominant role in the initial dust formation in AGB winds.}
        \label{fig:bond_energies}
    \end{figure}

    In oxygen-rich atmospheres $(\ch{C}/\ch{O}<1)$, carbon is predominantly locked-up in \ch{CO}, strongly inhibiting the formation of carbonaceous dust. \old{Alternatively, nucleation of pure oxygen is limited to \ch{O2} and \ch{O3} because no larger stable oxygen chains exist.} Highly stable molecules \rev{in an carbon-deficient gas} such \ch{CO}, \ch{N2}, and \ch{CN} only have a solid form (ice) at temperatures well below \SI{500}{\K}. \rev{Also solid oxygen only forms at extremely cold temperatures.} Hence, nucleation at high temperatures must proceed via hetero-atomic species such as composite metal\footnote{\label{foot:metal}We refer to the chemical use of \textit{metals} and not the astronomical one.} oxides. Monomers with high bond energies\footnote{Bond energy is a measure of the strength of a chemical bond.} are preferential candidates for first nucleation because higher energies generally allow for easier formation and more difficult destruction at higher temperatures. Therefore, bond energies of simple metal oxides give a hint for which molecules will play a predominant role. Considering the most abundant atomic metals in AGB winds, \ch{SiO}, \ch{TiO}, and \ch{AlO} are the metal oxides with the highest bond energy \fig{\ref{fig:bond_energies}}. Even though the amount of \ch{Ti} is almost a factor 40 and 400 lower than \ch{Al} and \ch{Si}, respectively, it can still be an important molecule due to its high bond energy. Similarly, \ch{MgO}, and \ch{FeO} have lower bond energies but the high atomic abundance of \ch{Mg} and \ch{Fe} can make them important nucleation candidates.\\\\
    Although the metal oxides hint at the engaged species, the most compelling evidence for nucleation building blocks comes from presolar grains. \rev{Considering all the presolar grains that originated from AGB stars, \ch{Al2O3} grains are the most frequently occurring oxygen-bearing ones. \citep{Hutcheon1994, Nittler1994, Choi1998, Nittler2008}} \old{\ch{Al2O3} grains are the most frequently occurring oxygen-bearing particles in presolar grains that originated from AGB stars \citep{Hutcheon1994, Nittler1994, Choi1998, Nittler2008}}. In these grains, \ch{Al2O3} is the basic building block (repeating formula unit) that forms the bulk grains with a variety of structural forms \citep{Stroud2004, Stroud2007}. The repetition of such a basic building block strengthens our assumption of homomolecular nucleation. The second most frequently found grains, roughly a factor $7$ less abundant, are the ones with \ch{MgAl2O4} as repeating formula unit \citep{Nittler1994, Choi1998, Nittler2008}. Additionally, there is some evidence for silicon and titanium oxides in presolar grains \citep{Nguyen2009, Bose2010a, Nittler2008, Bose2010}. However, as only little amount of this material is detected, it is unclear what the repeating basic building block is.\\\\
    Considering the occurrence in presolar grains, the atomic metal abundance, and the bond energy of simple metal oxides, we choose \ch{Al2O3} to be our primary nucleation candidate. Next, we do not consider \ch{MgAl2O4} as a candidate as this molecule consist of three different atoms, making it more complex to characterise its molecular features. We include \ch{MgO} as a candidate because it (and its clusters) might play a role in the formation of \ch{MgAl2O4} grains. Additionally, we take \ch{TiO2} as a nucleation candidate. Even though there is no substantial evidence for \ch{TiO2} to be the repeating formula unit in presolar grains containing titanium oxides, it is, however, the repeating basic building block in other commonly found titanium minerals on Earth (e.g. rutile and anatase). Lastly, we select \ch{SiO} as a candidate. Although there is no physical evidence in presolar grains that \ch{SiO} is the repeating formula unit, it does have the highest bond energy of the most abundant atomic metals and it most likely will play an important role in the formation of silicate grains. We exclude \ch{FeO} from this study because, so far, only one potential detection of \ch{FeO} in AGB circumstellar environment has been reported \citep{Decin2018}, nor has there been proof of \ch{FeO}-containing particles in pre-solar grains. Additionally, \ch{Fe}-containing nanoparticles can display various magnetic behaviours such as ferromagnetic, antiferromagnetic, ferrimagnetic, and nonmagnetic, and are therefore challenging to characterise.\\\\
    A typical interstellar dust grains of radius \SI{0.1}{\micron} contains \num{e9} monomer units with a typical radius of roughly \SI{0.1}{\nm} (Table~\ref{tab:cluster_info}). Hence, in order to construct a dust grain via reaction rate equations, one needs of the order of \num{e9} equations. As this is computationally impossible, we limit the maximum cluster size so the largest clusters roughly consist of \SIrange{20}{40}{} atoms, making it still feasible to perform high accuracy density functional theory calculations (Sec.~\ref{sec:quantum_chem}). We take the largest cluster to be \Ti{10}, \SiO{10}, \Mg{10}, and \Al{8}. Note that these cluster sizes are not necessarily the threshold from which the species can be considered as a macroscopic, solid dust grain (Sec.~\ref{sec:discussion}).
    
    \begin{table}
    \center
    \caption{Initial chemical composition. This is equal to the time-averaged mass fractions in the wind for a nucleosynthetic AGB evolutionary model  with an initial mass of 1 M$_{\odot}$ and metallicity Z=0.02 of \citet{Karakas2010}. The mass fraction of \ch{Ti} is that of solar abundance \citep{Asplund2009}.}
    \label{tab:init_composition}
        \begin{tabular}{lll}
            Element $i$ & Mass fraction $X_i$  & $n_i/n_H$\\ \hline
            \ch{He} & $3.11\cdot 10^{-1}$ & \SI{1.16e-1}{}\\
            \ch{C} & $2.63\cdot 10^{-3} $& \SI{3.26e-4}{}\\
            \ch{N} & $1.52\cdot 10^{-3} $& \SI{1.61e-4}{}\\
            \ch{O} & $ 9.60\cdot 10^{-3} $& \SI{8.92e-4}{}\\\\
            \ch{F} &  $4.06\cdot 10^{-7}$&  \SI{3.18e-8}{}\\
            \ch{Na} & $3.38\cdot 10^{-5} $& \SI{2.18e-6}{}\\
            \ch{Mg} &  $5.16\cdot 10^{-4} $& \SI{3.19e-5}{}\\
            \ch{Al} & $\SI{5.81e-5}{}$& \SI{3.20e-6}{} \\\\
            \ch{Si} &  $6.54\cdot 10^{-4} $& \SI{3.47e-5}{}\\
            \ch{P} & $8.17\cdot 10^{-6} $& \SI{3.92e-7}{}\\
            \ch{S} &  $3.97\cdot 10^{-4} $& \SI{1.84e-5}{}\\
            \ch{Ti} & $\SI{2.84e-6}{}$& \SI{8.44e-8}{} \\\\
            \ch{Fe} &  $1.17\cdot 10^{-3} $& \SI{3.16e-5}{}\\
            \ch{e-} & 0 & 0\\
            \ch{H} & $1 - \sum_i^N X_i $ & 1 \\ 
            &$=\SI{6.72e-1}{}$&\\
        \end{tabular}
    \end{table}
    
    \subsection{Closed nucleation networks}\label{sec:closed_model}
    A closed nucleation model corresponds to the evolution of a cluster system according to growth and destruction rate coefficients (Eqs.~\ref{eq:k_growth} and \ref{eq:k_destruction}) with the monomer as the smallest and the maximally considered cluster size as the largest allowed clusters. Such a model starts with an initial monomer abundance and follows the growth of this monomer over time at a fixed temperature. We construct a model grid in temperature and density that is primarily applicable to an AGB wind (but that is also valid in other environments) and evolve each model over a timescale of one year. The latter corresponds to the longest dynamically stable period (between pulsation-induced consecutive shocks), resulting in a roughly constant local temperature and density in that period. For the initial abundance of the monomer we assume all of the available atomic metal\textsuperscript{\ref{foot:metal}} to be locked-up in the monomer (Table~\ref{tab:init_composition}). For the available atomic metal abundance we choose the same composition as \citet{Boulangier2019} who take the time-averaged elemental mass fractions in the wind from \SI{1}{M_{\sun}} and $Z=0.02$ AGB evolution model of \citet{Karakas2010} (defined as $\left<X(i)\right>$ in \citet{Karakas2007}). For \ch{Ti} we take the solar abundance because this element is not considered in the nucleosynthesis networks of \citet{Karakas2010}\footnote{\rev{The abundance of \ch{Ti} is not affected by the slow neutron capture process because of low neutron capture cross sections for elements below iron, and burning temperatures are not high enough for higher burning processes to affect \ch{Ti}. Hence, $\left<X(i)\right>$ of Ti does not change between birth and death of low and intermediate mass stars.}}.

    \subsection{Comprehensive chemical nucleation network}\label{sec:comprehensive_model}
    A comprehensive nucleation model corresponds to the evolution of nucleation clusters in a large chemical network according to growth and destruction rate coefficients (Eqs.~\ref{eq:k_growth} and \ref{eq:k_destruction}) until a specified maximum cluster size. Such a model starts from the atomic composition rather than the initial monomer abundance which is used in a closed nucleation model (Sec.~\ref{sec:closed_model}). This is a more realistic prescription as is removes the assumption of the monomer being (abundantly) present. Moreover, it allows for more chemical interaction between species and the creation of other metal-bearing molecules besides the nucleation candidate clusters. In practice, the reaction network consists of the closed nucleation networks of \Ti{1}, \Mg{1}, \SiO{1}, and \Al{1} (Sec.~\ref{sec:closed_model}) extended with the reduced AGB wind network of \citet{Boulangier2019}. However, because their reduced network does not consider any \ch{Ti}, \ch{Al}, and only a few \ch{Mg} reactions, we have added all reactions that include these elements available in the literature. Additionally, where necessary and possible, we have included the reversed reaction based on the assumption of detailed balance\footnote{The reversed rate coefficient depends on the difference in Gibbs free energy of reactants and products (i.e. the Gibbs free energy of reaction). If there was insufficient data in the literature to calculate these energy values, we did not include the reversed reaction.}. As with the closed nucleation models, we compute the same grid of models in temperature and density over a one year period but with an initial atomic composition (Table~\ref{tab:init_composition}).
    
    \subsection{Justification of nucleation networks}
    \rev{It is instructive to investigate the nucleation of chemical species in a closed system with the assumption of an a priori monomer existence to gain insight in the efficiency of the nucleation process different species. Such preliminary nucleation investigations can already exclude candidates as viable AGB dust precursors based on inefficient nucleation at high temperatures. This pre-selection of nucleation candidates leads to a considerable reduction of the computational cost when coupling the reaction network to a hydrodynamical framework. Moreover, a closed nucleation investigation reduces the number of uncertainties when interpreting the nucleation process. For example, the nucleation of clusters in a large chemical network might not occur due to an insufficient or incorrect description of the gas-phase chemistry prior to the monomer formation rather than the nucleation process itself, which can be very effective. By ignoring the disentanglement between monomer formation and the nucleation process, the nucleation species can be wrongly discarded as a good dust candidate. Additionally, the closed nucleation system allows us to investigate the impact of using the improved nucleation description, such as monomer versus polymer nucleation and using molecular energies compared to bulk energies.}

    \subsection{Construction of nucleation networks}\label{sec:data_and_calcs}
    This section covers the additional chemical reactions, quantum mechanical properties and calculations needed to construct valuable nucleation reaction networks. The first section describes the addition of chemical reactions and the second section the collection and calculation of quantum mechanical properties of molecules and clusters necessary for certain reversed reactions.

    \subsubsection{Additional reactions}
    In order to construct a reaction network for the comprehensive nucleation models, reactions from atomic \ch{Ti}, \ch{Al}, \ch{Si}, and \ch{Mg} up to the corresponding nucleation monomer have to be included. Additionally, to increase the accuracy of chemical interactions, as many as possible other nucleation-related metal-bearing molecules should be added to the network with corresponding reactions. Even though some species or reactions might not be important and could be omitted, such filtering is beyond the scope of this paper because computation time is currently not an issue as we only perform grids of models rather than coupling it in real-time to a hydrodynamical framework.\\\\
    \ch{Ti}-bearing molecules are not well studied and corresponding reaction rate coefficients are lacking in astrochemical databases. We could only find 9 reactions of which only one had a reversed reaction. For the remaining 8 reversed reactions we assumed detailed balance. We did, however, ignore reactions for the \ch{Ti}-\ch{Cl}-\ch{H} system \citep{Teyssandier1998} due to the low abundance of both \ch{Cl} and \ch{Ti} in AGB stars.\\\\
    Apart from the SiO-nucleation reactions, \rev{just one} other \ch{Si}-reaction is added relative to \citet{Boulangier2019}, whose network is mainly constructed from the astrochemical databases UMIST \citep{McElroy2013} and KIDA \citep{Wakelam2012} in which \ch{Si}-bearing molecules are well-studied. \rev{The destruction of \ch{SiO2} by atomic hydrogen, calculated via detailed balance, is added to the chemical network to equilibrate the forward reaction. Previously, the only incorporated \ch{SiO2} destruction reaction was the collision of \ch{He+}, which requires very high temperatures.}\\\\
    Additionaly, 15 \ch{Mg}-related reactions are added. Only for 7 of them we added a reversed detailed balance reaction. However, due to a lack of quantum chemical data on \ch{MgO2}, \ch{MgO3}, and \ch{MgO4} no reversed reactions for reactions including such species are added. Reactions with ionised \ch{Mg}-bearing molecules can be found in the literature \citep{Whalley2010,Martinez-Nunez2010,Whalley2011} but are ignored because ionisation is unlikely at the low temperatures of our grid.\\\\
    In total 51 \ch{Al}-related reactions and their reversed detailed balance reactions are added, that mostly originate from combustion chemistry.

    \subsubsection{Quantum mechanical properties}\label{sec:quantum_chem}
    In order to calculate the reversed reaction rate coefficient under the assumption of detailed balance, one needs the Gibbs free energy (GFE) of all reactants and products, as a function of temperature at a specific pressure\footnote{One only needs to determine the GFE at a single pressure to be used in reversed rate coefficients. Often a standard pressure of $\SI{1}{\bar}=\SI{1e5}{\Pa}$ is used.} (Eq.~\eqref{eq:k_destruction} for nucleation and e.g. equations~$(73)-(76)$ in \citet{Grassi2014} in general). In principle, one can also use the difference in Gibbs free energy of formation (GFEoF) because the additional contribution of individual atoms cancels out (App.~\ref{app_sec:GFEoF}). On one hand, using the GFEoF has the advantage of being calculated for numerous species and being included in different databases, e.g. so-called NASA-polynomials\footnote{\url{http://garfield.chem.elte.hu/Burcat/burcat.html}} \citep{Burcat2005} and NIST-JANAF Thermochemical Tables\footnote{\url{https://janaf.nist.gov/}} \citep{Chase1998}. On the other hand, there are inconsistencies between both databases such as the same species having different GFEoF values. By benchmarking, \citet[]{Tsai2017} also came to this conclusion and assign the discrepancies between the databases to a differently defined reference level that corresponds to zero energy. Another reason might be that the GFEoF values rely on experimentally determined values of quantities at room temperature which can have large error bars. Moreover, the details of the calculations or experiments are often unclear as these have been performed decades ago and frequently lack detailed descriptions. For consistency, we use (and strongly encourage to use) GFE rather than GFEoF. Because the GFE is an intrinsic property of a species, it does not rely on any experimental value at a reference temperature (e.g. room temperature) but can be calculated from first principles with absolute zero as a reference point (App.~\ref{app_sec:GFE}). In short, to calculate the GFE as a function of temperature, one only needs the total partition function and the electronic potential energy at zero Kelvin (Eq.\,\ref{app:G_Z_1}).\\\\
    We calculate the GFE of all clusters of the four nucleation species \nucspec by first gathering the most recent structural information (i.e. atomic coordinates) of the lowest energy isomers, i.e. the so-called global minima (Table\,\ref{tab:cluster_info}). Subsequently, using \gaussian \citep{Frisch2013}, we perform density functional theory (DFT) calculations including a vibrational analysis to determine the GFE. For consistency, we always use the same functional and basis set, namely the B3LYP functional \citep{Becke1993} and 6-311+G*\footnote{This basis set is spanned by 6 primitive Gaussians, includes diffusion(+) and polarisation(*).} basis set. Other functionals and/or basis sets might be more accurate for specific properties or species, yet B3LYP is well established and suitable for inorganic oxides \citep{Cora2005}, and 6-311+G* is a good compromise between accuracy and computation time.\\\\
    For all non-cluster species participating in reversed reactions, we have collected the electronic potential energies when available (Table~\ref{tab:quantum_data}). All energies originate from DFT calculations by the Computational Chemistry Comparison and Benchmark DataBase\footnote{\url{https://cccbdb.nist.gov/}} \citep[CCCBDB,][]{Johnson2018}. For consistency we always use results of the same functional and basis set, namely B3LYP and 6-31+G**\footnote{CCCBDB does not contain calculations with 6-311+G*, the one we used for the nucleation clusters. The 6-31+G** basis set is slightly smaller but also includes diffusion and polarisation, and most closely resembles 6-311+G*}. We perform DFT calculations for the species of which no electronic potential energies are present in any database, using the same DFT setup as for the nucleation clusters (Table~\ref{tab:quantum_data}).\\\\
    When possible, we have gathered partition functions\footnote{Note that this excludes the translational part because that depends on the number of particles and the pressure for which one wants to calculate the total partition function.} of the non-cluster species participating in reversed reactions (Table~\ref{tab:quantum_data}). These values originate from detailed calculations and/or experiments. If no literature partition functions could be found, we have calculated them from internal energy levels (rotational, vibration, and electronic\footnote{The number of electronic energy levels is truncated to be valid below $\sim\SI{10000}{K}$, which is more than sufficient for the purpose of this paper.}) found in the CCCBDB (App.~\ref{app_sec:GFE}). Note that this method is less precise due to approximations such as considering the species as a rigid rotor and harmonic oscillator. Again, when no energy levels were available in the literature, we have calculated them via a vibrational analysis as a follow-up on the DFT calculations (Table~\ref{tab:quantum_data}).\\\\

\section{Results}\label{sec:results}
    This section presents the simulation results of the two main model setups, one with closed nucleation networks and one with a comprehensive chemical nucleation network. The closed nucleation network setup considers four nucleation species, \nucspec. Additionally, each of these sub-setups will use the monomer nucleation (MN) and the polymer nucleation (PN) approach. The comprehensive chemical nucleation network setup will encompass all four mentioned nucleation species but only use the polymer nucleation approach.\\\\
    Because our results include four parameters (temperature, gas density, cluster number density, and time), we reduce the dimensionallity to analyse the outcome. The analysis of the cluster size distributions in $(T, \rho)$-space is limited to the end of the simulation, i.e. after one year. Subsequently, to infer temporal effects, we choose a benchmark constant total mass density of \SI{1e-9}{\kg\per\m\cubed}, which is a typical value we expect in an AGB wind \citep[fig.~10]{Boulangier2019}. Note that we use the total mass density of the gas as a parameter since this value remains constant as compared to the total number density.
    
    \subsection{Closed nucleation networks}\label{sec:results_closed_ntw}
    This section covers the evolution of four nucleation species \nucspec for a closed nucleation network setup with both the monomer nucleation (MN) and the polymer nucleation (PN) description. To ensure the overview, we mainly discuss the largest clusters because they are most interesting to understand formation of macroscopic dust grains. Additional figures for all clusters can be found in Appendix~\ref{app:res_closed_ntw}. 
        
        \begin{figure*}
            \centering
            \includegraphics[width=0.9\columnwidth]{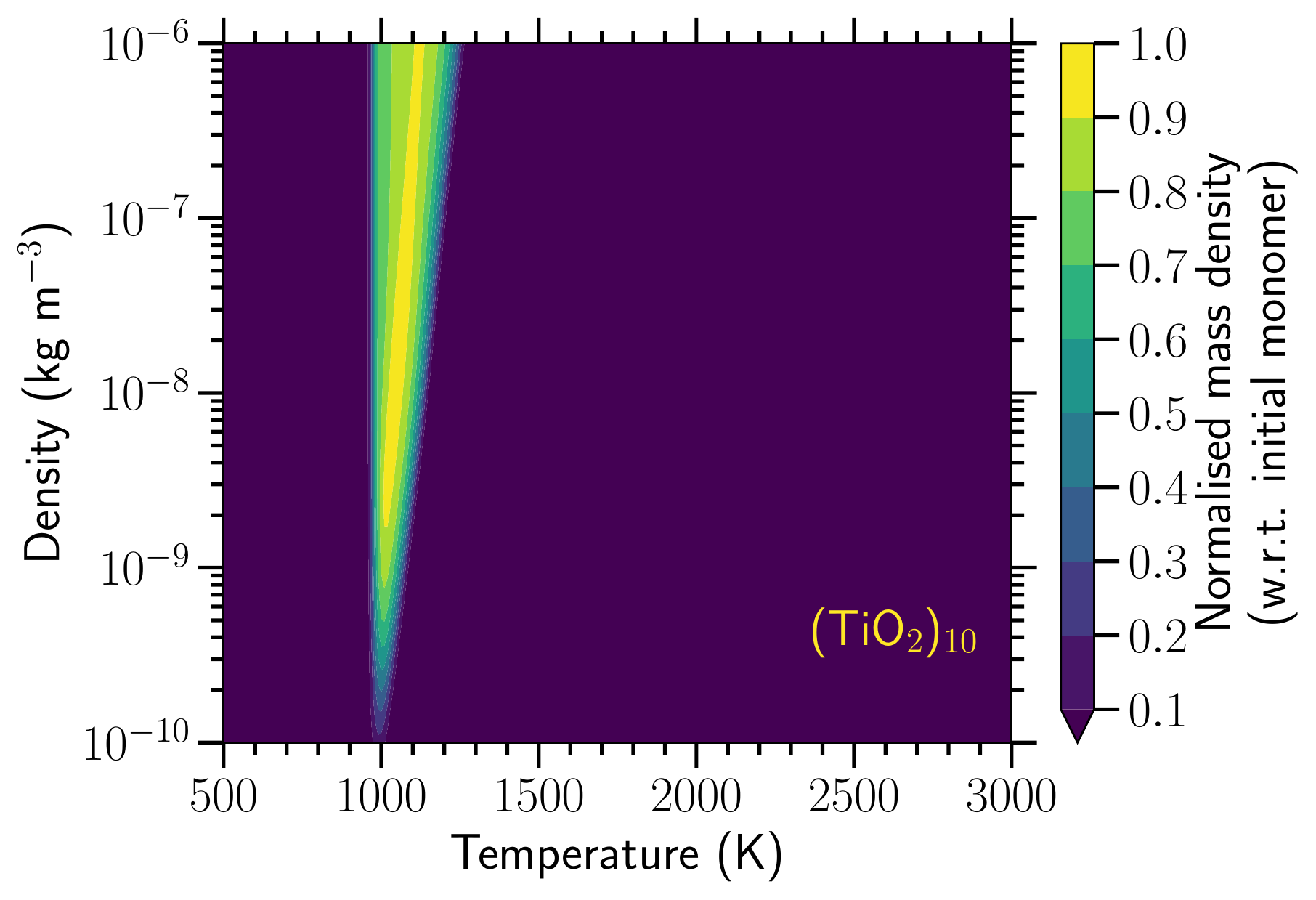}
            \includegraphics[width=0.9\columnwidth]{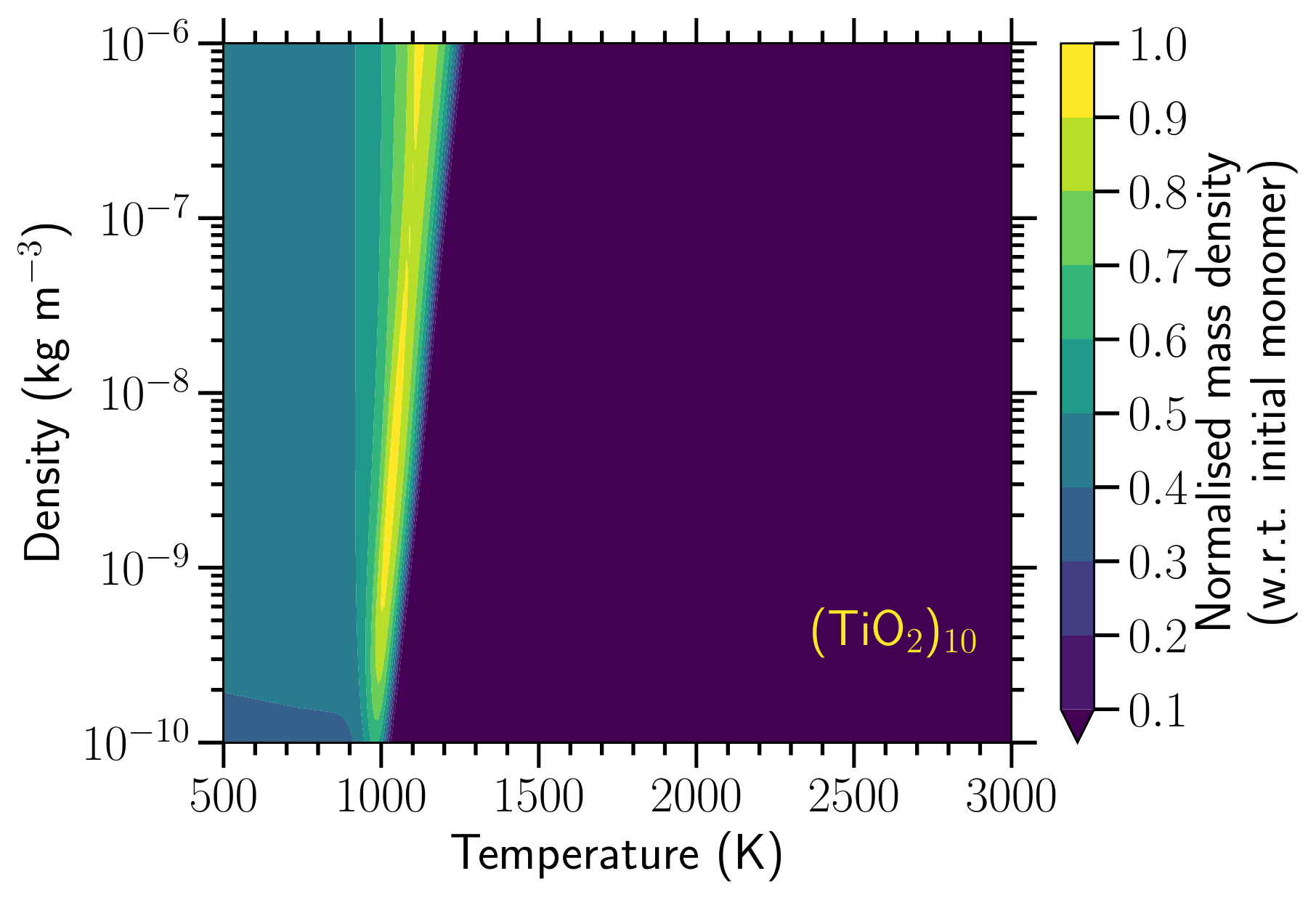}
            \includegraphics[width=0.9\columnwidth]{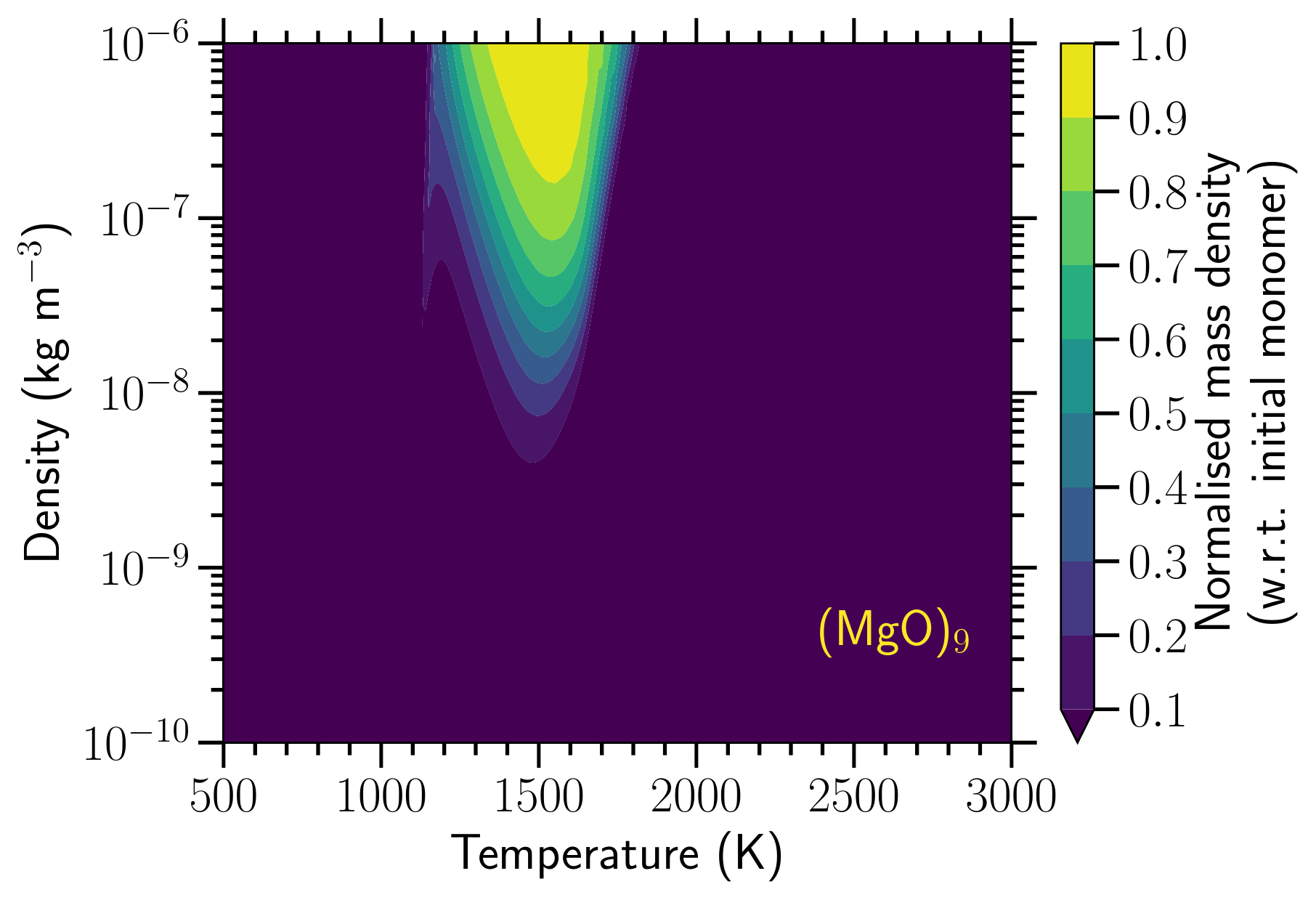}
            \includegraphics[width=0.9\columnwidth]{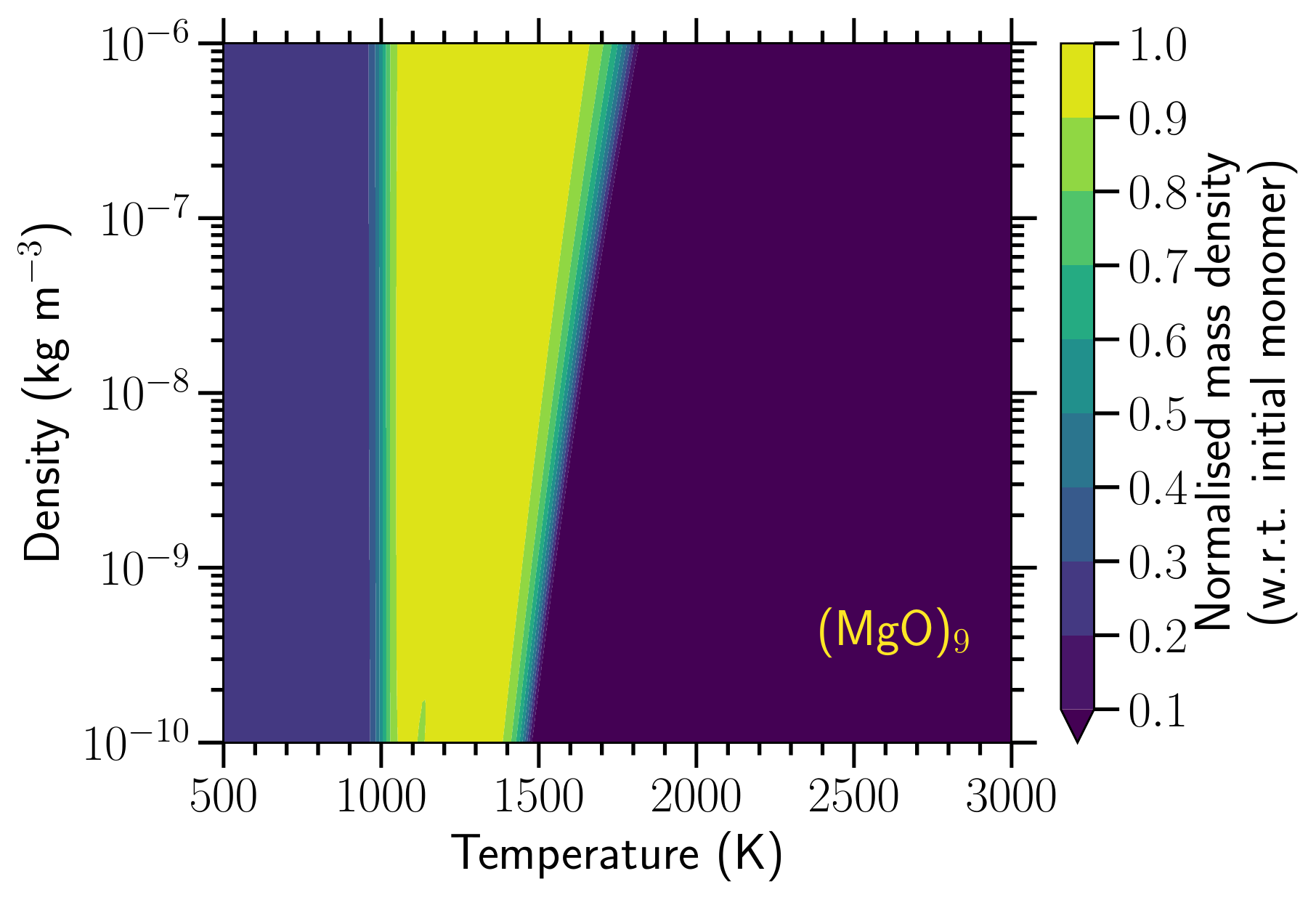}
            \includegraphics[width=0.9\columnwidth]{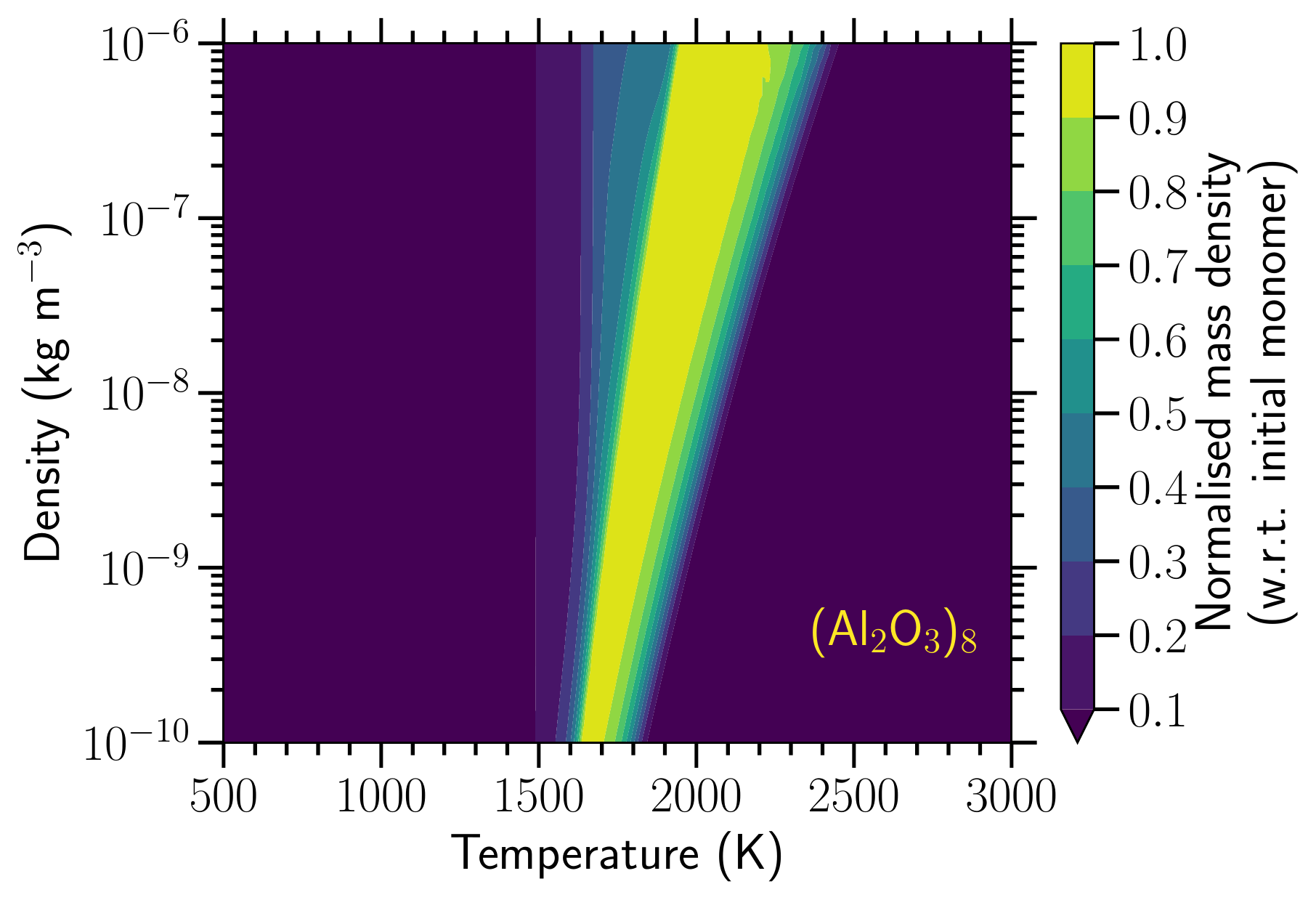}
            \includegraphics[width=0.9\columnwidth]{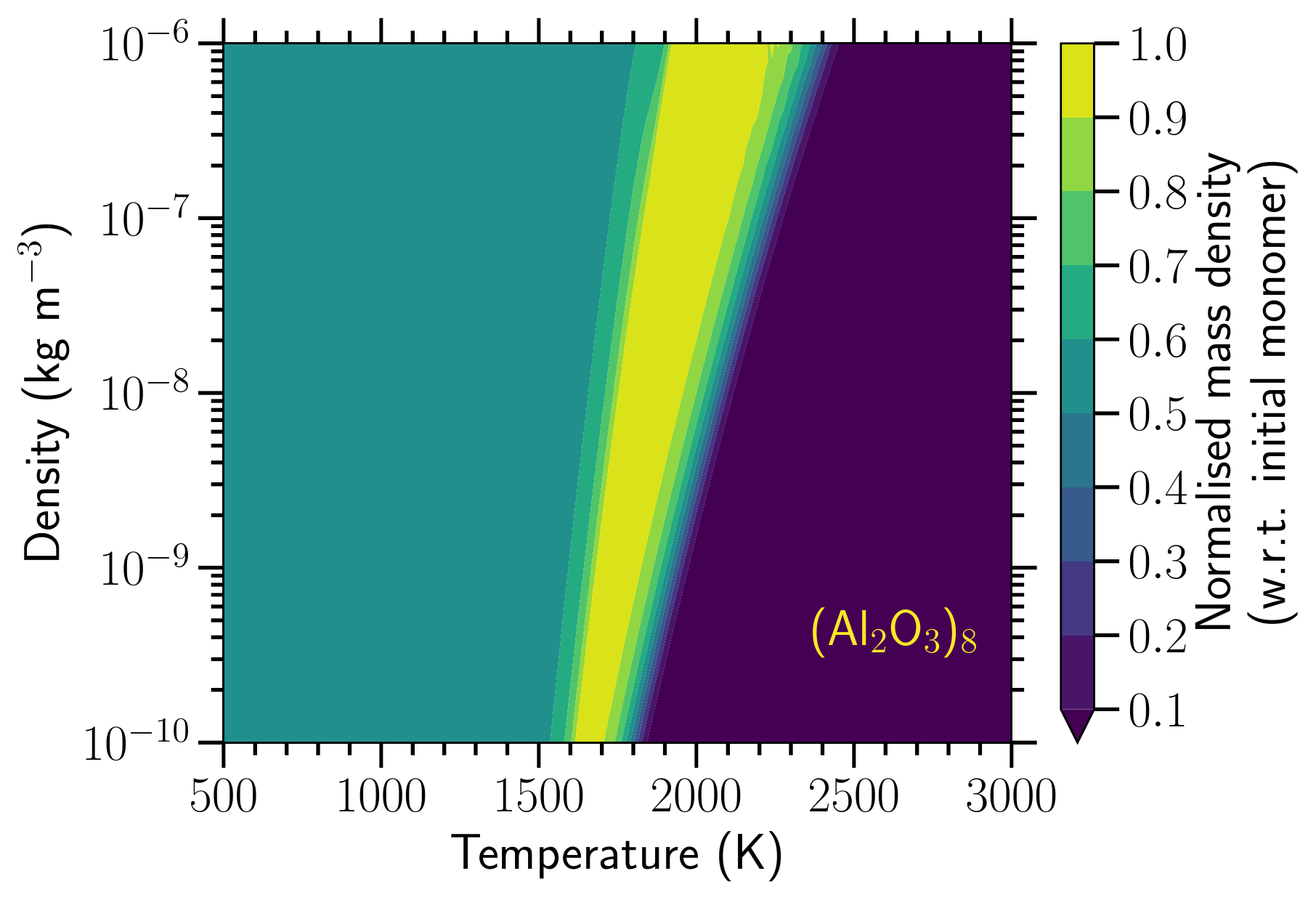}
            \caption{Normalised mass density (or mass fraction) w.r.t. the initially available monomers after one year of \protect\Ti{10}, \protect\Mg{9}, and \protect\Al{8} for the closed nucleation models with left monomer and right polymer nucleation description. We refrain from showing \protect\SiO{10} since its abundance is zero in the entire parameter space. Note that \protect\Mg{9} is the second largest cluster, but most stable and more abundant one.  Monomer nucleation under predicts the amount of large clusters at low temperature, as compared to polymer nucleation. This under prediction is due to the limitation of growth-by-monomers in the monomer nucleation description. In the most favourable nucleation conditions, more than 90 per cent of the initial monomers end up in the largest cluster. \protect\Al{1}-clusters are the primary candidate for first dust precursors because \protect\Al{8} forms at the highest temperature as compared to the other candidates. Normalised number densities w.r.t. the initially available monomers can easily be found by dividing the normalised mass density by the cluster size, i.e. divide by 8 in the case of \protect\Al{8}. An overview of all clusters of all candidates can be found in Appendix \ref{app:res_closed_ntw} with an in-depth analysis in Sections \ref{sec:results_closed_ntw} and \ref{sec:impact_fullntw}.}
            \label{fig:CN_all_clusters_normalised}
        \end{figure*}

        \begin{figure*}
            \centering
            \includegraphics[width=0.9\columnwidth]{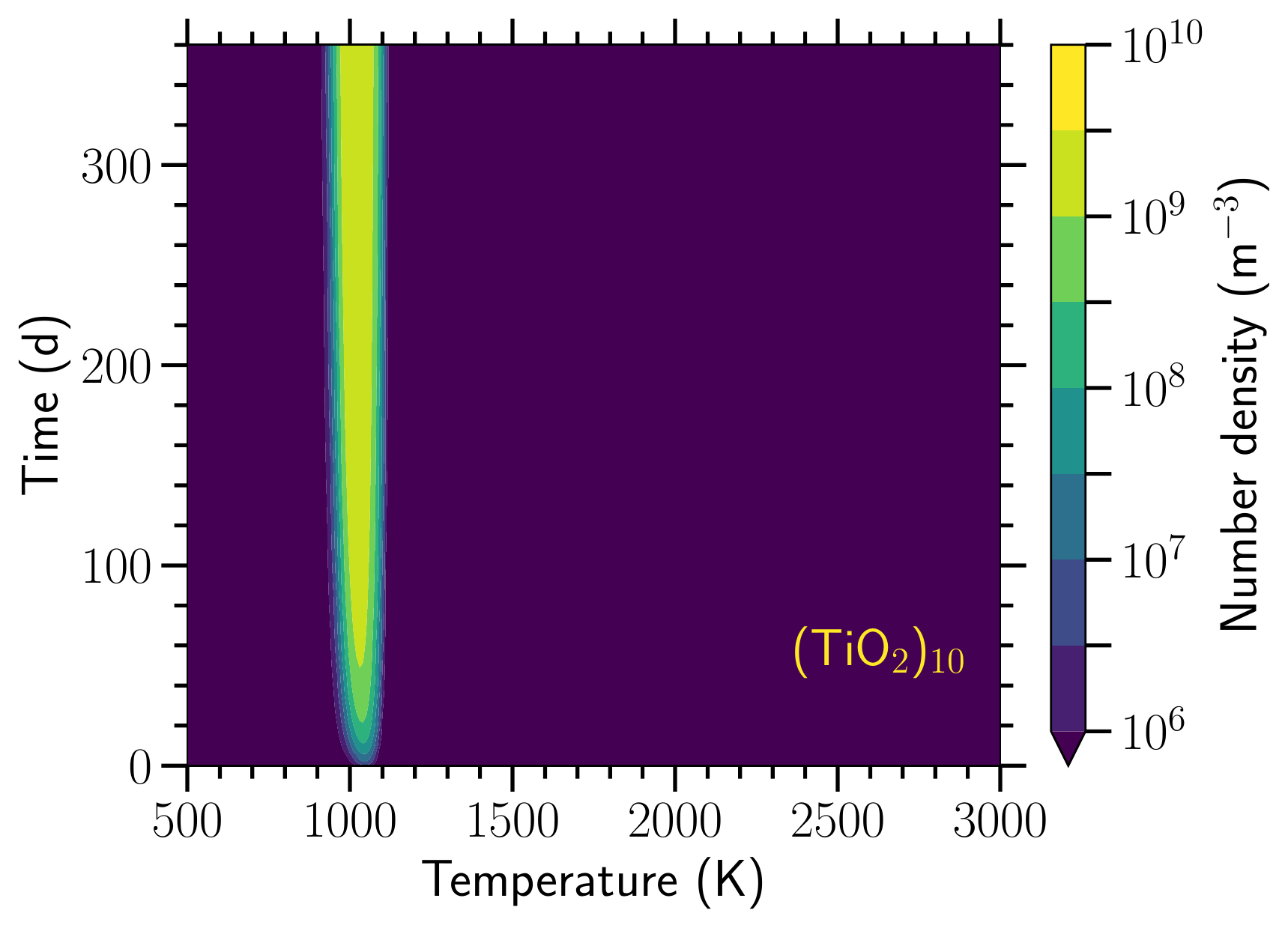}
            \includegraphics[width=0.9\columnwidth]{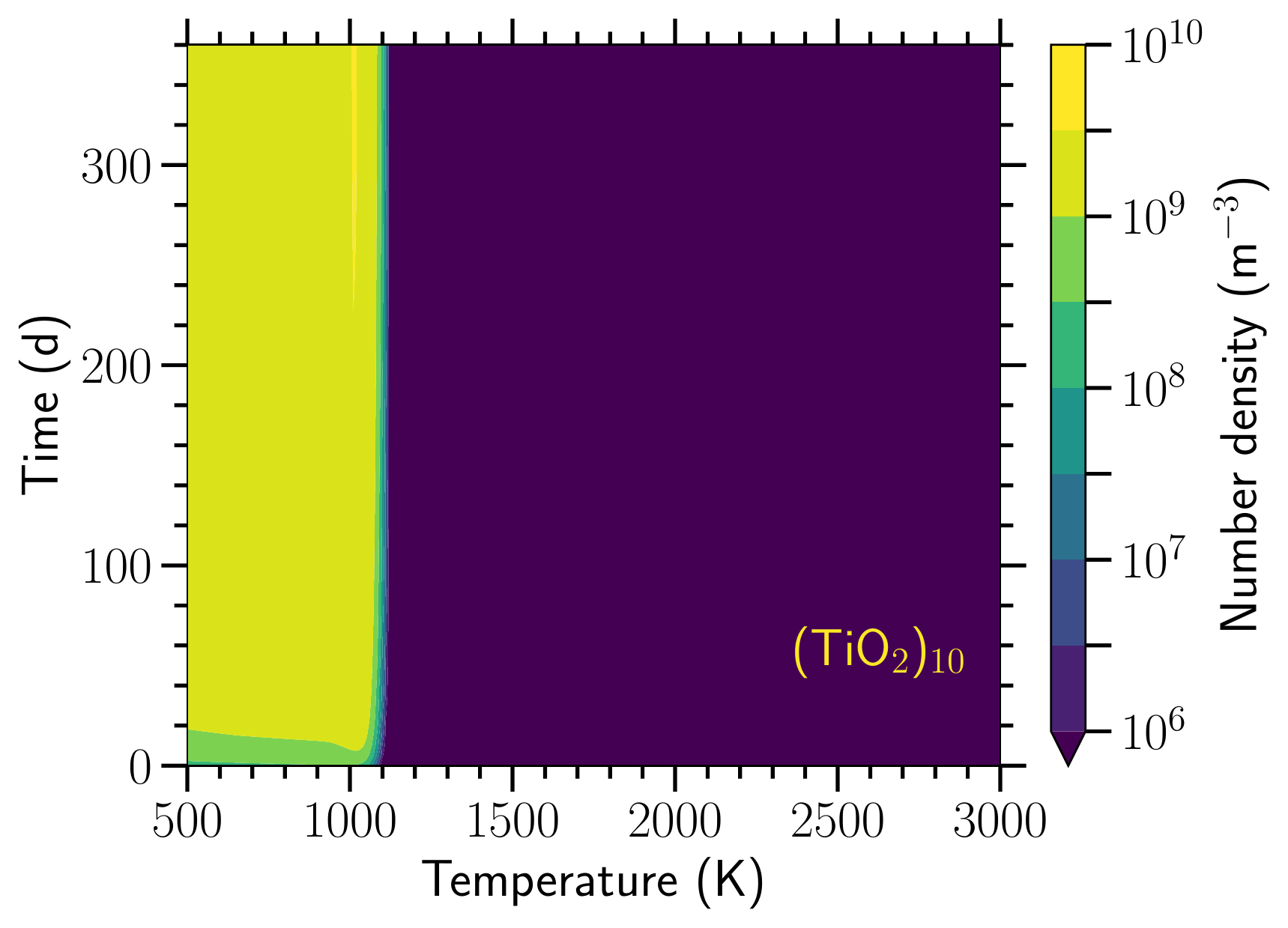}
            \includegraphics[width=0.9\columnwidth]{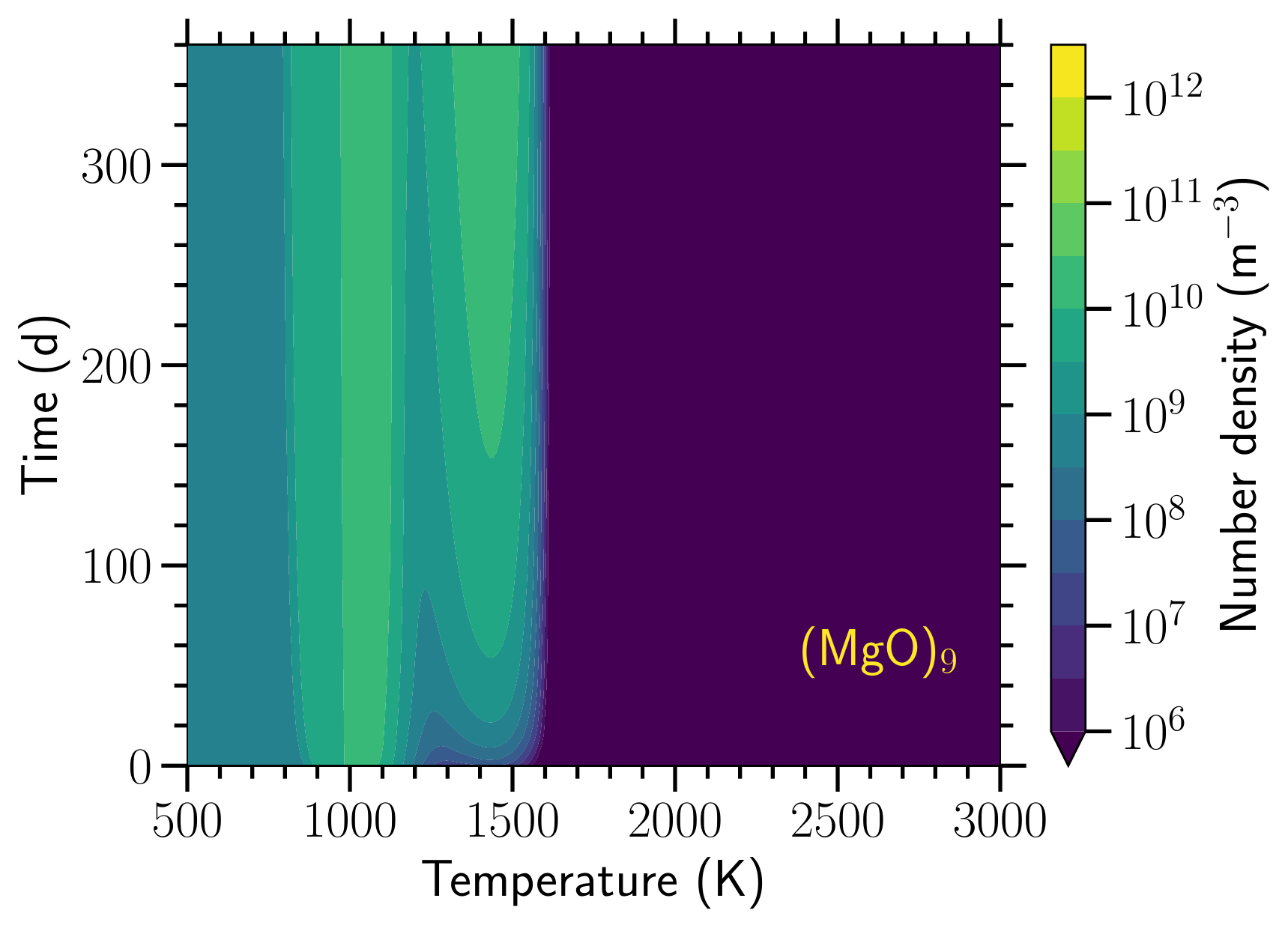}
            \includegraphics[width=0.9\columnwidth]{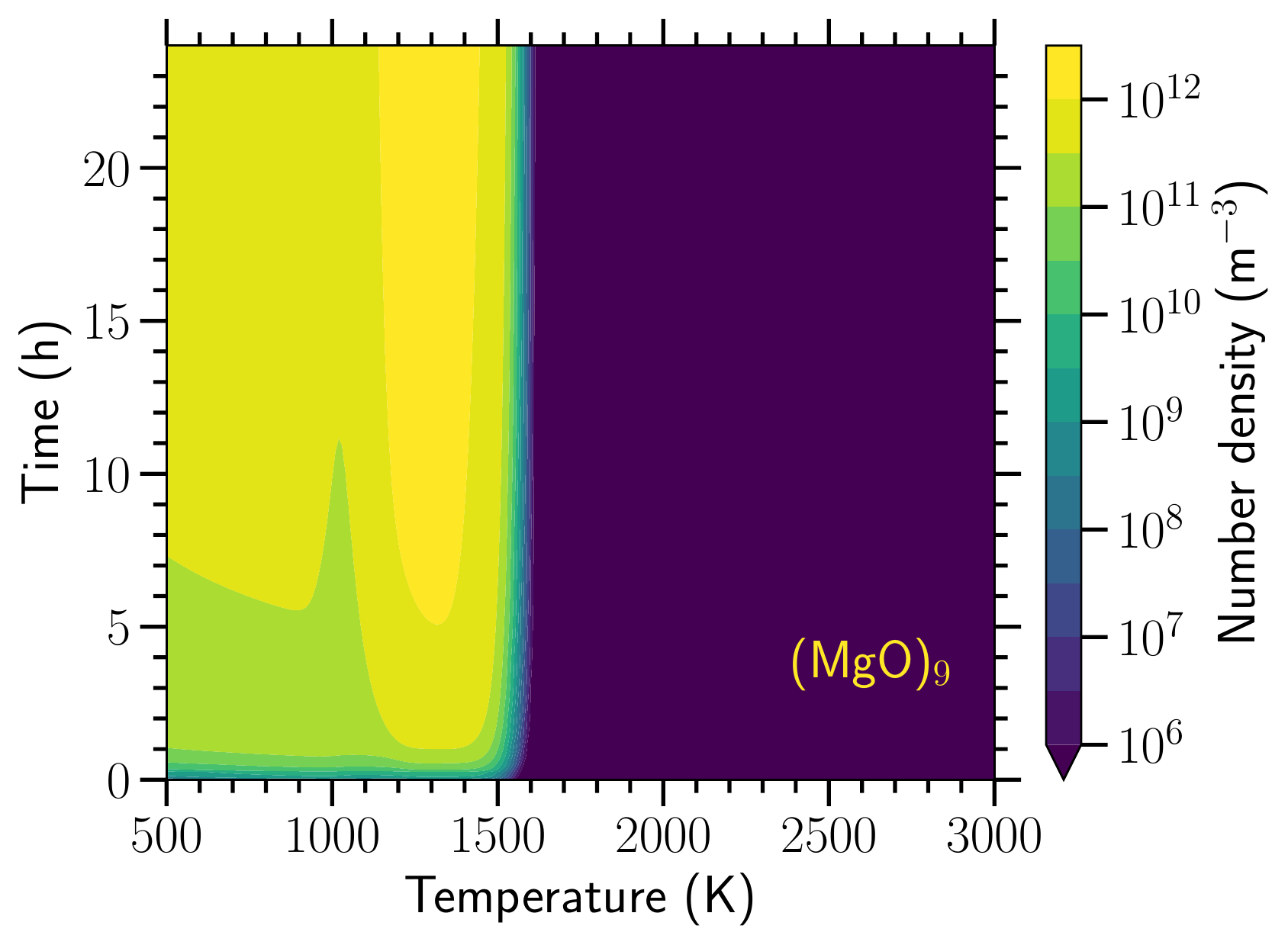}
            \includegraphics[width=0.9\columnwidth]{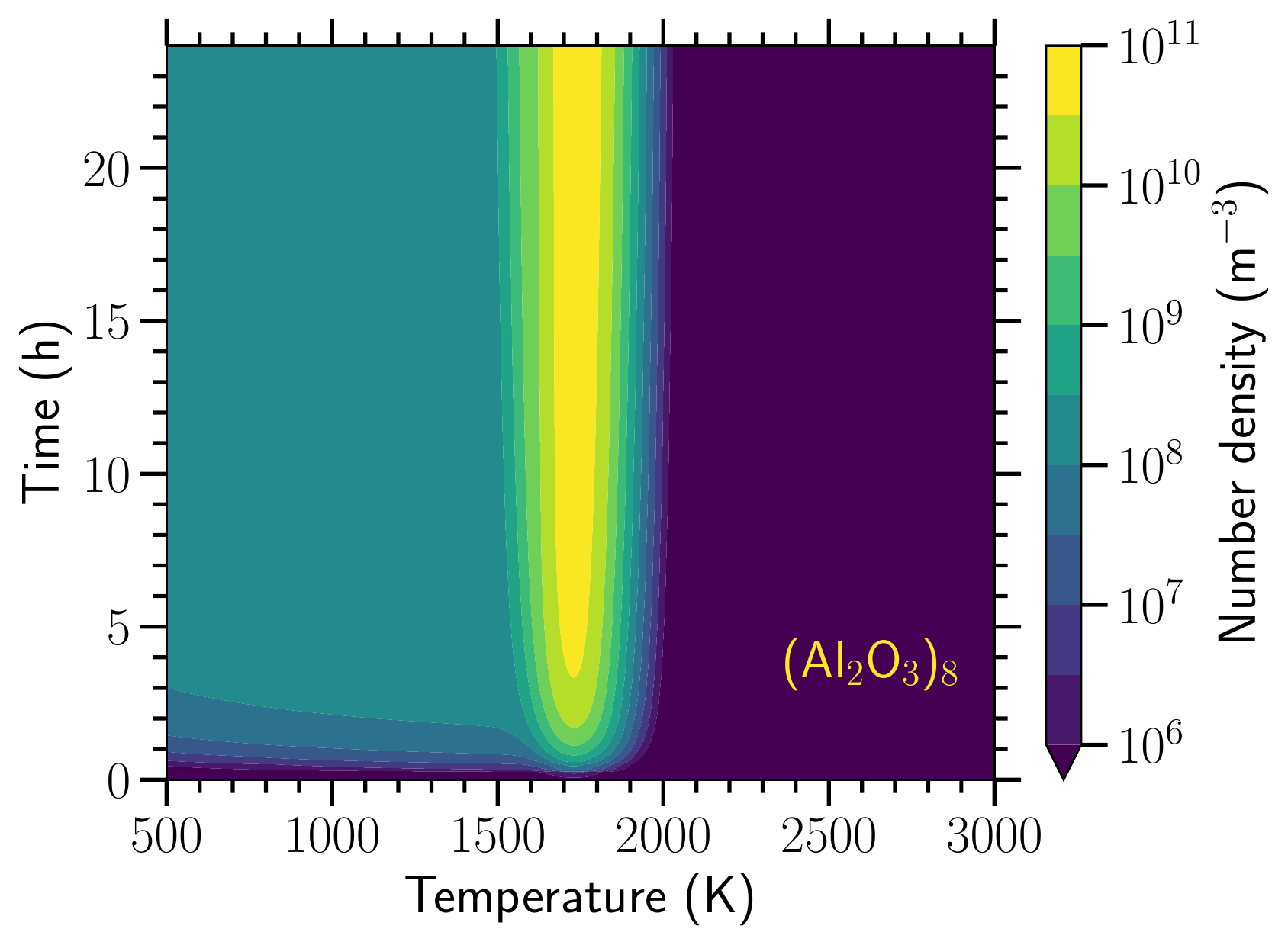}
            \includegraphics[width=0.9\columnwidth]{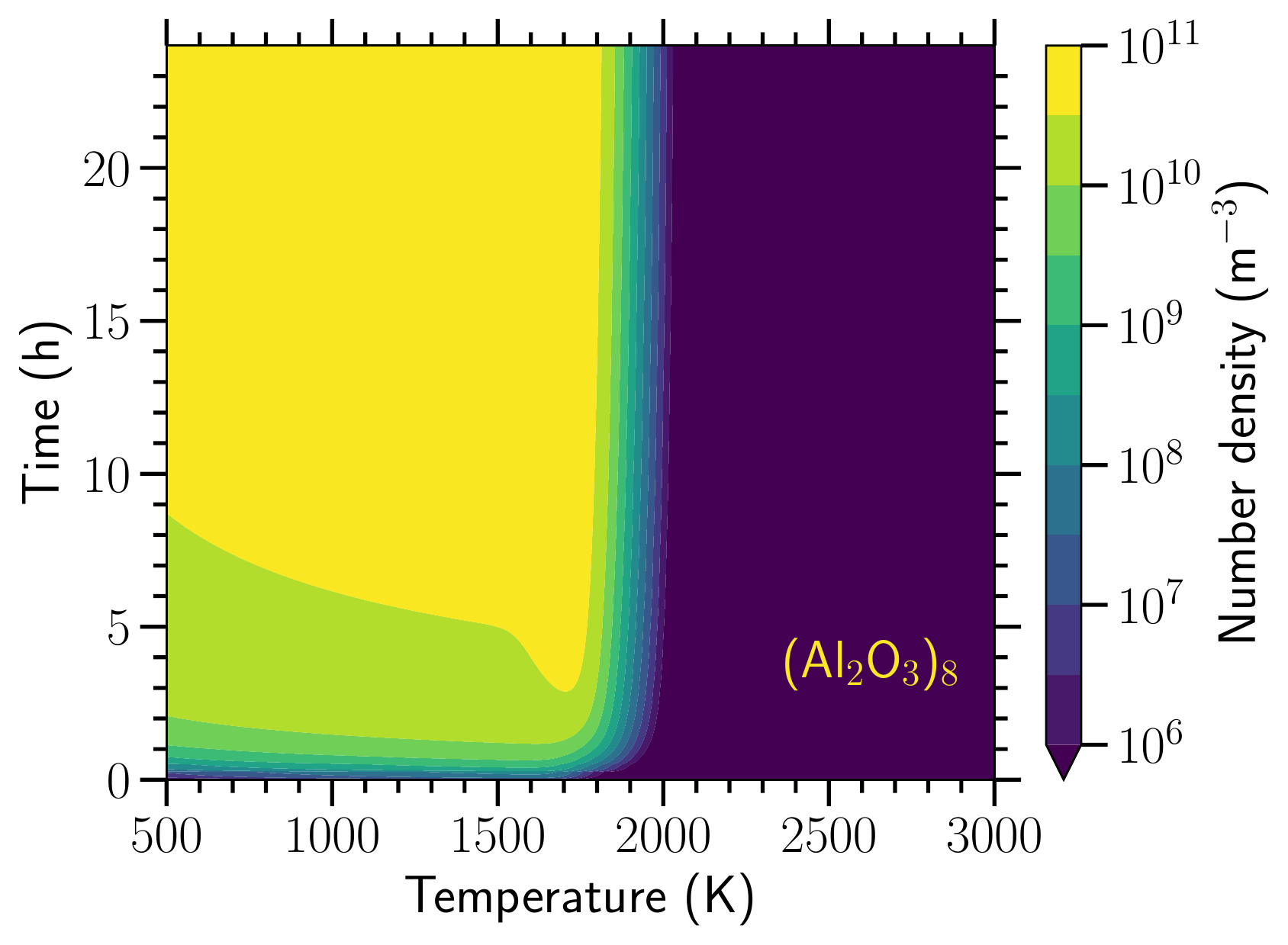}
            \caption{Temporal evolution of the absolute number density of \protect\Ti{10}, \protect\Mg{9}, and \protect\Al{8} at the benchmark total gas density $\rho=\SI{1e-9}{\kg\per\m\cubed}$ with left monomer and right polymer nucleation description. Be aware of the different time scales between species. Overall, convergence with monomer nucleation description takes slightly longer than using the polymer nucleation one. It can also yield vastly different final abundances which is most noticeable for \protect\Mg{9}. We refrain from showing \protect\SiO{10} since its abundance is zero in the entire parameter space. An overview of all clusters of all candidates can be found in Appendix \ref{app:res_closed_ntw} with an in-depth analysis in Secs. \ref{sec:results_closed_ntw} and \ref{sec:impact_fullntw}.}
            \label{fig:CN_all_clusters_time_evolution}
        \end{figure*}
    
        \subsubsection{\ch{TiO2}}\label{sec:resTiO2}
        \Ti{10} forms when the temperature drops below the sharp threshold at \SIrange{1000}{1200}{\K}, where the low (high) temperature threshold is for the lowest (highest) densities \fig{\ref{fig:CN_all_clusters_normalised}}. At temperatures above \SI{1200}{\K}, its abundance drops orders of magnitude \figs{\ref{fig:TiO2_clusters_monomer_norm_same_scale}}{\ref{fig:TiO2_clusters_general_norm_same_scale}}. As expected, a higher density leads to more collisions facilitating nucleation at higher temperatures. Between roughly \SI{950}{K} and the upper temperature threshold for both MN and PN, almost all of the available monomers end up in \Ti{10} (> 80 per cent). However, using MN or PN yields vastly different results at low temperatures. In this regime, roughly below \SI{950}{\K}, the abundance of \Ti{10} drops orders of magnitude in the case of MN in contrast to PN, where its abundance is nearly identical and accounts for \SIrange{40}{50}{} per cent of the available titanium. The low abundance in the MN case is caused by a relatively rapidly developing lack of monomers in this temperature range, because, by design, growth is only allowed by the addition of monomers.  Once the bulk of the material is clustered in $N=$\,\SIrange{2}{4}{} chains, the monomer population becomes depleted and further growth is quenched \fig{\ref{fig:TiO2_clusters_monomer_norm_same_scale}}. At our benchmark density of \SI{1e-9}{\kg\per\m\cubed}, this typically happens in less than a day. This bottleneck does not occur in the case of PN since, by design, all clusters are allowed to participate in the growth process \fig{\ref{fig:TiO2_clusters_general_norm_same_scale}}. Therefore, even in the case of a lack of monomers other small clusters can interact and form larger clusters. In this low temperature regime, this occurs so efficiently that the small clusters ($N=$\,\SIrange{2}{4}{}) are completely depleted and turned into large clusters. The fact that clusters of intermediate ($N>6$) size are still present is somewhat artificial since they are only allowed to grow by addition of smaller ones due to the limitation of a maximum size of $N=10$. As these small clusters are already depleted, the intermediate growth is quenched. In reality clusters of size $N=6$ and $N=7$ can interact to form an $N=13$ sized cluster.\\\\
        Using MN, the abundance of the largest molecules converges\footnote{This happens over the entire temperature range unless stated otherwise.} slowest, roughly after \SIlist{20;60}{\day} for \Ti{9} and \Ti{10}, respectively. All smaller molecules roughly converge after \SI{20}{\day} or less \fig{\ref{fig:TiO2_clusters_monomer_time_evolution}}. Using PN, the convergence of \Ti{10} occurs faster, already after \SI{20}{\day}, even in less than \SI{1}{\day} for the slightly smaller clusters. All small clusters are also formed within \SI{1}{\day} but continue to steadily grow into larger ones \fig{\ref{fig:TiO2_clusters_general_time_evolution}}.
        
        \subsubsection{\ch{MgO}}\label{sec:resMgO}
        Unlike for \Ti{1} clusters, the conditions that determine the presence of the largest \Mg{1} cluster differ strongly between the different nucleation descriptions, being more complex in the MN case. Yet both nucleation descriptions reveal that the second largest cluster \Mg{9}, rather than the largest cluster \Mg{10}, is the most stable and therefore most abundant cluster \figs{\ref{fig:MgO_clusters_monomer_norm_same_scale}}{\ref{fig:MgO_clusters_general_norm_same_scale}}. Hence, we mainly discuss \Mg{9}. In the MN case, between \SIlist{1100;1500}{\K} and at the highest densities all available monomers end up in \Mg{9}  \fig{\ref{fig:CN_all_clusters_normalised}}. But, within this temperature range, this amount strongly decreases with decreasing density where at \SI{1e-8}{\kg\per\m\cubed} just 10 per cent ends up in \Mg{9} and at the lowest densities this amount reduces to \SI{0.01}{} per cent \fig{\ref{fig:MgO_clusters_monomer_norm_same_scale}}. Note that below \SI{1100}{\K} \Mg{9} clusters can also exist but maximally take up 1 per cent of the available monomers. In the PN case, \Mg{9}-clusters already form below \SIrange{1500}{1700}{\K} and above \SI{1000}{\K} they contain over 90 per cent of the available monomers \fig{\ref{fig:CN_all_clusters_normalised}}. Below \SI{1000}{\K}, they are less abundant but still encompass between \SIrange{20}{30}{} per cent of the monomers. Note that below \SI{1000}{\K}, there is more \Mg{10} than \Mg{9}, making the largest cluster the most stable one at low temperatures \fig{\ref{fig:MgO_clusters_general_norm_same_scale}}. Similar to the other nucleation candidates, the lack of large \Mg{1}-clusters at low temperatures, below \SI{1000}{\K}, in the MN case is due to the construction of this nucleation description, that limits growth by addition of monomers. It is also interesting to note that in both nucleation cases and above \SI{1000}{\K}, cluster sizes $N=$~\SIlist{2;4;6;9}{} are more abundant than their direct size-neighbours. This is a consequence of the energetic stability of these \ch{MgO} cluster sizes. This phenomenon would not arise when using extrapolated bulk properties for the clusters (i.e. classical nucleation), but only when calculating the energy on a microscopic level (i.e. quantum mechanically). \\\\
        Determining the time scale of abundance convergence for \Mg{1}-clusters is problematic, due to the complex behaviour in temperature-space. We give a rough convergence time scale below and above \SI{1000}{\K}. Below \SI{1000}{\K} and in the case of MN, all clusters converge in just a few hours \fig{\ref{fig:MgO_clusters_monomer_time_evolution_short}}. In the case of PN, the largest clusters do converge in a few hours but smaller clusters form in less than a few hours and then gradually get destroyed again over the course of a few days before reaching convergence \fig{\ref{fig:MgO_clusters_general_time_evolution_short}}. \Mg{5} stands out as its abundance still gradually changes on time scales of \SIrange{10}{100}{\day} \fig{\ref{fig:MgO_clusters_general_time_evolution}}. Because the evolution above \SI{1000}{\K} is less straightforward, we limit the analysis to the largest most stable cluster \Mg{9}, and refer the reader to figures \ref{fig:MgO_clusters_monomer_time_evolution}, \ref{fig:MgO_clusters_monomer_time_evolution_short} and \ref{fig:MgO_clusters_general_time_evolution}, \ref{fig:MgO_clusters_general_time_evolution_short} for more details on all clusters. In the case of MN, the abundance of \Mg{9} converges after roughly \SI{180}{\day}, whereas using PN this happens in only a few hours \fig{\ref{fig:CN_all_clusters_time_evolution}}.
        
        \subsubsection{\ch{SiO}}\label{sec:resSiO}
        In both nucleation cases, the largest \SiO{1}-clusters do not form significantly in our $(T,\rho)$-range \figs{\ref{fig:SiO_clusters_monomer_norm_same_scale}}{\ref{fig:SiO_clusters_general_norm_same_scale}}. Between \SIrange{500}{700}{\K} most monomers end up in \SiO{3} and remain in the monomer above this temperature. Note that sizes $N=$~\SIrange{5}{9}{} do not form at all. Since no large clusters form in our $(T,\rho)$-range, we refrain from analysing any time dependence.
        
        \subsubsection{\ch{Al2O3}}\label{sec:resAl2O3}
        For both nucleation descriptions, the largest \Alclusters already form at temperatures as high as \SIrange{1800}{2400}{\K}, depending on the total gas density \fig{\ref{fig:CN_all_clusters_normalised}}, i.e. in hotter regimes than any of the other nucleation candidates. Moreover, between \SIrange{1600}{1700}{\K} and \SIrange{1900}{2200}{\K} more than 90 per cent of the available monomers are locked-up in the largest cluster \Al{8}. Between the lower limits and \SI{1500}{\K}{}, \Al{8} encompasses between 10 and 90 per cent of the available material for the MN description. Below \SI{1500}{\K}, MN again impedes a subsequent growth because the monomers are depleted once small clusters have formed, resulting in a pile-up of small clusters unable to continue to grow \fig{\ref{fig:Al2O3_clusters_monomer_norm_same_scale}}. PN does not have this limitation and \Al{8} contains more than 50 per cent of the available monomers in the entire temperature range below the formation threshold. Additionally PN growth is so efficient that the bulk of the material grows to sizes above $N=5$, removing all smaller clusters \fig{\ref{fig:Al2O3_clusters_general_norm_same_scale}}.\\\\
        In both nucleation cases, the formation of \Al{8} happens so fast that it is invisible on a time scale of days \figs{\ref{fig:Al2O3_clusters_monomer_time_evolution}}{\ref{fig:Al2O3_clusters_general_time_evolution}}. Refining the time sampling reveals that, in both nucleation cases, convergence of the abundance of \Al{8} already occurs after roughly \SIrange{5}{10}{\hour} \fig{\ref{fig:CN_all_clusters_time_evolution}}. For MN, convergence happens even faster for smaller clusters \fig{\ref{fig:Al2O3_clusters_monomer_time_evolution_short}}. For PN, however, there is a gradual creation and destruction of the smaller clusters, on a time scale of hours \fig{\ref{fig:Al2O3_clusters_general_time_evolution_short}}. Even on a time scale of \SI{100}{\day}, the smallest clusters do not converge but gradually get converted to larger ones \fig{\ref{fig:Al2O3_clusters_general_time_evolution}}.
        
        \subsubsection{Comparison with equilibrium compositions}\label{sec:equi_ratios}
        \rev{The equilibrium abundance ratio of two clusters with different sizes w.r.t to the equilibrium abundance ratio of two other cluster sizes can be calculated via Eq.~\eqref{eq:equi_ratios}. Such ratio of ratios can be used to more quantitatively discuss if clusters distributions have reached the equilibrium composition. Since it is most meaningful to compare ratios if nucleation is feasible, the ratios of two smaller clusters w.r.t. the ratio of the two largest clusters are discussed in the favourable temperature range. The results, shown for comparison with the equilibrium abundances, correspond to the closed PN models for the benchmark total gas density $\rho = \SI{1e-9}{\kg\per\m\cubed}$ at the final time step (one year). Note that if the number density of any of the four clusters species is below the numerical solver accuracy of \SI{1e-20}{\per\cm\cubed}, the ratios are not shown.\\\\
        The relative abundances ratios of \ch{TiO2}- and \ch{MgO}-clusters do not reach the equilibrium ratios in the entire temperature range (Figs.~\ref{fig:equ_ratios_TiO2} and \ref{fig:equ_ratios_MgO}). At the highest temperatures, at which the nucleation is feasible, the model results correspond to the equilibrium ratios. However, at lower temperatures, the clusters need more time to reach the equilibrium ratios since the interaction probability is lower. This transition is visible between \SIrange{900}{1000}{\K} and \SIrange{1000}{1300}{\K} for the \ch{TiO2}- and \ch{MgO}-clusters, respectively. The fact that the clusters have not yet reached equilibrium ratios is also visible from the temporally changing abundances in Figs.~(\ref{fig:TiO2_clusters_general_time_evolution}) and (\ref{fig:MgO_clusters_general_time_evolution}). The relative abundance ratios of \ch{Al2O3}-clusters deviate more from the equilibrium ratios (Fig.~\ref{fig:equ_ratios_Al2O3}). Due to the large variation in number densities of the clusters in different temperature regimes (order of magnitude), it is often impossible to compare ratios of the \ch{Al2O3}-clusters. This variation is more clearly visible in Fig.~(\ref{fig:Al2O3_clusters_general_time_evolution}). \SiO{1}-clusters are not discussed since they do not significantly form in the temperature range of interest.
        }

    \subsection{Comprehensive chemical nucleation network}\label{sec:resAllchem}
    This section covers the evolution of the four nucleation species \nucspec for a comprehensive chemical nucleation network with the polymer nucleation (PN) description. To ensure the overview, we mainly discuss the species that also contain the cluster metals (\ch{Ti}, \ch{Mg}, \ch{Si}, and \ch{Al}) because they are most interesting to understand formation of macroscopic dust grains. In analogy with Section~\ref{sec:results_closed_ntw}, only the temporal evolution of the nucleation clusters is presented. Additional figures for all species of interest can be found in Appendix~\ref{app:res_full_ntw}. 
    
        \subsubsection{\ch{TiO2}}\label{sec:resAllTiO2}
        The formation of \Ti{10} occurs at the same temperature and density conditions as in the closed nucleation model with the PN approach, i.e. when the temperature drops below the sharp threshold at \SIrange{1000}{1200}{\K} \fig{\ref{fig:full_ntw_TiO2_max_normalised}}. Above this threshold, \ch{Ti} resides in either \Ti{1}, \ch{TiO}, or remains atomic, with the atomic state preferred at the highest temperatures (above \SI{2000}{\K} or higher for higher densities). \fig{\ref{fig:full_ntw_Ti-molecules_norm_same_scale}} The convergence of \Ti{10} happens within roughly \SI{40}{\day}, similar to the closed nucleation model with PN \fig{\ref{fig:full_ntw_TiO2_max_normalised}}. The convergence of other \Ticlusters is also similar to the closed PN case \fig{\ref{fig:full_ntw_Ti-molecules_time_evolution}}.
        
        \subsubsection{\ch{MgO}}\label{sec:resAllMgO}
        All available \ch{Mg} remains atomic. Neither \Mg{1}, nor the \Mg{1}-clusters, nor any \ch{Mg}-bearing molecules are formed. Hence we refrain from showing the abundance figures.
        
        \subsubsection{\ch{SiO}}\label{sec:resAllSiO}
        The abundance evolution of all \SiO{1}-clusters,  in temperature, density, and time, is the same as for the closed nucleation PN model, i.e. the large clusters do not form in the considered temperature-density range and the smallest clusters only form at the lowest temperatures \fig{\ref{fig:full_ntw_Si-molecules_norm_same_scale}}. Above roughly \SI{700}{\K}, all \ch{Si} is locked-up in the \ch{SiO2} molecule (except at the highest temperatures and lowest densities, which is due to time constraints of the simulation). This finding is somewhat in contrast to the higher binding energies of \ch{SiO} compared to \ch{SiO2} (Section\,\ref{sec:limChem}). Below \SI{700}{\K}, the most abundant molecules are \ch{SiO} and \SiO{3}. Note that in the entire $(T,\rho)$-grid, \ch{Si} does not remain atomic.
        
        \subsubsection{\ch{Al2O3}}\label{sec:resAllAl2O3}
        Most of the \ch{Al} remains atomic except for some specific $(T, \rho)$-combinations. Overall creation of \ch{Al}-molecules is up to maximally 1 per cent of the total available \ch{Al}, except at the lowest temperatures for both extremes in the considered density range where it can be up to roughly 50 per cent \fig{\ref{fig:full_ntw_Al-molecules_norm_same_scale}}. The most abundant molecules are \ch{AlO}, \ch{AlH}, \ch{Al(OH)2}, and \ch{Al(OH)3}. Their formation regimes can be recovered in the abundance figure of \ch{Al}, and  only \ch{AlO} forms in the entire temperature range. Note that the figures of less abundant \ch{Al}-bearing molecules are only shown in Appendix\,\ref{app:res_full_ntw} since their abundance never exceeds the chosen threshold \fig{\ref{fig:full_ntw_Al-molecules}}.
        
        \begin{figure}
            \centering
            \includegraphics[width=\columnwidth]{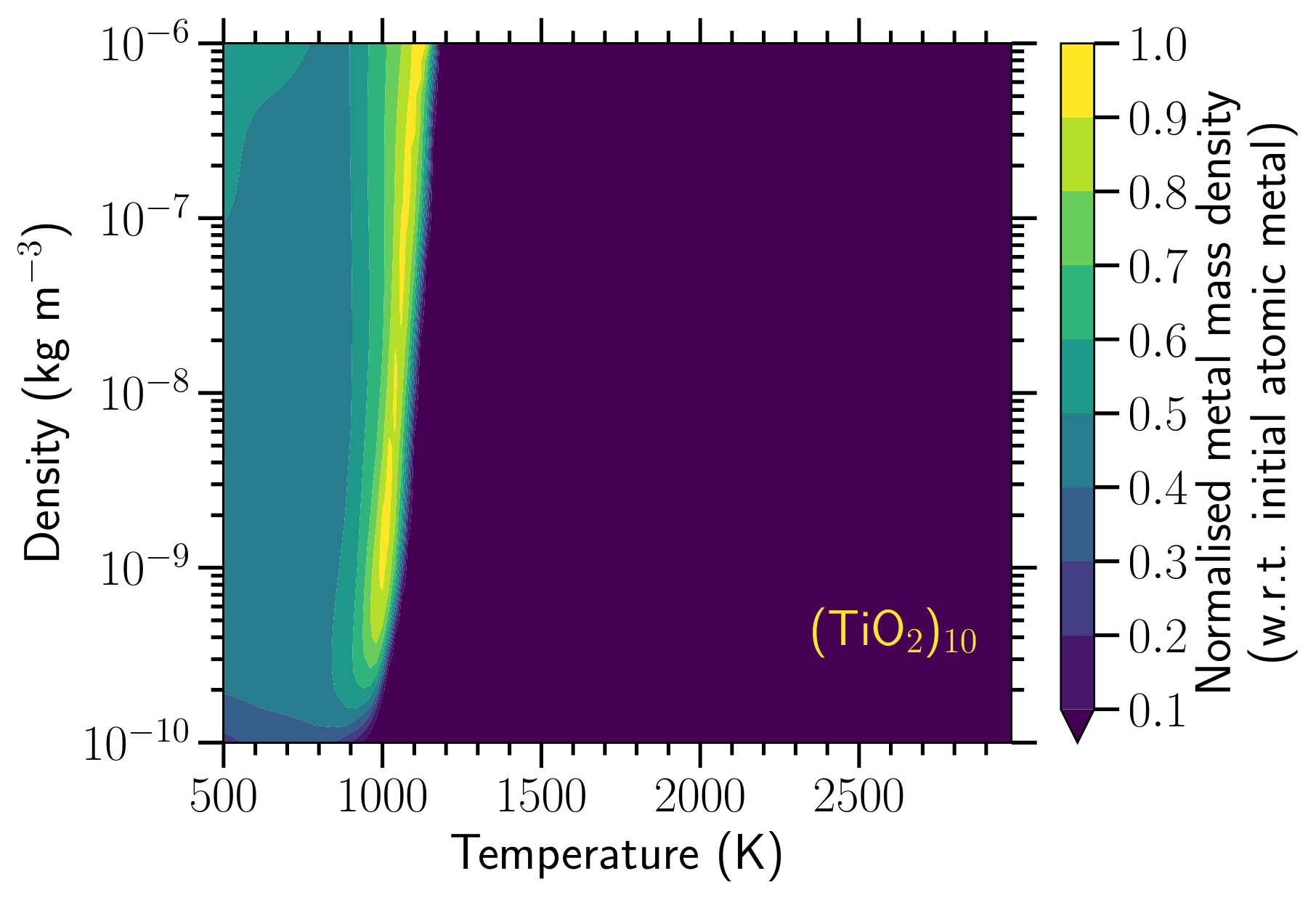}
            \includegraphics[width=\columnwidth]{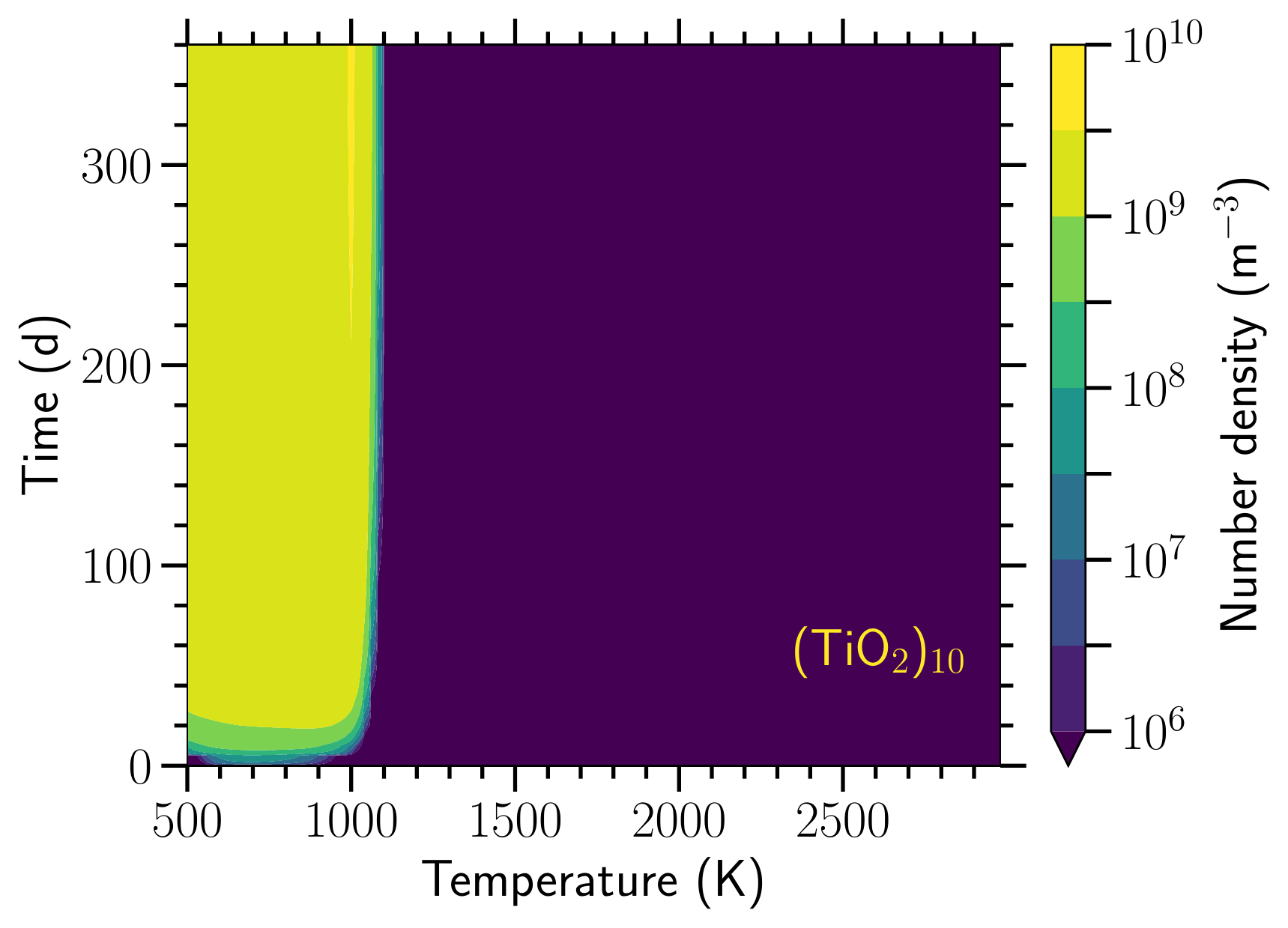}
            \caption{Normalised mass density after one year (top) and temporal evolution of the absolute number density at the benchmark total gas density $\rho=\SI{1e-9}{\kg\per\m\cubed}$ (bottom) of \protect\Ti{10} for the comprehensive chemical nucleation models using the polymer nucleation description. The results are similar to the closed nucleation model \fig{\ref{fig:CN_all_clusters_normalised}} where \protect\Ti{10} forms from \SIrange{1000}{1200}{\K} and converges within roughly \SI{20}{\day}. The largest cluster encompasses more than 90 per cent of the available \ch{Ti}, in the most favourable nucleation conditions. This implies that all atomic \ch{Ti} quickly forms \protect\Ti{1} which subsequently starts to nucleate, in favourable conditions. An overview of all \protect\ch{Ti}-bearing molecules can be found in Appendix \ref{app:res_full_ntw} with an in-depth analysis in Sections \ref{sec:resAllchem} and \ref{sec:impact_fullntw}. }
            \label{fig:full_ntw_TiO2_max_normalised}
        \end{figure}
        
        \begin{figure*}
            \begin{flushleft}
            \includegraphics[width=0.32\textwidth]{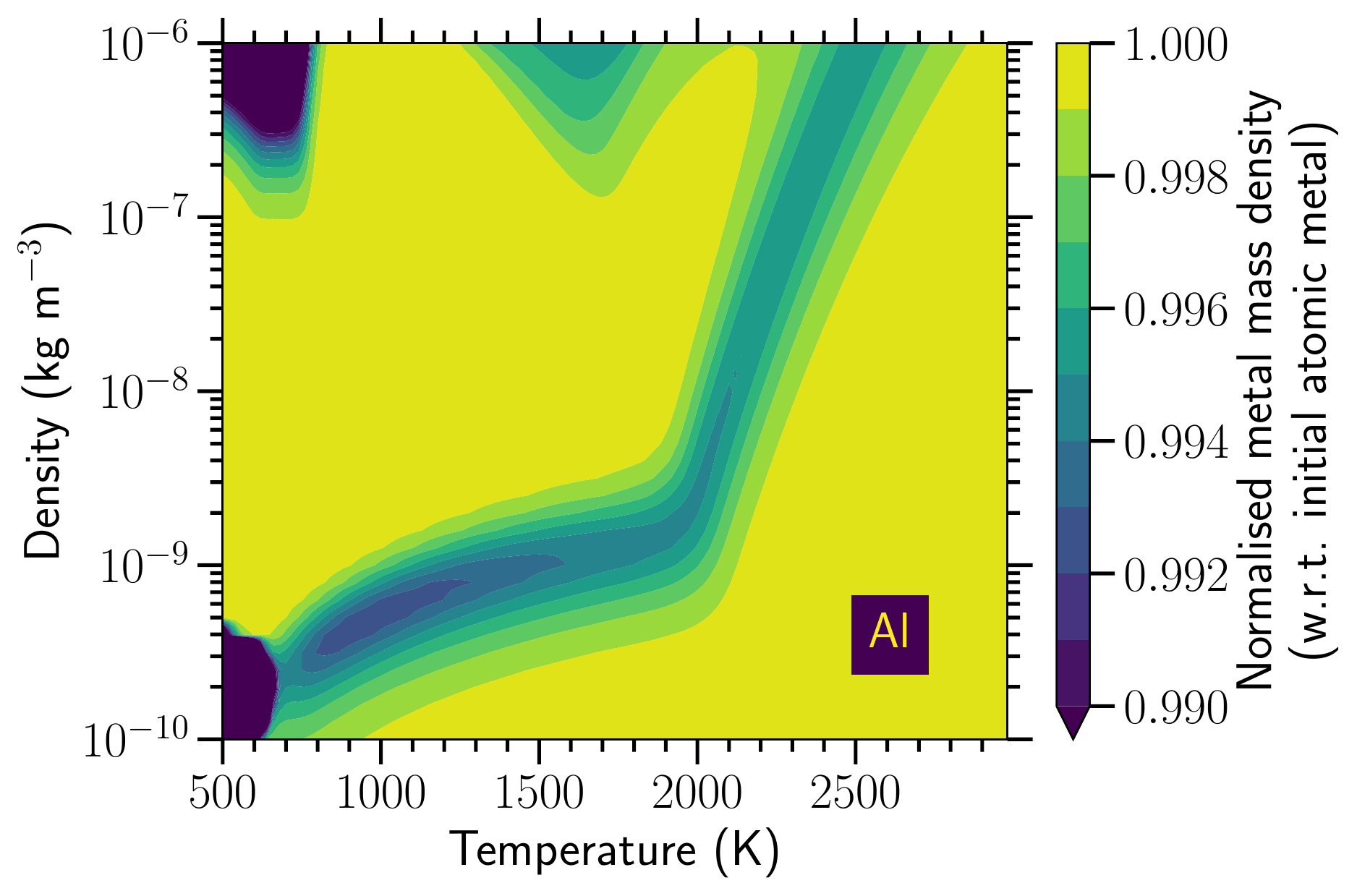}
            \includegraphics[width=0.32\textwidth]{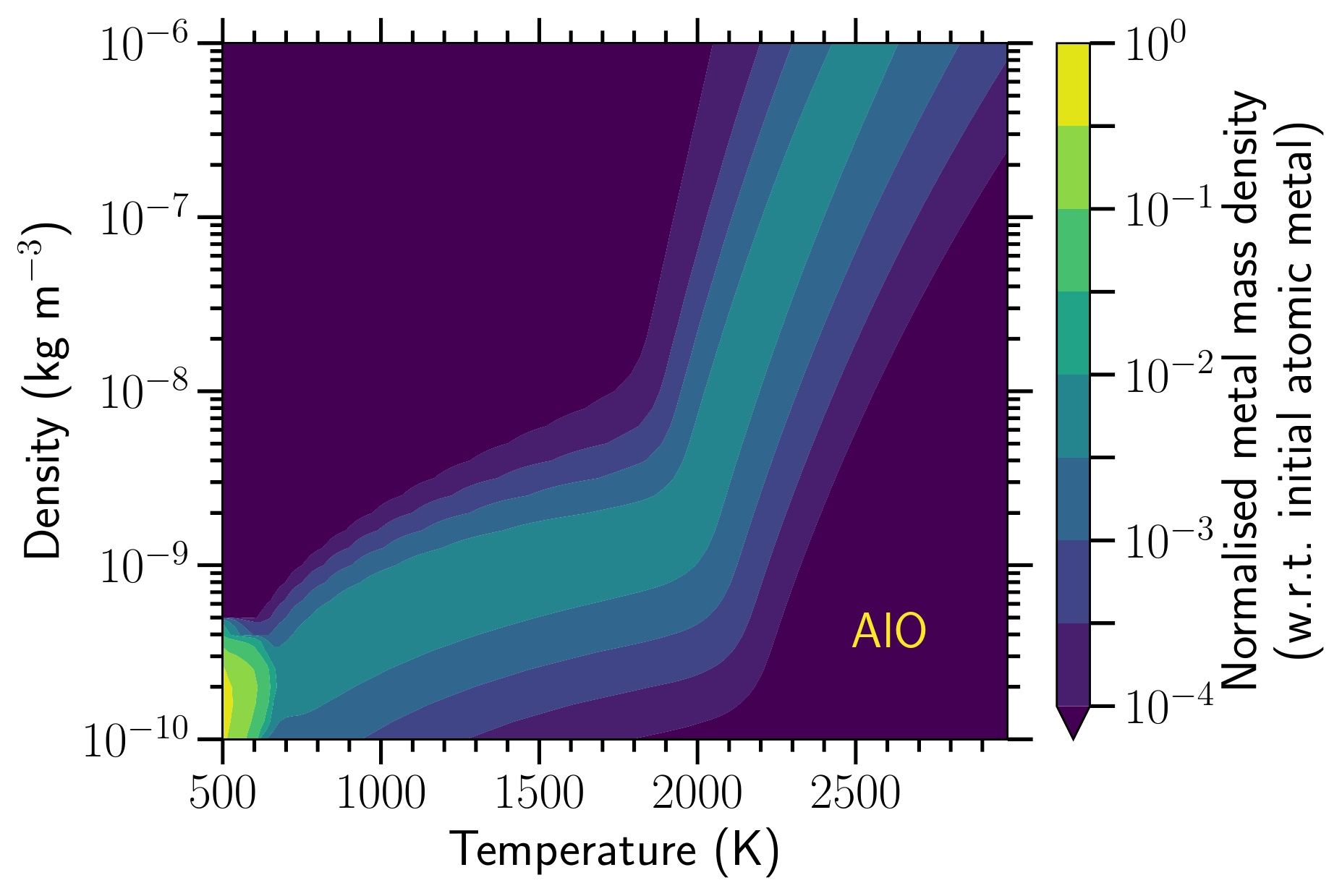}
            \includegraphics[width=0.32\textwidth]{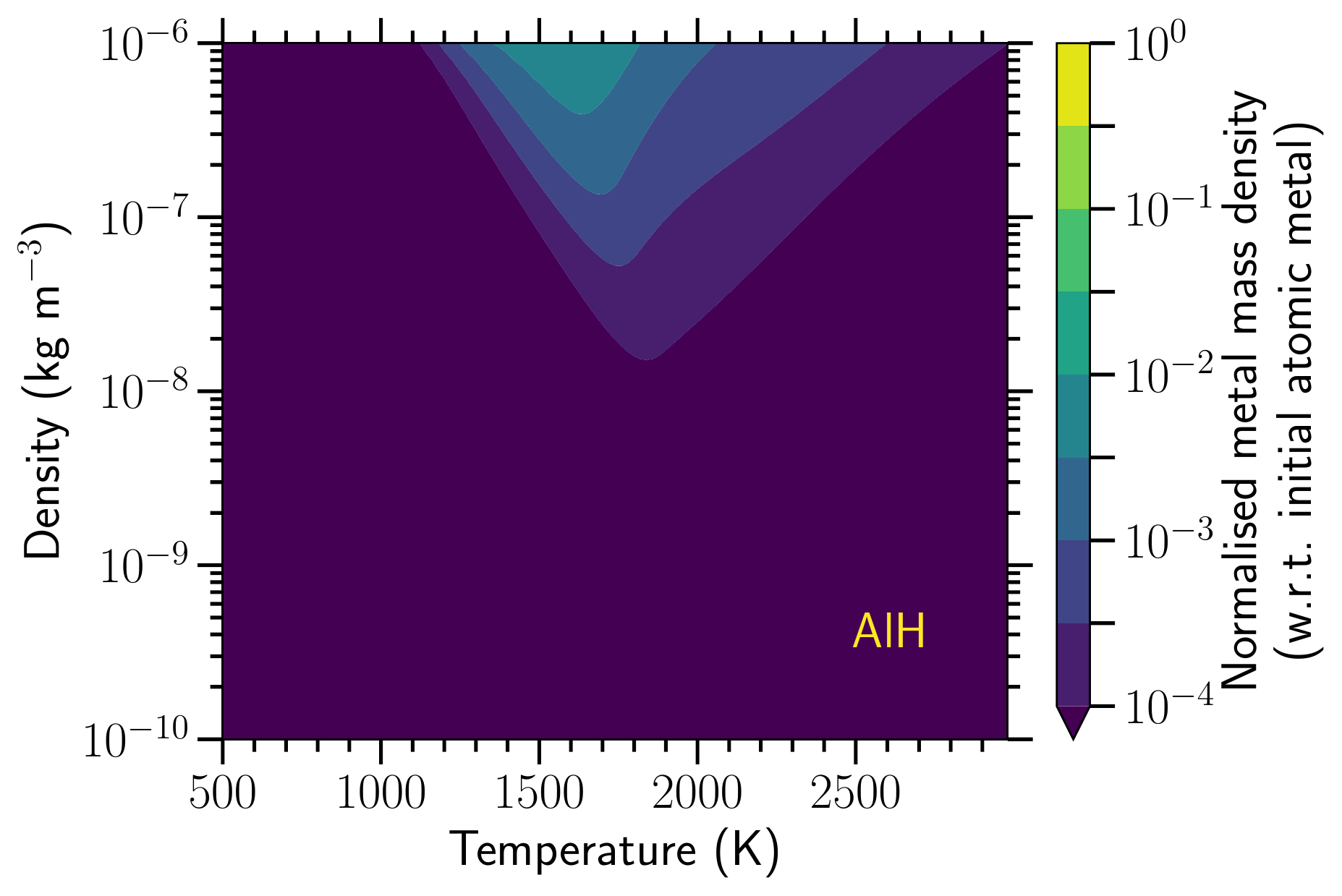}
            \includegraphics[width=0.32\textwidth]{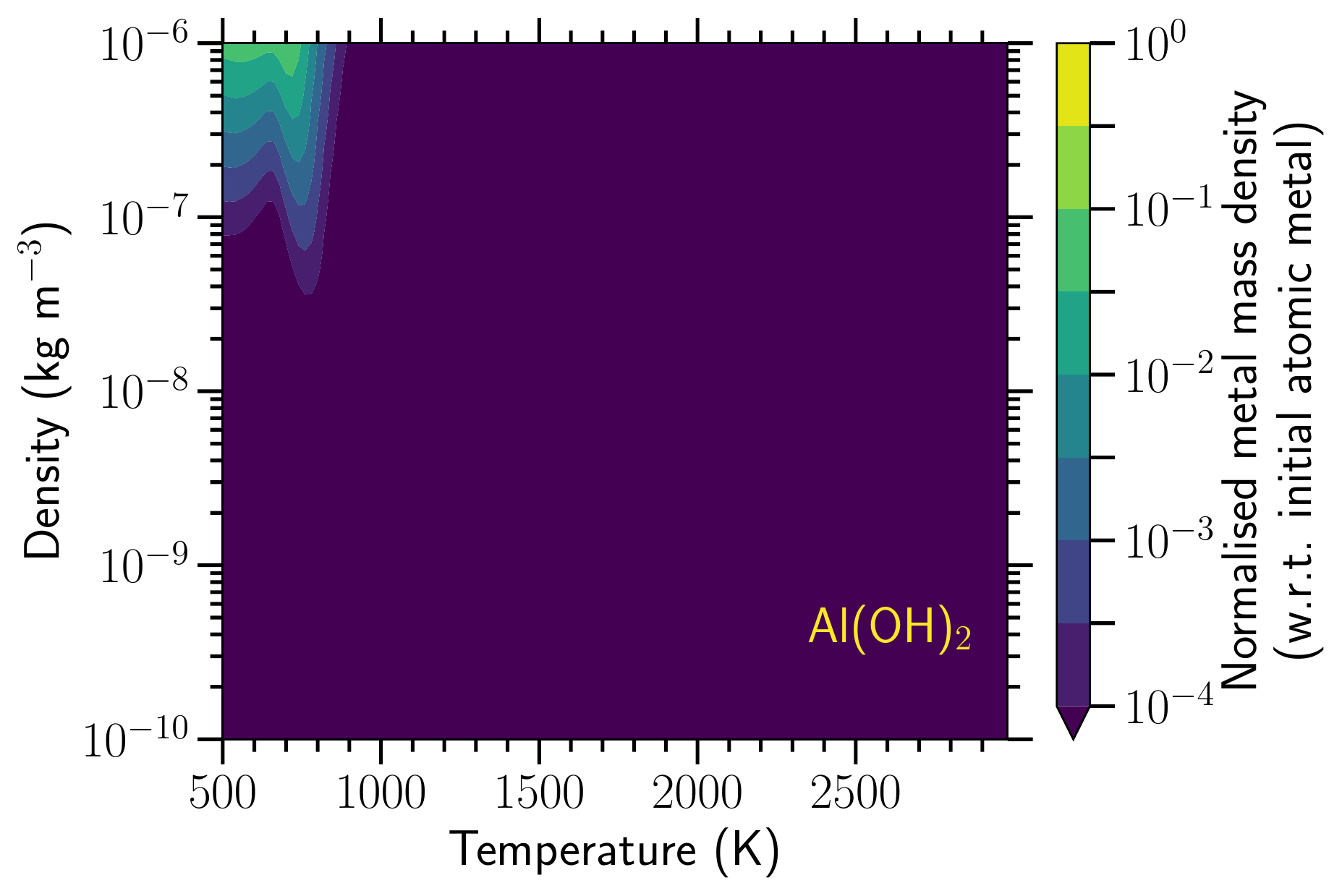}
            \includegraphics[width=0.32\textwidth]{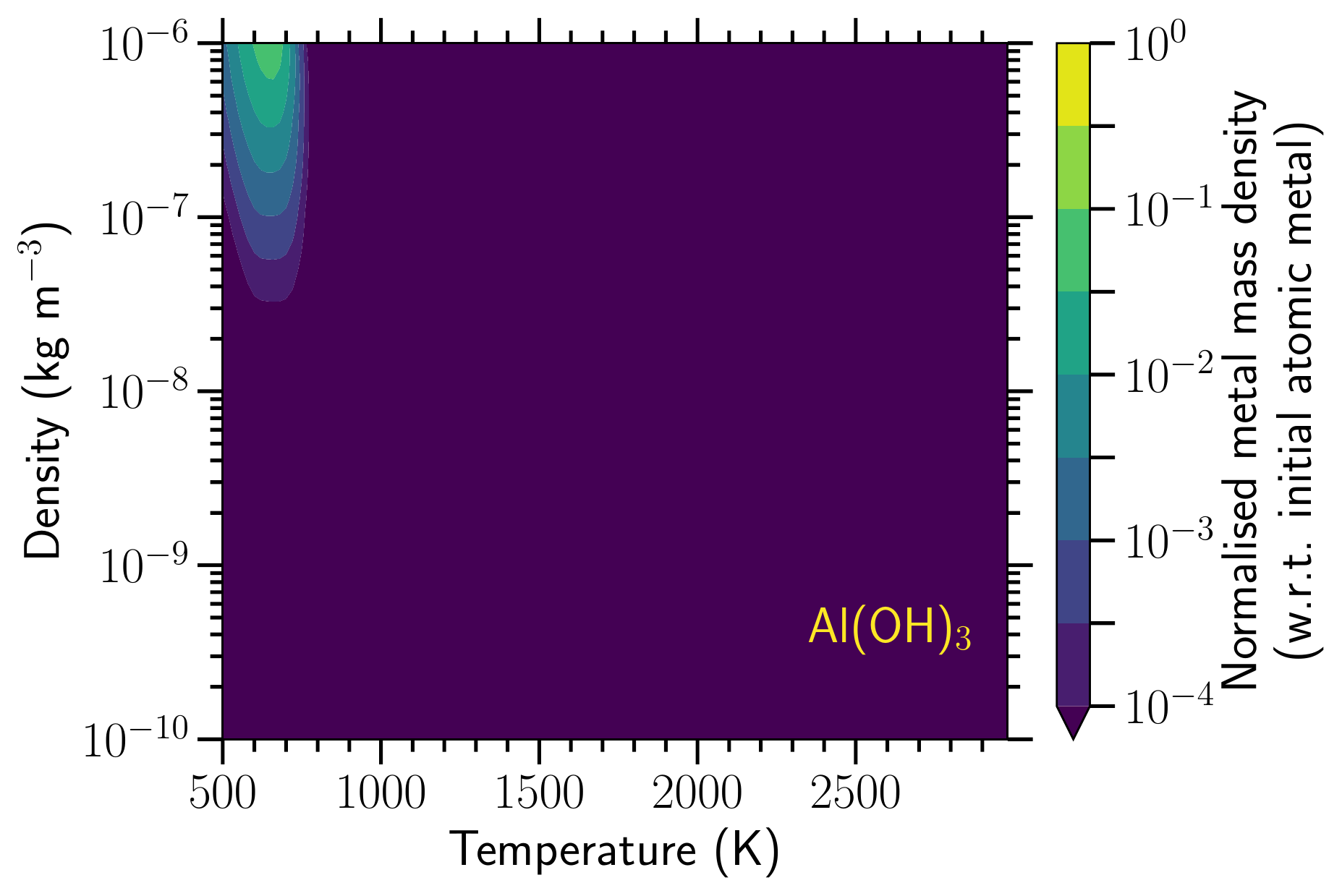}
            \end{flushleft}
            \caption{Normalised mass density after one year of the most abundant \protect\ch{Al}-bearing molecules for the comprehensive chemical nucleation models using the polymer nucleation description. Most \ch{Al} remains atomic with up to 1 per cent in \ch{Al}-bearing molecules. \protect\Al{1}, nor its precursors \ch{Al2O2}, \ch{AlO2} are able to form anywhere in the considered $(T,\rho)$-grid. Hence, no \protect\Al{1}-clusters can form either. We believe this issue is due to incomplete rate coefficients of \ch{Al}-molecule formation reactions. An overview of all \protect\ch{Al}-bearing molecules plus a temporal evolution of \ch{Al} and \ch{AlO} can be found in Appendix \ref{app:res_full_ntw} with an in-depth analysis in Sections \ref{sec:resAllchem} and \ref{sec:impact_fullntw}.  }\label{fig:full_ntw_Al-molecules_norm_same_scale}
        \end{figure*}
        
\section{Implications of results}\label{sec:impact}
This section interprets the nucleation model results and what they implicate for AGB dust precursors. Be aware that conclusions drawn from closed nucleation networks are based on the underlying assumption that the monomer exists and that all of the nucleation-related metal is turned into the monomer. \rev{The reader should be cautious when using these results as they are not necessarily physical. They are, however, useful in their own right to investigate the efficiency of individual nucleation species and the improved nucleation description.}
    
    \subsection{Closed nucleation networks}\label{sec:inter_res_closed}
    The most prominent result is that large \Alclusters can form fast $(< \SI{1}{\day})$ at high temperatures (around \SIrange{1800}{2400}{\K}). This makes \ch{Al2O3} the favoured candidate to become the first dust particles in the inner AGB wind. The second favoured candidates are \Mgclusters, which can form fast $(< \SI{1}{\day})$ around \SI{1500}{\K}. We find, that \Mg{9} forms more easily than the largest considered cluster \Mg{10} thanks to its higher stability. This is a consequence of the used non-classical nucleation description that relies on the Gibbs free energy of the clusters, which is lower for \Mg{9} than for \Mg{10}, making the former more energetically stable. Another consequence of the non-classical description is the preferred cluster sizes $N=$ \SIlist{2;4;6;9}{}, a situation that would never occur when using a classical nucleation theory (also noted by \citealt{Kohler1997}). The third preferred dust candidates are \Ticlusters, which only form below \SI{1000}{\K} at a relatively slow rate (time scale of tens of days compared to hours for \Mg{1}- and \Alclusters). Finally, we discard \Siclusters to be important as first dust species as their growth  requires conditions that are too cold and too dense compared to the conditions in an inner AGB wind. \Siclusters might form dust grains further out in the wind, where the temperature is below \SI{500}{\K}.\\\\ 
    Using the monomer or polymer nucleation description can result in substantial differences in typical formation times of the nucleation products, hence in their abundances after one year \figs{\ref{fig:CN_all_clusters_normalised}}{\ref{fig:CN_all_clusters_time_evolution}}. The most striking difference is the absence of large clusters at low temperatures when using the MN description. This can have profound implications while the wind is cooling down, underestimating the total number of large clusters. Using the abundance of the largest clusters as a gauge of dust formation, the MN description will yield less dust, which can delay or even hamper wind-driving. The formation time of large clusters can be several times larger when using the MN description. E.g. the convergence of \Ti{10} takes \SI{60}{\day} as compared to less than \SI{20}{\day} when using the PN description. For \Mg{9} the difference is \SI{180}{\day} compared to a few hours \fig{\ref{fig:CN_all_clusters_time_evolution}}. Additionally, at our benchmark density of \SI{e-9}{\kg\per\m\cubed} the abundance of \Mg{9} converges to roughly \SI{e12}{\per\m\cubed} in mere hours in the polymer case whereas in the monomer case it takes almost \SI{200}{\day} to converges to only \SI{e10}{\per\m\cubed} \fig{\ref{fig:CN_all_clusters_time_evolution}}.\\\\
    \rev{Although the abundance of some clusters converges, this does not happen for all clusters over the entire temperature regime. This result implies that no all clusters have reached equilibrium abundances yet. Hence, the assumption of a steady state nucleation is generally not valid in the entire temperature range. Therefore, it is necessary to use a time dependent nucleation description to accurately trace the nucleation process.}
    
    \subsection{Comprehensive chemical nucleation network}\label{sec:impact_fullntw}
    Although \Al{1}-clusters are the primary dust precursor candidate  according to the closed nucleation models (Sec~\ref{sec:inter_res_closed}), no \Al{1}-clusters form in the comprehensive nucleation models since the smallest building block, the monomer, cannot be created. Most \ch{Al} remains atomic, though up to maximally 1 per cent can form molecules (\ch{AlO}, \ch{AlH}, \ch{Al(OH)2}, and \ch{Al(OH)3}, Fig.~\ref{fig:full_ntw_Al-molecules_norm_same_scale}). The second favoured candidates, \Mg{1}-clusters, do not exist either because all the available \ch{Mg} remains atomic. The third favoured candidates according to the closed nucleation model, \Ti{1}-clusters, form equally efficient in the comprehensive nucleation model. Lastly, as in the closed nucleation models, \SiO{1}-clusters are discarded as first dust precursors in the considered temperature-density regime.\\\\
    These results suggest that, of the considered candidates, \Ti{1}-clusters are the only possible dust precursors. However, firstly there is ample evidence that pre-solar AGB grains mainly encompass \Al{1}-grains rather than \Ti{1}-grains \citep{Hutcheon1994, Nittler1994, Choi1998, Nittler2008, Bose2010a}. Secondly, dust has been observed to exist close to AGB stars,  at $\sim 1.5 R_\star$ for R\,Dor \citep{Khouri2016}, at $< 2R_\star$ for R\,Dor, W\,Hya and R\,Leo \citep{Norris2012}, and at $< 2R_\star$ for W\,Hya \citep{Zhao-Geisler2015, Ohnaka2016}. The temperature corresponding to those spatial regions is roughly \SIrange{1500}{2000}{\K}, which is higher than the formation temperature of \Ti{10}, that is around \SIrange{1000}{1200}{\K} \figs{\ref{fig:CN_all_clusters_normalised}}{\ref{fig:full_ntw_TiO2_max_normalised}}. Large \Mg{1} and \Al{1}-clusters, however, are able to form at such high temperatures \fig{\ref{fig:CN_all_clusters_normalised}}. Both observational arguments question the viability of \Ti{1}-clusters as first dust species and favour \Al{1}-clusters, yet our comprehensive model does not predict this. This discrepancy indicates that our current model lacks chemical reaction physics to form \Al{1} monomers. Since we cannot form any of the two \ch{Al2O3} precursors either (\ch{Al2O2} and \ch{AlO2}, Tab.~\ref{tab:Al2O3-formation-pathways}), we believe that the current reaction rate coefficients involving \ch{Al}-oxides are incorrect and need revision or that alternative small \ch{Al2O3}-cluster formation pathways are missing.

    \begin{table}
        \centering
        \begin{tabular}{ll}
            $^a$ \ch{AlO + AlO + M-> Al2O2 + M}&  \\
            $^a$ \ch{Al + AlO2 + M -> Al2O2 + M}& \\
            $^a$ \ch{Al2O + O + M -> Al2O2 + M}&  \\\\
            $^b$ \ch{AlO + O + M -> AlO2 + M}&  \\
            $^b$ \ch{AlO + O2 -> AlO2 + O}& \\
            $^a$ \ch{AlO + CO2 -> AlO2 + CO}& \\\\
			$^a$ \ch{Al2O2 + O + M-> Al2O3 + M}\\
			$^a$ \ch{AlO2 + AlO + M-> Al2O3 + M}\\\hline
             \multicolumn{2}{p{\columnwidth}}{Rate coefficients are determined by: a) Reversed of \citet{Catoire2003, Washburn2008} via detailed balance. b) \citet{Sharipov2012}.}
        \end{tabular}
        \caption{Formation of \ch{Al2O3} can only occur via \ch{Al2O2} or \ch{AlO2}, according to the reactions available in the literature. \ch{M} is by convention a third body which can be any chemical species.}
        \label{tab:Al2O3-formation-pathways}
    \end{table}

\section{Discussion}\label{sec:discussion}
This section discusses the limitations of our models (Sec.~\ref{sec:limitations}) and compares our model results with other literature studies (Sec.~\ref{sec:compLit}). 

    \subsection{Limitations}\label{sec:limitations}
    This section focuses on the limitations of the improved nucleation theory (Sec~\ref{sec:limKNT}), the used chemical reactions (Sec~\ref{sec:limChem}), and the inference of dust properties (Sec~\ref{sec:limDust}).
    
        \subsubsection{Nucleation theory}\label{sec:limKNT}
\begin{figure}
           \centering
           \includegraphics[width=\columnwidth]{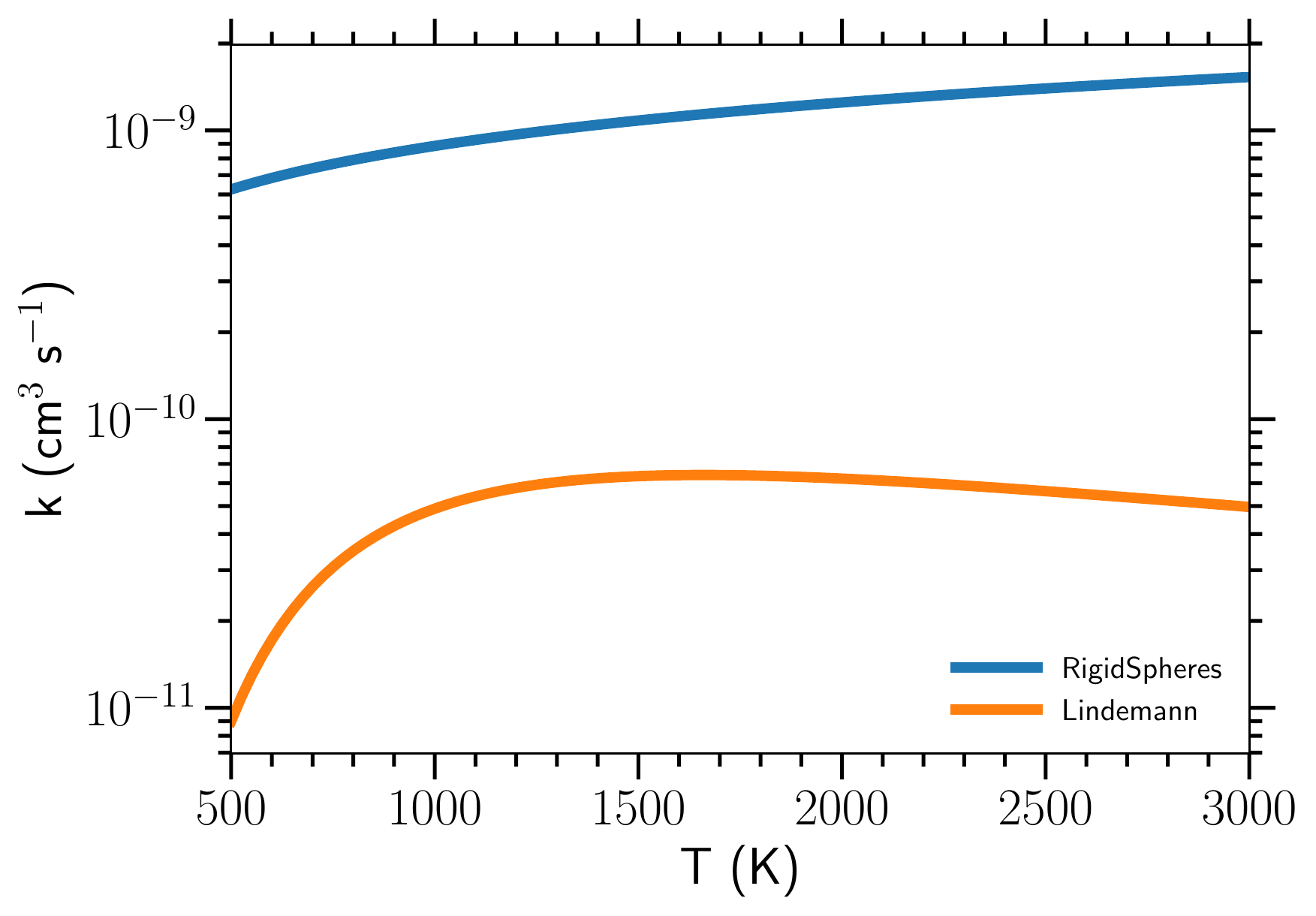}
           \caption{\rev{The reaction rate coefficients of \ch{Al2O3 + Al2O3 -> (Al2O3)_2} with the approximation of a collision of rigid spheres, used in this work, and calculated with Rice-Ramsperger-Kassel-Marcus theory plus a Lindemann fit by \citet{Sharipov2018}. As this latter also depends on the total number density, we have chosen a typical value for the inner AGB wind, $n_\text{tot} = \SI{e18}{\per\m\cubed}.$ Our approximation over predicts the dimerisation by roughly an order of magnitude compared to the more realistic coefficient using the Lindemann fit. Therefore, using a rigid sphere approximation, as in this work, will most likely overestimate the efficiency of the nucleation process.}}
           \label{fig:Al2O3-dimerisation}
        \end{figure}

        Our non-classical, non-equilibrium nucleation theory has some limitations. The most prominent one is most likely that it describes the growth of clusters as an inelastic collision between rigid spheres. This assumption does not account for the shape of the clusters nor mutual interaction forces. Using detailed chemical reaction coefficients for each cluster reaction\rev{, which account for possible energy barriers,} would be a large improvement. Unfortunately, such information does not yet exist. \rev{Recently, \citet{Sharipov2018} have calculated rate coefficients of the dimerisation of \Al{1} based on Rice-Ramsperger-Kassel-Marcus (RRKM) theory, which is a more realistic apprximation than the using the rigid spheres. We show both approximations as an example on how much the coefficients can differ (Fig.~\ref{fig:Al2O3-dimerisation}). Similarly, \citet{Suh2001} and \citet{Bromley2016} have determined \ch{SiO}-clustering rate coefficients with RRMK theory.} 
        Additionally, in the cluster growth coefficient (Eq.~\ref{eq:k_growth}), we write the radius of each cluster as a function of the monomer radius. However, since we know the shape of each cluster, it is possible to calculate an effective radius for each cluster, yielding a more correct geometrical cross-section between cluster collisions. Another limitation is set by using spontaneous clusters destruction reactions that rely on detailed balance. Incorporating chemical or collisionally induced destruction reactions would increase the accuracy of the model. Furthermore, the entire nucleation process is assumed to be homomolecular. There is, however, no good reason that it cannot be heteromolecular. Heteromolecular nucleation is most likely necessary to create \ch{MgAl2O4}-clusters, which are abundant in pre-solar AGB grains, or \ch{Mg}-containing silicates \citep{Goumans2012}. Including heteromolecular nucleation will increase the number of possible reactions exponentially and will increases the amount of detailed quantum mechanical calculation and data needed for those reaction rate coefficients.\\\\
        \rev{The assumption that nucleation starts with the formation of the monomer is not yet established. Small clusters might be formed via pathways which bypass the monomer molecule. This could possibly solve the issue of not forming \Al{1}-monomer in our models. Additionally, the fact that nucleation occurs via the addition of monomer-multiples with a fixed stoichiometry is not established either. Clusters could possibly grow via the addition of other stoichiometric ratios, as investigated by \citet{Patzer2005} for small aluminium oxide clusters.}\\\\
        \rev{Note that the used nucleation description considers the process as a statistical ensemble of particles which all have the same mean temperature. However, as this is a process of molecular interactions, the notion of `temperature' can become unclear. In reality, the particles have a temperature distribution around a mean kinetic temperature. Molecular dynamics simulations, which do not rely on a mean temperature, reveal that small temperature fluctuations amongst particles initiate the nucleation process \citep{Tanaka2011,Diemand2013,Toxvaerd2015}.}\\\\
        \rev{A last limitation is the artificial maximum cluster size. In reality, the clusters would continue to grow to form solid material. This material can then, on its turn, sublimate and return nucleation species to the gas phase. Whether the sublimation process returns small clusters, monomers, atoms, or simple molecules is unclear. Additionally, to estimate the sublimation rate one needs the binding energies of the surface layer of the solid material. However, the phase transition process to a solid dust grain is often described by one fast reaction \citep[e.g.][]{Huang2009,Bojko2014}. A better approach would be to evolve the nucleation of clusters until a chosen maximal cluster size is reached, after which it can be considered as a solid particle and can grow via grain-grain interactions such a coagulation.}
        
        \subsubsection{Chemical reactions}\label{sec:limChem}
        To infer abundances of the largest nucleation clusters, it is crucial to correctly predict the creation of its fundamental building block, the monomer. Hence, the chemical reaction path ways from atoms to monomers have to be accurate. However, astrochemical databases lack the necessary monomer formation reactions. Yet, there are individual studies that provide some reactions, but they are scarce depending on the nucleation candidate. To determine the AGB dust precursors, we believe that \ch{Ti} and \ch{Al} reactions are the most pressing. There are hardly any \ch{Ti}-related reactions (App.~\ref{app:chem_network}) 
        and most \ch{Al}-related reactions have extremely high reaction barriers. The latter mainly originate from combustion studies and are therefore often only determined in the high density limit. Moreover, most \ch{Al}-related reaction rate coefficients are determined from destruction of larger molecules, which is the opposite of what is actually needed. Therefore, the growth coefficients rely on the assumption of detailed balance.\\\\
        It is important that the entire chemical network contains sufficient reactions with accurate rates. As pointed out by \citet{Boulangier2019}, we are largely dependent on the astrochemical databases which do not contain all the reactions that are necessary. 
        \old{An example of the limited availability of reaction rate coefficients in the literature is the surprising abundance of \ch{SiO2} (Sec.\,\ref{sec:resAllSiO}, Fig.\,\ref{fig:full_ntw_Si-molecules_norm_same_scale}). The lower binding binding energies of \ch{SiO2} (6.5\,eV) compared to \ch{SiO} (8.3\, eV) should result in the formation of \ch{SiO2} at lower temperatures than \ch{SiO}, which is not the case. However, there is only one \ch{SiO2} formation reaction available in both UMIST and KIDA, which is even constant in temperature (\ch{SiO + OH -> SiO2 + H}). According to both databases, the destruction of this molecule is only possible via collisions with ions (not abundant) or photons (shielded from \rev{an external radiation field} due to high density), thus inefficient. There is also the NIST Chemical Kinetics Database\footnote{\label{foot:nist_chem}\url{https://doi.org/10.18434/T4D303}}, which we only used for reactions related to the nucleation monomers. This database does provide a temperature dependent rate coefficient for \ch{SiO + OH -> SiO2 + H}, and has one collisional \ch{SiO2} destruction rate. The latter, however, is just an experimental value at room temperature and thus again temperature independent. \ch{SiO2} is just an example molecule which we happen to discuss here, but the limitations hold for all species.} 
        Due to the lack of reactions rate coefficients and especially the unknown temperature dependence, caution is advised when interpreting chemical evolution results and the existence of certain molecules based on the gas temperature.
        
        \subsubsection{Inference of dust properties from clusters}\label{sec:limDust}
        This work focuses on nucleation clusters to infer AGB dust properties such as abundance, composition and formation times. However, the largest clusters that we consider are only a fraction of the size of a dust grain nor do they resemble the bulk geometry. The largest clusters' radii range from \SIrange{0.16}{0.71}{\nm} whereas dust grains can be as large as a few micron. \citet{Lamiel-Garcia2017} predict that \Ti{1}-clusters only resemble the bulk geometry from $N \ge 125$. For \rev{highly ionically bonded materials} such as \Mg{1}-clusters this can already be at $N=20$ due to the strong electrostatic interactions between atoms. Therefore, one has to be careful when using nucleation clusters as a gauge for dust grains. Yet, due to computational constraints a small cross-over size, from clusters to dust, has to be chosen. From this cross-over size, the particles should not be considered as molecular clusters any more but as tiny grains which can numerically be binned in size \rev{and can grow via various physical processes} \citep[e.g.][]{Jacobson2013, Grassi2017, McKinnon2018, Sluder2018}. \rev{Because of our artificial maximum cluster-size, one has to be cautious when interpreting the abundances of the largest clusters in this work since in reality these will most likely continue to grow to actual dust grains.}

    \subsection{Comparison with literature}\label{sec:compLit}
    This section compares our model results with other nucleation models (Sec.~\ref{sec:disc_nucl}), with seed particle requirements of dynamical wind models (Sec.~\ref{sec:disc_dynamical_models}), and with molecular observations of AGB stars (Sec.~\ref{sec:disc_observations}).
    
        \subsubsection{Nucleation models}\label{sec:disc_nucl}
\begin{figure}
           \centering
           \includegraphics[width=\columnwidth]{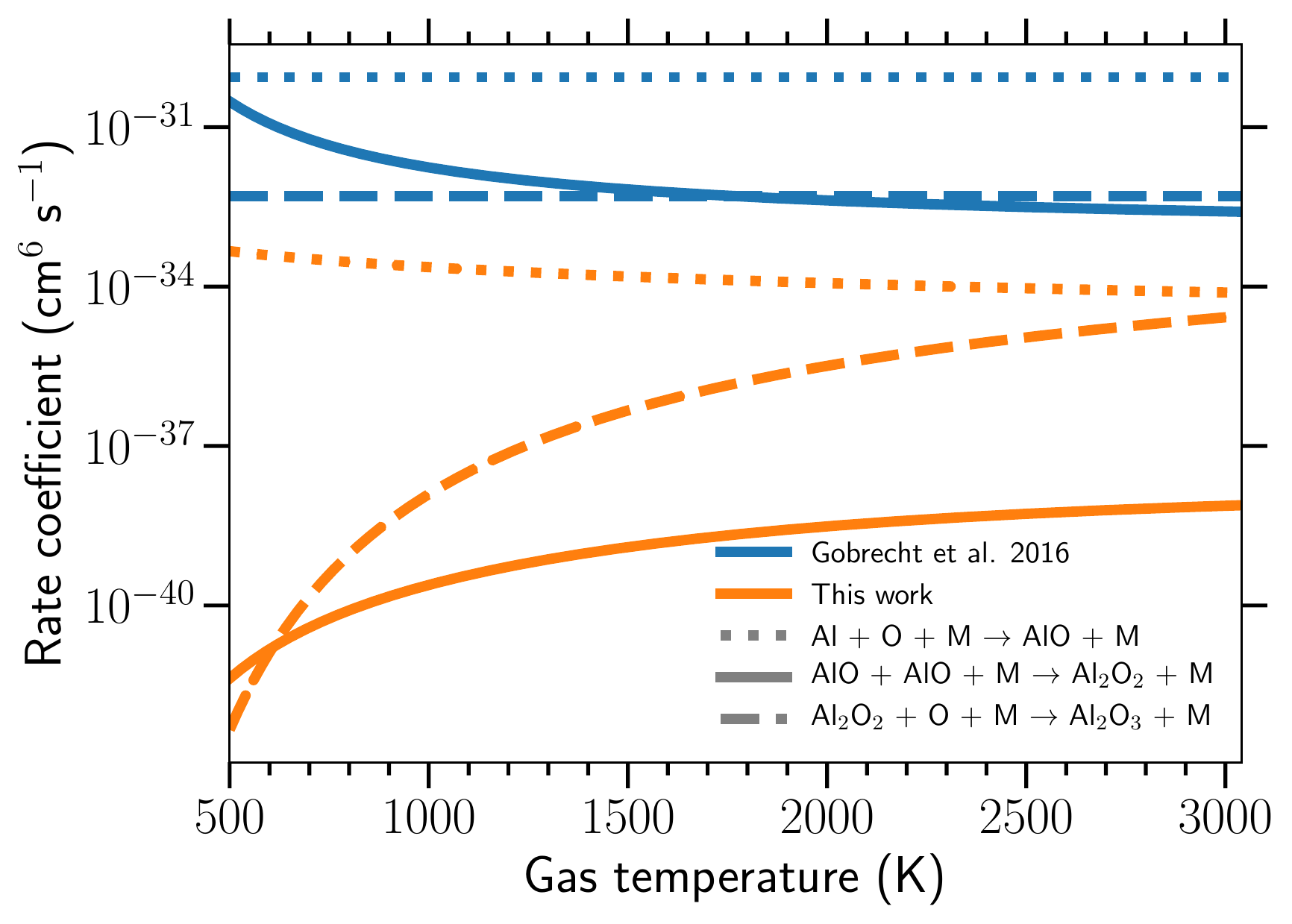}
           \caption{The reaction rate coefficients of some key \ch{Al2O3} formation reaction used by \citet{Gobrecht2016} are \SIrange{2}{10}{} orders of magnitude higher than the ones used is this work. These large differences could explain why \citet{Gobrecht2016} form \ch{Al2O3} and we do not. Moreover, they estimate the barrierless three-body reactions of the type \ch{A + B + M -> AB + M} to be temperature independent, hampering an investigating of the temperature dependence for \ch{Al2O3}-cluster formation.}
           \label{fig:Al-rates}
        \end{figure}

        In contrast to \citet{Gobrecht2016}, our most complete model (comprehensive network with polymer nucleation) does not produce any \Al{1}-clusters. However, unlike this work, \citet{Gobrecht2016} used a simplified formulation to determine reversed formation rates for \ch{Al}-bearing molecules resulting in a temperature independent rate coefficients. Some key formation reactions reveal that the used rate coefficients can differ by up to 10 orders of magnitude (e.g. \ch{AlO + AlO + M -> Al2O2 + M}, Fig.~\ref{fig:Al-rates}. Note that \citet{Sluder2018} use an even higher rate coefficient for this reaction.). Such large differences could explain different results of \citet{Gobrecht2016}, as compared to this work. Moreover, we give a more realistic rate description by incorporating a temperature dependence in addition to the strong density dependence which is crucial \rev{to} investigate the existence of large clusters and dust grain as a function of temperature (e.g. \ch{Al2O2 + O + M -> Al2O3 + M}, Fig.~\ref{fig:Al-rates}). Compared to observations, \citet{Gobrecht2016} overpredict the abundance of \ch{Al}-bearing molecules (\ch{AlO} and \ch{AlOH}), whereas our models agree better with the most recent observations (Sec.~\ref{sec:disc_observations}).\\\\
        \rev{An approach similar to this work has recently been used by \citet{Savelev2018} who investigated the nucleation of \Al{1}-clusters up to a cluster size of 75 during the combustion of aluminized fuels. Similarly, they also model the nucleation kinetically with a set a chemical reactions. Their nucleation reactions, however, only consider monomer interactions. They do consider much larger clusters than we do. However, the authors rely on estimates (interpolations) of the Gibbs free energies for $N=5 - 75$ and do not perform DFT calculations of the global minima candidates. The authors do not provide the geometries of these larger sized \ch{Al2O3}-clusters. Therefore, we cannot verify these isomers with the lowest-energy structures used in the present study. Moreover, it is difficult to compare results since their environment has a density of several orders of magnitude higher making the nucleation occur on milli- and microsecond time scales. See \citet{Starik2015} for a recent review of modelling aluminium nanoparticles in the fuel combustion community.}\\\\
        The nucleation efficiency of species is often determined by the steady state nucleation rate, $J_*/n_{\ch{H}}$, which represents the number of dust seed particles formed per second per total number of hydrogen. However, this rate relies on two main assumptions. Firstly, growth of clusters only occurs via addition of monomers. Secondly, the system of clusters is in a steady state, i.e. the number densities of all clusters remain constant over time, ergo chemical equilibrium. This latter implies that the net formation of all clusters is the same and size independent. Detailed derivations for the steady state nucleation rate can be found in \citet{Patzer1998} but the notation used by \citet{Bromley2016} is clearer. The latter explicitly shows that $J_*/n_{\ch{H}}$ solely depends on the amount of monomers\footnote{Since a steady state is assumed, this refers to the number of monomers at chemical equilibrium. A detail which is usually overlooked.} and all rate coefficients between clusters. To determine this equilibrium abundance, one has to know the Gibbs free energy of the lowest energy configuration for all cluster sizes (App.~\ref{app:minGFE}). This data is unavailable for large clusters. It is often unclear how this abundance is determined in nucleation papers, either from the vapour pressure of the monomer and the solid form \citep[as explained by][]{Patzer1998, Helling2006}\footnote{\rev{Determining the equilibrium monomer abundance from the phase equilibrium with the bulk material via the vaporisation pressure inherently assumes that the bulk material exists. However, since we are investigating the existence of bulk material can actually happen in certain conditions, such assumption should not be made.}} or by chemical equilibrium calculations of the gas without considering the clusters \cite[e.g.][]{Jeong2003, Lee2015}\footnote{Though we could not confirm which of these two methods \citet{Jeong2003} used, we note that if the vapour pressure was used than the nucleation of \ch{Al2O3} should be higher than that of \ch{TiO2} since the former has a lower vapour pressure. According to that method, this means less monomers thus all material is in the solid form. However, they find a smaller $J_*/n_{\ch{H}}$ for \ch{Al2O3} than for \ch{TiO2}.}. Because the steady state monomer nucleation description differs significantly from ours \rev{and requires knowing the equilibrium abundance of the monomer}, it is difficult to compare with. Using $J_*/n_{\ch{H}}$, it is often claimed that only \Ti{1} nucleates efficiently enough to form the first dust precursor. We limit the comparison to our monomer nucleation description since the steady state one also \rev{assumes nucleation by monomers.} \old{make this assumption.} When comparing our results with \citet[fig.~1]{Jeong2003}, we note that both predictions of \Ti{2}-clusters have a steep cut-off around \SI{1000}{\K} \fig{\ref{fig:CN_all_clusters_normalised}}. However, our time dependent description does not yield the high nucleation that the steady state one does at low temperatures since the availability of monomers decreases quickly hereby quenching the growth process. Additionally, the assumption of steady state is invalid since there is a clear time dependence in cluster growth \fig{\ref{fig:CN_all_clusters_time_evolution}}. \citet{Jeong2003} exclude \Al{1}-clusters to be a primary dust precursor due to the low 
        $J_*/n_{\ch{H}}$. One should be careful with interpreting this result since, as they point out, this is due to the low equilibrium abundance of the monomer and not necessarily the capability of nucleating \Al{1}-clusters. They do not discuss the efficiency of \Al{1} vs \Ti{1}-nucleation based on stability of the clusters. We find that, if \Al{1}-monomers could exist, they will nucleate at much higher temperatures than \Ti{1} \fig{\ref{fig:CN_all_clusters_normalised}}. However, we are also unable to form the \Al{1}-monomers with an initial atomic gas (Sec.~\ref{sec:resAllAl2O3}).\\\\
        Our results indicate that \Al{1}-nucleation is dominant at high temperatures but the formation of the monomer via chemical reactions is unattainable with currently available data. \rev{Moreover, there is experimental evidence that small \Al{1}-clusters do exist when vaporising the solid material \citep{vanHeijnsbergen2003,Demyk2004,Sierka2007}} This is a clear incentive for the scientific community to investigate rate coefficients of \ch{Al}-bearing reactions at high temperatures. Without this data, it will remain unclear which species forms the first dust precursors in AGB winds. 
        
        \subsubsection{Dynamical models}\label{sec:disc_dynamical_models}
        \citet{Hofner2016} show that the minimal normalised number of \Al{1} dust seed particles (assumed to be clusters of size $N=1000$) for driving an AGB wind is of the order of $n_s/n_{\ch{H}} \sim 10^{-16}$, with $n_s$ the seed particle number density. For comparison, we do a rough extrapolation of our results by assuming that all the largest \Al{1}-clusters get turned into clusters of size $N=1000$. This is in line with rapid formation of the largest clusters and depletion of the smallest ones (Sec.~\ref{sec:resAl2O3}). Since our largest cluster has roughly size $N=10$, the number of seed particles of $N=1000$ would be 100 times smaller. If we compare with \Al{1}-clusters and assume that 1 per cent of the available \ch{Al} turns into \Al{1} (Sec~\ref{sec:resAllAl2O3} and \ref{sec:impact_fullntw}), then the total number of largest clusters is roughly 10 per cent of the initial number of monomers. This translates to $n_{\Al{1000}}/n_{\ch{Al}} \approx 10^{-5}$. Using $n_{\ch{Al}}/n_{\ch{H}}$ from Table~\ref{tab:init_composition}, this yields a normalised number of seed particles $n_{\Al{1000}}/n_{\ch{H}} \approx 3 \cdot 10^{-11}$, which is already \num{100000} times more than needed according to the models of \citet{Hofner2016}. We can also compare this with the number of \Ti{10}-clusters. Here no assumption on the number of monomers has to be made because the comprehensive network model with PN already predicts the amount of \Ti{10}. This is roughly 10 per cent of the available number of \ch{Ti}. Again assuming that the number of \Ti{1000}-clusters is 100 times smaller than \Ti{10} and using the initial $n_{\ch{Ti}}/n_{\ch{H}}$ from Table~\ref{tab:init_composition}, yields $n_{\Ti{1000}}/n_{\ch{H}} \approx 8 \cdot 10^{-11}$. This is in line with the \Al{1000}-cluster extrapolation.\\\\
        
        \subsubsection{Observations}\label{sec:disc_observations}
\begin{figure}
            \centering
            \includegraphics[width=\columnwidth]{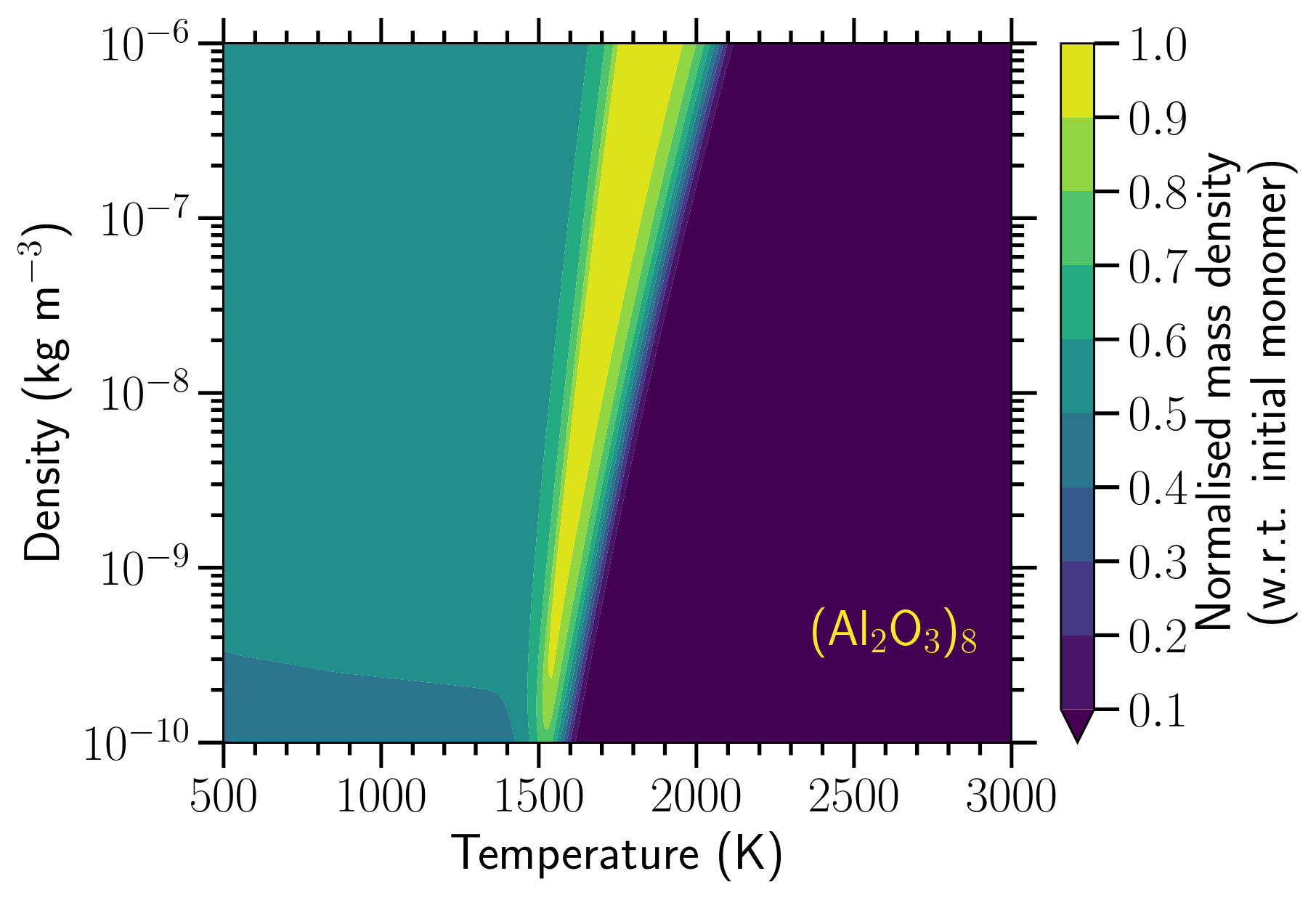}
            \includegraphics[width=\columnwidth]{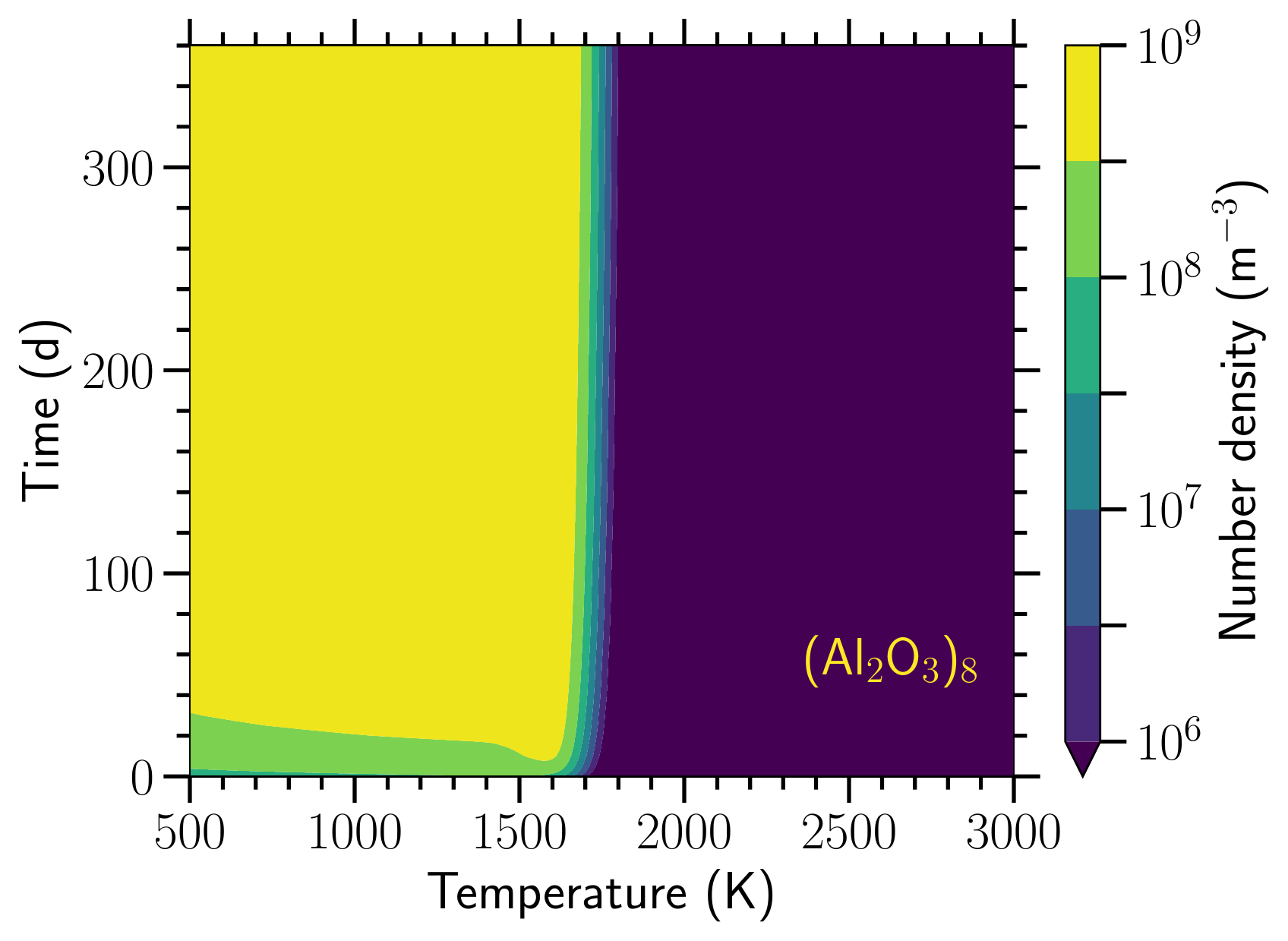}
            \caption{Normalised mass density after one year (top) and temporal evolution of the absolute number density at the benchmark total gas density $\rho=\SI{1e-9}{\kg\per\m\cubed}$ (bottom) of \protect\Al{8} for the closed nucleation model with an initial \protect\Al{1} abundance of 1 per cent of the available \ch{Al} using the polymer nucleation description. The results are similar to the closed nucleation model with all \ch{Al} turned into \protect\Al{1} \figs{\ref{fig:CN_all_clusters_normalised}}{\ref{fig:CN_all_clusters_time_evolution}}. Due to the lower amount of species the formation threshold is slightly lower at \SIrange{1600}{2100}{\K} and convergence takes a little longer, roughly \SI{20}{\day}.} 
            \label{fig:Al2O3_low_init_general_norm_same_scale_and_time_evolution}
        \end{figure}

        Our prediction of \Ti{1}-clusters \fig{\ref{fig:full_ntw_TiO2_max_normalised}} agrees with \citet{Kaminski2017} who state that there is no solid \Ti{1} close to the star ($T>1000$\,K). They also claim that \ch{TiO} and \ch{TiO2} are abundantly present in the extended envelope (\SIrange{170}{500}{\K}) and therefore \Ti{1}-clusters should not significantly exist to aid in wind driving. However, according to models of \citet[]{Hofner2016}, a tiny faction of seed particles ($n_s/n_H \sim 10^{-16}$) can be sufficient to aid in wind driving (Sec.~\ref{sec:disc_dynamical_models}). The lower left corner of our $(T, \rho)$-grid most closely resembles the extended envelope regime (i.e. cold and sparse), which shows that the \Ti{1} molecule and \Ti{1}-clusters can simultaneously be present \fig{\ref{fig:full_ntw_Ti-molecules_norm_same_scale}}. When intuitively extrapolating to lower temperatures and lower densities, as if moving further out into the extended envelope, we expect a higher \Ti{1} and \ch{TiO} abundance and less \Ti{1}-clusters.\\\\
        \citet{Khouri2018} observe that for the oxygen-rich AGB star o\,Cet 4.5 per cent of the atomic \ch{Ti} is locked-up in \Ti{1}. It is however difficult to compare with our model grid since the presence of the molecule is extremely sensitive to gas temperature and its abundance ranges from \SIrange{0}{100}{} per cent of the intitial atomic \ch{Ti} \fig{\ref{fig:full_ntw_Ti-molecules_norm_same_scale}}. As it is unclear what the temperature coverage of the observation is, the derived abundance is most likely an average in a certain temperature range. \citet{Kaminski2016} discovered \ch{AlO}, \ch{AlOH}, and \ch{AlH} in o\,Cet but could only determine the abundance of \ch{AlO}. They find $n_{\ch{AlO}}/n_{\ch{H}} = 10^{-9} - 10^{-7}$, which agrees with our model predictions that maximally 1 per cent of all \ch{Al} is turned into molecules, with \ch{AlO} the most abundant molecule $\sim n_{\ch{AlO}}/n_{\ch{H}} < 10^{-8}$. \citet{Kaminski2016} do state that \ch{AlOH} is present in a gas temperature of \SI{1960 \pm 170}{\K}, and that \ch{AlH} is detected between \SIrange{2.5}{4}{R_{\star}}. Both observational constraints comply with our model predictions \fig{\ref{fig:full_ntw_Al-molecules}}. Additionally, \citet{Decin2017} find that for AGB stars IK\,Tau and R\,Dor the amount of \ch{AlO}, \ch{AlOH}, and \ch{AlCl} accounts for maximally 2 per cent of the total aluminium budget. Both observations are in line with our prediction that maximally 1 per cent of all \ch{Al} is turned into molecules \fig{\ref{fig:full_ntw_Al-molecules_norm_same_scale}}. The amount of detected \ch{AlOH} in R\,Dor only accounts for roughly 0.02 per cent, yet this is still significantly more than our models predict \fig{\ref{fig:full_ntw_Al-molecules}}. Lastly, \citet{Khouri2018} also deduce that less than 0.1 per cent of the atomic \ch{Al} is converted into \ch{AlO}. In conclusion, all three observational studies agree with our prediction that maximally 1 per cent of all \ch{Al} is turned into molecules. Our prediction also better supports the recent observations than the significantly higher abundances of \ch{Al}-bearing molecules predicted by models of \citet{Gobrecht2016}.\\\\
        As both observations and our comprehensive model agree that maximally 1 per cent of all atomic \ch{Al} turns into a molecule (Sec.~\ref{sec:resAllAl2O3}), it is interesting to analyse the results of a closed nucleation model with only 1 per cent of the available \ch{Al} as initial \Al{1} abundance. We choose to only use the polymer nucleation description. Compared to a 100 per cent initial abundance, the temperature formation threshold of \Al{8} has slightly lowered to \SIrange{1600}{2100}{\K} \fig{\ref{fig:Al2O3_low_init_general_norm_same_scale_and_time_evolution}}. This is expected as a lower density produces less collisions therefore making it more difficult to form clusters at higher temperatures. Similarly, \Al{8} converges only after roughly \SI{20}{\day} which is significantly longer than the \SIrange{5}{10}{\hour} for the 100 per cent initial abundance model \fig{\ref{fig:Al2O3_low_init_general_norm_same_scale_and_time_evolution}}. Besides the temporal effects, the results are analogous to the 100 per cent case where \Al{8} also contains more than 90 per cent of the available monomers at the highest formation temperatures.
        
\section{Summary and prospects}\label{sec:summary}
In this paper, we have constructed and investigated an improved nucleation theory by abandoning the assumption of chemical equilibrium, dropping the restriction of cluster growth by only monomers, and using accurate quantum mechanical properties of molecular clusters. We have examined the viability of \nucspec as candidates of the first dust precursors in oxygen-rich AGB winds. The choice of candidates is based on rigorous theoretical and observational evidence (Sec.~\ref{sec:nucl_candidates}).\\\\
This work consists of two main nucleation descriptions, one that only allows cluster growth via monomers and one that allows polymer interaction. Both assume the nucleation processes to be homogeneous and homomolecular. With these descriptions, two main types of systems are evolved in a grid of temperature and density that is typical for AGB winds: a closed nucleation system and a comprehensive chemical nucleation system. The former considers the growth of one nucleation candidate species with the monomer as the smallest building block and assumes that all available atomic metal is locked-up in the monomer. The latter allows chemical interaction between species in a gas mixture which includes all nucleation species and starts with an atomic composition. The former provides insight in the nucleating efficiency of each candidate in temperature and density space, and the latter yields a more complete chemical nucleation model by removing the assumption of the a priori existence of the monomer (Sec.~\ref{sec:modelsetup}).\\\\
Constructing the nucleation reaction networks required quantum mechanical data of all clusters, which we calculated with high precision density functional theory. Since such calculations exponentially increase with cluster size, we limit the maximal size to roughly $N=10$.  The comprehensive chemical reaction network is constructed by adding relevant chemical reactions to an already carefully designed reduced network for AGB winds \citep{Boulangier2019}. The extension includes all relevant and available reactions to form the nucleation monomers. Since a significant amount of reversed reactions is not present in the literature, quantum mechanical data for the participating species is needed to calculate those reaction rate coefficients. We have gathered as much as possible data from the literature and performed density functional theory calculations when this was unavailable (Sec.~\ref{sec:data_and_calcs}).\\\\
Overall, using the monomer nucleation description as compared to the polymer one, will underestimate the abundance and overestimate the formation time of the large clusters. Using the abundance of the largest clusters as a gauge of dust formation, the monomer nucleation scenario would underestimate the amount of dust and overestimate its formation time. This can lead to less efficient wind-driving or even the absence of a wind in theoretical simulations. The monomer description also inhibits the formation of large clusters at low temperatures due to a rapidly developing lack of monomers, which by design is the only growth mechanism. The polymer description does not suffer from this limitation and is therefore more realistic. \rev{Comparison with equilibrium abundance ratios reveals that the assumption of equilibrium is not valid over the entire temperature range for a period of one year. Hence, a time-dependent description in necessary to investigate the nucleation process in AGB winds.}\\\\
The closed nucleation models, which assume that the nucleation monomers are present, predict that \Al{1} is the primary candidate to be the first AGB dust precursor. These clusters rapidly form at much higher temperatures than any other cluster, around \SIrange{1800}{2400}{\K} and in less than a few days. Rapid dust formation at high temperatures will aid in driving the AGB wind, since the wind is cooling down from hot shocks \citep[$\sim$\SI{10000}{\K},][]{Boulangier2019}. At around \SIrange{1500}{1700}{\K}, large \Mg{1}-clusters can form and only at \SIrange{1000}{1200}{\K} large \Ti{1}-clusters arise. Formation of \SiO{1}-clusters is not favourable in the considered temperature range but requires colder conditions. Note  that the above conclusions are drawn on the underlying assumption that the monomer exists (Sec.~\ref{sec:results_closed_ntw}).\\\\
The comprehensive chemical nucleation model yields different results from the closed nucleation ones. Firstly, it does not predict any \Al{1}-clusters, nor its monomer, nor its molecular precursors (\ch{Al2O2} and \ch{AlO2}) but most \ch{Al} remains atomic with maximally 1 per cent in \ch{Al}-bearing molecules which is mainly \ch{AlO}. Secondly, all available \ch{Mg} remains atomic and no \Mg{1}-clusters can exist. Hence, the most favoured nucleation candidates, according to the closed models, are non-existent. Only \Ti{1}-clusters exist in the comprehensive model, with similar formation conditions as in the closed model. \SiO{1}-clusters are again discarded due to their low formation temperature (Sec.~\ref{sec:resAllchem}).\\\\
The results from the comprehensive nucleation model suggest that \Ti{1} is the only possible AGB dust precursor of the considered nucleation candidates. However, this contradicts the substantial amount of \Al{1}-favouring evidence. Firstly, the number of \Al{1}-clusters found in pre-solar AGB grains far exceeds the amount of \Ti{1}-clusters. Secondly, numerous AGB dust observations indicate that dust already exists \rev{close to the star and thus} at temperatures as high as \SIrange{1500}{2000}{\K}, a regime in which, according to our model results, only \Al{1}-clusters can exist. \Ticlusters require temperatures below \SIrange{1000}{1200}{\K}. We believe that this discrepancy suggests that our current chemical reaction network is incomplete. \rev{Additionally, since there is experimental evidence that gaseous small \Al{1}-clusters can exists,} we believe that either the current reaction rate coefficients involving \ch{AlO}-bearing molecules are not accurate enough and need to be re-evaluated, or that alternative small \ch{Al2O3}-cluster formation pathways are missing. Moreover, most \ch{Al}-molecule formation rate coefficients are unavailable in the literature and rely on the assumption of detailed balance with their corresponding destruction process. We therefore urge the scientific community to investigate rate coefficients of formation reactions of \ch{Al}-bearing molecules at high temperatures. Without this data, it will remain unclear which species will form the initial dust precursors in AGB winds.\\\\
This paper has constructed and investigated an improved nucleation theory for more accurate modelling of the formation of dust. The improved description is time-dependent, allows growth by polymers, and considers quantum mechanical molecular properties. This procedure is universal and can be applied to any astrophysical environment, where this paper focuses on AGB winds. This work serves as a initial model \rev{which will be extended with macroscopic dust formation processes such as gas accretion, gas sputtering, dust coagulation, dust shattering, and dust evaporation in a future paper.} \old{to extend with macroscopic dust formation processes such as gas accretion, gas sputtering, dust coagulation and dust shattering.} It is the second in a series where we strive for increased self-consistency regarding chemistry, dust creation, and dynamics. The developed and improved chemical nucleation description can be incorporated into a hydrochemical model such as the first paper in this series \citep{Boulangier2019}. Currently, the results indicate which species, how much, how fast, and under which conditions they nucleate in an AGB wind.

\section*{Acknowledgements}
J.B., D.G., and L.D. acknowledge support from the ERC consolidator grant 646758 AEROSOL.
This research made use of Matplotlib \citep{Hunter2007}, NumPy \citep{Oliphant2006}, and Astropy \citep{Robitaille2013,TheAstropyCollaboration2018}, which are community-developed core Python packages for science and astronomy.



\bibliographystyle{mnras}
\bibliography{manual_ref_revision} 



\appendix


    \section{Equilibrium composition in the dilute limit}\label{app:minGFE}
    This section describes, step by step, how to determine the equilibrium composition of a gas mixture. This allows to determine the equilibrium ratio of two species, which in needed in equation \eqref{eq:kdestruction}. We focus on a nucleating system as this is the main purpose of this work. Because the number densities of nucleating molecules are small compared to the total gas number density, a nucleating system can be considered as a dilute solution where the bulk gas is the solvent and the nucleation molecules are the solutes. The Gibbs free energy of a pure solvent of $N_A$ particles $A$, is just $N_A$ times the chemical potential,
    \begin{equation}
        G = N_A \mu_A(T,P),
    \end{equation}
    where $\mu_A(T,P)$ is the chemical potential of the pure solvent, that is a function of temperature and pressure, $T$ and $P$. Imagine, adding a single $B$ particle to this system while holding the temperature and pressure fixed. This changes the Gibbs free energy by
    \begin{equation}
        \dd G = \dd U + P \dd V -T \dd S,
    \end{equation}
    where $U$ is the internal energy, $V$ the volume, and $S$ the entropy of the system. Note that $\dd U$ nor $P\dd V$ depend on $N_A$ but on how the $B$ particle interacts with its nearby neighbours, regardless of the total number of $A$ particles. $\dd S$ is partly independent of $N_A$, but part comes from the freedom of choosing where to put this $B$ particle. As this is proportional to the total number of $A$ particles, the entropy changes as
    \begin{equation}
        \dd S = k \ln N_A + (\text{terms independent of } N_A).
    \end{equation}
    We drop the $B$ subscript of the Boltzmann constant to avoid confusion with the $B$ particle.
    The total change in Gibbs free energy can then be written as
    \begin{equation}
        \dd G = G_B(T,P) - kT \ln N_A,
    \end{equation}
    where $G_B(T,P)$ is a function of temperature and pressure but independent of $N_A$. We shall call this the intrinsic Gibbs free energy of particle $B$. Generalising to adding $N_B$ particles results in a change
    \begin{equation}
        \dd G = N_B G_B(T,P) - N_BkT \ln N_A+kT \ln (N_B!)
    \end{equation}
    where the last term is introduced because all $B$ particles are identical and interchanging them does not result in a distinct state. Because $N_B \gg 1$, Stirling's approximation can be used to get rid of the factorial, leading to
    \begin{equation}
        \dd G = N_B G_B(T,P) - N_BkT \ln N_A+ N_B kT \ln N_B -N_B kT .
    \end{equation}
    Generalising this to adding $\mathcal{M}-1$ different particles (so that $\mathcal{M}$ includes the solvent particle), the total Gibbs free energy of the system is given by
    \begin{equation}\label{appA:totalGFE}
        G = N_A\mu_A(T,P) + \sum_{i=2}^{\mathcal{M}} N_i G_i(T,P) -N_ikT\ln N_A + N_ikT \ln N_i -N_i kT.
    \end{equation}
    Note that this expression is only valid in the limit $N_i \ll N_A$, that is when the solution is dilute. If not, then all $i$ particles would also interact with each other and the volume occupied by the particles will matter in the total Gibbs free energy determination \citep{Lepinoux2006}. In order to determine the equilibrium composition of the system, its Gibbs free energy (Eq.\,\ref{appA:totalGFE}) has to be minimised.\\\\
    In general, when optimising a multivariate function $f(x_1,\ldots,x_n)$ with $m$ number of constraints $g_k(x_1,\ldots,x_n) = 0$ with $k \in \{1,\ldots,m\}$, the Lagrangian that needs to be optimised (to each variable) takes the form
    \begin{equation}
        \mathcal{L}(x_1,\ldots,x_n,\lambda_1,\ldots,\lambda_m) = f(x_1,\ldots,x_n) - \sum_{k=1}^m\lambda_kg_k(x_1,\ldots,x_n),
    \end{equation}
    where each $\lambda_k$ is called a Lagrangian multiplier. Minimising the total Gibbs free energy of the nucleating system, eq. \eqref{appA:totalGFE}, can be achieved under the constraint that the total number of atoms in the system is constant
    \begin{equation}
        \sum_{i=1}^{\mathcal{M}}\sum_{j=1}^{\mathcal{A}}N_i x_{ij} = C,
    \end{equation}
    where $C$ is the total number of atoms, $\mathcal{A}$ is the number of different atoms, and $x_{ij}$ is the number of $j$ atoms in molecule $i$. Since this one constraint is sufficient, $g_k=g_1=g$ and $\lambda_k = \lambda_1 = \lambda$. Rewriting the constraint gives
    \begin{equation}
        g(N_1,\ldots,N_\mathcal{M}) = \sum_{i=1}^{\mathcal{M}}\sum_{j=1}^{\mathcal{A}}N_i x_{ij} - C = 0,
    \end{equation}
    with $N_1=N_A$ (the solvent). The Lagrangian of the system can then be written as
    \begin{equation}
        \mathcal{L}(N_1,\ldots,N_\mathcal{M}) = G(N_1,\ldots,N_\mathcal{M}) - \lambda \left(\sum_{i=1}^{\mathcal{M}}\sum_{j=1}^{\mathcal{A}}N_i x_{ij} - C\right).
    \end{equation}
    Minimising this Lagrangian to each variable leads to the set of $\mathcal{M}+1$ equations
    \begin{align}
        \frac{\partial\mathcal{L}}{\partial N_A} &= \mu_A(T,P) - \frac{kT}{N_A}\sum_{i=2}^{\mathcal{M}}N_i -\lambda\sum_{j=1}^\mathcal{A}x_{Aj} &= 0\\
        \frac{\partial\mathcal{L}}{\partial N_i} &= G_i(T,P) + kT \ln\left(\frac{N_i}{N_A}\right) - \lambda\sum_{j=1}^\mathcal{A}x_{ij} &= 0 \label{app:GFEminNi}\\
        \frac{\partial\mathcal{L}}{\partial \lambda} &= \sum_{i=1}^{\mathcal{M}}\sum_{j=1}^{\mathcal{A}}N_i x_{ij} - C &= 0
    \end{align}
    where eq. \eqref{app:GFEminNi} is valid for all $i \in \{2,\ldots,\mathcal{M}\}$. Solving this matrix will result in the equilibrium distribution of all molecules.\\\\
    For our purpose (making use of detailed balance, eq. \ref{eq:kdestruction}), we are interested in the ratio between molecules and by rewriting equation \eqref{app:GFEminNi} the number of solute molecules compared to the solvent is given by
    \begin{equation}\label{app:GFEminiNiNA}
        \frac{N_i}{N_A} = \exp\left(\frac{-G_i + \lambda\sum_j^\mathcal{A}x_{ij}  }{kT}\right).
    \end{equation}
    Note this represents the equilibrium values but we omit the "eq" superscript for clarity. As we consider a nucleating system, this equation can be simplified, because the number of atoms in a cluster scales linearly with the size of the cluster. Let $X$ be the number of atoms in the monomer, then for a cluster of size $n$:
    \begin{equation}
        \sum_j^\mathcal{A}x_{nj} = n\sum_j^\mathcal{A}x_{1j}=nX.
    \end{equation}
    Then, according to equation \eqref{app:GFEminiNiNA}, the number fraction of an $n$-sized cluster is given by
    \begin{equation}\label{app:GFEminiNiNAL}
        \frac{N_n}{N_A} = \exp\left(\frac{-G_n + \lambda n X  }{kT}\right).
    \end{equation}
    Consequently, the ratio of two different cluster sizes $n$ and $m$ is
    \begin{equation}
        \frac{N_n}{N_m} = \exp \left( \frac{-G_n + G_m + (n-m)\lambda X}{kT} \right).
    \end{equation}
    Introducing an $(n-m)$-sized cluster, with $n>m$, removes the $\lambda X$ term. I.e. using Eq. \eqref{app:GFEminiNiNAL}, one can write
    \begin{equation}
        \frac{N_{n-m}}{N_A} = \exp\left(\frac{-G_{n-m} + (n-m)\lambda X }{kT}\right),
    \end{equation}
    and hence
    \begin{equation}\label{app:(n-m)LambdaX}
        (n-m)\lambda X = kT\ln\left( \frac{N_{n-m}}{N_A} \right)  + G_{n-m}.
    \end{equation}
    Substitution Eq. \eqref{app:(n-m)LambdaX} into equation \eqref{app:GFEminiNiNAL}, results in the ratio,
    \begin{equation}\label{app:N_ratio}
        \frac{N_n}{N_m} = \frac{N_{n-m}}{N_A}\exp \left( \frac{-G_n + G_m + G_{n-m}}{kT} \right).
    \end{equation}\\\\
    Remember that each $G_i = G_i(T,P)$ is temperature and pressure dependent. For convenience these values are often calculated at a so-called standard pressure of $P\st=1 \si{\bar}$ $(= \SI{1e5}{\Pa} = \SI{1e6}{dyne\per\cm\squared})$. The superscript $\st$ refers to a quantity at this standard pressure. The Gibbs free energy of a particle at any pressure can be written as a function of the standard one,
    \begin{equation}\label{app:G_Gst}
        G = G\st -kT\ln\left(\frac{P\st}{P}\right),
    \end{equation}
    because only the translational partition function is a pressure dependent term (Eqs.\,\ref{app:Z_tr}--\ref{app:G_Z_1}),
    \begin{equation}
        Z_t = Z_t\st\frac{P\st}{P}.
    \end{equation}
    Using the standard Gibbs free energy and substituting equation \eqref{app:G_Gst} in equation \eqref{app:N_ratio}, the ratio of cluster sizes becomes,
    \begin{align}
        \frac{N_n}{N_m} &= \frac{N_{n-m}}{N_A}\exp \left( \frac{-G_n\st + G_m\st + G_{n-m}\st}{kT} \right)\frac{P}{P\st} \nonumber\\
                        &= N_{n-m}\frac{kT}{P\st V}\exp \left( \frac{-G_n\st + G_m\st + G_{n-m}\st}{kT} \right)
    \end{align}
    Hence, in equilibrium, the ratio of number densities of two clusters of sizes $N$ and $M$, with $N>M$, is described by,
    \begin{equation}\label{app:ratioEnd}
        \frac{n\eq_{N}}{n\eq_M} = n\eq_{N-M}\frac{kT}{P\st}\exp \left( \frac{-G_N\st + G_M\st + G_{N-M}\st}{kT} \right).
    \end{equation}

\section{Gibbs free energy}\label{app_sec:GFE}
    The Gibbs free energy of a system is defined as
    \begin{equation}\label{app:G}
        G = H -TS,
    \end{equation}
    where $H$ is the enthalpy, $S$ is the entropy, and $T$ is the temperature of the system. The enthalpy is defined as
    \begin{equation}\label{app:H}
        H = U + PV,
    \end{equation}
    where $U$ is the internal energy of the system, $P$ is the pressure of the system, and $V$ is the volume of the system.
    Both entropy and internal energy depend on the configurational freedom of the particles in the system. This configurational freedom or statistical properties of a particle is described by its partition function. When dealing with a system of $N$ non-interacting particles, the system's partition function is given by
    \begin{equation}\label{app:Z_N}
        Z_N = \frac{1}{N!}Z_1^N,
    \end{equation}
    where $Z_1$ is the partition function of a single particle.\\\\
    The entropy for a system consisting of $N$ particles is defined as
    \begin{align}\label{app:S_N}
        S_N &= \left.\frac{\pd kT\ln Z_N}{\pd T}\right|_{V,N} \nonumber\\
            &= k\ln Z_N + kT\left.\frac{\pd \ln Z_N}{\pd T}\right|_{V,N}
    \end{align}
    Substituting $Z_N$ using equation~\eqref{app:Z_N} yields,
    \begin{align}
        S_N &= Nk\ln Z_1 - k\ln(N!) + kT\left.\frac{\pd N\ln Z_1 - k\ln(N!)}{\pd T}\right|_{V,N} \nonumber\\
            &= Nk\ln Z_1 + NkT\left.\frac{\pd\ln Z_1}{\pd T}\right|_{V}-k\ln(N!) \label{app:S_N_Z-1__2}\\
            &=N S_1 -k\ln(N!) \nonumber \\
            &\approx N S_1 -Nk\ln N +k N,\nonumber
    \end{align}
    where the last transition uses Stirling's approximation which is valid for $N\gg1$. As this quantity is often calculated for one mole (\num{6.022140758e23} particles), this is a valid approximation.\\\\
    The internal energy of a system consisting of $N$ particles is defined as
    \begin{equation}
        U_N = \left.kT^2\frac{\pd\ln Z_N}{\pd T}\right|_{V,N}
    \end{equation}
    Again, substituting $Z_N$ with equation \eqref{app:Z_N}, this reduces to
    \begin{align}\label{app:U_N_or}
        U_N &= \left.NkT^2\frac{\pd\ln Z_1}{\pd T}\right|_{V}\nonumber\\
            &= N U_1.
    \end{align}
    Typically, the partition function is calculated with respect to the bottom of the particle's energy well (Sec.\,\ref{app:Z_e}) Therefore this energy value, $U_0$\footnote{\label{note:U_0}$U_0$ is the sum of the electronic ground state and nuclear-nuclear repulsion energies, isolated in vacuum, without vibration at \SI{0}{\K}.}, is separated from the partition function and equation \eqref{app:U_N_or} becomes,
    \begin{align}
        U_N &= \left.NkT^2\frac{\pd\ln Z_1}{\pd T}\right|_{V} + N U_0\label{app:U_N__2}\\
            &= N (U_1 + U_0).\nonumber
    \end{align}
    Substituting equations \eqref{app:H}, \eqref{app:S_N_Z-1__2}, and \eqref{app:U_N__2} into \eqref{app:G}, combined with the ideal gas law, yields the Gibbs free energy of a system of $N$ particles,
    \begin{equation}\label{app:G_Z_1}
        G_N = NU_0 - N k T \ln Z_1 + Nk T \ln N,
    \end{equation}
    which only depends on the total partition function of a single particle and $U_0$ of that particle.
    
    \subsection{Partition functions of one particle}
    According to the Born-Oppenheimer approximation rotational, vibrational, and electronic energies are independent of each other, and the partition function of one particle can be written as the product of separate contributors namely translational, rotational, vibrational, and electronic degrees of freedom, $Z_1=Z\tr Z\rot Z\vib Z\elec$. This section contains a summary of all different partition function for the most general case of a non-linear poly atomic ideal gas, a linear poly atomic ideal gas, and a mono atomic ideal gas.
        
        \subsubsection{Translation}
        
            The translational part is always given by
            \begin{align}\label{app:Z_tr}
                Z\tr &=\left(\frac{2\pi m k T}{h^2}\right)^{3/2}V \nonumber\\ 
                     &=\left(\frac{2\pi m k T}{h^2}\right)^{3/2}\frac{Nk T}{P},
            \end{align}
            where $m$ is the mass of the particle and $h$ is the Planck constant Note that $V$ is the volume of the embedding system meaning that $N$ is the total number of particles of the system in which this one particle resides.
            
        \subsubsection{Rotation}
        \begin{enumerate}[(I)]
            \item Non-linear poly atomic
            \begin{equation}
                Z\rot = \frac{1}{\sigma}\left(\frac{\pi T^3}{\Theta_x \Theta_y \Theta_z}\right)^{1/2},
            \end{equation}
            where $\sigma$ is the molecule's symmetry number\footnote{A molecule's symmetry number is the number of different but indistinguishable views of the molecule to correct for counting equivalent views.}, and $\Theta_i$ the rotational temperature related to the moments of inertia, $I_x, I_y, I_z$, via
            \begin{equation}
                \Theta_i = \frac{\hbar^2}{2 I_i k} \qquad i \in \{x,y,z\}
            \end{equation}
            \item Linear poly atomic
            \begin{equation}
                Z\rot = \frac{T\Theta\rot}{\sigma},
            \end{equation}
            where $\Theta\rot$ is the rotational temperature related to the moment of inertia, $I$ via
            \begin{equation}
                \Theta = \frac{\hbar^2}{2 I k}
            \end{equation}
            \item Mono atomic
            \begin{equation}
                Z\rot = 0
            \end{equation}
            \end{enumerate}
            Note that this is a high temperature approximation which is valid when the temperature is much larger than rotational temperature, which is the case in all our simulations. 
            
        \subsubsection{Vibration}
        A molecules consisting of $N$ atoms has $3N$ degrees of freedom, where the factor "3" corresponds to the possible movements of a particle in three-dimensional space. In the most general case, a molecule has $3N - 3 - 3=3N-6$ vibrational degrees of freedom where the "$-3$" terms are the translational and rotational degrees of freedom of the molecule. We choose the zero-energy reference point as the bottom of the potential well and not the vibrational ground state.
        \begin{enumerate}[(1)]
            \item Non-linear poly atomic
            \begin{equation}
                Z\vib = \prod_{\Theta_v \in \mathcal{T}_v} \frac{e^{-\Theta_v/2T}}{1-e^{-\Theta_v/T}},
            \end{equation}
            where $\Theta_v$ is the vibrational temperature related to a vibrational frequency $\nu$ of the molecule via
            \begin{equation}
                \Theta_v=\frac{h\nu}{k}
            \end{equation}
            when assuming that the vibrational modes of the molecule behave like harmonic oscillators. $\mathcal{T}_v$ is the set of all $3N-6$ vibrational modes of the molecule.\\
            \item Linear poly atomic
            \begin{equation}
                Z\vib = \prod_{\Theta_v \in \mathcal{T}_v} \frac{e^{-\Theta_v/2T}}{1-e^{-\Theta_v/T}}.
            \end{equation}
            Note that $\mathcal{T}_v$ only contains $3N-5$ vibrational modes due to a rotational symmetry of the molecule.\\
            \item Mono atomic
            \begin{equation}
                Z\vib = 0
            \end{equation}
            \end{enumerate}
        \subsubsection{Electronic}\label{app:Z_e}
        The electronic part is always given by
        \begin{equation}
            Z\elec = \sum_{i=0}^{N_e} g_i e^{-\epsilon_i/kT}
        \end{equation}
        with $\epsilon_i$ the $i$th electronic energy level w.r.t. the bottom of the electronic potential well, $g_i$ the degeneracy of the $i$th level due to spin splitting and $N_e$ the number of energy levels. Each energy level can be scaled by choosing the bottom of the well to be 0\footnote{This energy difference should be added again in the total internal energy of the molecule (Eq.\,\ref{app:U_N__2}.)}, giving $\varepsilon_i=\epsilon_i-\epsilon_0$. The number of levels can also be limited to the one where $\varepsilon_{N_{\text{lim}}} \gg kT$.
        \begin{equation}
            Z\elec = g_0 + \sum_{i=1}^{N_{\text{lim}}} g_i e^{-\varepsilon_i/kT}
        \end{equation}

\section{Gibbs free energy of formation}\label{app_sec:GFEoF}
Generally, the standard Gibbs free energy of formation (GFEoF), rather than the Gibbs free energy (GFE), is used to determine reversed reaction rate coefficients under the assumption of detailed balance. Although both can be used, we opt for GFE for reason explained in the main text (Sec.\,\ref{sec:quantum_chem}) but explain GFEoF for completeness and comparison. The GFEoF of a compound is the change in GFE that occurs when one mole of the compound is formed from its component elements in their most thermodynamically stable states under standard conditions (pressure of 1 bar = \SI{1e5}{\Pa}). Note that this state, depending on the components can be gaseous, solid, or liquid.\\\\
Consider a molecule $m$ consisting of $N$ unique atoms with each atom $a$ occurring $v_a$ times in the molecule. Then, the set with unique atoms is defined as $\mathcal{A} = \{a_1,a_2, \ldots, a_N\}$. For an example molecule $m=\ch{H2O}$, this gives $N=2$, $\mathcal{A} = \{H, O\}$, $v_H = 2$, and $v_O=1$. Following the documentation of \gaussian \citep{Frisch2013}, the standard GFEoF of molecule $m$ at a given temperature $T$,  $\xform{G}{T}{m}$, is described by
\begin{equation}\label{app:GFEf_T}
    \xform{G}{T}{m} = \xform{H}{T}{m} - T\left(\xquan{S}{T}{m} - \sum_{a\in \mathcal{A}} v_a \xquan{S}{T}{a} \right),
\end{equation}
where $\xform{H}{T}{m}$ is the standard enthalpy of formation\footnote{The standard enthalpy of formation of a compound is the change of enthalpy during the formation of one mole of that substance from its constituent elements, with all substances in their standard states. For an atom, this is the standard enthalpy of phase transition w.r.t. the phase in its standard state, i.e. the energy that must be supplied as heat at constant pressure per mole to convert from one phase to the other.} of molecule $m$ at a given temperature, $\xquan{S}{T}{m}$ and $\xquan{S}{T}{a}$ are the entropy at a given temperature of molecule $m$ and atom $a$, respectively. The $\st$ notation refers to the quantity at standard pressure of 1 bar (= \SI{1e5}{\Pa}). The standard enthalpy of formation of molecule $m$ at temperature $T$ is described by
\begin{equation}\label{app:Hf_T}
    \xform{H}{T}{m} = \xform{H}{0}{m} + \xquan{H}{T}{m} - \xquan{H}{0}{m} - \sum_{a\in \mathcal{A}} v_a \left(\xquan{H}{T}{a} - \xquan{H}{0}{a}\right),
\end{equation}
where $H\st_T$ denotes the standard (thermal) enthalpy (Eq.\,\ref{app:H}) which excludes the electronic potential energy $U_0$\textsuperscript{\ref{note:U_0}} of the species. The standard enthalpy of formation of a molecule at absolute zero is given by
\begin{equation}\label{app:Hf_0}
    \xform{H}{0}{m} = U_{0,m} + U_{\text{zpve},m} - \sum_{a\in \mathcal{A}} v_a \left(U_{0,a} - \xform{H}{0}{a} \right),
\end{equation}
where $U_{\text{zpve},m}$ is the zero point vibration energy of a molecule, which is the lowest vibrational energy (ground state) at \SI{0}{\K}. Note that this is not the bottom of the vibrational potential well (when representing this as harmonic oscillator potential). Combining equations \ref{app:GFEf_T}, \ref{app:Hf_T}, and \ref{app:Hf_0}, and rearranging some terms, the standard GFEoF is given by
\begin{align}
    \xform{G}{T}{m} &= \xquan{H}{T}{m} -\xquan{H}{0}{m} + U_{0,m} + U_{\text{zpve},m} - T\xquan{S}{T}{m}\nonumber\\ 
                    &- \sum_{a\in \mathcal{A}}v_a \left( \xquan{H}{T}{a} -\xquan{H}{0}{a} + U_{0,a} - \xform{H}{0}{a} - T\xquan{S}{T}{a}\right).
\end{align}
When realising that $\xquan{H}{0}{m} = U_{\text{zpve},m}$ for a molecule and $\xquan{H}{0}{a} = 0$ for an atom, the standard GFEoF reduces to
\begin{align}
    \xform{G}{T}{m} &= \xquan{H}{T}{m} + U_{0,m} - T\xquan{S}{T}{m}\nonumber\\ 
    &- \sum_{a\in \mathcal{A}}v_a \left( \xquan{H}{T}{a} + U_{0,a} - \xform{H}{0}{a} - T\xquan{S}{T}{a}\right).
\end{align}
\section{Quantum mechanical data}\label{app:quantum_data}
This section contains an overview of all quantum mechanical data that was collected and calculated (Table~\ref{tab:cluster_info} for the nucleation species and Table~\ref{tab:quantum_data} for all other species). All gathered data has been homogenised and is available as a JSON file. A collection of used literature input files (raw and cleaned versions), reference files, and info files is also available online\footnote{Zenodo: \zenodo}. All this data was used to calculate Gibbs free energies, which are also available online for the temperature range on our interest \SIrange{500}{3000}{\K} at standard pressure of \SI{1}{\bar}, which also have been included in KROME. Adaptations of these tables can easily be produced with our open-source repository\footnote{\repothermo} and the provided data.

    \begin{table}
        \centering
        \caption{Nucleation cluster specifications. All quantum mechanical properties of these clusters are calculated in this work ($U_0$, $Z_1/Z\tr$, $\Theta\rot, \Theta\vib$). }
        \label{tab:cluster_info}
        \begin{tabular}{llll}
            Cluster &  Sizes & Global minimum & $r_\text{monomer}$ (nm)\\ \hline
            \Ti{1}  & 1-10 & \citet{Lamiel-Garcia2017} & 0.162$^a$\\
            \SiO{1} & 1-10 & \citet{Bromley2016} & 0.075765$^b$\\
            \Mg{1} & 1-10 & \citet{Chen2014} & 0.0865$^c$\\
            \Al{1} & 1-7 & \citet{Li2012} & 0.3304$^d$\\
                    & 8 & \citet{Gobrecht2018} & \\\hline
            \multicolumn{4}{p{\columnwidth}}{\textbf{Notes:} a) Inter atomic \ch{Ti-O} distance from \citet{Jeong2000}. b) Half a \ch{Si-O} bond length from \citet{Bromley2016}. c) Half a \ch{Mg-O} bond length from \citet{Farrow2014}. d) Inter atomic distance \ch{O-Al-O} (linear geometry) from \citet{Archibong1999}. All used monomer radii can be more accurate by accounting for the geometry of the non-linear molecules and using our re-evaluated structures (Sec.\,\ref{sec:limKNT}).}
        \end{tabular}
    \end{table}

    \begin{table*}
        \centering
        \caption{Overview of the sources of all quantum mechanical data, either gathered or calculated, as defined in Appendix~\ref{app_sec:GFE}.}
        \label{tab:quantum_data}
        \begin{tabular}{llllll}
            Species & Global minimum & $U_0$ & $Z_1/Z\tr$ & $\Theta\rot, \Theta\vib$ & $\varepsilon_i$\\ \hline
            \ch{TiO}    & -  & CCCBDB & \citet{Kurucz1992}$^a$  & - & \citet{Phillips1971}$^c$  \\
            \ch{CO2}    & -  & CCCBDB & \citet{Rothman2010}$^a$  & - & \citet{Herzberg1966}$^e$  \\
            \ch{OH}    & -  & CCCBDB & \citet{Rothman2010}$^a$  & - & \citet{Huber1979}$^e$  \\
            \ch{AlO}    & -  & CCCBDB & \citet{Patrascu2015}$^a$  & - & -  \\
            \ch{AlH}    & -  & CCCBDB & \citet{Yurchenko2018}$^a$  & - & -  \\
            \ch{NO}    & -  & CCCBDB & \citet{Wong2017}$^a$  & - & \citet{Huber1979}$^e$  \\
            \ch{CO}    & -  & CCCBDB & \citet{Li2015}$^a$  & - & -  \\
            \ch{SO}    & -  & CCCBDB & \citet{Gamache2017}$^b$  & - & \qmark$^c$  \\
            \ch{SO2}    & -  & CCCBDB & \citet{Gamache2017}$^b$  & - & \citet{Herzberg1966}$^e$  \\
            \ch{HO2}    & -  & CCCBDB & \citet{Gamache2017}$^b$  & - & -  \\
            \ch{H2O2}    & -  & CCCBDB & \citet{Gamache2017}$^b$  & - & -  \\
            \ch{O2}    & -  & CCCBDB & \citet{Gamache2017}$^b$  & - & \citet{Huber1979}$^e$  \\
            \ch{N2}    & -  & CCCBDB & \citet{Gamache2017}$^b$  & - & -  \\
            \ch{N2O}    & -  & CCCBDB & \citet{Gamache2017}$^b$  & - & \citet{Herzberg1966}$^e$  \\
            \ch{NO2}    & -  & CCCBDB & \citet{Gamache2017}$^b$  & - & \qmark$^e$  \\
            \ch{H2O}    &  -  & CCCBDB  & \citet{Furtenbacher2016}$^f$  & -  & - \\
            \ch{H2}    & -  & CCCBDB & \citet{Popovas2016}  & - & \citet{Huber1979}$^e$ \\
            \ch{AlC} &  - & CCCBDB  &  \checkmark  &  CCCBDB & - \\
            \ch{AlH2} &  - & CCCBDB  &  \checkmark  &  CCCBDB & - \\
            \ch{AlH3} &  - & CCCBDB  &  \checkmark  &  CCCBDB & - \\
            \ch{HCO} &  - & CCCBDB  &  \checkmark  &  CCCBDB & \citet{Johns1963}$^{e,g}$ \\
            \ch{HO2} &  - & CCCBDB  &  \checkmark  &  CCCBDB & \citet{Becker1978}$^{e,h}$\\
            \ch{MgO} &  - & CCCBDB  &  \checkmark  &  CCCBDB & \citet{Bauschlicher2017}$^{i}$, \citet{Huber1979}$^e$\\
            \ch{MgOH} &  - & CCCBDB  &  \checkmark  &  CCCBDB & - \\
            \ch{Mg(OH)2} &  - & CCCBDB  &  \checkmark  &  CCCBDB & - \\
            \ch{MgCO3} &  - & CCCBDB  &  \checkmark  &  CCCBDB & - \\
            \ch{O3} &  - & CCCBDB  &  \checkmark  &  CCCBDB & - \\
            \rev{\ch{SiO2}} &  - & CCCBDB  &  \checkmark  &  CCCBDB & - \\
            \ch{AlO2}    & \citet{Patzer2005}   & \checkmark  &  \checkmark &  \checkmark & - \\
            \ch{Al2O}    & \citet{Patzer2005}   & \checkmark  &  \checkmark &  \checkmark & - \\
            \ch{Al2O2}    & \citet{Patzer2005}   & \checkmark  &  \checkmark &  \checkmark & - \\
            \ch{AlOH}    & \checkmark   & \checkmark  &  \checkmark &  \checkmark & - \\
            \ch{AlO2H}    & \checkmark   & \checkmark  &  \checkmark &  \checkmark & - \\
            \ch{Al(OH)2}    & \checkmark   & \checkmark  &  \checkmark &  \checkmark & - \\
            \ch{Al(OH)3}    & \checkmark   & \checkmark  &  \checkmark &  \checkmark & - \\
            \ch{H}    &  \xmark & CCCBDB  &  \xmark &  \xmark & - \\
            \ch{C}    &  \xmark & CCCBDB  &  \xmark &  \xmark & \citet{Haris2017,Beckmann1975}$^d$\\
            \ch{Mg}    & \xmark  & \checkmark  &  \xmark &  \xmark & - \\
            \ch{N}    &  \xmark &  \checkmark  &  \xmark &  \xmark & - \\
            \ch{O}    & \xmark  & \checkmark  &  \xmark &  \xmark &  \citet{Moore1993}$^d$\\
            \ch{Si}    & \xmark  & \checkmark  &  \xmark &  \xmark  & \citet{Martin1983}$^d$\\
            \ch{Al}    & \xmark  & \checkmark &  \xmark &  \xmark & \citet{Martin1979}$^d$\\
            \ch{Ti}    & \xmark  & \checkmark & \xmark &  \xmark & \citet{Saloman2012}$^d$\\
             &   &   &   &   & \\
            \multicolumn{6}{p{2\columnwidth}}{\textbf{Legend}: \checkmark: This work, \xmark: Not applicable, -: Unnecessary, \qmark: No references provided, CCCBDB: NIST Computational Chemistry Comparison and Benchmark Database \citep{Johnson2018}.}\\ \hline
            \multicolumn{6}{p{2\columnwidth}}{ a) via ExoMol (\url{http://exomol.com/}). b) via HITRAN \citep{Gordon2017}. c) via NIST chemistry WebBook (\url{https://doi.org/10.18434/T4D303}). d) via NIST Atomic Spectra Database \citep{Kramida2018}. e) via CCCBDB \citep{Johnson2018} f) Uses $g=1$ and $g=3$ as para-ortho degeneracy which is preferred over using $g=1/4$ and $g=3/4$ like \citet{Vidler2000}. g) The most likely reference of list of the references provided by NIST chemistry WebBook (\url{https://doi.org/10.18434/T4D303}). h) Unclear reference for the second energy level. i) First level: improved theoretical value over the theoretical one of \citet{Huber1979}. }\\
        \end{tabular}
    \end{table*}

\newpage
\section{Results}\label{app:results}
    This appendix encompasses additional figures of the nucleation models.
    Figures which are not shown in this appendix are either already present in the main body or provide
    no added value.
    
    \subsection{Closed nucleation networks}\label{app:res_closed_ntw}
    This section contains a more thorough overview of all closed nucleation models of all nucleation clusters results.
    
        \subsubsection{Monomer nucleation}
\begin{figure*}
        \begin{flushleft}
        \includegraphics[width=0.32\textwidth]{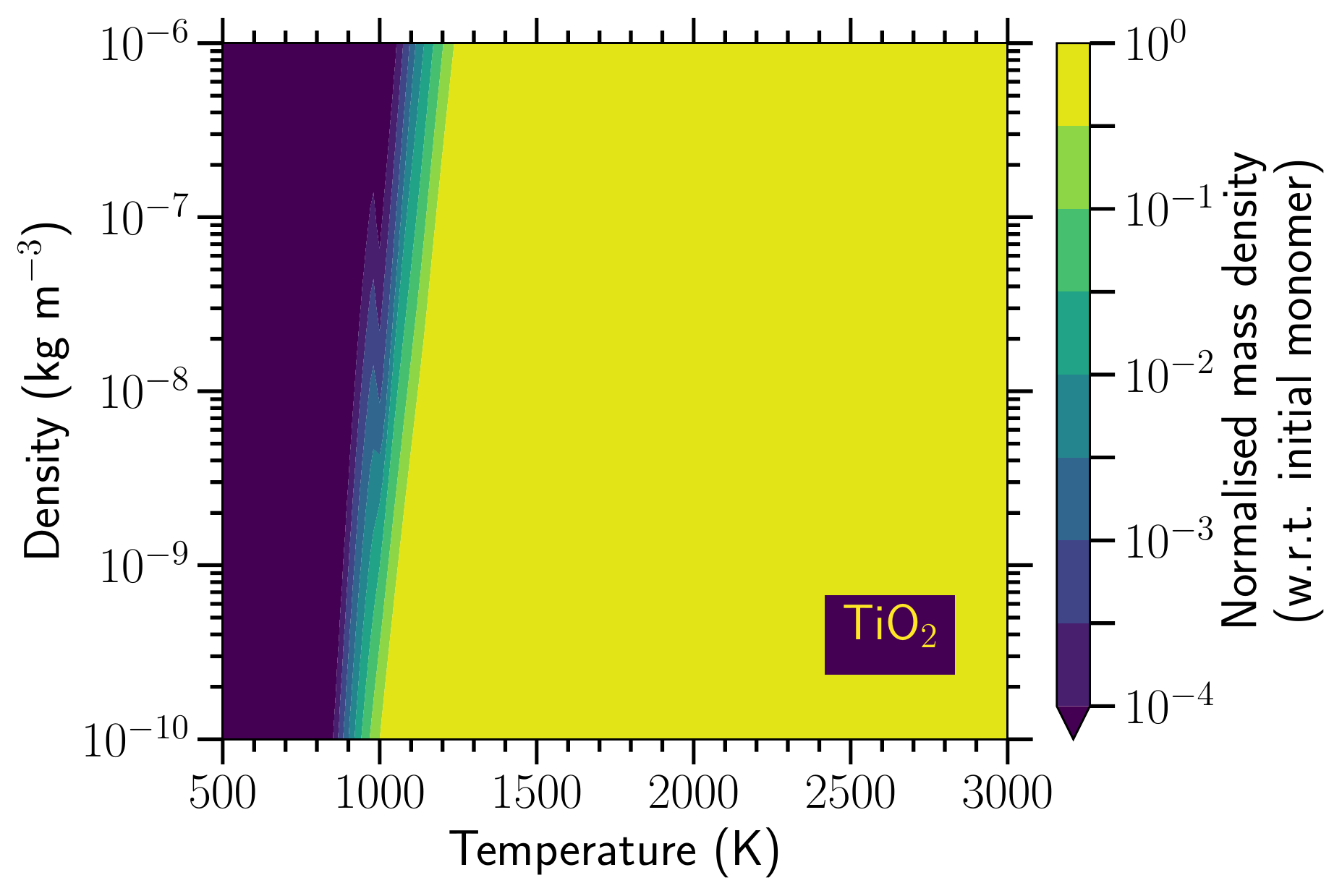}
        \includegraphics[width=0.32\textwidth]{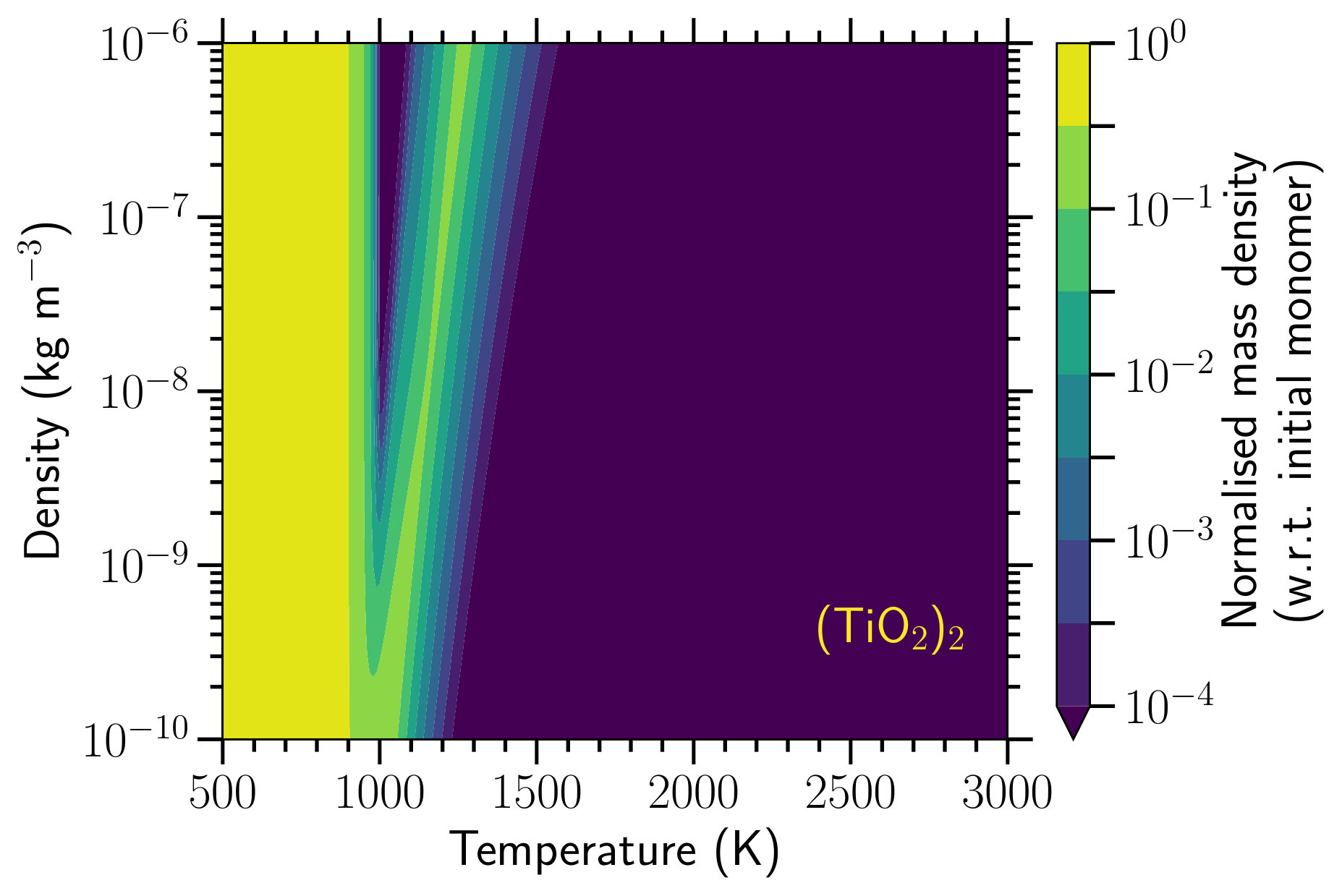}
        \includegraphics[width=0.32\textwidth]{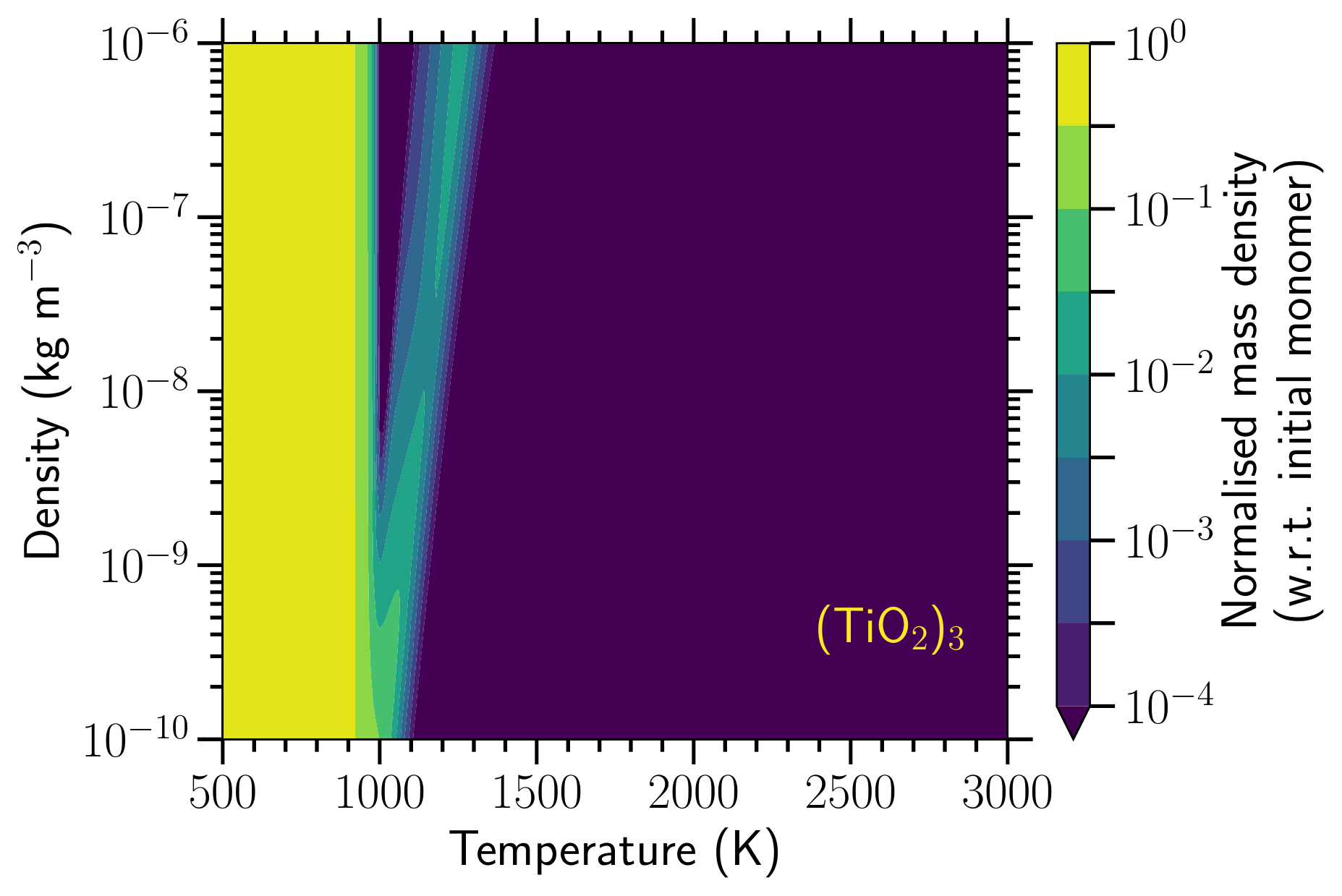}
        \includegraphics[width=0.32\textwidth]{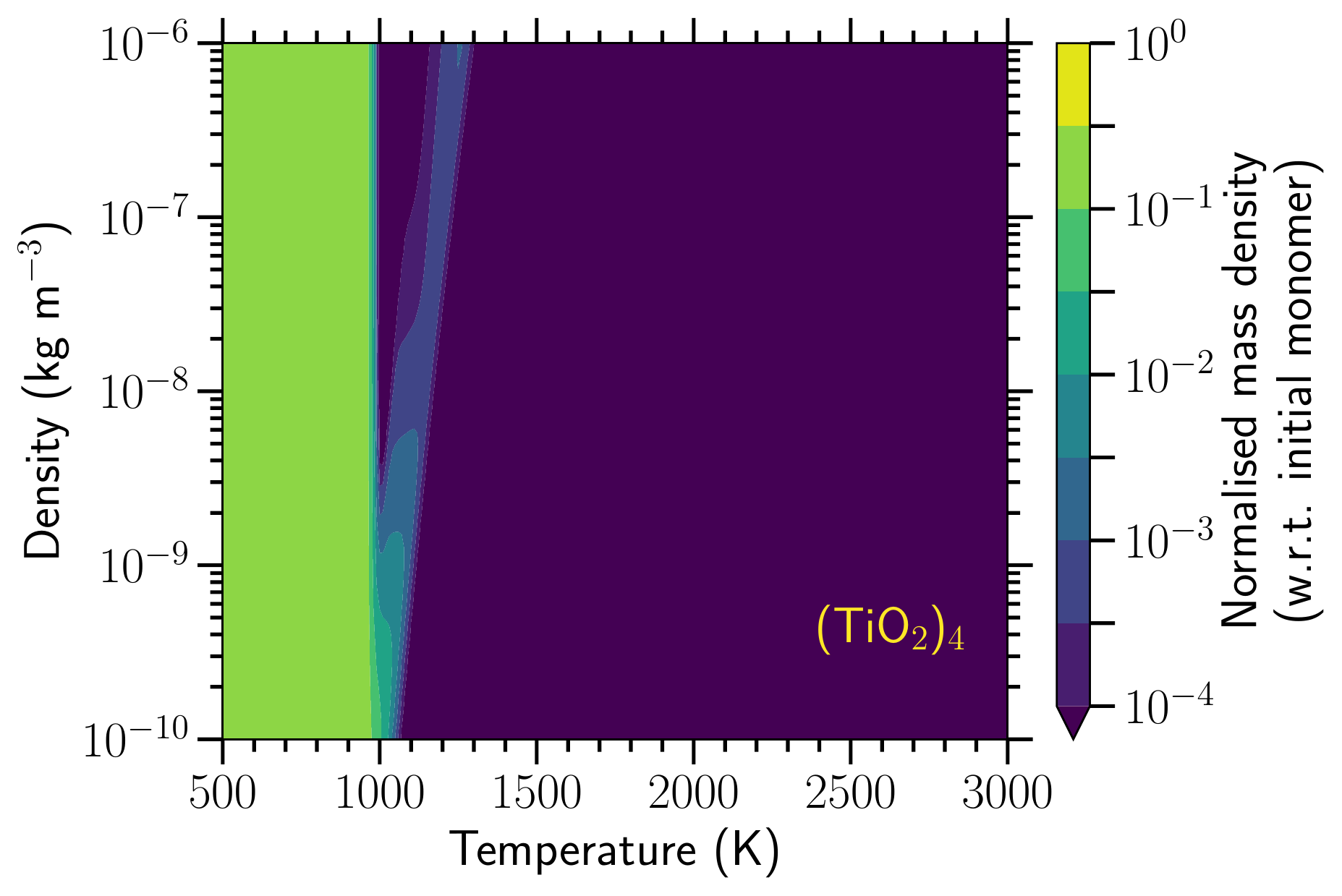}
        \includegraphics[width=0.32\textwidth]{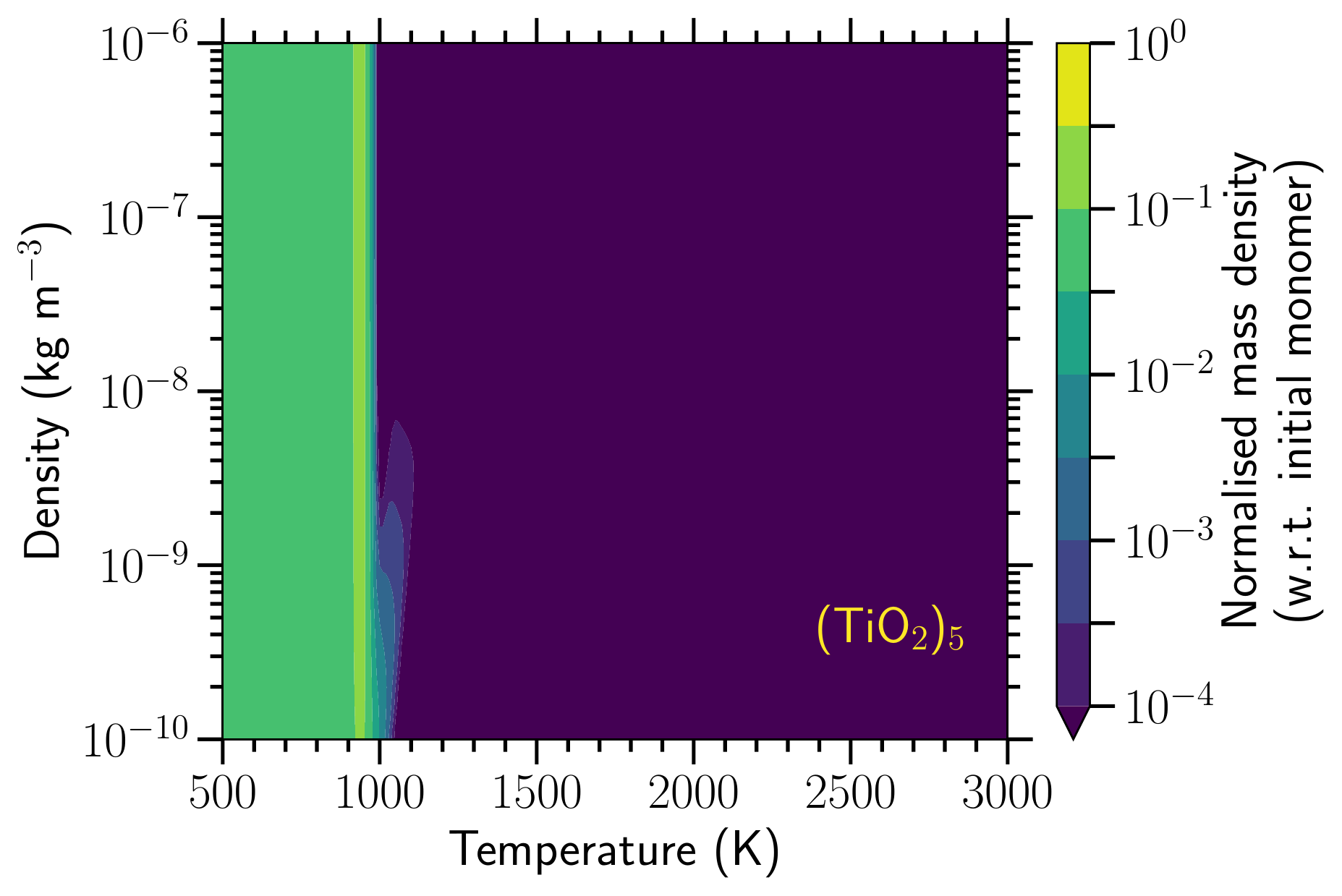}
        \includegraphics[width=0.32\textwidth]{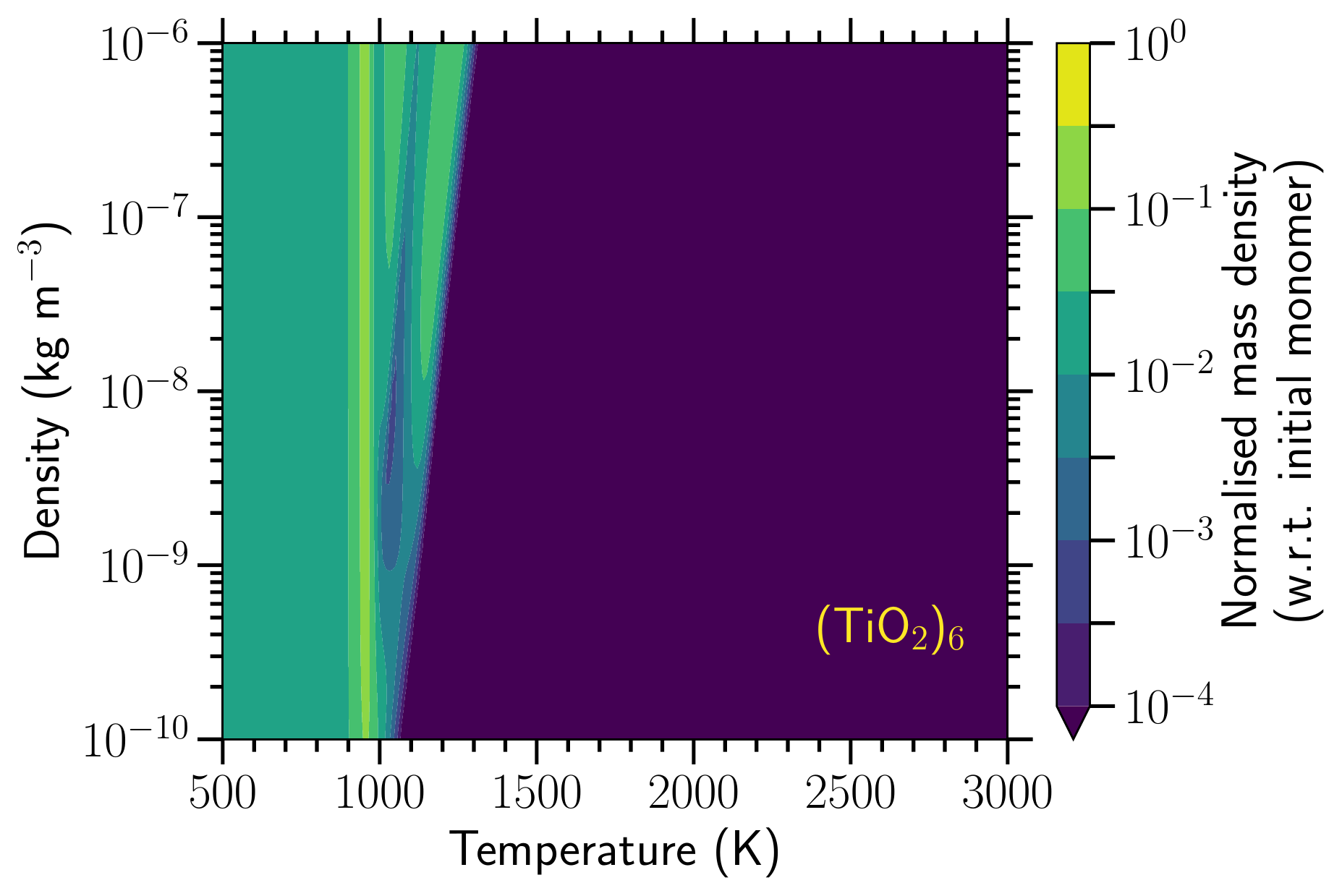}
        \includegraphics[width=0.32\textwidth]{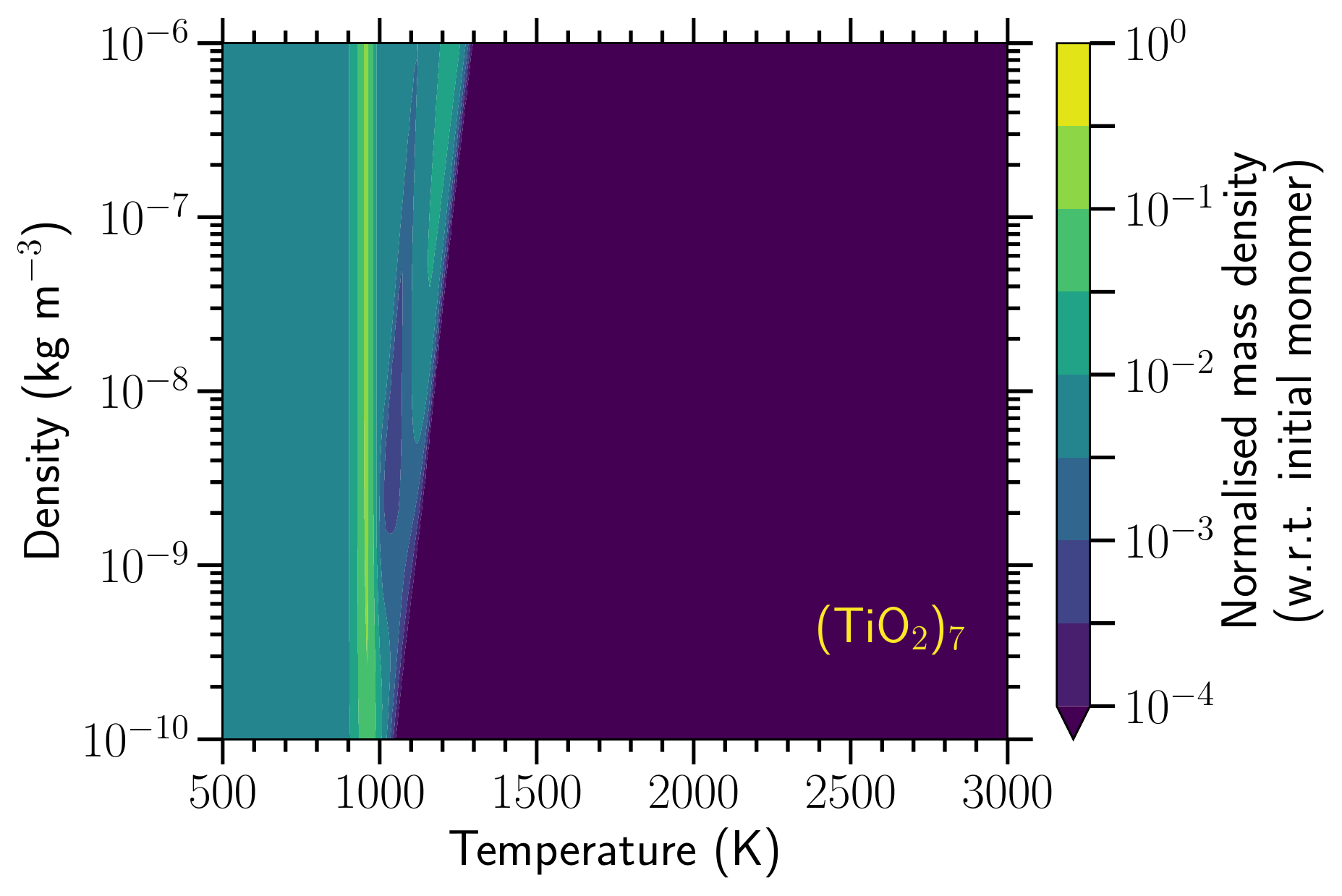}
        \includegraphics[width=0.32\textwidth]{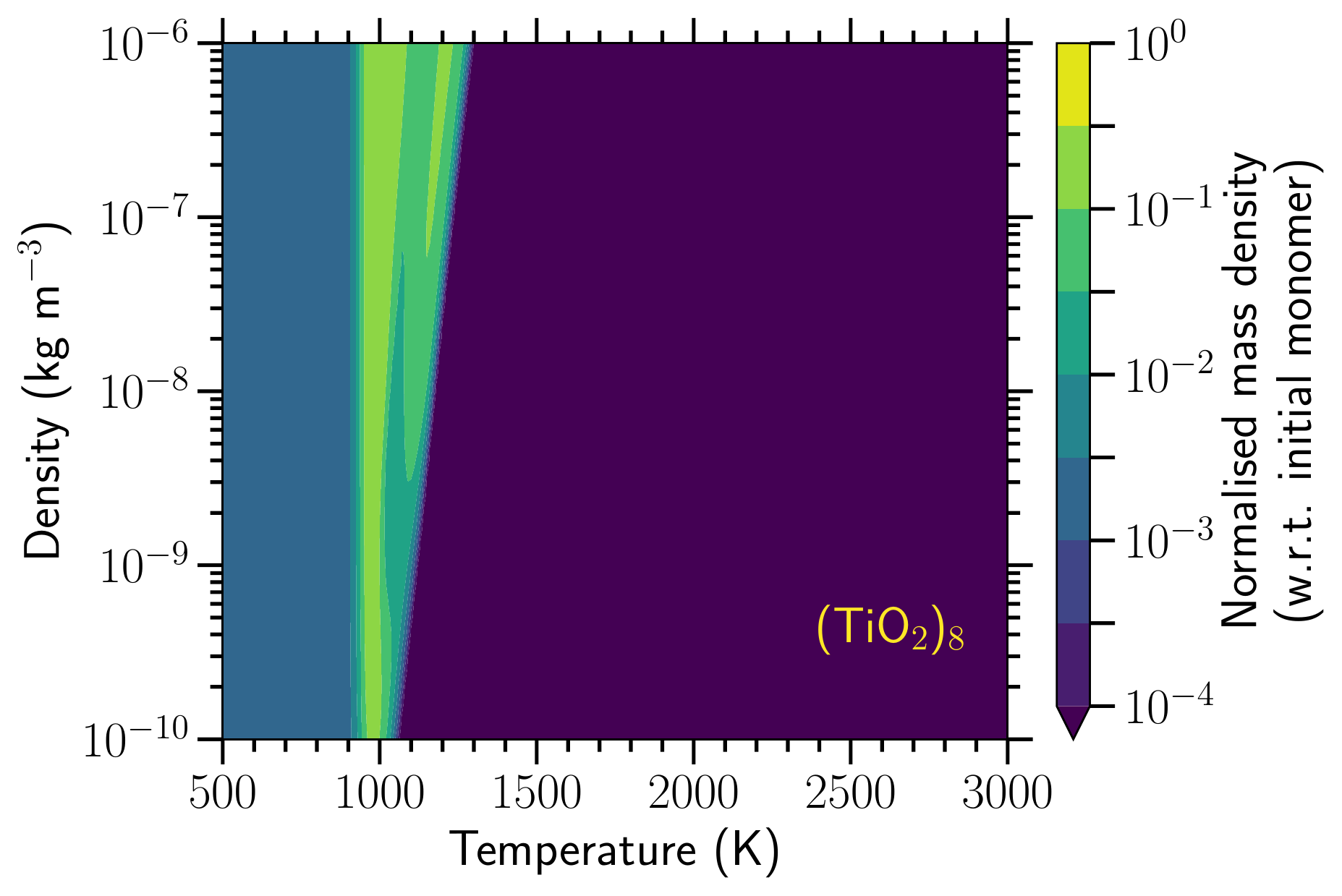}
        \includegraphics[width=0.32\textwidth]{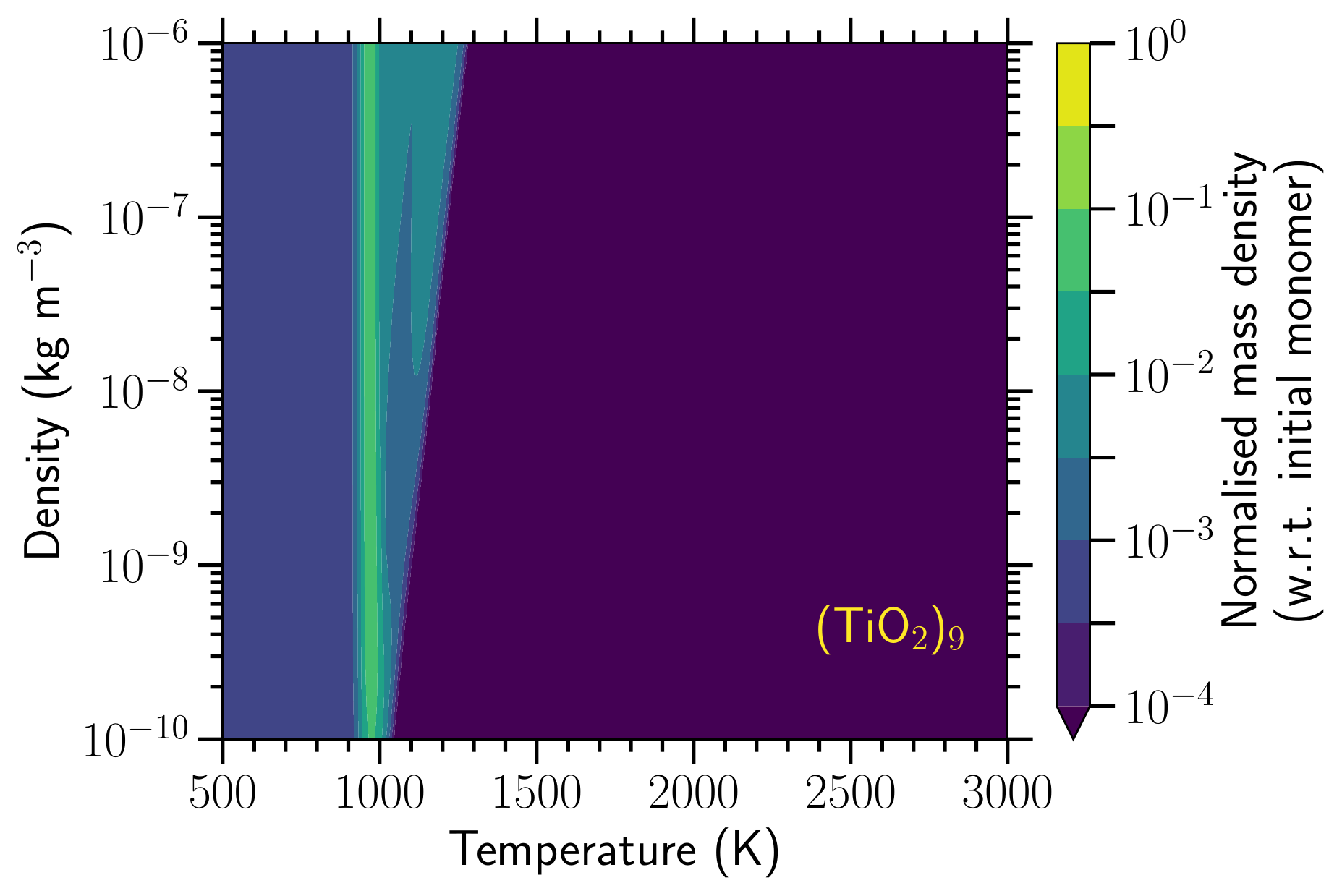}
        \includegraphics[width=0.32\textwidth]{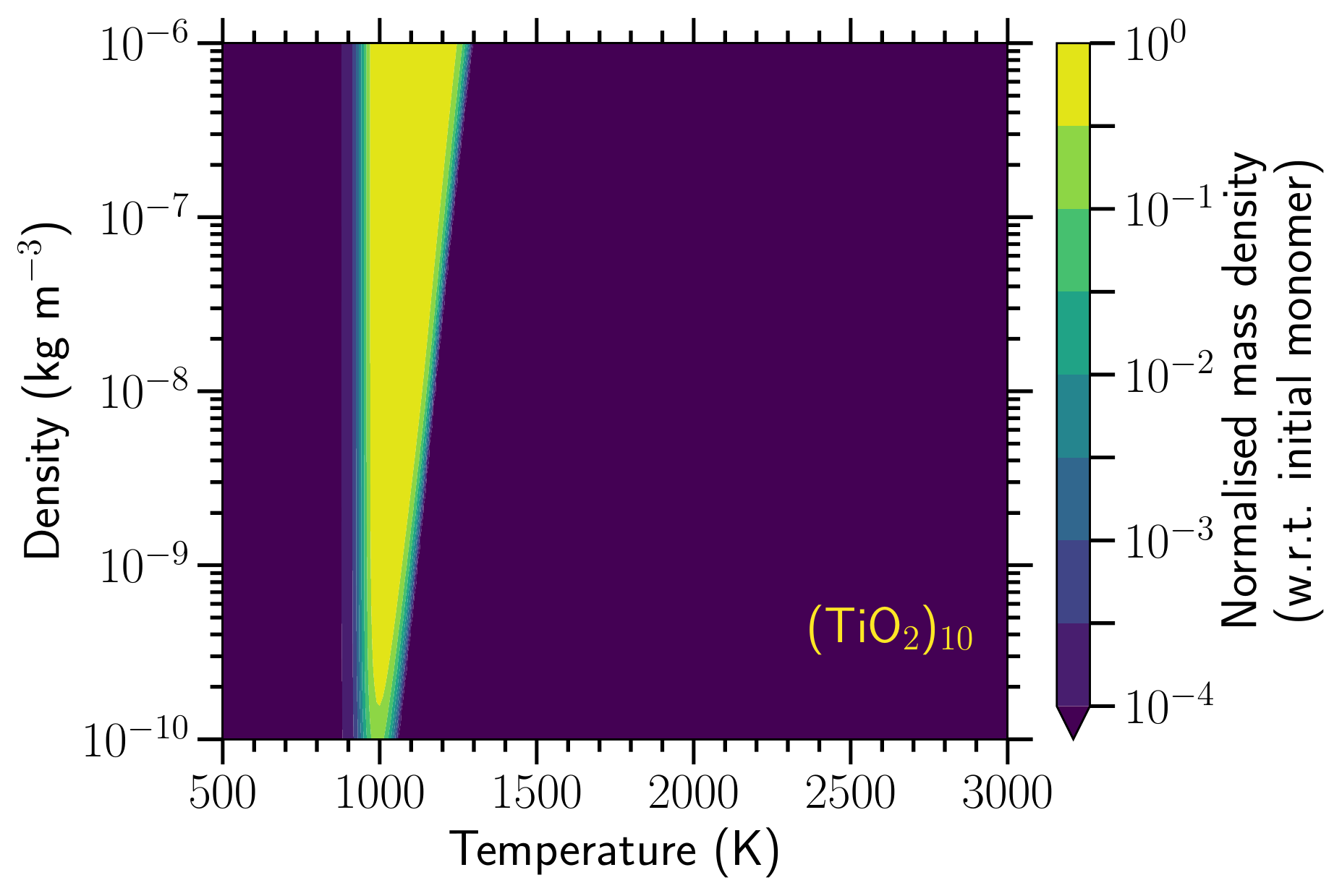}
        \end{flushleft}
        \caption{Overview of the normalised mass density after one year of all \protect\Ti{1}-clusters for a closed nucleation model using the monomer nucleation description.}
        \label{fig:TiO2_clusters_monomer_norm_same_scale}
    \end{figure*}

\begin{figure*}
        \begin{flushleft}
        \includegraphics[width=0.32\textwidth]{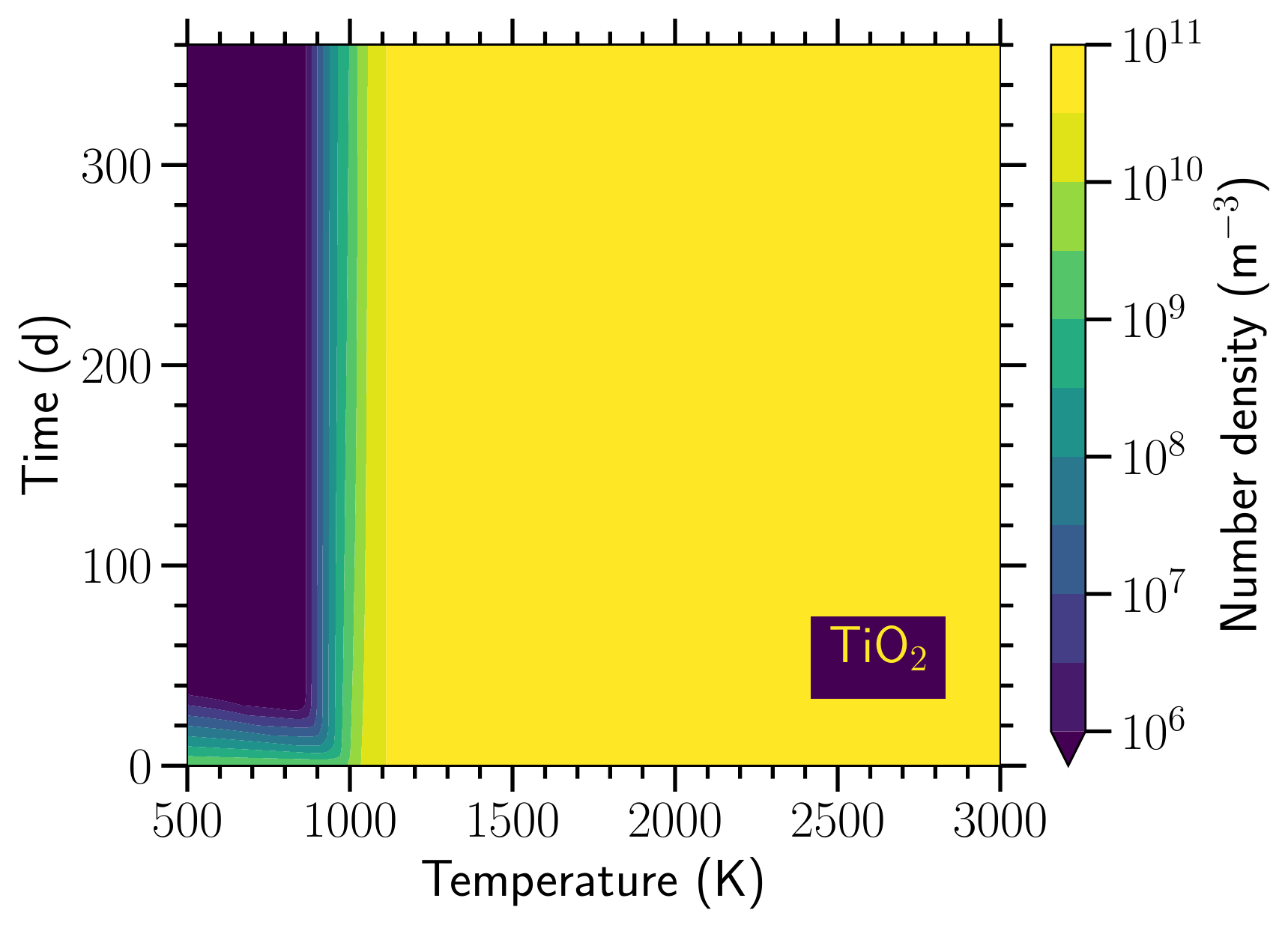}
        \includegraphics[width=0.32\textwidth]{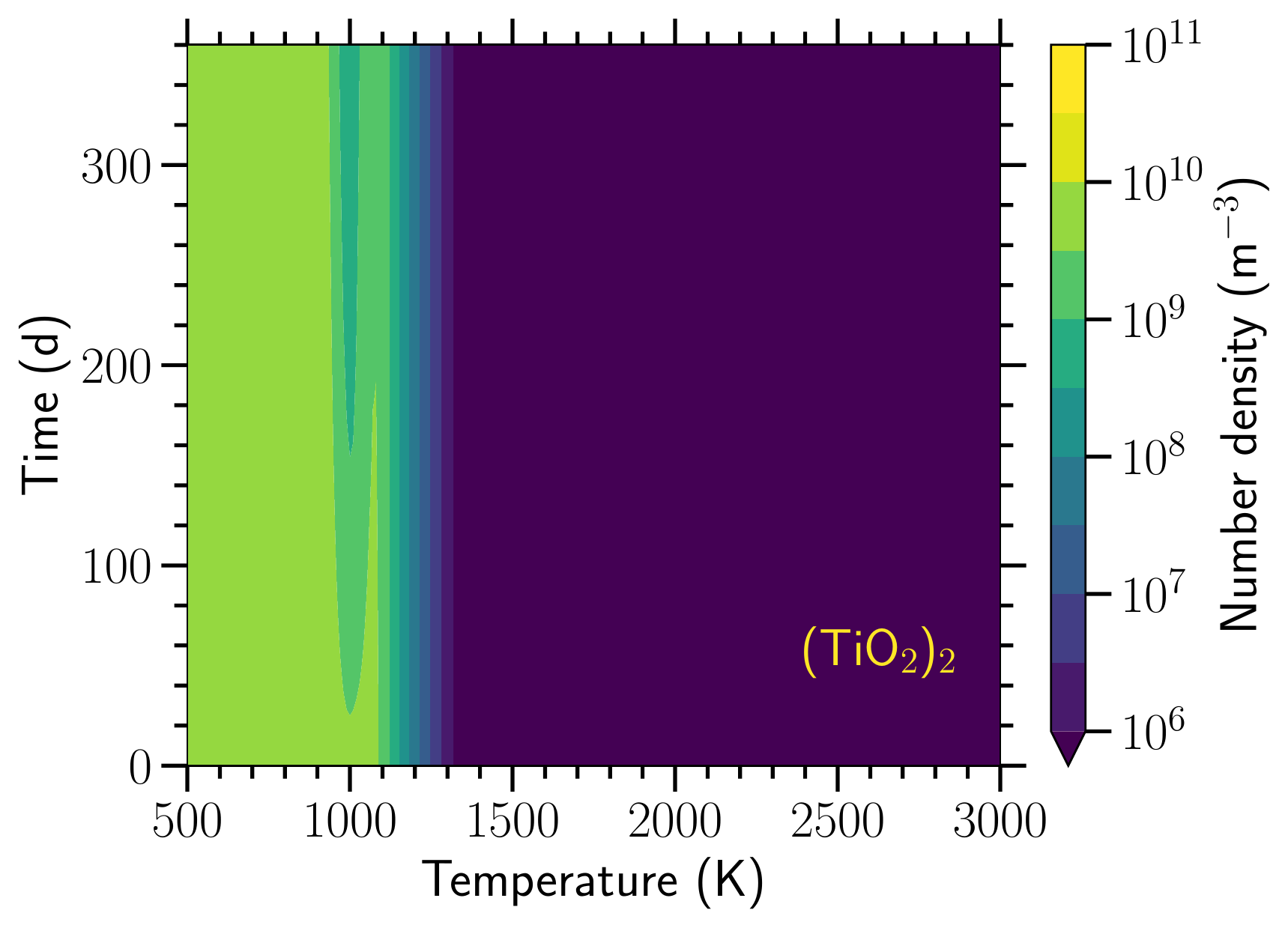}
        \includegraphics[width=0.32\textwidth]{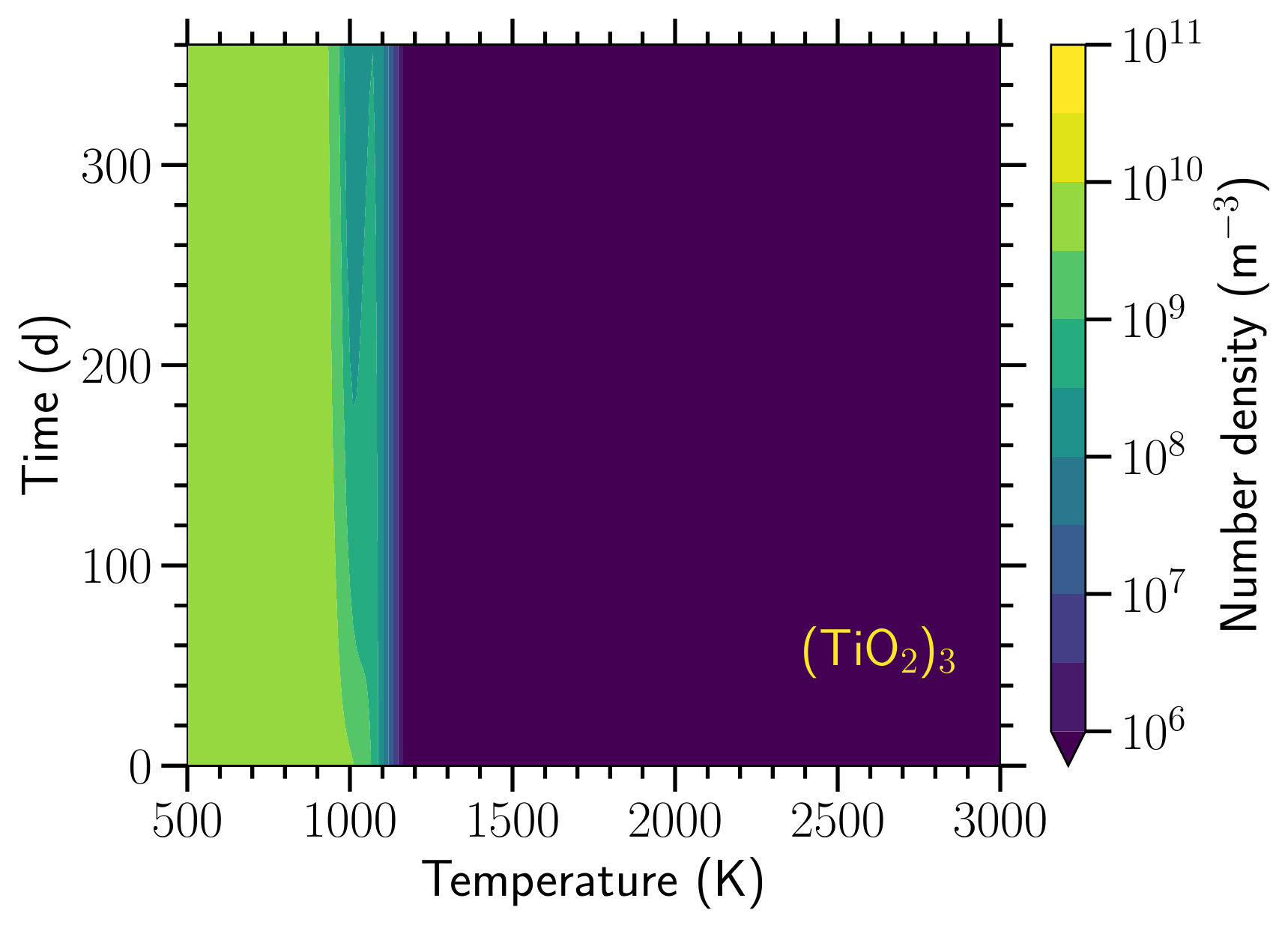}
        \includegraphics[width=0.32\textwidth]{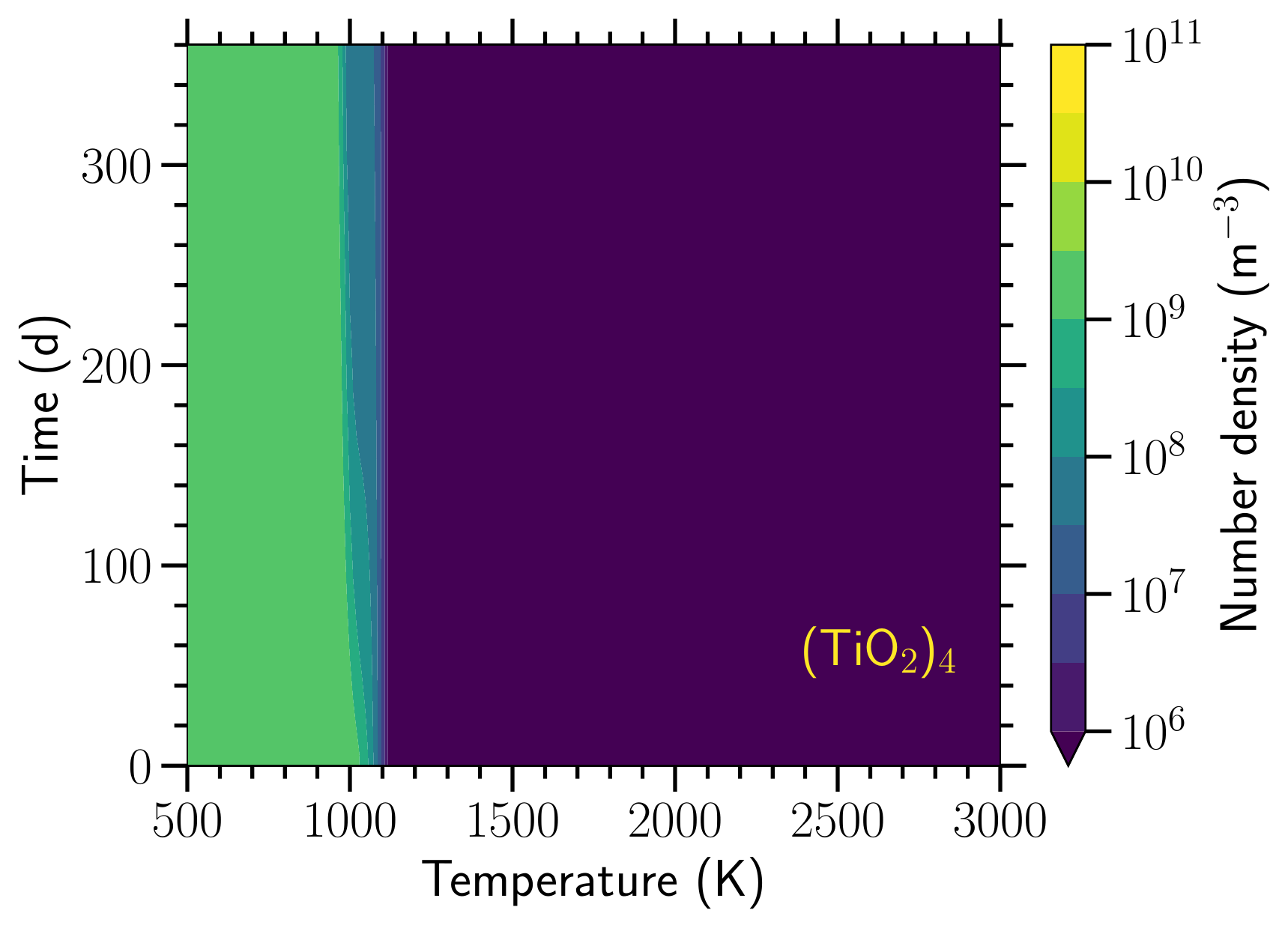}
        \includegraphics[width=0.32\textwidth]{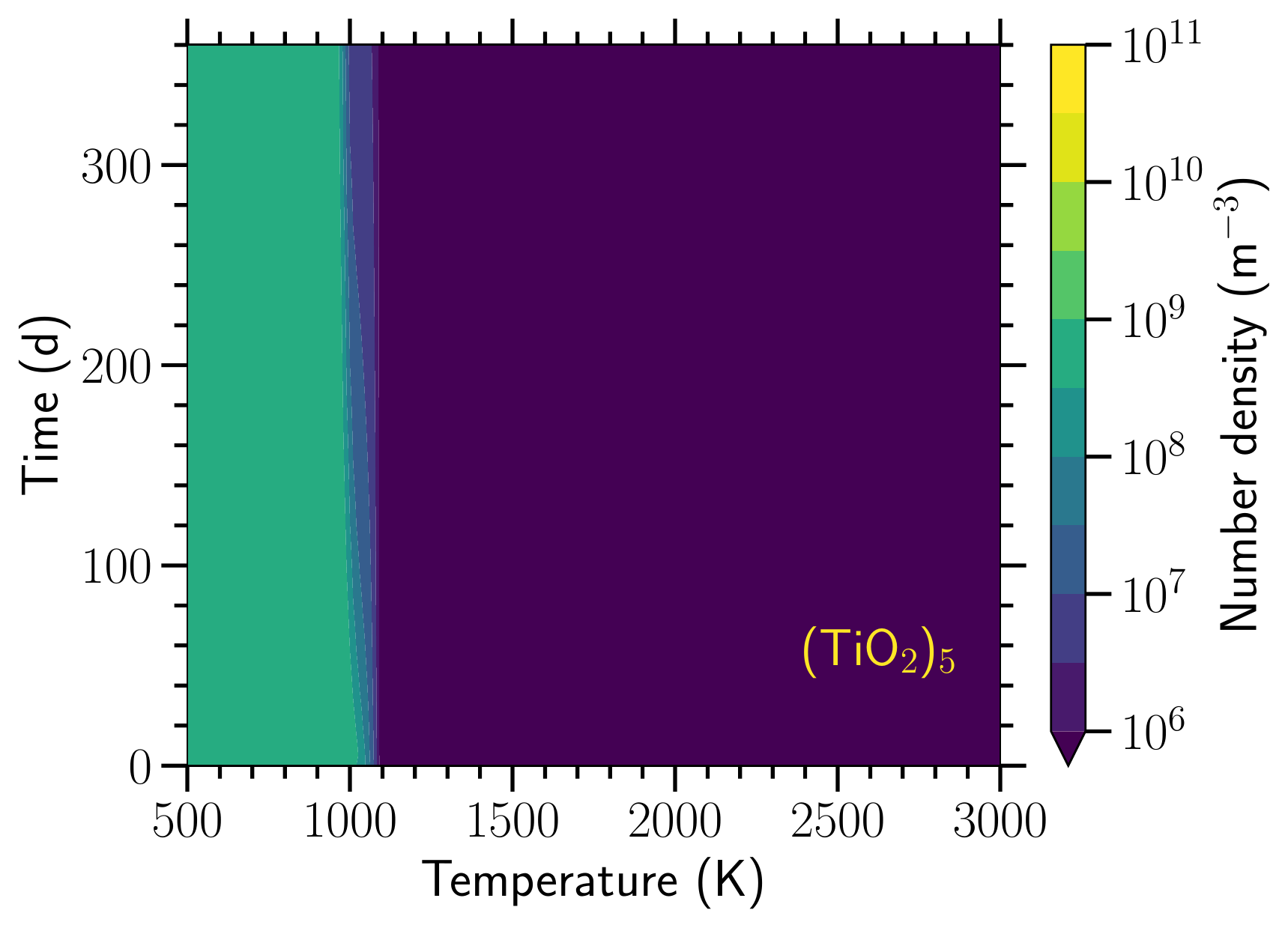}
        \includegraphics[width=0.32\textwidth]{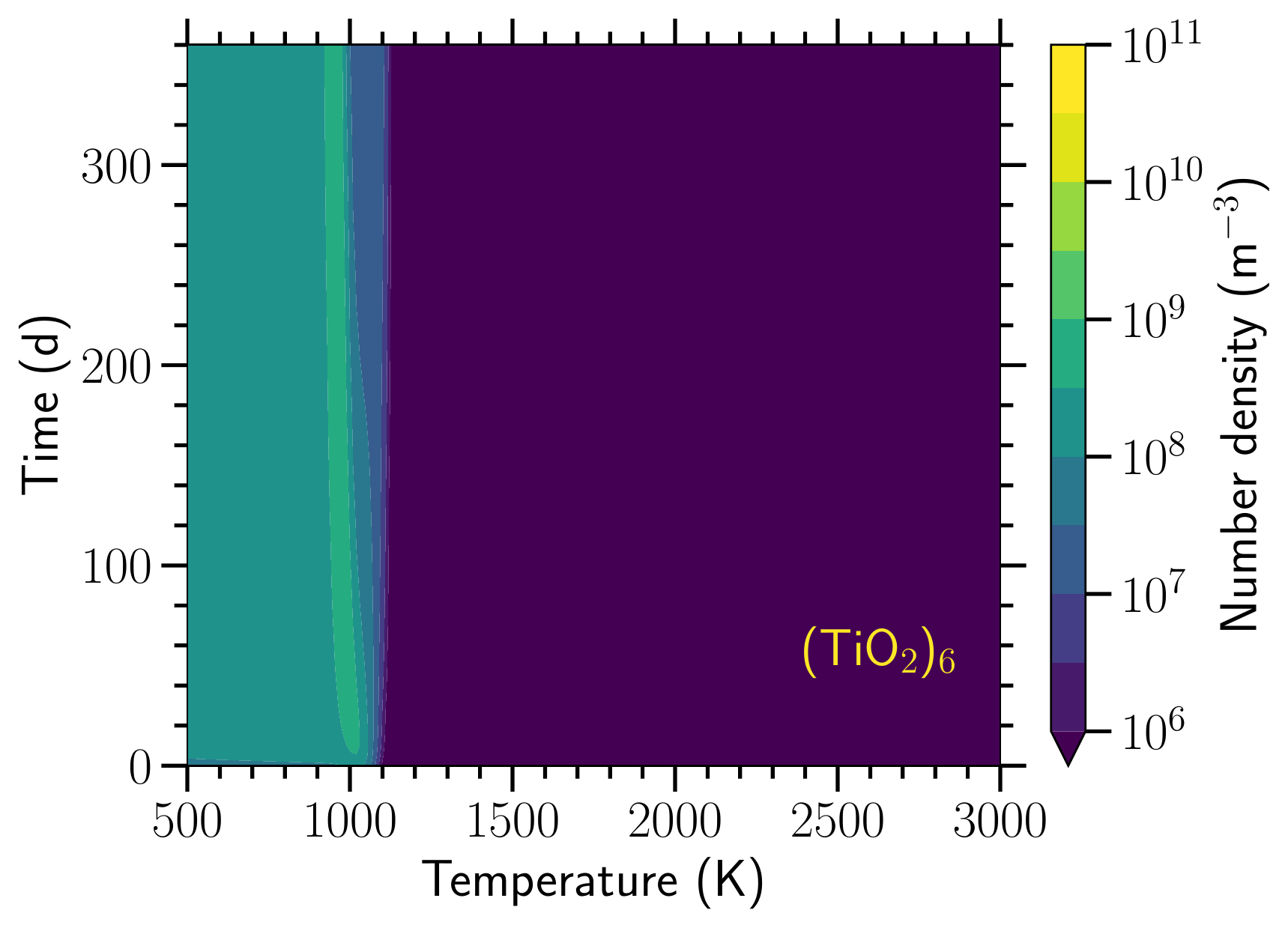}
        \includegraphics[width=0.32\textwidth]{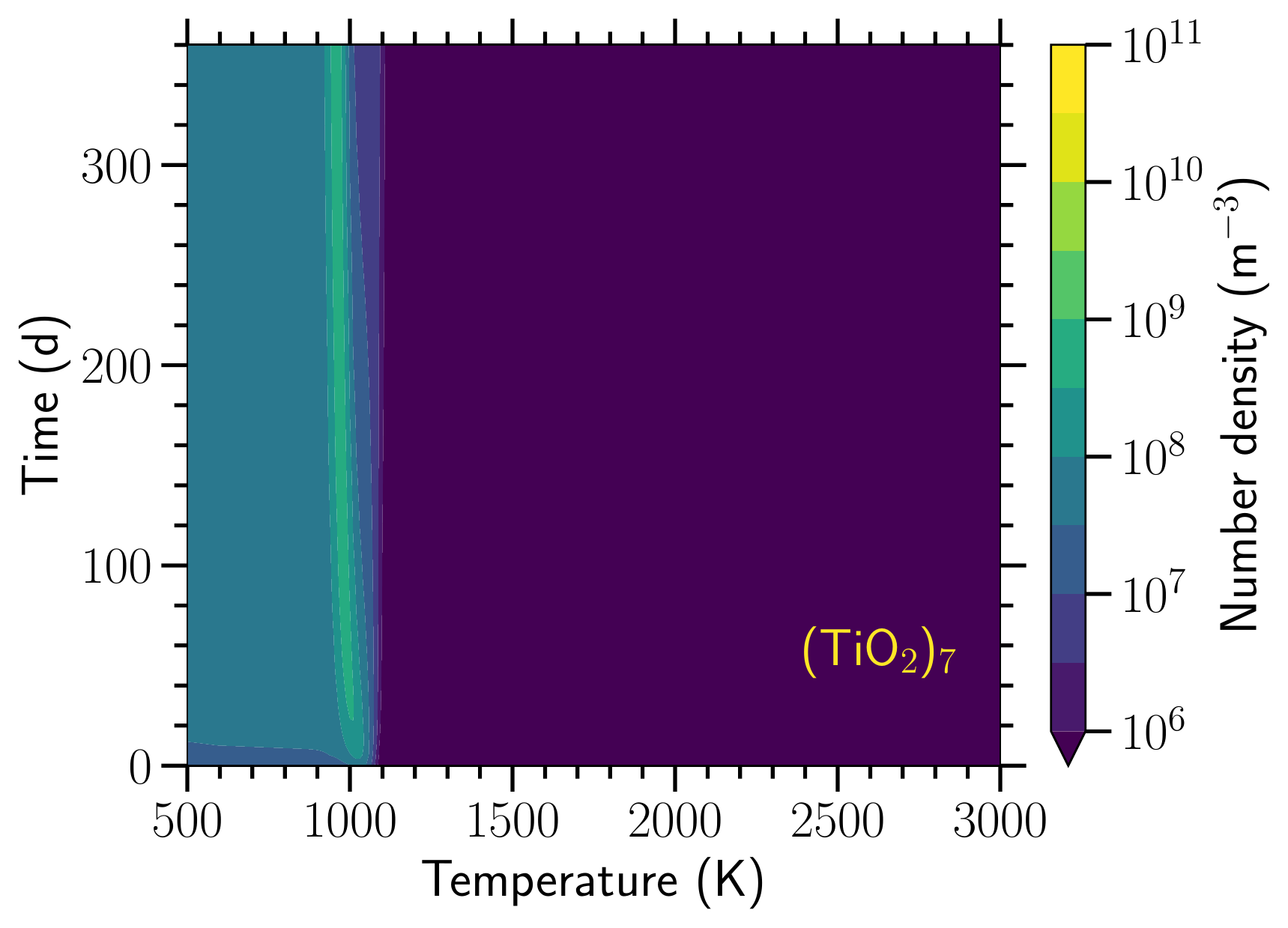}
        \includegraphics[width=0.32\textwidth]{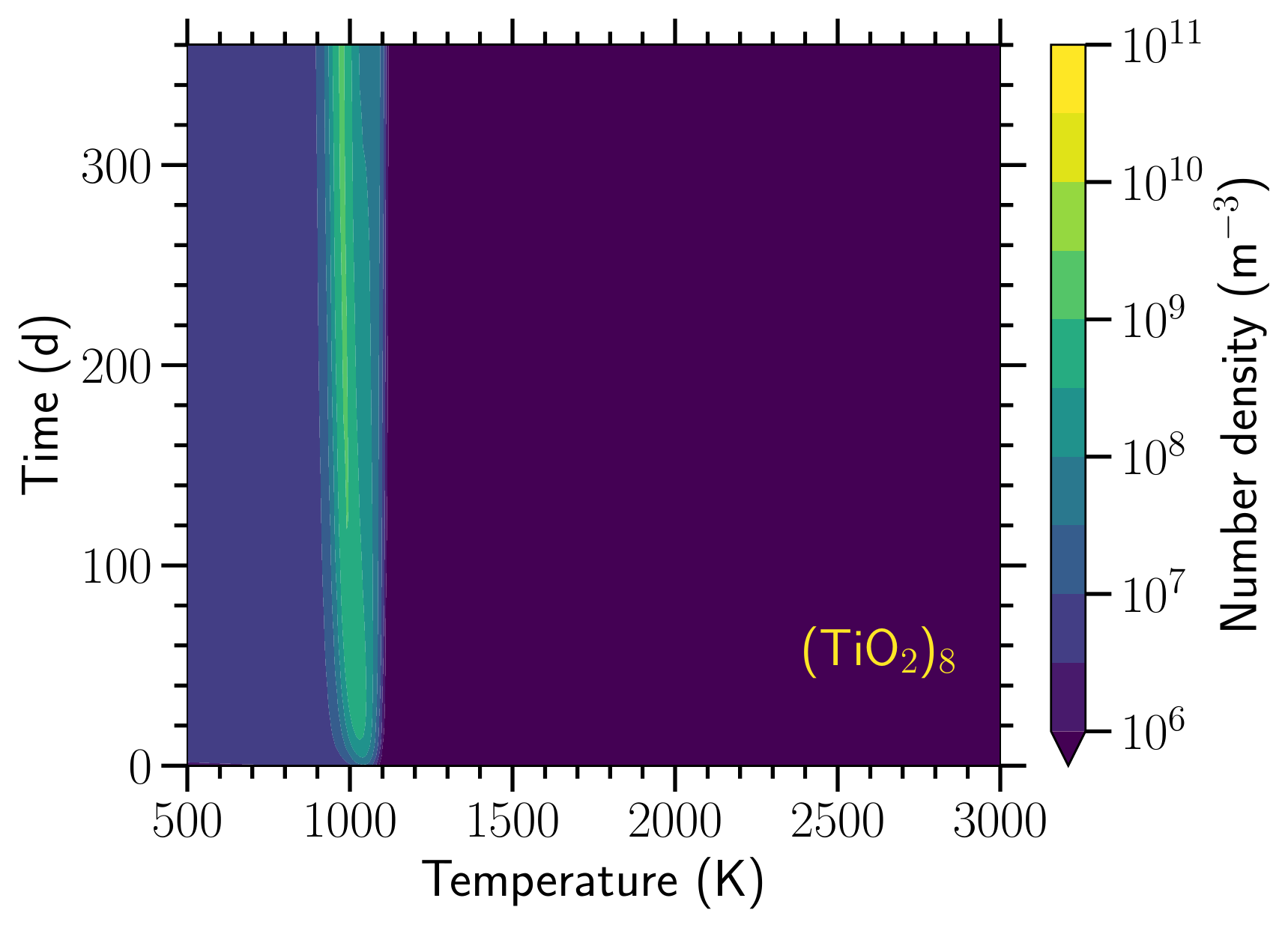}
        \includegraphics[width=0.32\textwidth]{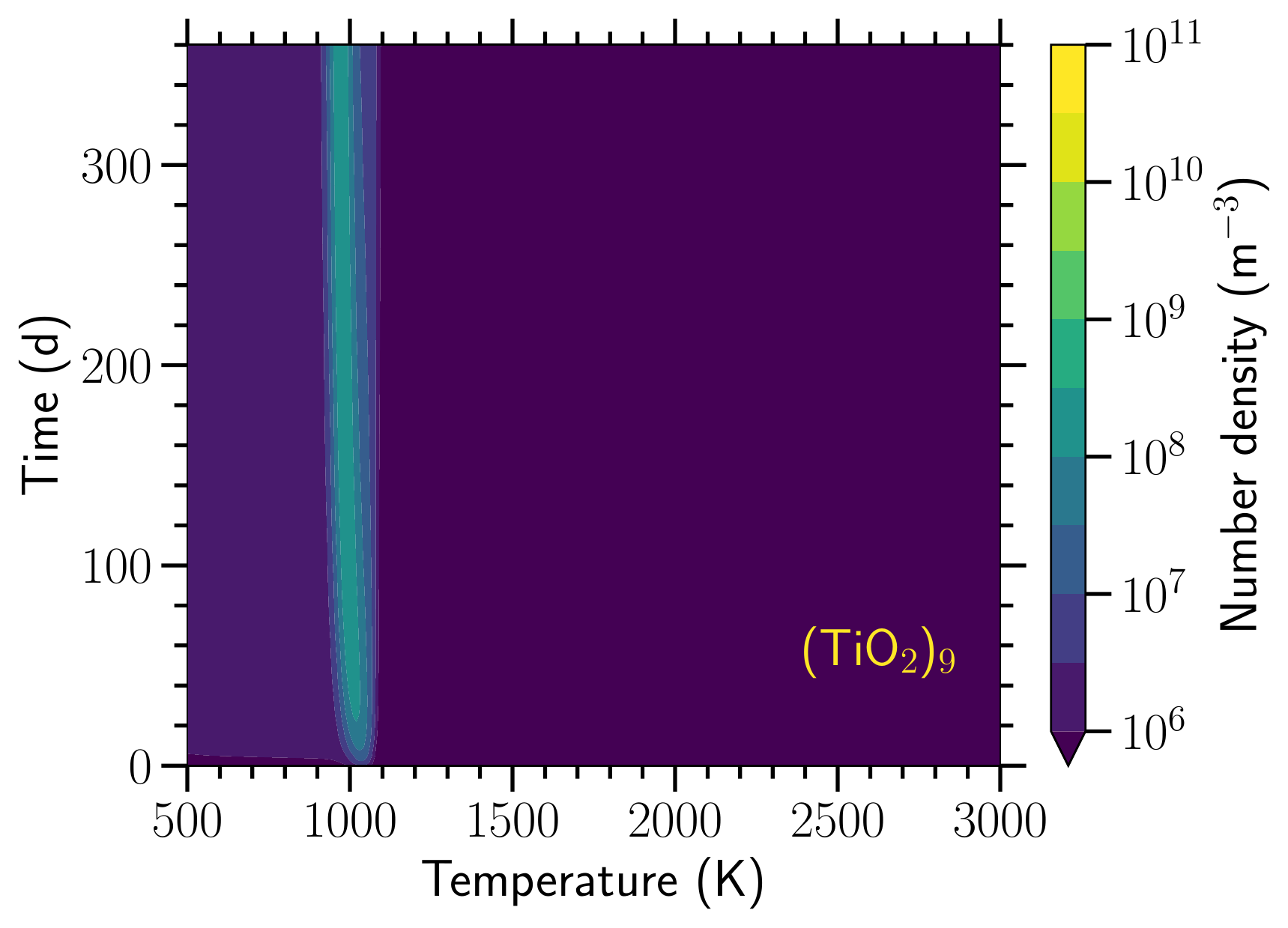}
        \includegraphics[width=0.32\textwidth]{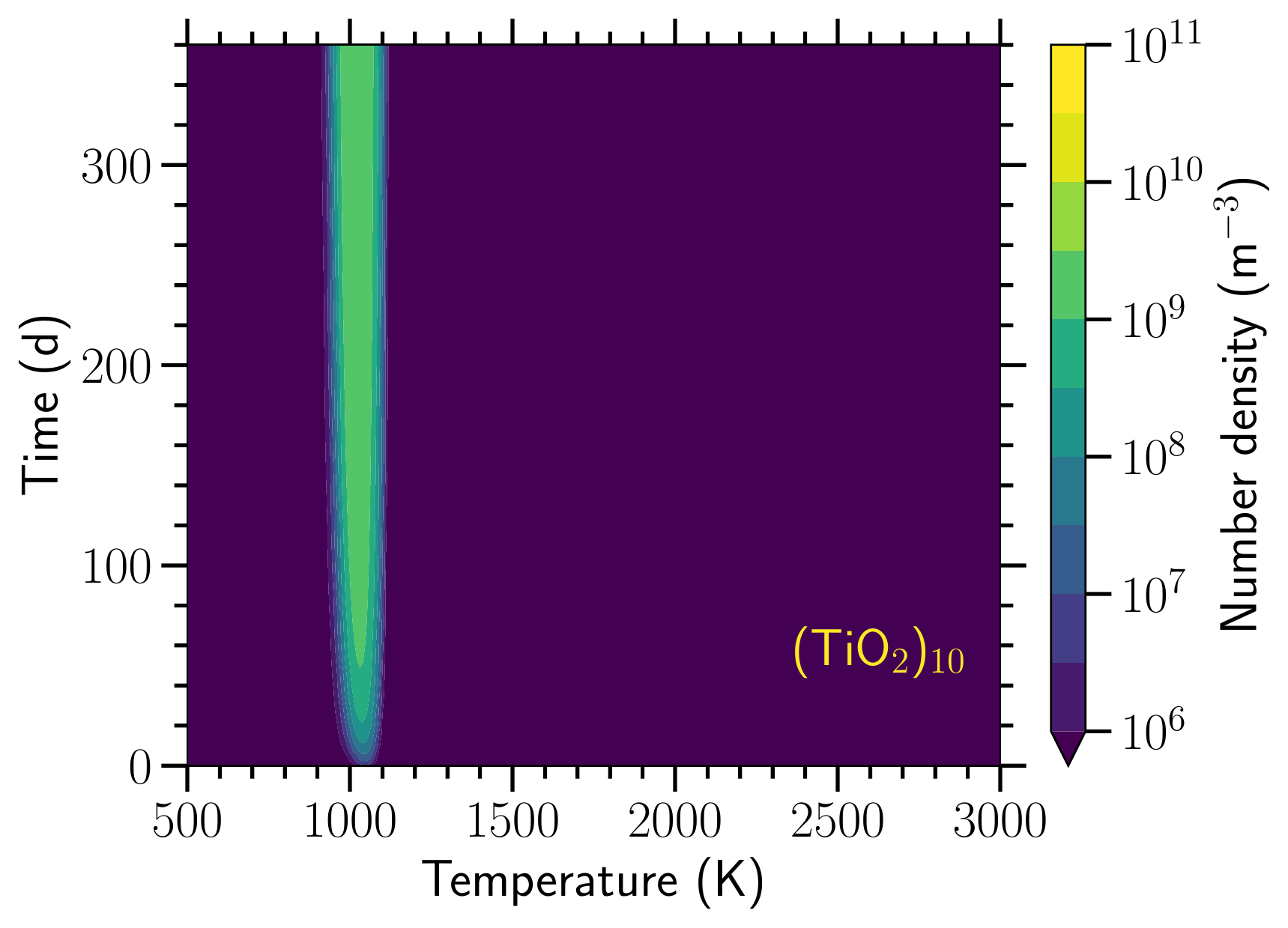}
        \end{flushleft}
        \caption{Temporal evolution of the absolute number density of all \protect\Ti{1}-clusters at the benchmark total gas density $\rho=\SI{1e-9}{\kg\per\m\cubed}$ for a closed nucleation model using the monomer nucleation description.}
        \label{fig:TiO2_clusters_monomer_time_evolution}
    \end{figure*}

\begin{figure*}
        \begin{flushleft}
        \includegraphics[width=0.32\textwidth]{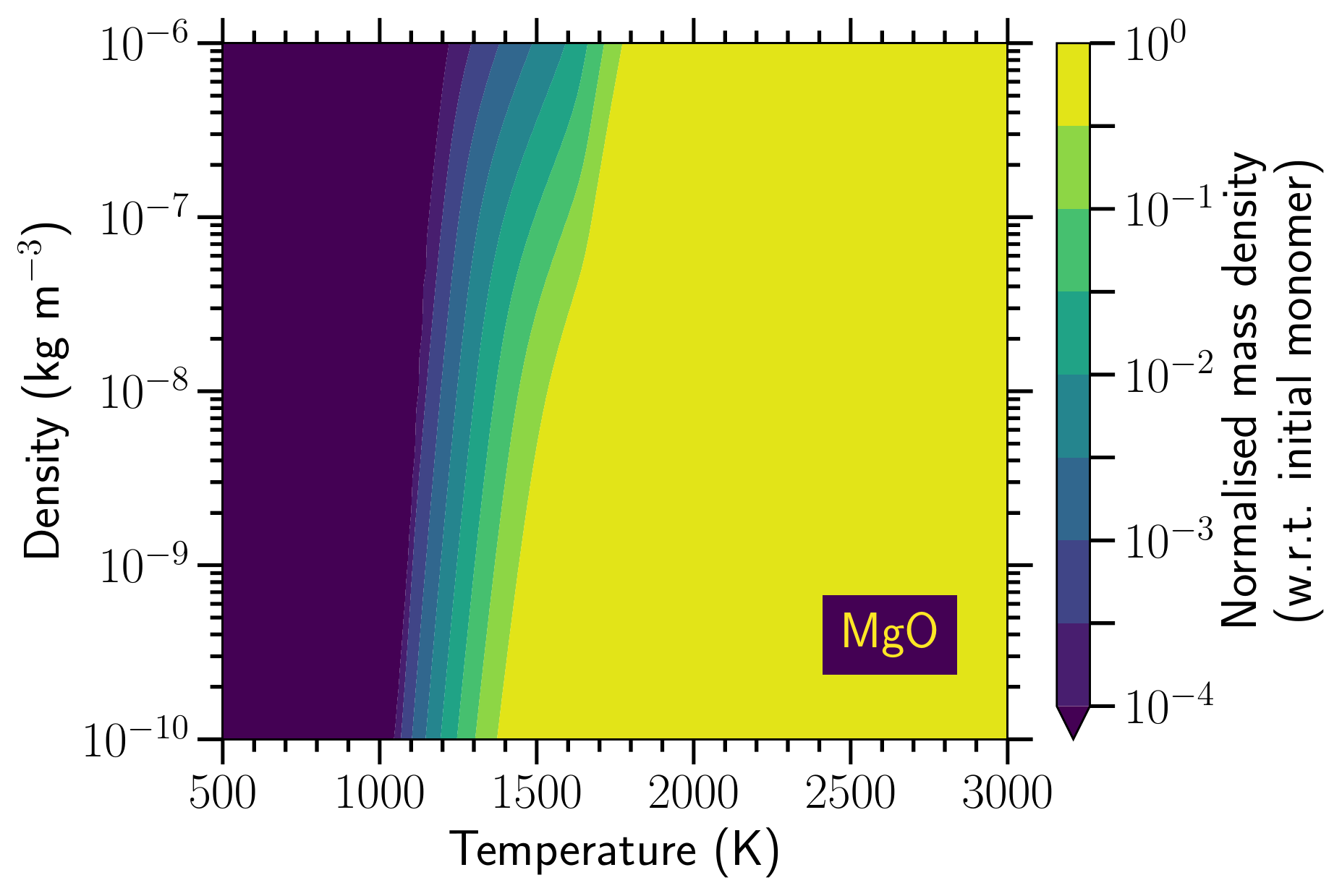}
        \includegraphics[width=0.32\textwidth]{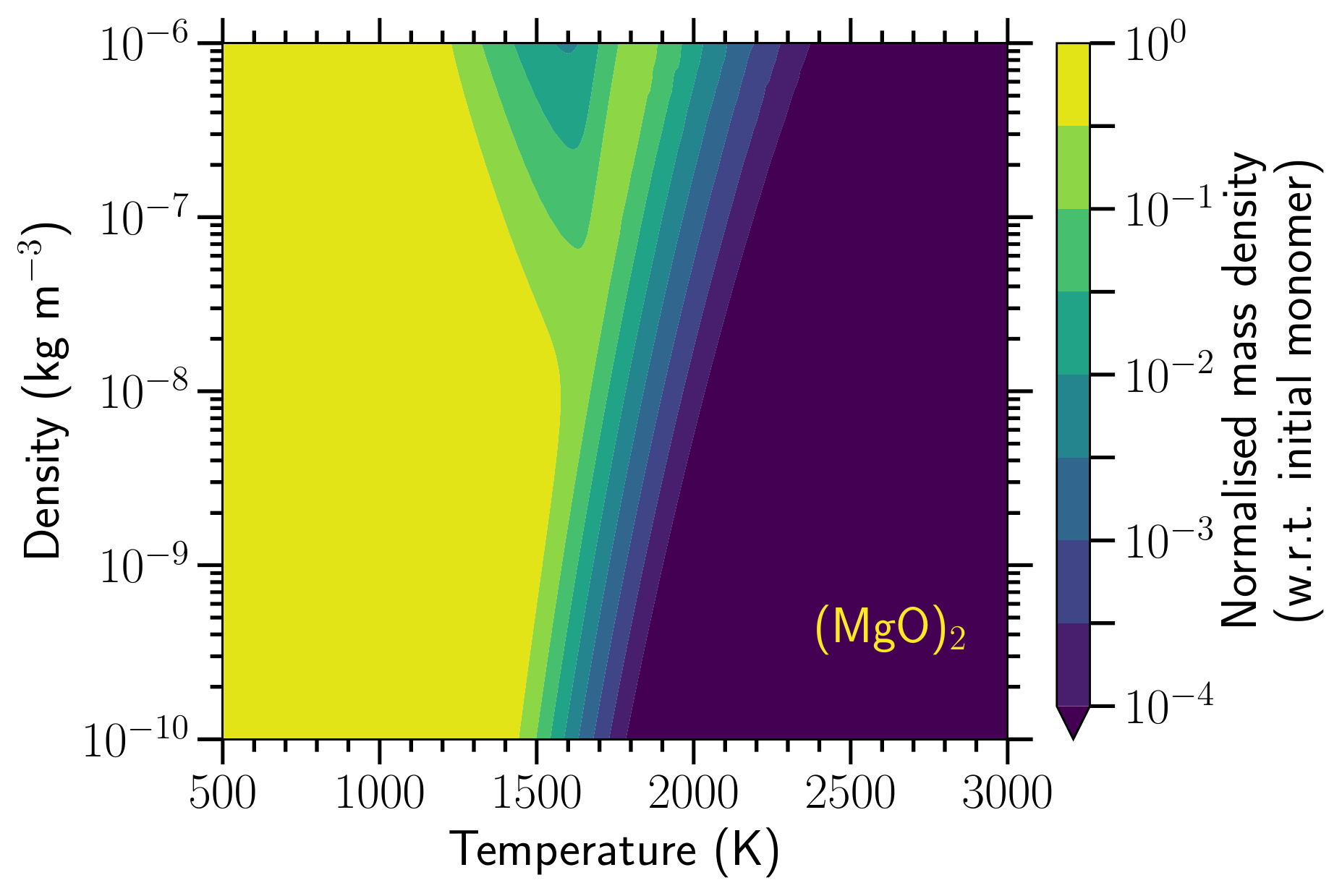}
        \includegraphics[width=0.32\textwidth]{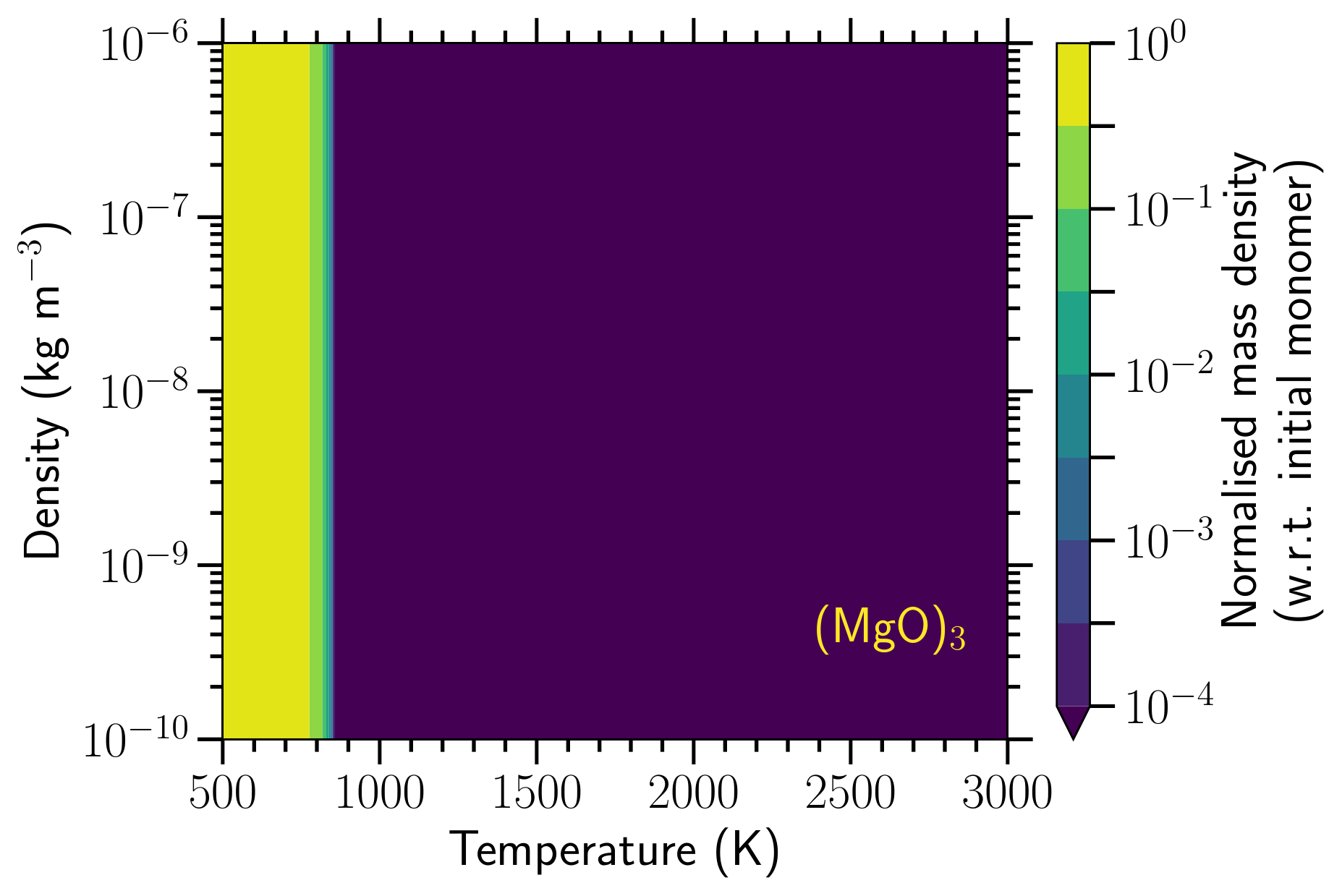}
        \includegraphics[width=0.32\textwidth]{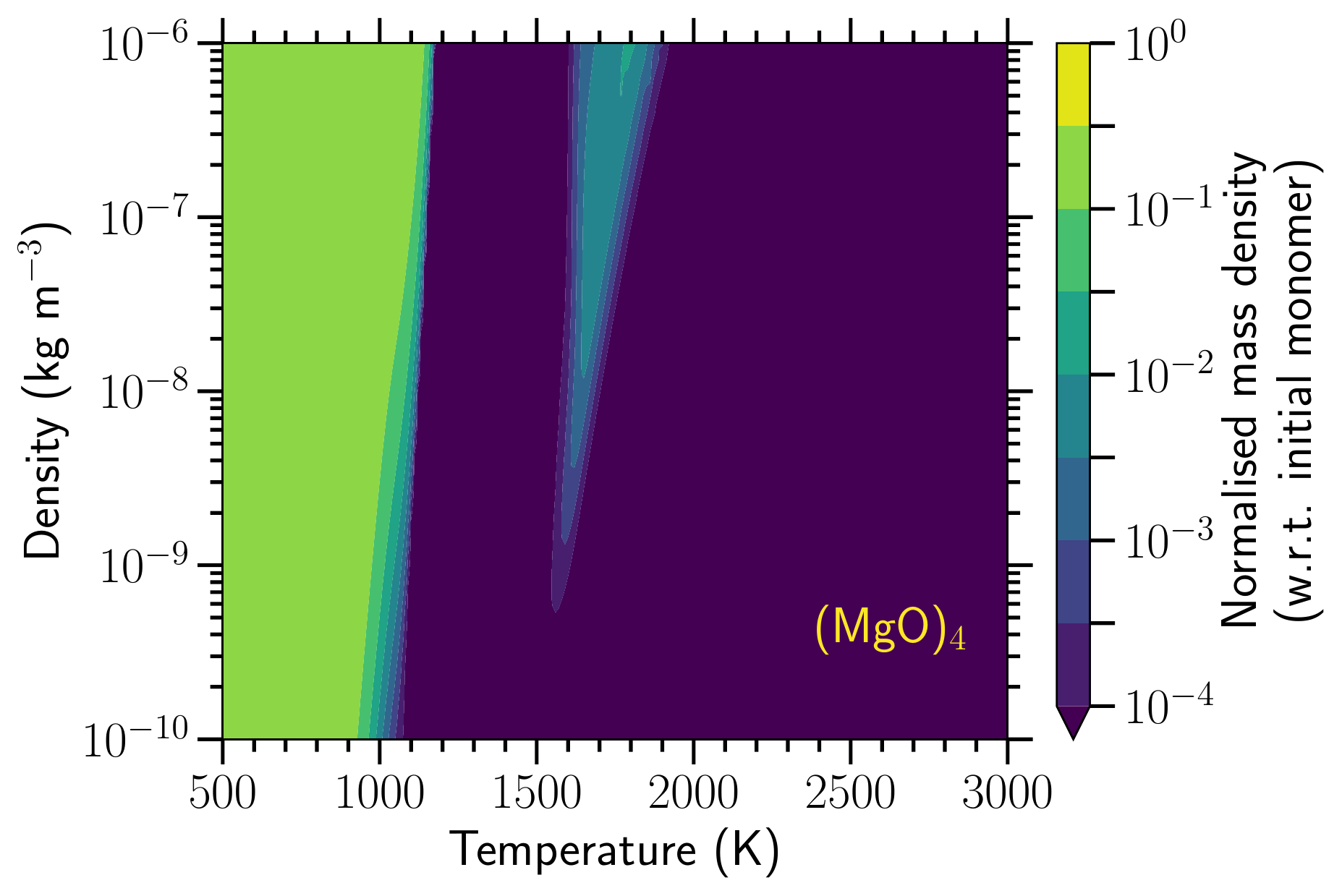}
        \includegraphics[width=0.32\textwidth]{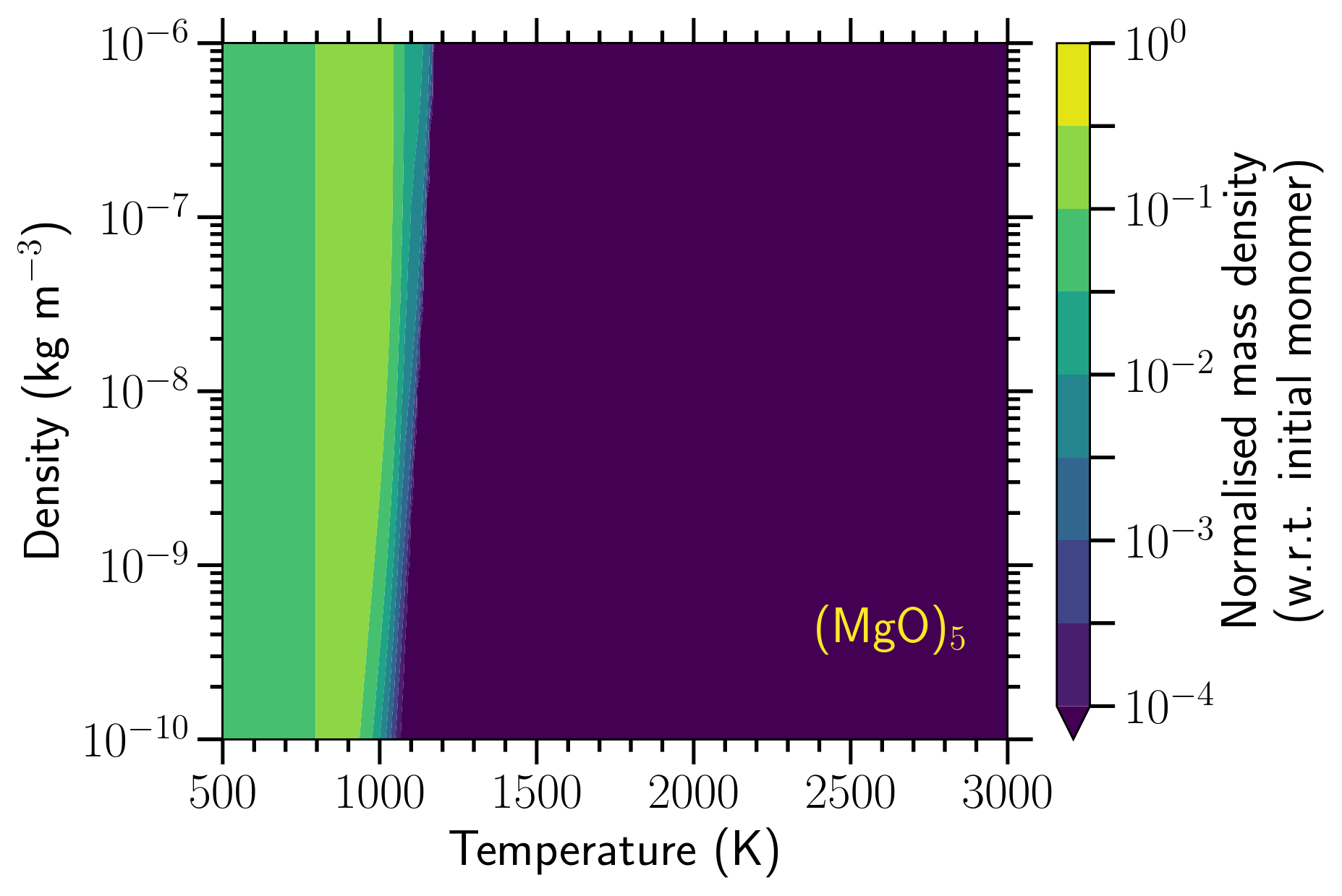}
        \includegraphics[width=0.32\textwidth]{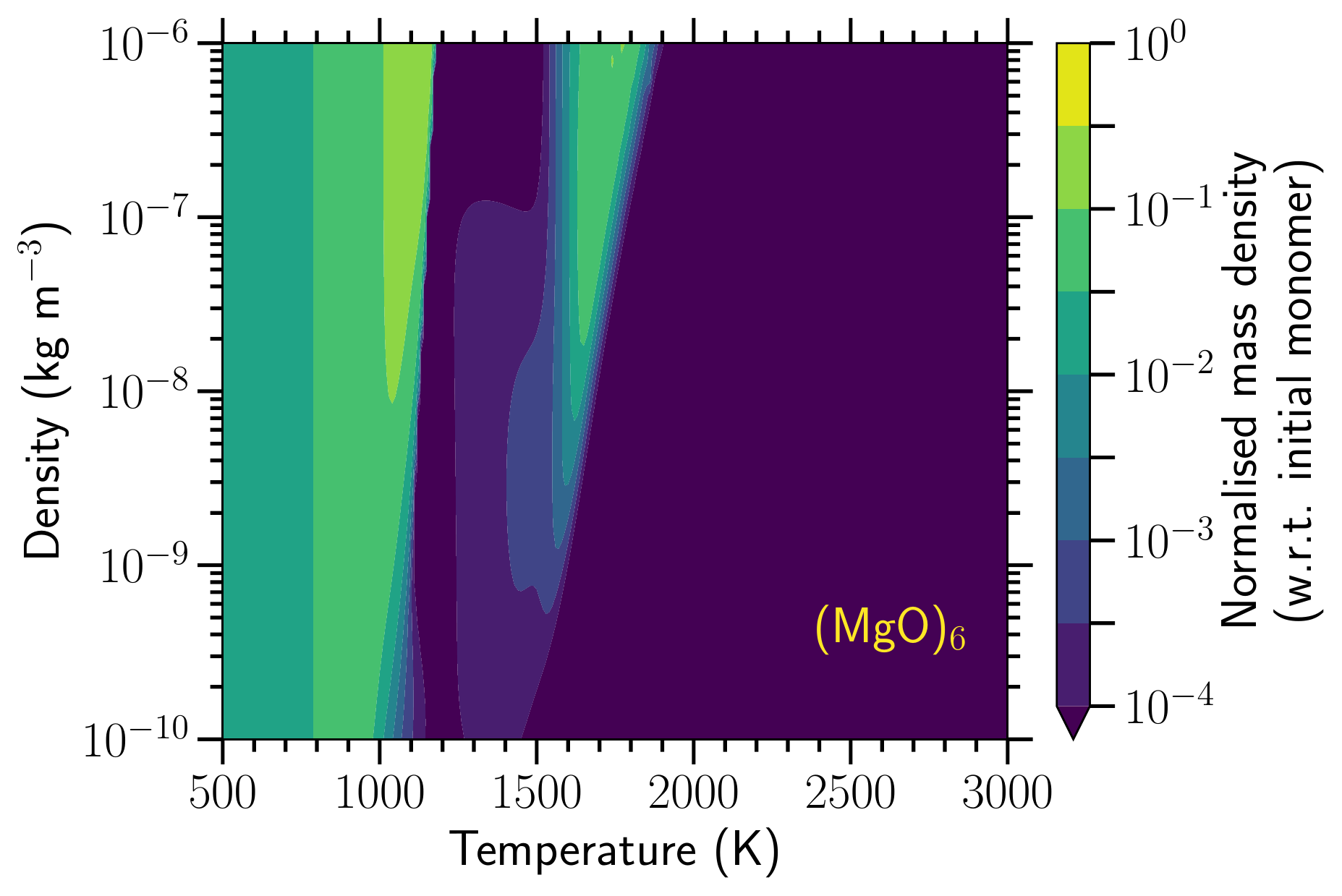}
        \includegraphics[width=0.32\textwidth]{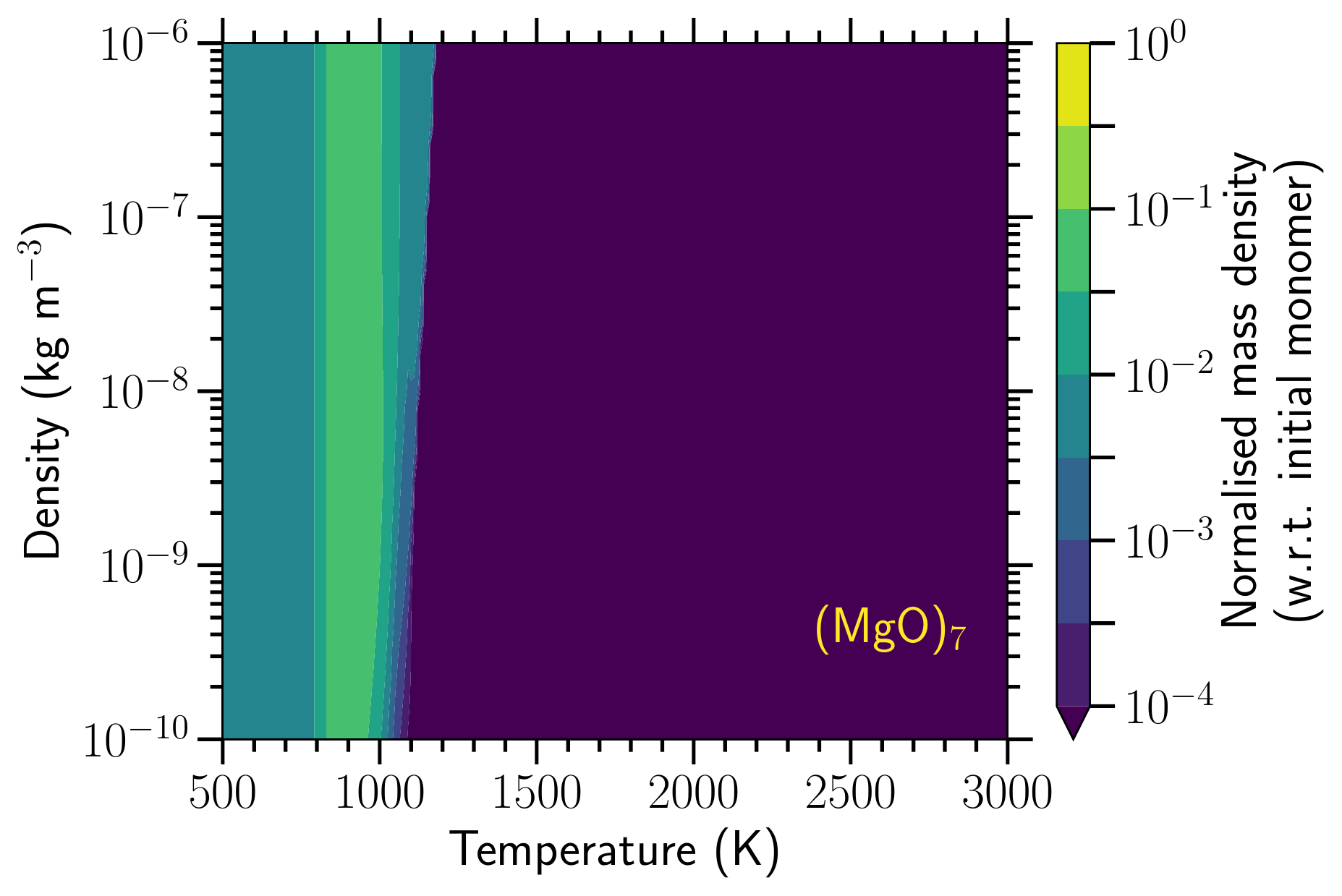}
        \includegraphics[width=0.32\textwidth]{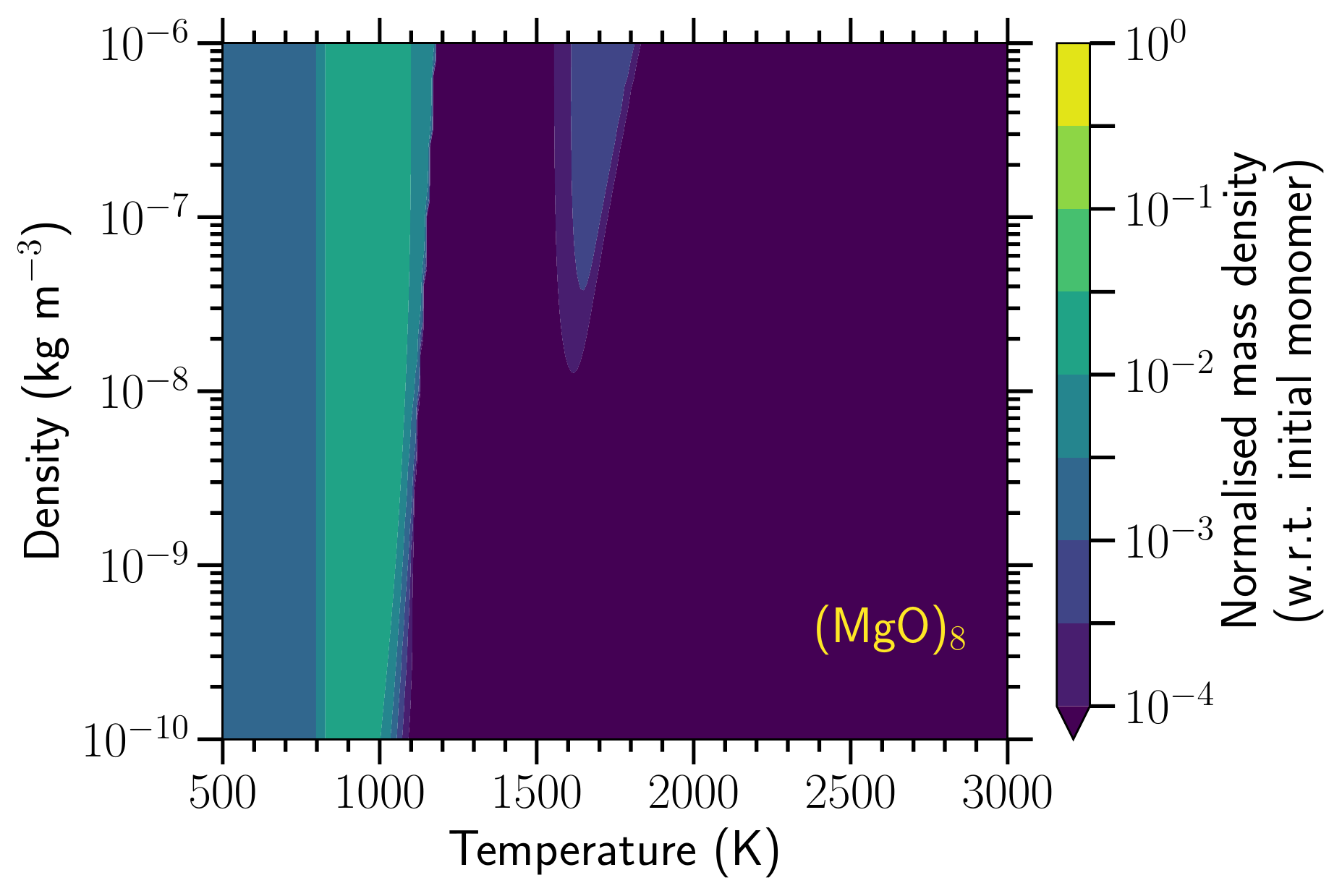}
        \includegraphics[width=0.32\textwidth]{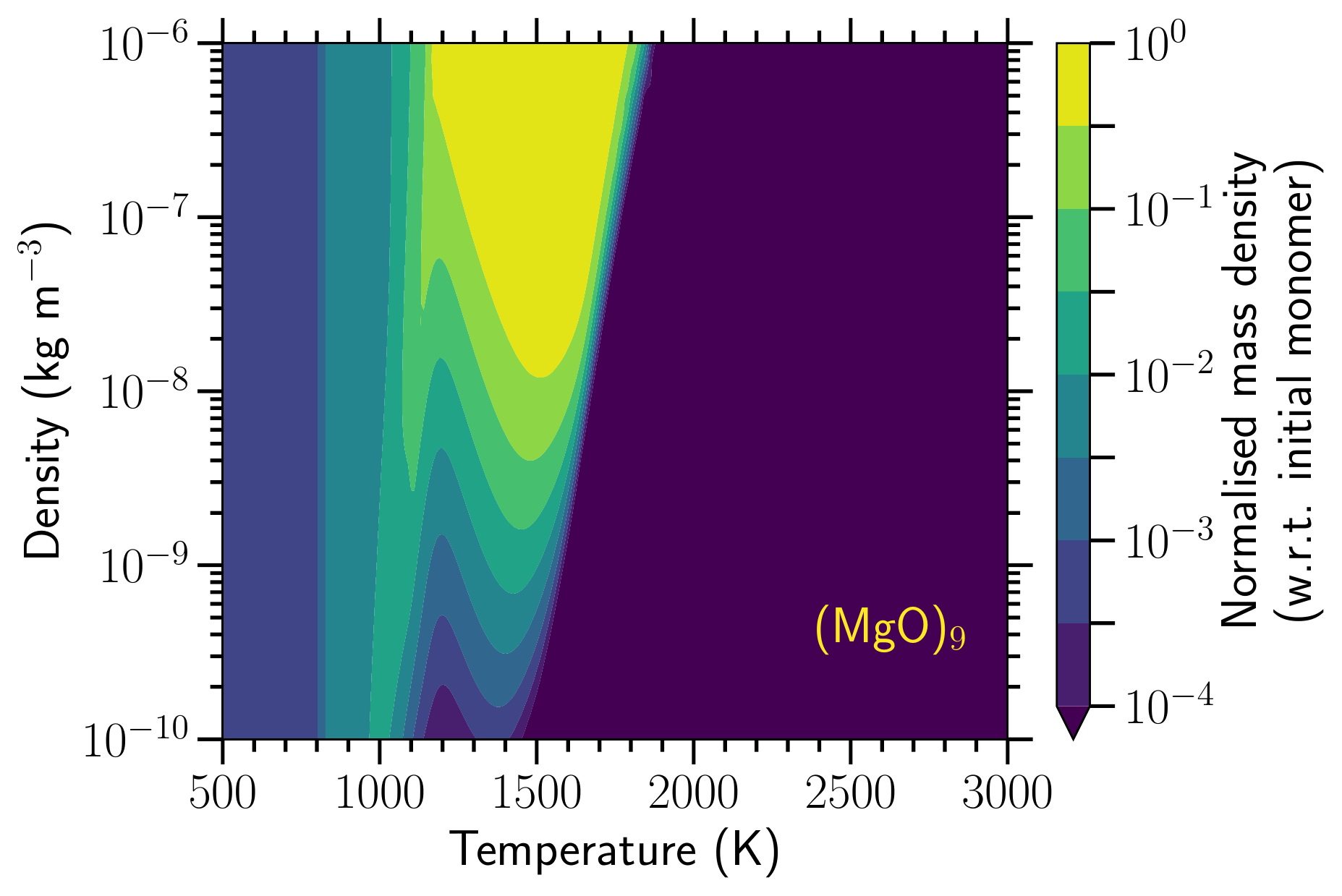}
        \includegraphics[width=0.32\textwidth]{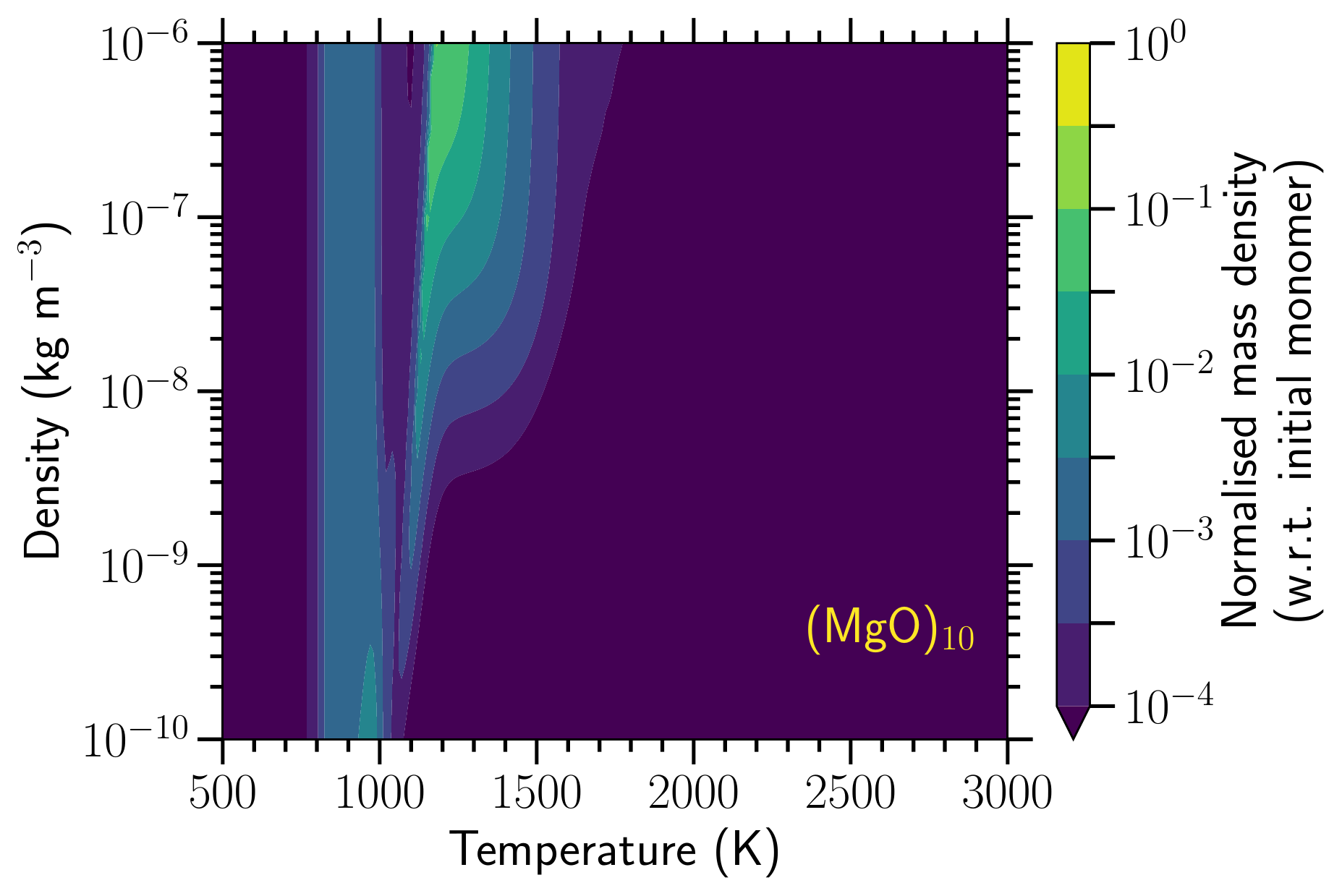}
        \end{flushleft}
        \caption{Overview of the normalised mass density after one year of all \protect\Mg{1}-clusters for a closed nucleation model using the monomer nucleation description.}
        \label{fig:MgO_clusters_monomer_norm_same_scale}
    \end{figure*}

\begin{figure*}
        \begin{flushleft}
        \includegraphics[width=0.32\textwidth]{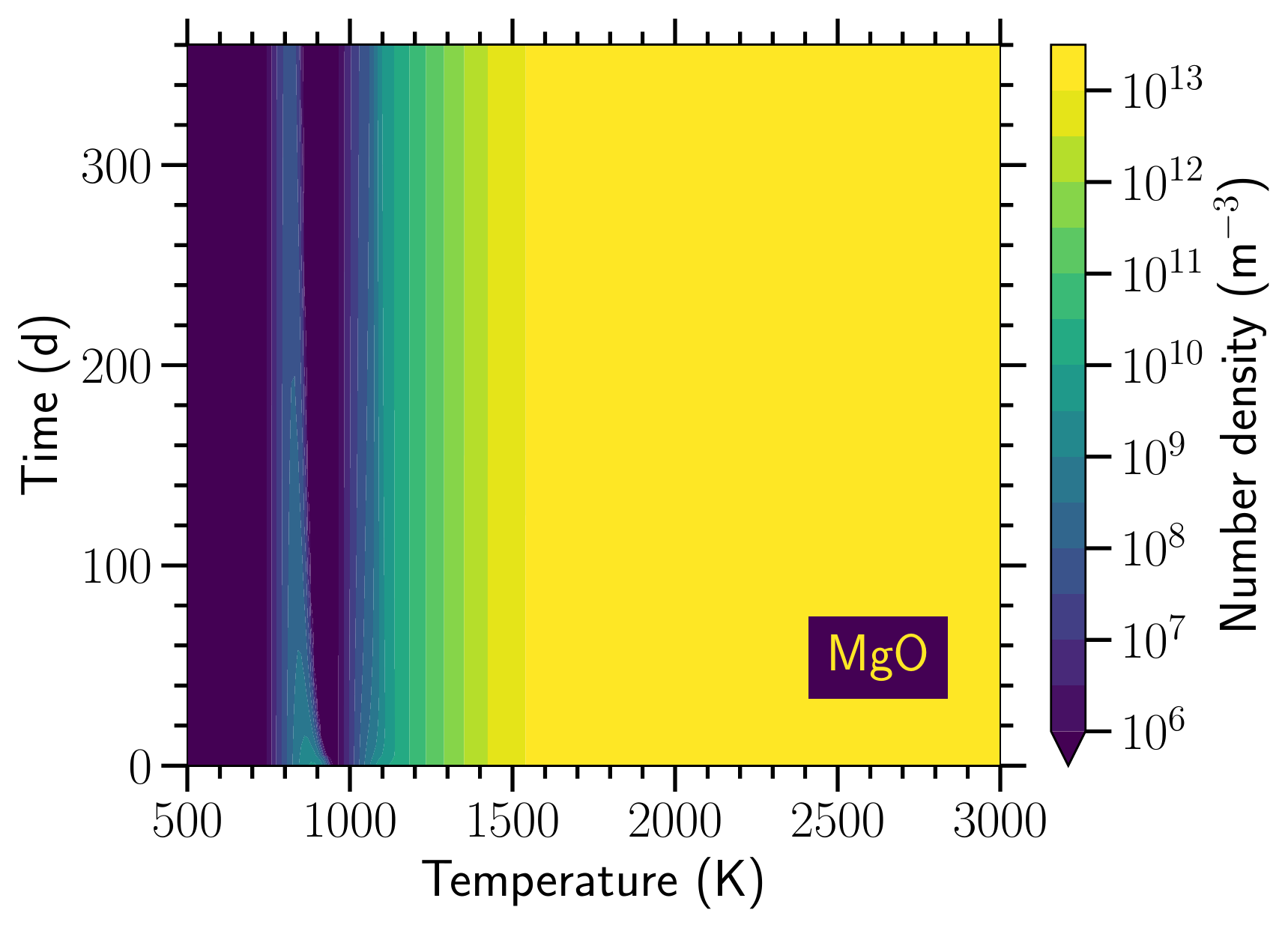}
        \includegraphics[width=0.32\textwidth]{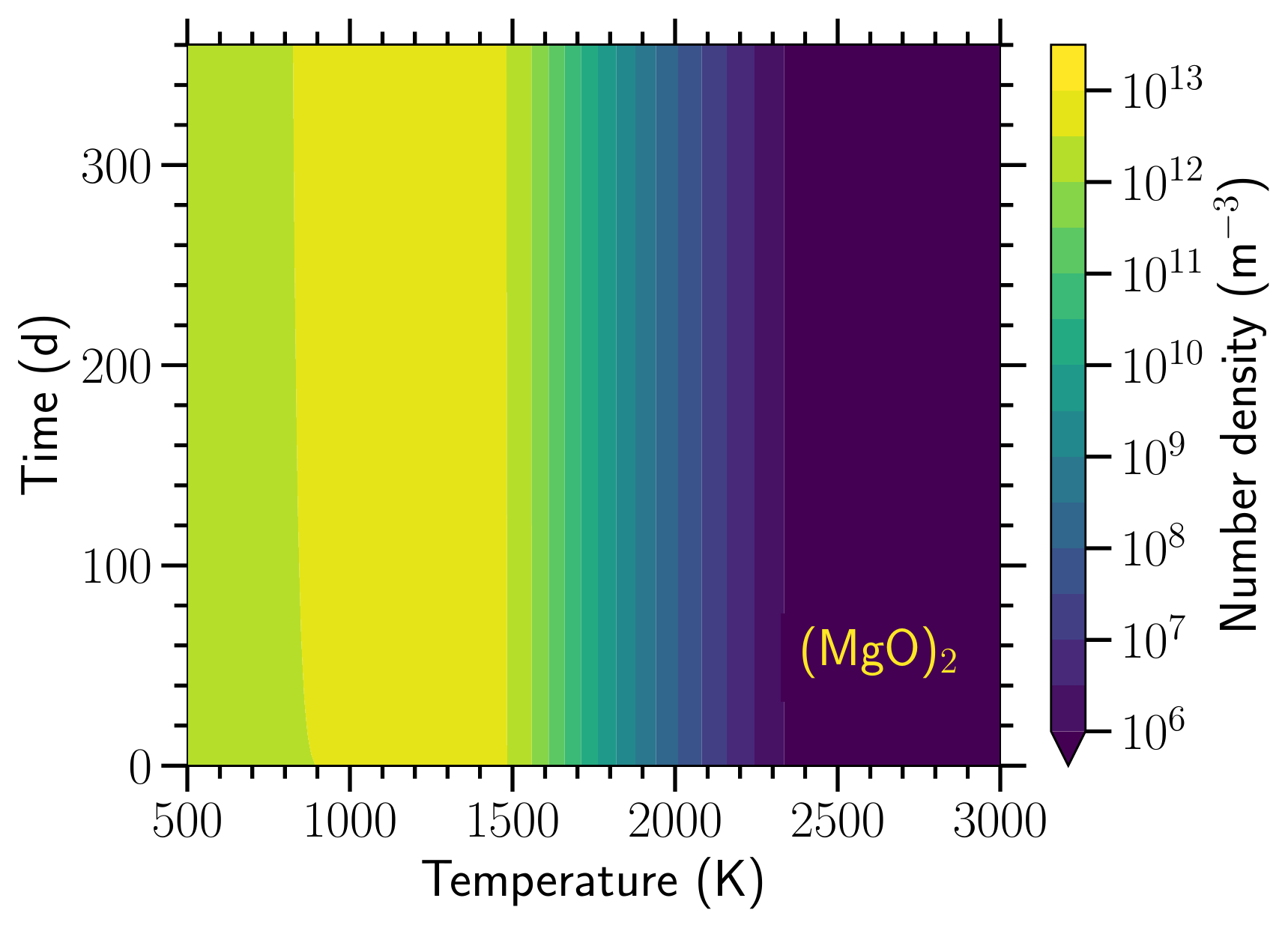}
        \includegraphics[width=0.32\textwidth]{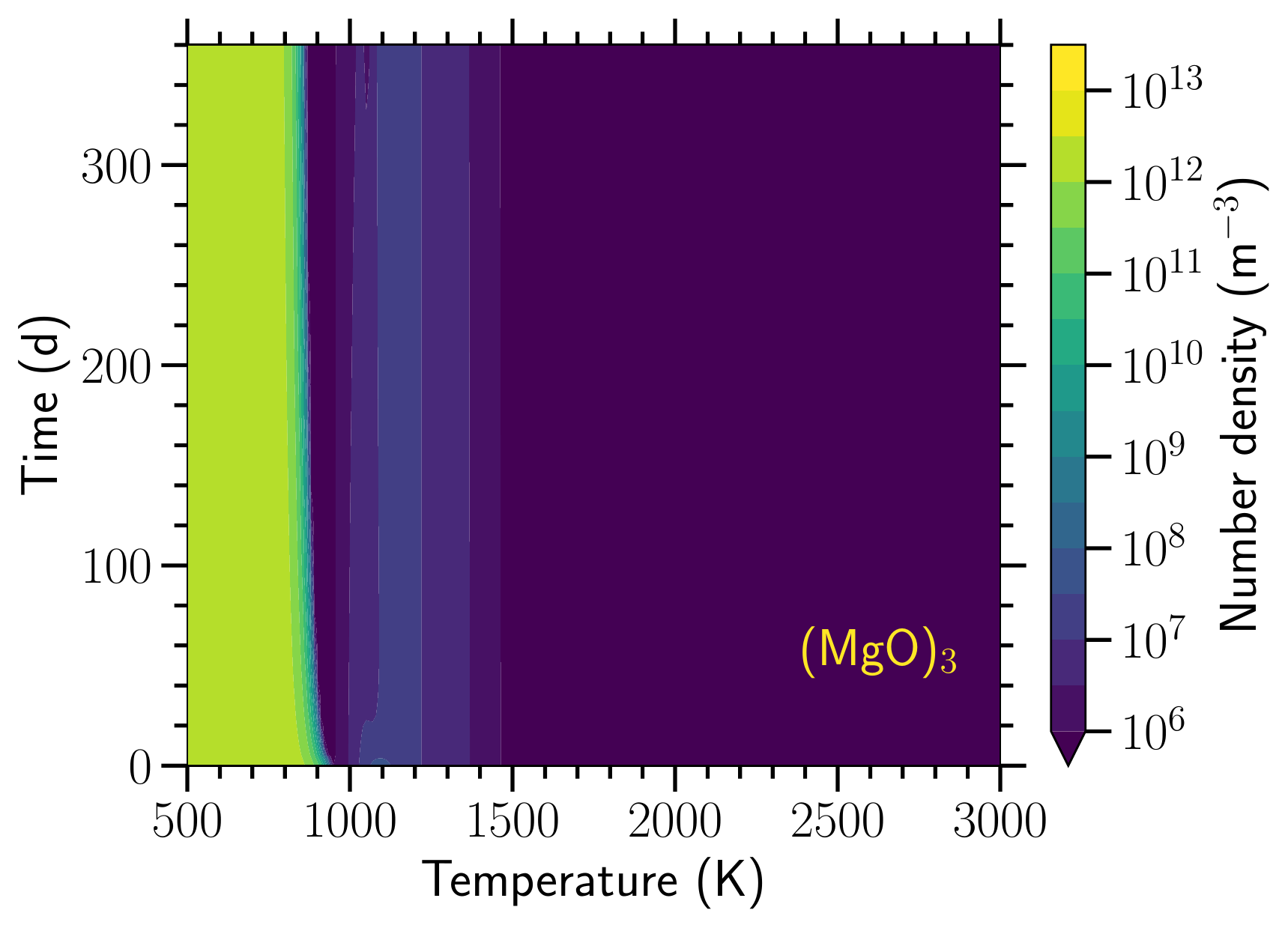}
        \includegraphics[width=0.32\textwidth]{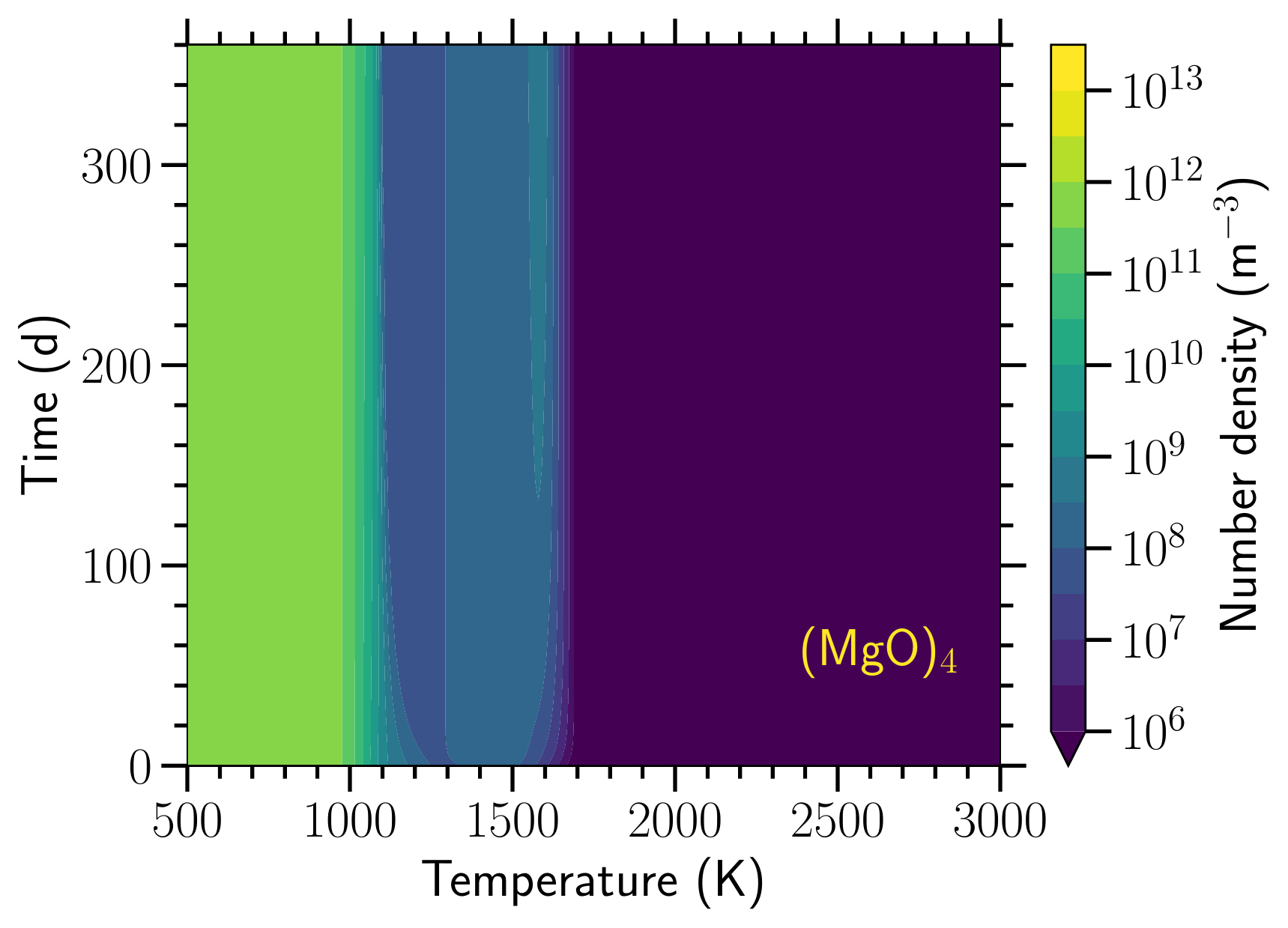}
        \includegraphics[width=0.32\textwidth]{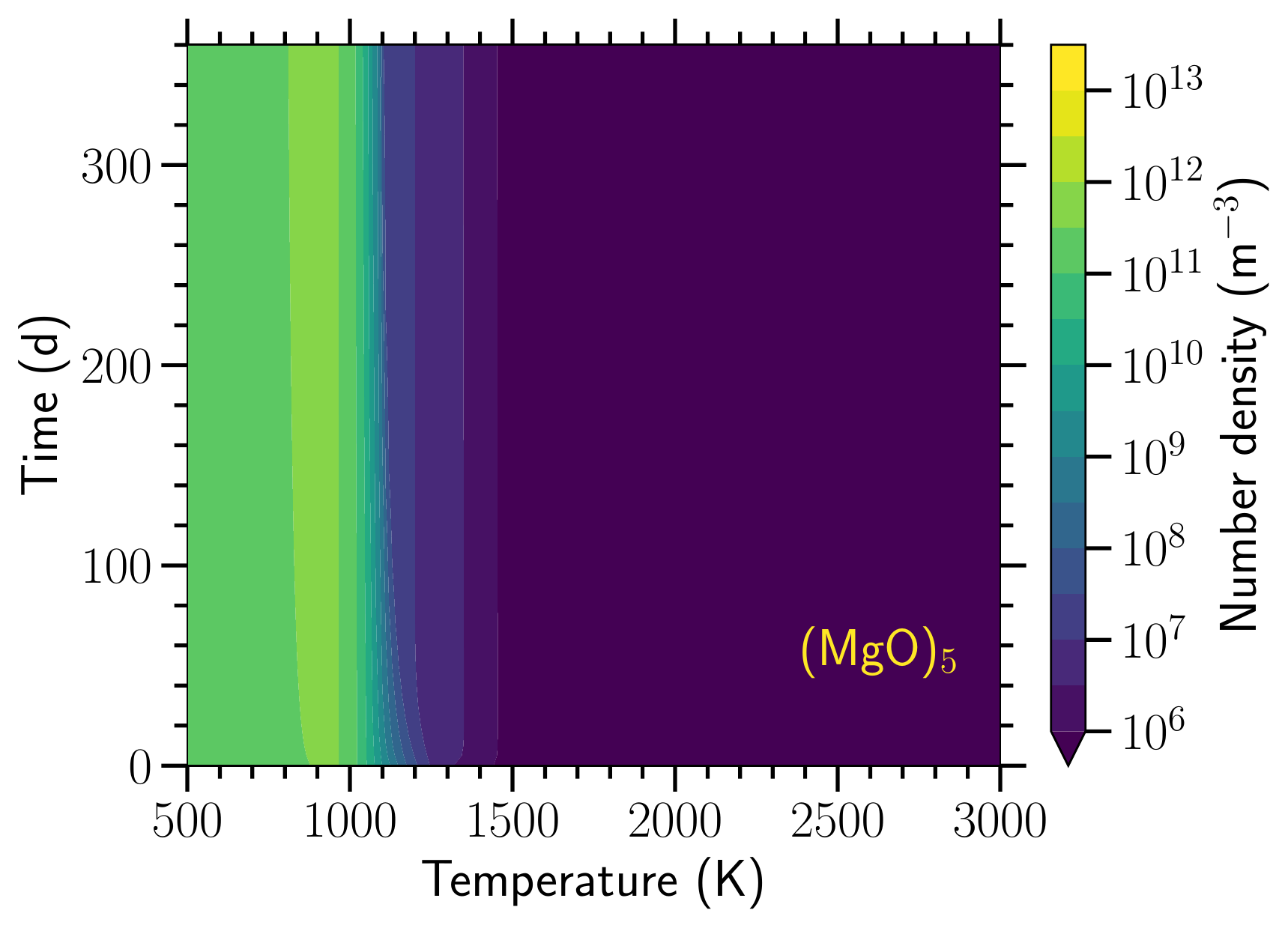}
        \includegraphics[width=0.32\textwidth]{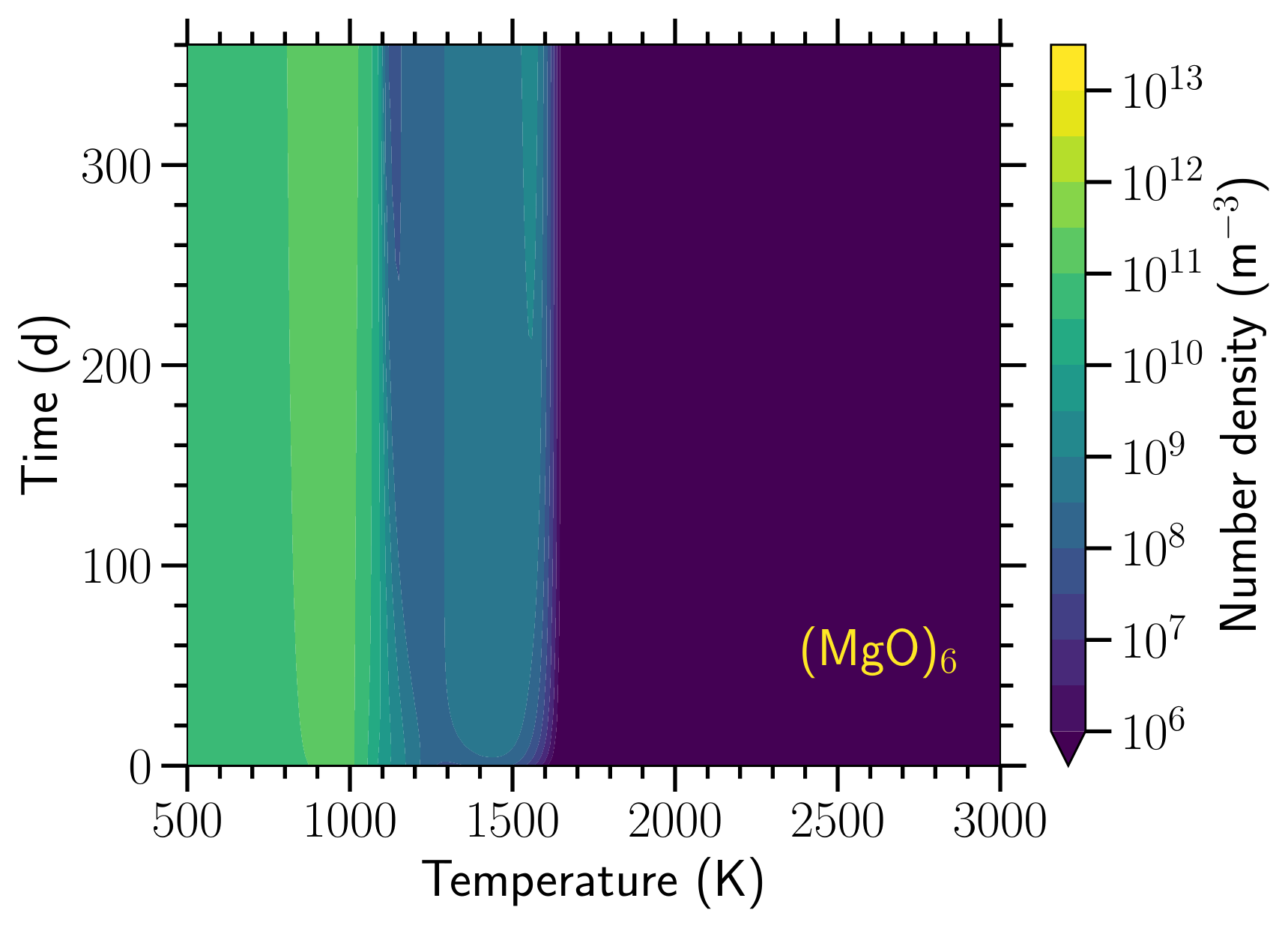}
        \includegraphics[width=0.32\textwidth]{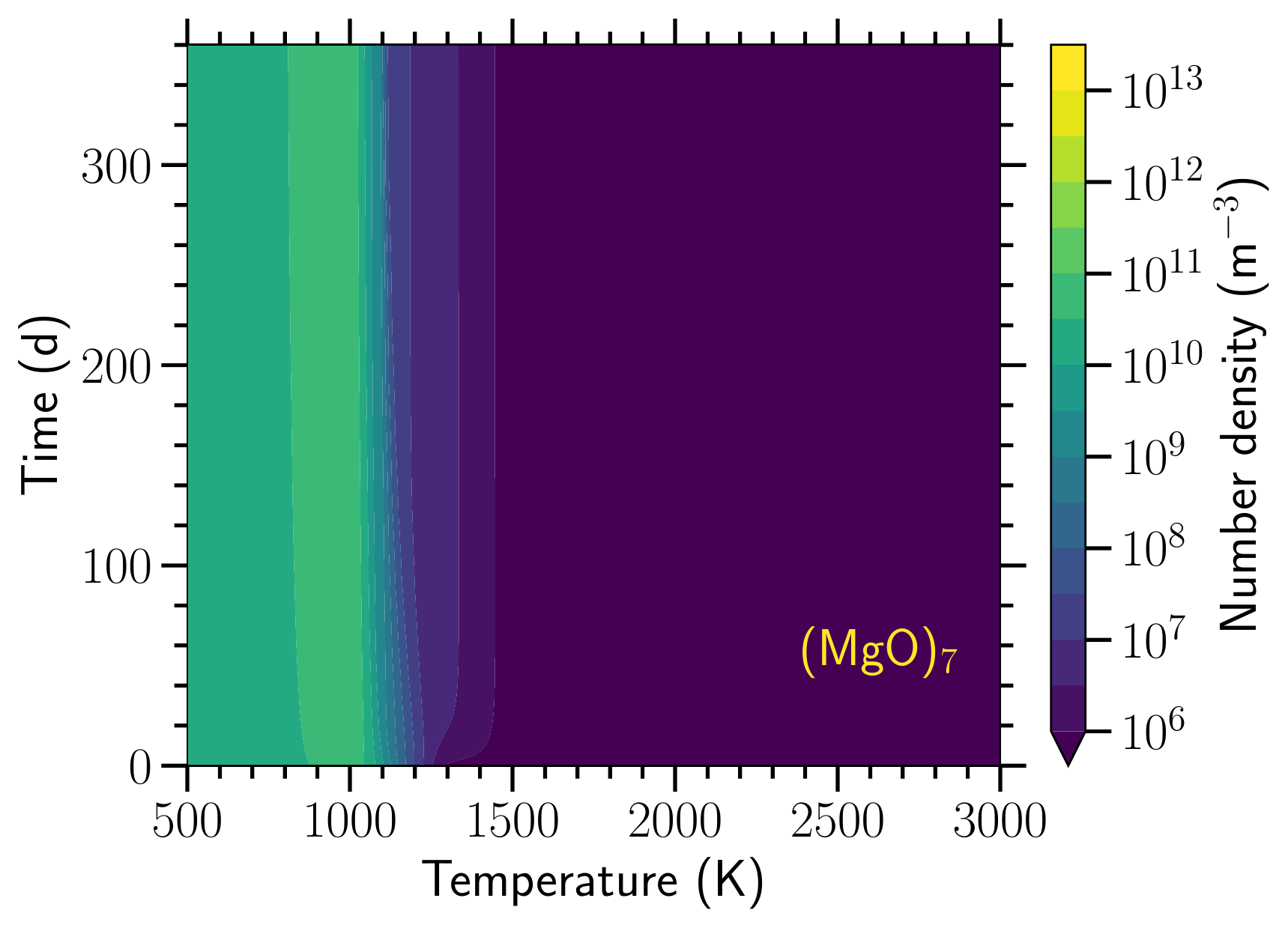}
        \includegraphics[width=0.32\textwidth]{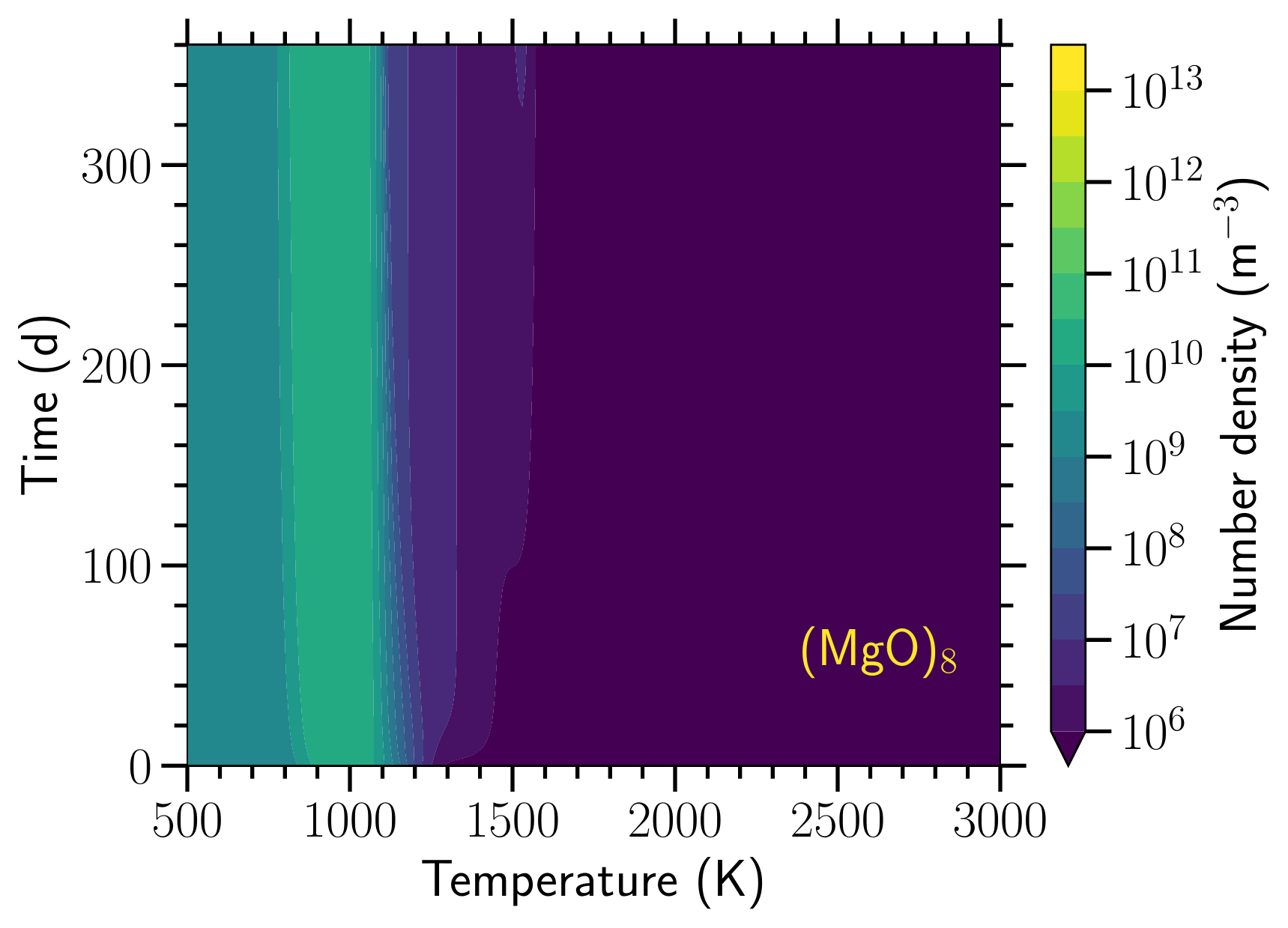}
        \includegraphics[width=0.32\textwidth]{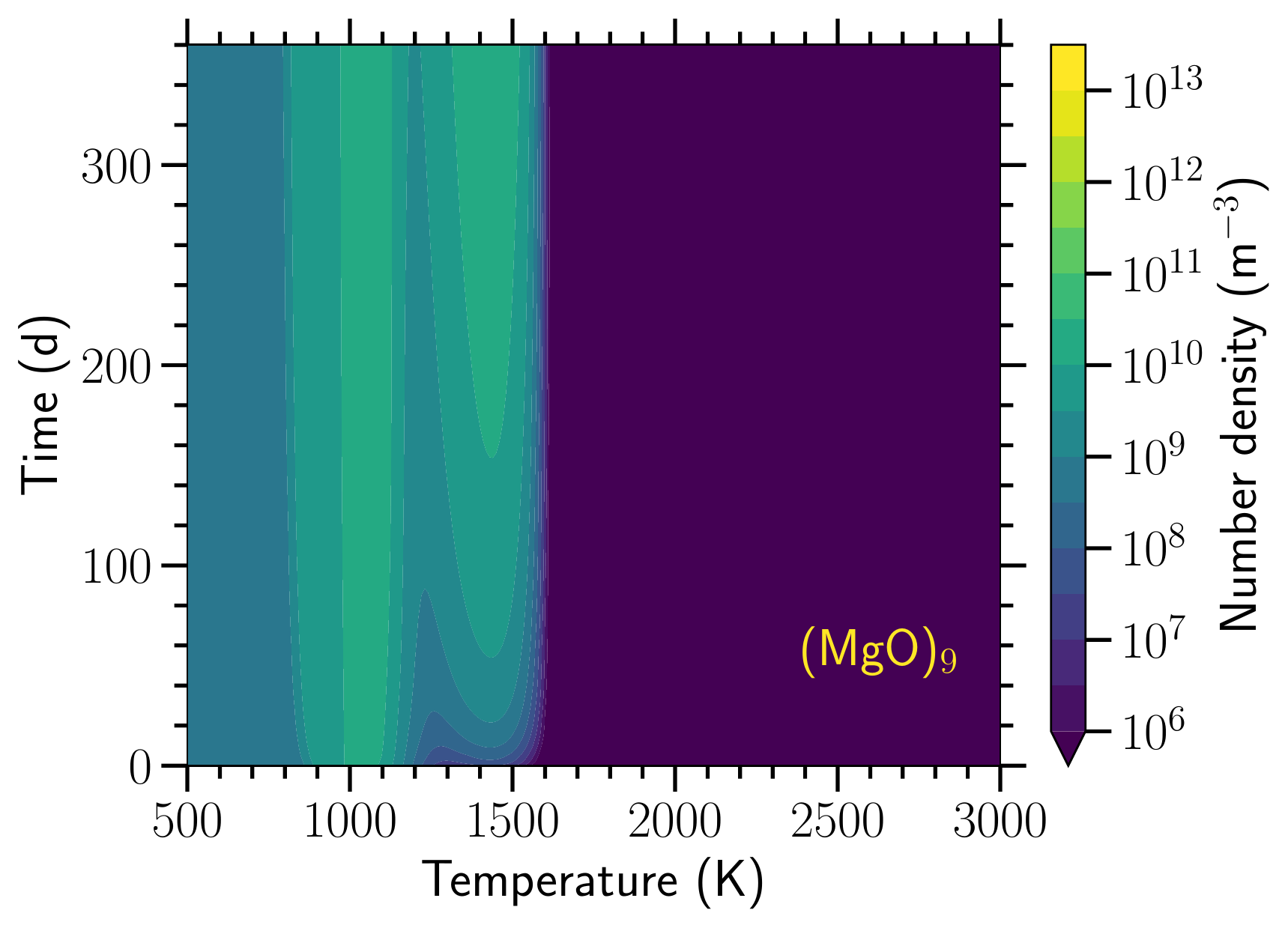}
        \includegraphics[width=0.32\textwidth]{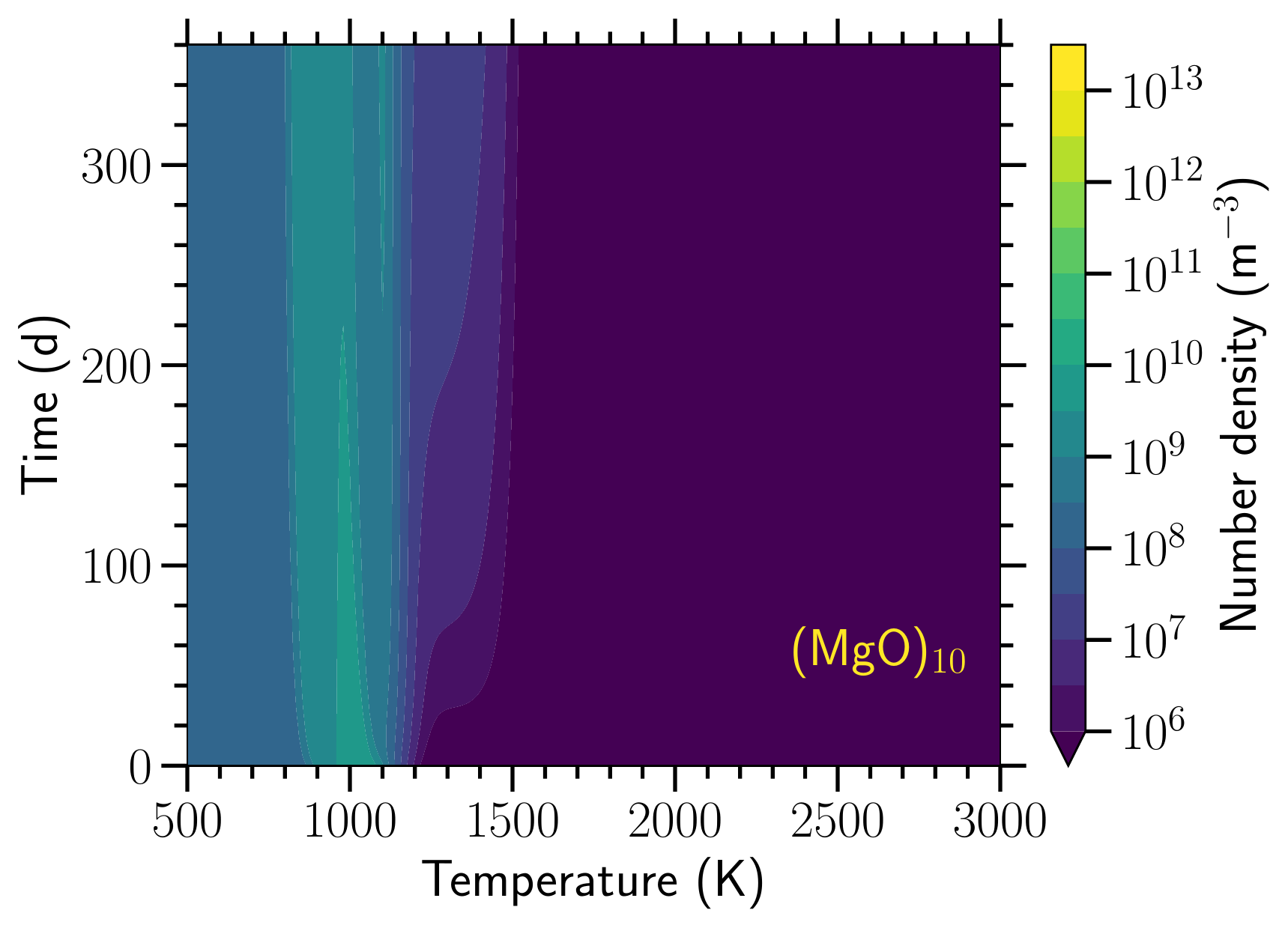}
        \end{flushleft}
        \caption{Temporal evolution of the absolute number density of all \protect\Mg{1}-clusters at the benchmark total gas density $\rho=\SI{1e-9}{\kg\per\m\cubed}$ for a closed nucleation model using the monomer nucleation description.}
        \label{fig:MgO_clusters_monomer_time_evolution}
    \end{figure*}

\begin{figure*}
        \begin{flushleft}
        \includegraphics[width=0.32\textwidth]{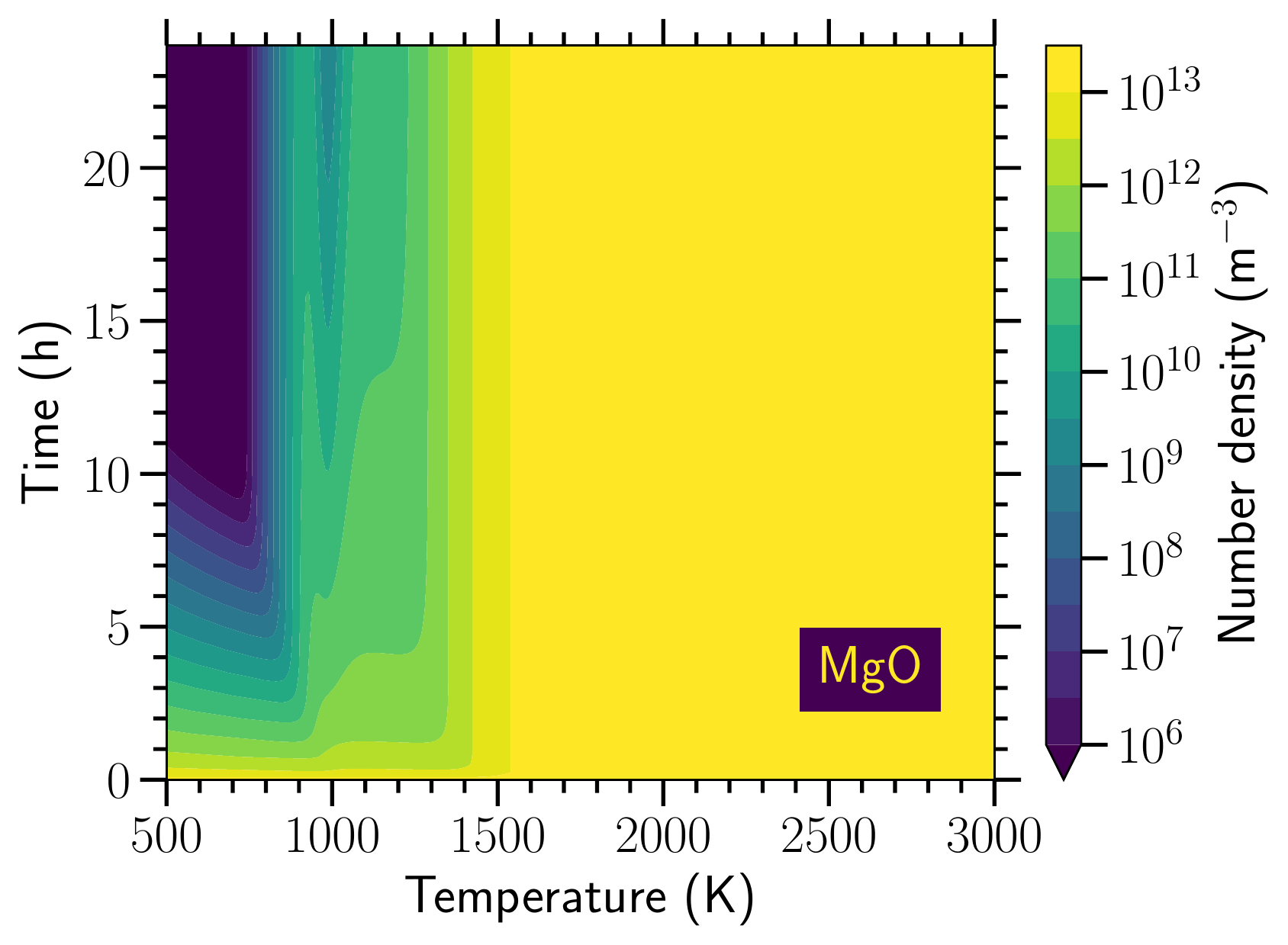}
        \includegraphics[width=0.32\textwidth]{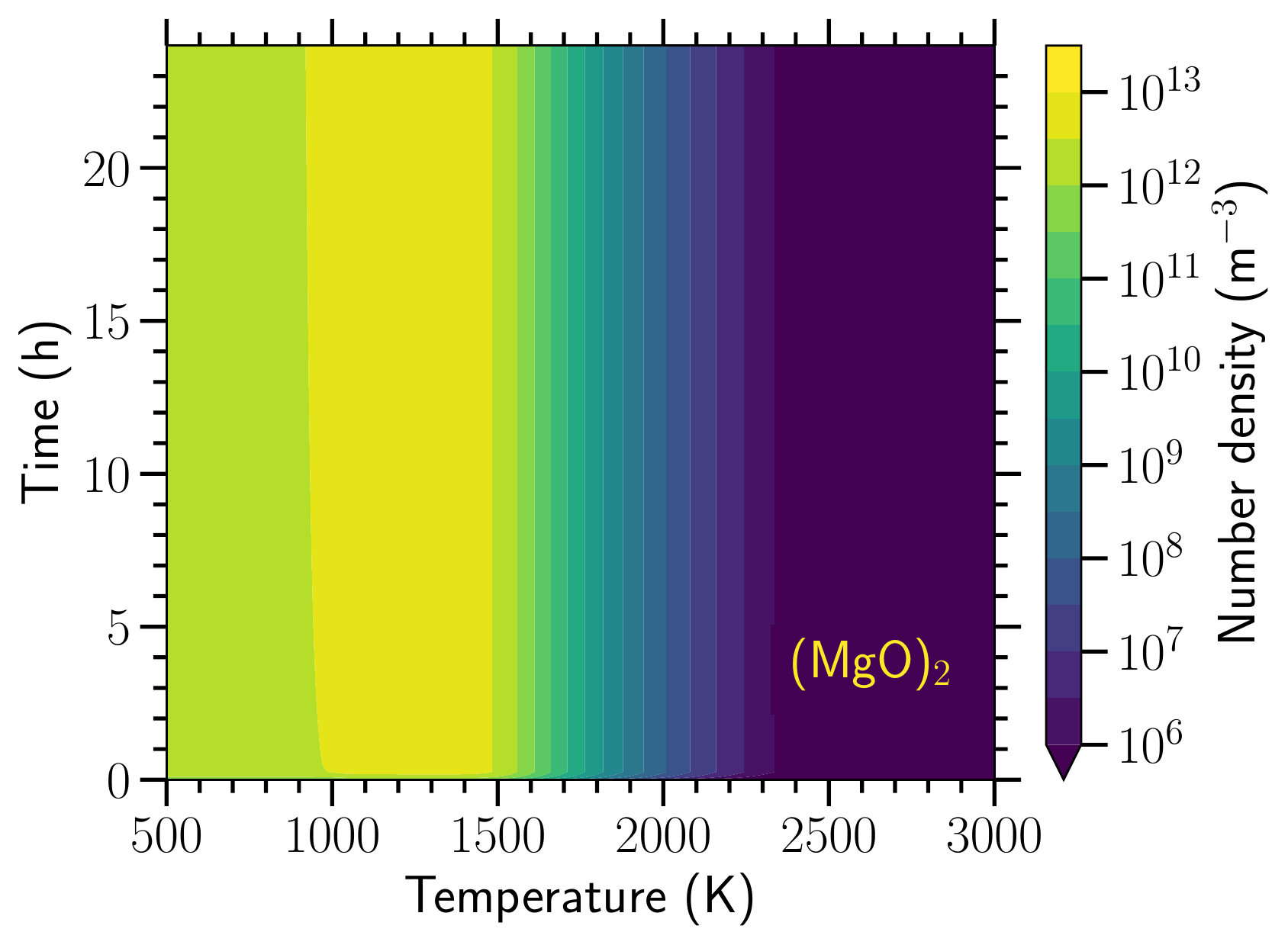}
        \includegraphics[width=0.32\textwidth]{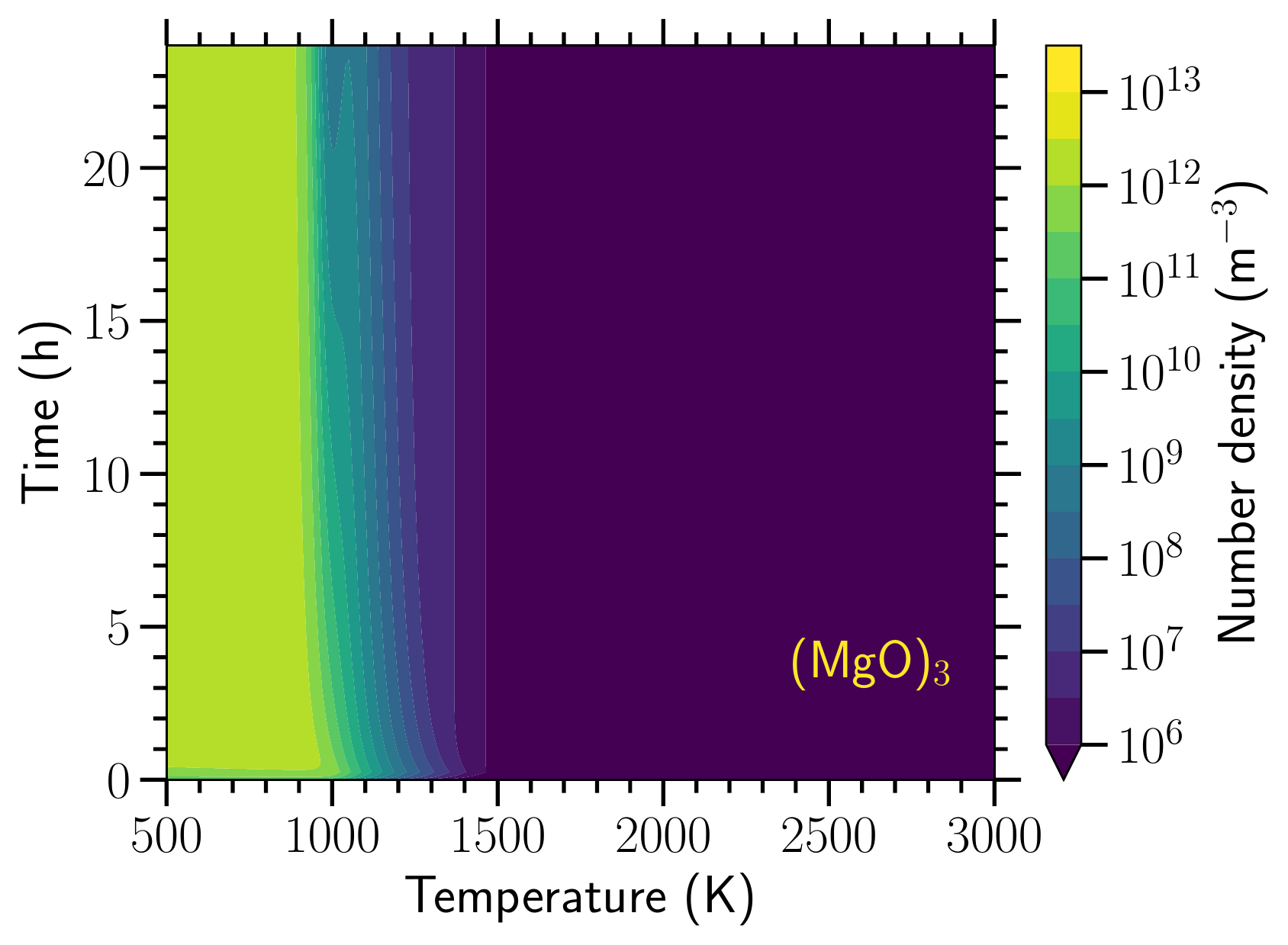}
        \includegraphics[width=0.32\textwidth]{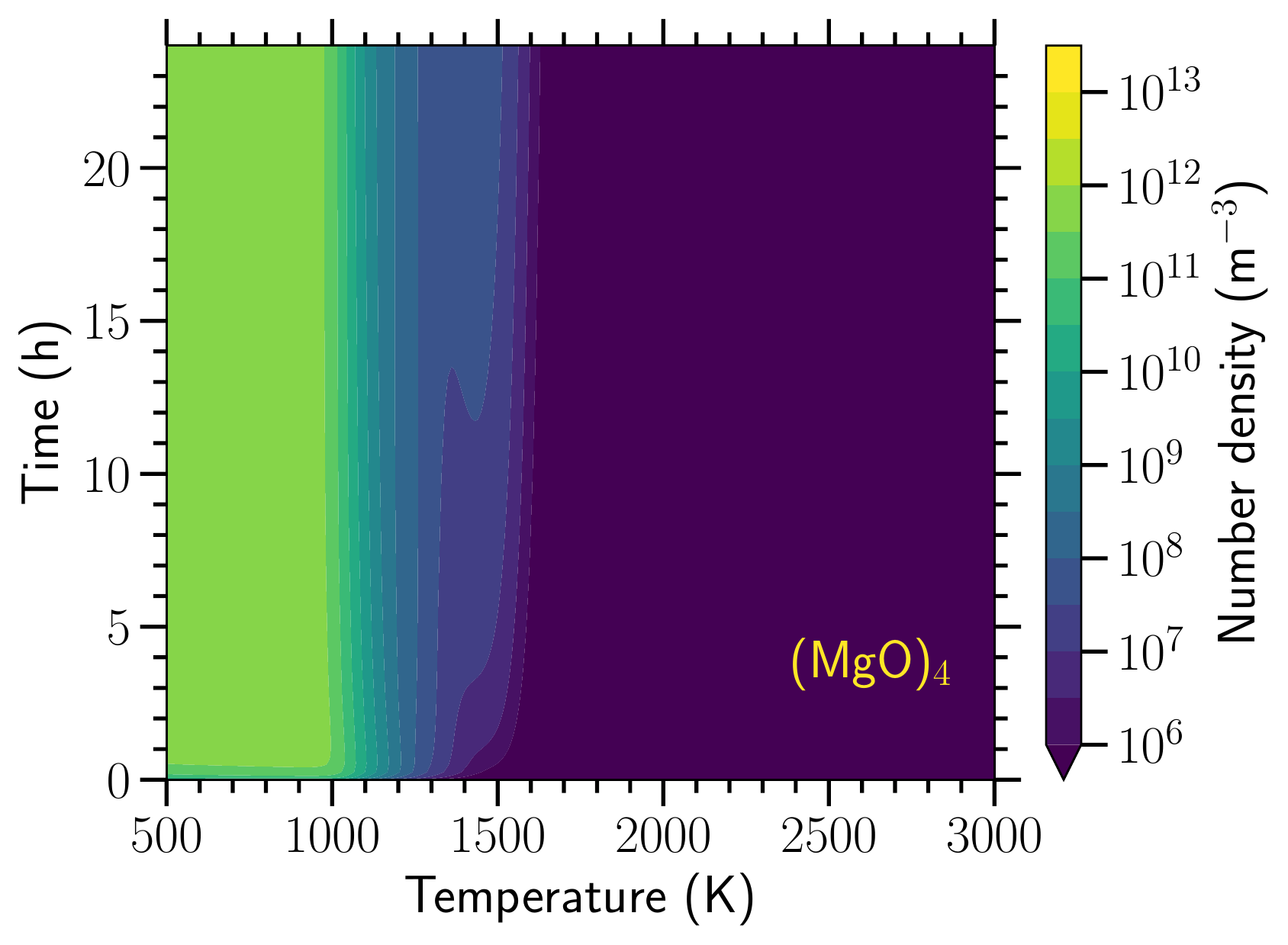}
        \includegraphics[width=0.32\textwidth]{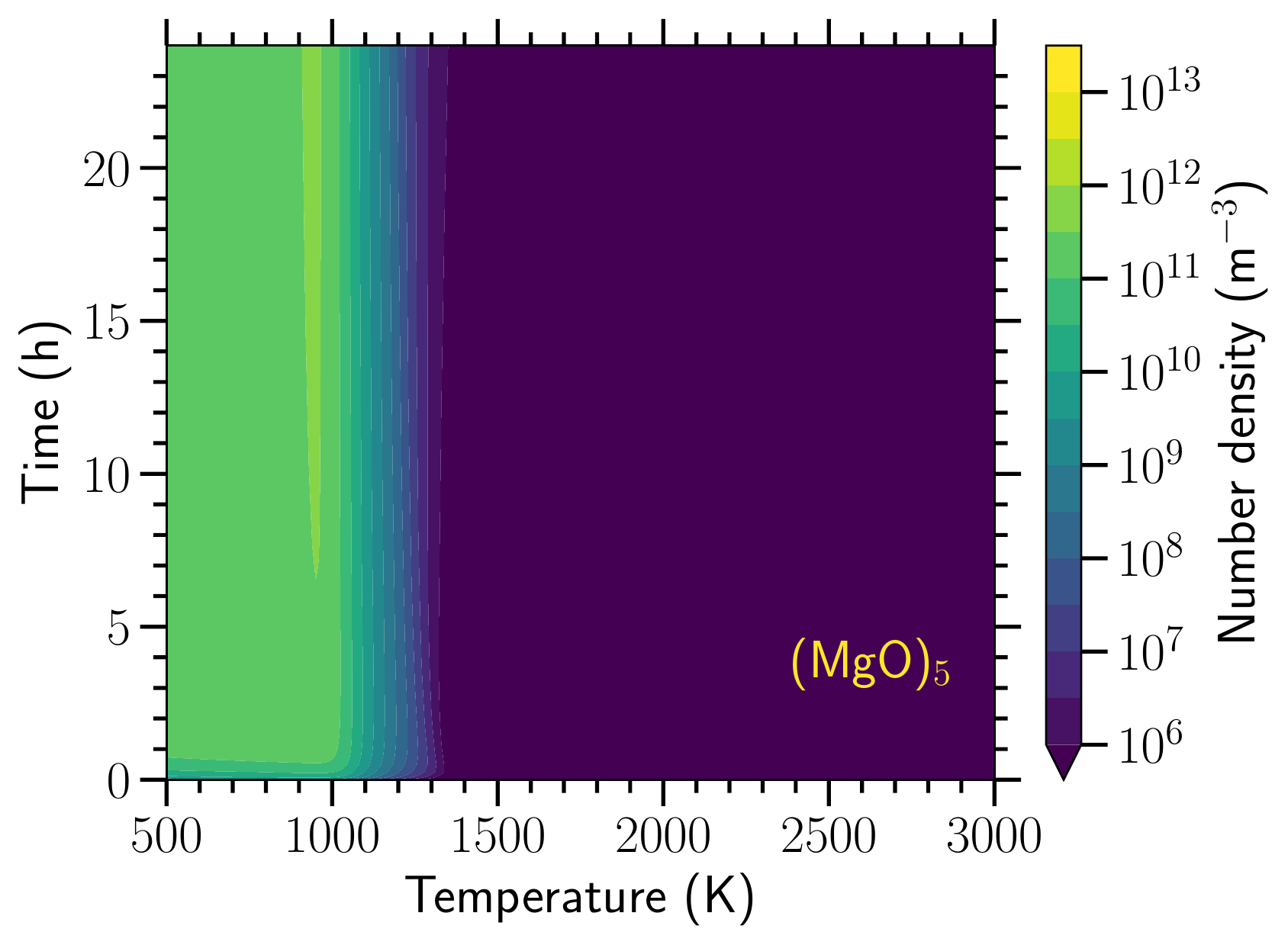}
        \includegraphics[width=0.32\textwidth]{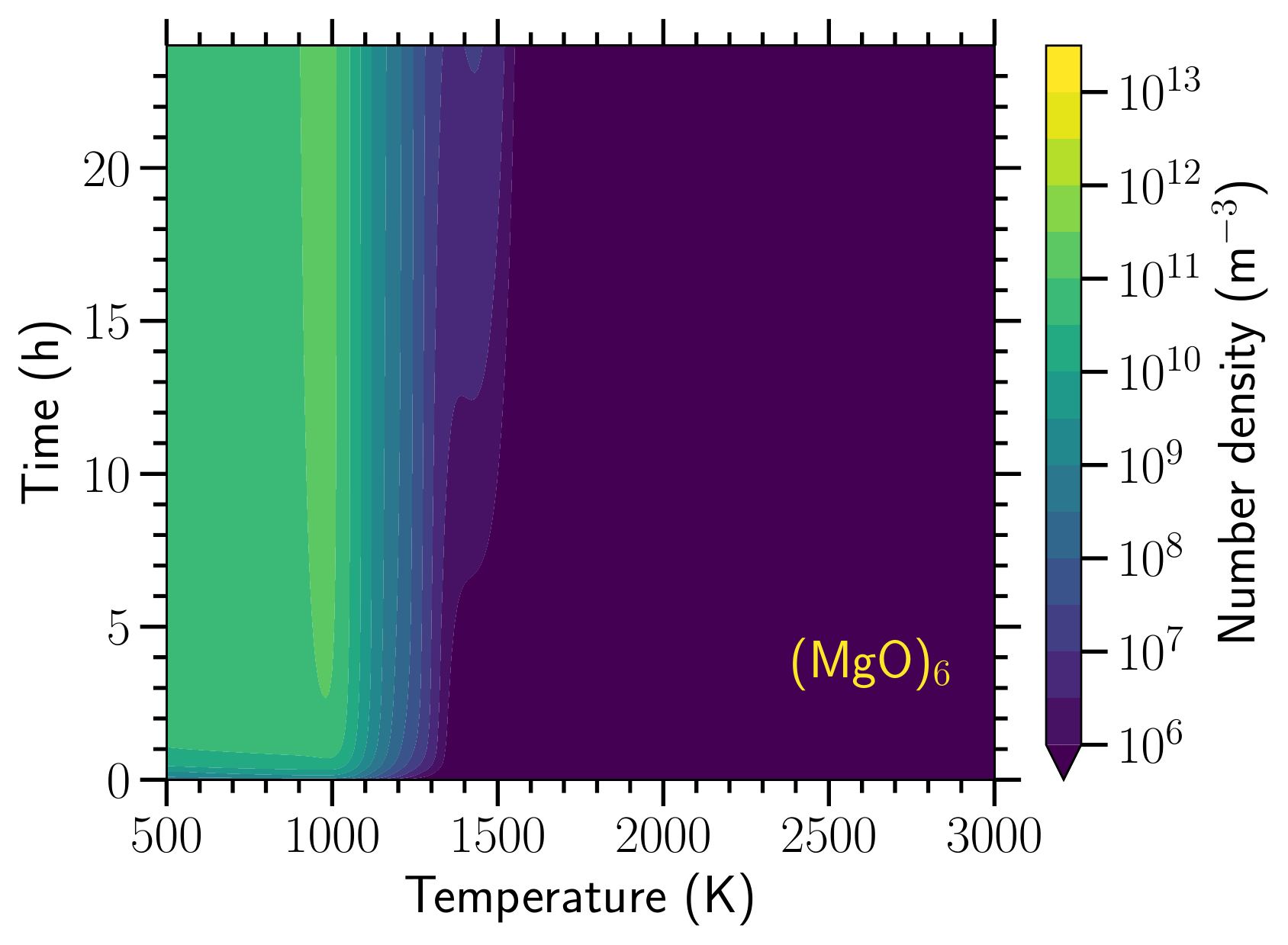}
        \includegraphics[width=0.32\textwidth]{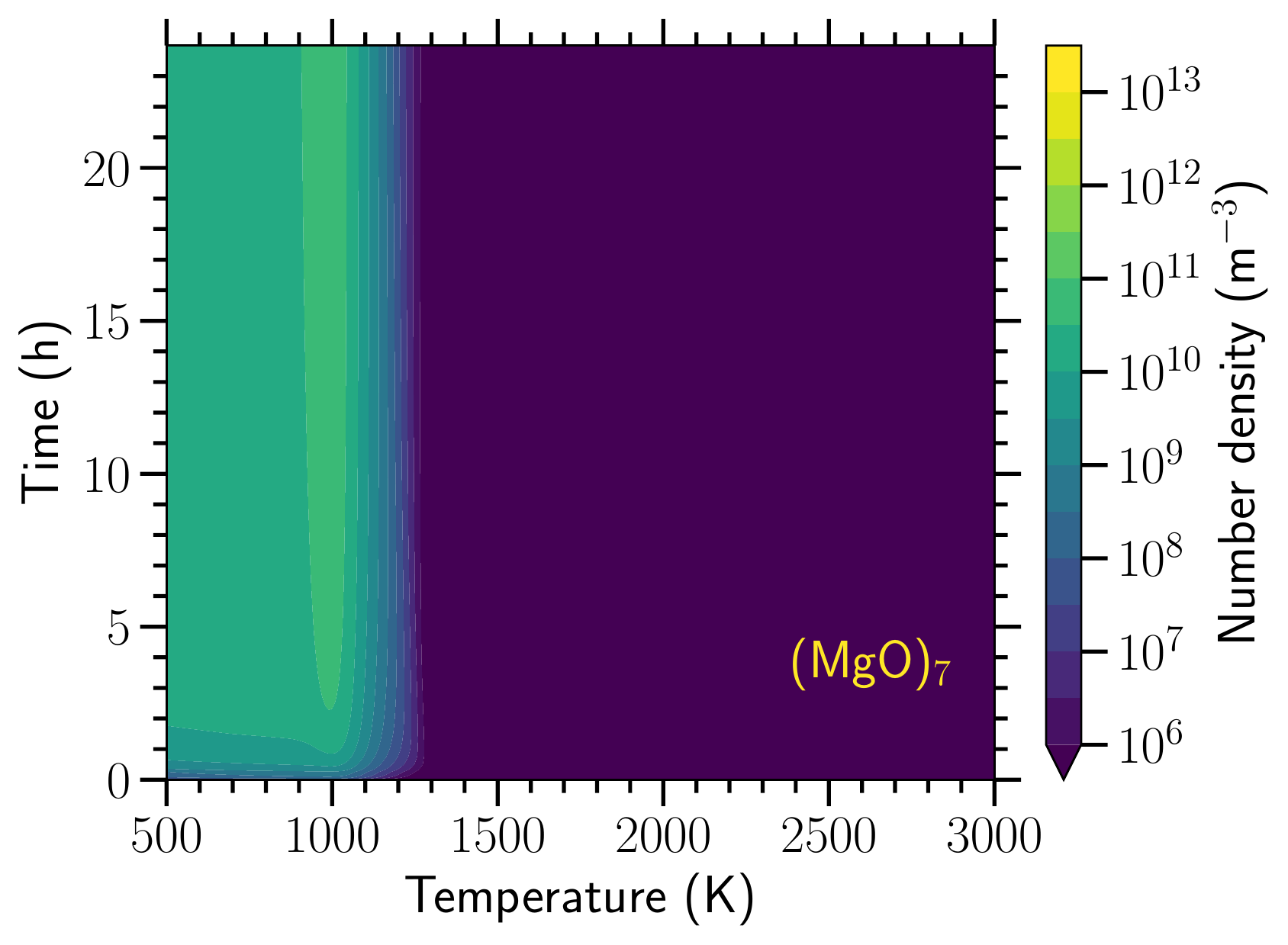}
        \includegraphics[width=0.32\textwidth]{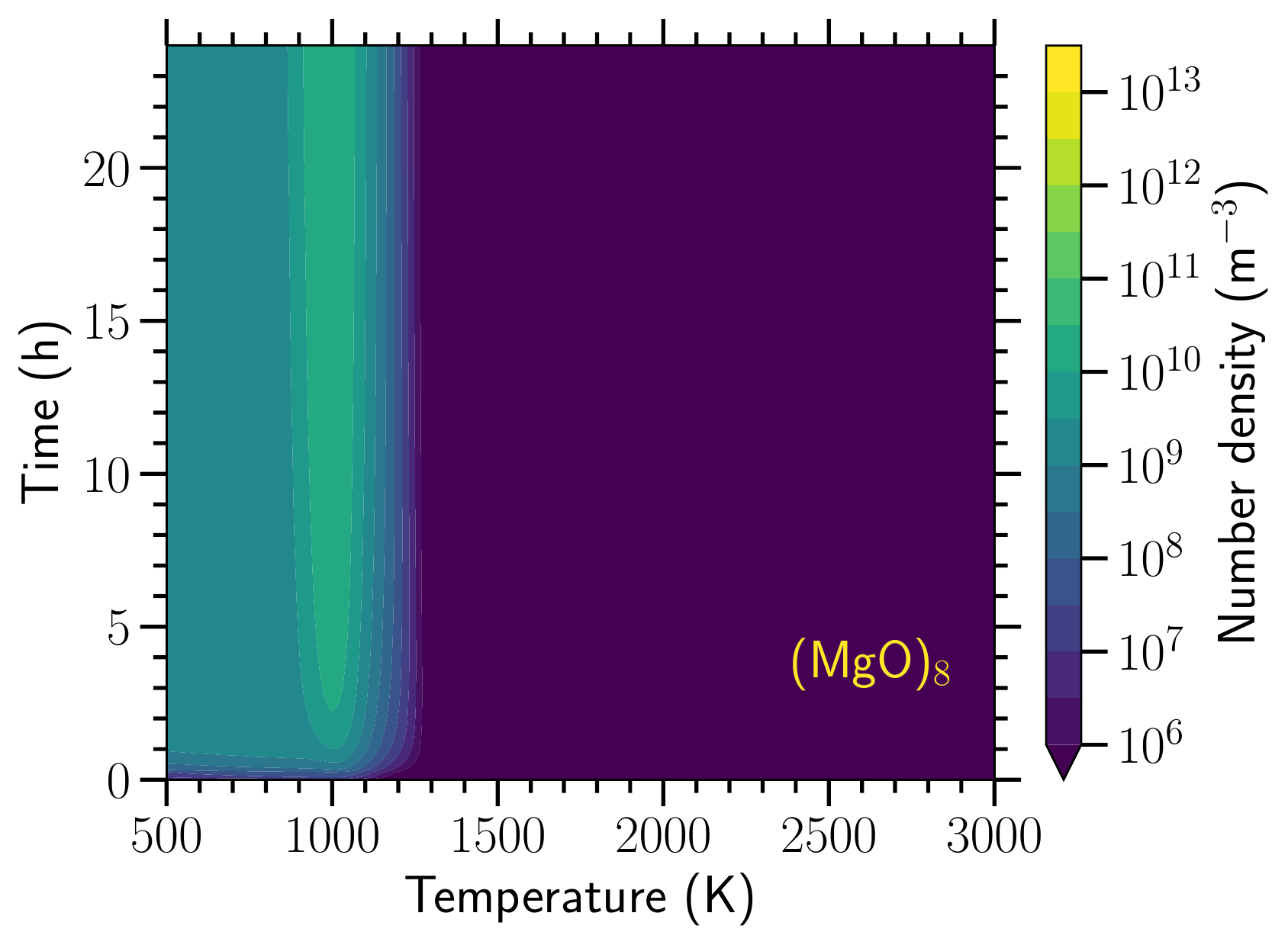}
        \includegraphics[width=0.32\textwidth]{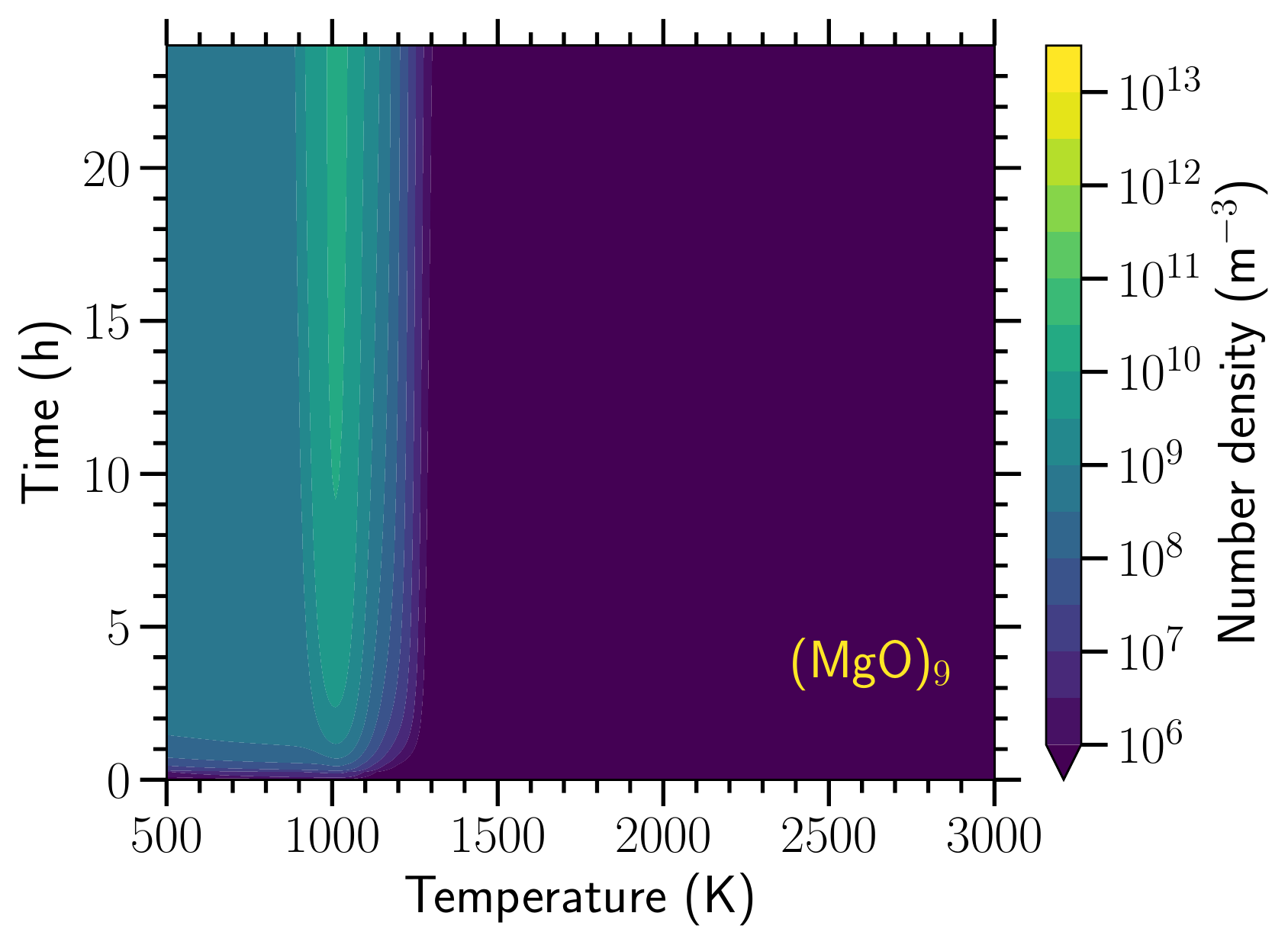}
        \includegraphics[width=0.32\textwidth]{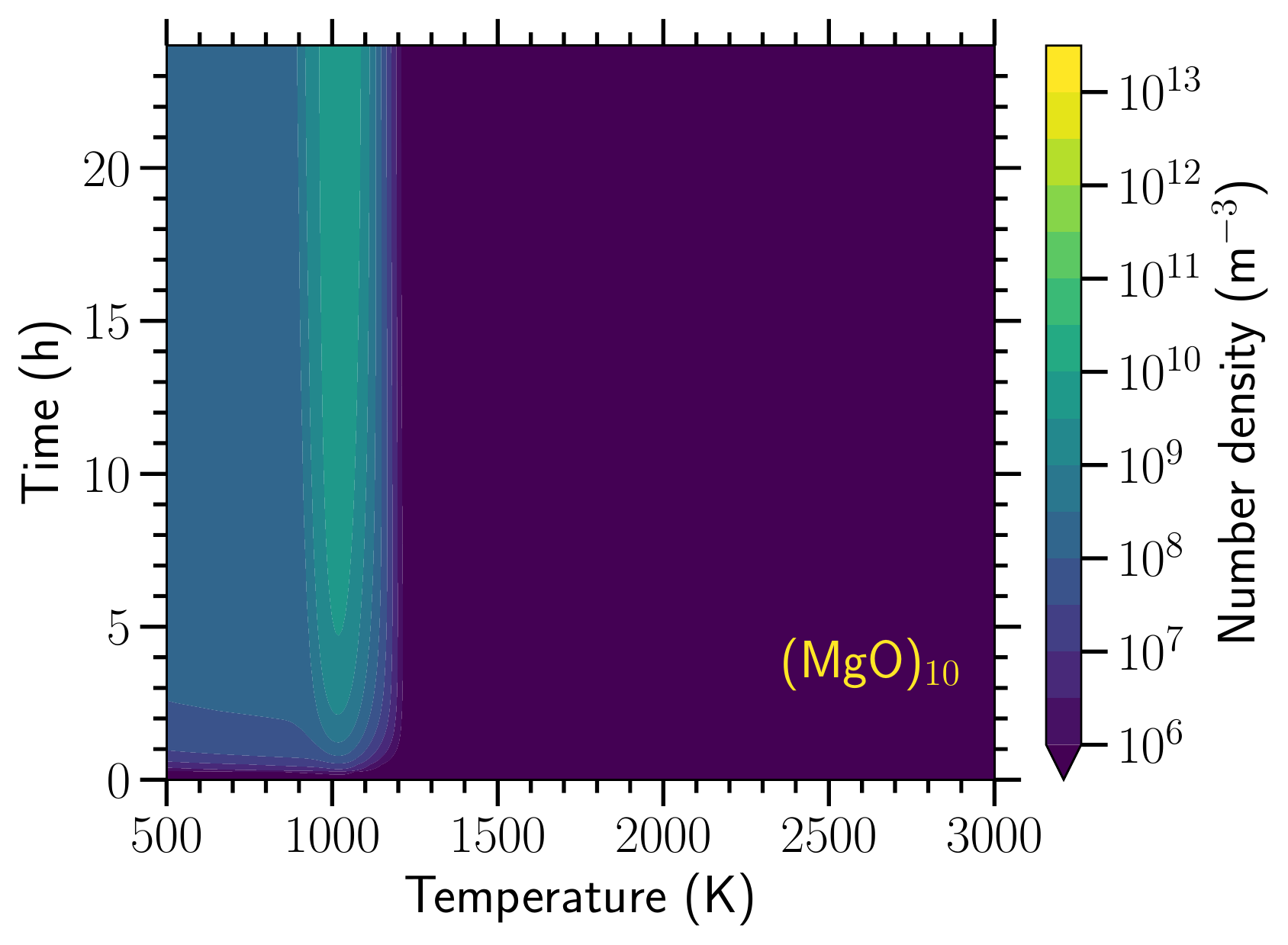}
        \end{flushleft}
        \caption{Refined temporal evolution of the absolute number density of all \protect\Mg{1}-clusters at the benchmark total gas density $\rho=\SI{1e-9}{\kg\per\m\cubed}$ for a closed nucleation model using the monomer nucleation description.}
        \label{fig:MgO_clusters_monomer_time_evolution_short}
    \end{figure*}

\begin{figure*}
        \begin{flushleft}
        \includegraphics[width=0.32\textwidth]{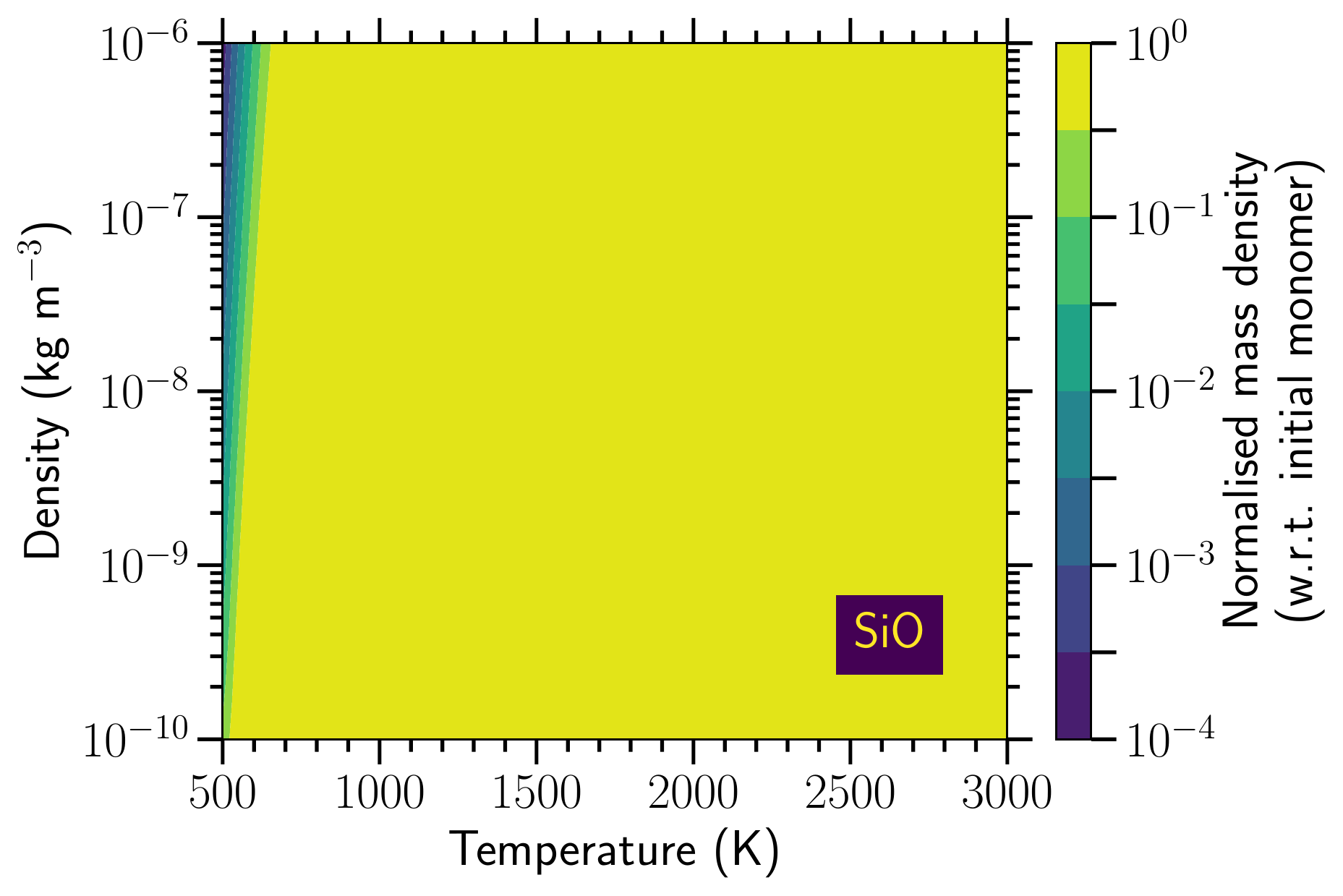}
        \includegraphics[width=0.32\textwidth]{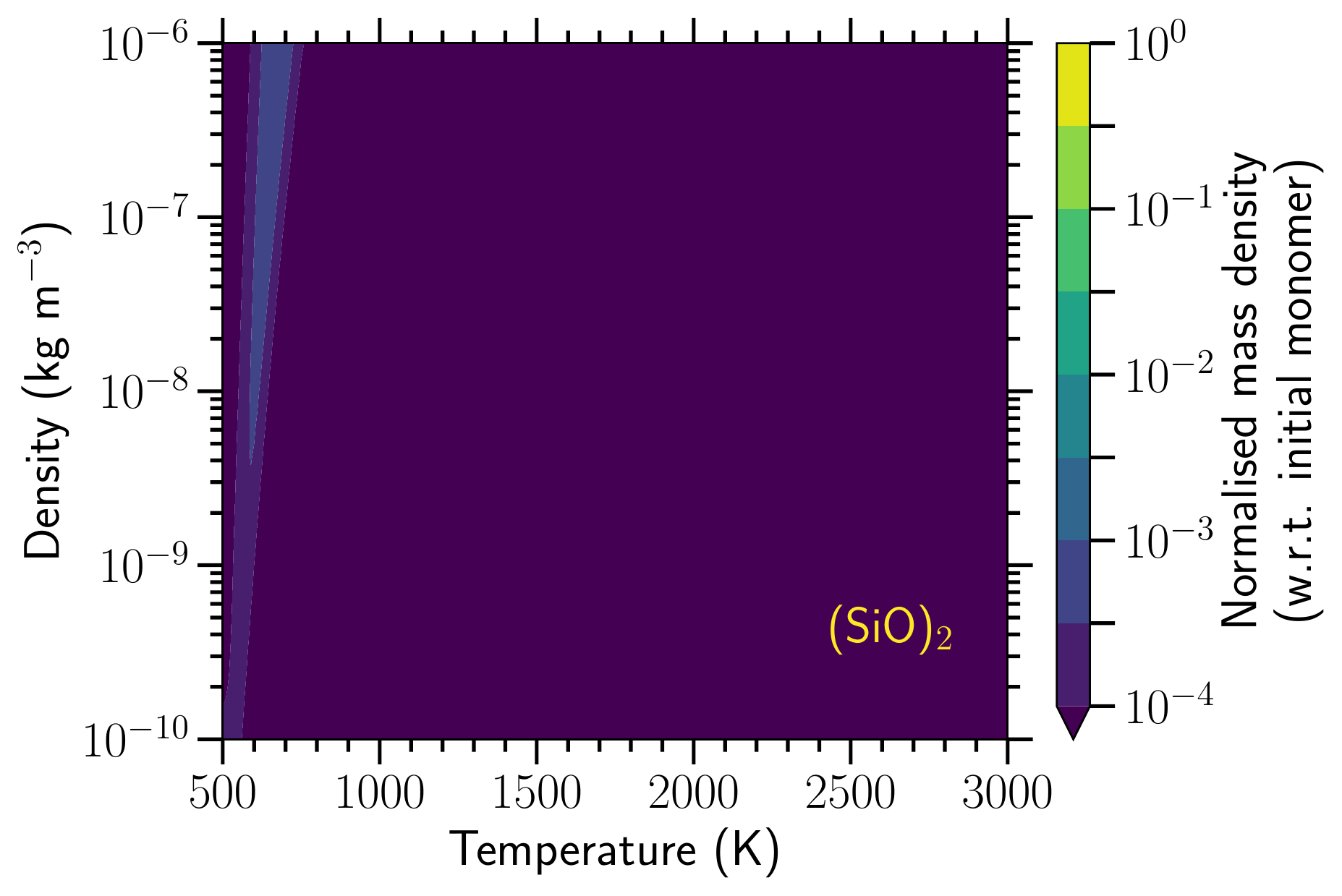}
        \includegraphics[width=0.32\textwidth]{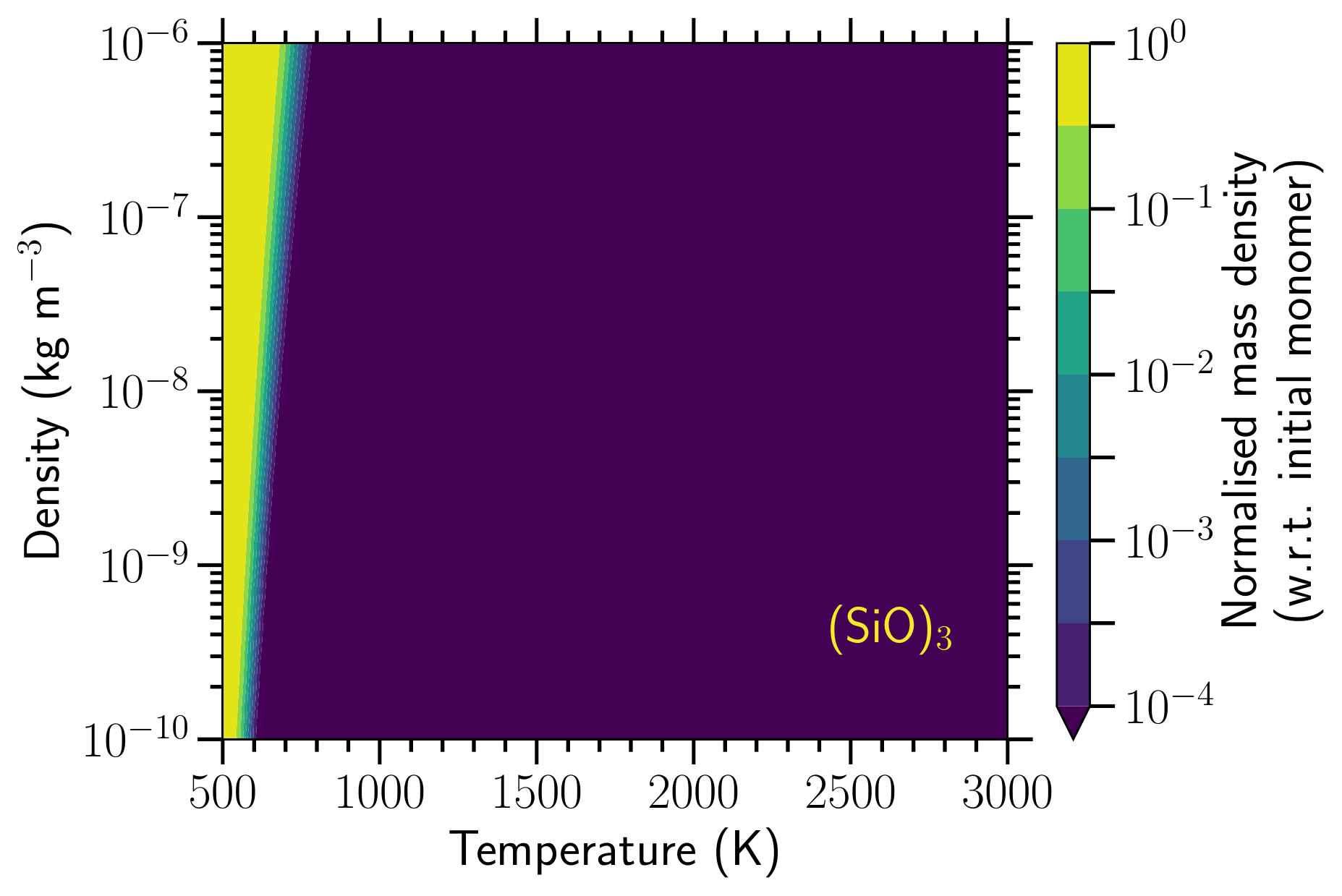}
        \includegraphics[width=0.32\textwidth]{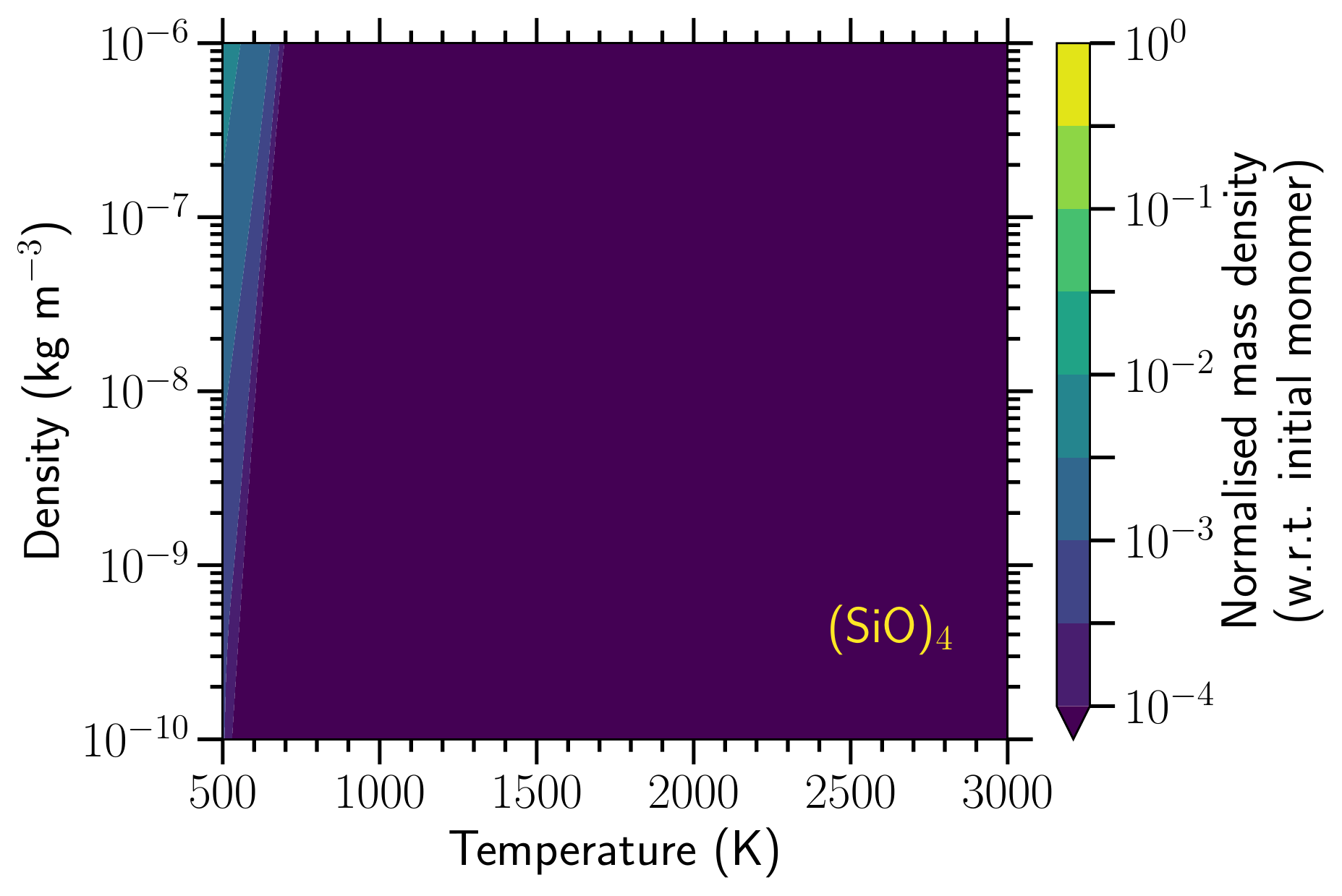}
        \includegraphics[width=0.32\textwidth]{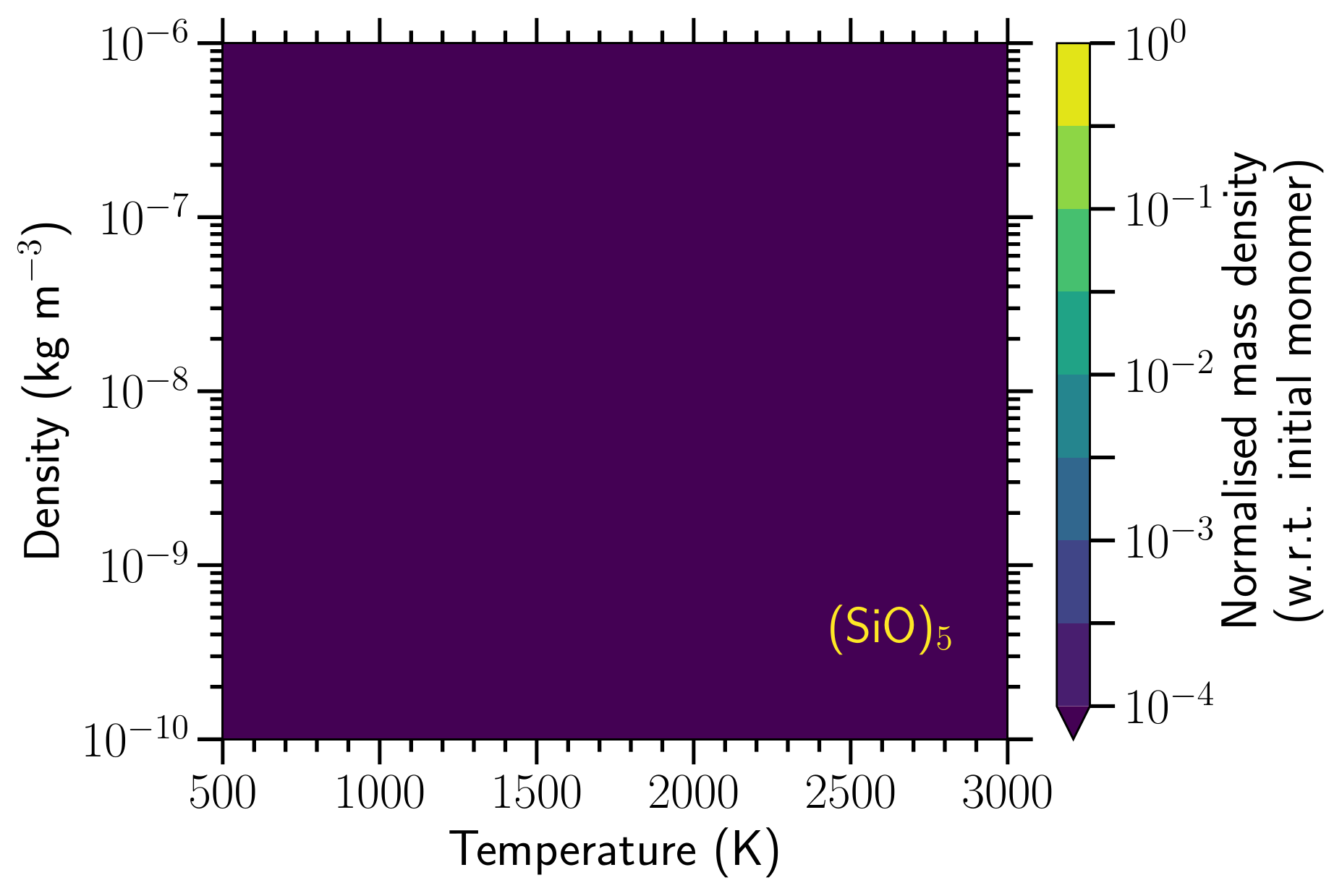}
        \includegraphics[width=0.32\textwidth]{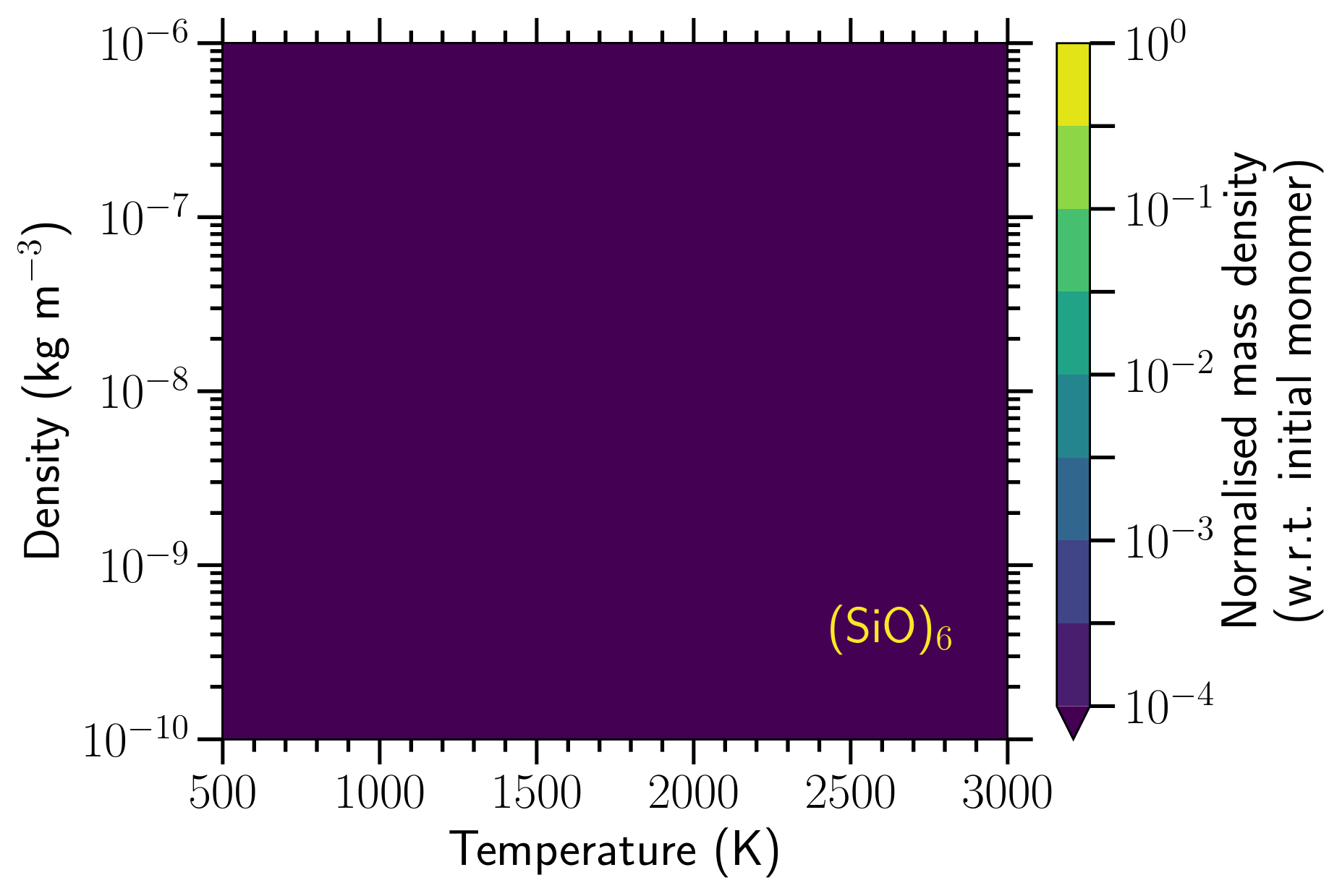}
        \includegraphics[width=0.32\textwidth]{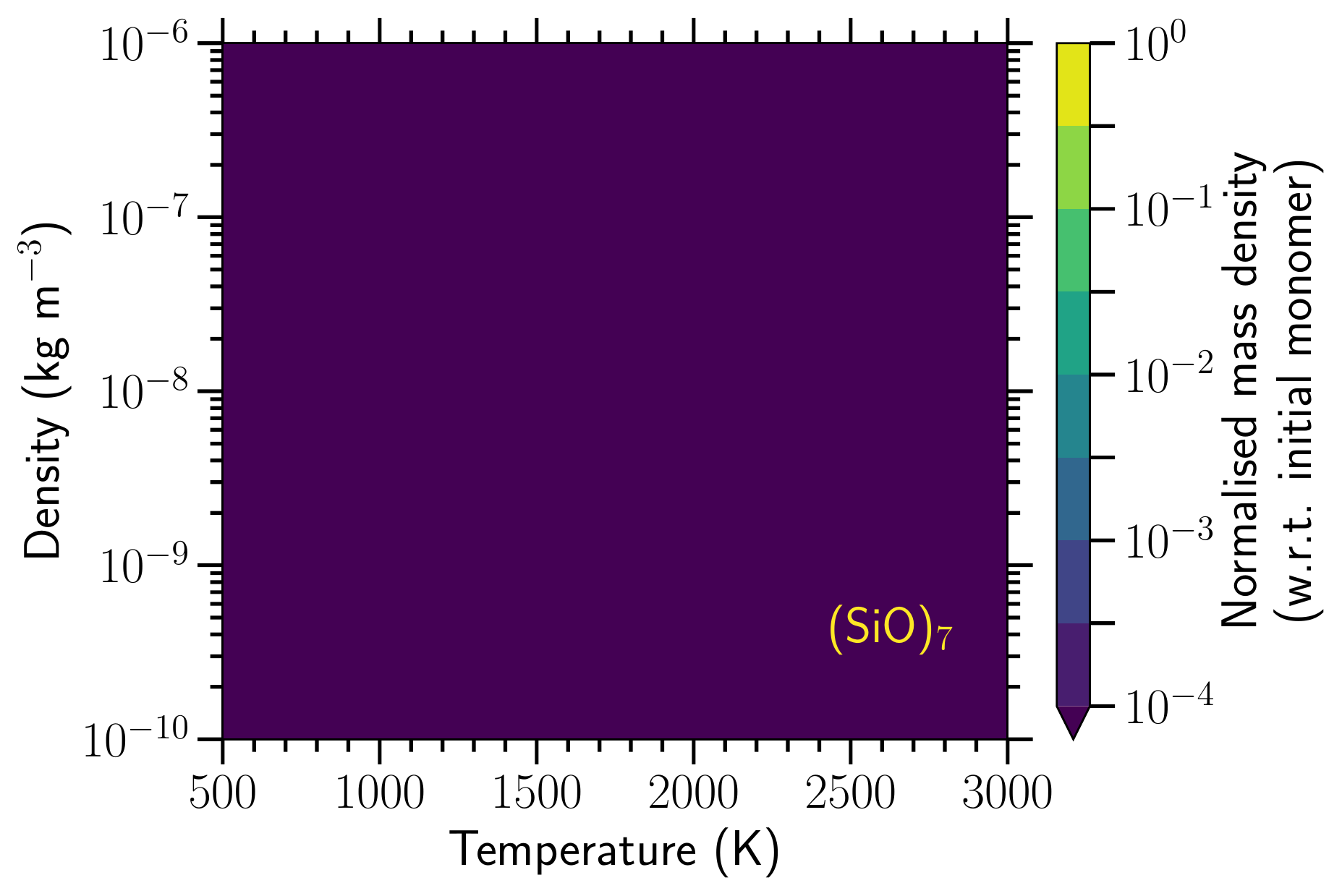}
        \includegraphics[width=0.32\textwidth]{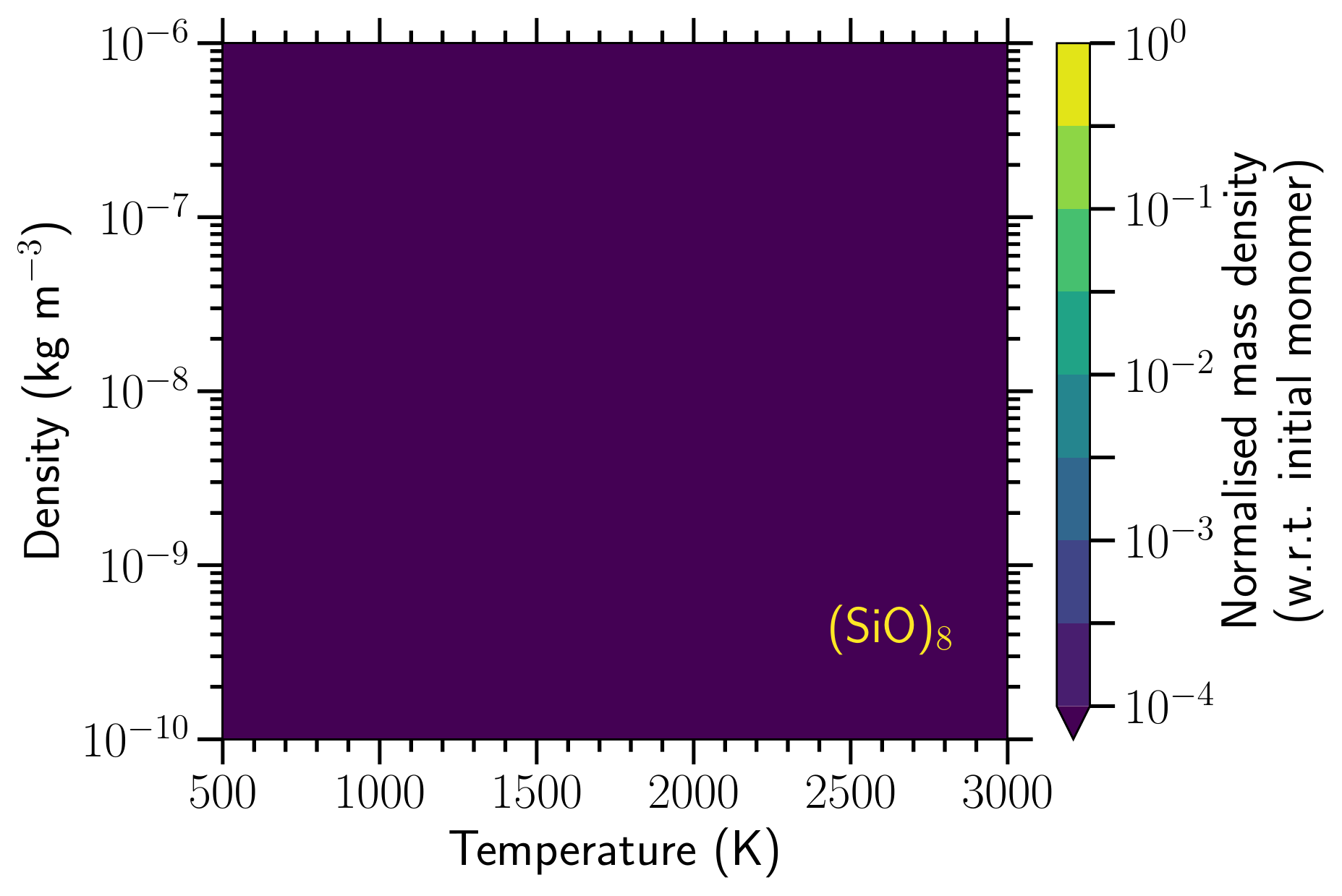}
        \includegraphics[width=0.32\textwidth]{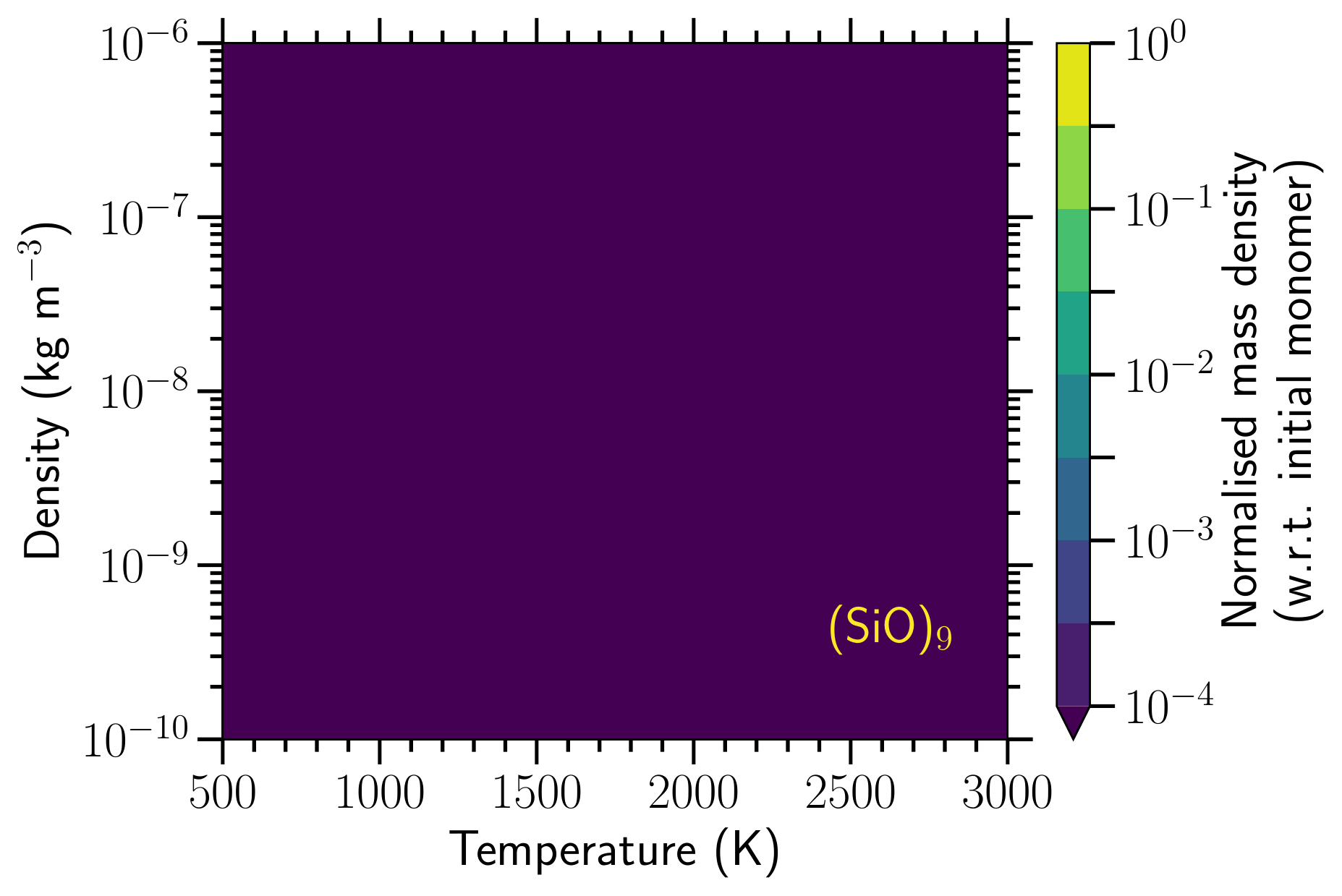}
        \includegraphics[width=0.32\textwidth]{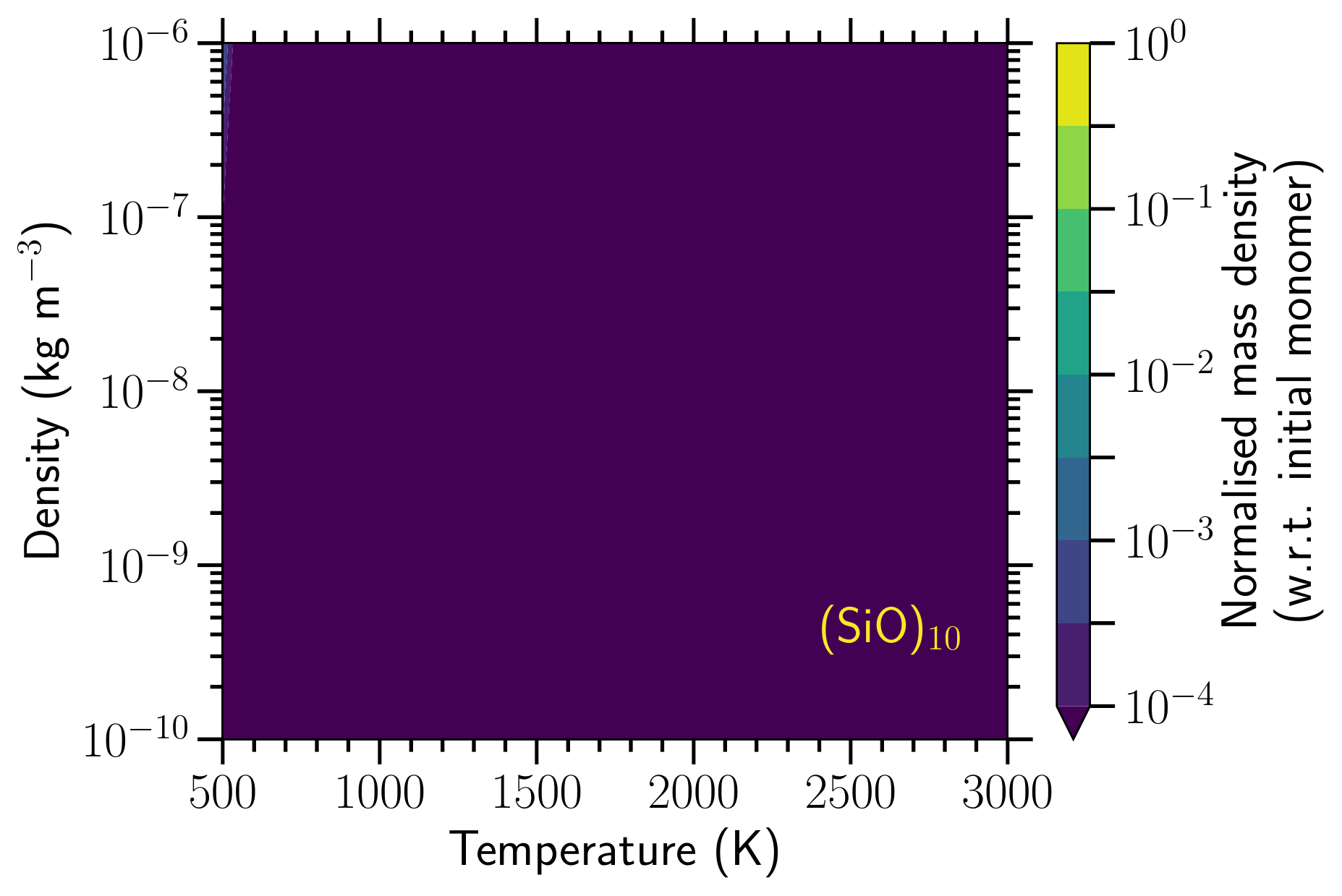}
        \end{flushleft}
        \caption{Overview of the normalised mass density after one year of all \protect\SiO{1}-clusters for a closed nucleation model using the monomer nucleation description.}
        \label{fig:SiO_clusters_monomer_norm_same_scale}
    \end{figure*}

\begin{figure*}
        \begin{flushleft}
        \includegraphics[width=0.32\textwidth]{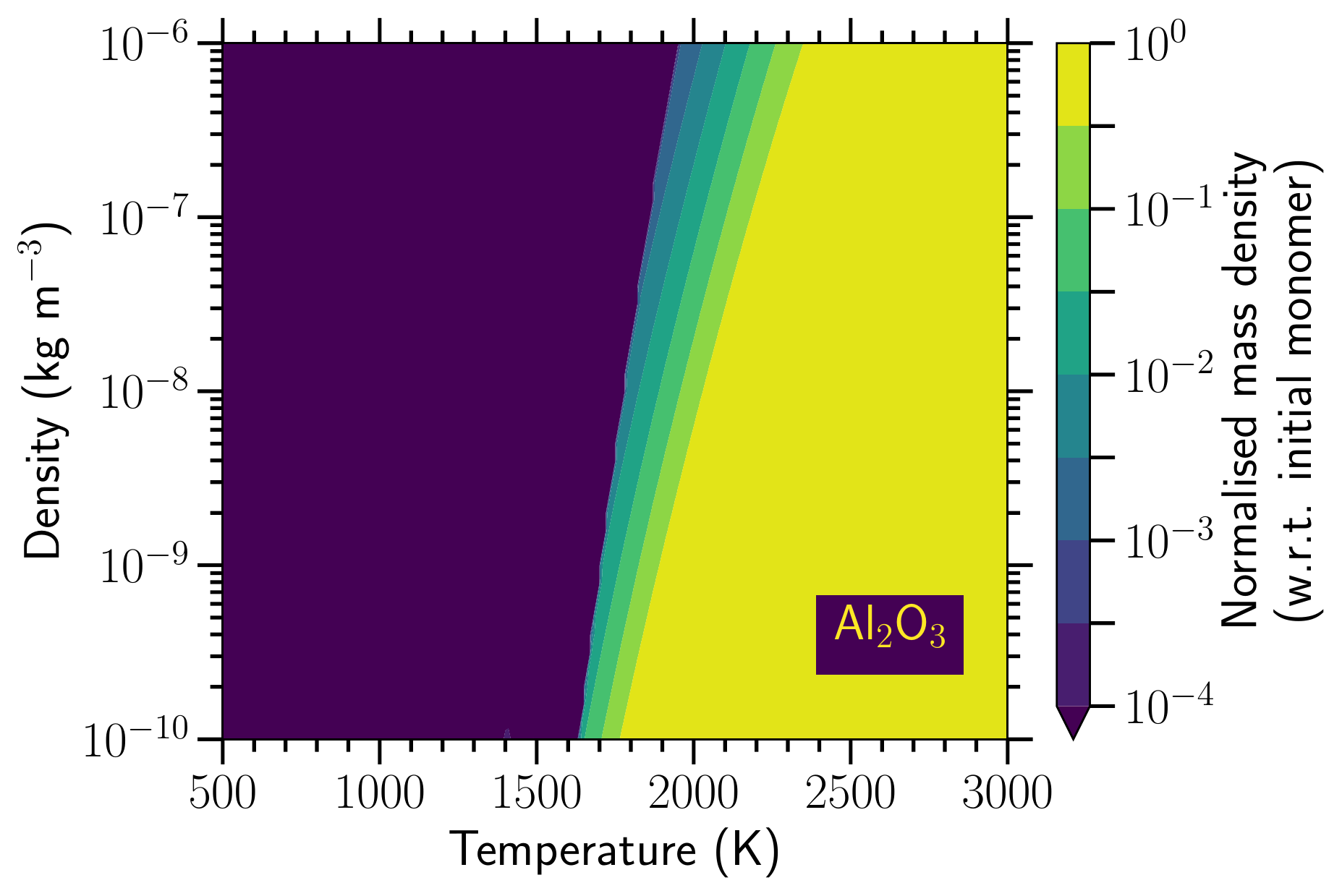}
        \includegraphics[width=0.32\textwidth]{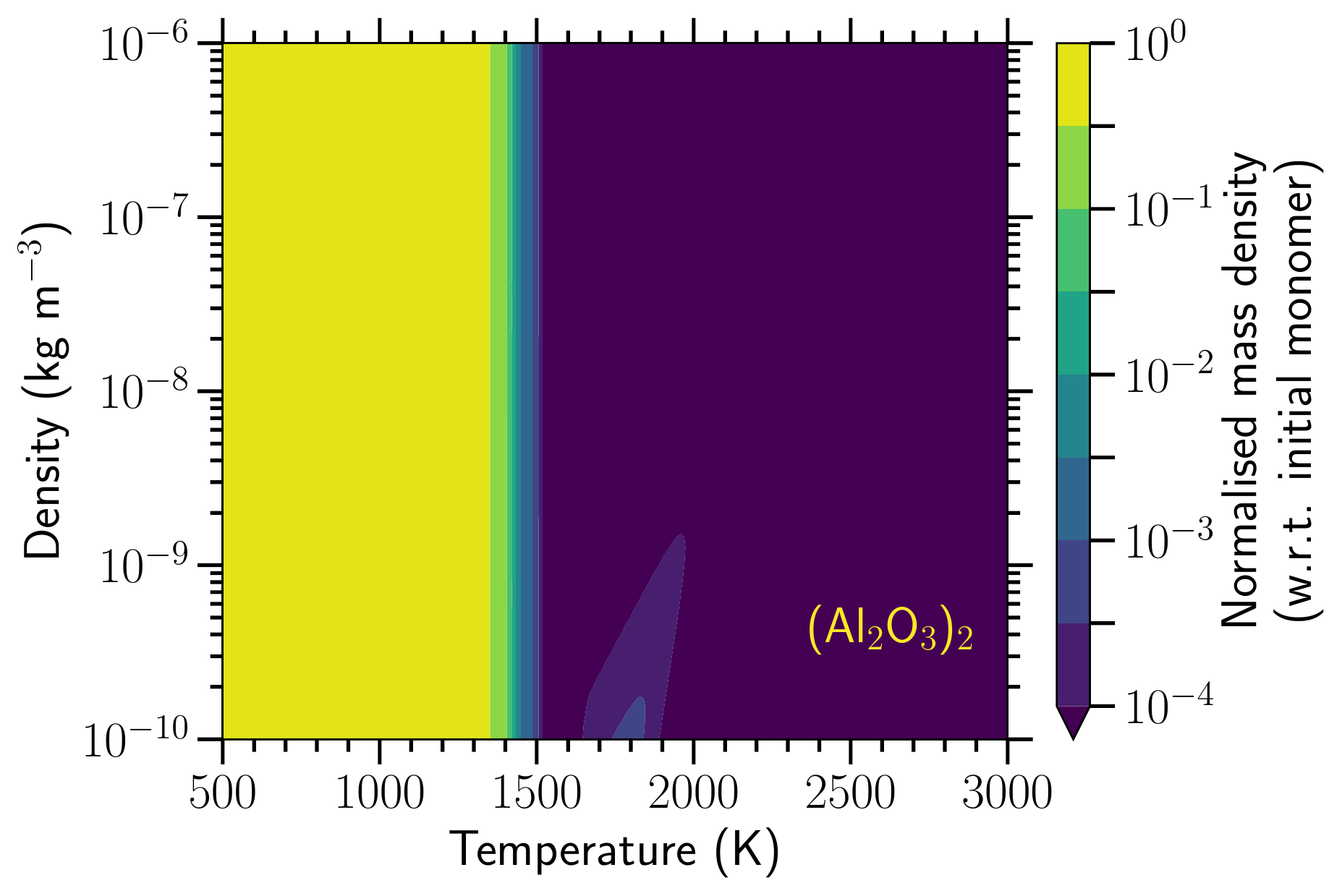}
        \includegraphics[width=0.32\textwidth]{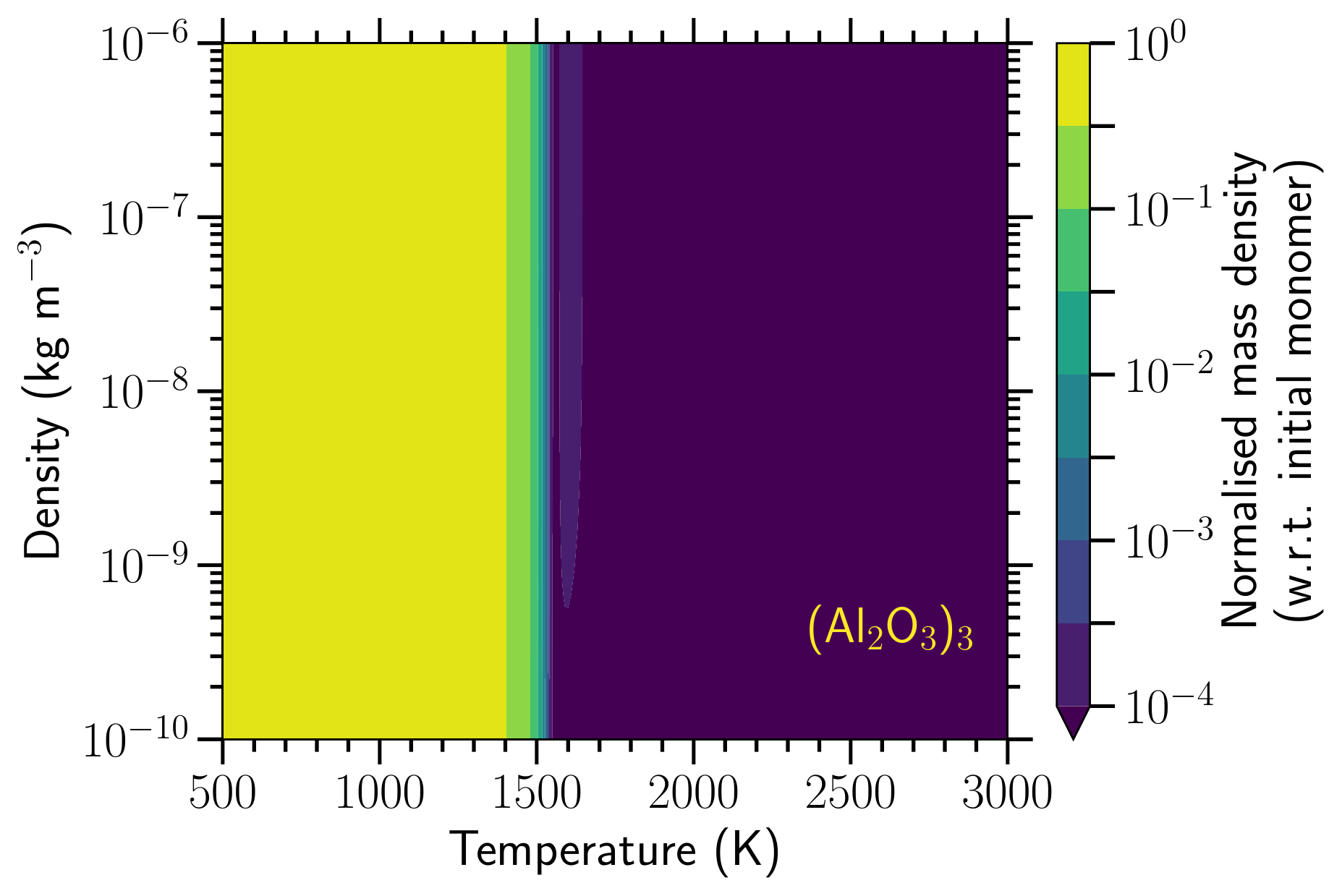}
        \includegraphics[width=0.32\textwidth]{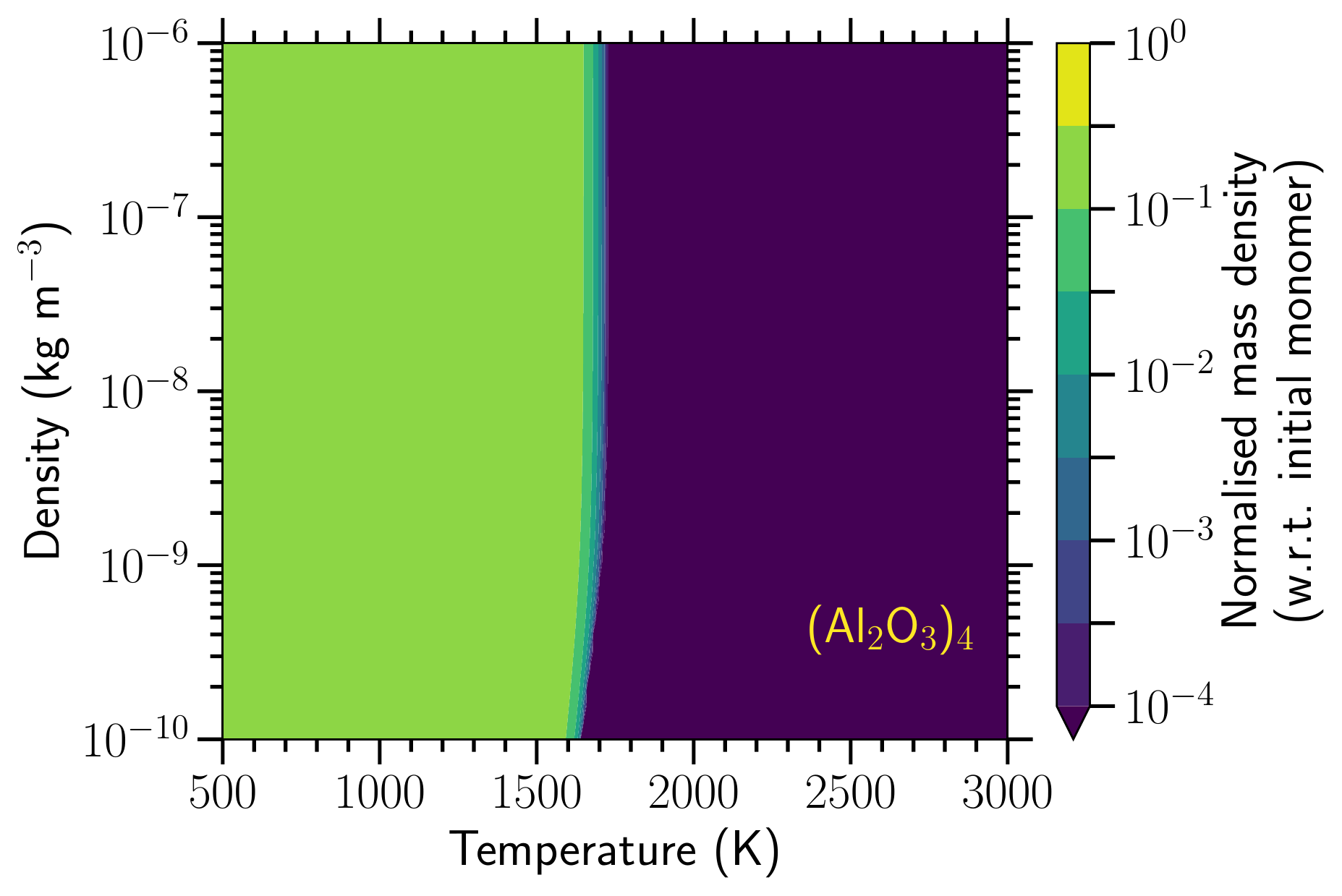}
        \includegraphics[width=0.32\textwidth]{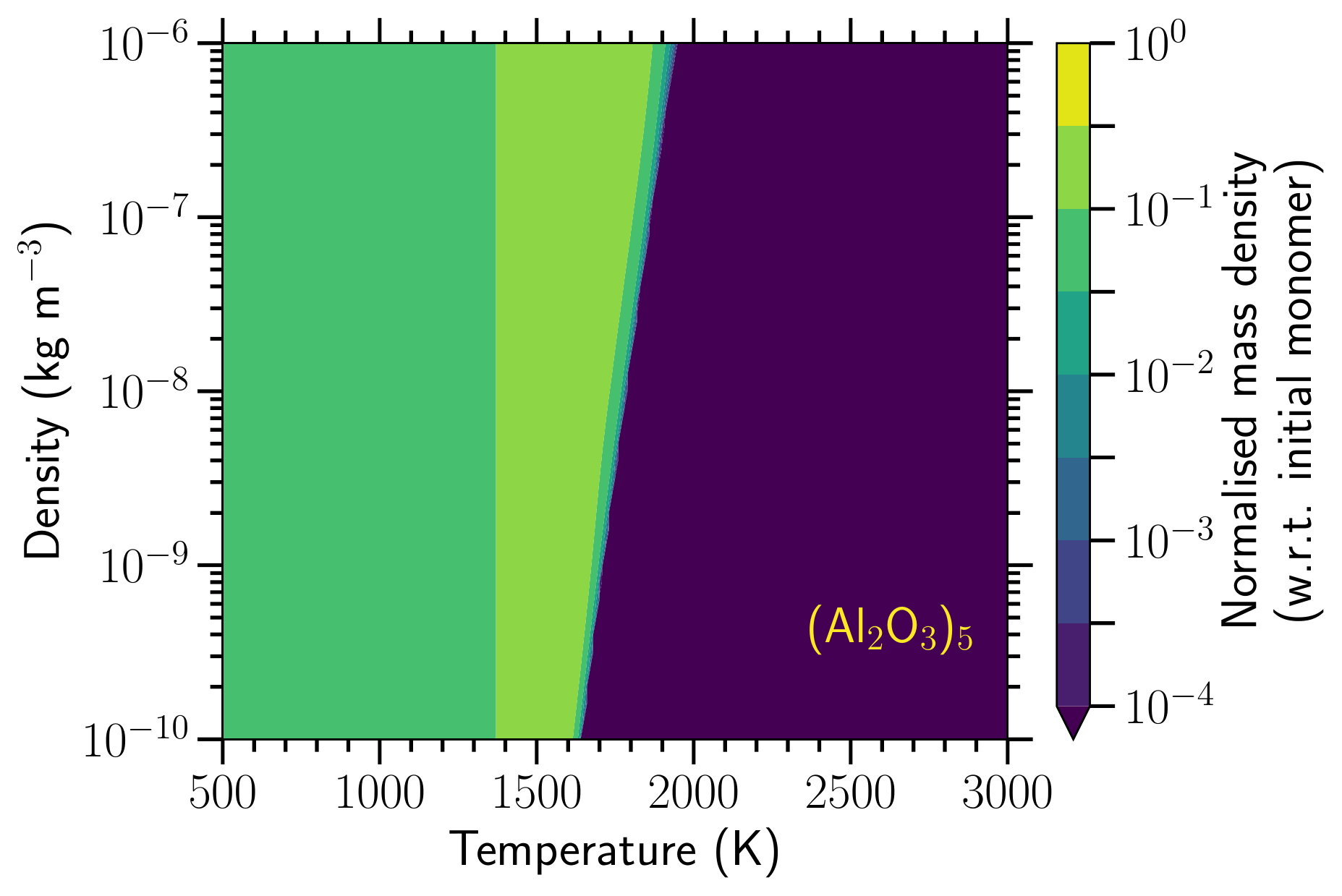}
        \includegraphics[width=0.32\textwidth]{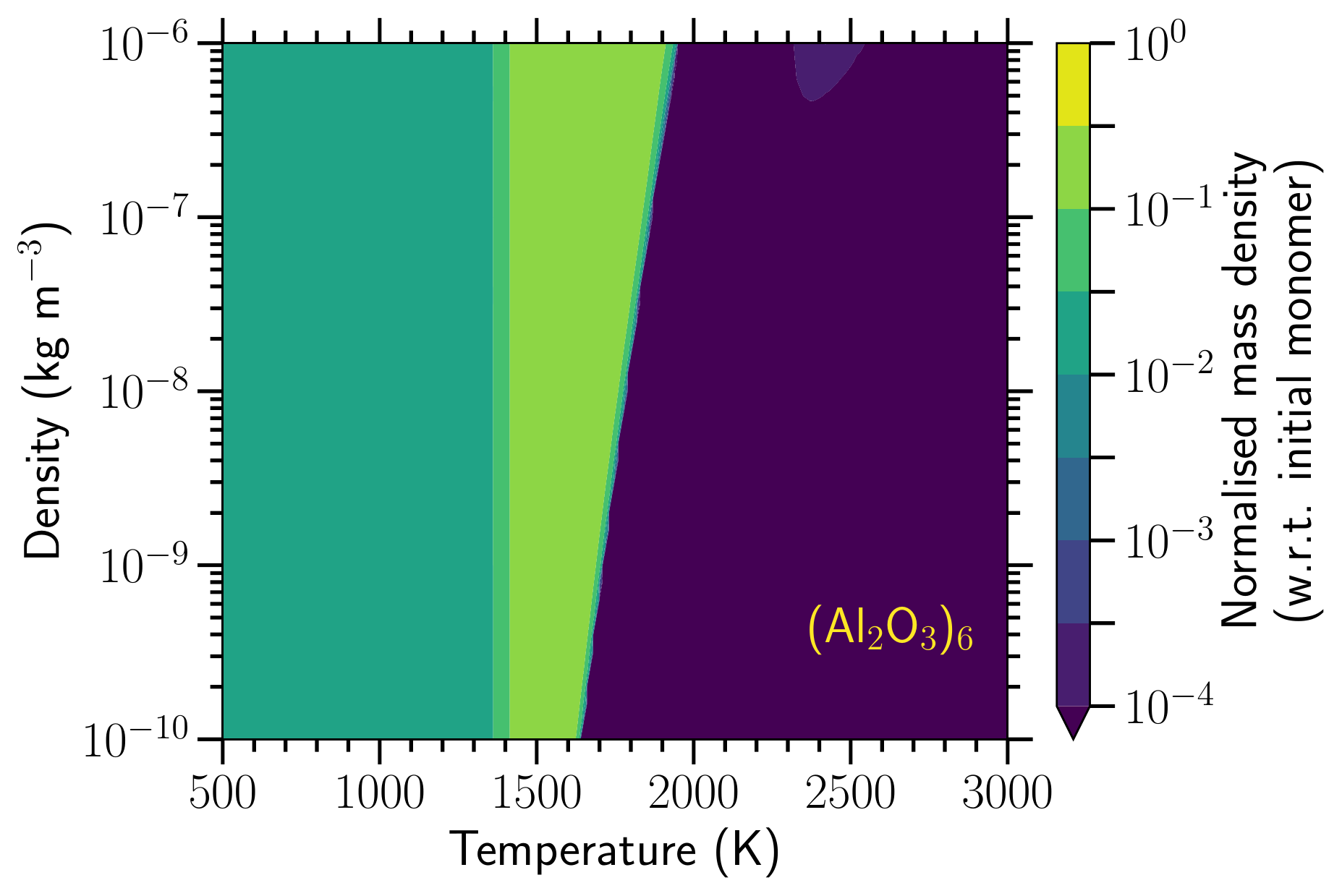}
        \includegraphics[width=0.32\textwidth]{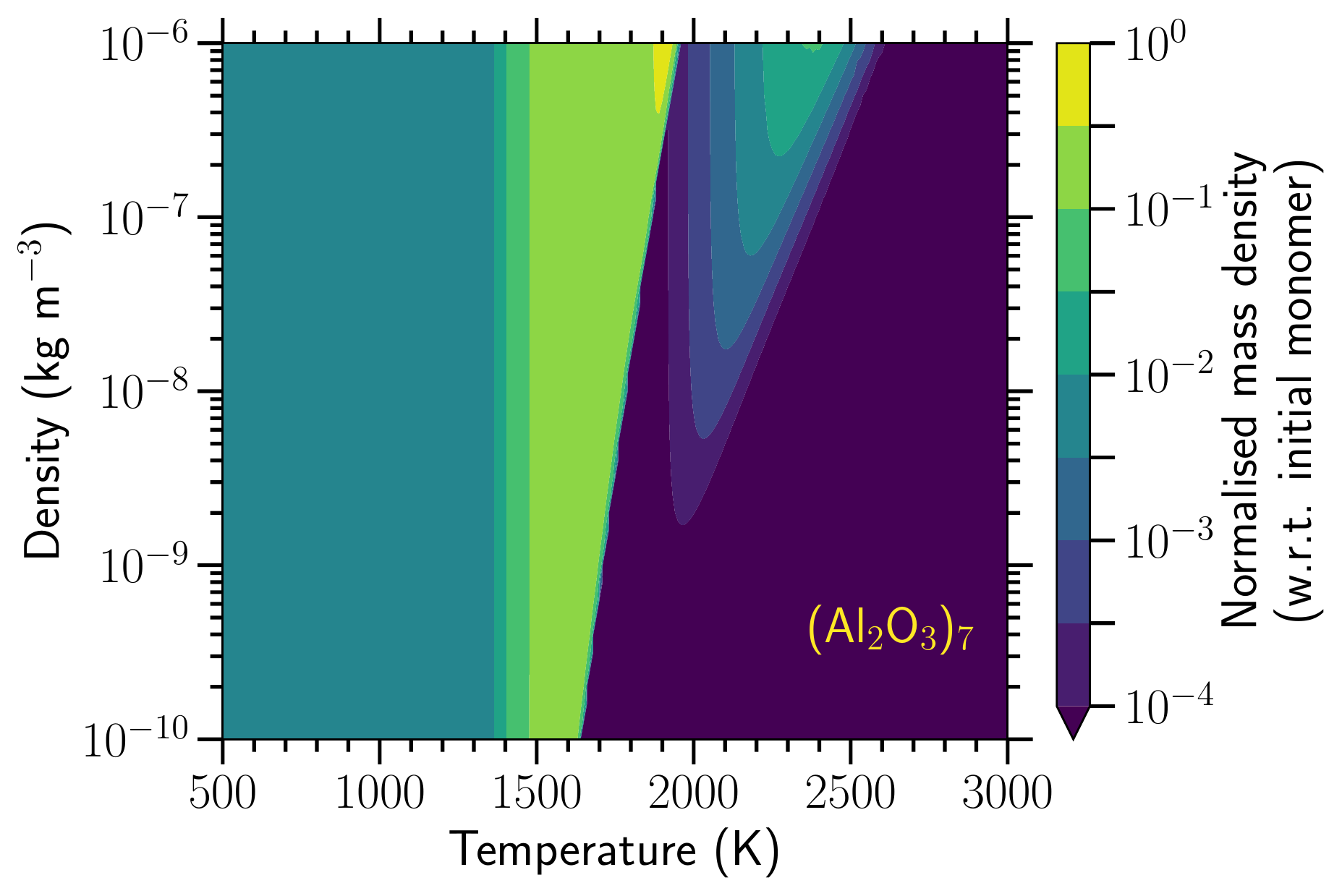}
        \includegraphics[width=0.32\textwidth]{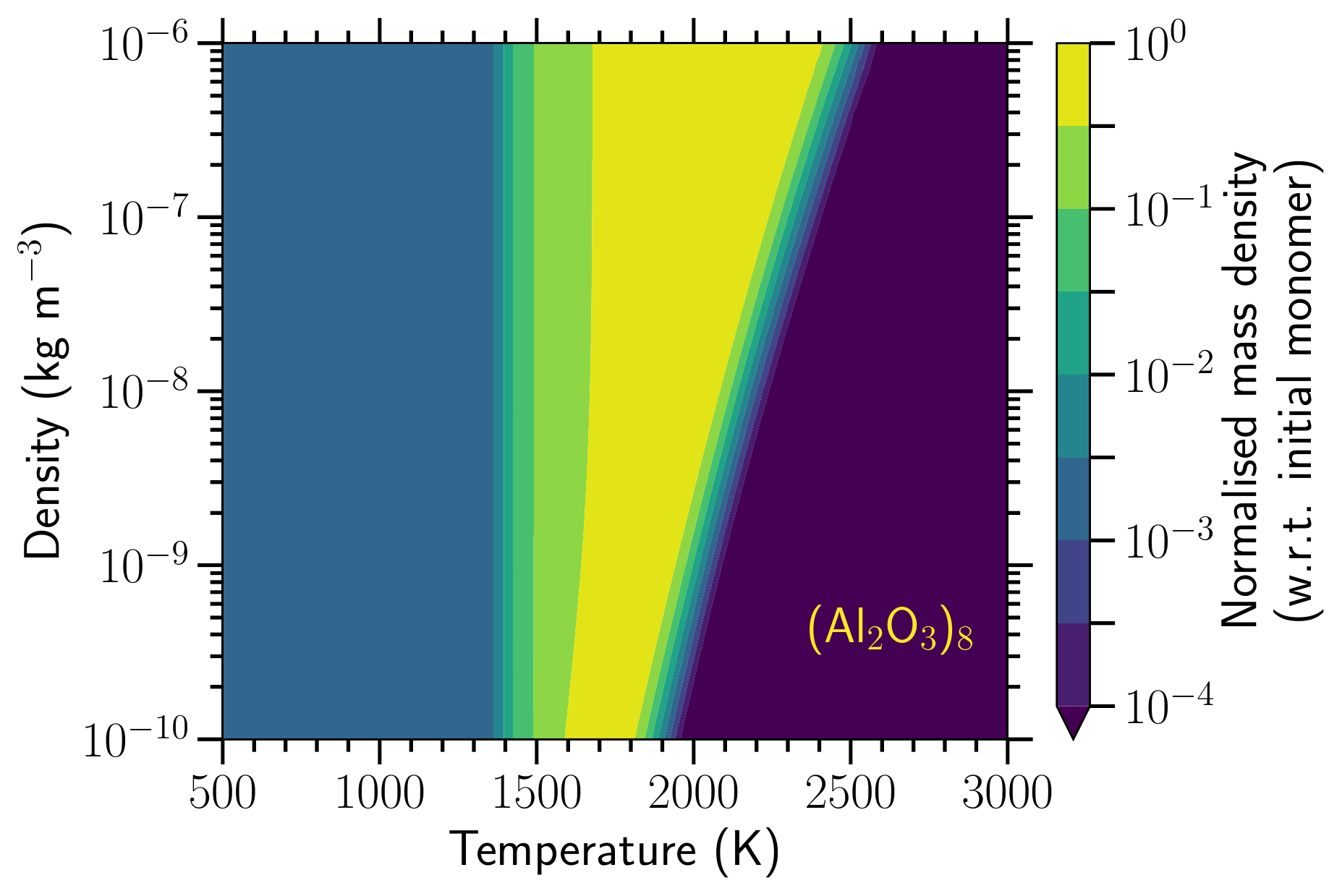}
        \end{flushleft}
        \caption{Overview of the normalised mass density after one year of all \protect\Al{1}-clusters for a closed nucleation model using the monomer nucleation description.}
        \label{fig:Al2O3_clusters_monomer_norm_same_scale}
    \end{figure*}

\begin{figure*}
        \begin{flushleft}
        \includegraphics[width=0.32\textwidth]{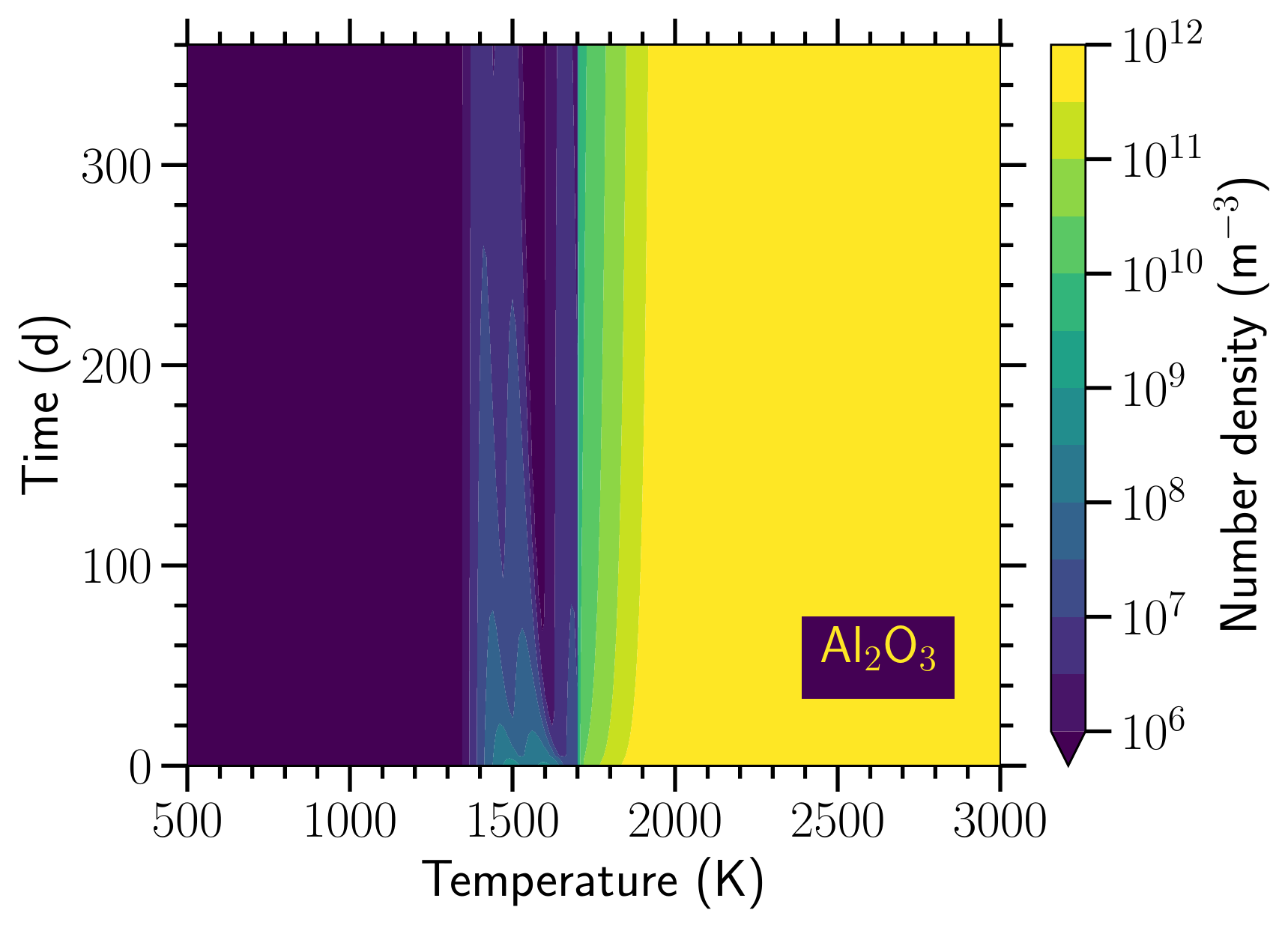}
        \includegraphics[width=0.32\textwidth]{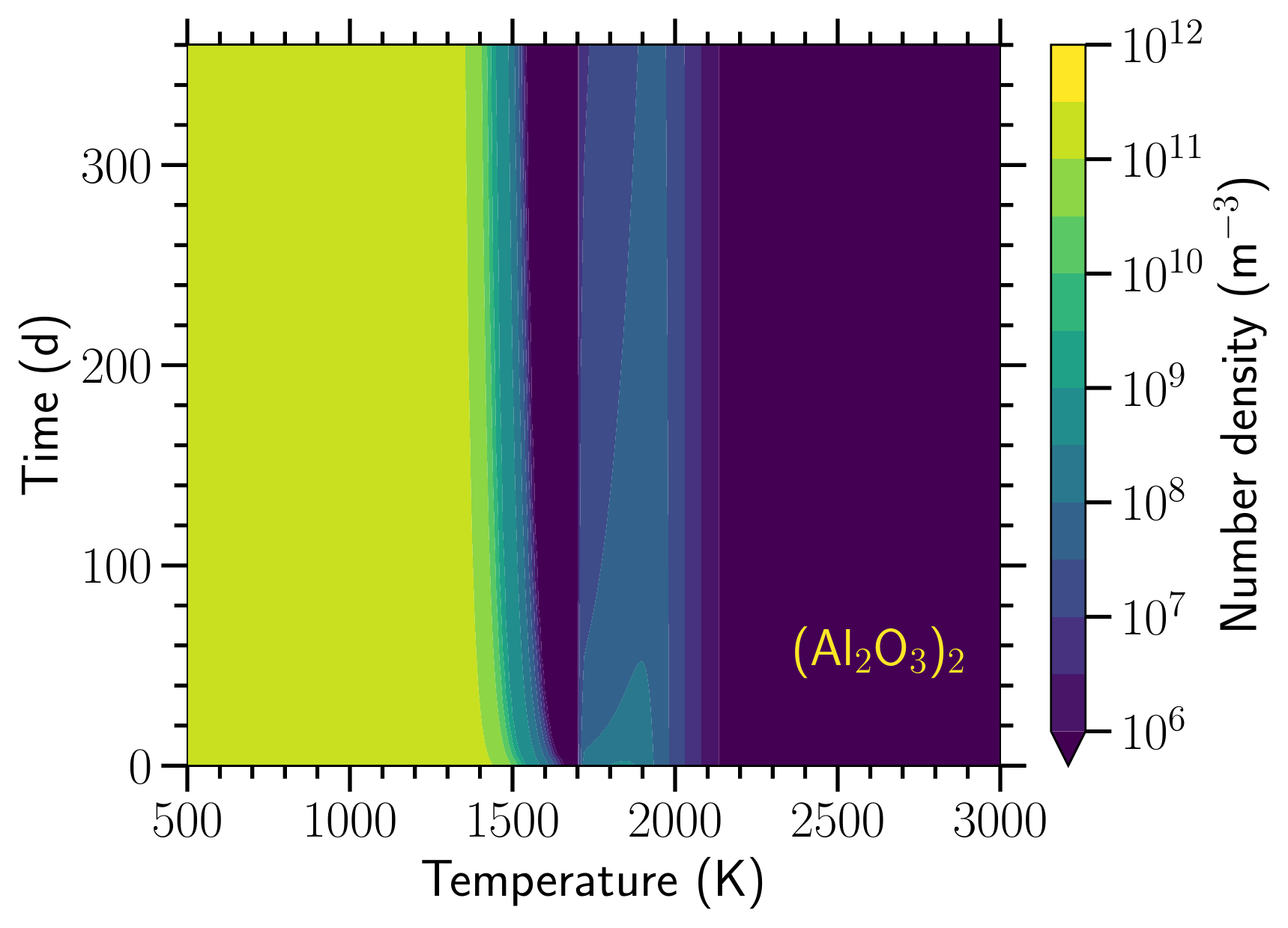}
        \includegraphics[width=0.32\textwidth]{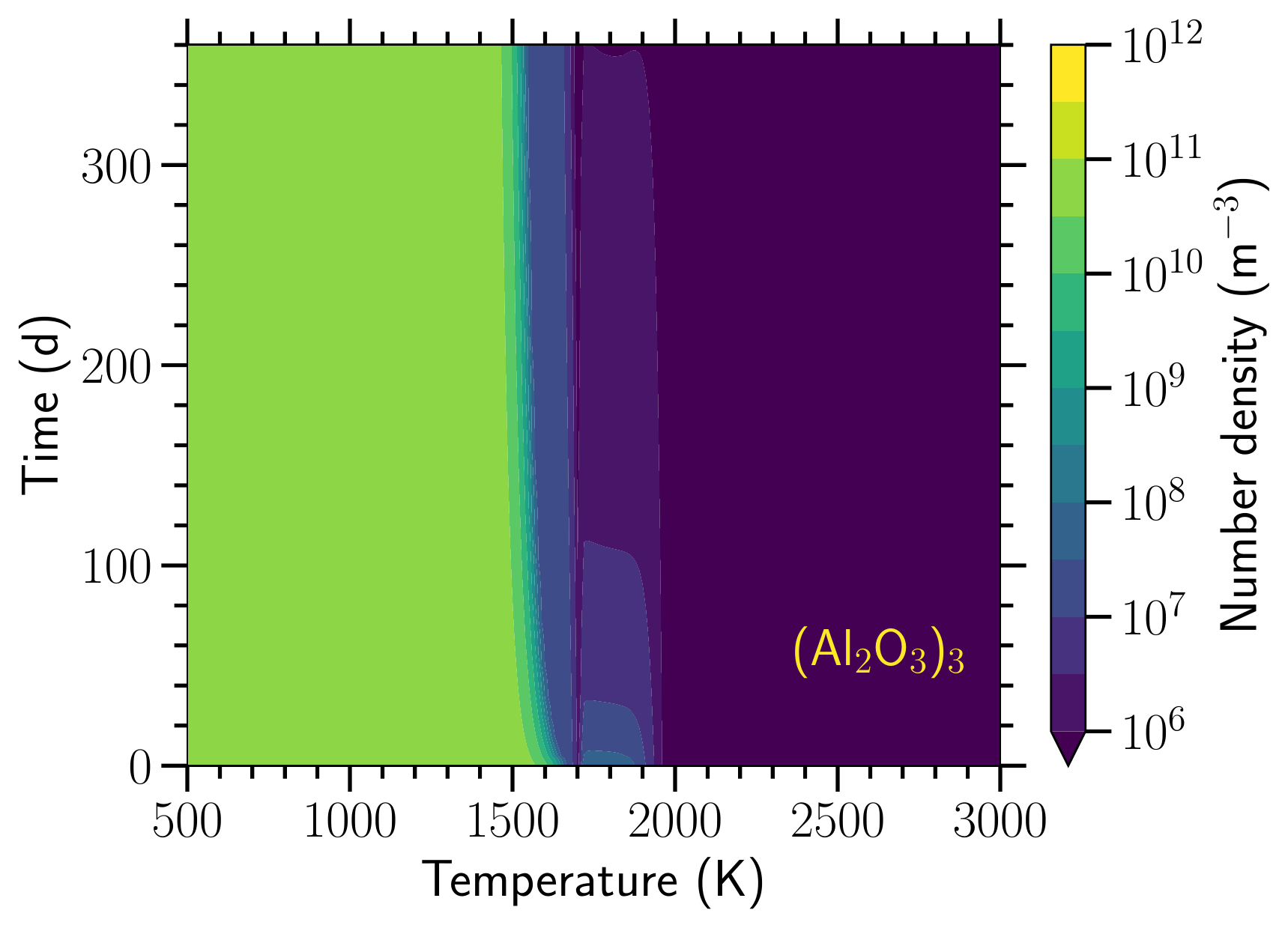}
        \includegraphics[width=0.32\textwidth]{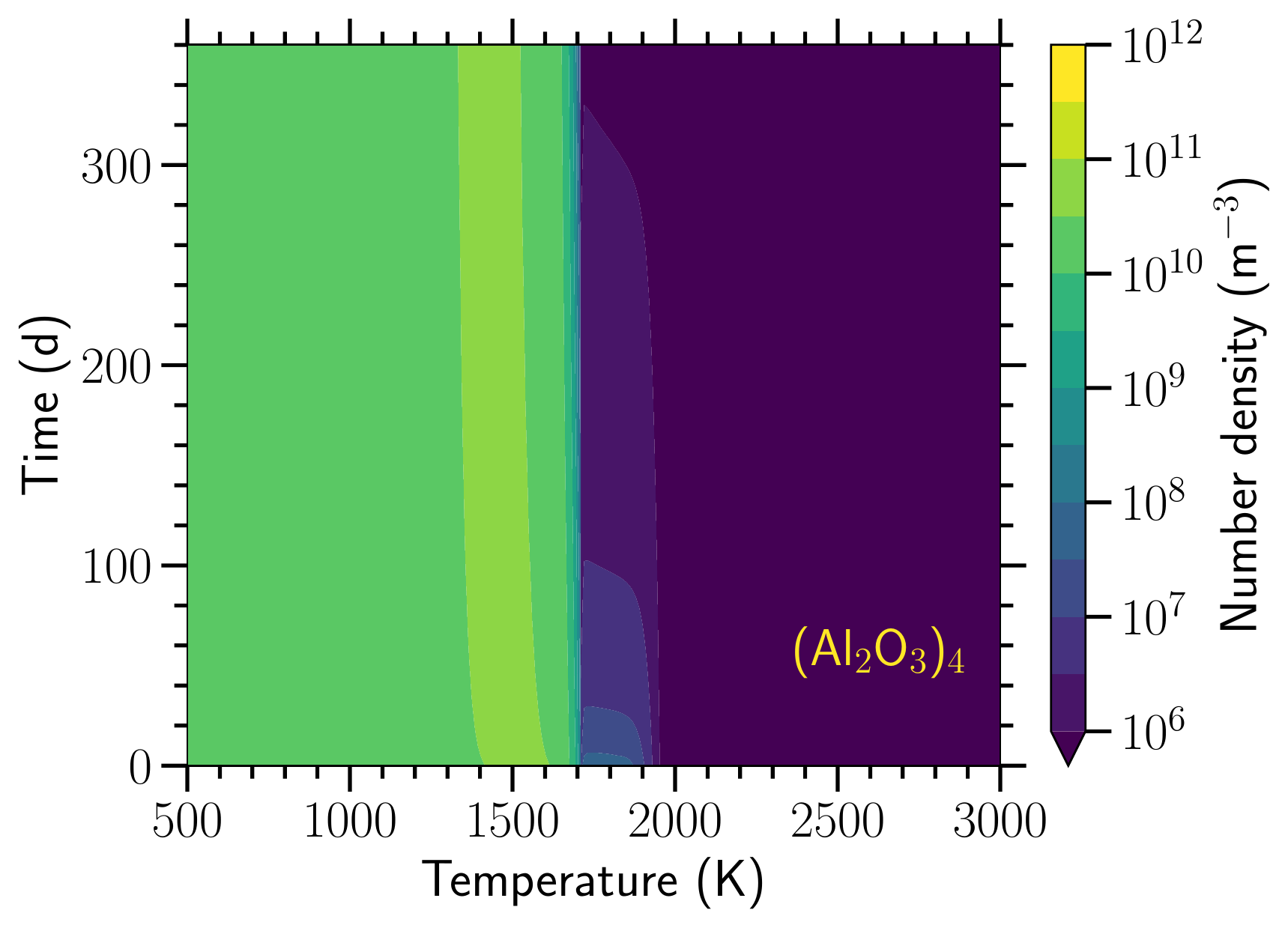}
        \includegraphics[width=0.32\textwidth]{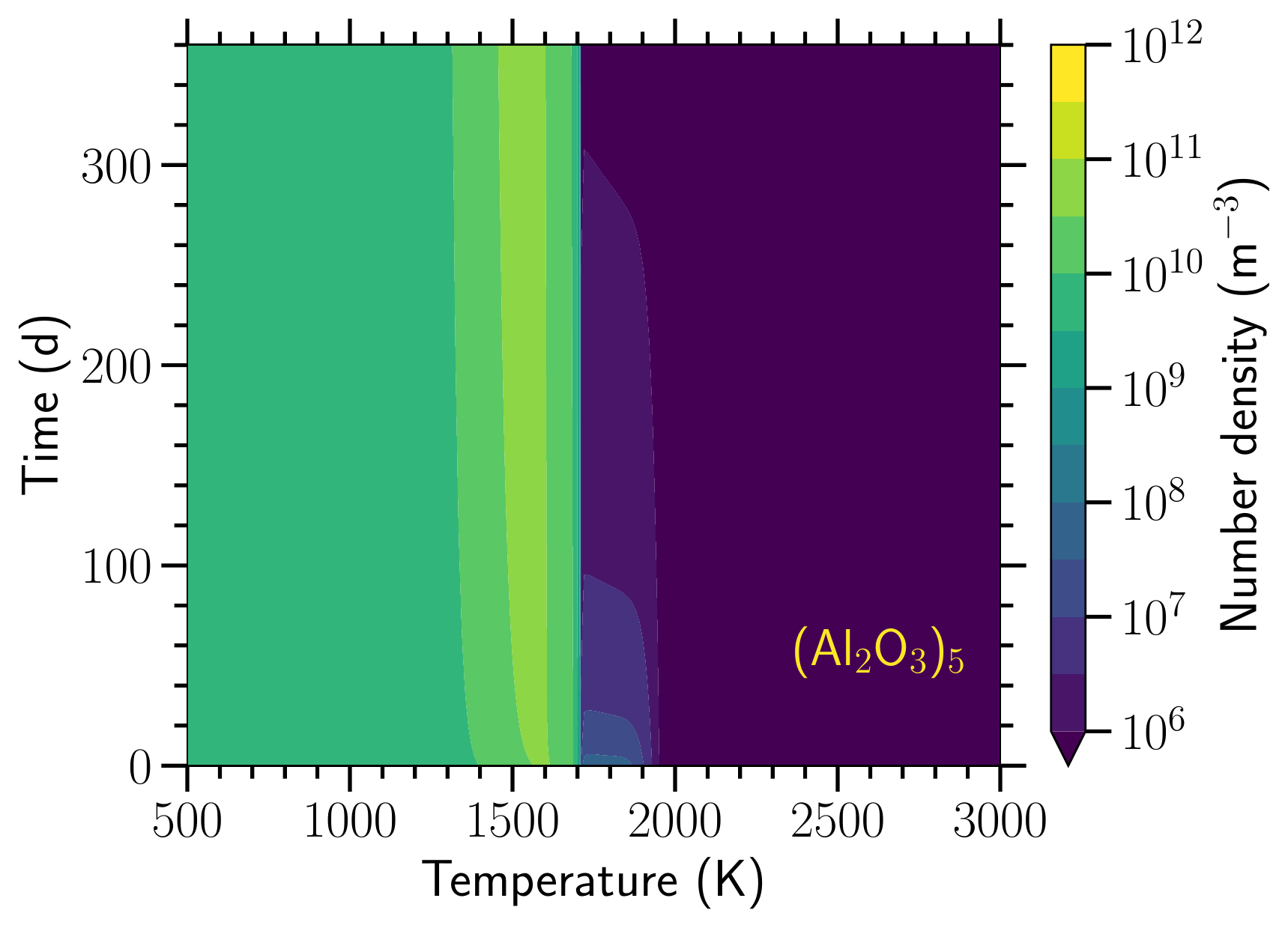}
        \includegraphics[width=0.32\textwidth]{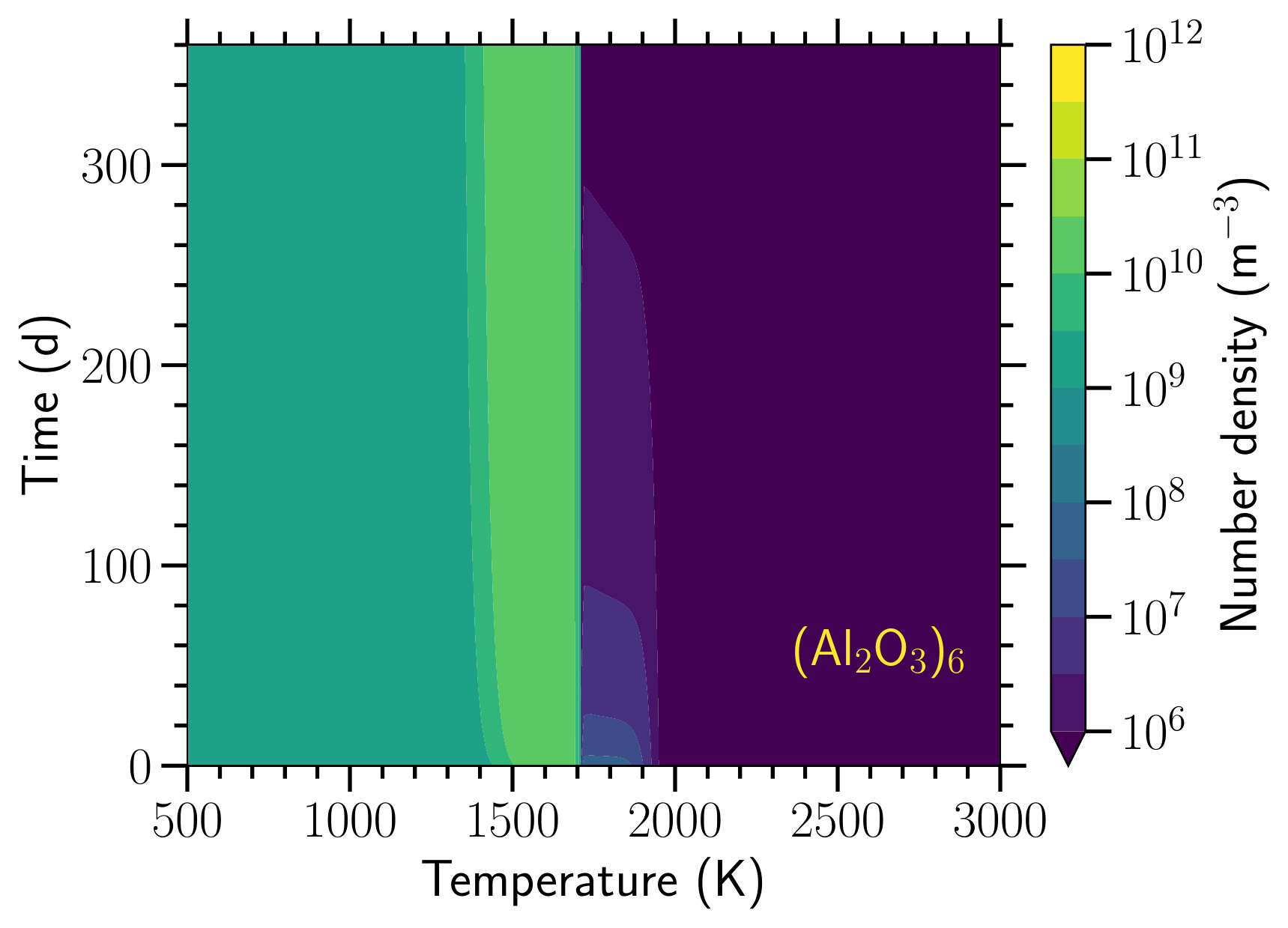}
        \includegraphics[width=0.32\textwidth]{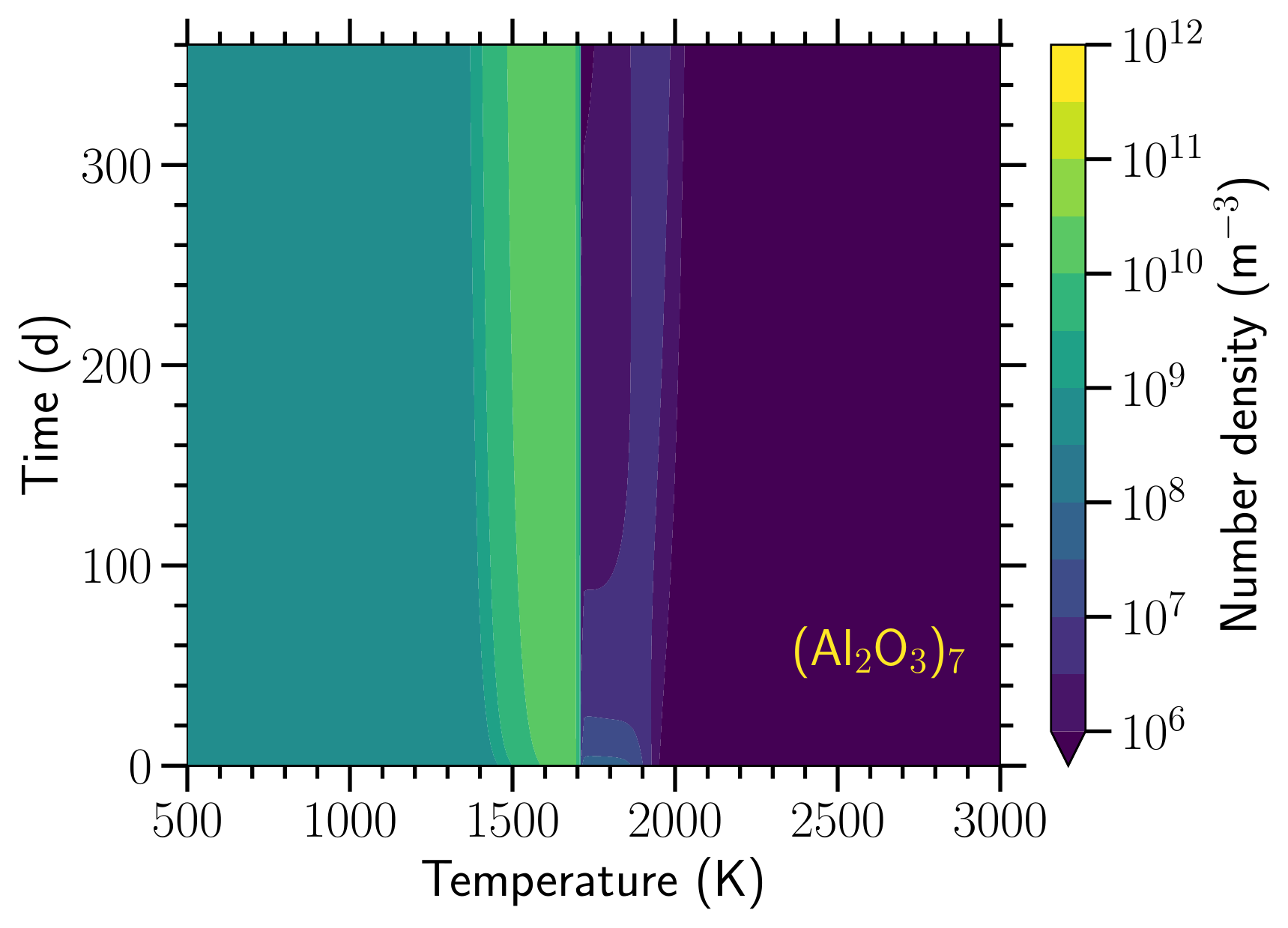}
        \includegraphics[width=0.32\textwidth]{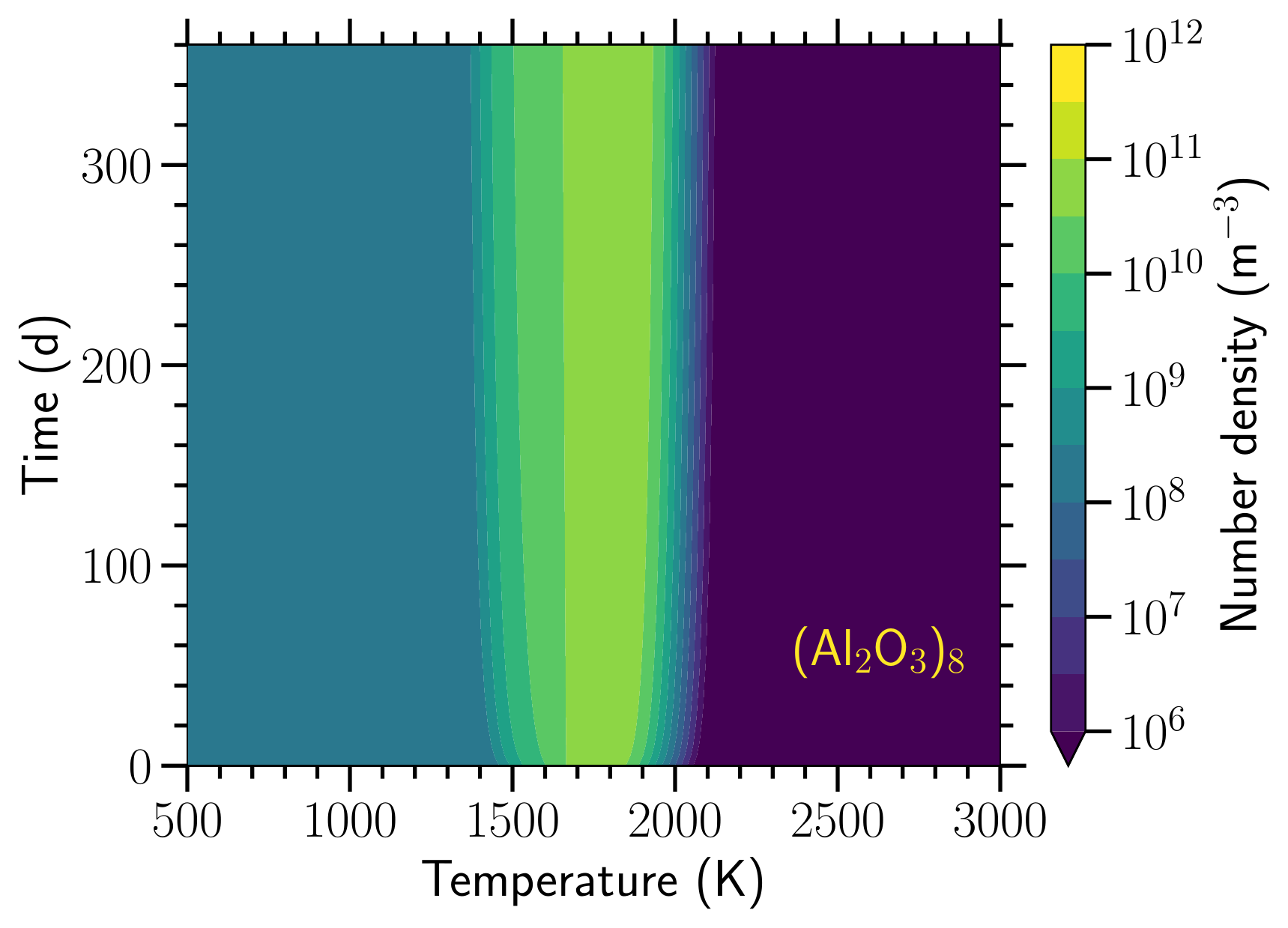}
        \end{flushleft}
        \caption{Temporal evolution of the absolute number density of all \protect\Al{1}-clusters at the benchmark total gas density $\rho=\SI{1e-9}{\kg\per\m\cubed}$ for a closed nucleation model using the monomer nucleation description.}
        \label{fig:Al2O3_clusters_monomer_time_evolution}
    \end{figure*}

\begin{figure*}
        \begin{flushleft}
        \includegraphics[width=0.32\textwidth]{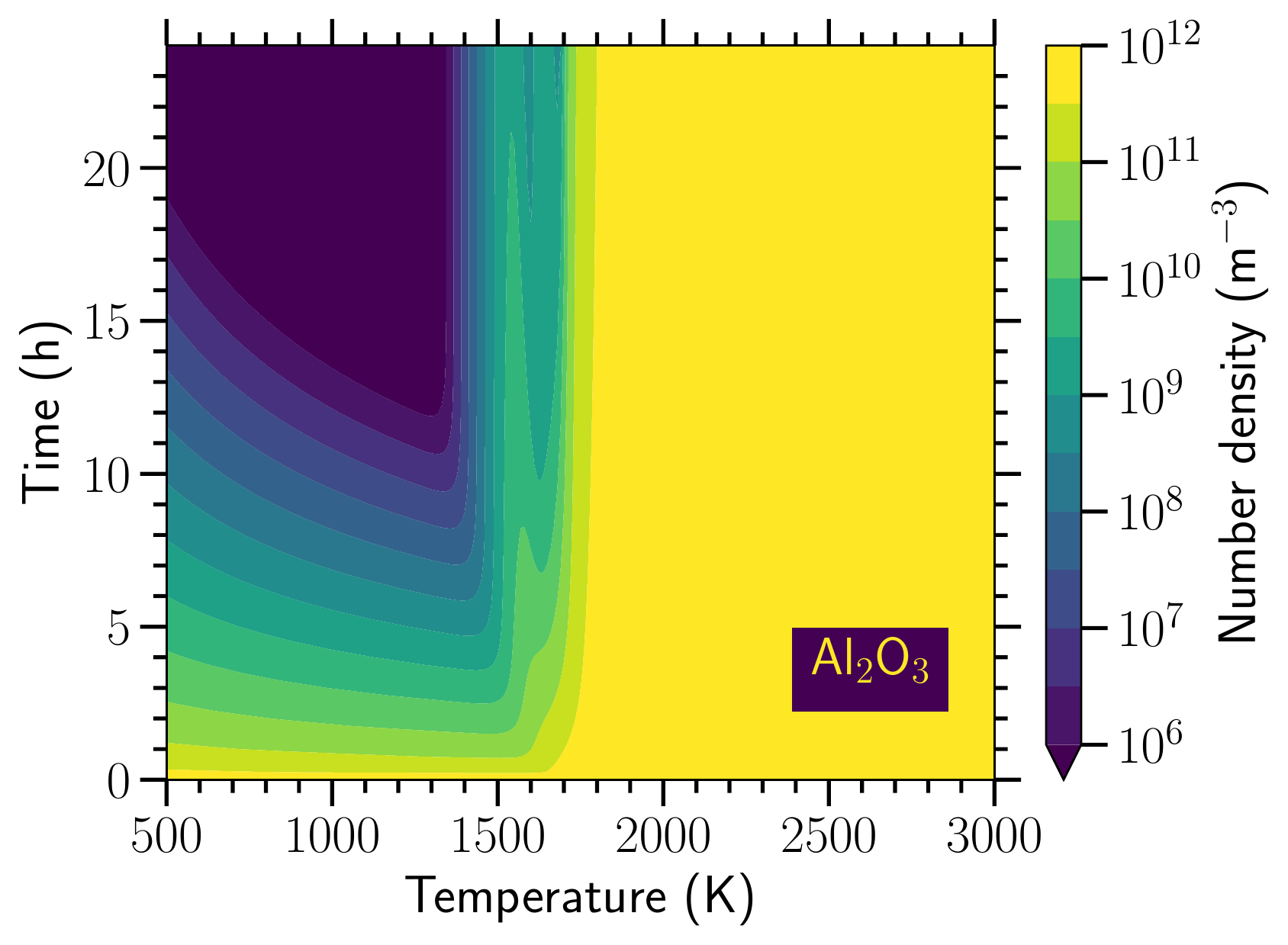}
        \includegraphics[width=0.32\textwidth]{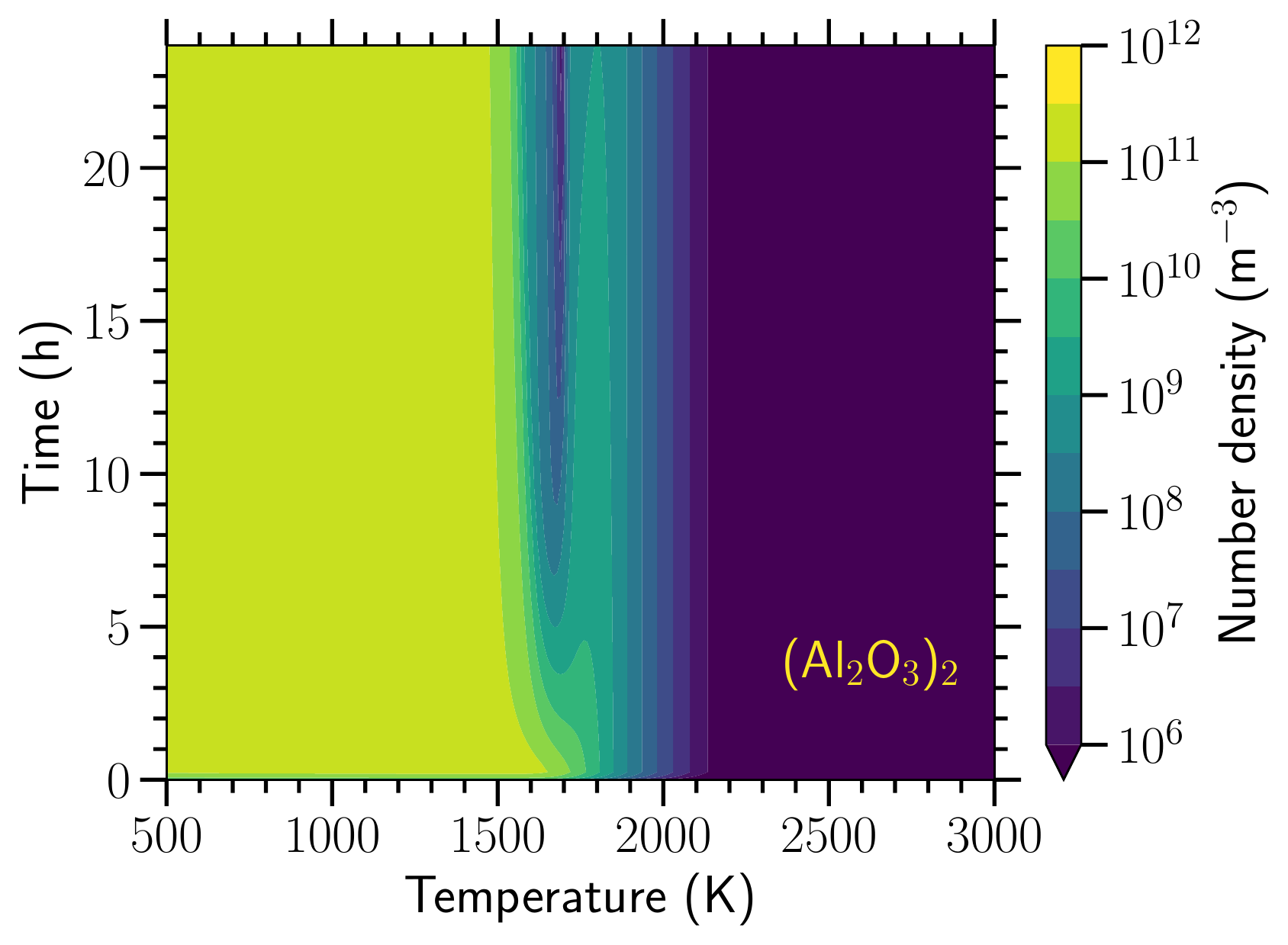}
        \includegraphics[width=0.32\textwidth]{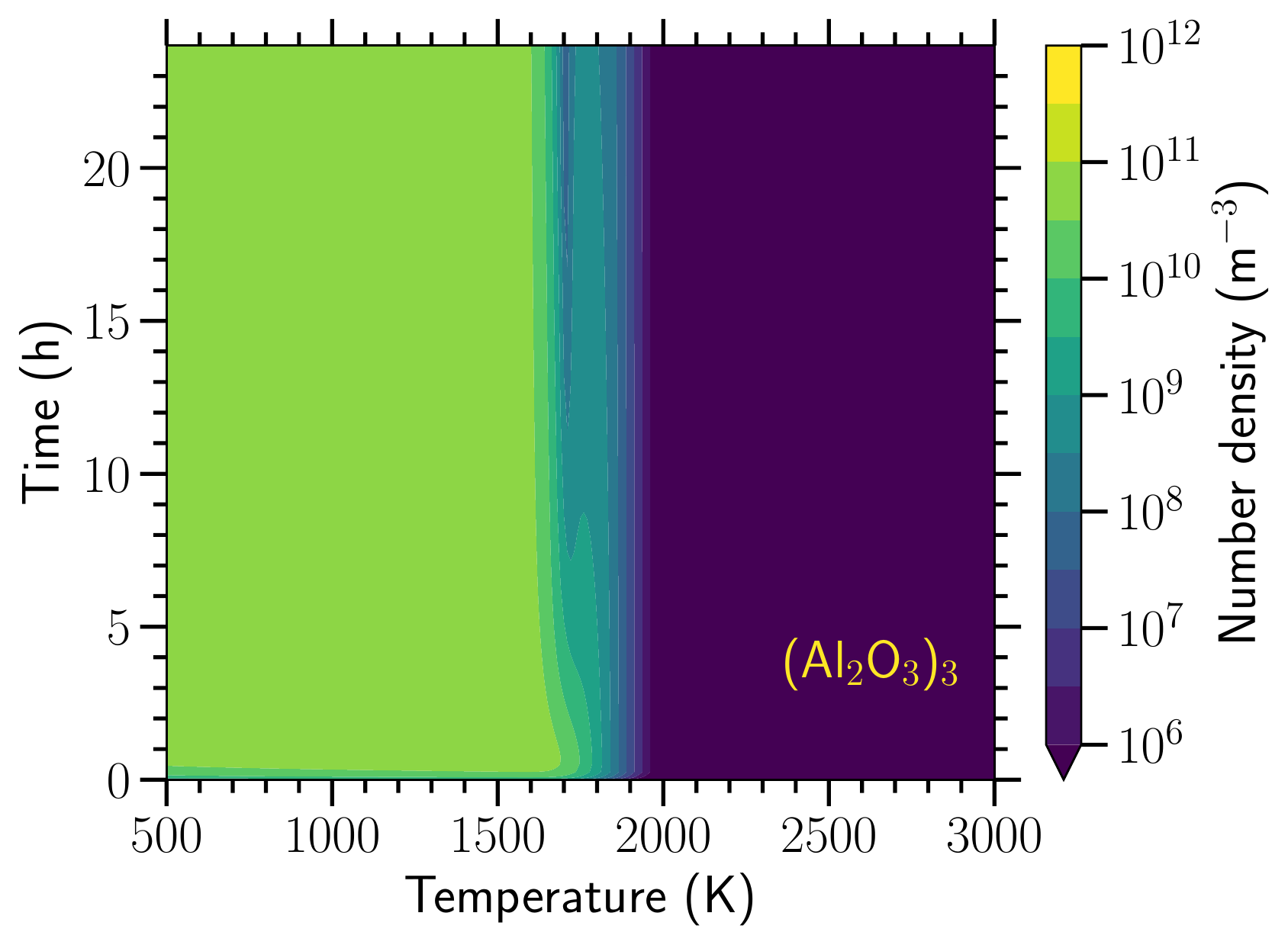}
        \includegraphics[width=0.32\textwidth]{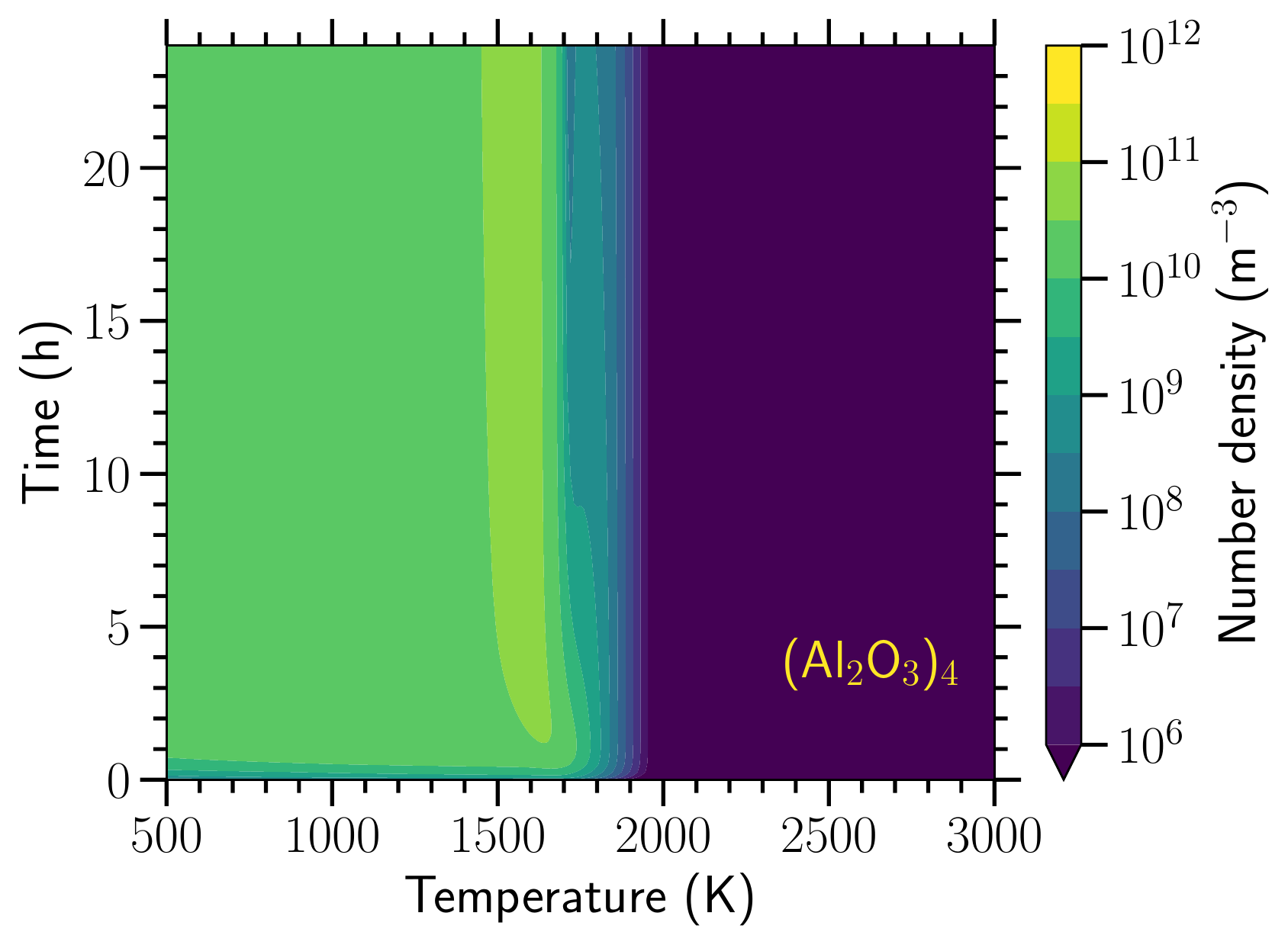}
        \includegraphics[width=0.32\textwidth]{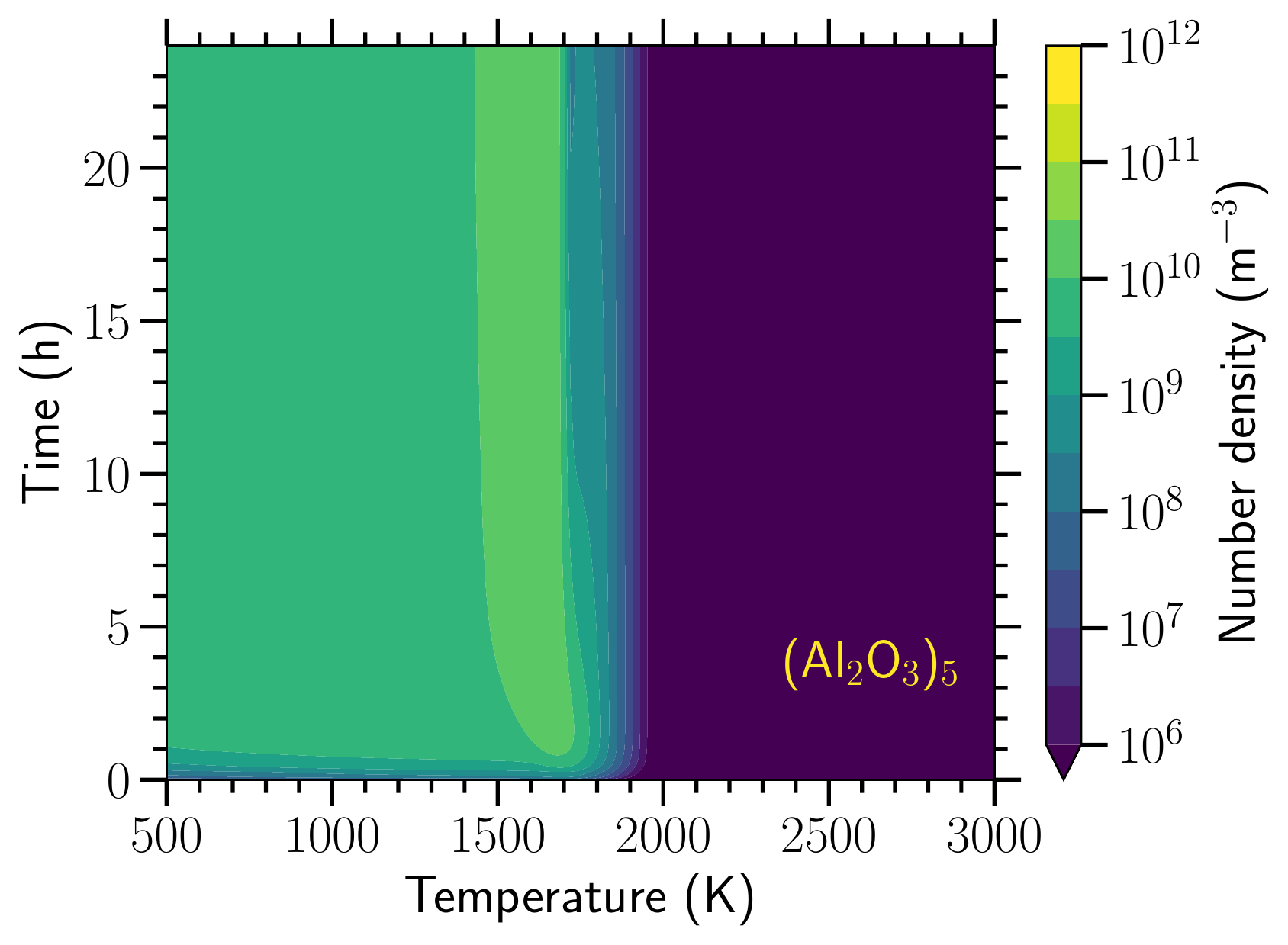}
        \includegraphics[width=0.32\textwidth]{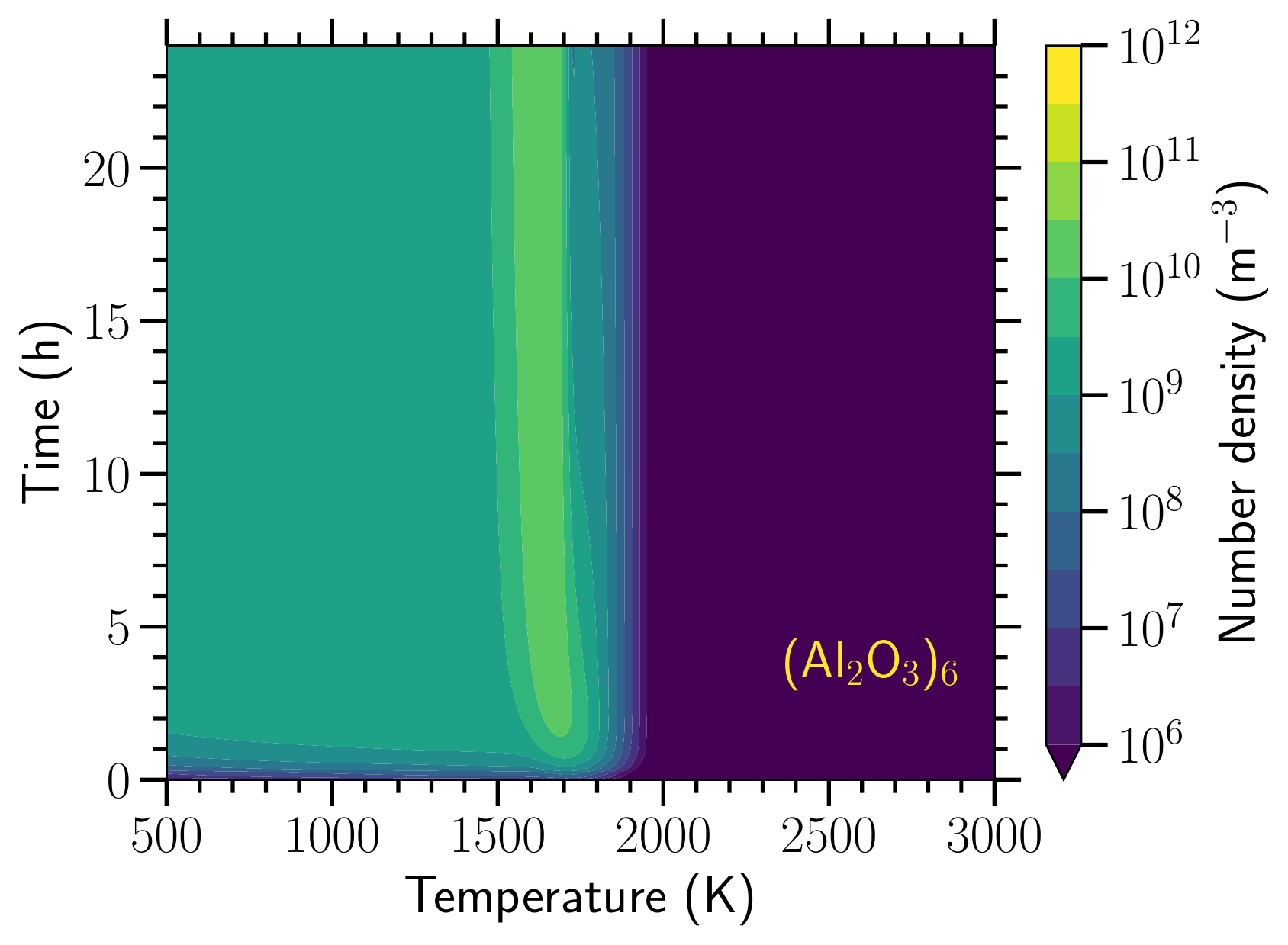}
        \includegraphics[width=0.32\textwidth]{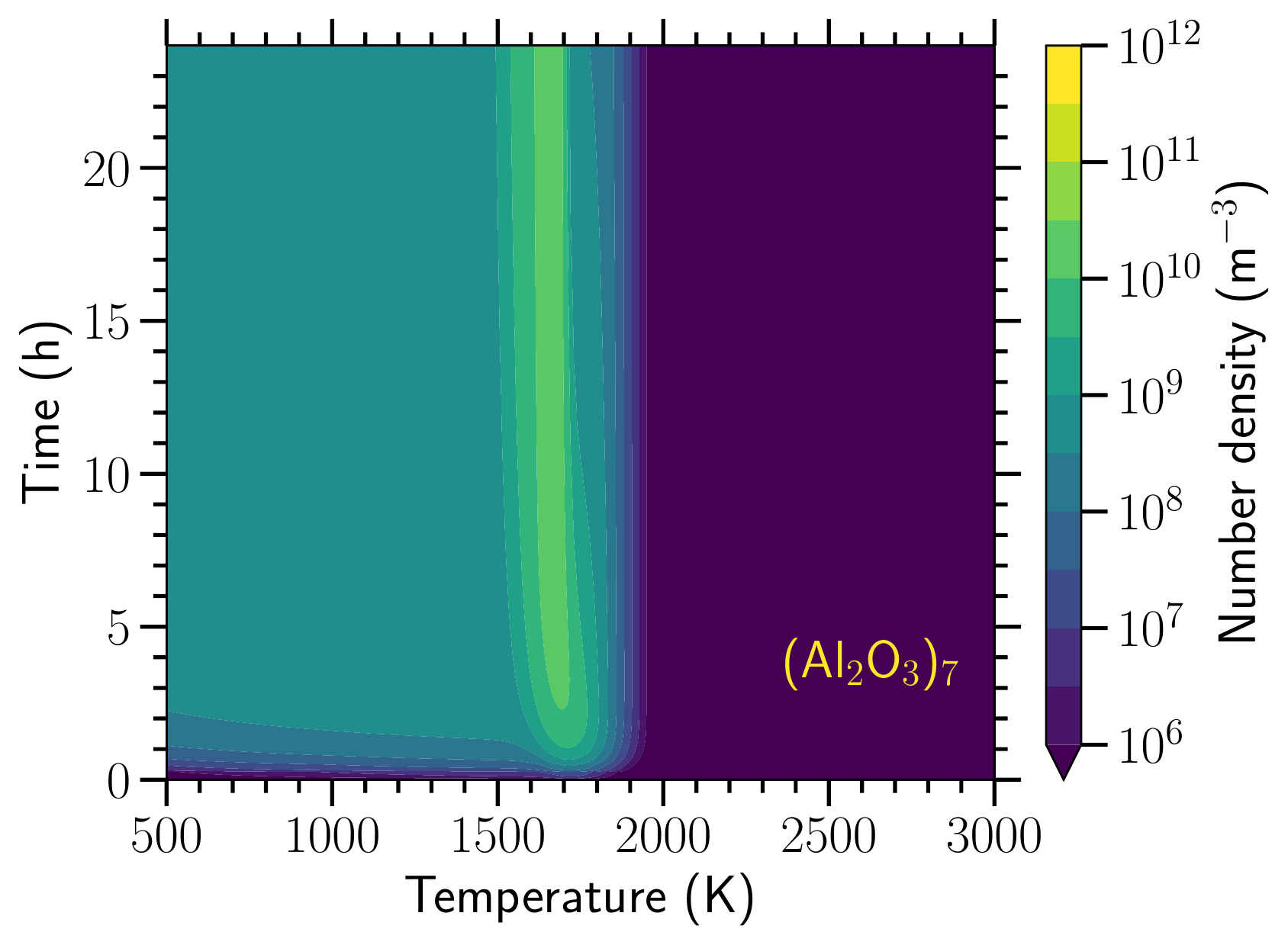}
        \includegraphics[width=0.32\textwidth]{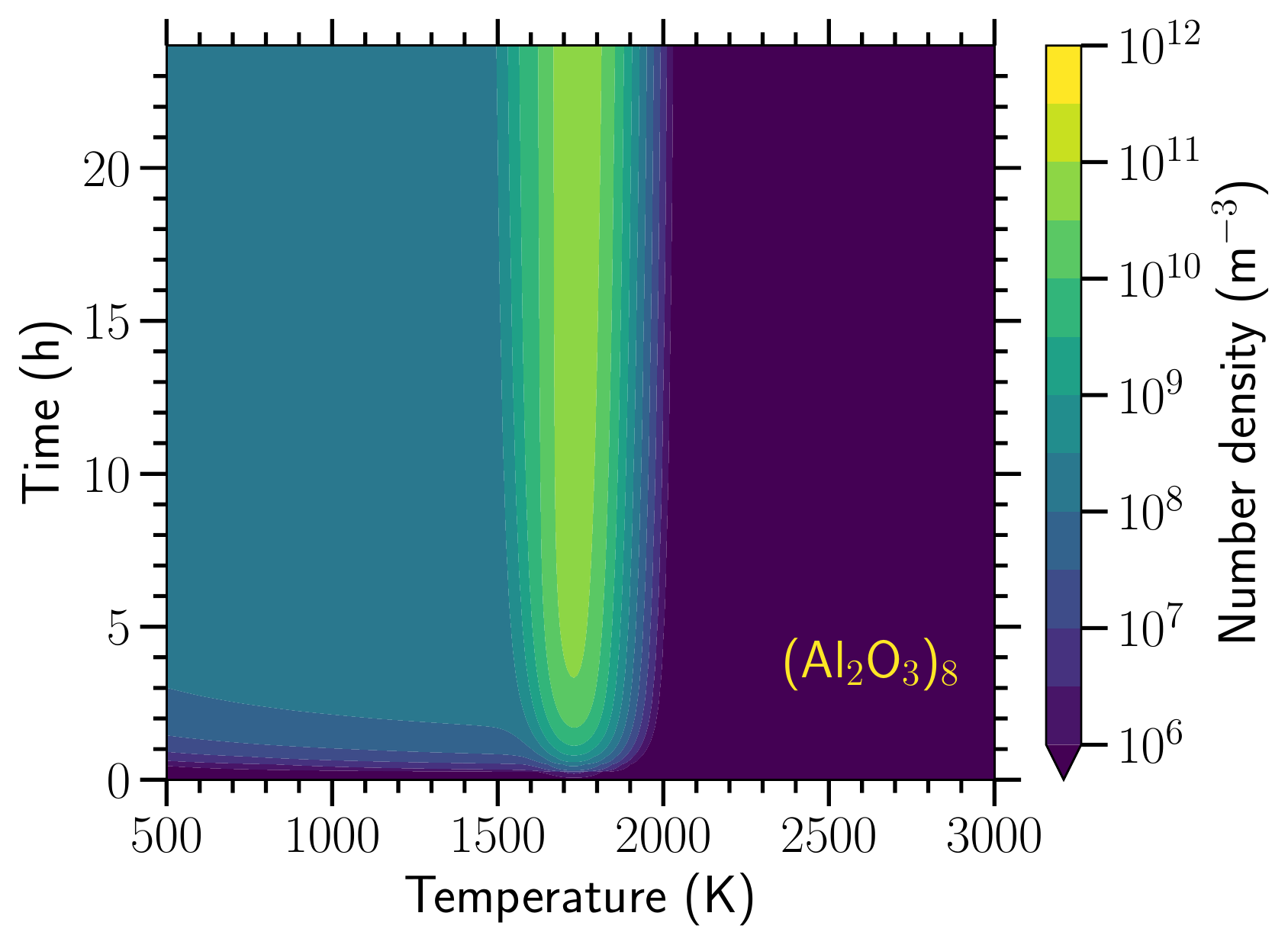}
        \end{flushleft}
        \caption{Refined temporal evolution of the absolute number density of all \protect\Al{1}-clusters at the benchmark total gas density $\rho=\SI{1e-9}{\kg\per\m\cubed}$ for a closed nucleation model using the monomer nucleation description.}
        \label{fig:Al2O3_clusters_monomer_time_evolution_short}
    \end{figure*}

        This section contains a more complete overview of the closed nucleation models using the monomer nucleation description of all nucleation clusters results.\\\\
        \Ti{1}-clusters: Figs.\,\ref{fig:TiO2_clusters_monomer_norm_same_scale} to \ref{fig:TiO2_clusters_monomer_time_evolution}\\
        \Mg{1}-clusters: Figs.\,\ref{fig:MgO_clusters_monomer_norm_same_scale} to \ref{fig:MgO_clusters_monomer_time_evolution_short}\\
        \SiO{1}-clusters: Fig.\,\ref{fig:SiO_clusters_monomer_norm_same_scale}\\
        \Al{1}-clusters: Figs.\,\ref{fig:Al2O3_clusters_monomer_norm_same_scale} to \ref{fig:Al2O3_clusters_monomer_time_evolution_short}
        \subsubsection{Polymer nucleation}
\begin{figure*}
        \begin{flushleft}
        \includegraphics[width=0.32\textwidth]{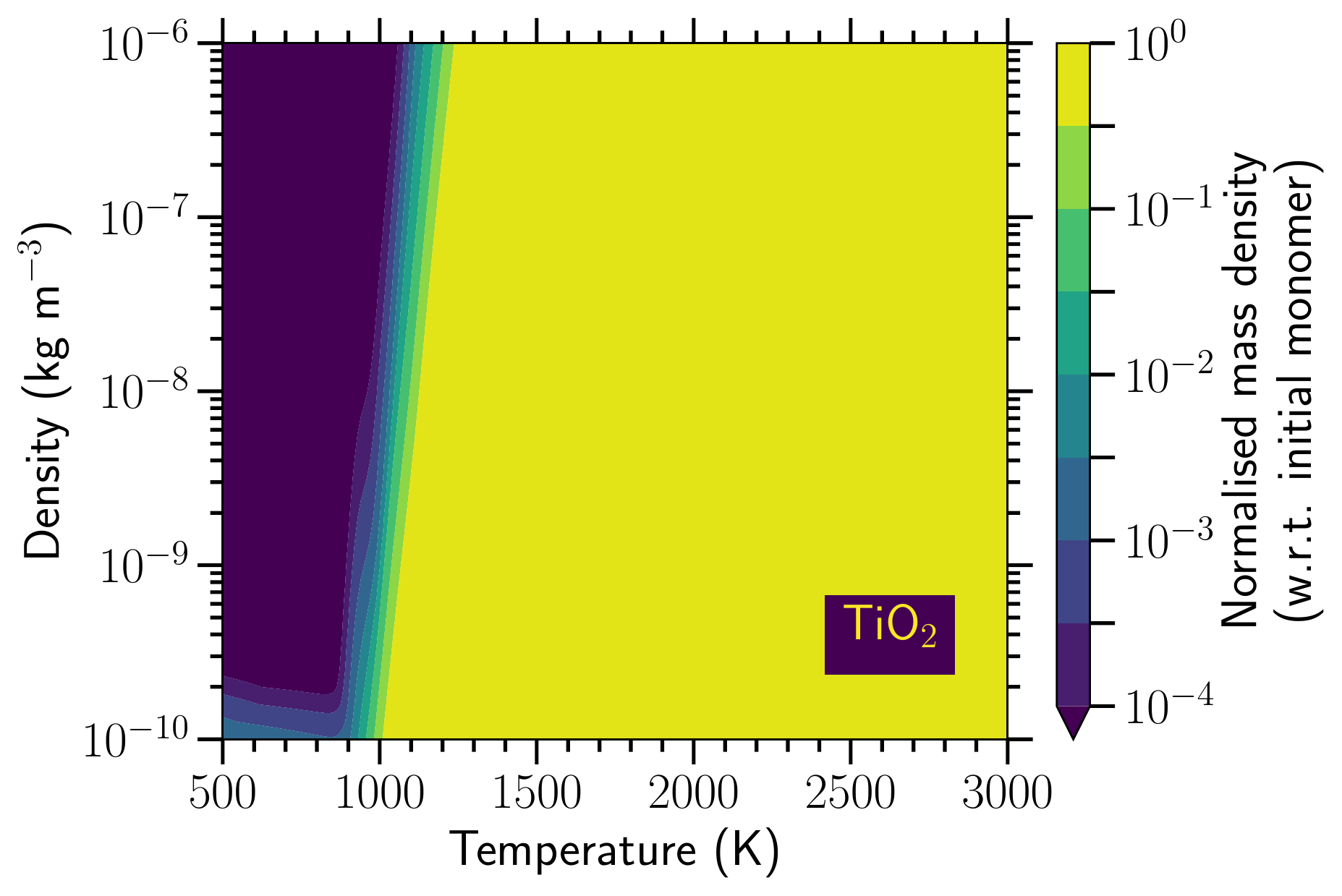}
        \includegraphics[width=0.32\textwidth]{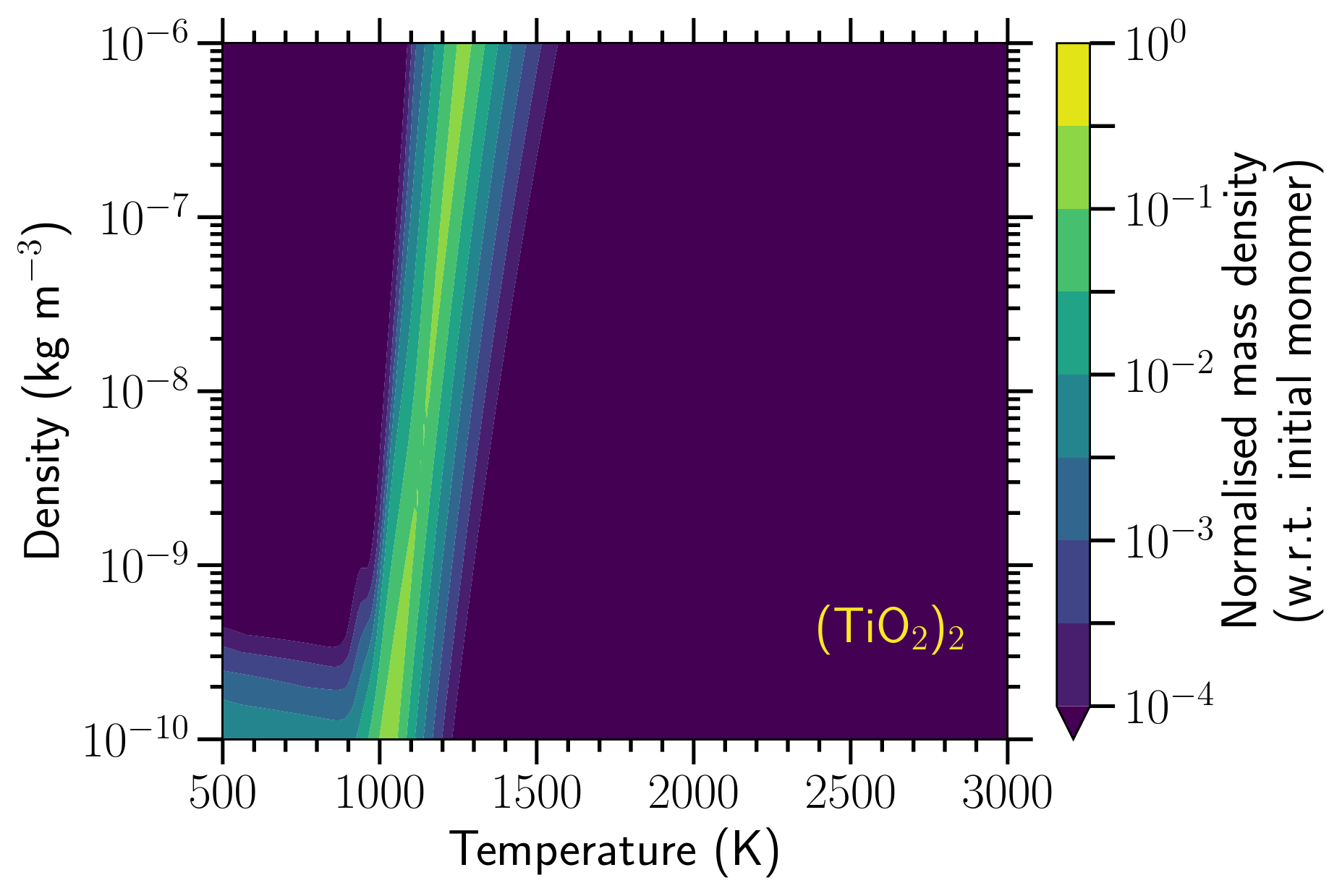}
        \includegraphics[width=0.32\textwidth]{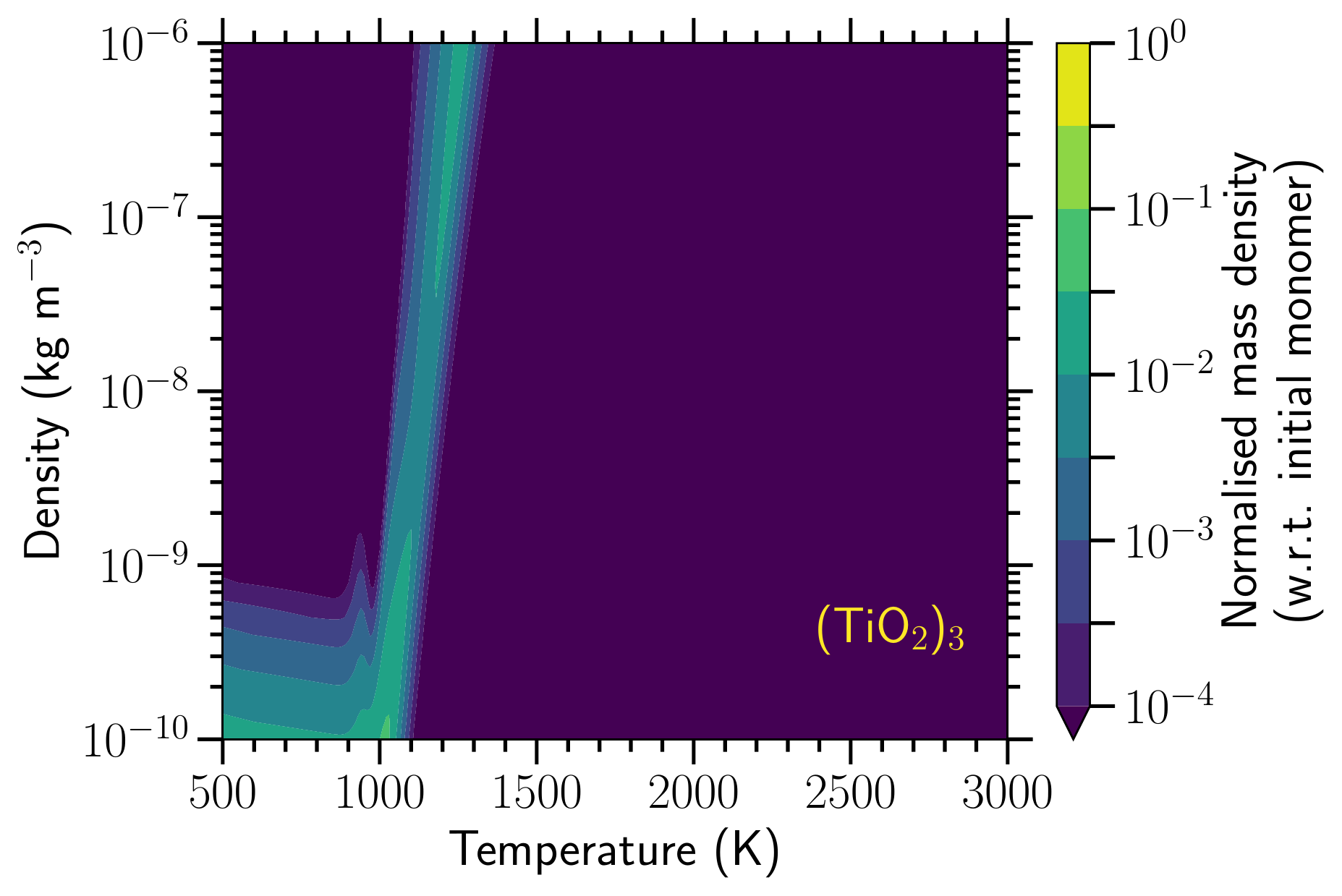}
        \includegraphics[width=0.32\textwidth]{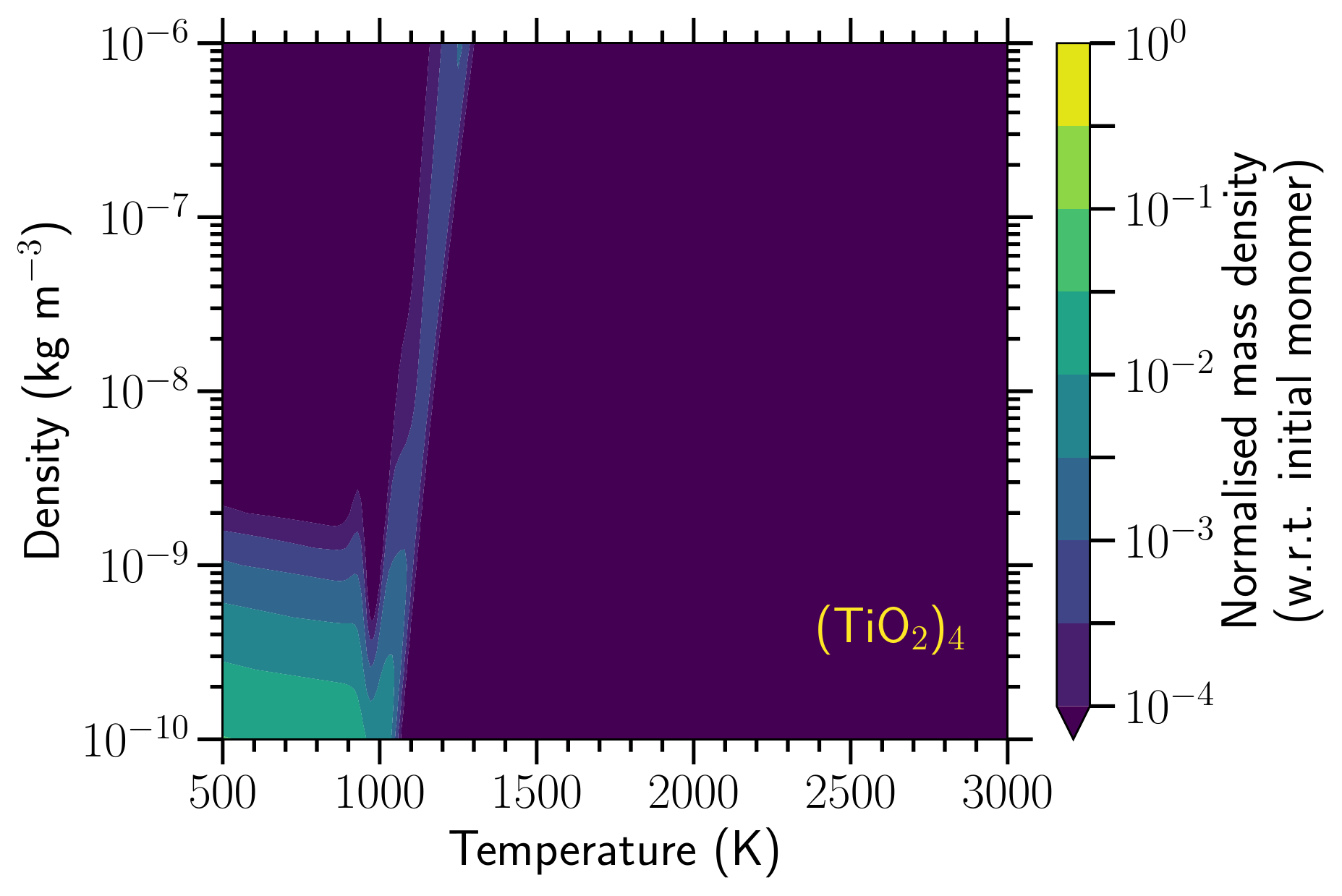}
        \includegraphics[width=0.32\textwidth]{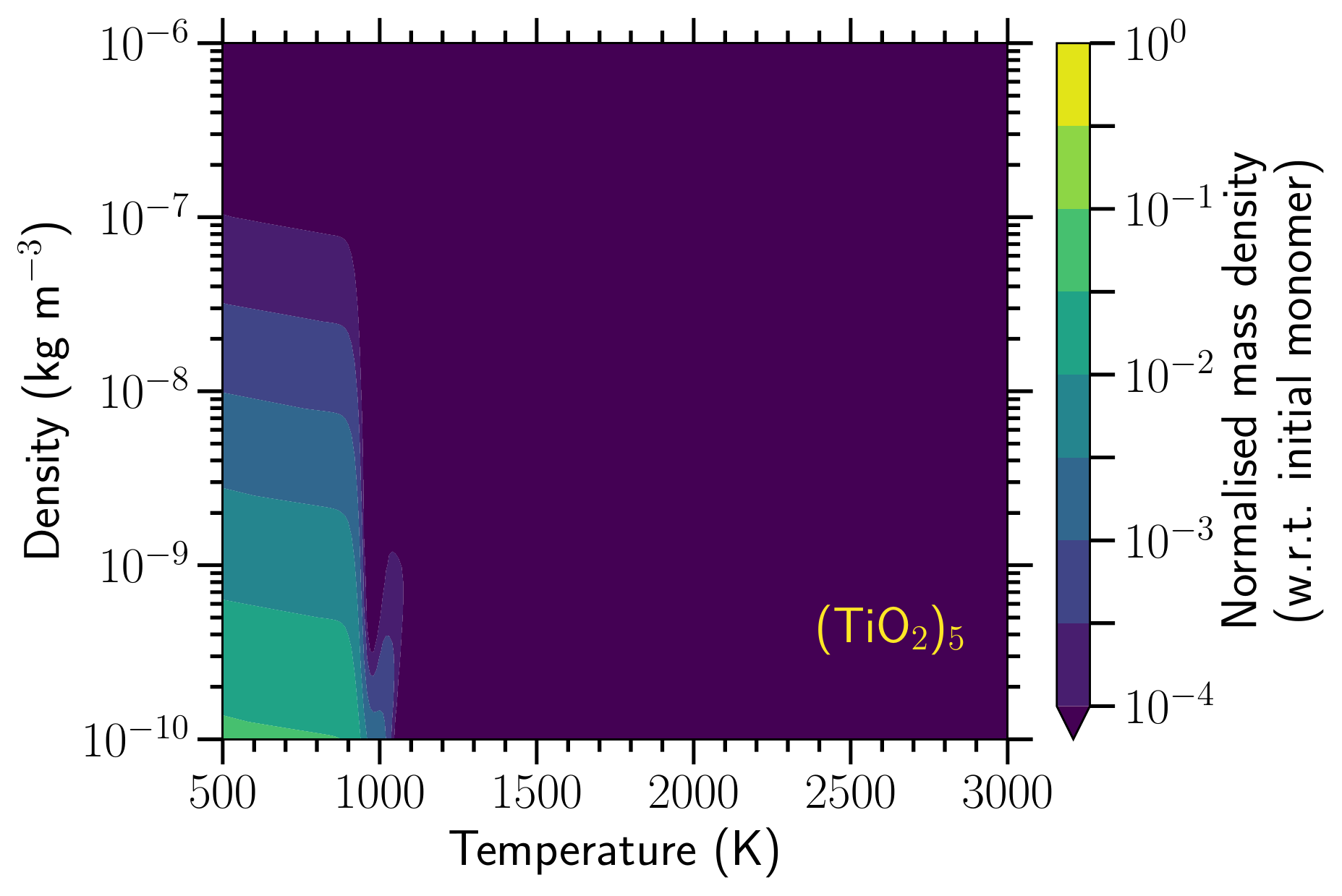}
        \includegraphics[width=0.32\textwidth]{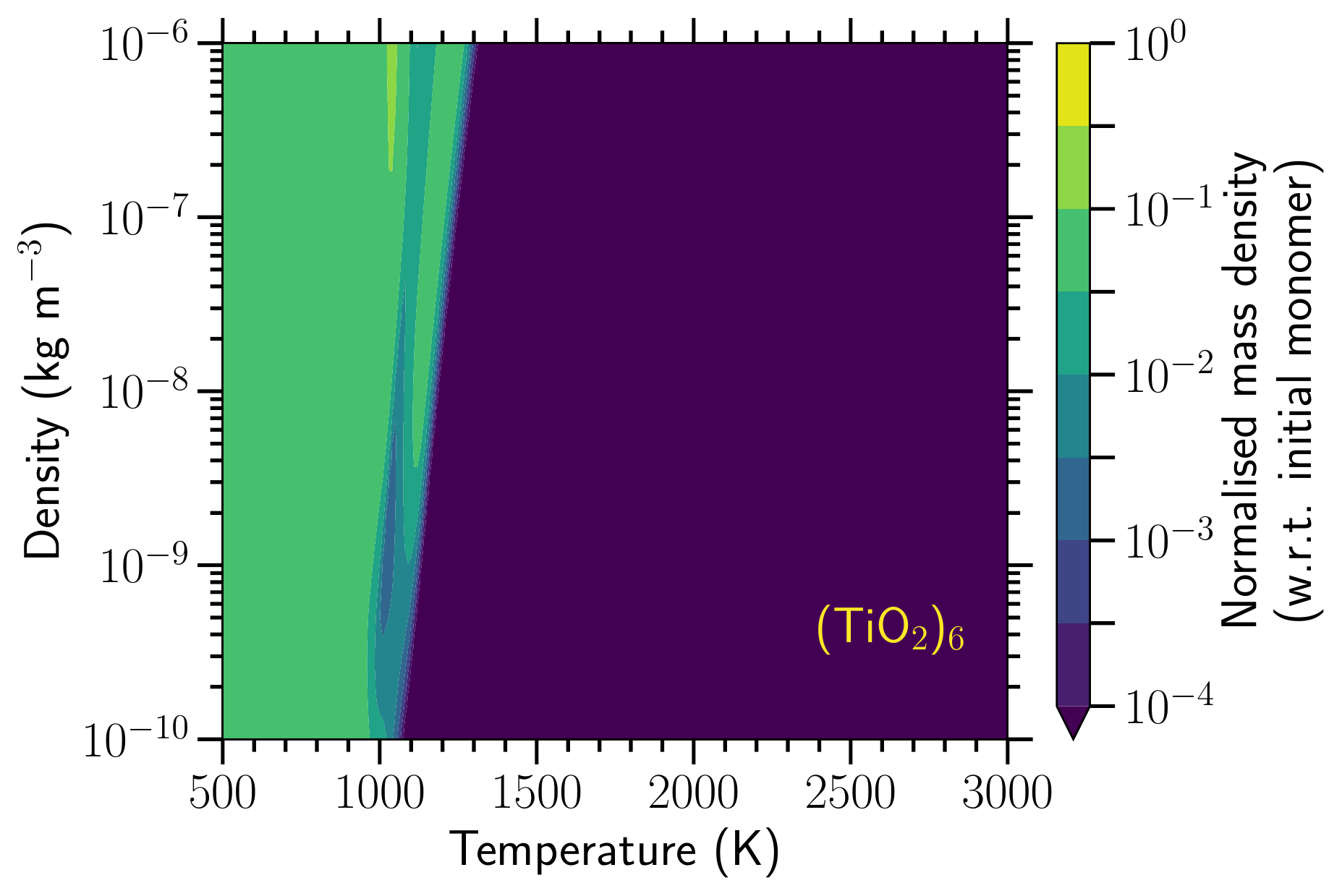}
        \includegraphics[width=0.32\textwidth]{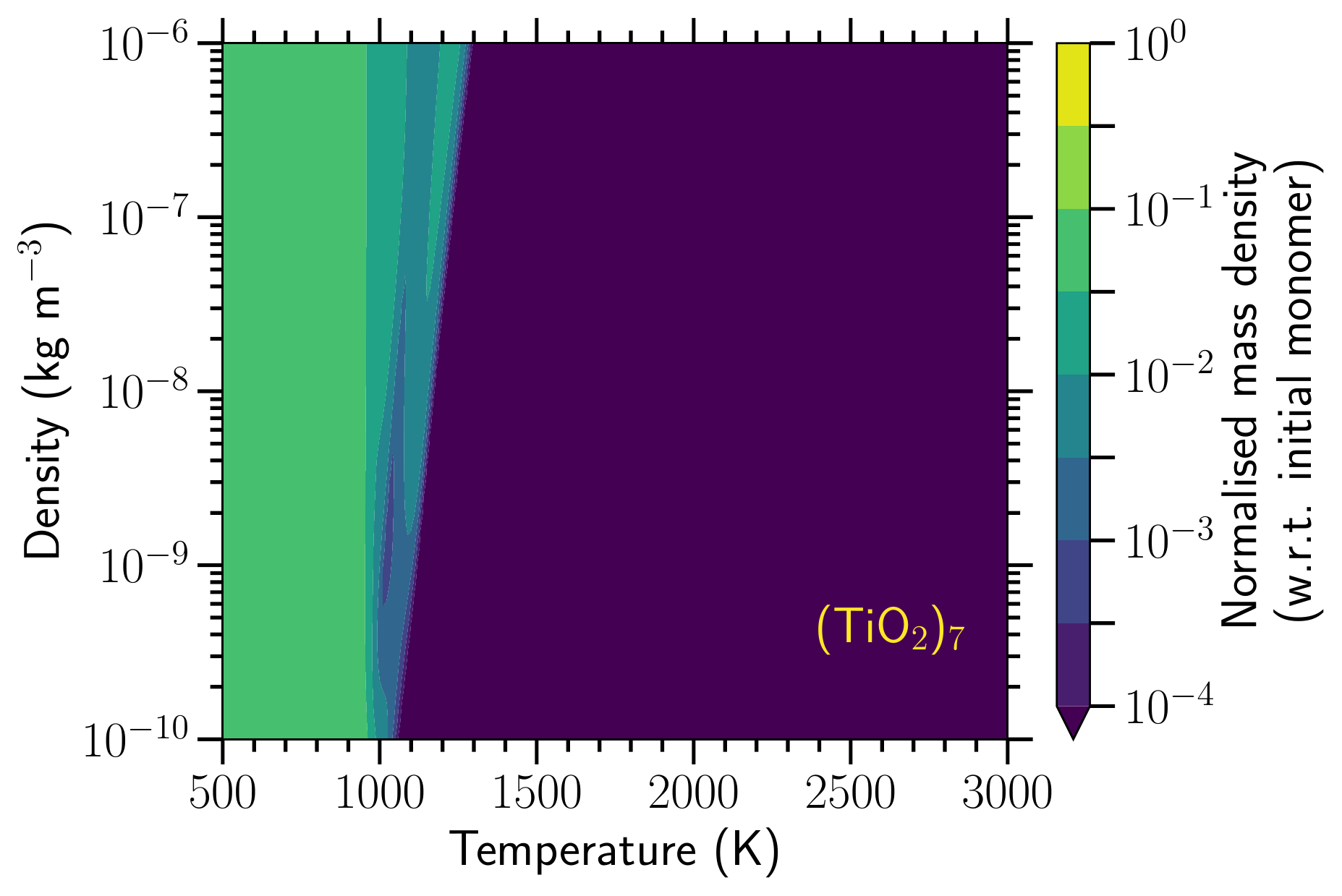}
        \includegraphics[width=0.32\textwidth]{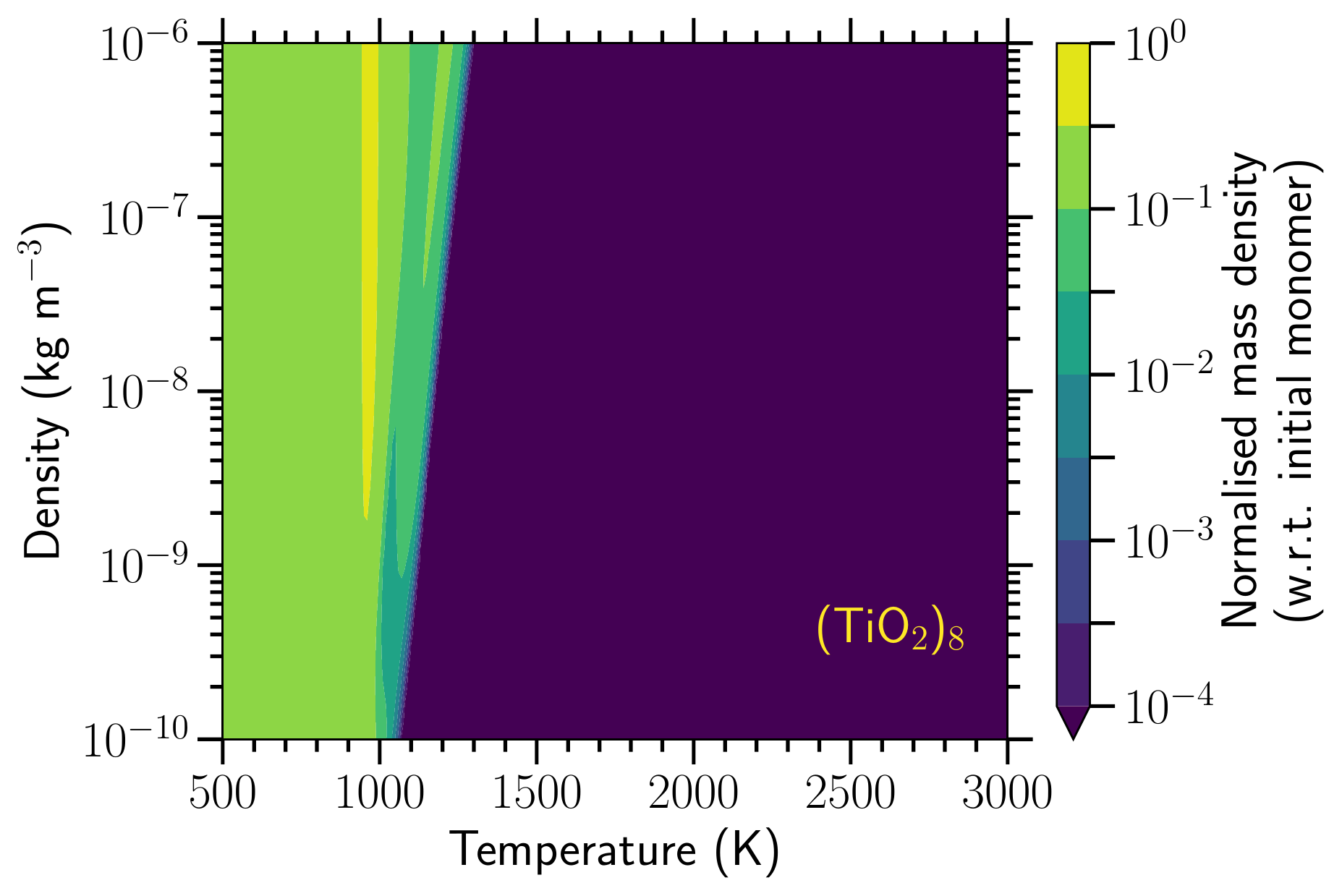}
        \includegraphics[width=0.32\textwidth]{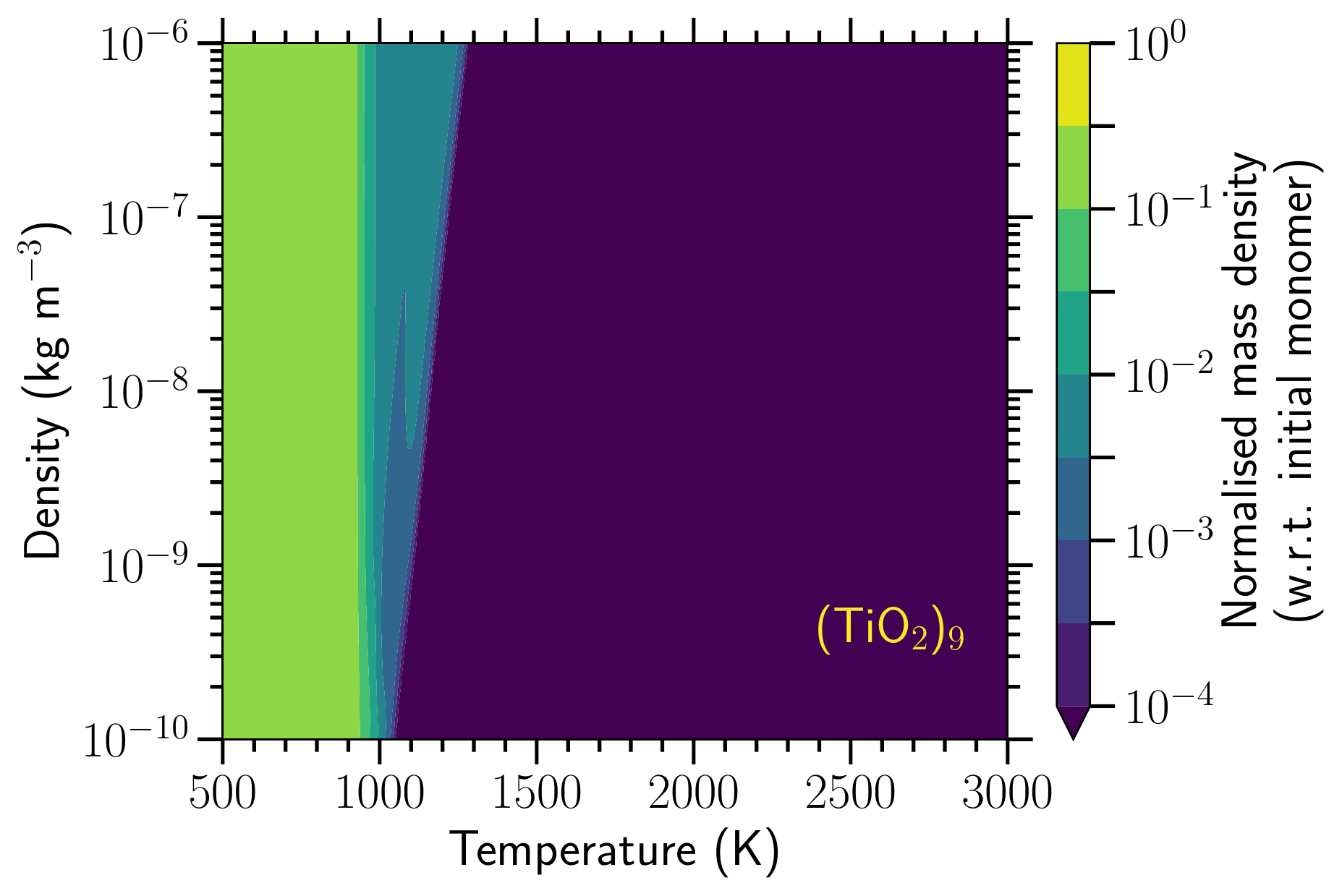}
        \includegraphics[width=0.32\textwidth]{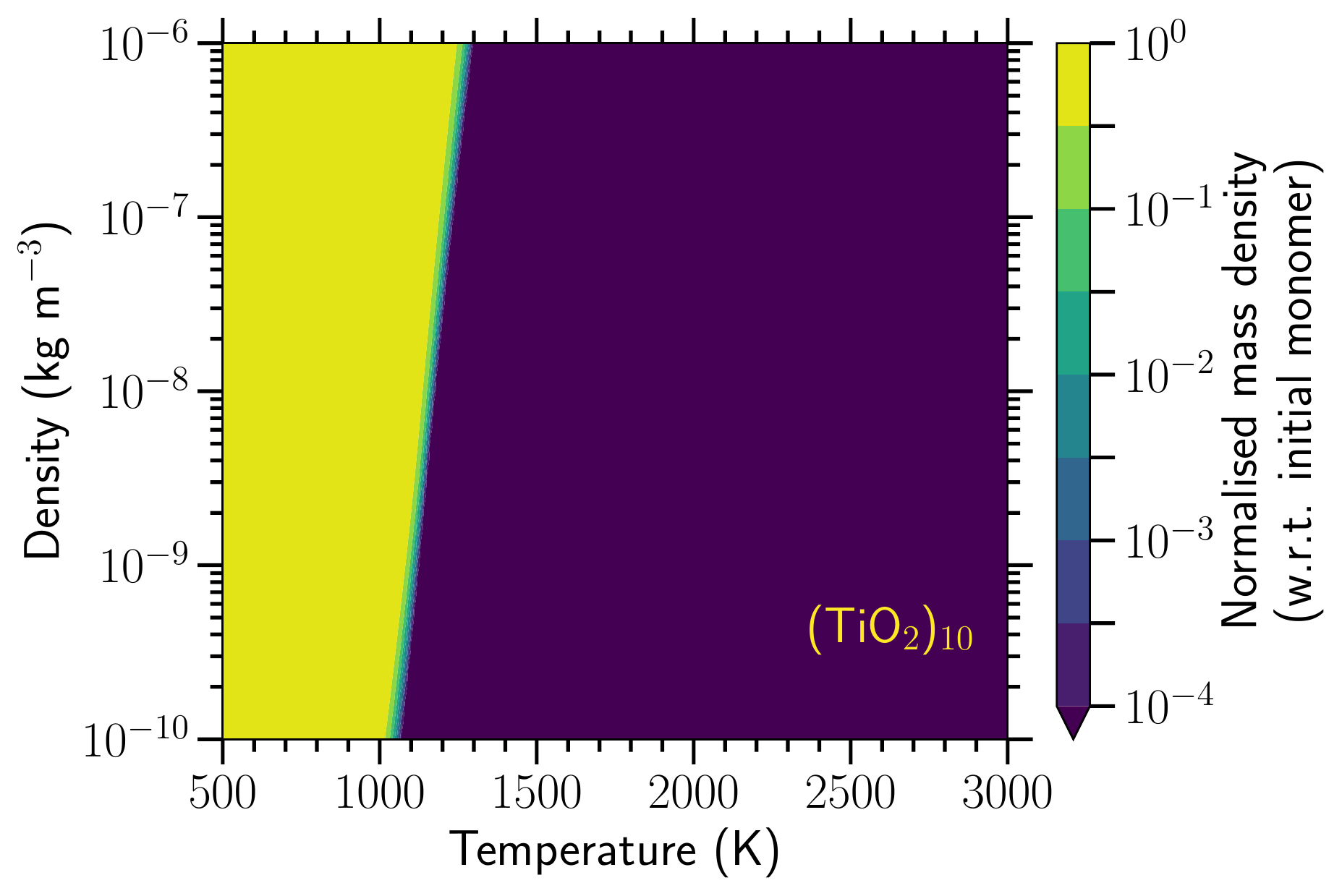}
        \end{flushleft}
        \caption{Overview of the normalised mass density after one year of all \protect\Ti{1}-clusters for a closed nucleation model using the polymer nucleation description.}
        \label{fig:TiO2_clusters_general_norm_same_scale}
    \end{figure*}

\begin{figure*}
        \begin{flushleft}
        \includegraphics[width=0.32\textwidth]{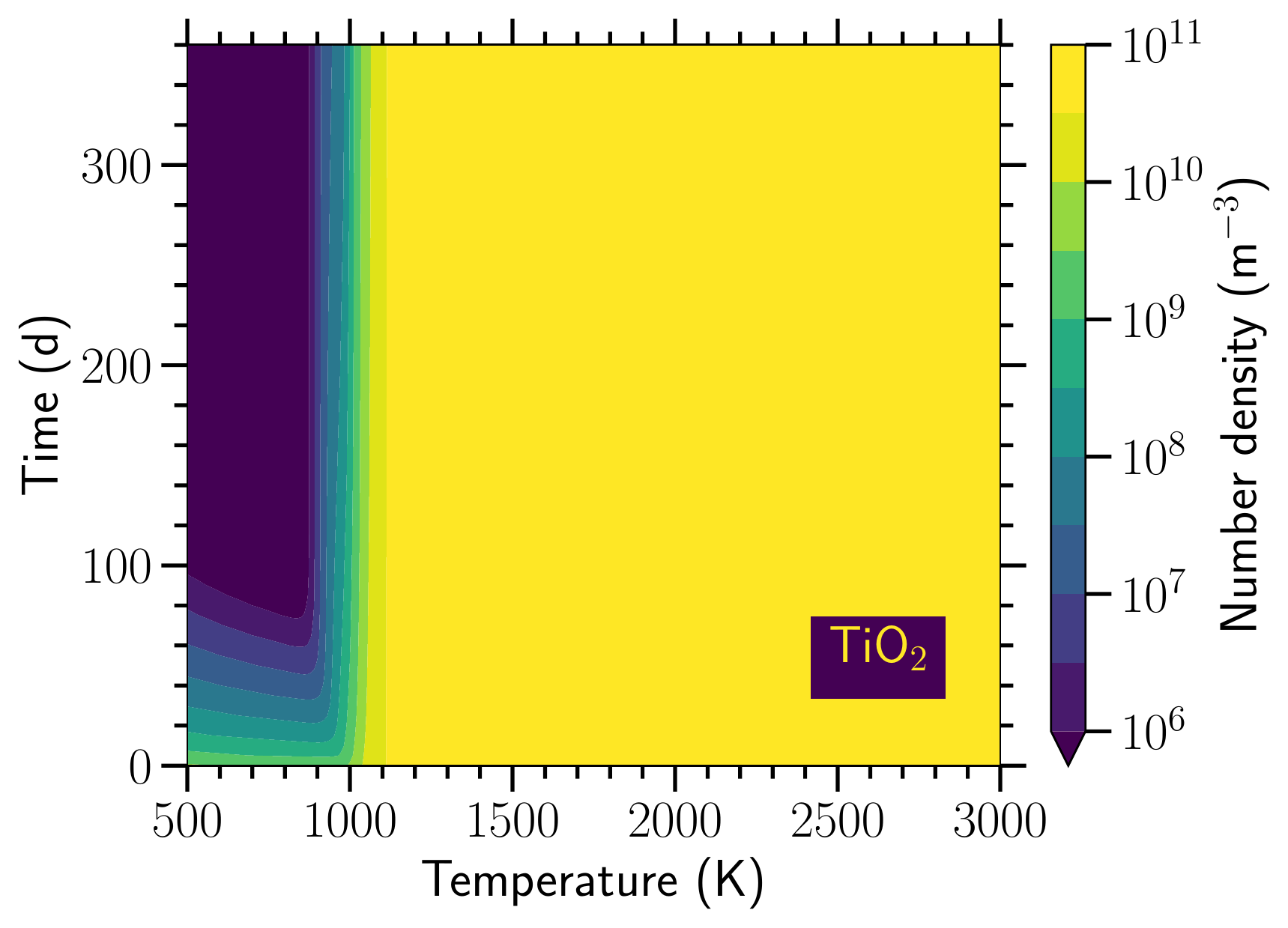}
        \includegraphics[width=0.32\textwidth]{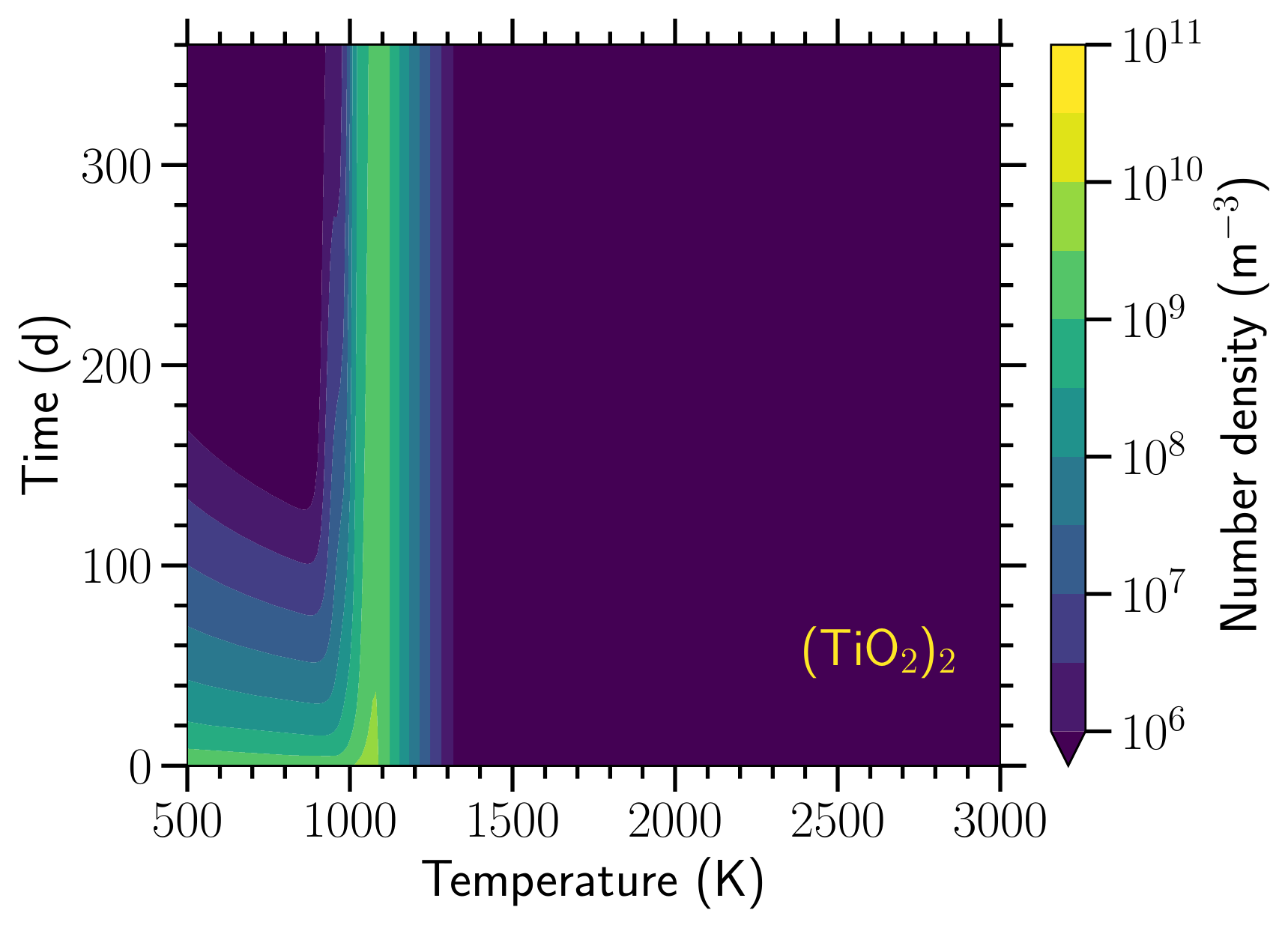}
        \includegraphics[width=0.32\textwidth]{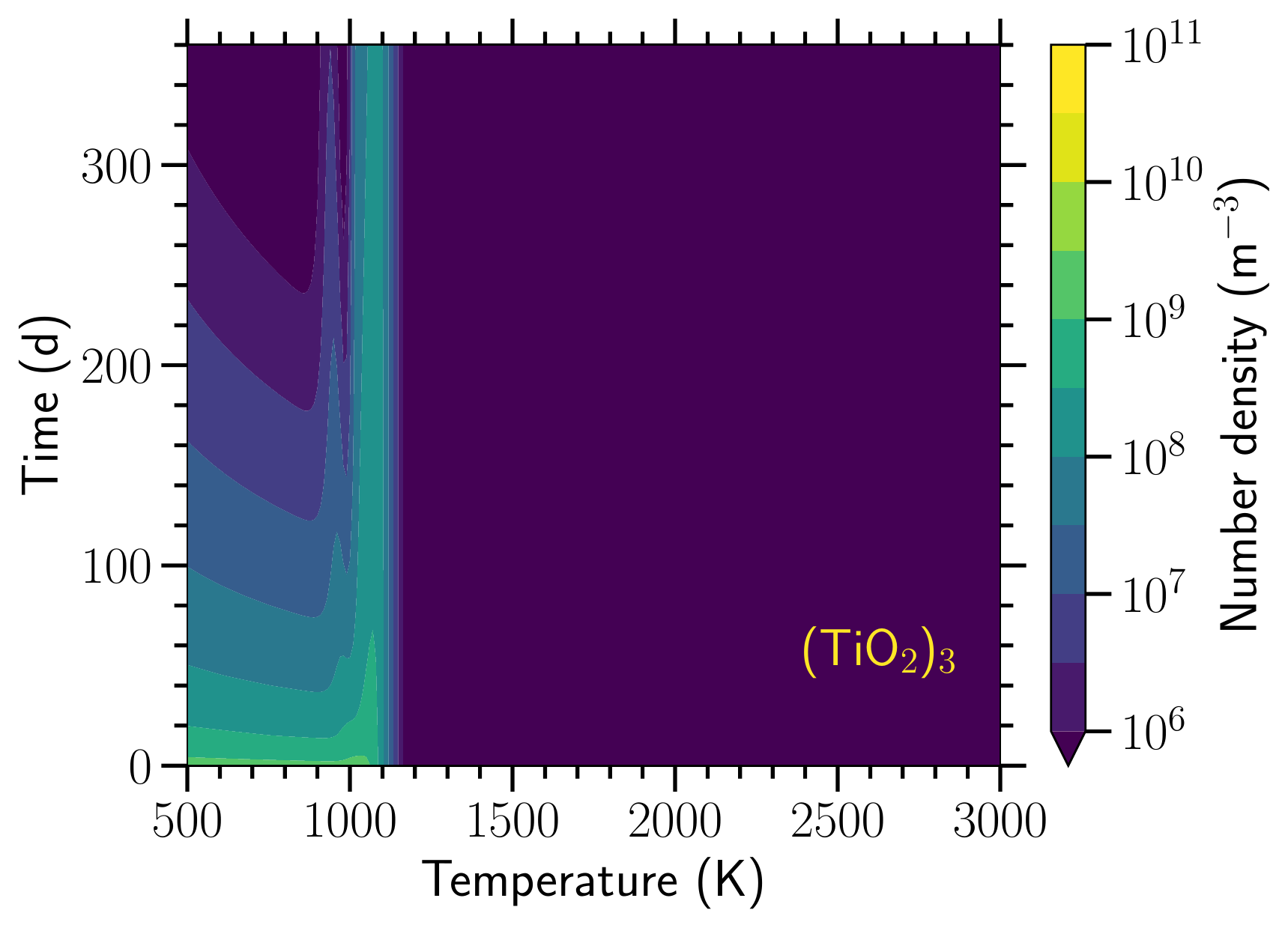}
        \includegraphics[width=0.32\textwidth]{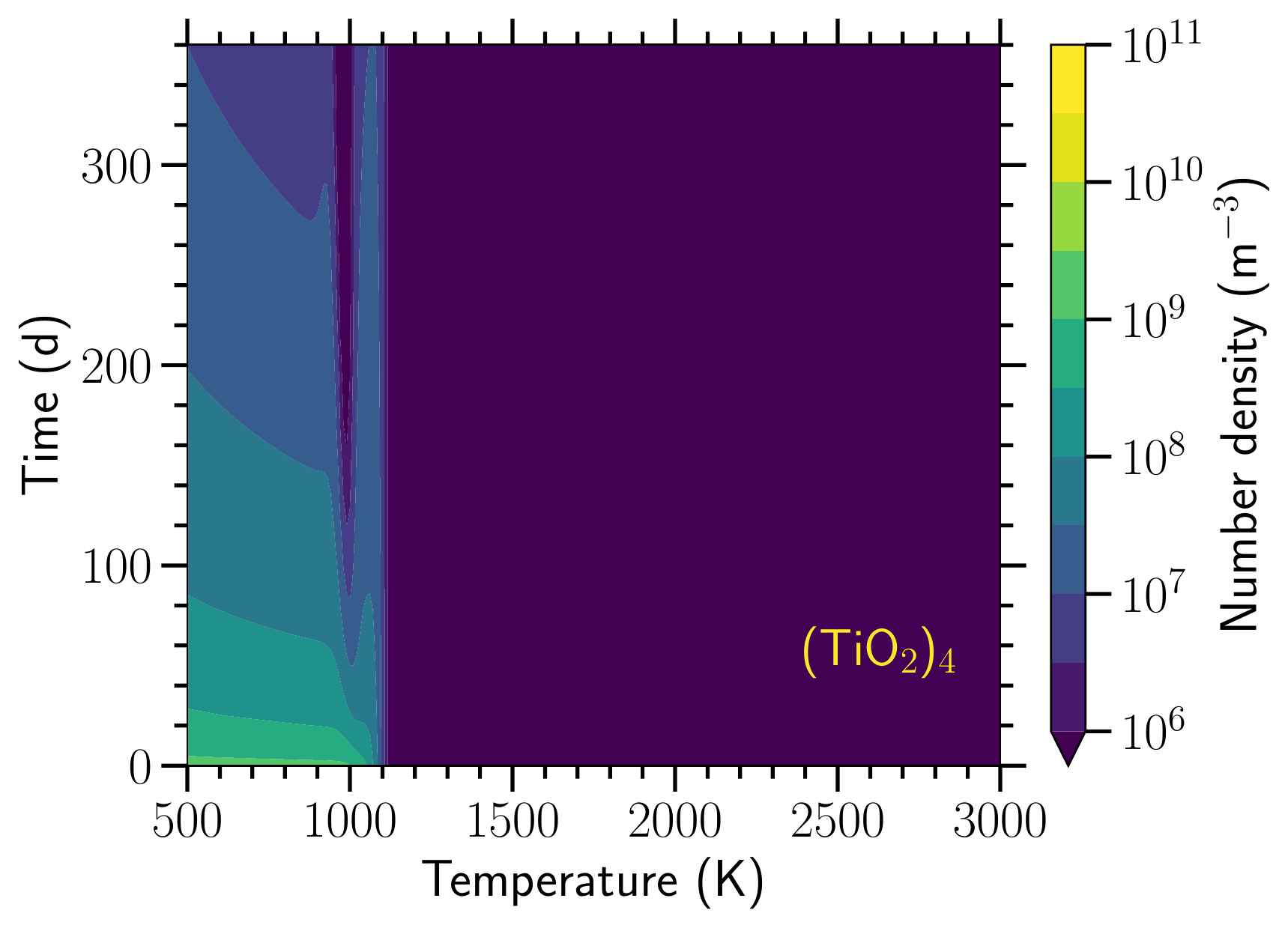}
        \includegraphics[width=0.32\textwidth]{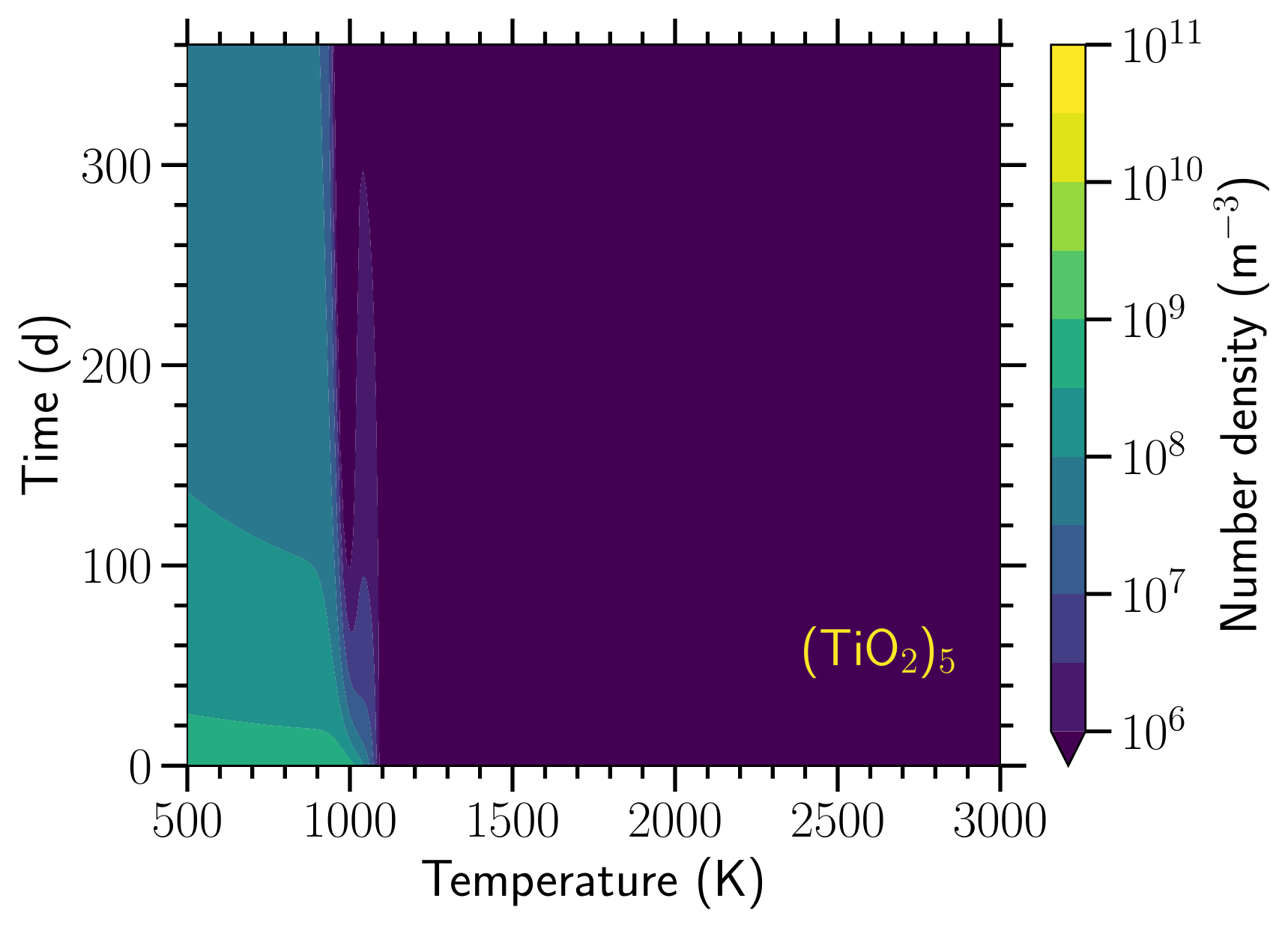}
        \includegraphics[width=0.32\textwidth]{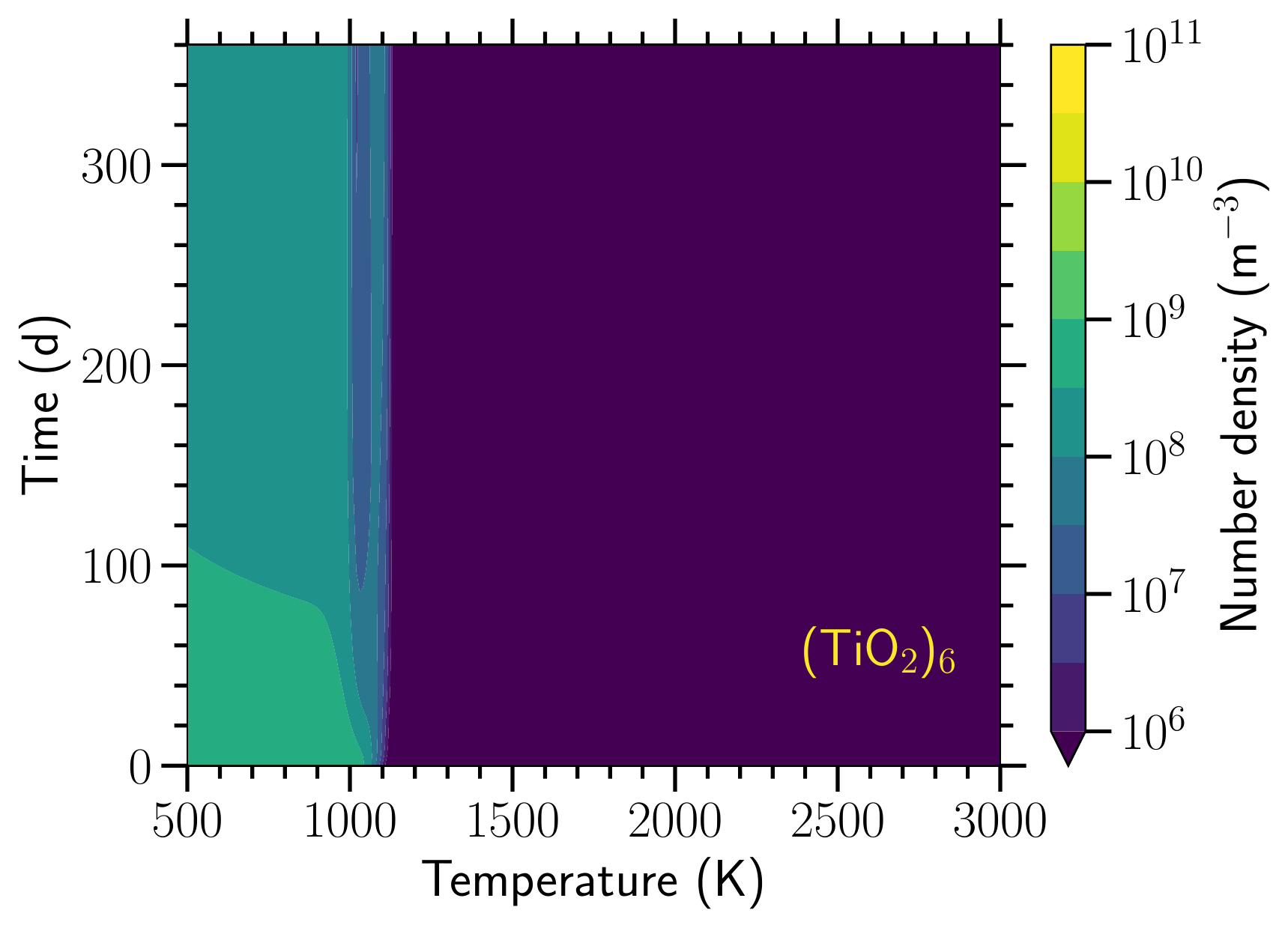}
        \includegraphics[width=0.32\textwidth]{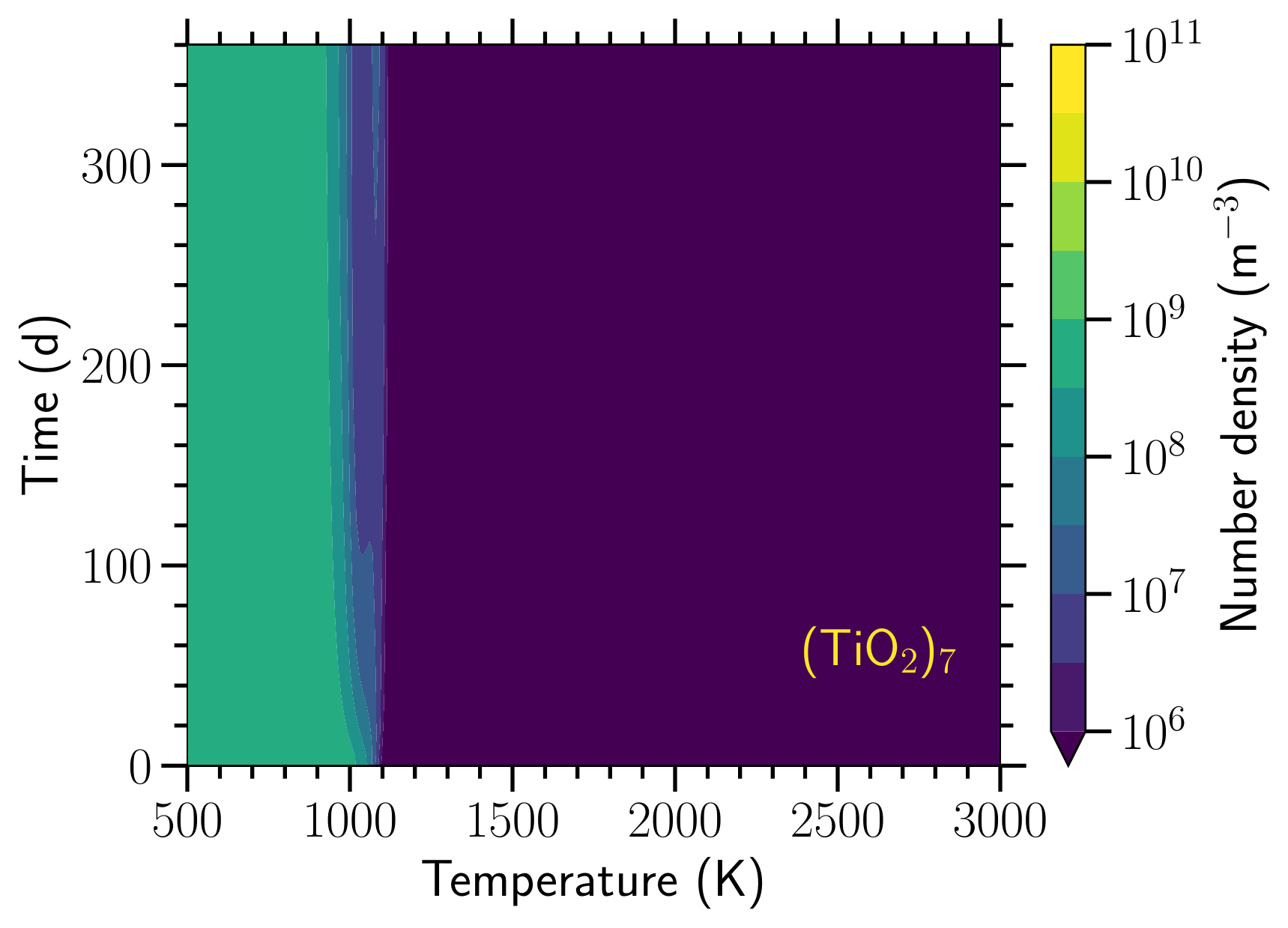}
        \includegraphics[width=0.32\textwidth]{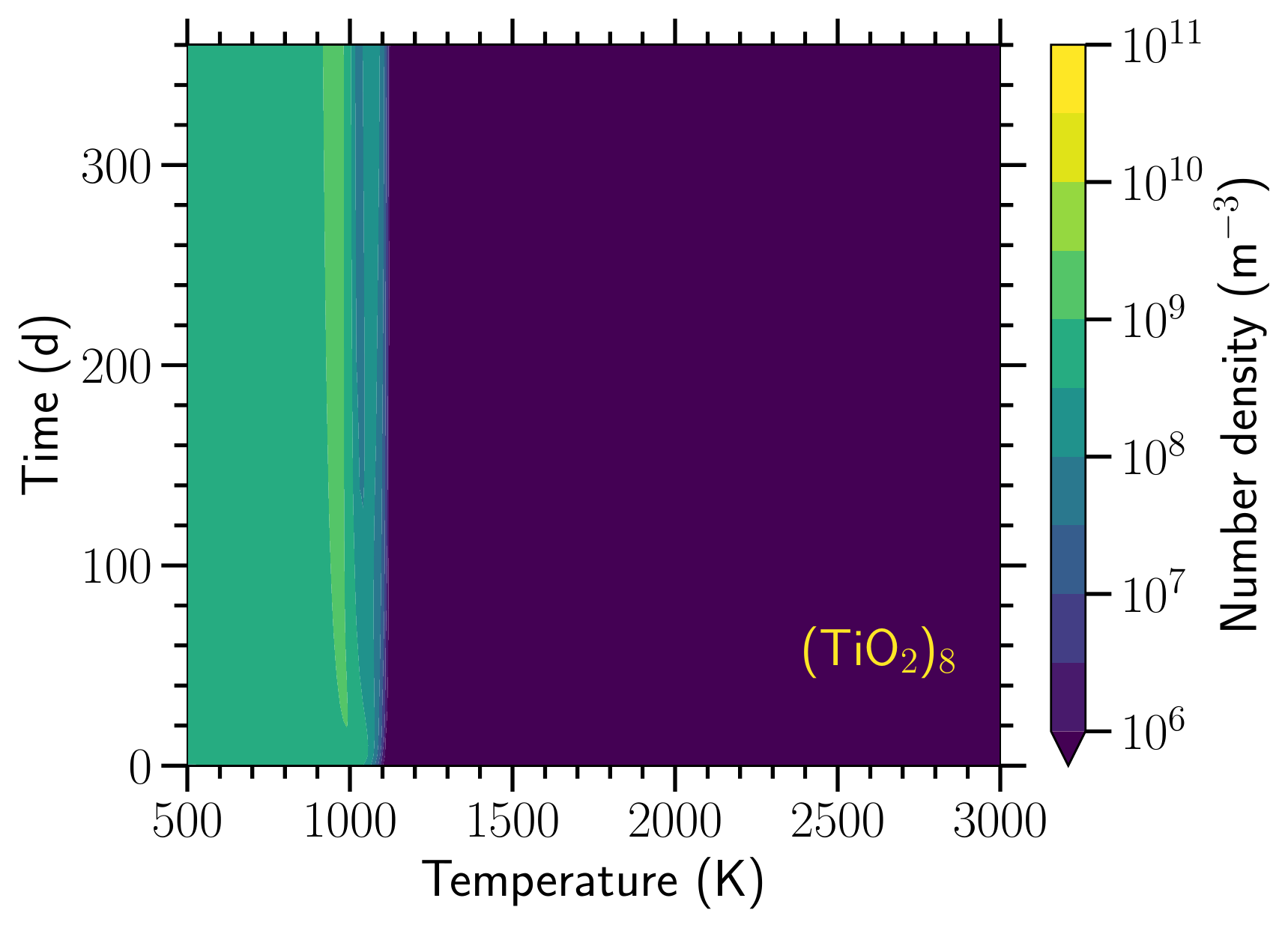}
        \includegraphics[width=0.32\textwidth]{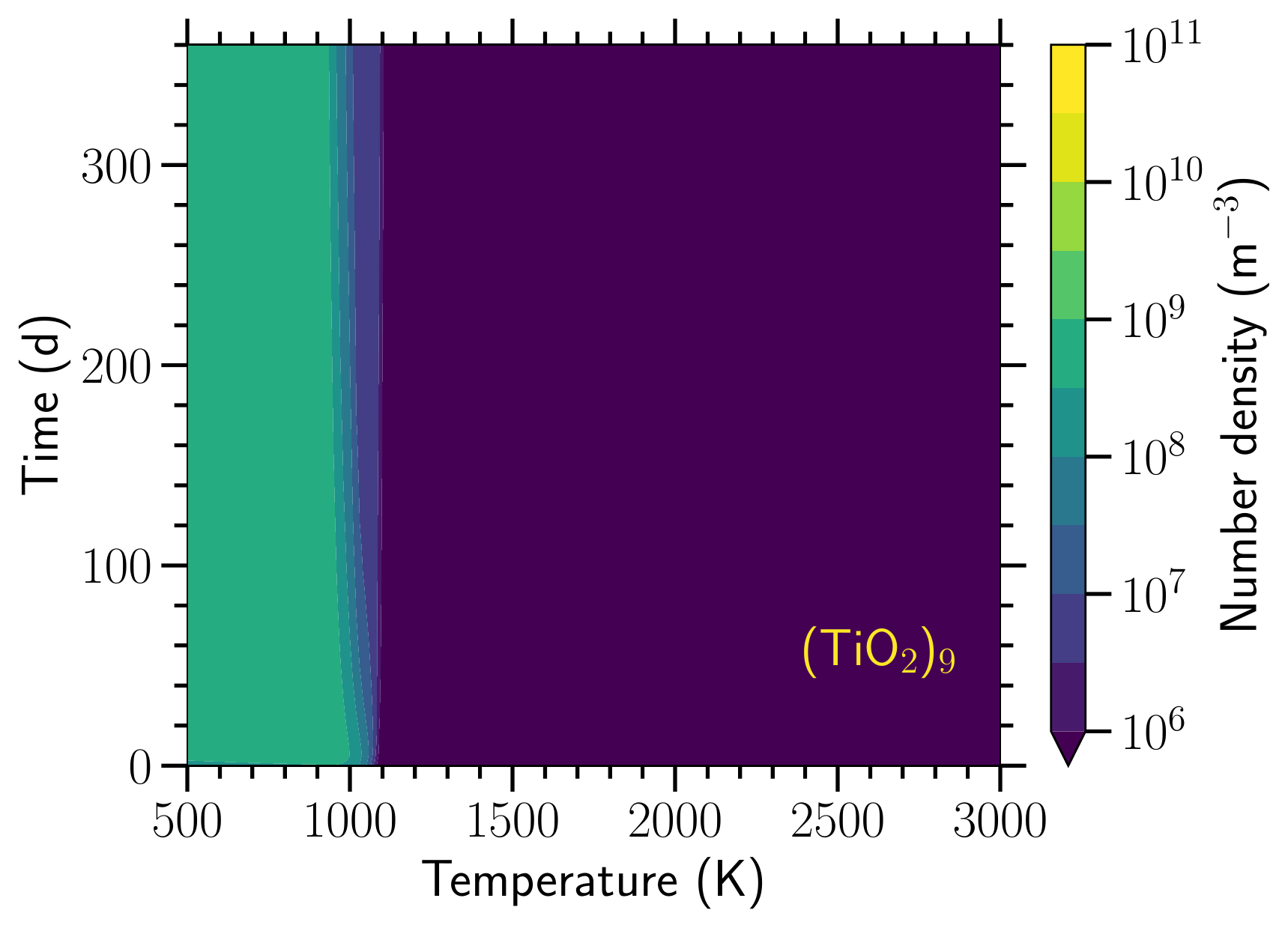}
        \includegraphics[width=0.32\textwidth]{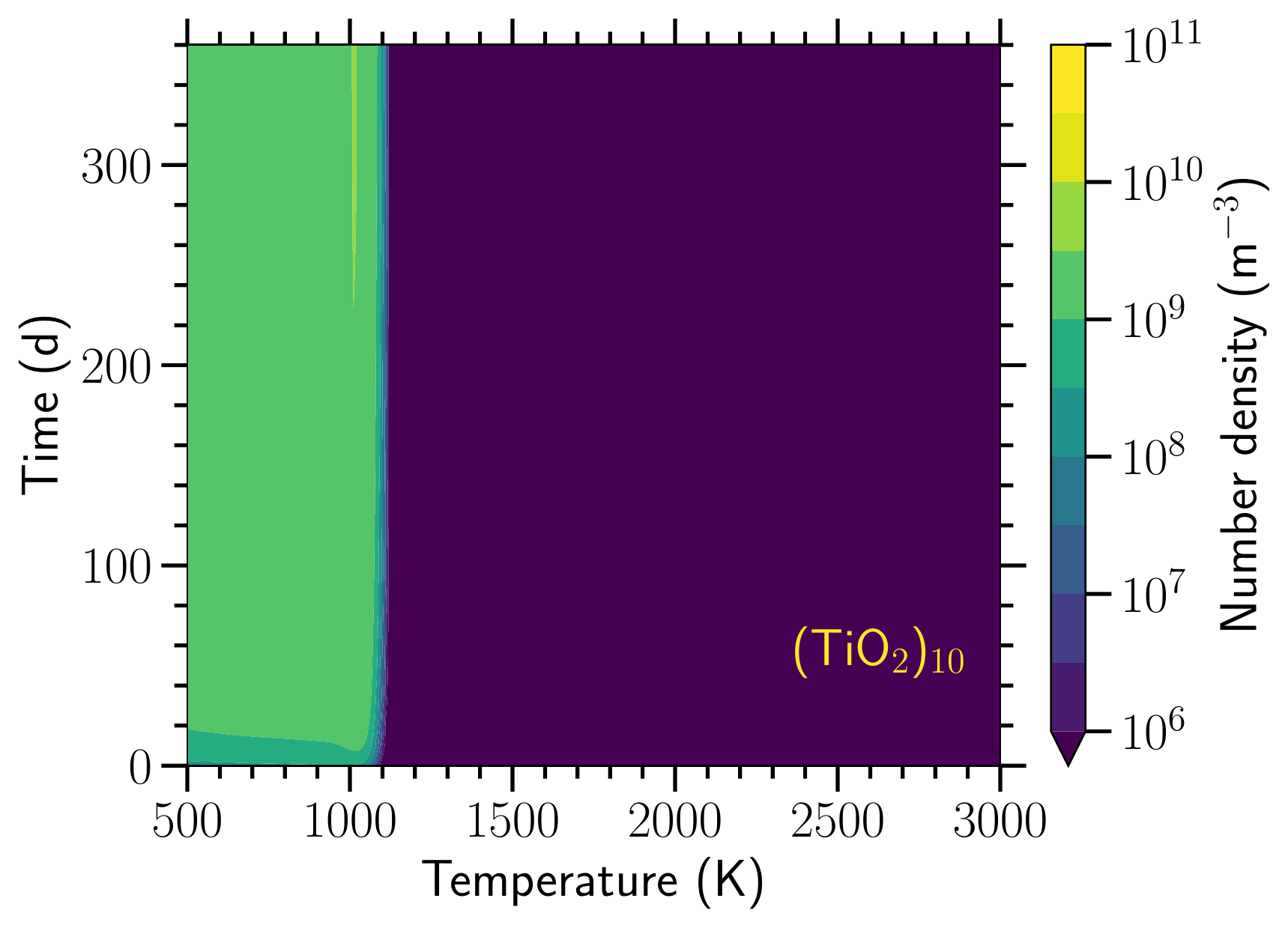}
        \end{flushleft}
        \caption{Temporal evolution of the absolute number density of all \protect\Ti{1}-clusters at the benchmark total gas density $\rho=\SI{1e-9}{\kg\per\m\cubed}$ for a closed nucleation model using the polymer nucleation description.}
        \label{fig:TiO2_clusters_general_time_evolution}
    \end{figure*}

\begin{figure*}
        \begin{flushleft}
        \includegraphics[width=0.32\textwidth]{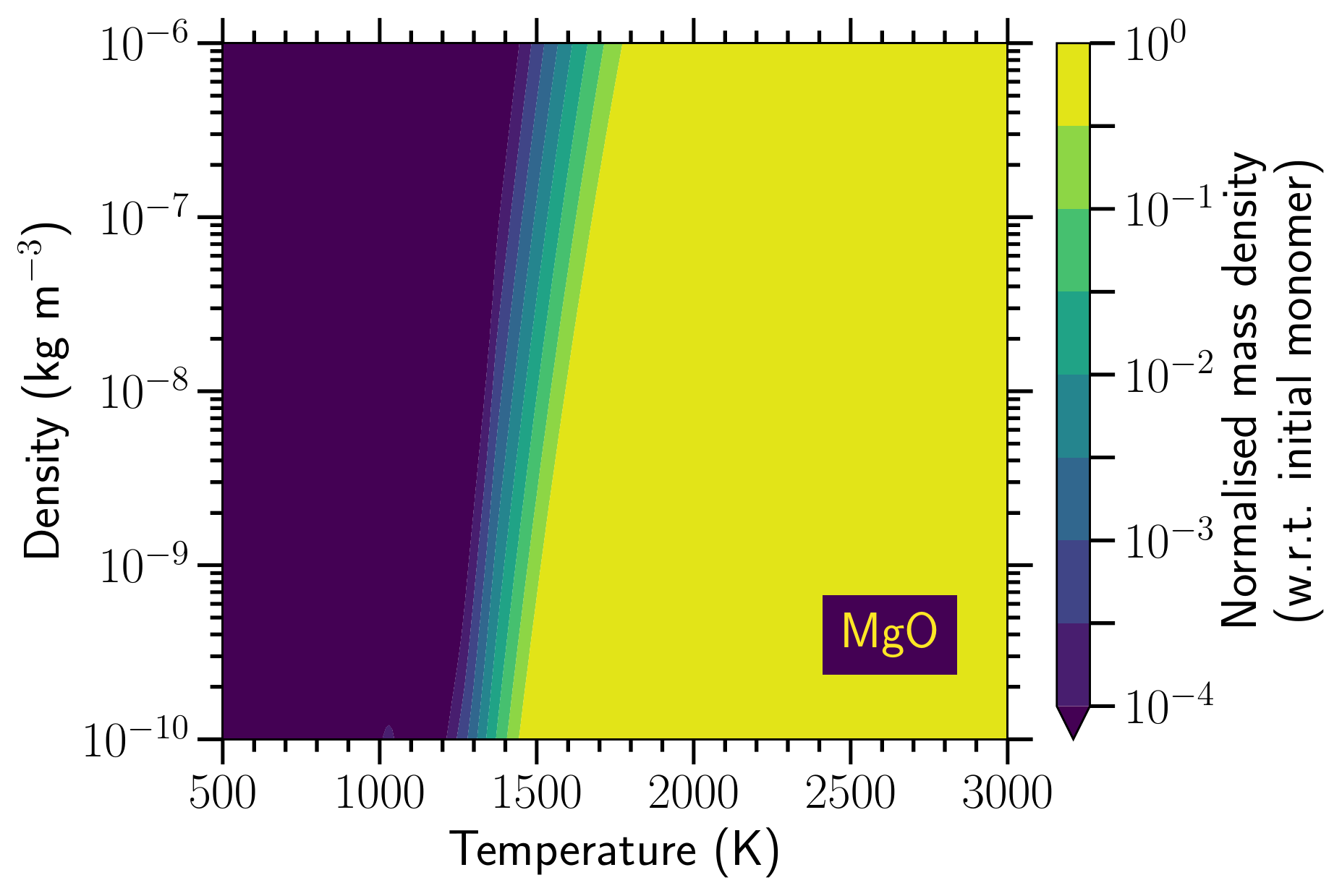}
        \includegraphics[width=0.32\textwidth]{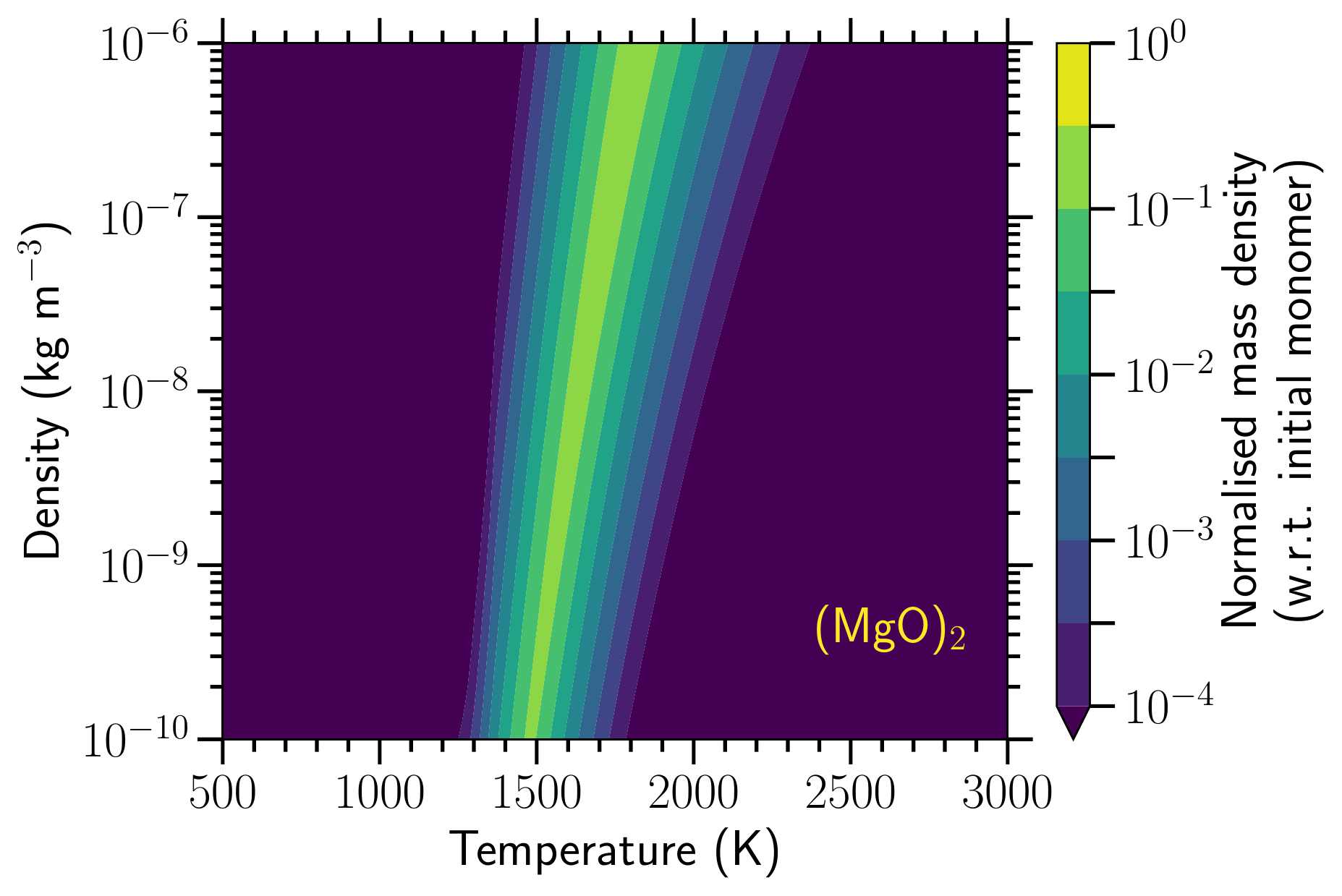}
        \includegraphics[width=0.32\textwidth]{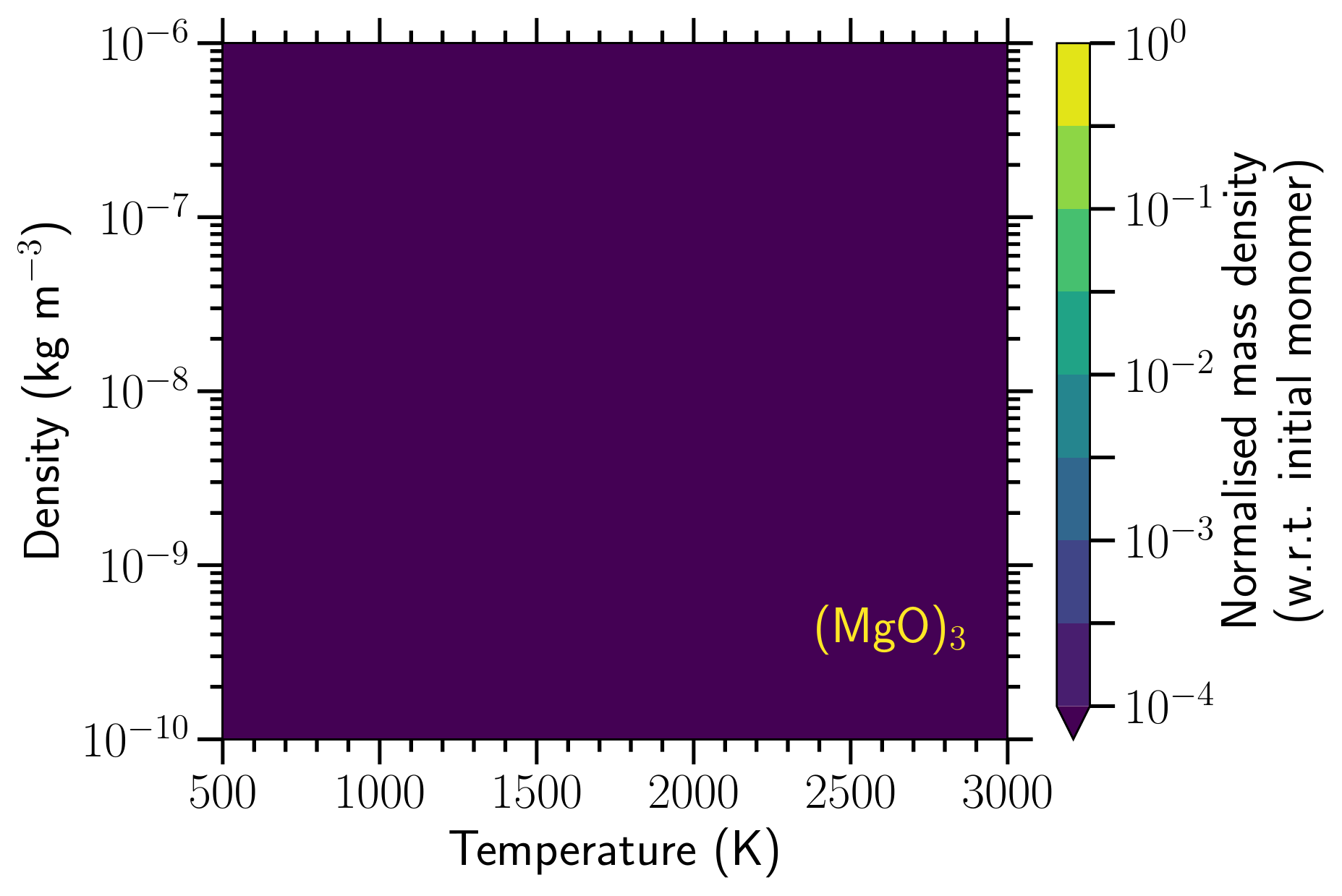}
        \includegraphics[width=0.32\textwidth]{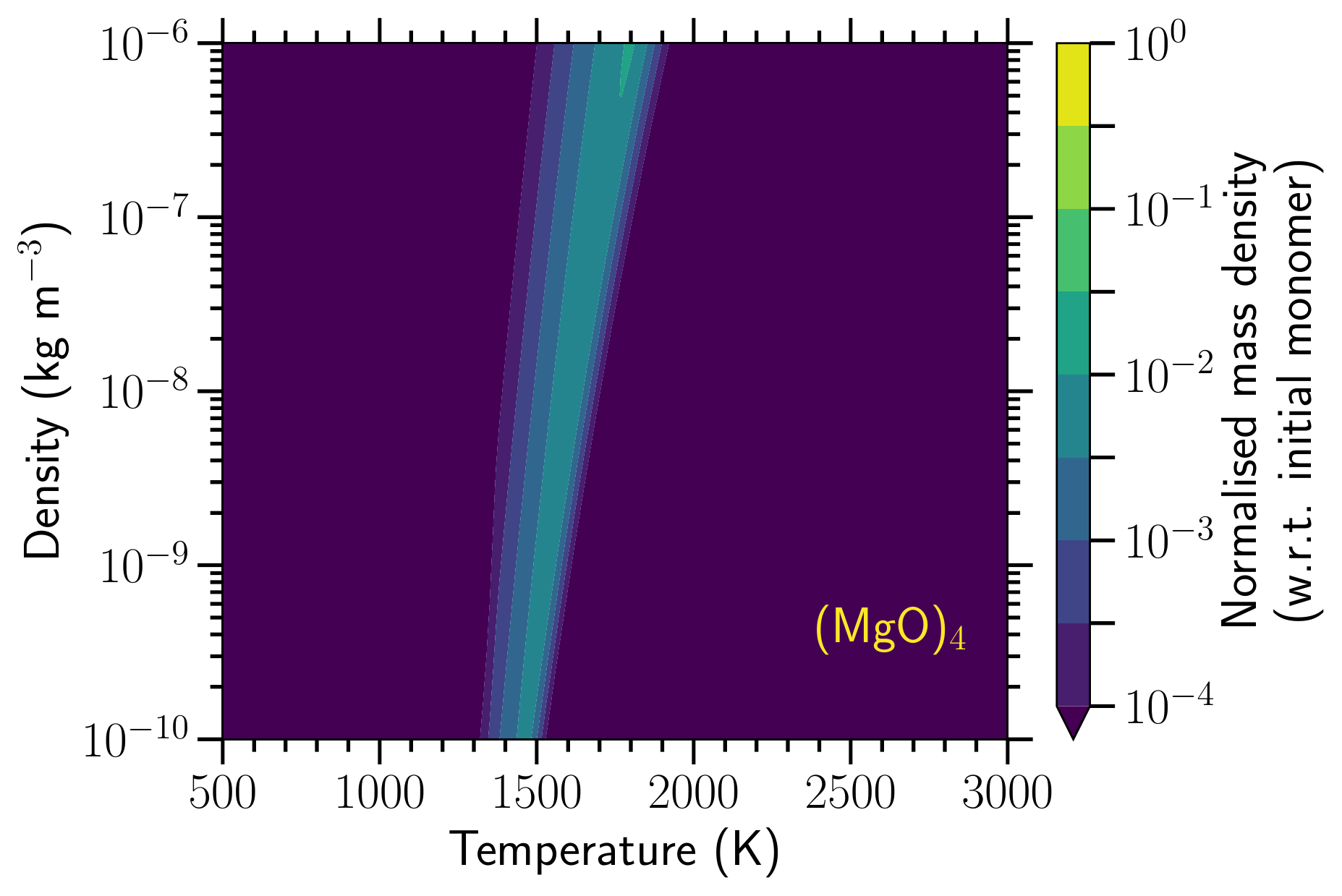}
        \includegraphics[width=0.32\textwidth]{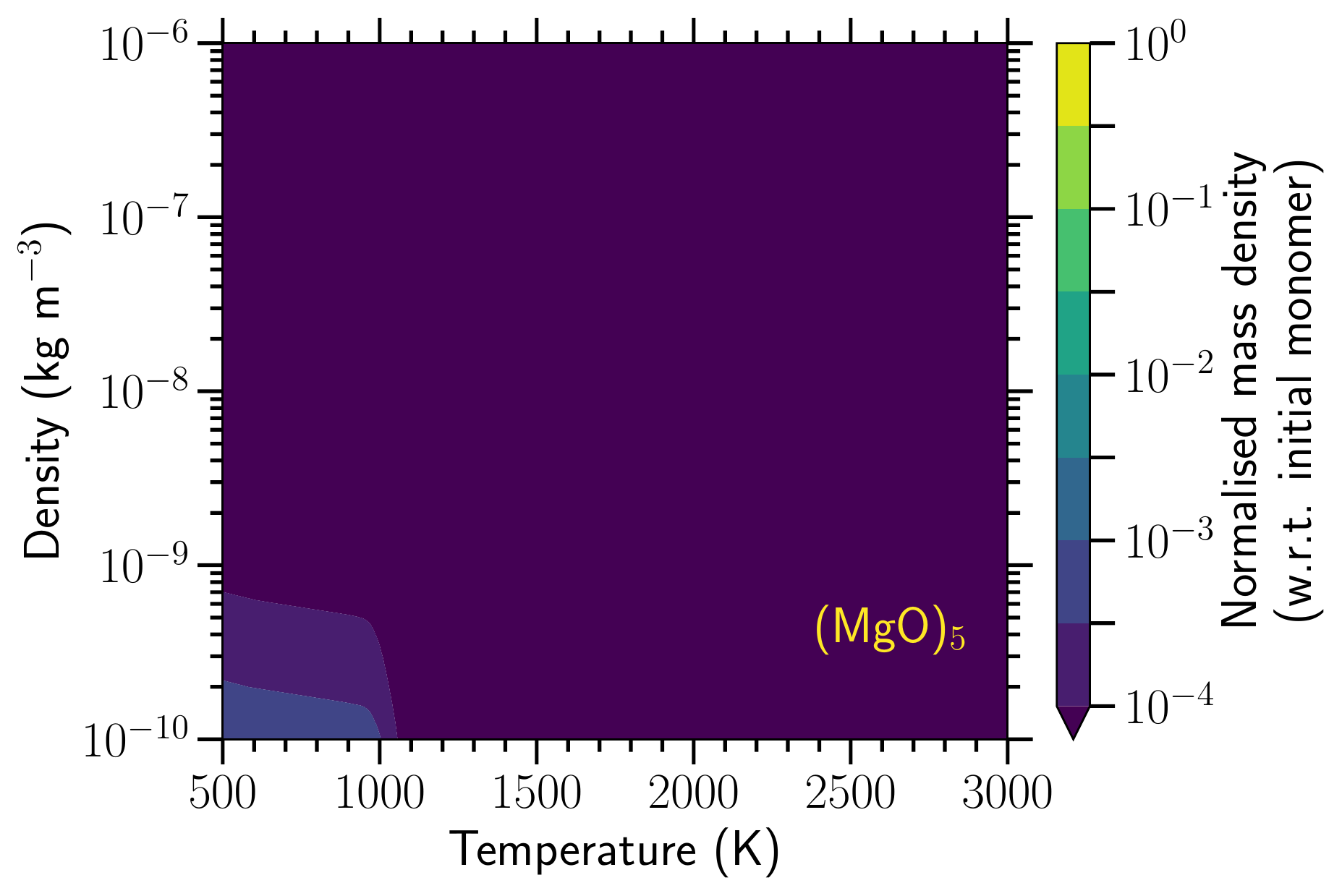}
        \includegraphics[width=0.32\textwidth]{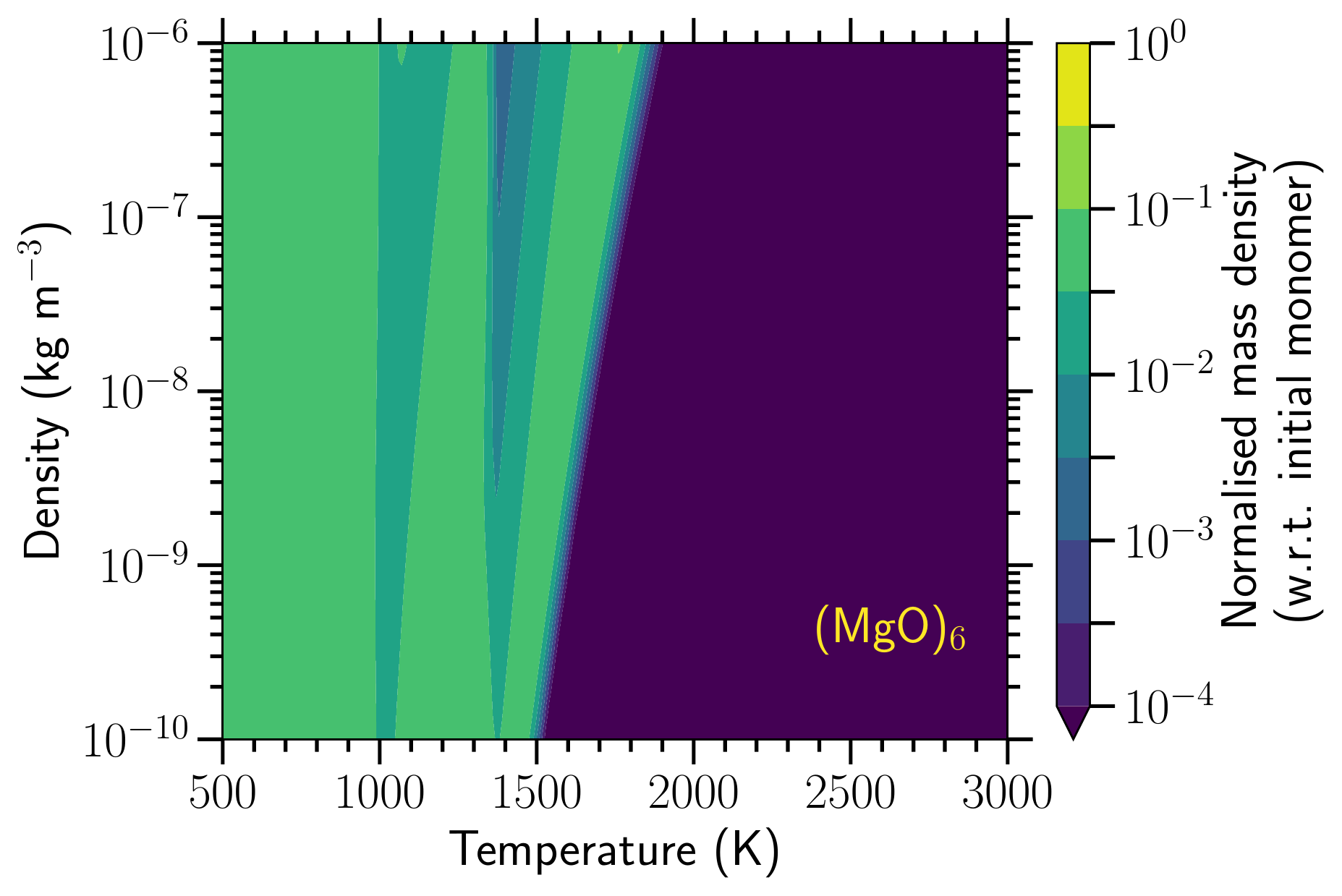}
        \includegraphics[width=0.32\textwidth]{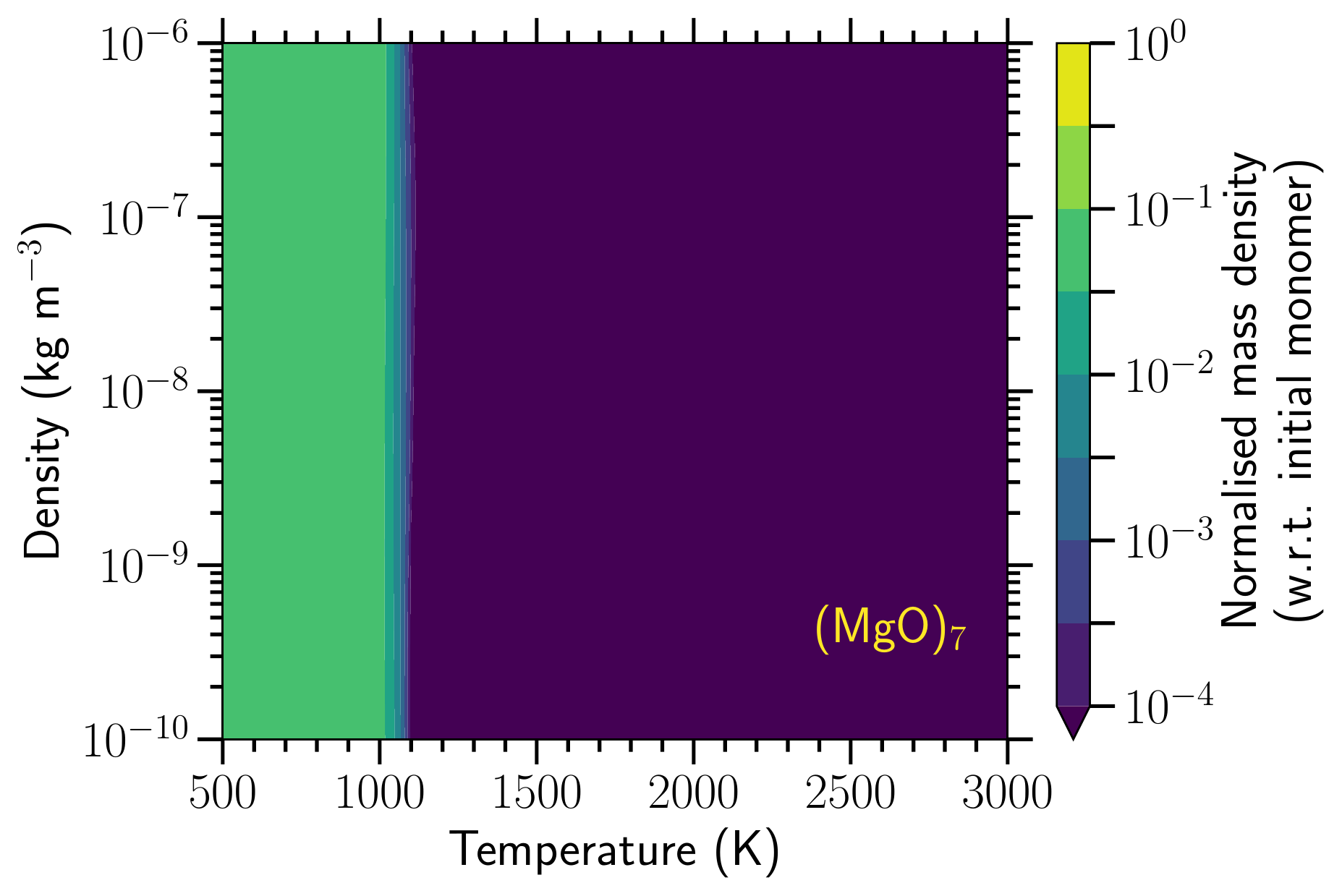}
        \includegraphics[width=0.32\textwidth]{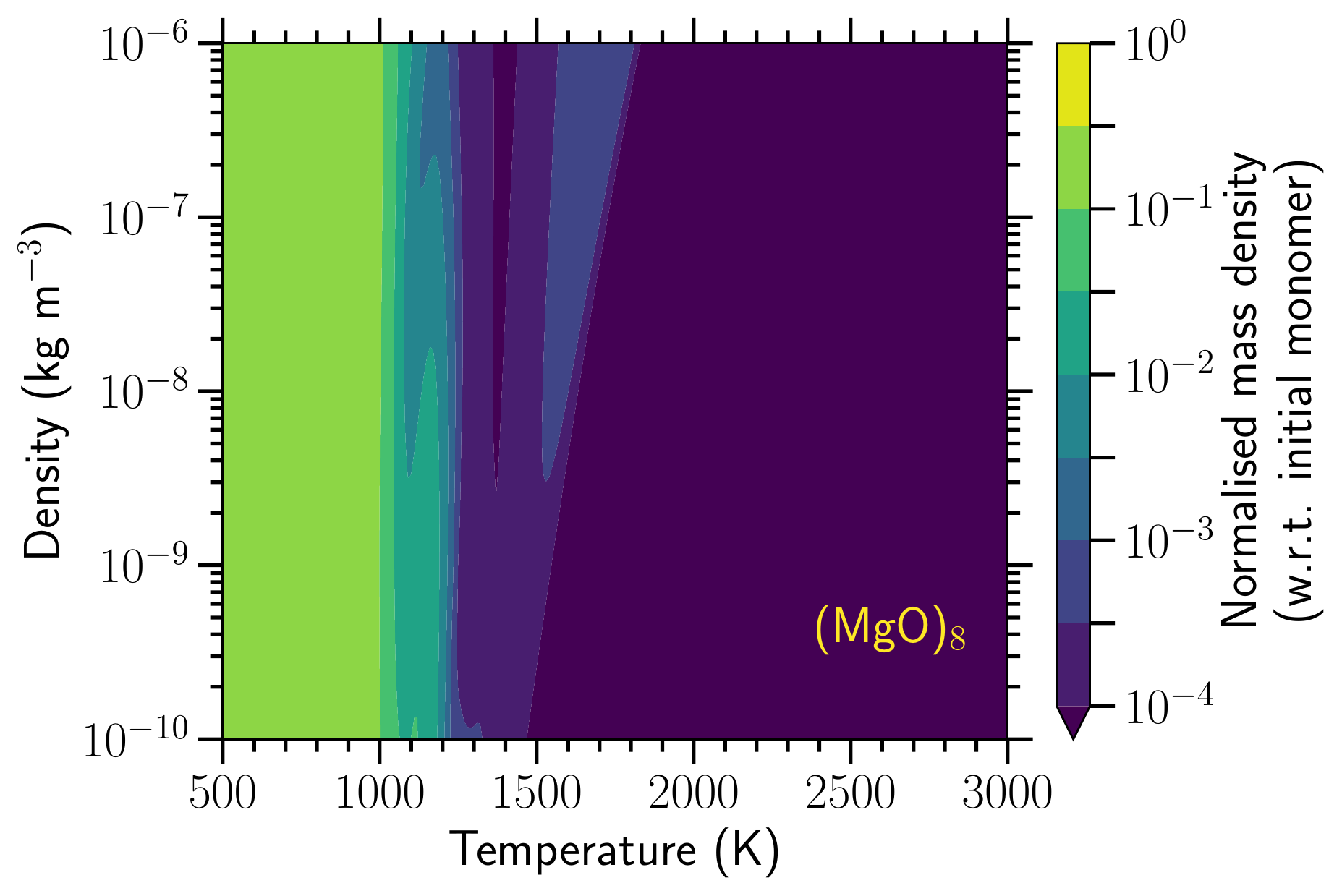}
        \includegraphics[width=0.32\textwidth]{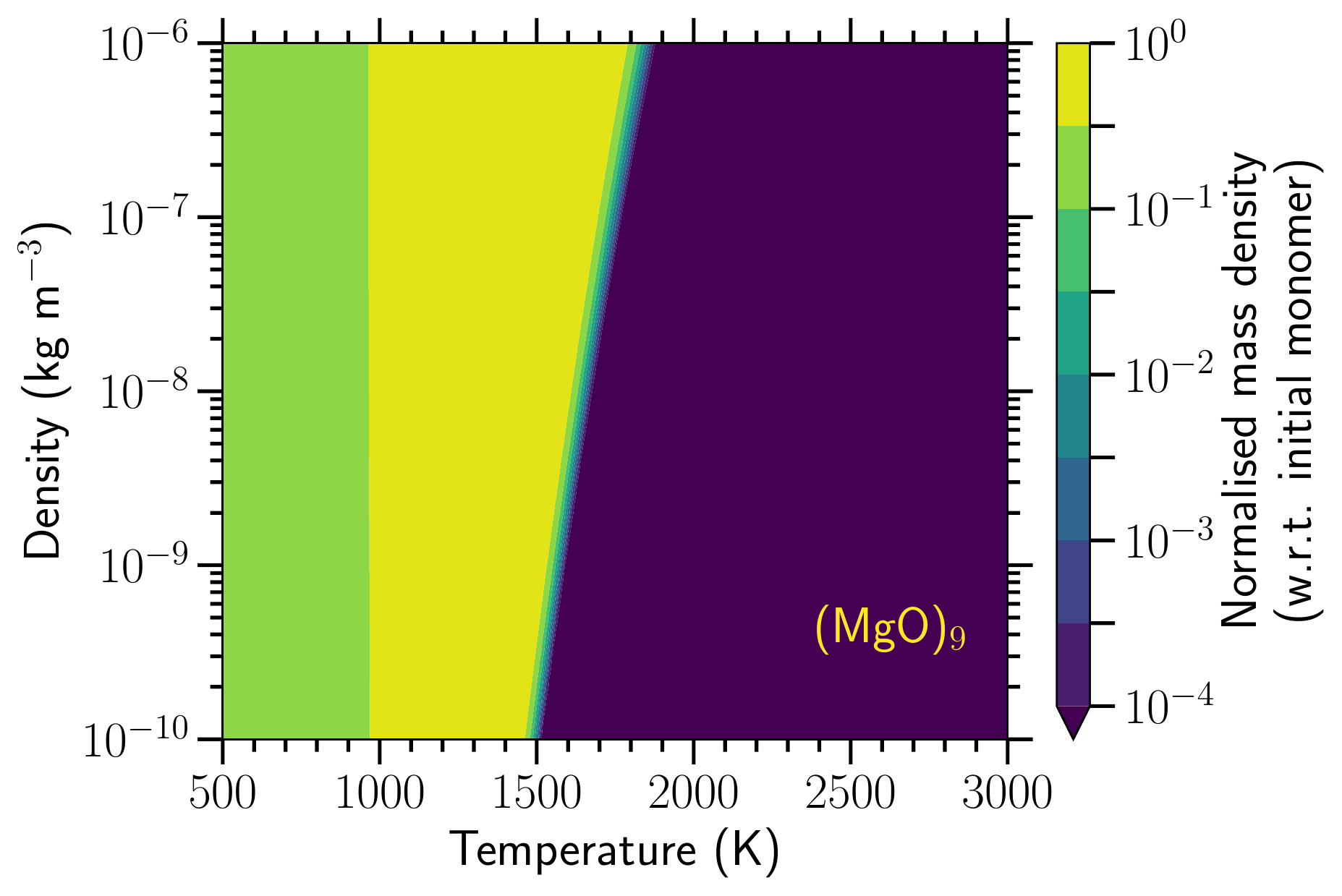}
        \includegraphics[width=0.32\textwidth]{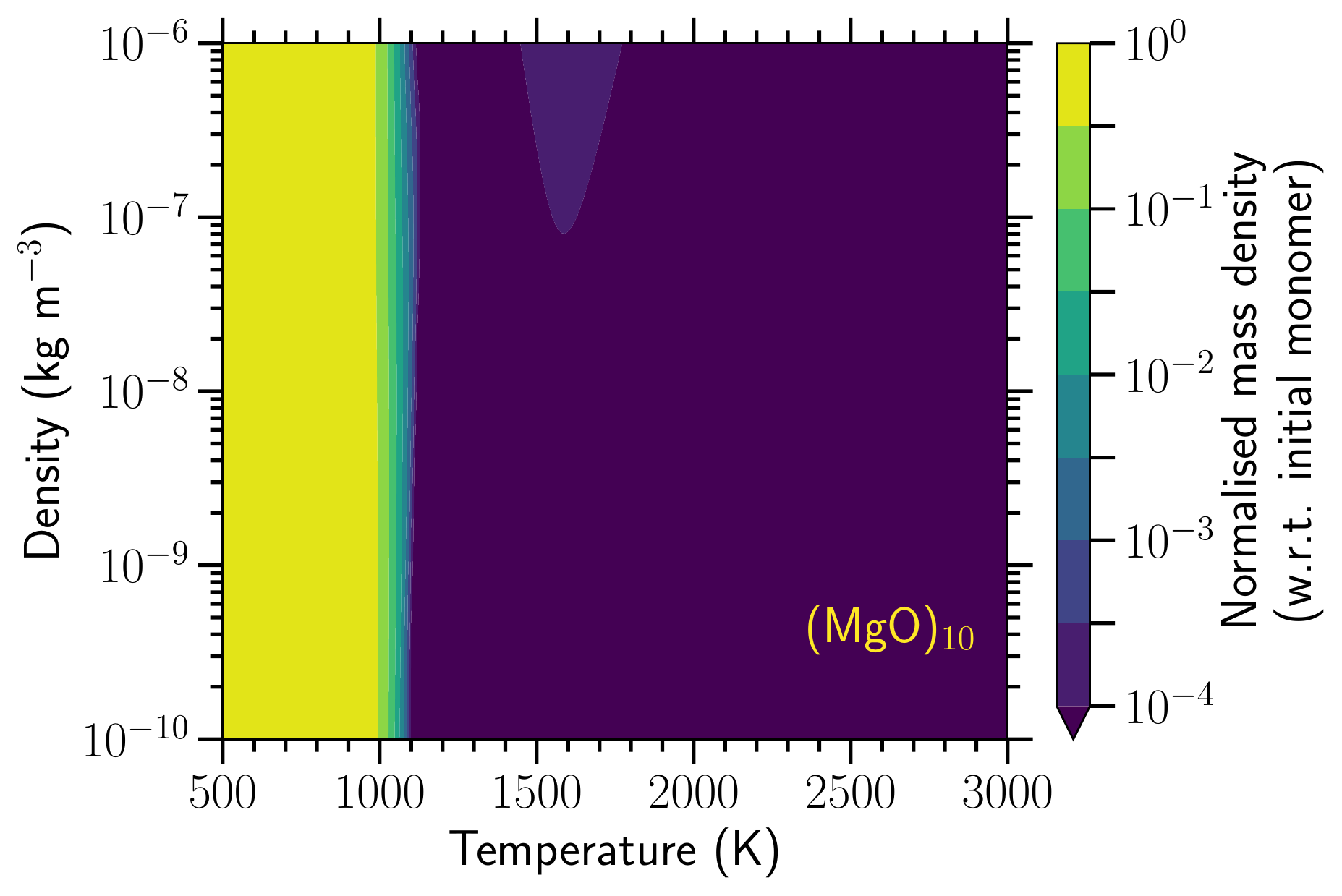}
        \end{flushleft}
        \caption{Overview of the normalised mass density after one year of all \protect\Mg{1}-clusters for a closed nucleation model using the polymer nucleation description.}
        \label{fig:MgO_clusters_general_norm_same_scale}
    \end{figure*}

\begin{figure*}
        \begin{flushleft}
        \includegraphics[width=0.32\textwidth]{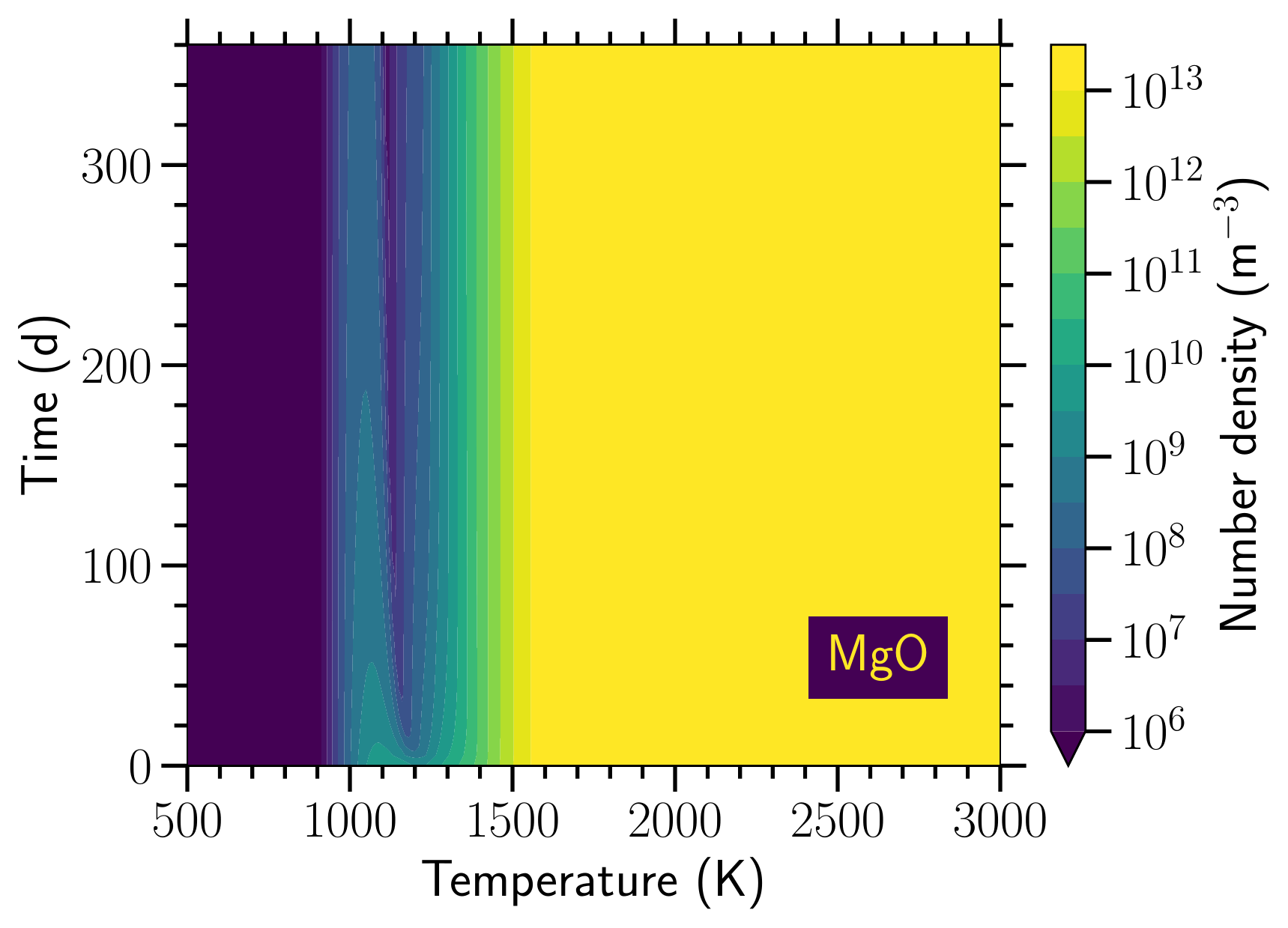}
        \includegraphics[width=0.32\textwidth]{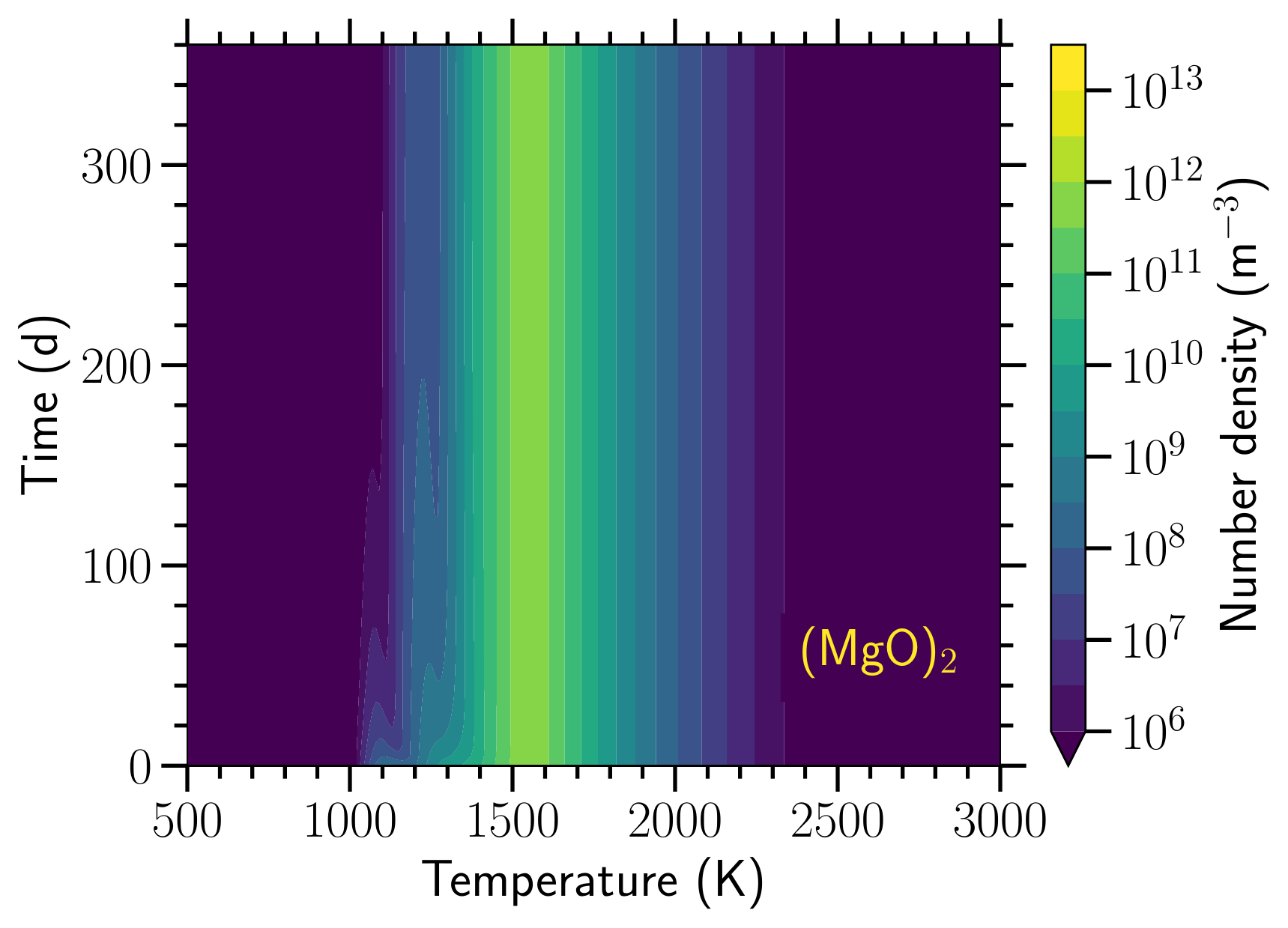}
        \includegraphics[width=0.32\textwidth]{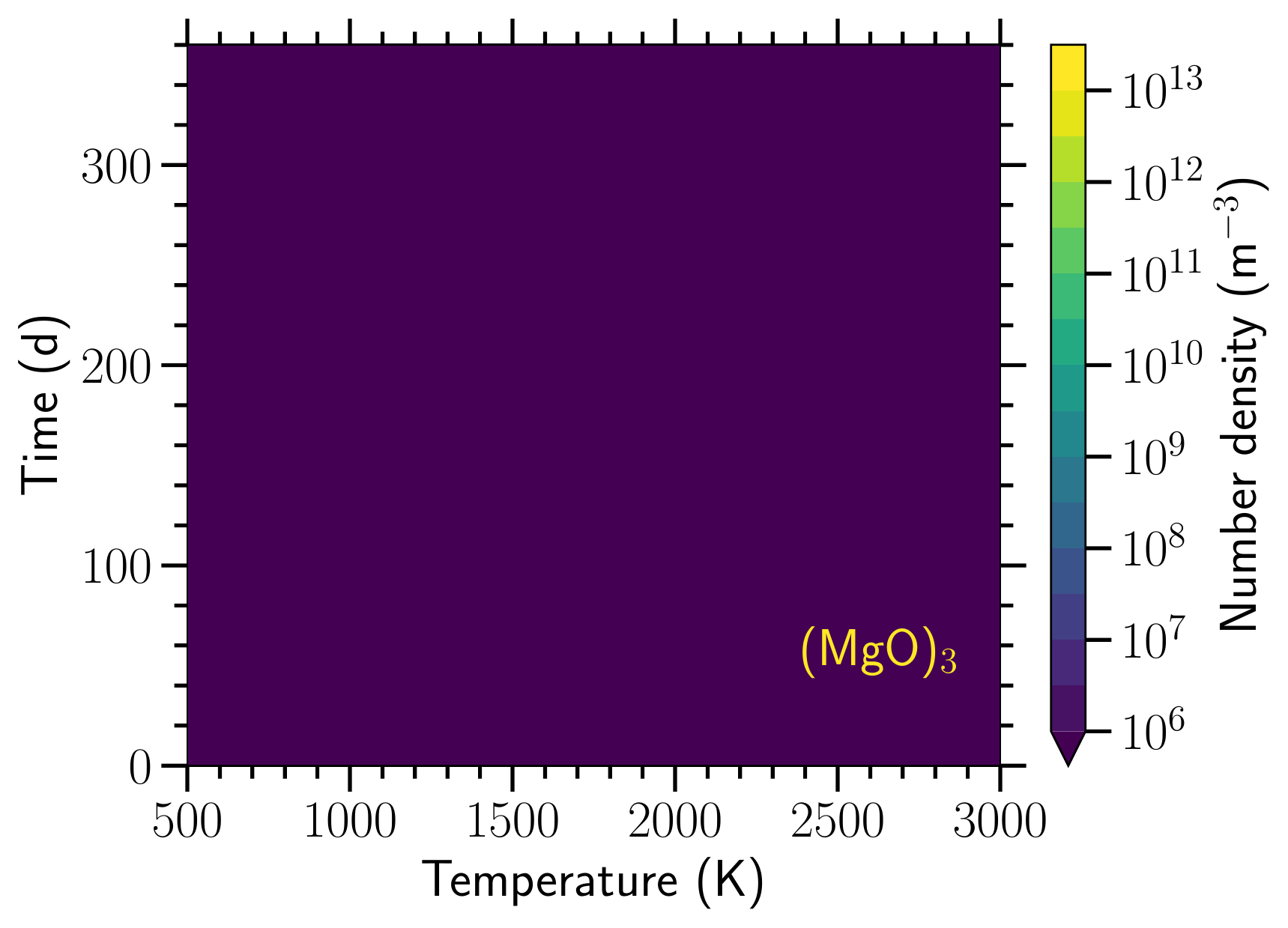}
        \includegraphics[width=0.32\textwidth]{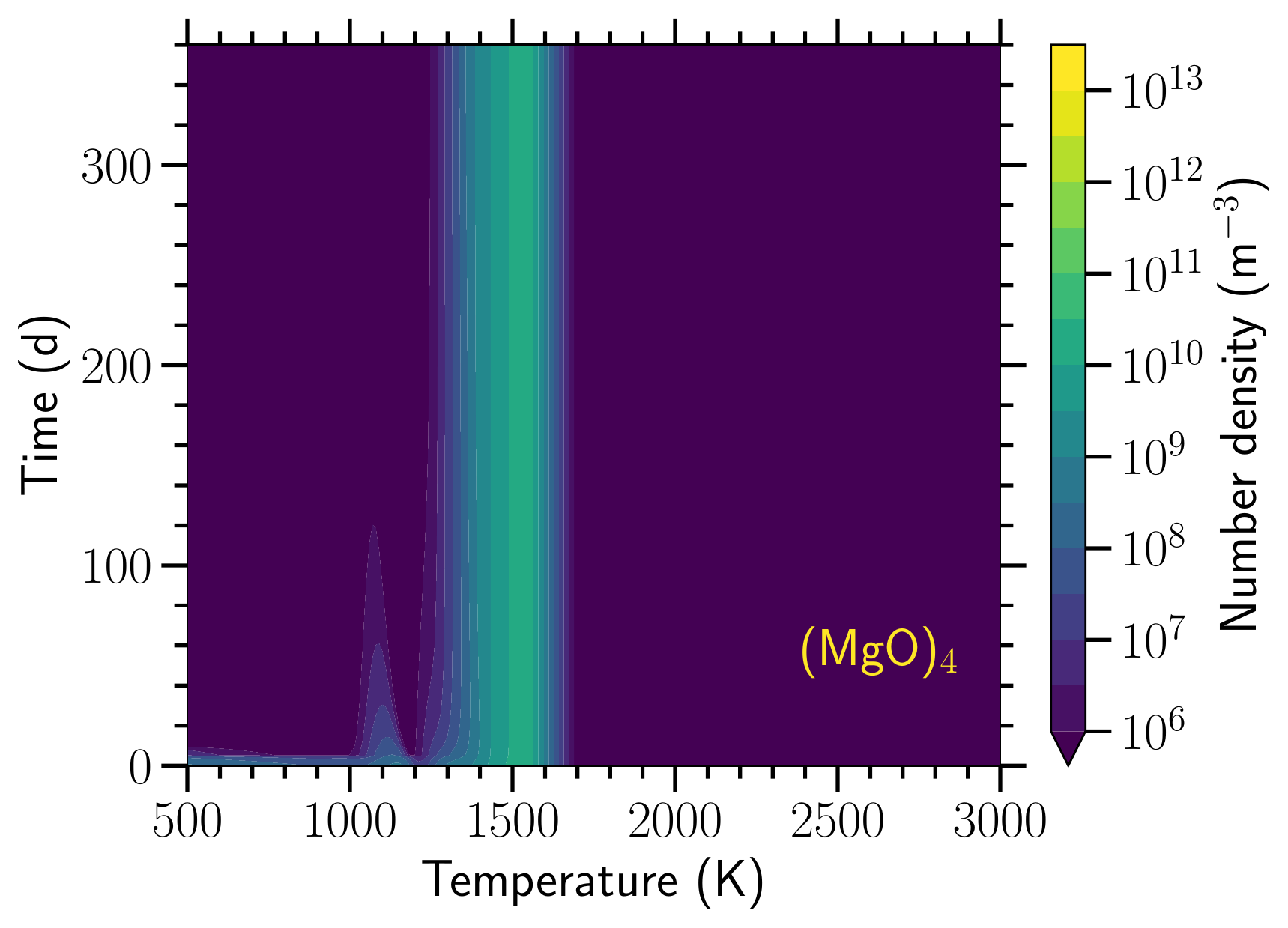}
        \includegraphics[width=0.32\textwidth]{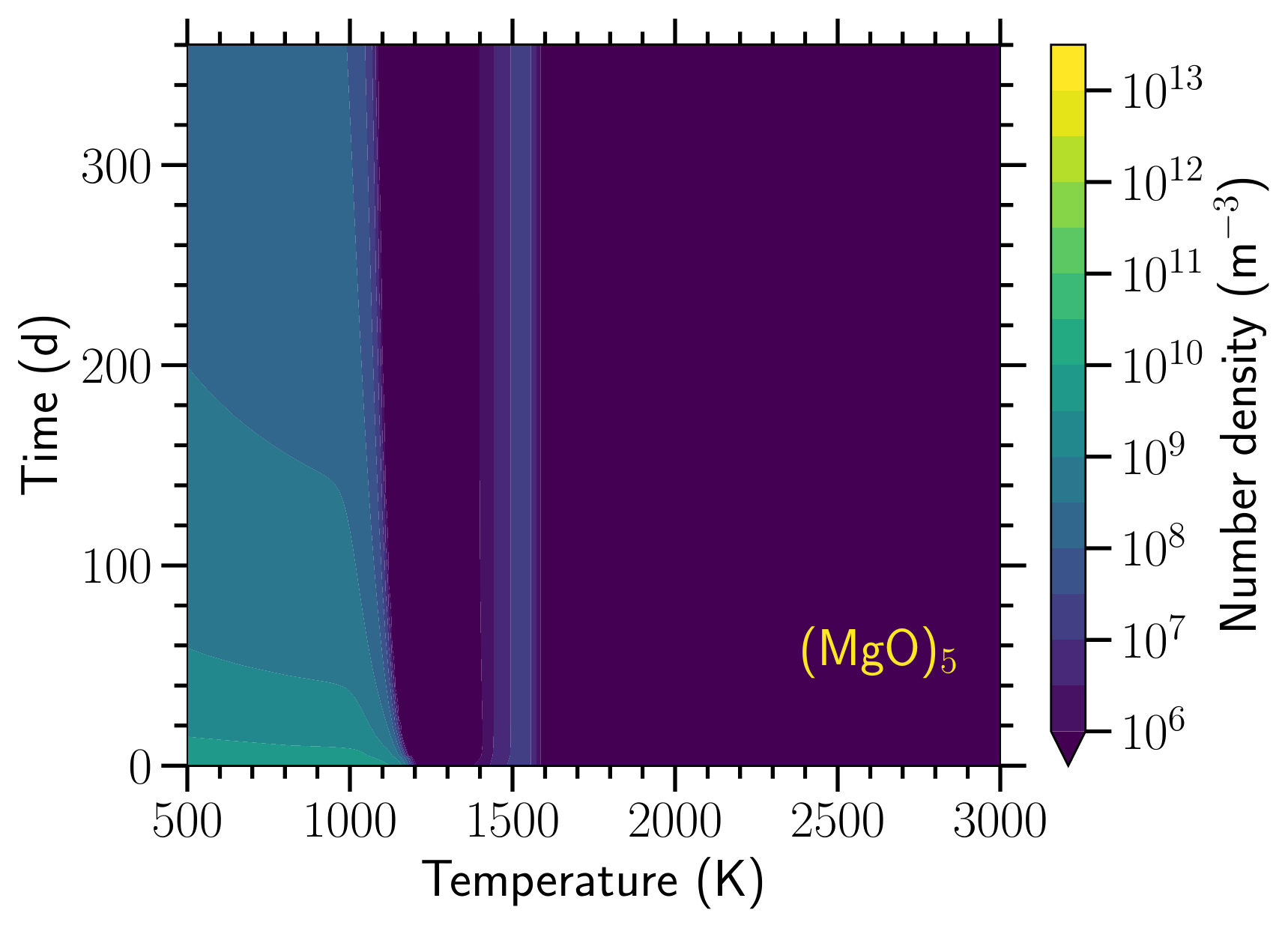}
        \includegraphics[width=0.32\textwidth]{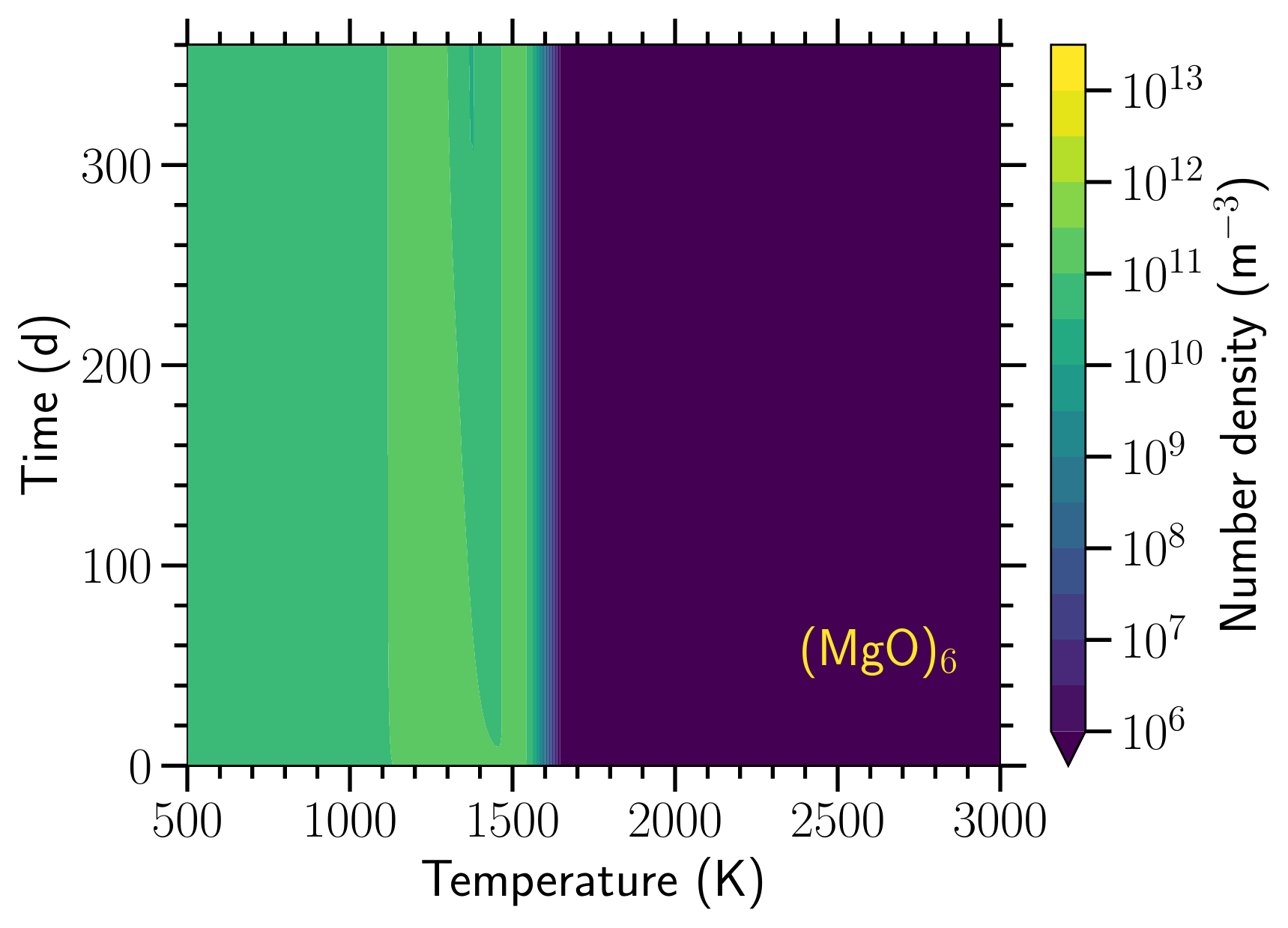}
        \includegraphics[width=0.32\textwidth]{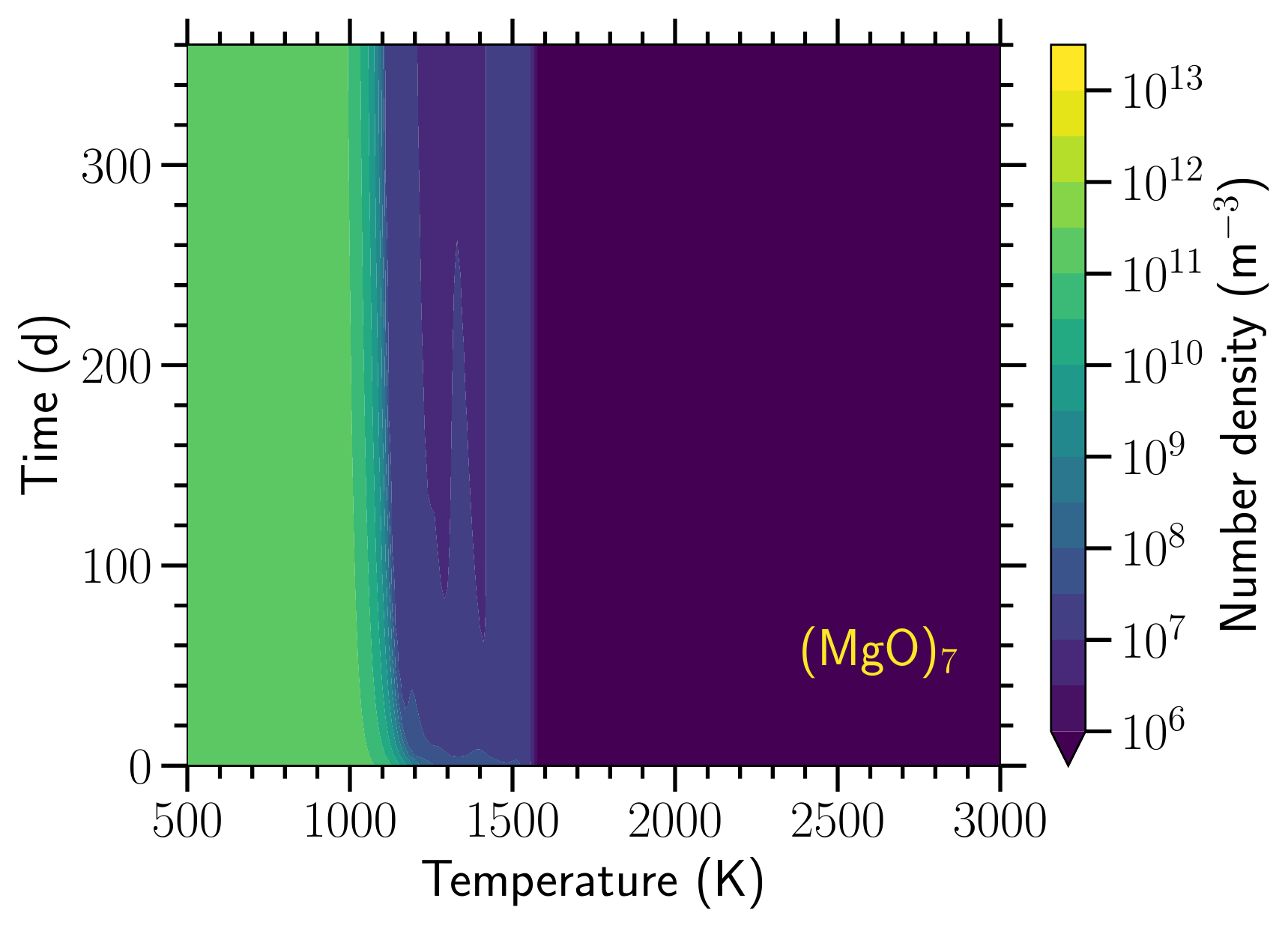}
        \includegraphics[width=0.32\textwidth]{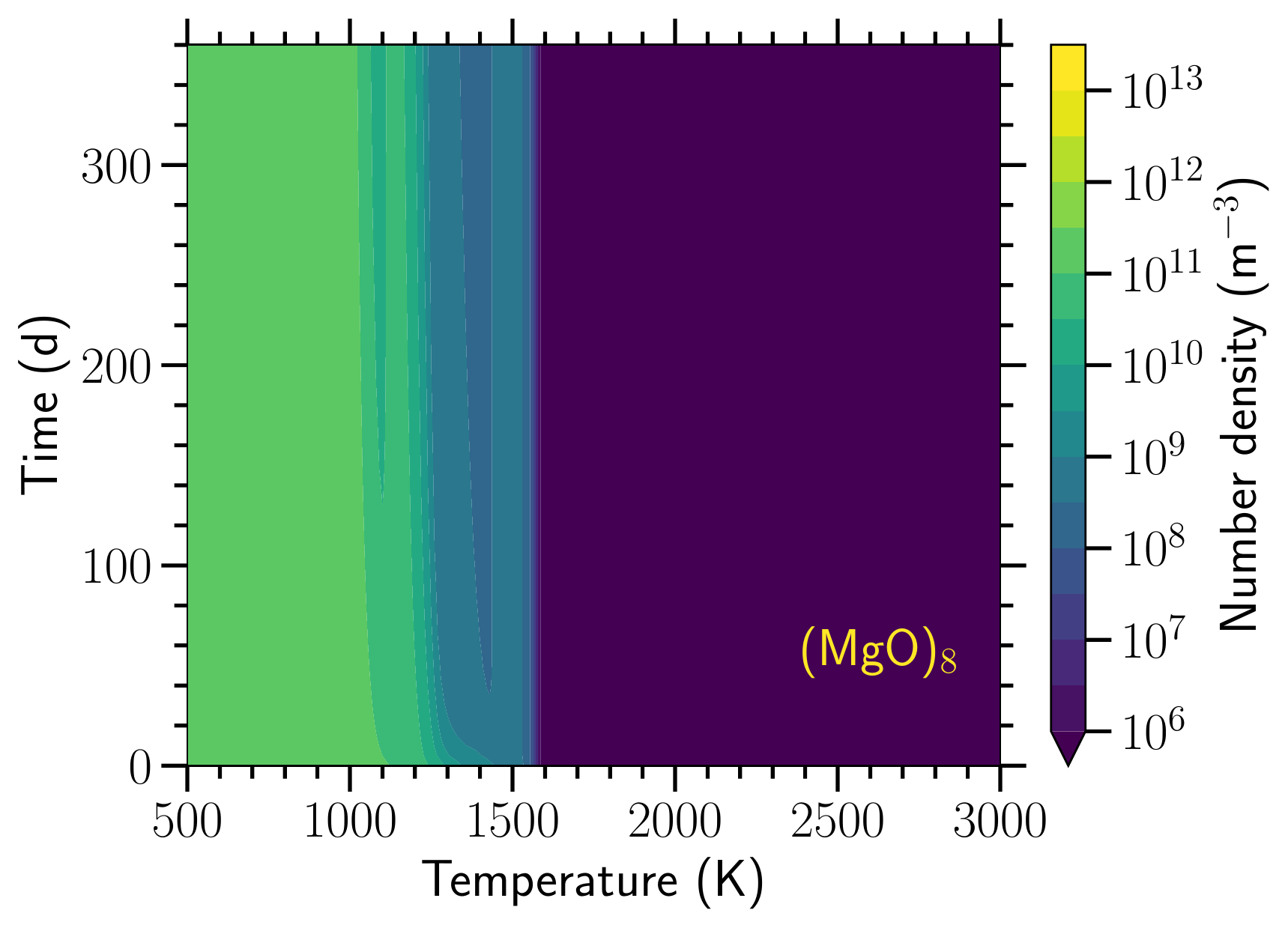}
        \includegraphics[width=0.32\textwidth]{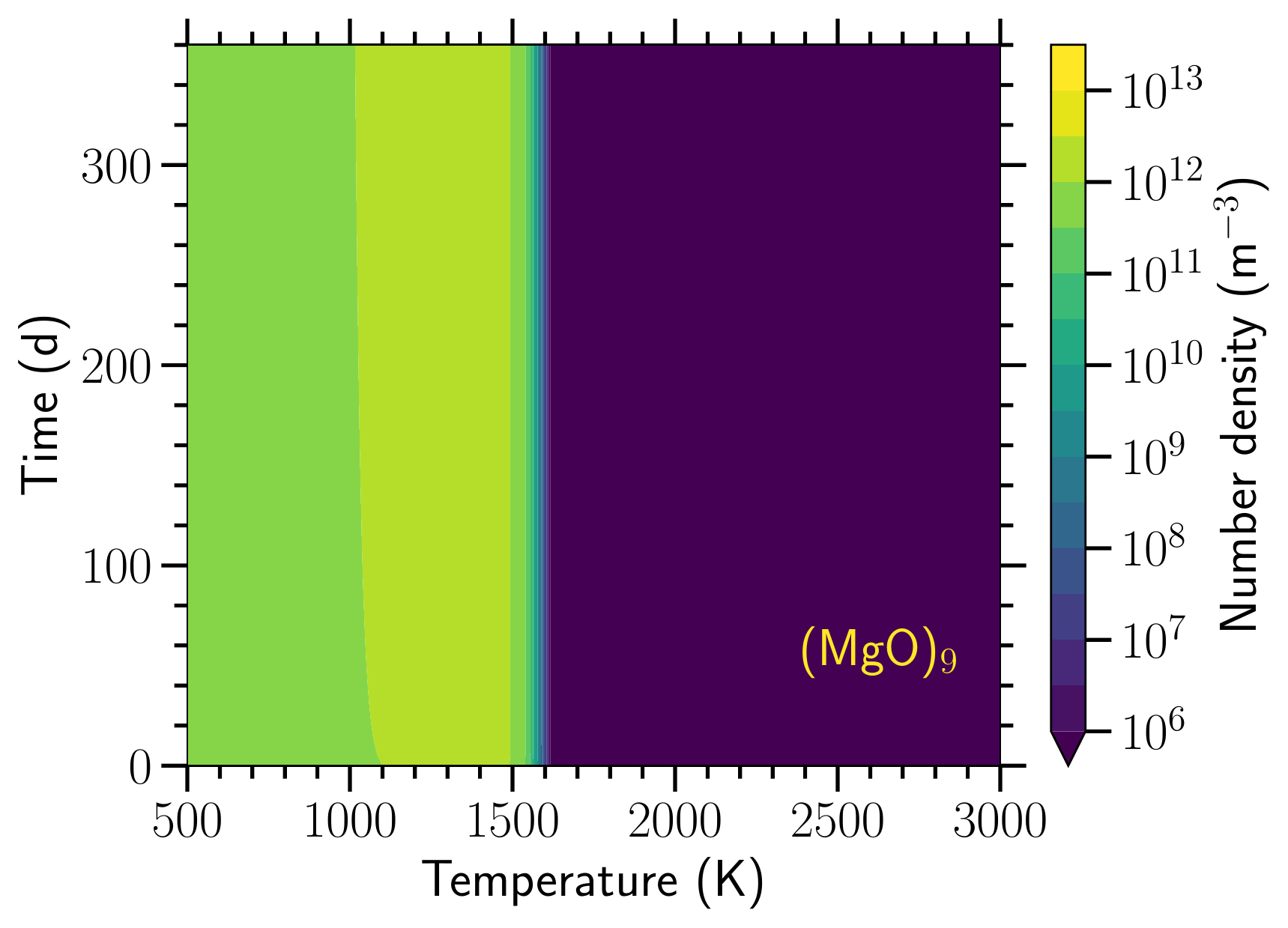}
        \includegraphics[width=0.32\textwidth]{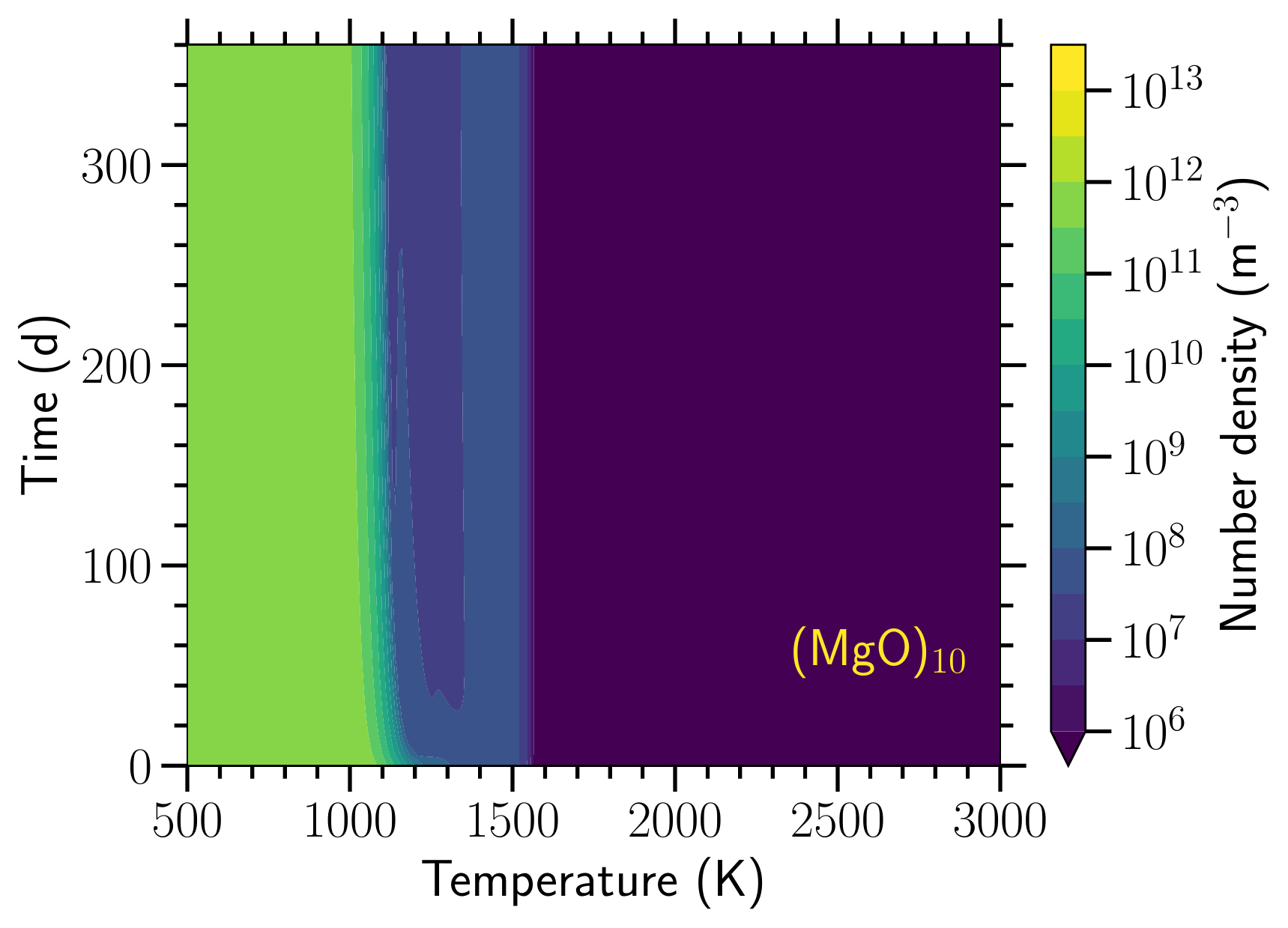}
        \end{flushleft}
        \caption{Temporal evolution of the absolute number density of all \protect\Mg{1}-clusters at the benchmark total gas density $\rho=\SI{1e-9}{\kg\per\m\cubed}$ for a closed nucleation model using the polymer nucleation description.}
        \label{fig:MgO_clusters_general_time_evolution}
    \end{figure*}

\begin{figure*}
        \begin{flushleft}
        \includegraphics[width=0.32\textwidth]{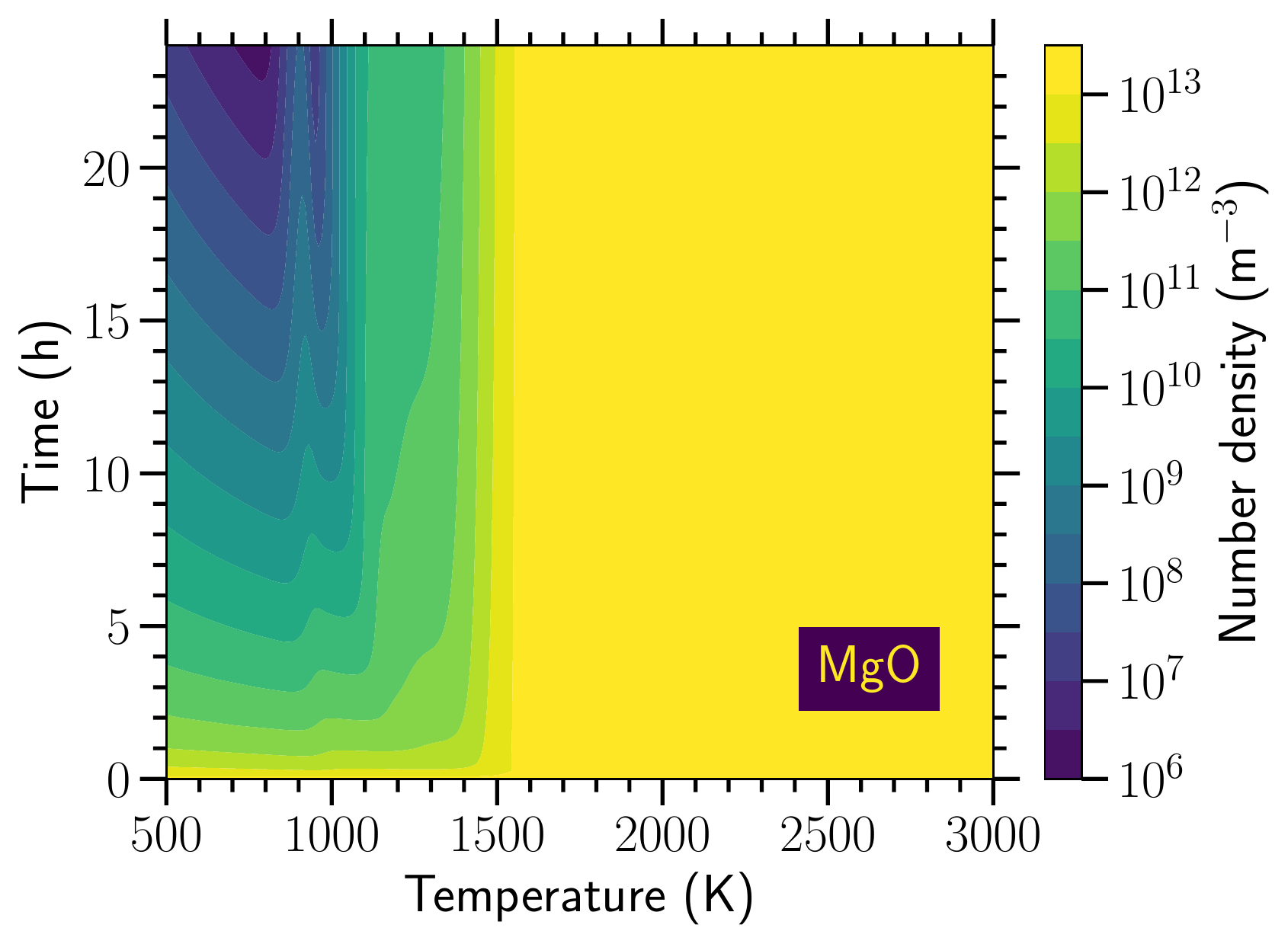}
        \includegraphics[width=0.32\textwidth]{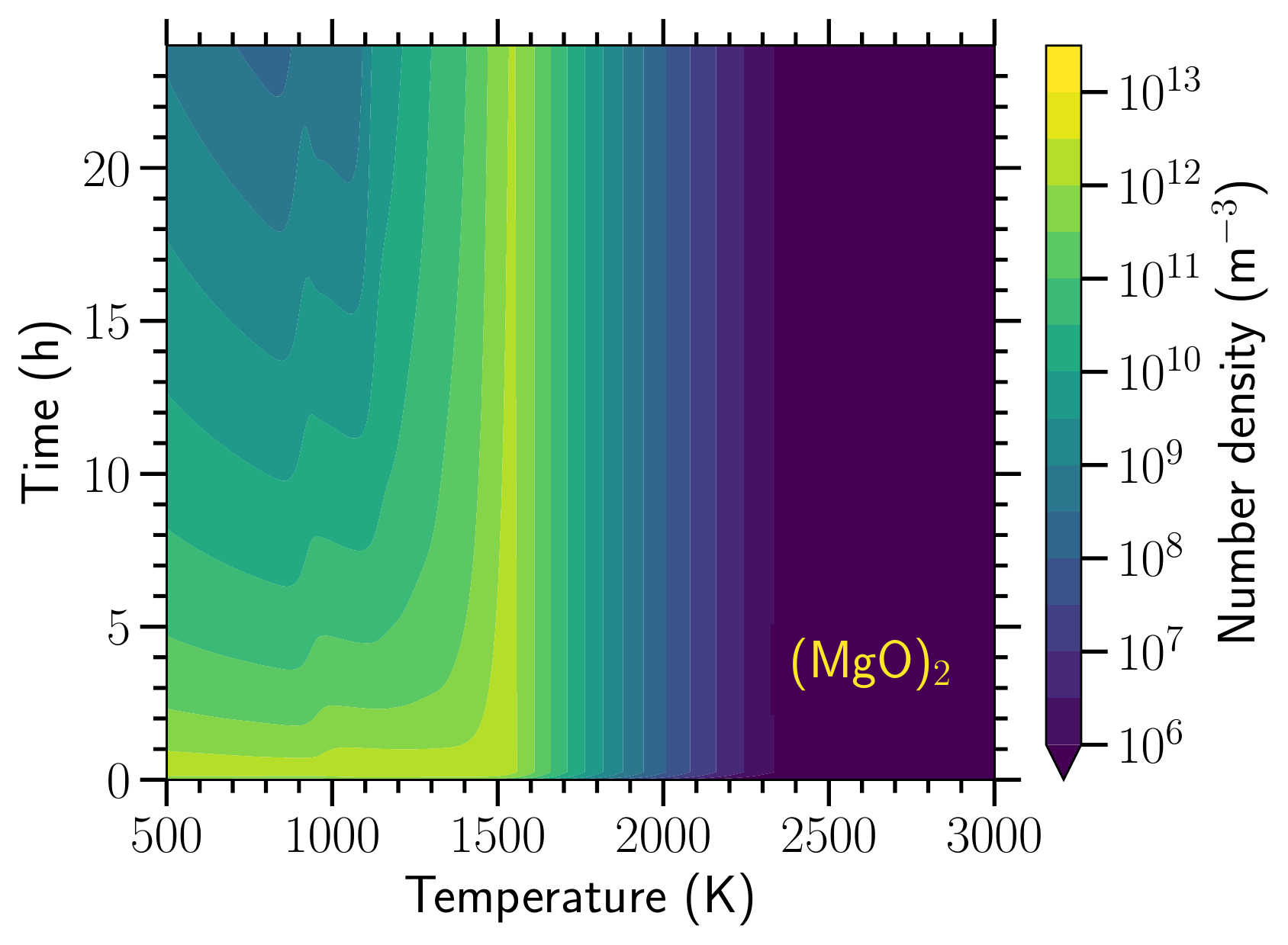}
        \includegraphics[width=0.32\textwidth]{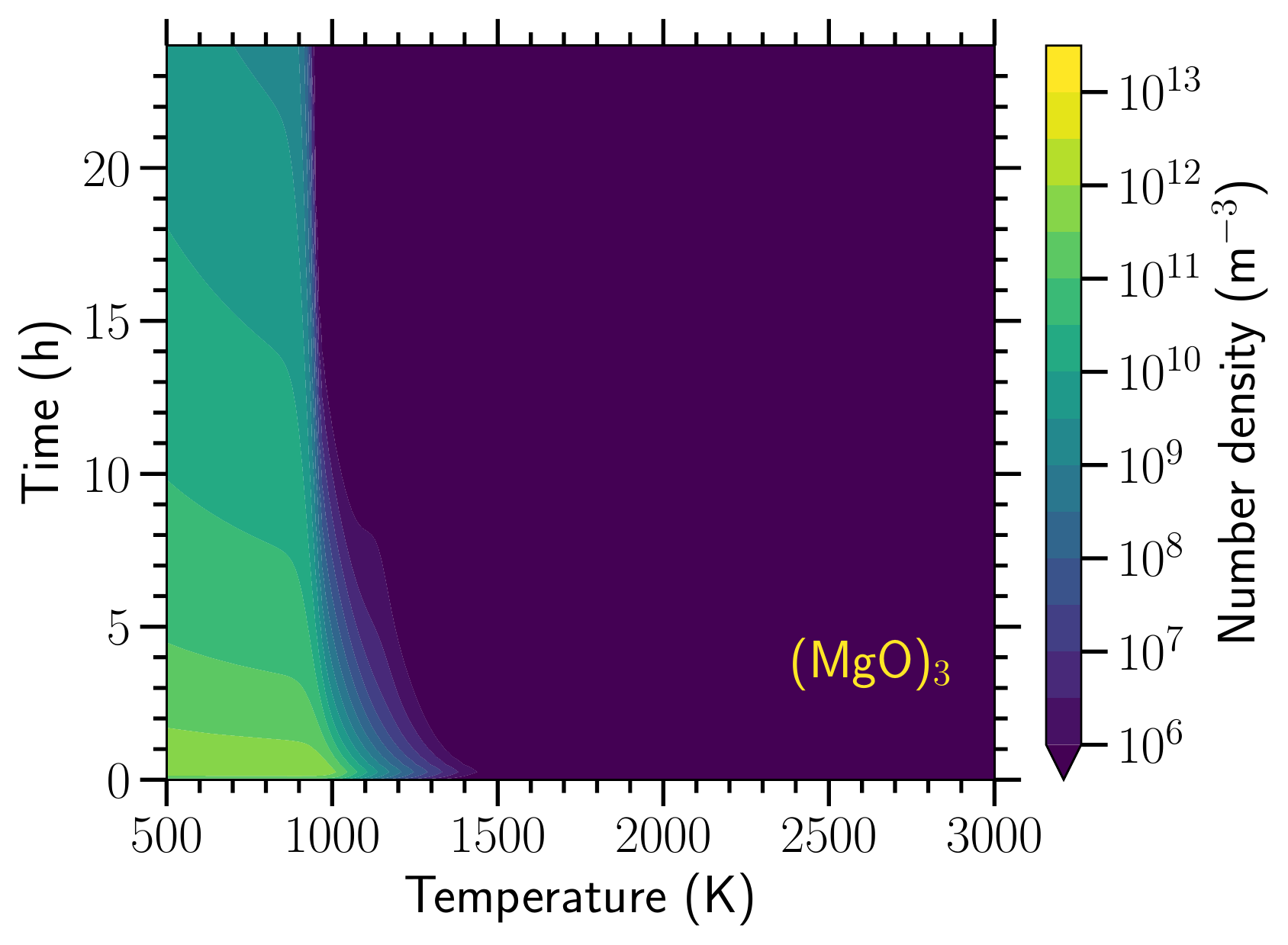}
        \includegraphics[width=0.32\textwidth]{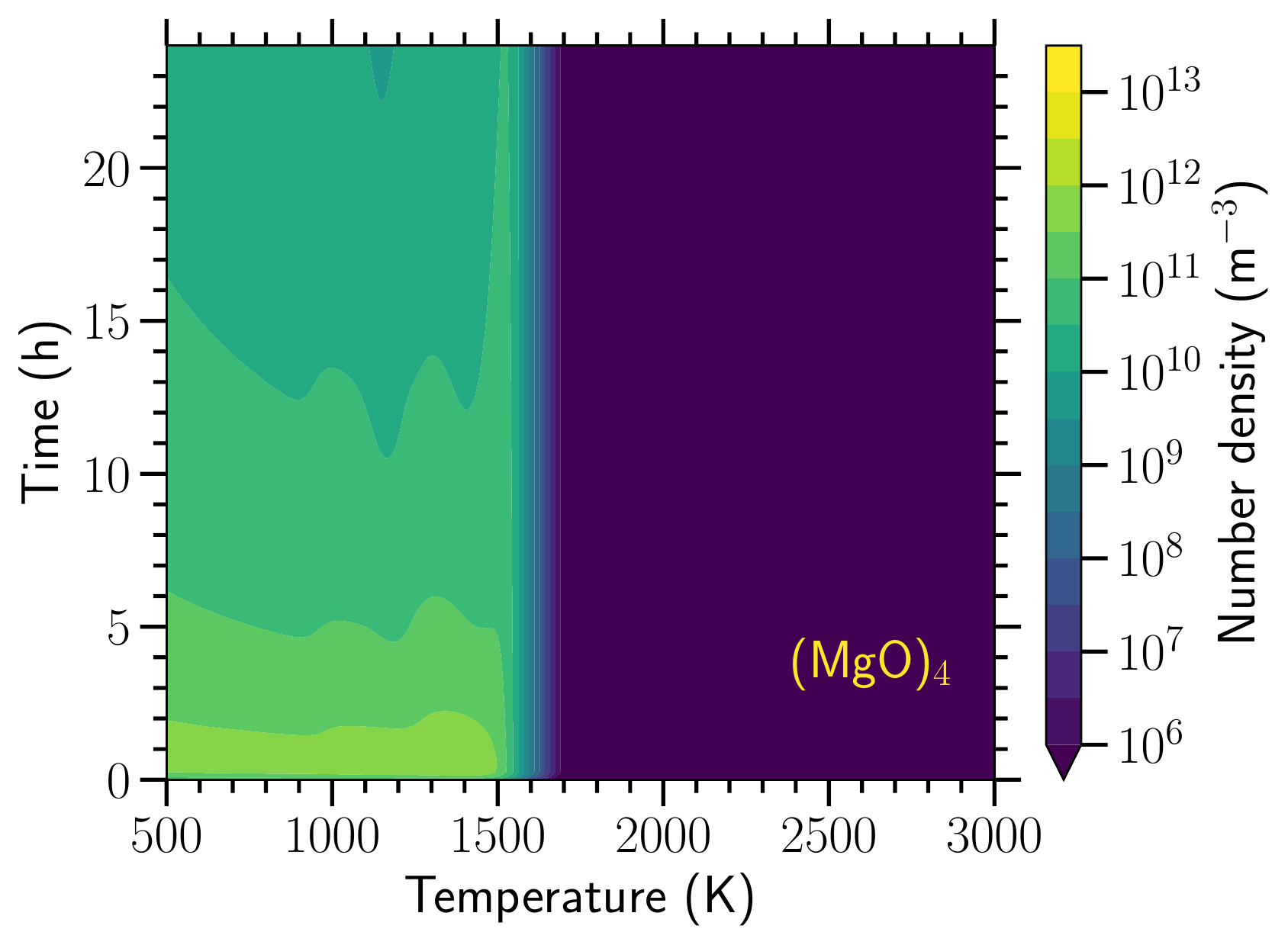}
        \includegraphics[width=0.32\textwidth]{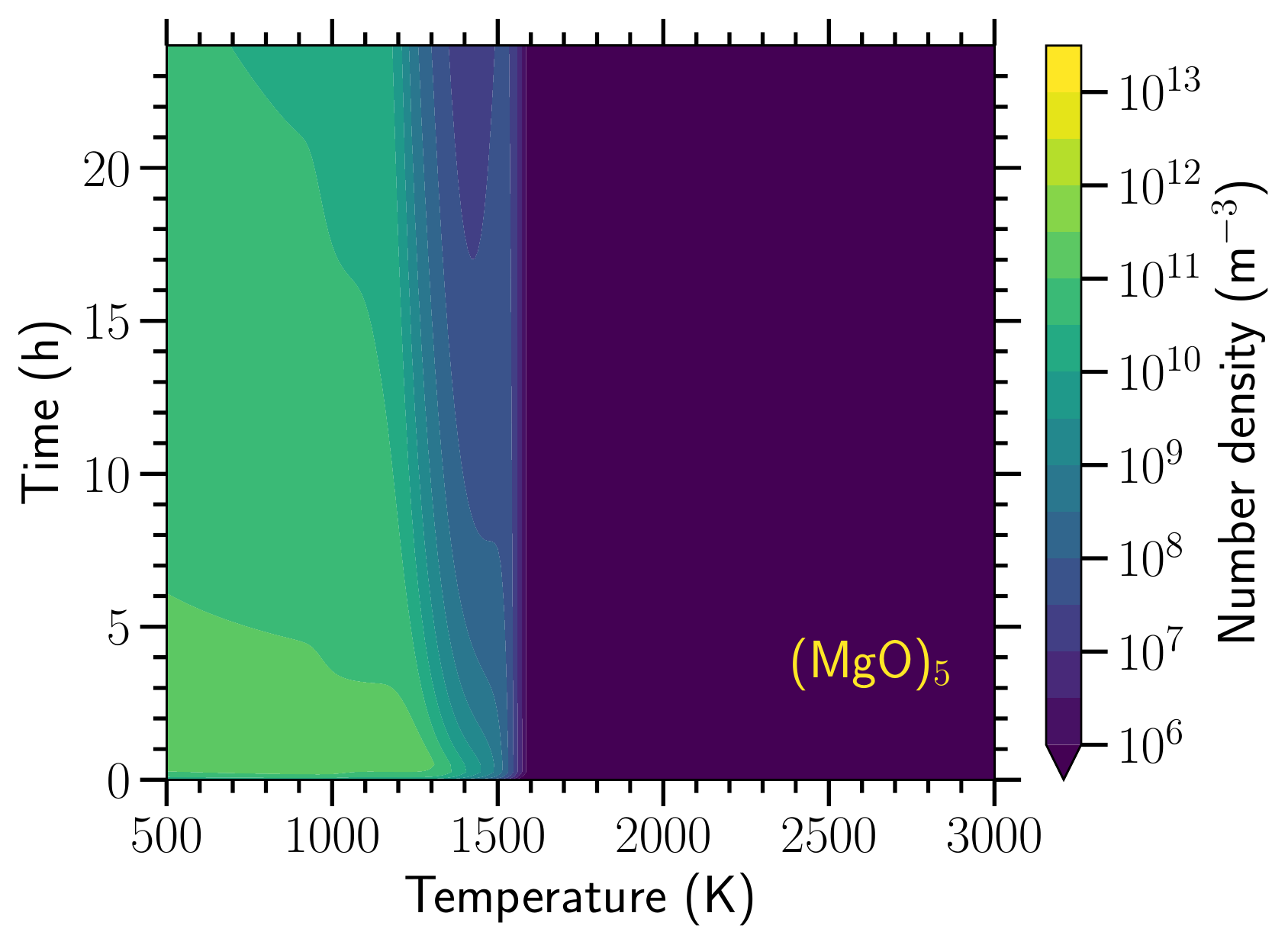}
        \includegraphics[width=0.32\textwidth]{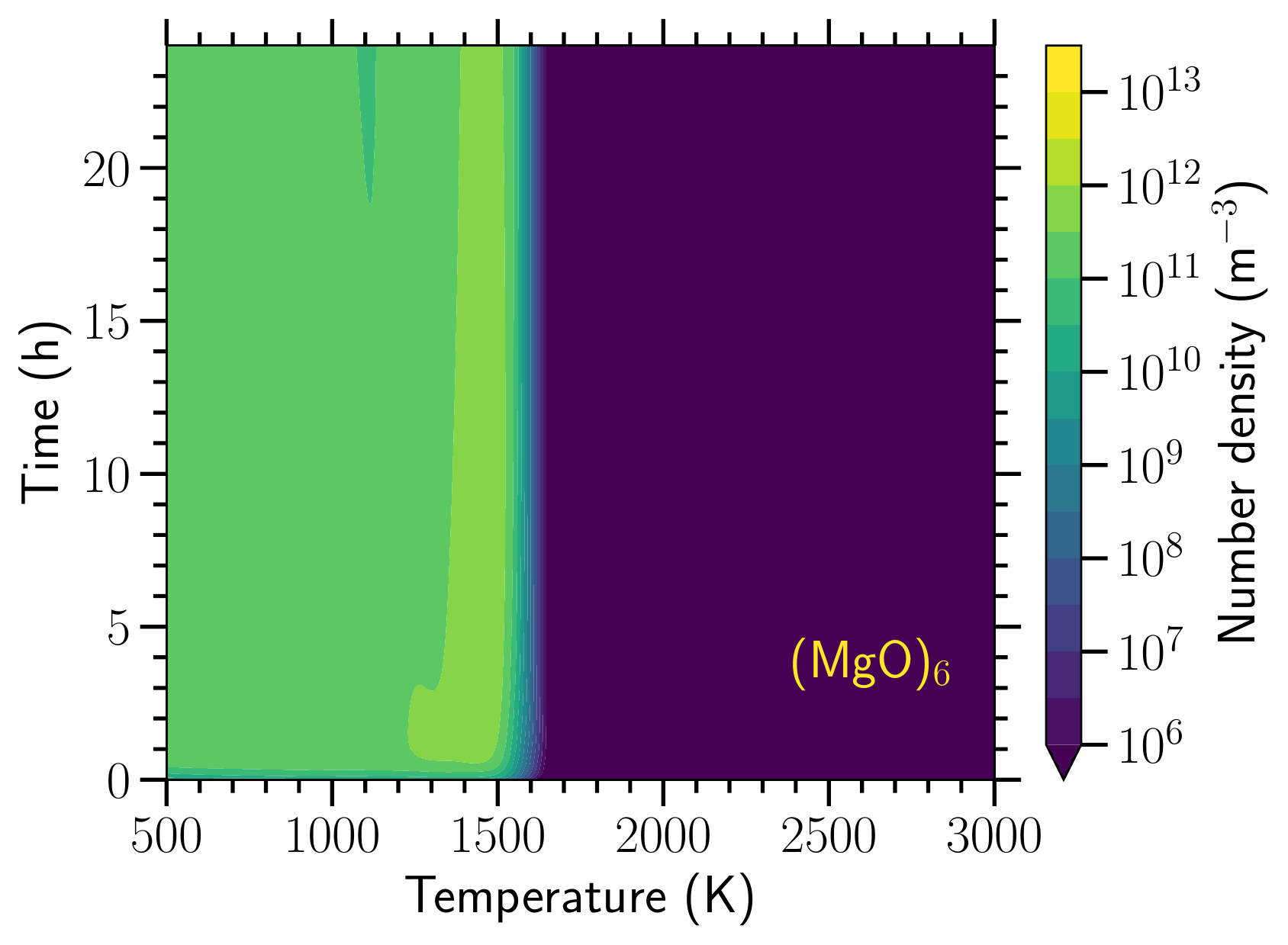}
        \includegraphics[width=0.32\textwidth]{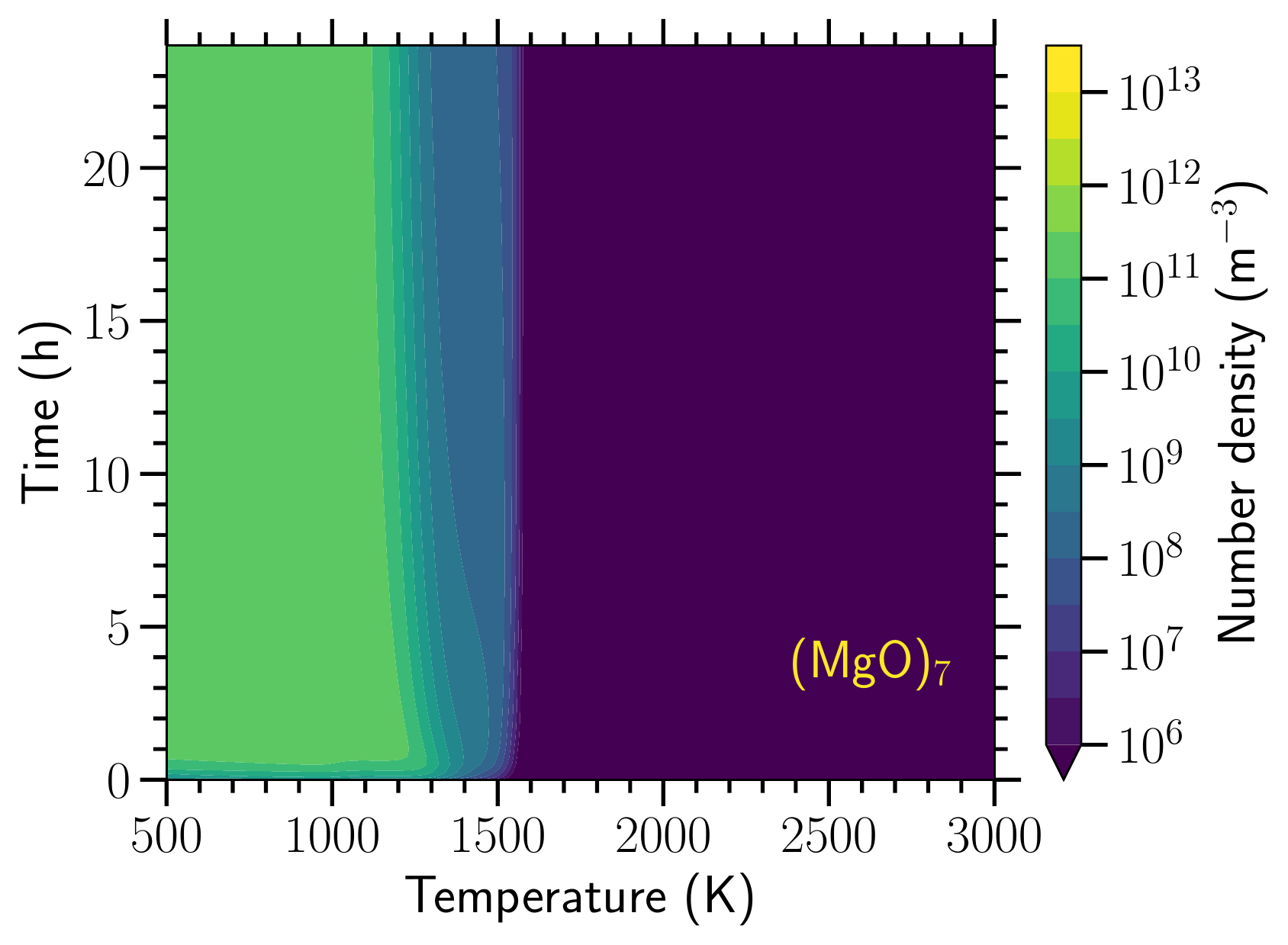}
        \includegraphics[width=0.32\textwidth]{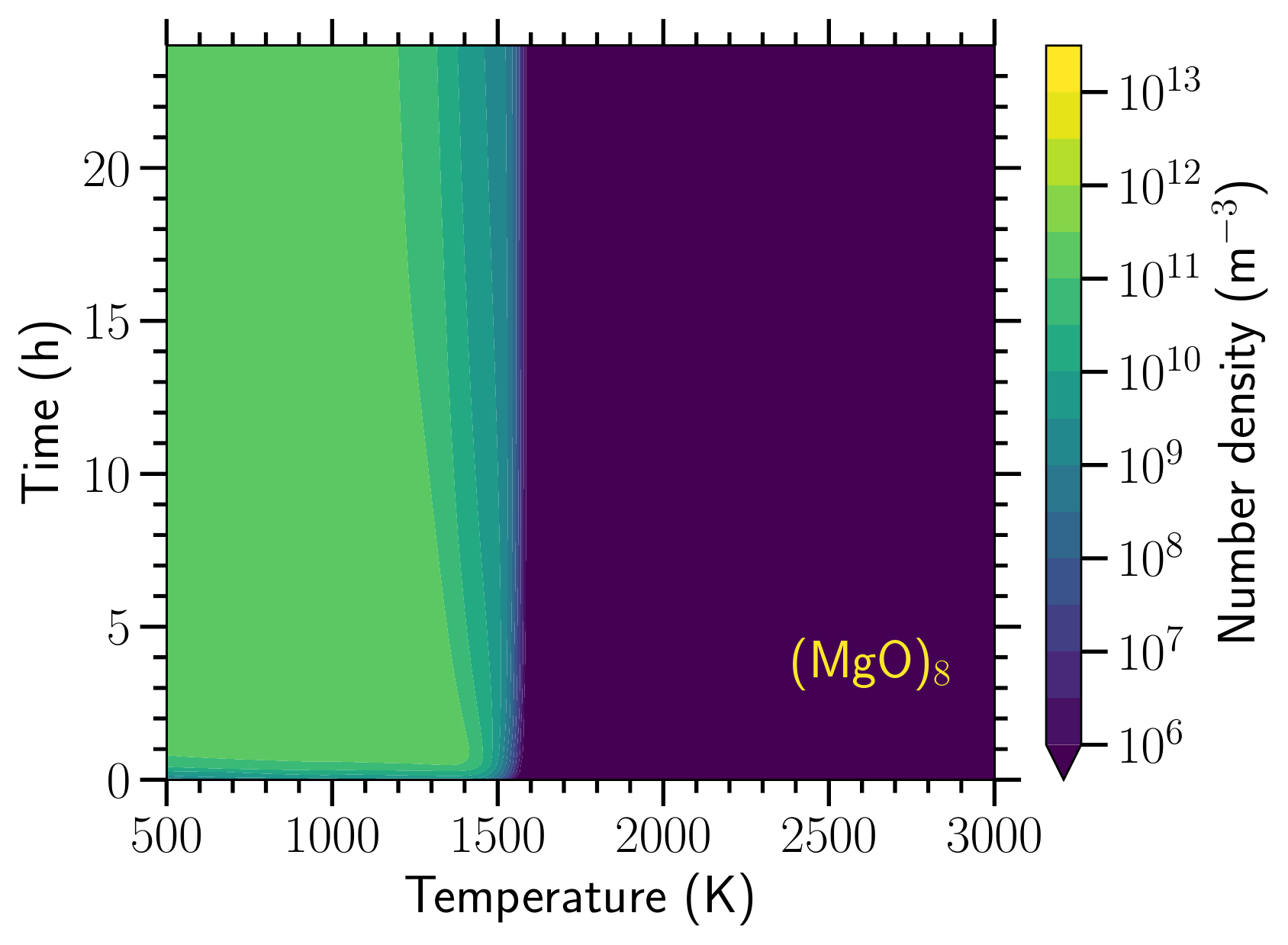}
        \includegraphics[width=0.32\textwidth]{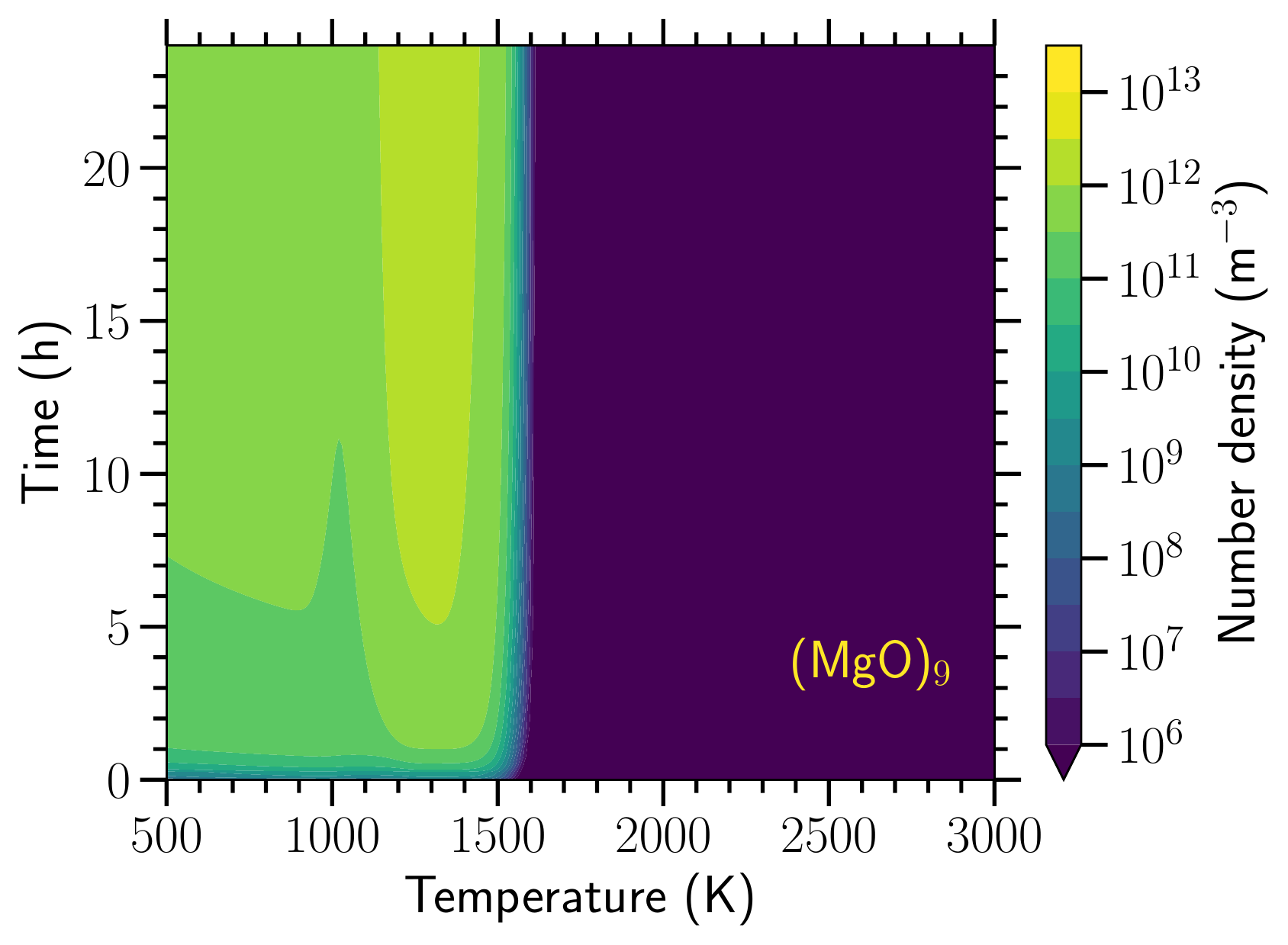}
        \includegraphics[width=0.32\textwidth]{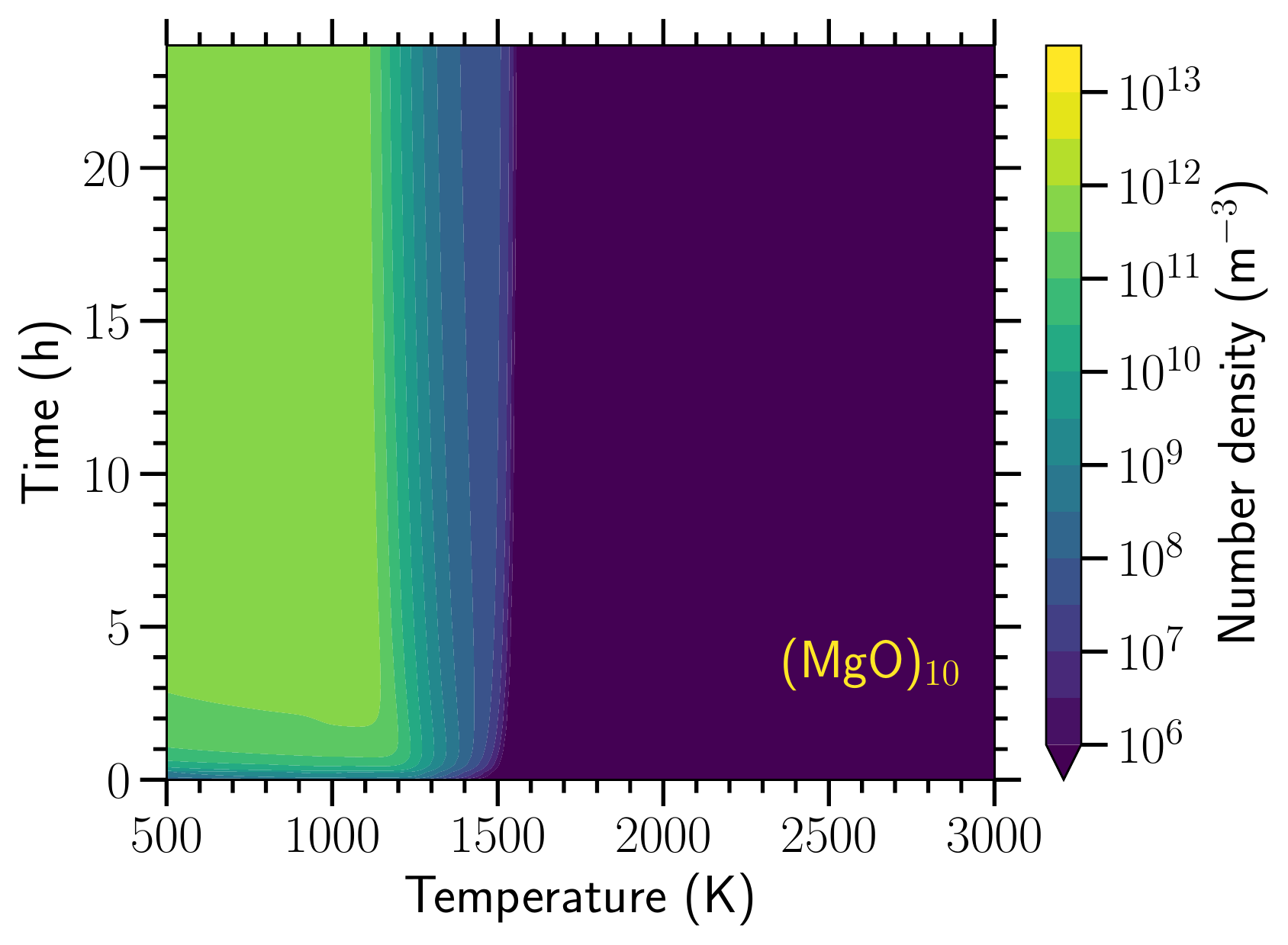}
        \end{flushleft}
        \caption{Refined temporal evolution of the absolute number density of all \protect\Mg{1}-clusters at the benchmark total gas density $\rho=\SI{1e-9}{\kg\per\m\cubed}$ for a closed nucleation model using the polymer nucleation description.}
        \label{fig:MgO_clusters_general_time_evolution_short}
    \end{figure*}

\begin{figure*}
        \begin{flushleft}
        \includegraphics[width=0.32\textwidth]{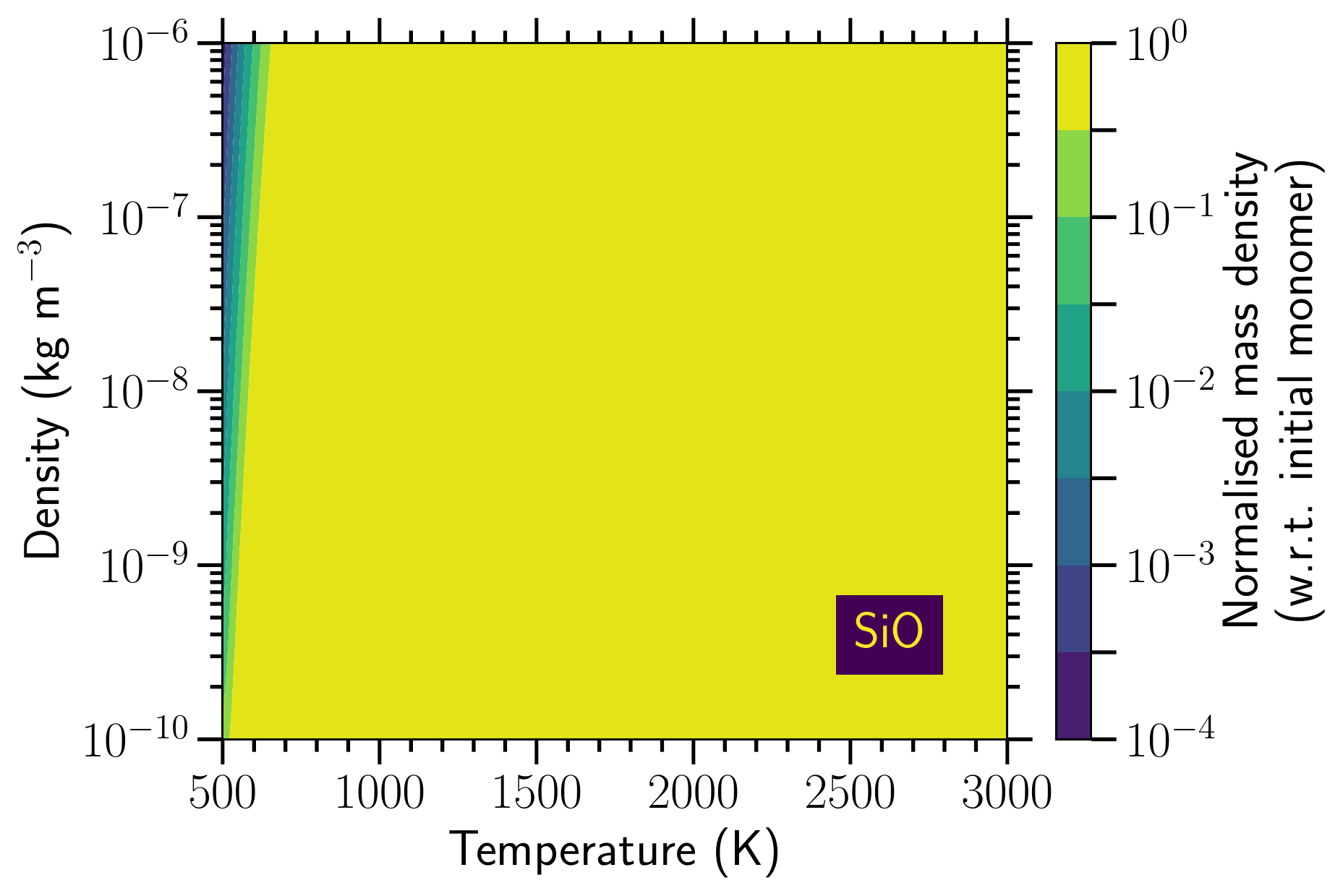}
        \includegraphics[width=0.32\textwidth]{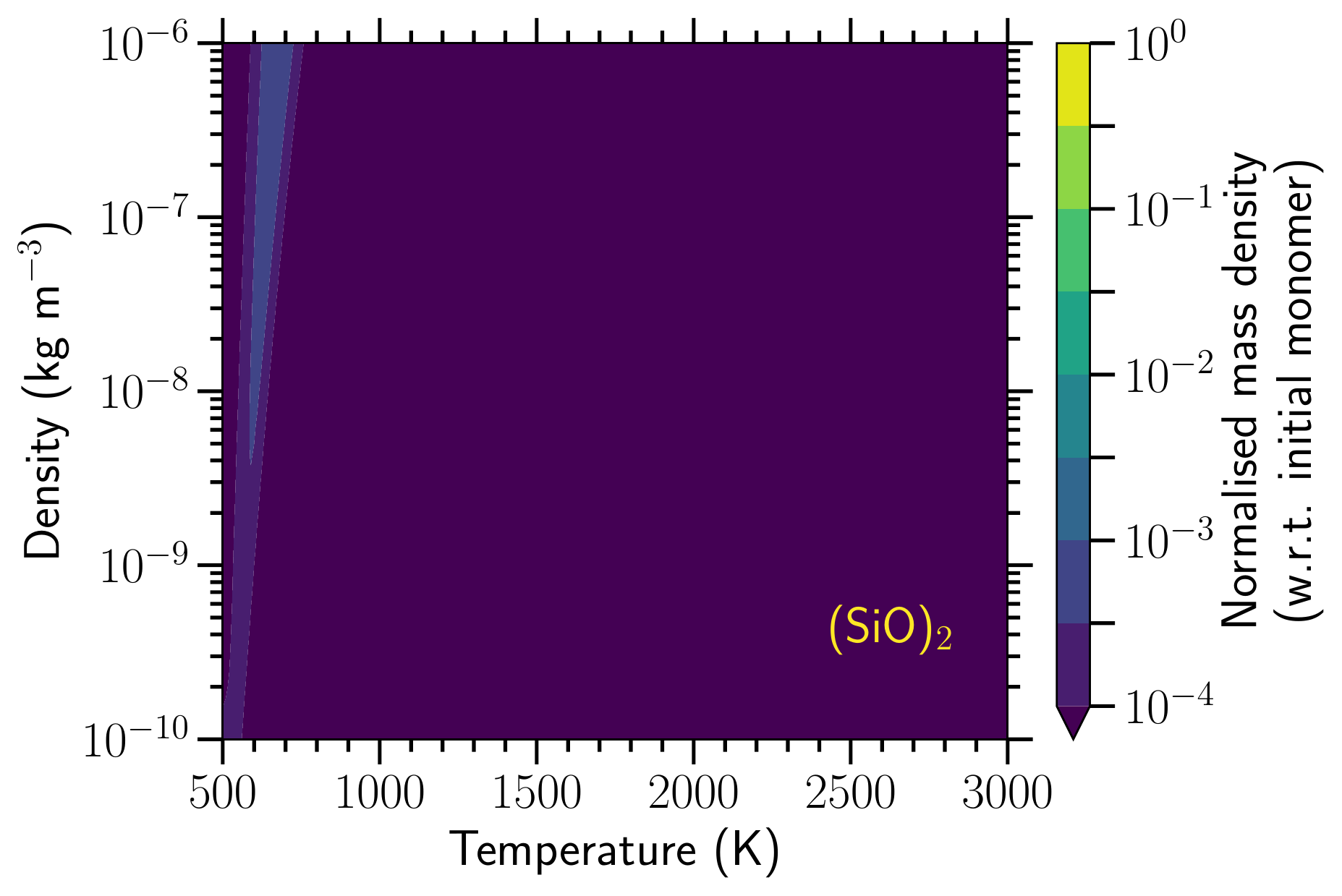}
        \includegraphics[width=0.32\textwidth]{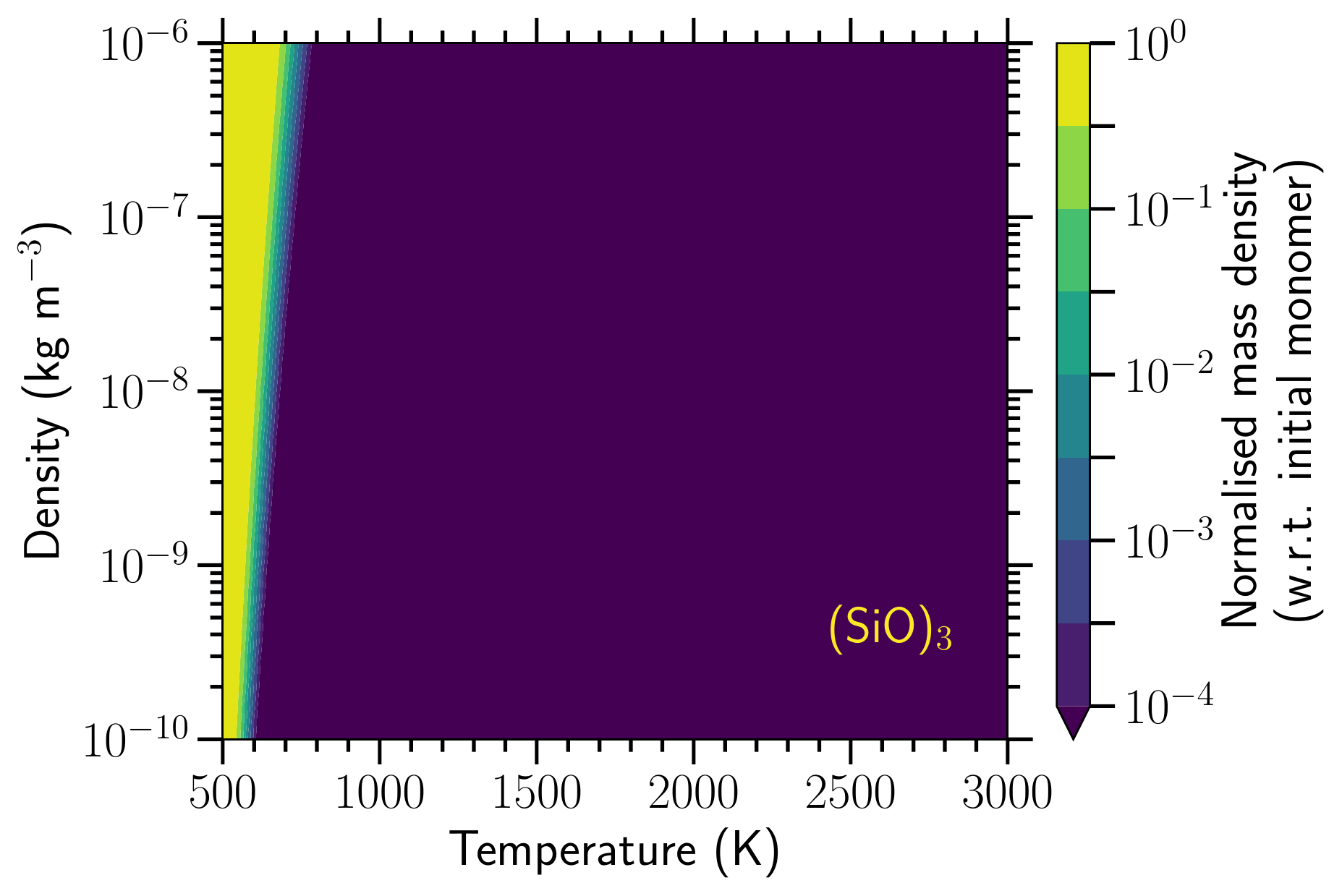}
        \includegraphics[width=0.32\textwidth]{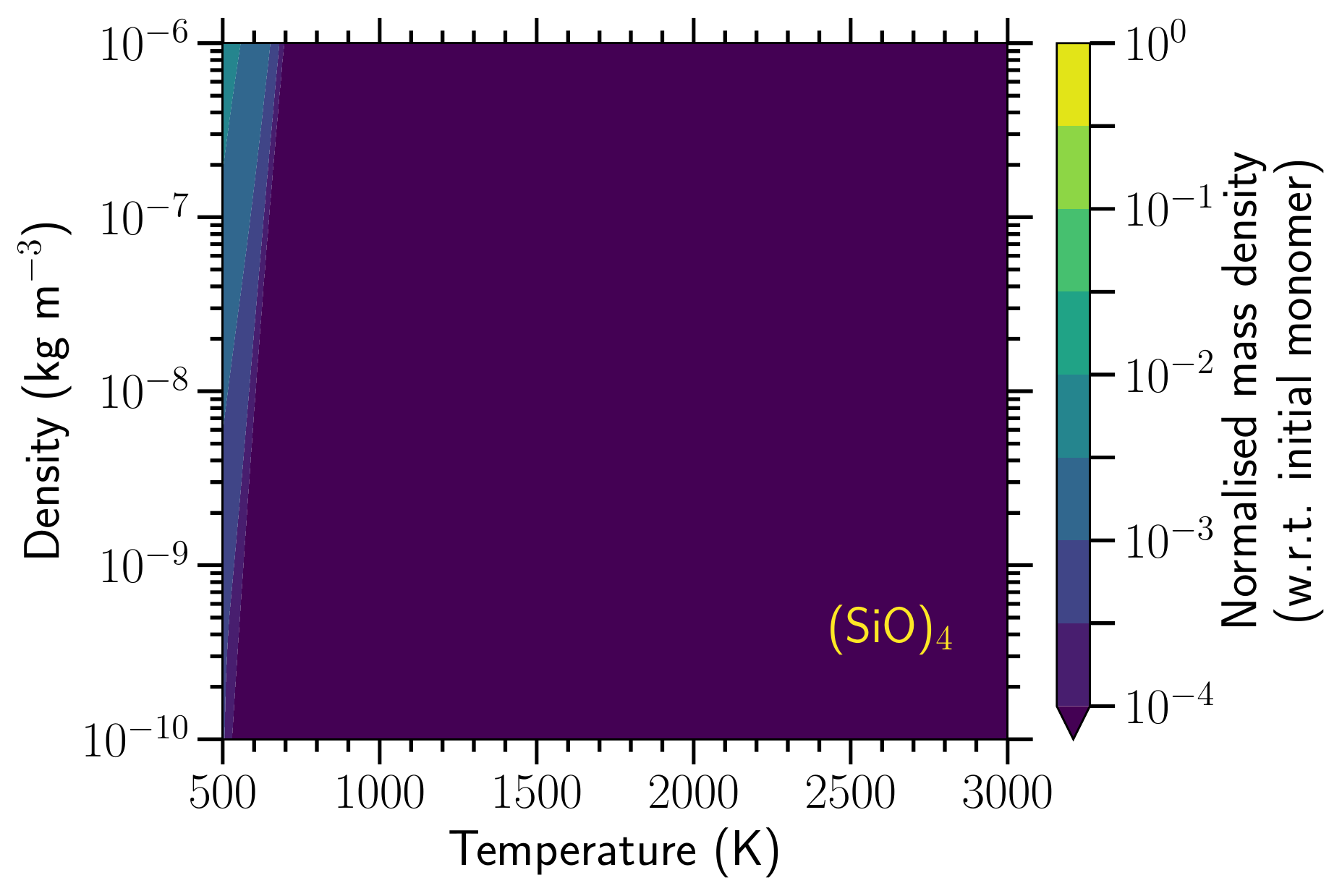}
        \includegraphics[width=0.32\textwidth]{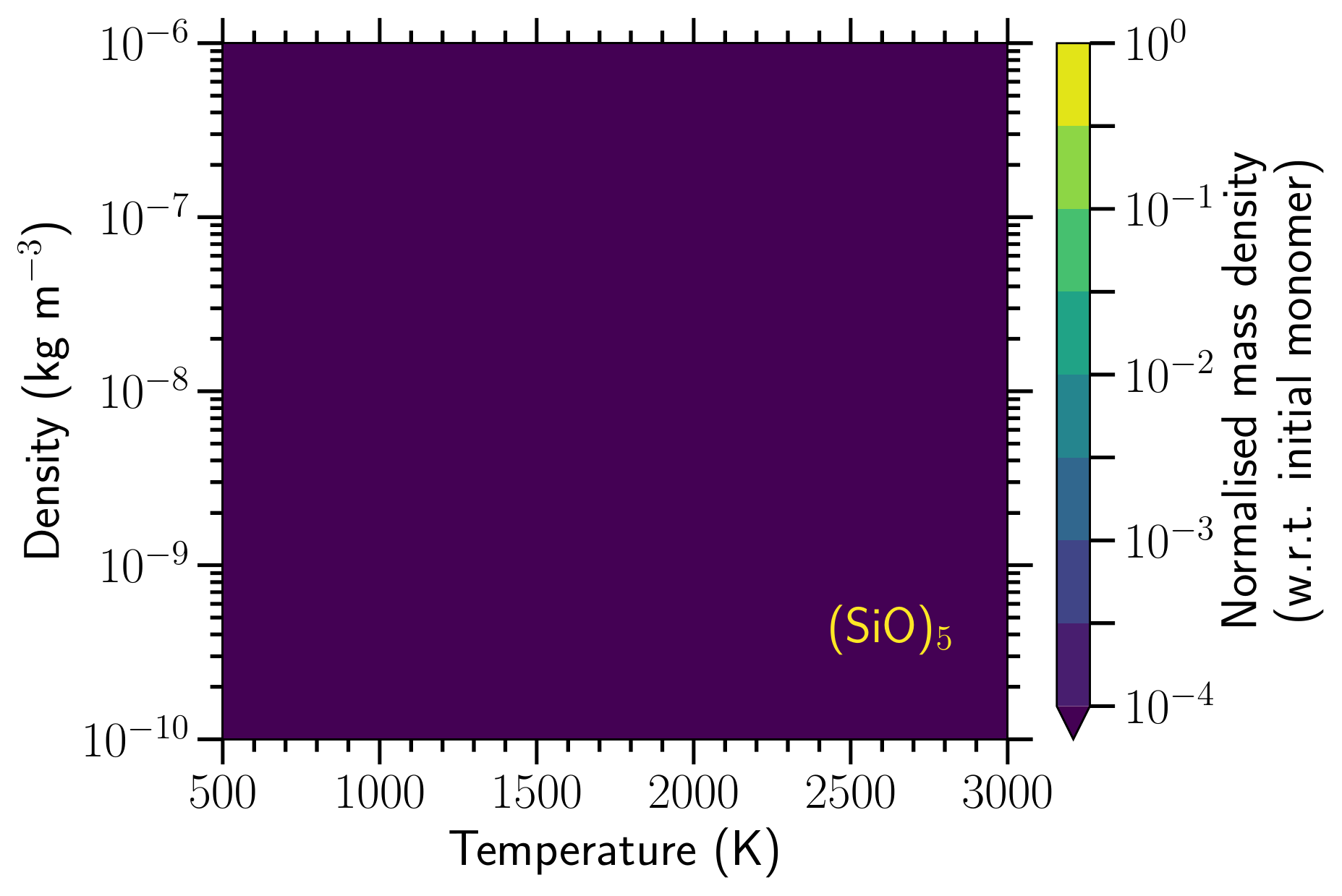}
        \includegraphics[width=0.32\textwidth]{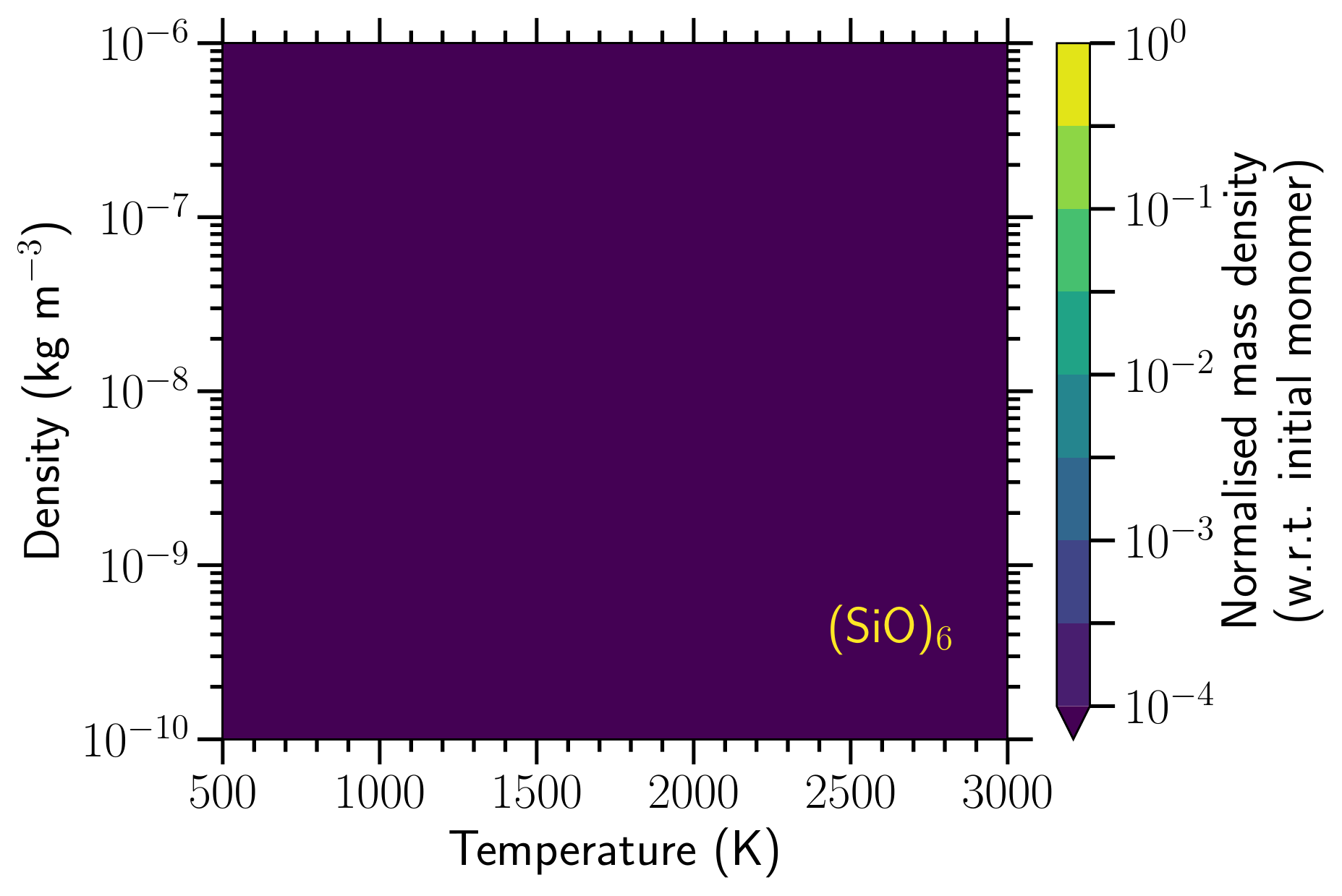}
        \includegraphics[width=0.32\textwidth]{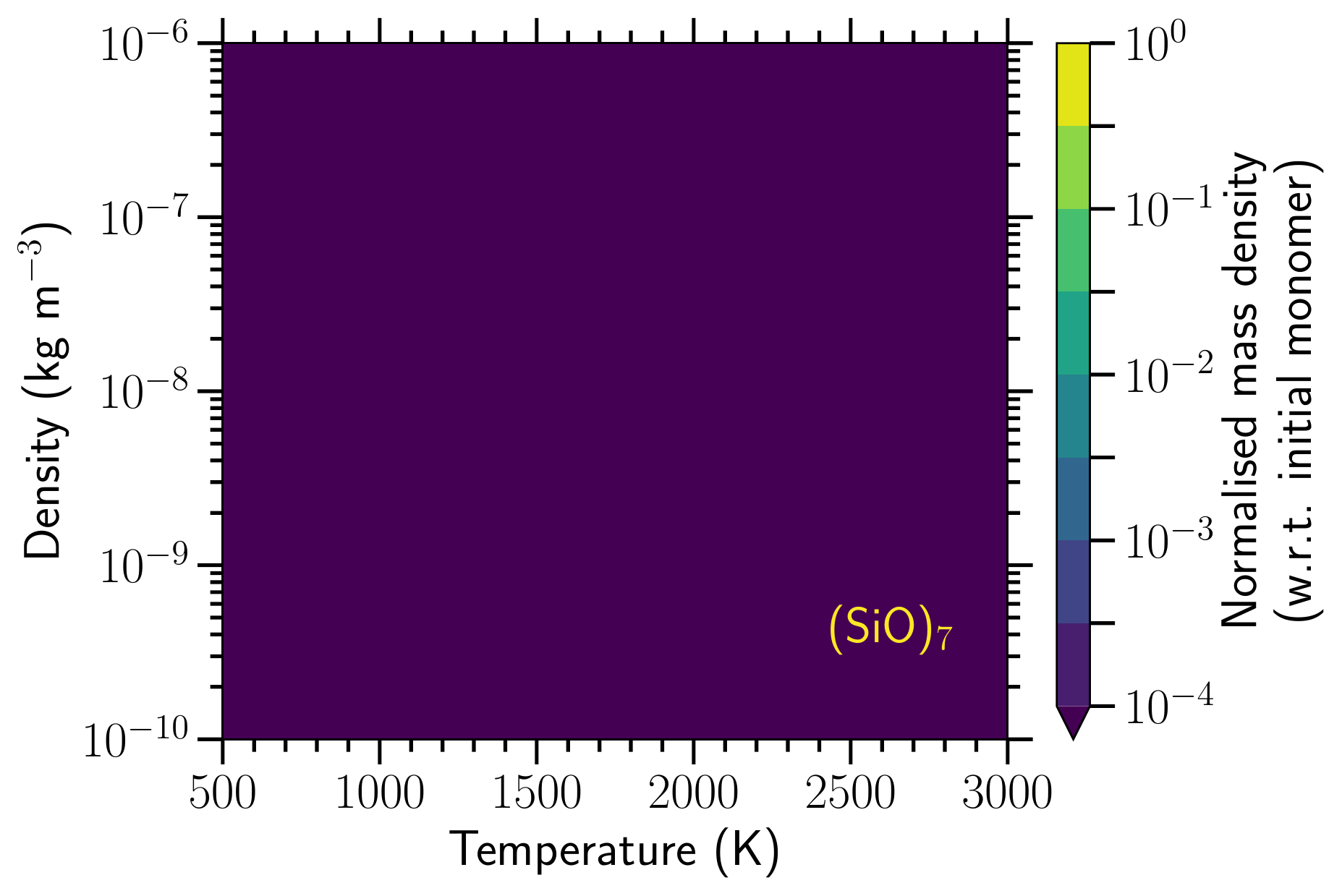}
        \includegraphics[width=0.32\textwidth]{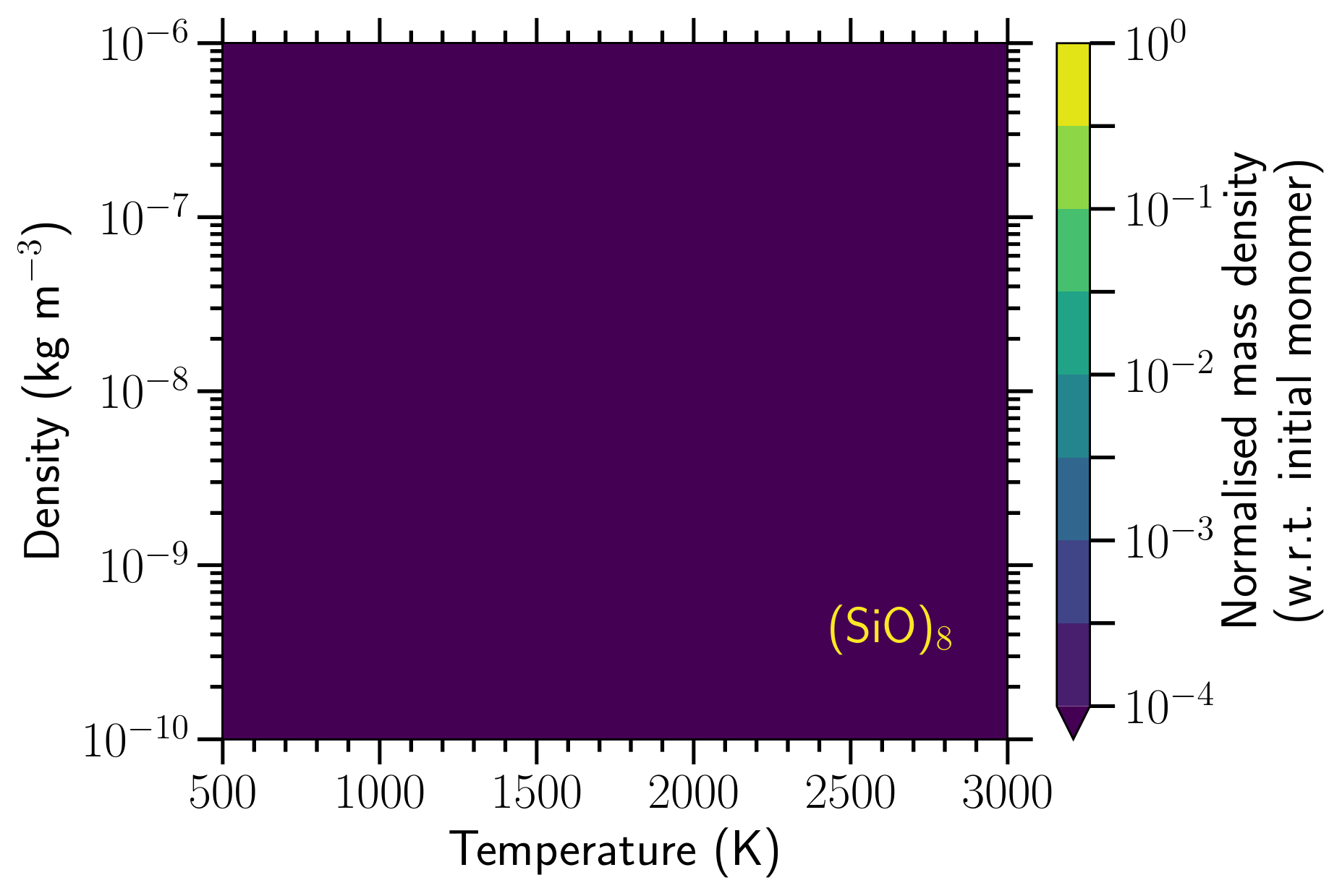}
        \includegraphics[width=0.32\textwidth]{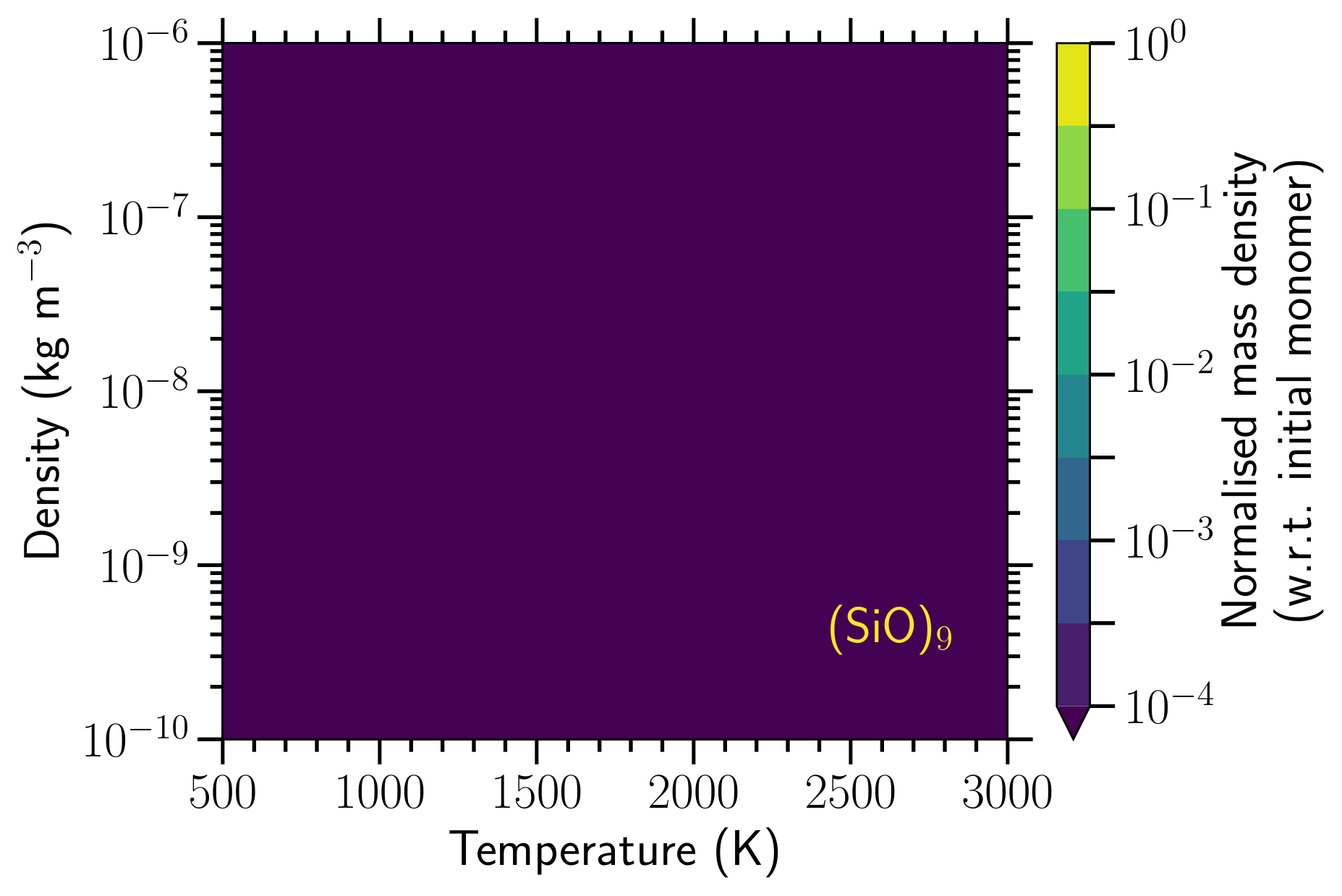}
        \includegraphics[width=0.32\textwidth]{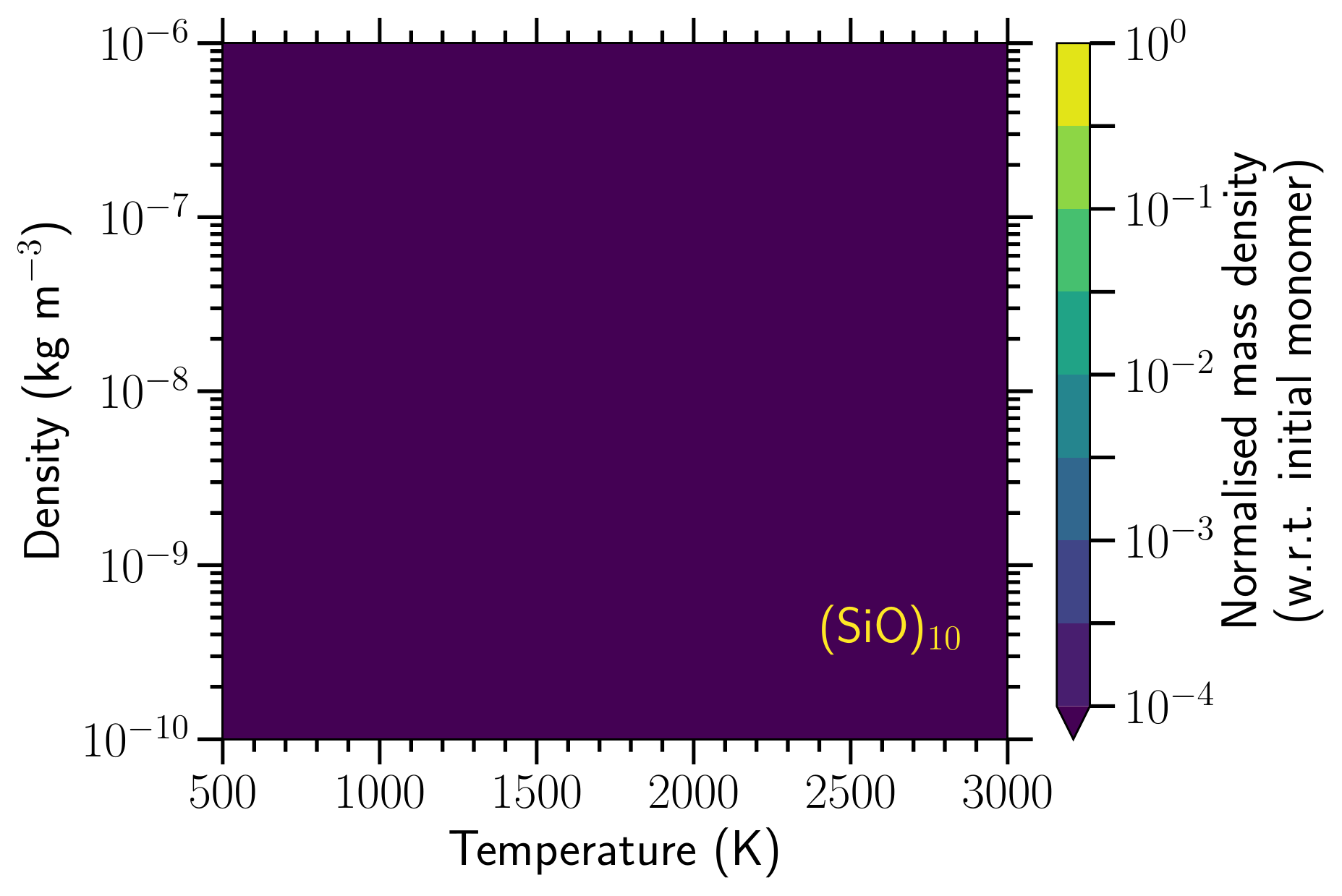}
        \end{flushleft}
        \caption{Overview of the normalised mass density after one year of all \protect\SiO{1}-clusters for a closed nucleation model using the polymer nucleation description.}
        \label{fig:SiO_clusters_general_norm_same_scale}
    \end{figure*}

\begin{figure*}
        \begin{flushleft}
        \includegraphics[width=0.32\textwidth]{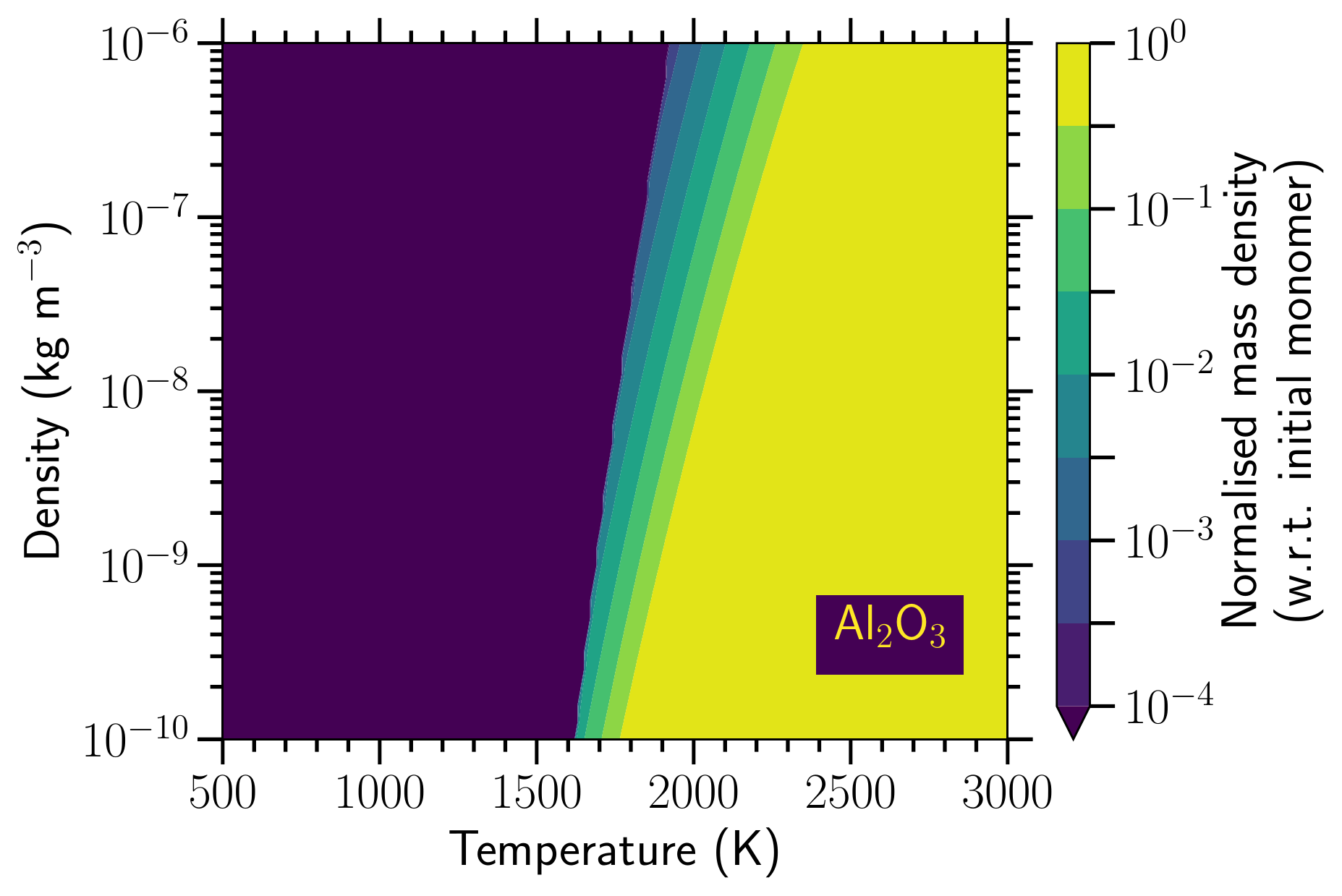}
        \includegraphics[width=0.32\textwidth]{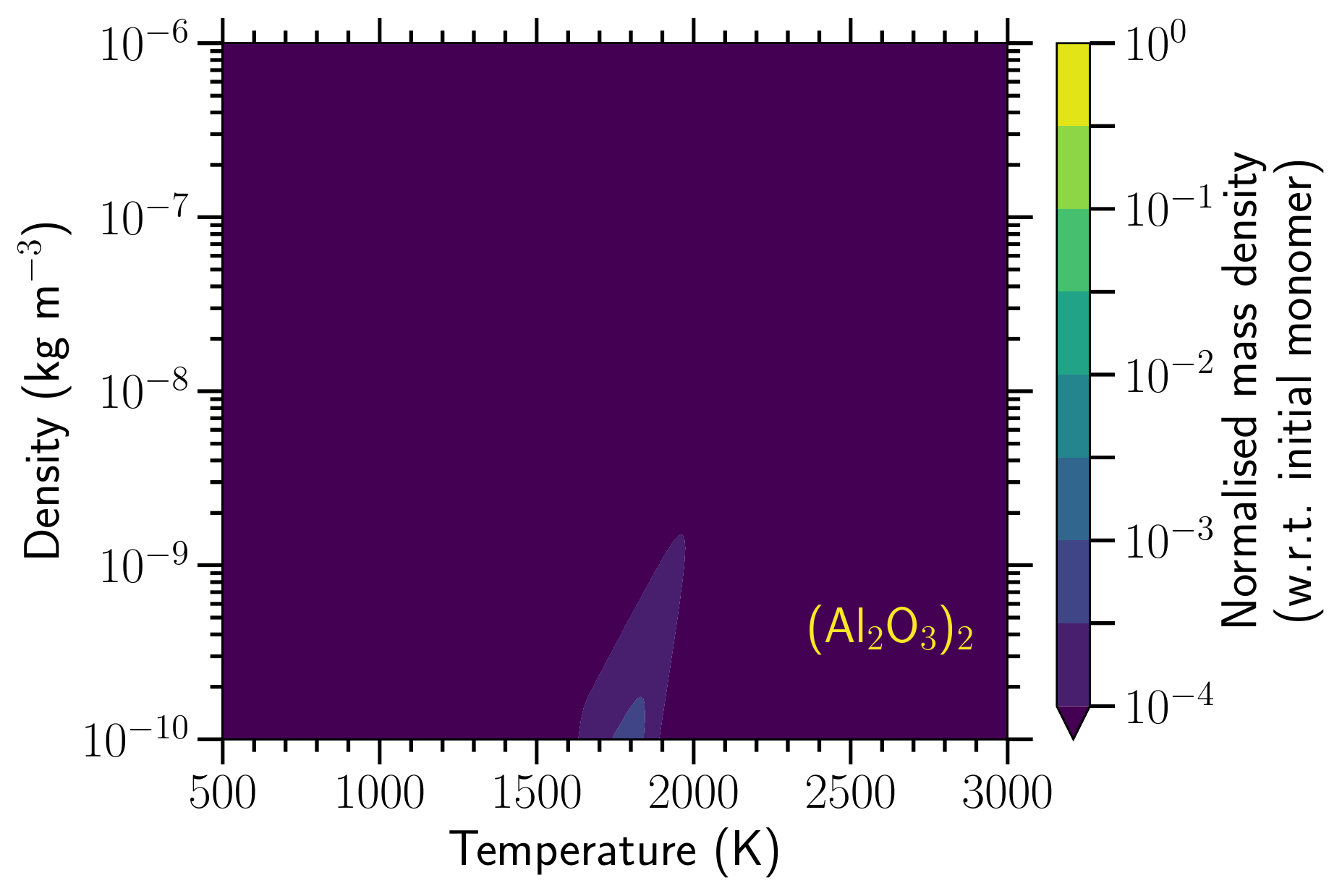}
        \includegraphics[width=0.32\textwidth]{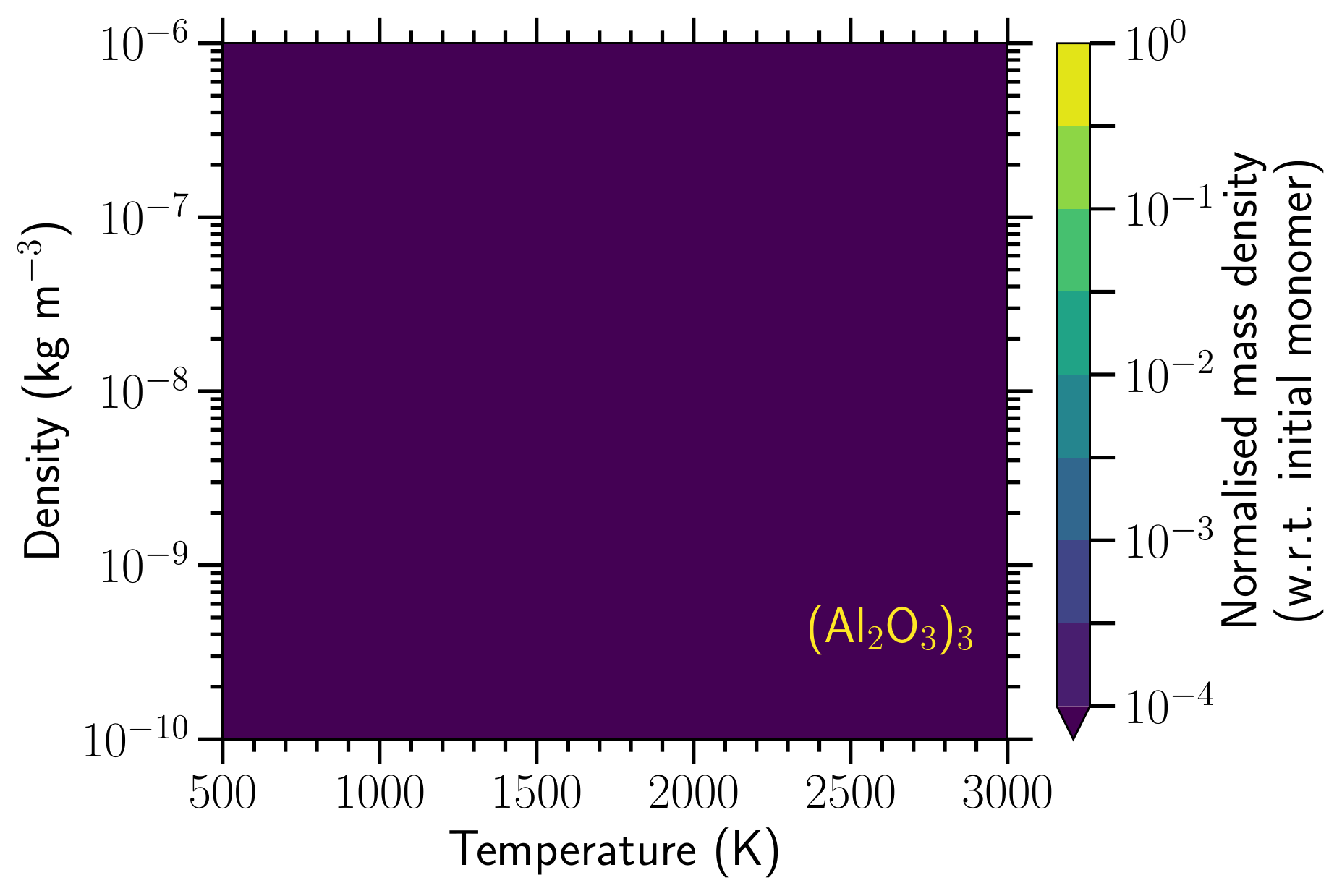}
        \includegraphics[width=0.32\textwidth]{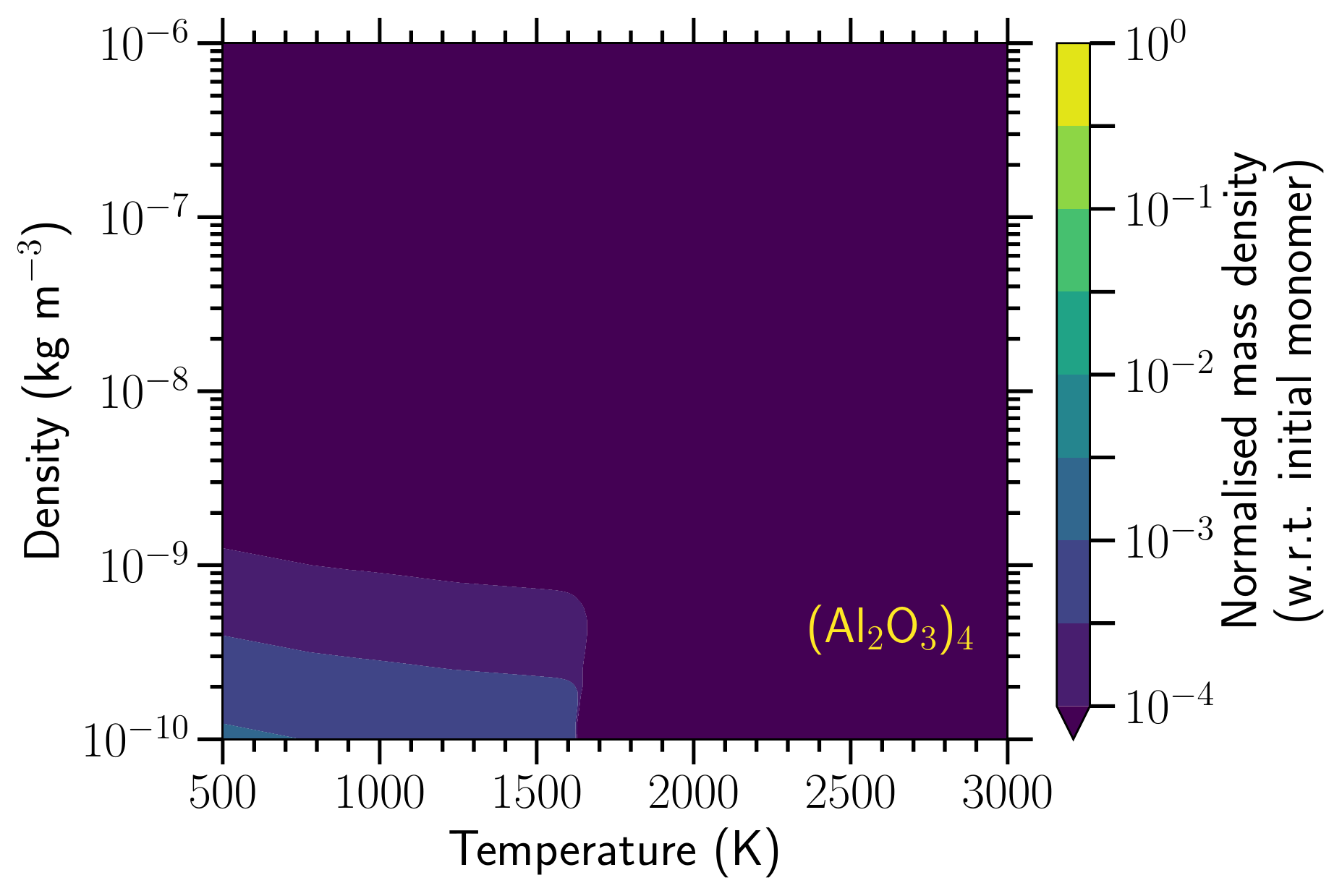}
        \includegraphics[width=0.32\textwidth]{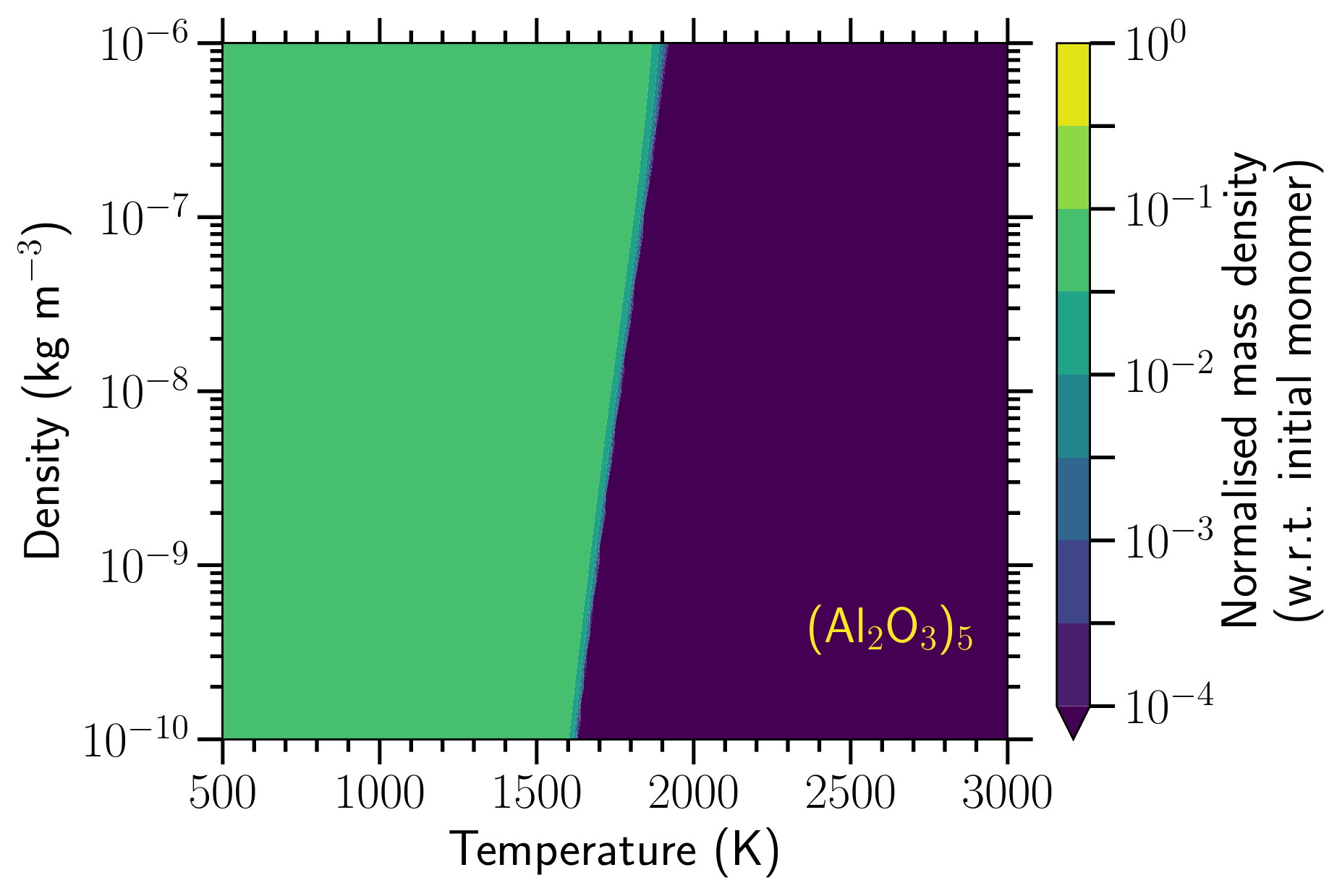}
        \includegraphics[width=0.32\textwidth]{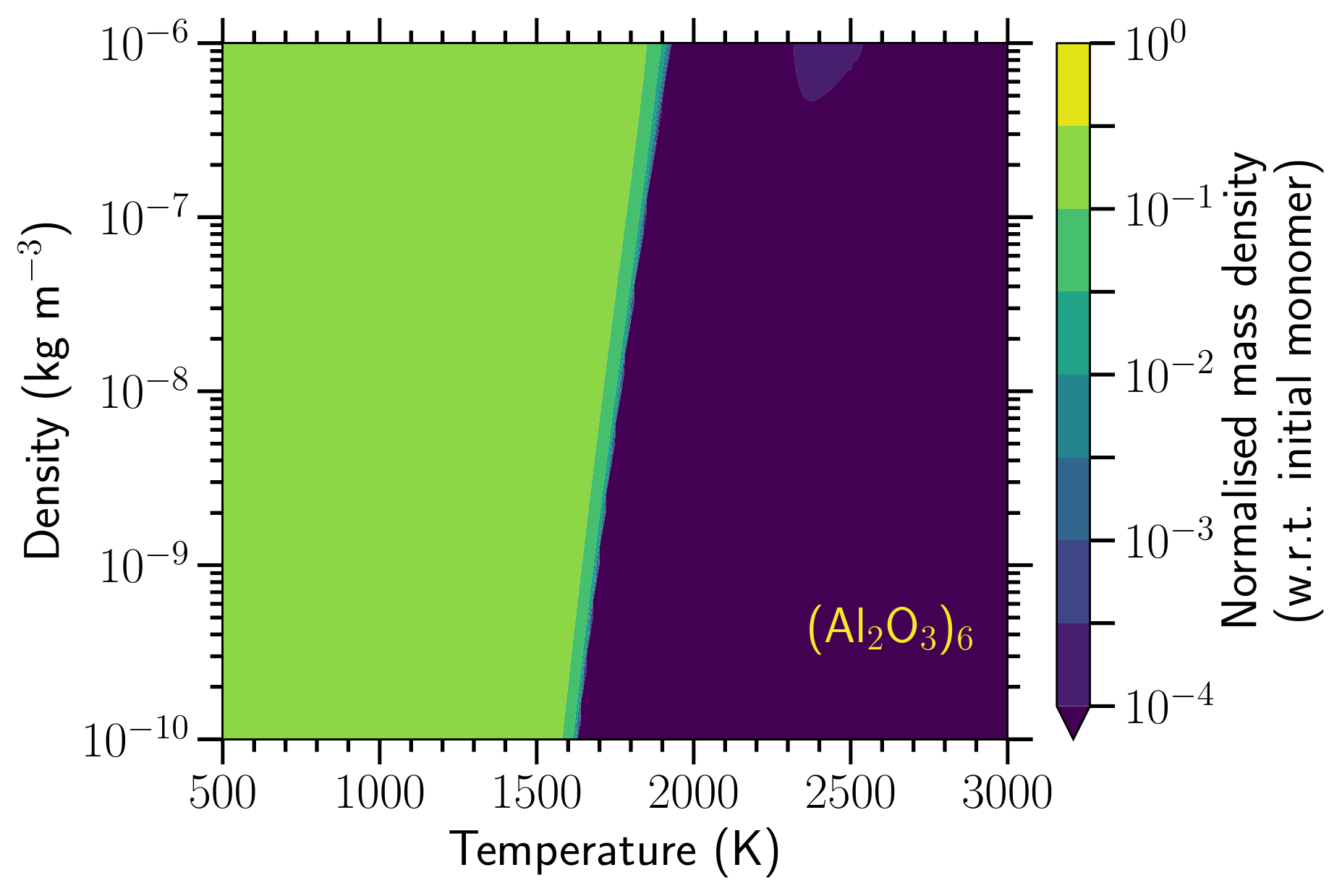}
        \includegraphics[width=0.32\textwidth]{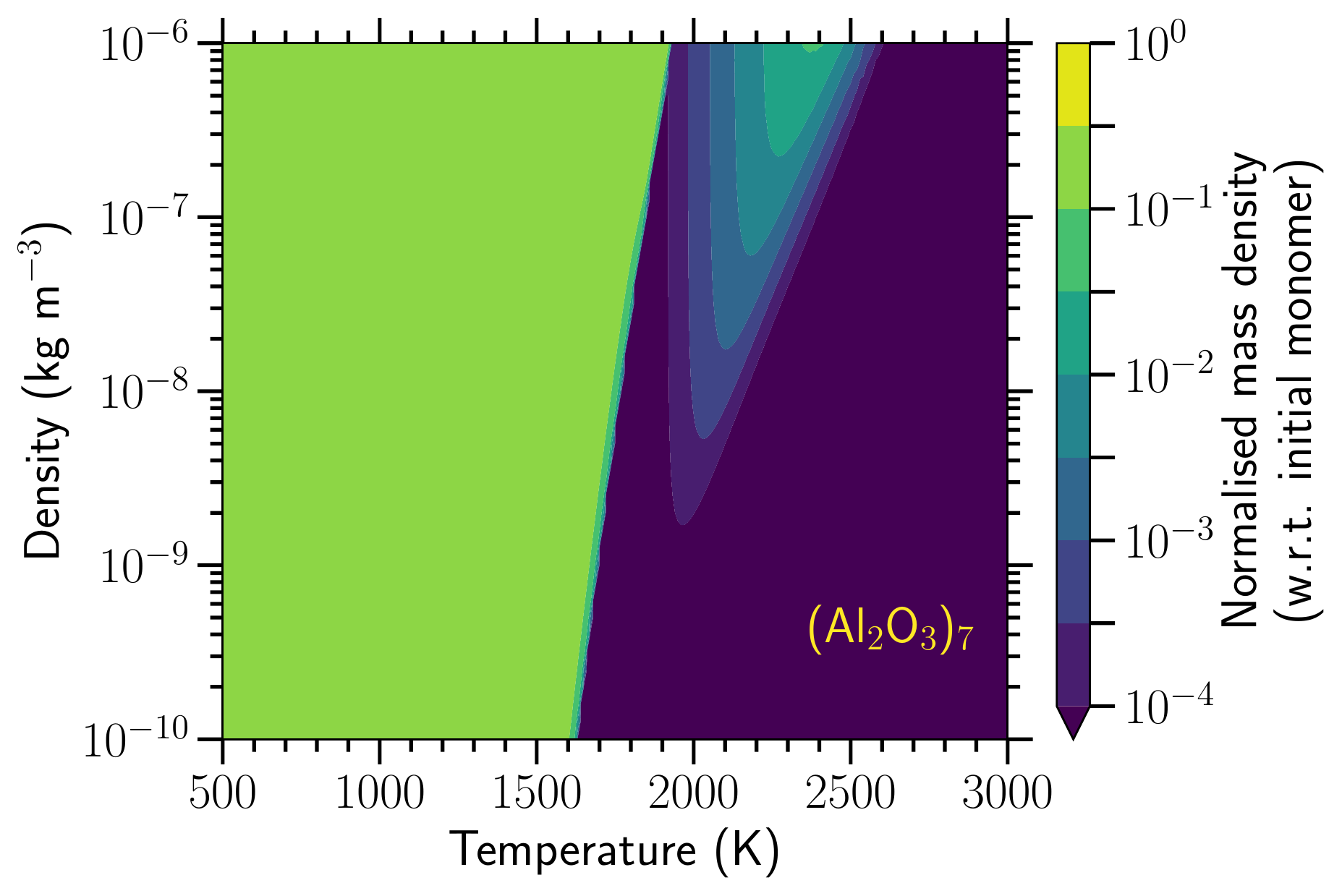}
        \includegraphics[width=0.32\textwidth]{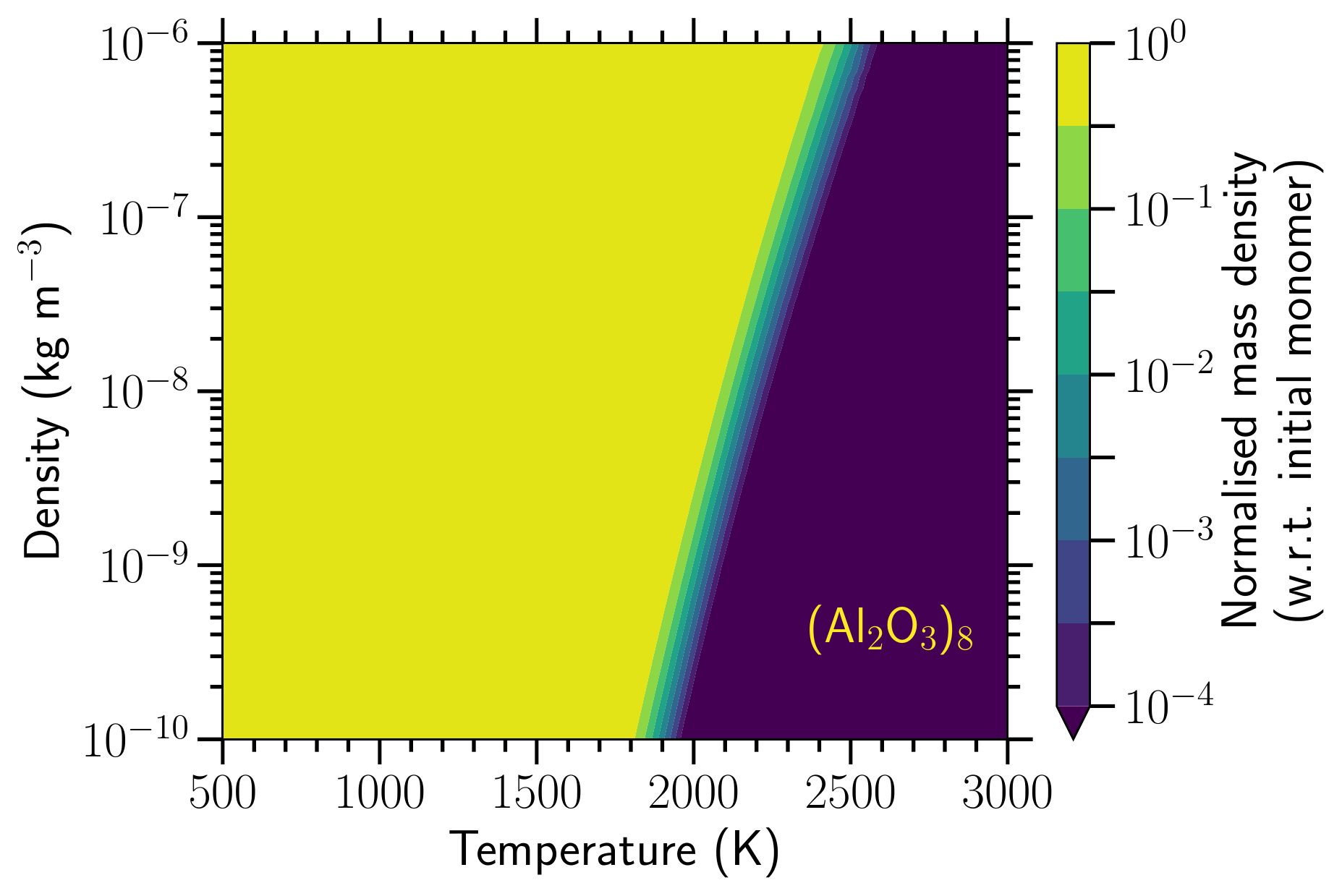}
        \end{flushleft}
        \caption{Overview of the normalised mass density after one year of all \protect\Al{1}-clusters for a closed nucleation model using the polymer nucleation description.}
        \label{fig:Al2O3_clusters_general_norm_same_scale}
    \end{figure*}

\begin{figure*}
        \begin{flushleft}
        \includegraphics[width=0.32\textwidth]{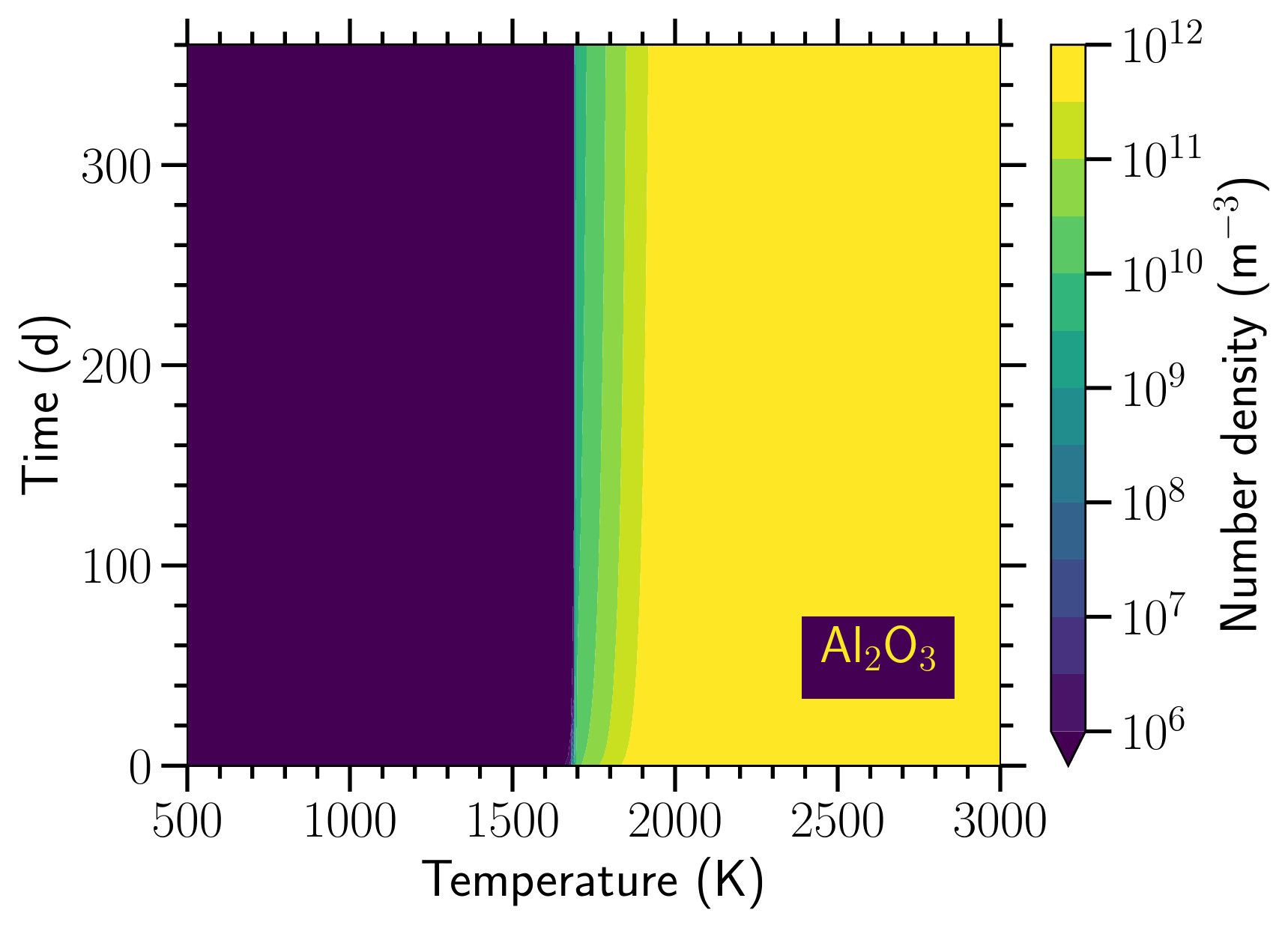}
        \includegraphics[width=0.32\textwidth]{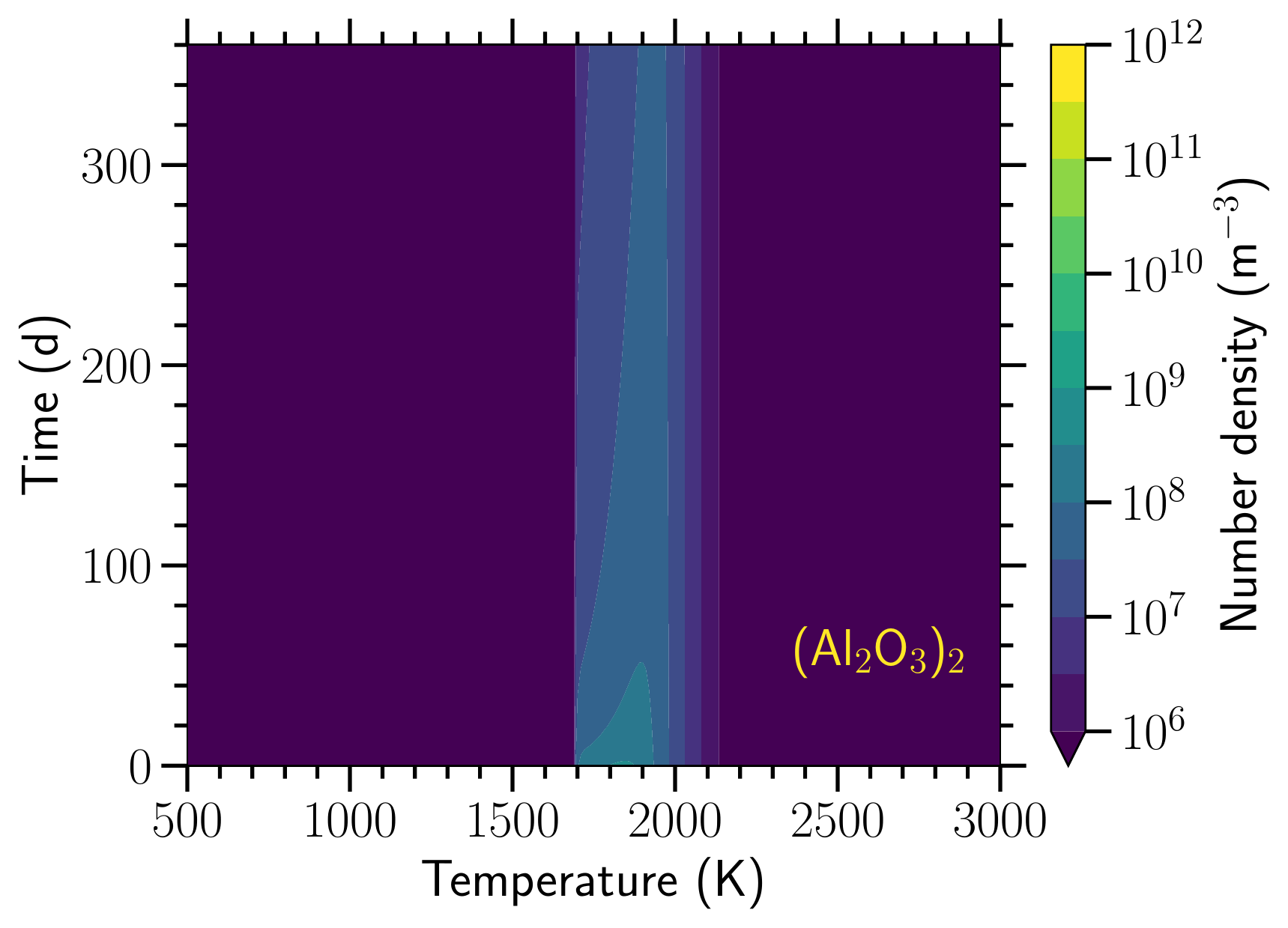}
        \includegraphics[width=0.32\textwidth]{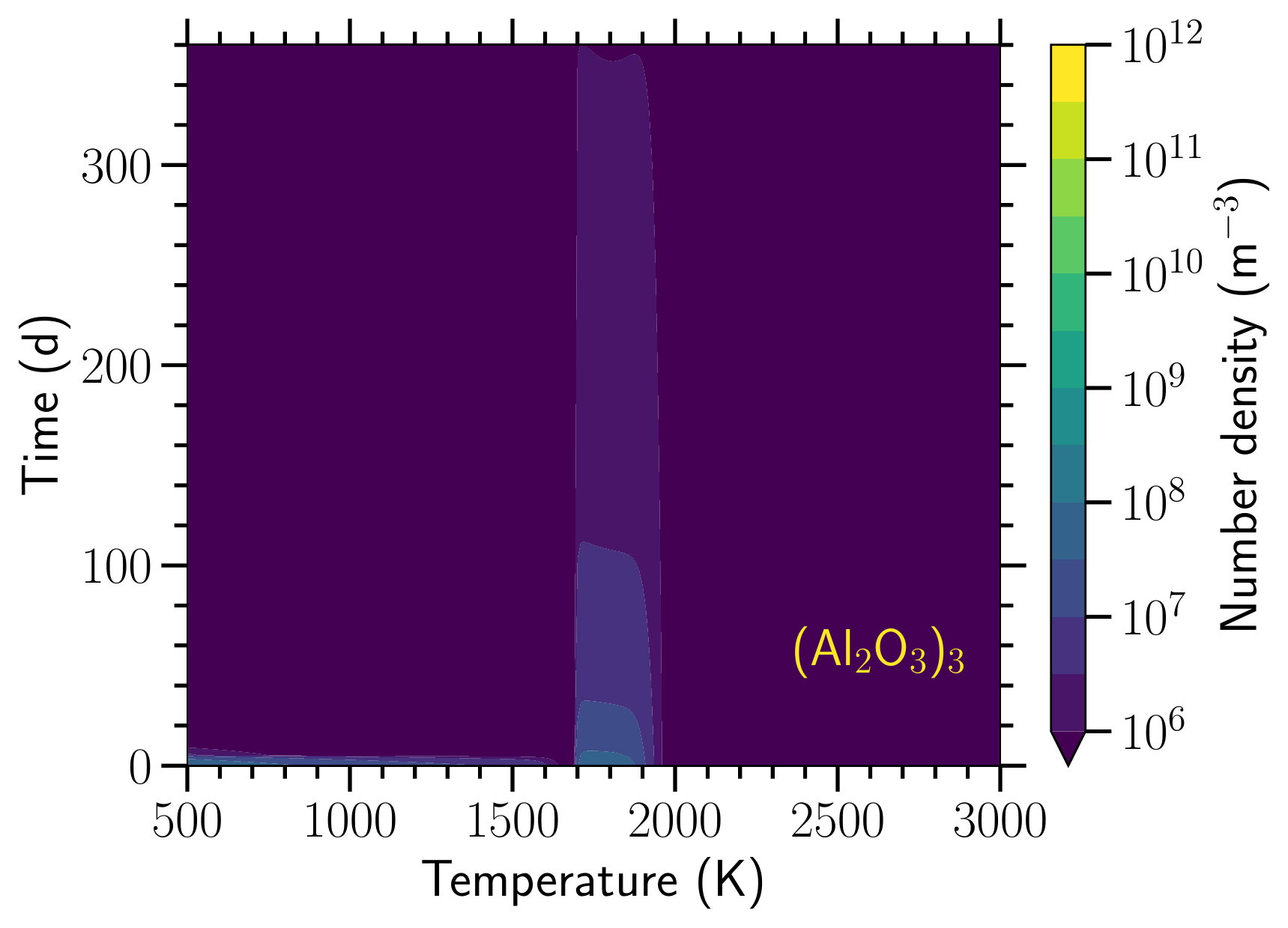}
        \includegraphics[width=0.32\textwidth]{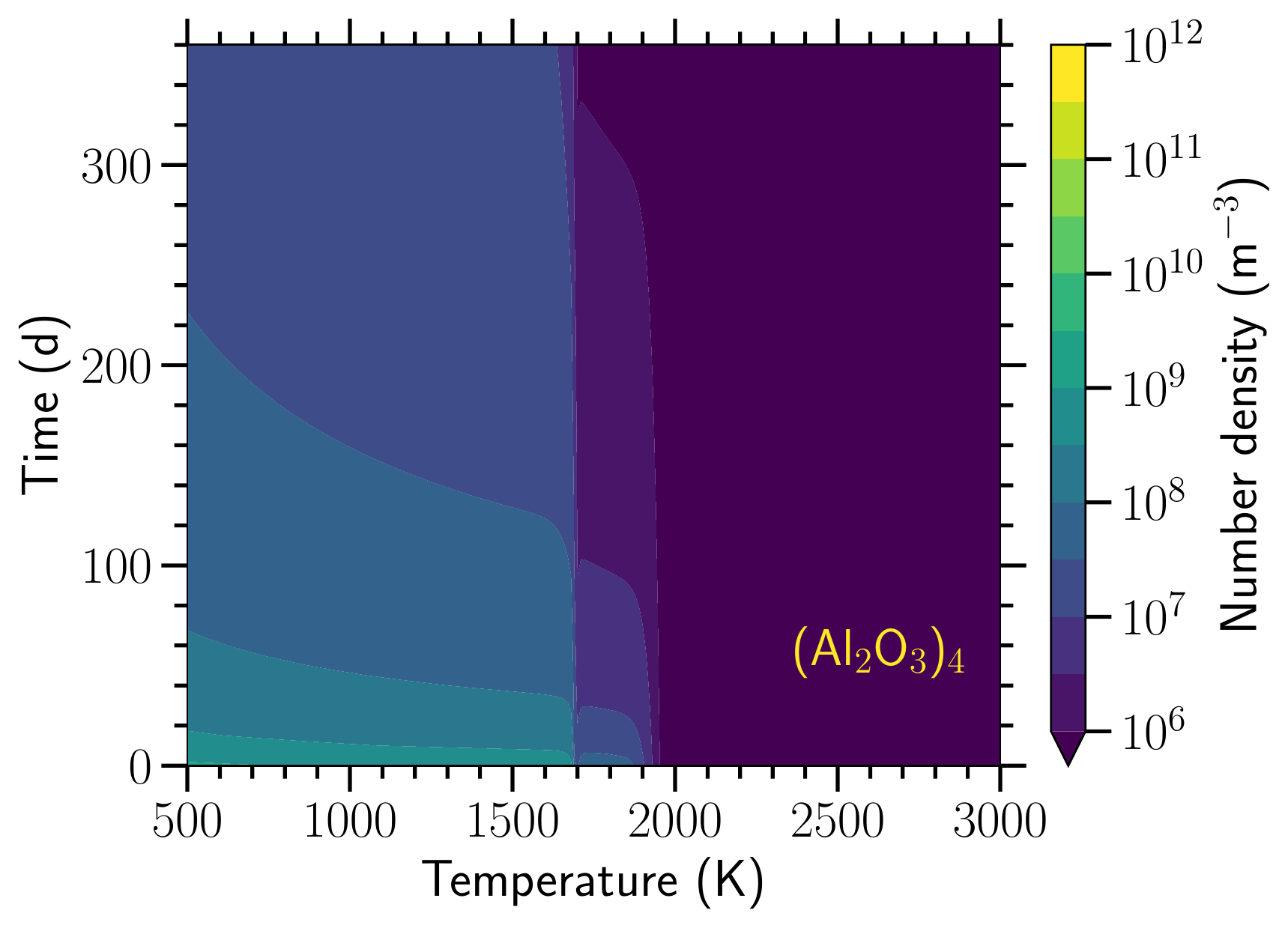}
        \includegraphics[width=0.32\textwidth]{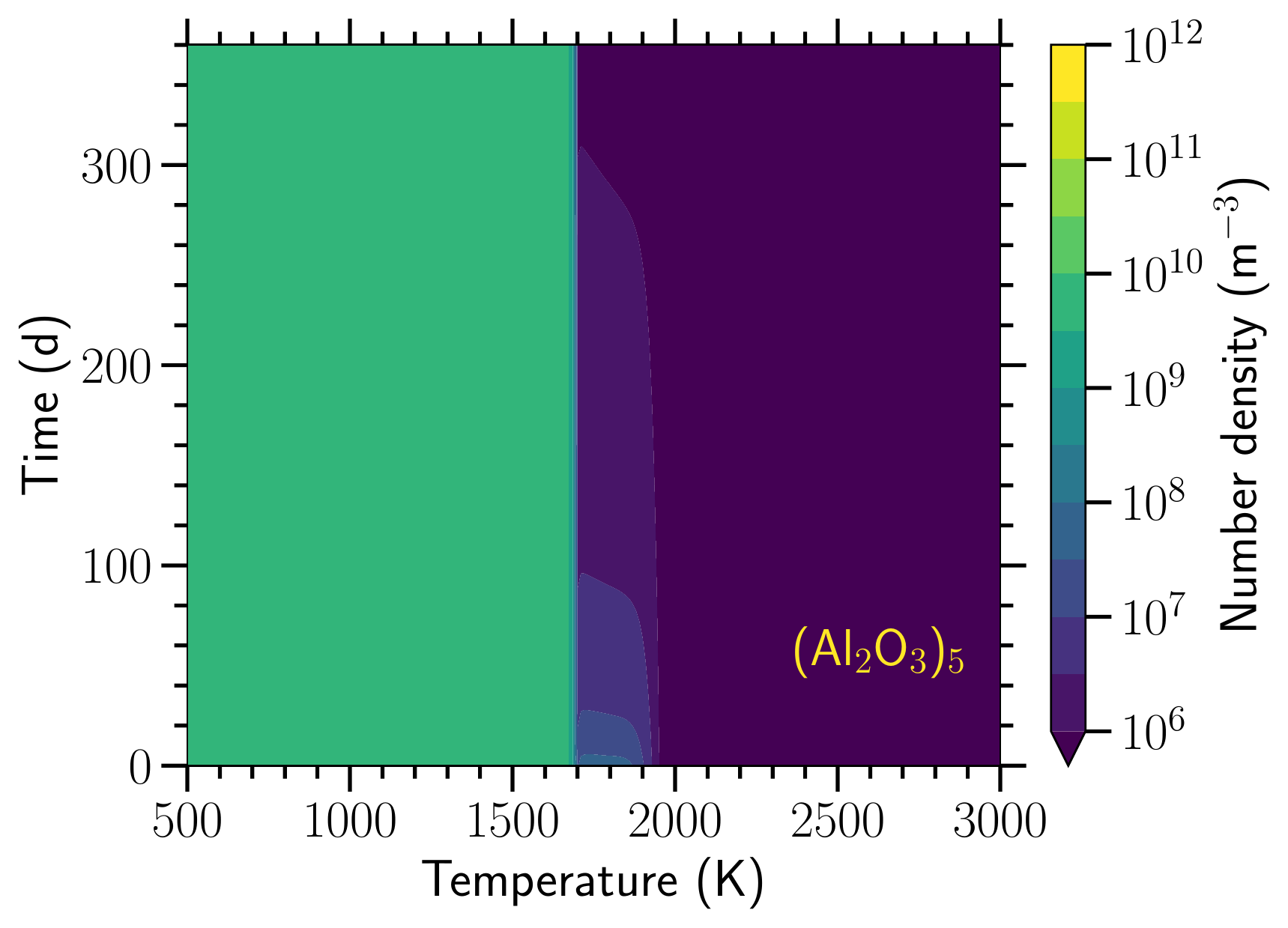}
        \includegraphics[width=0.32\textwidth]{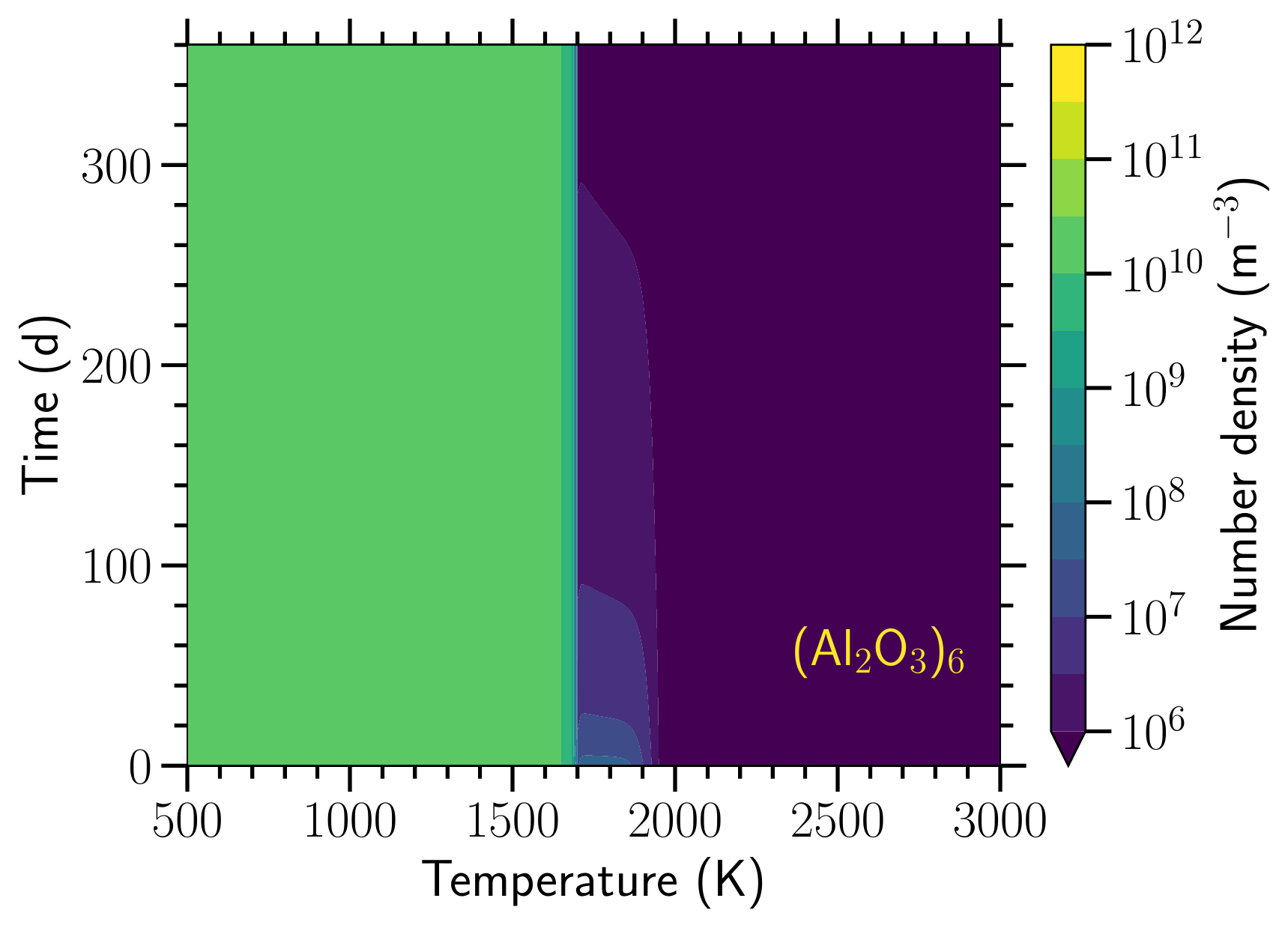}
        \includegraphics[width=0.32\textwidth]{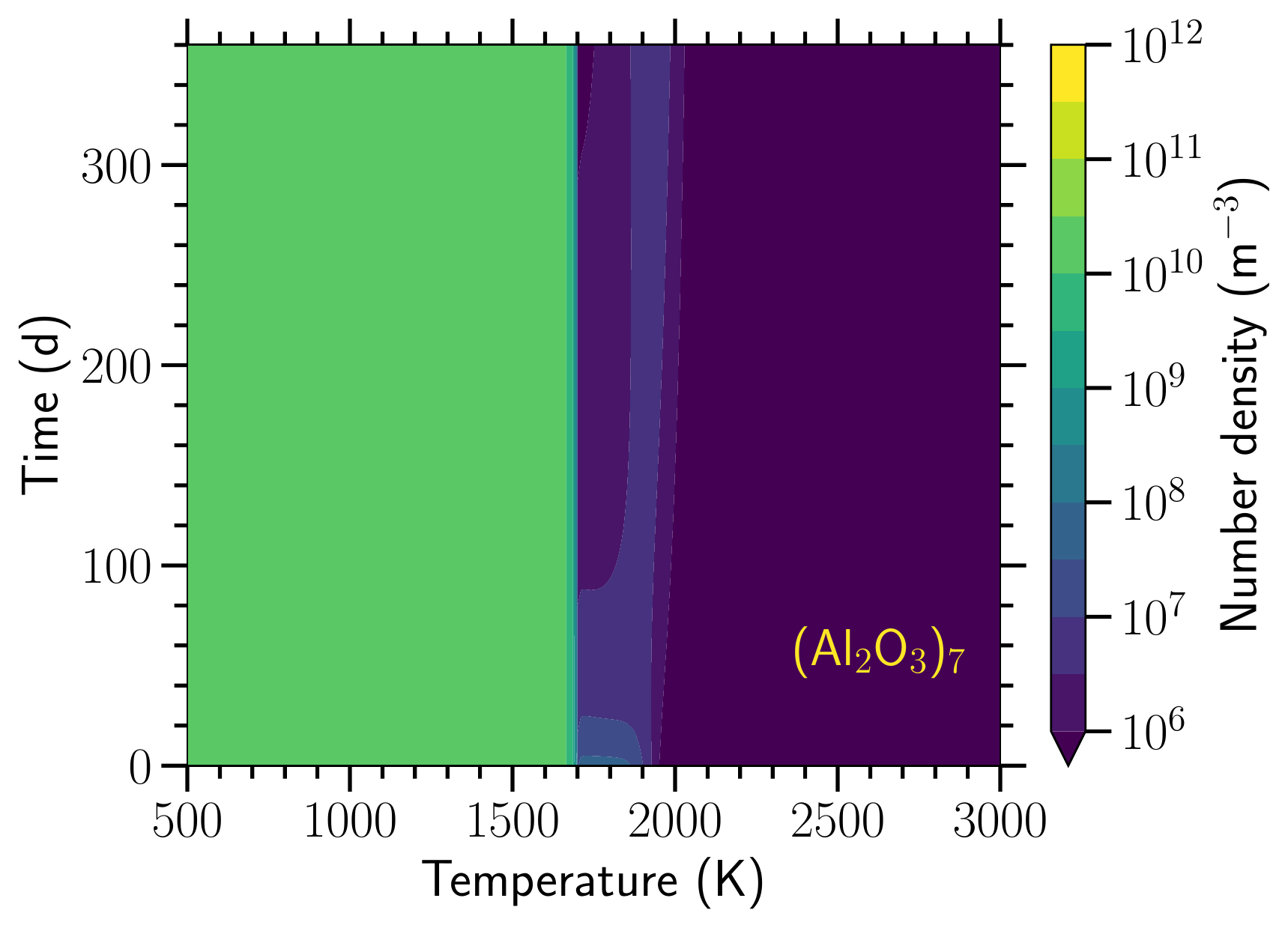}
        \includegraphics[width=0.32\textwidth]{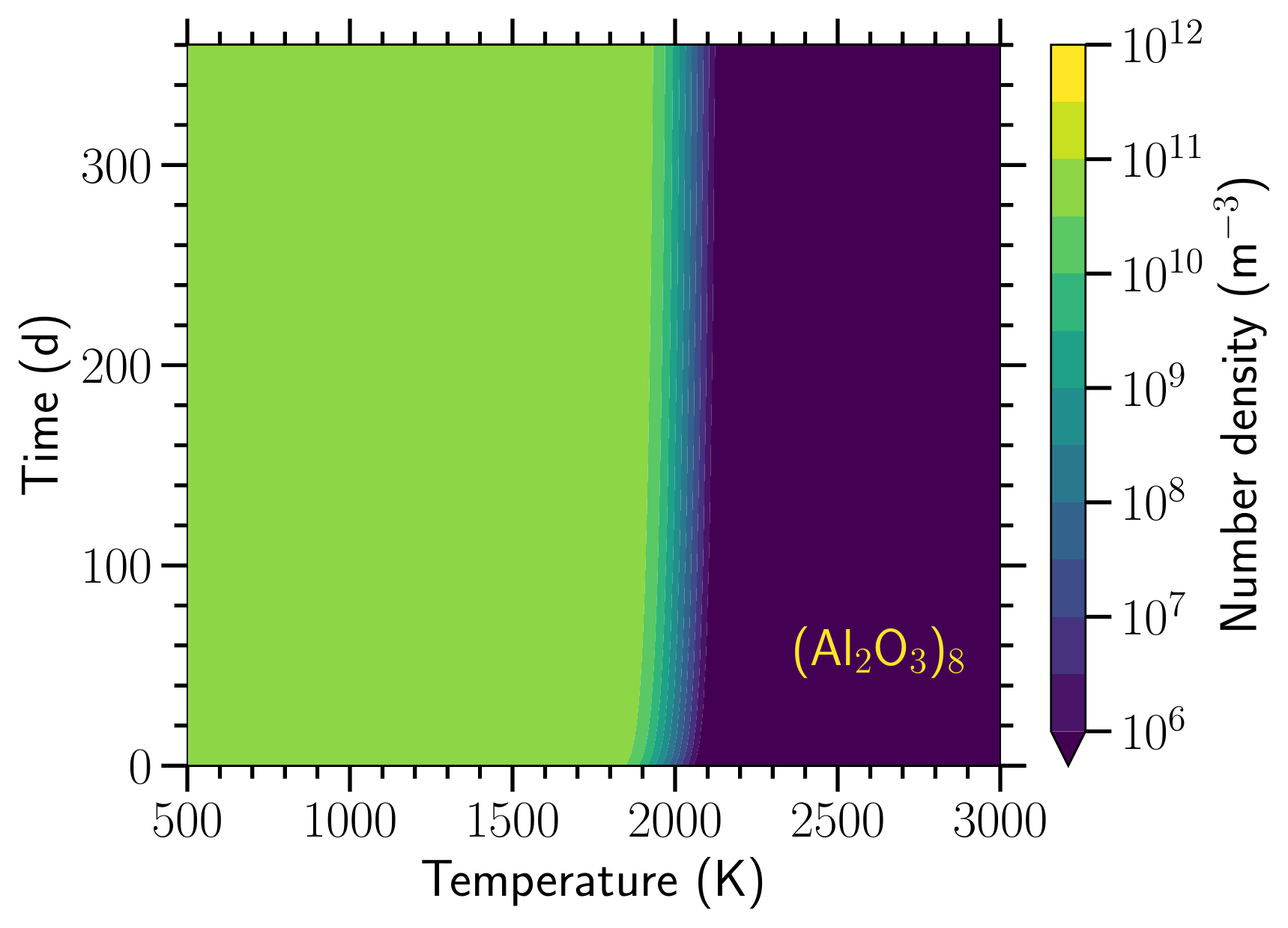}
        \end{flushleft}
        \caption{Temporal evolution of the absolute number density of all \protect\Al{1}-clusters at the benchmark total gas density $\rho=\SI{1e-9}{\kg\per\m\cubed}$ for a closed nucleation model using the polymer nucleation description.}
        \label{fig:Al2O3_clusters_general_time_evolution}
    \end{figure*}

\begin{figure*}
        \begin{flushleft}
        \includegraphics[width=0.32\textwidth]{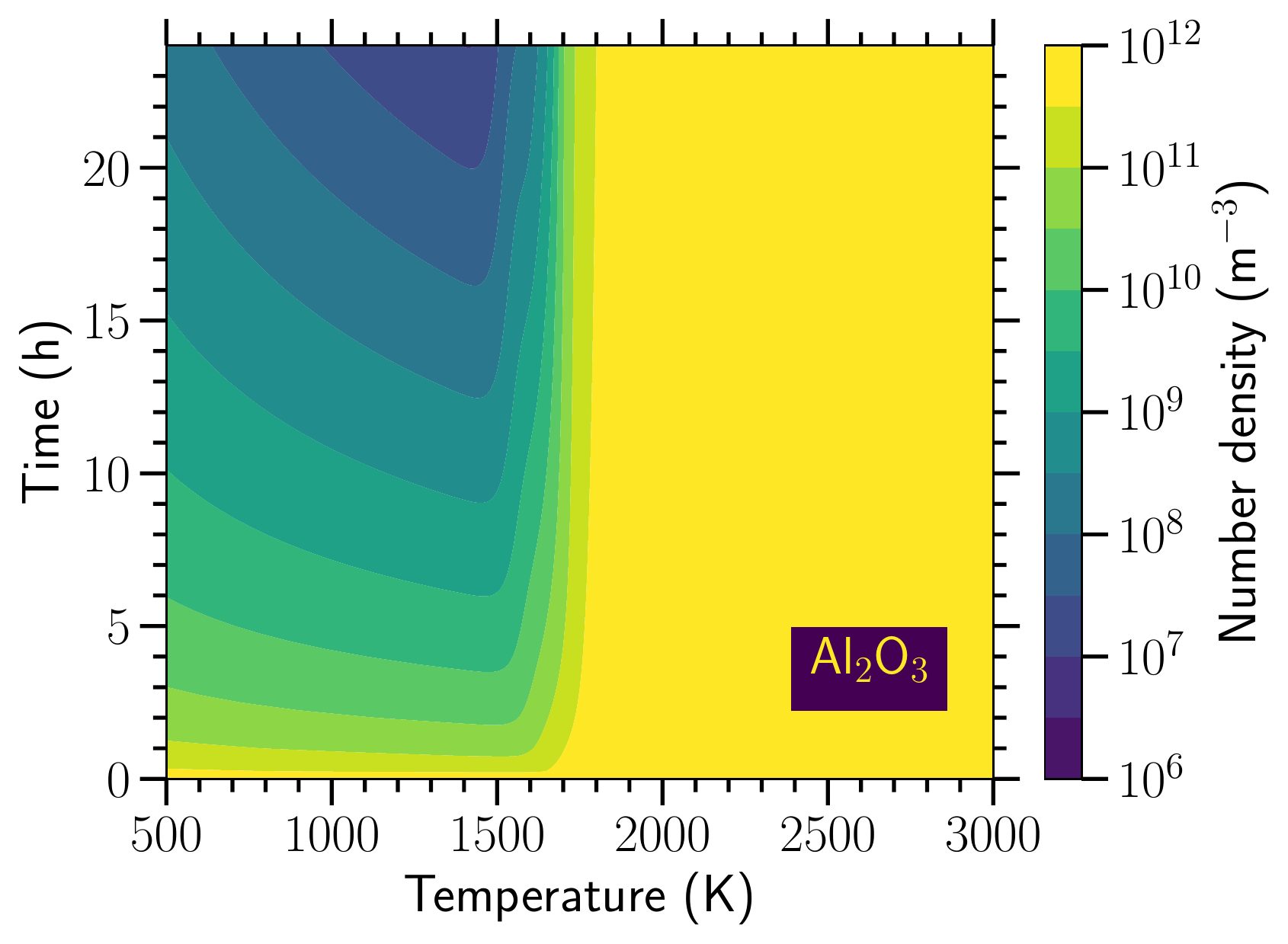}
        \includegraphics[width=0.32\textwidth]{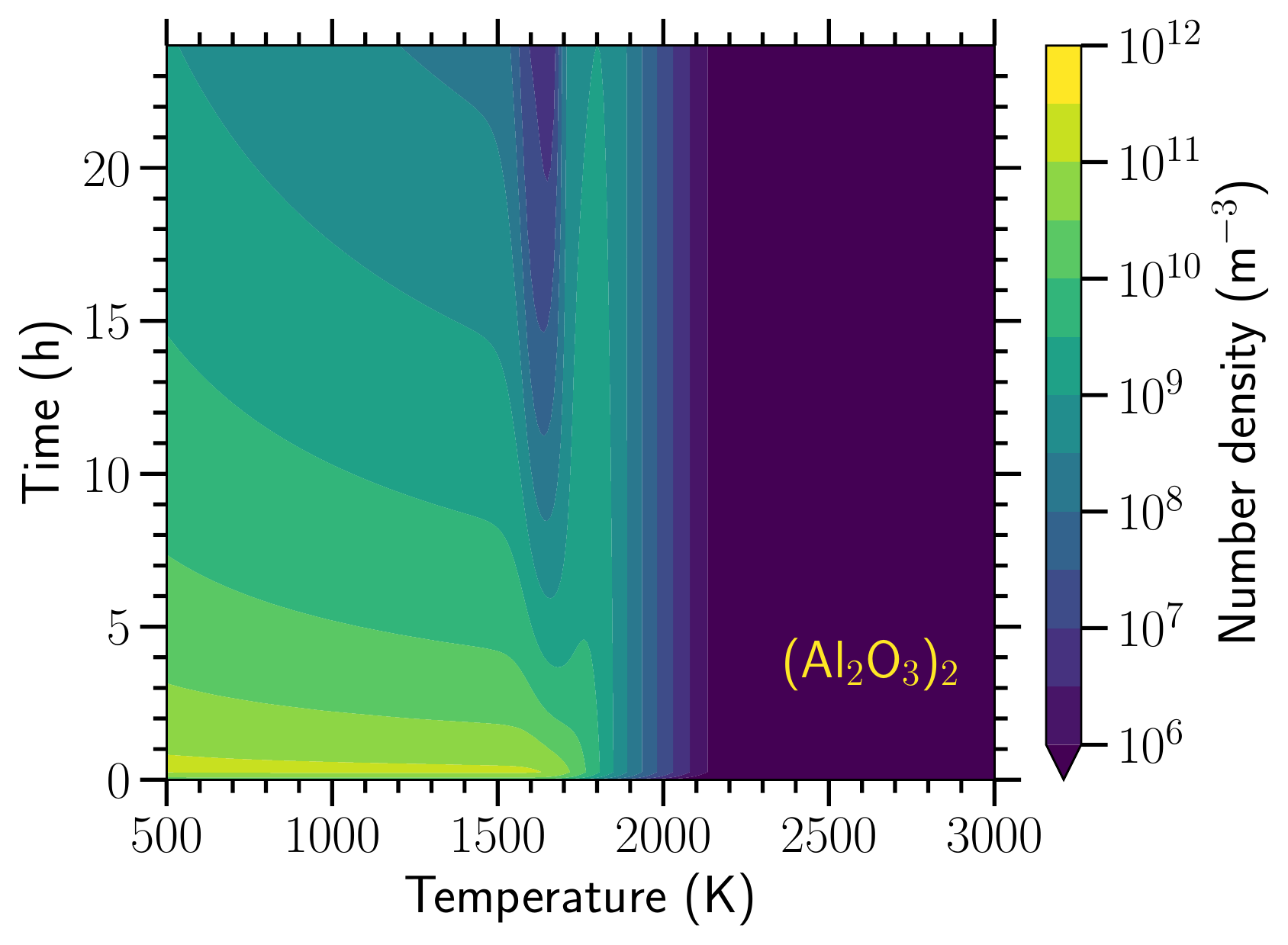}
        \includegraphics[width=0.32\textwidth]{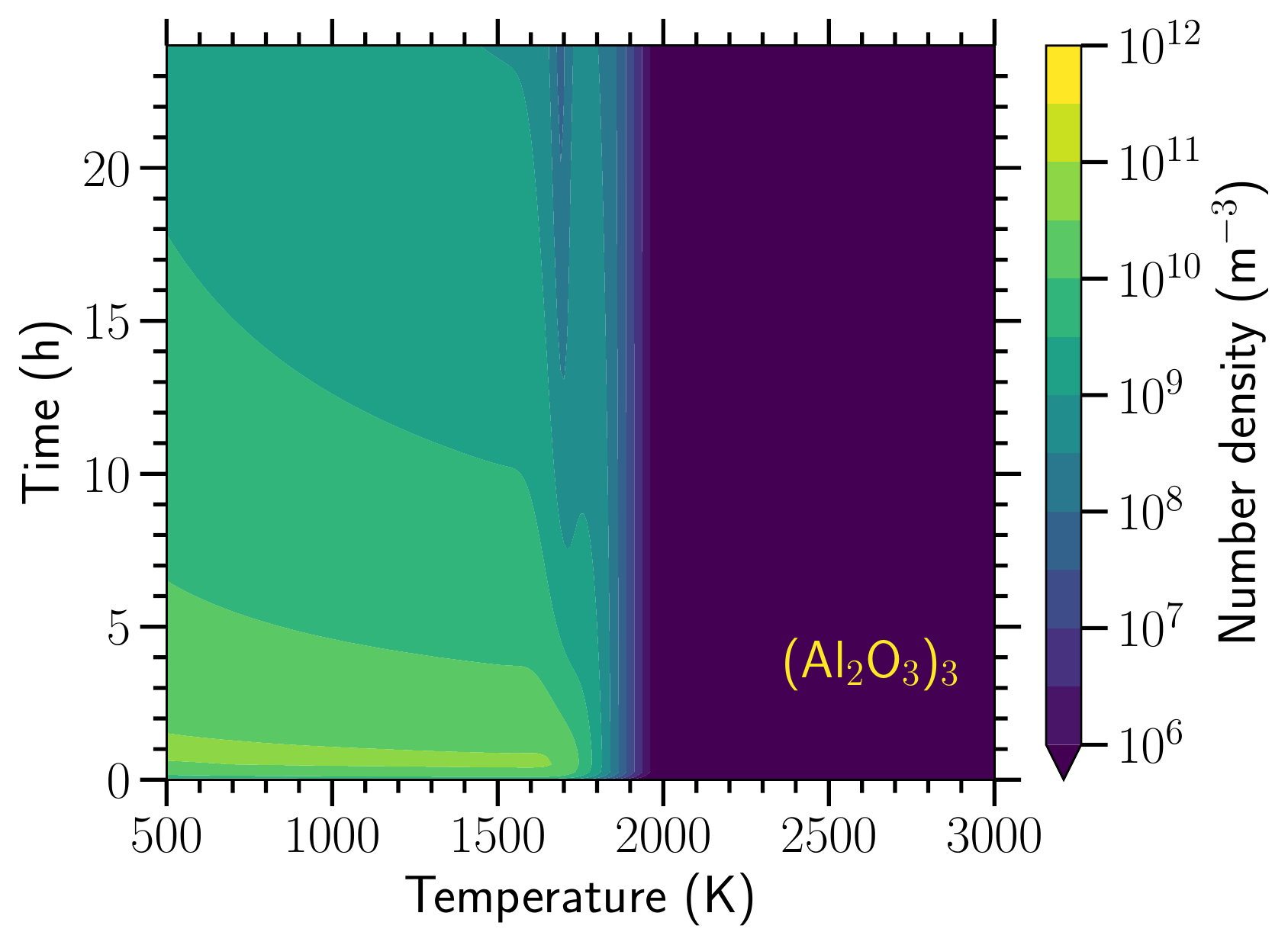}
        \includegraphics[width=0.32\textwidth]{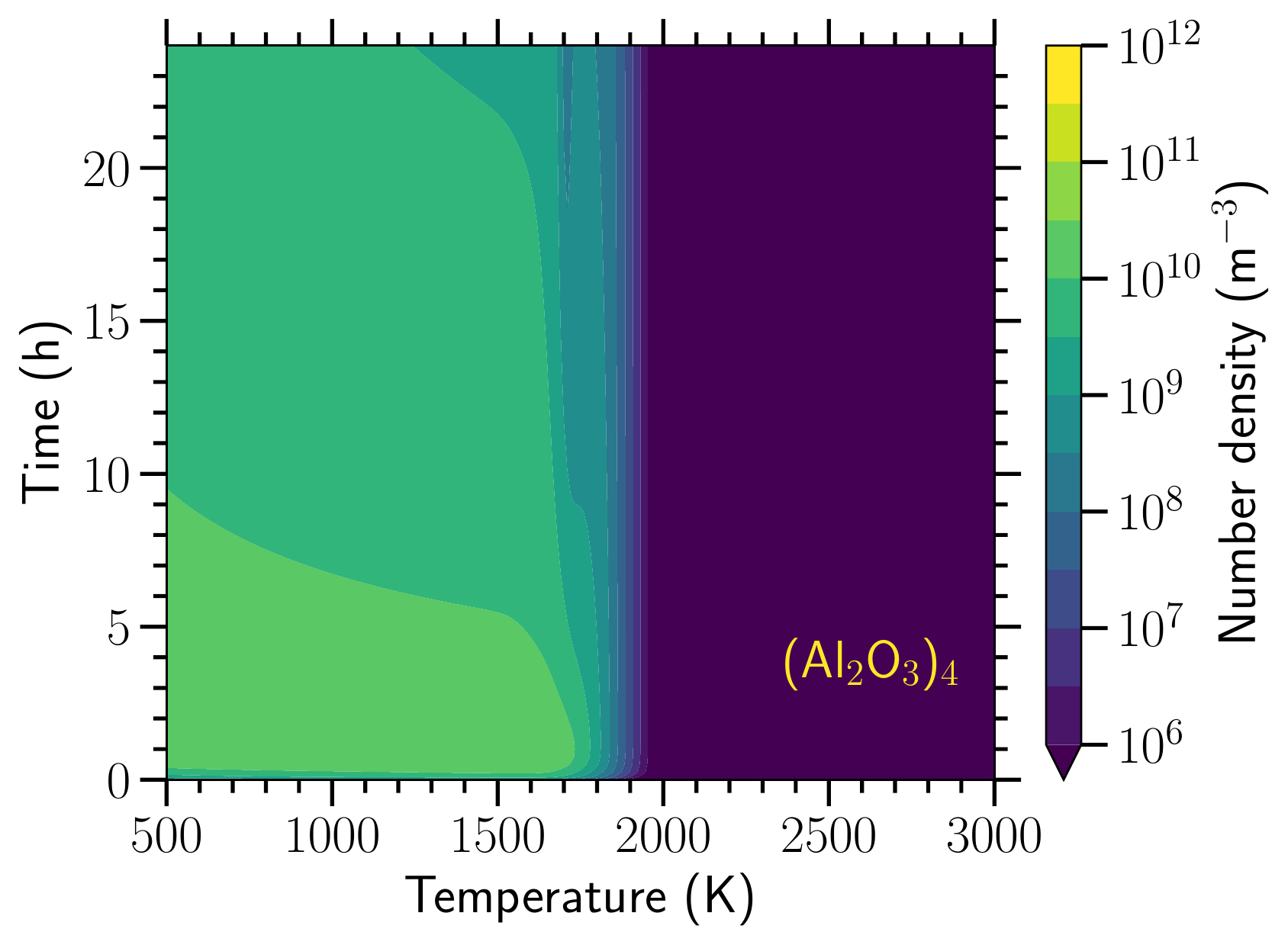}
        \includegraphics[width=0.32\textwidth]{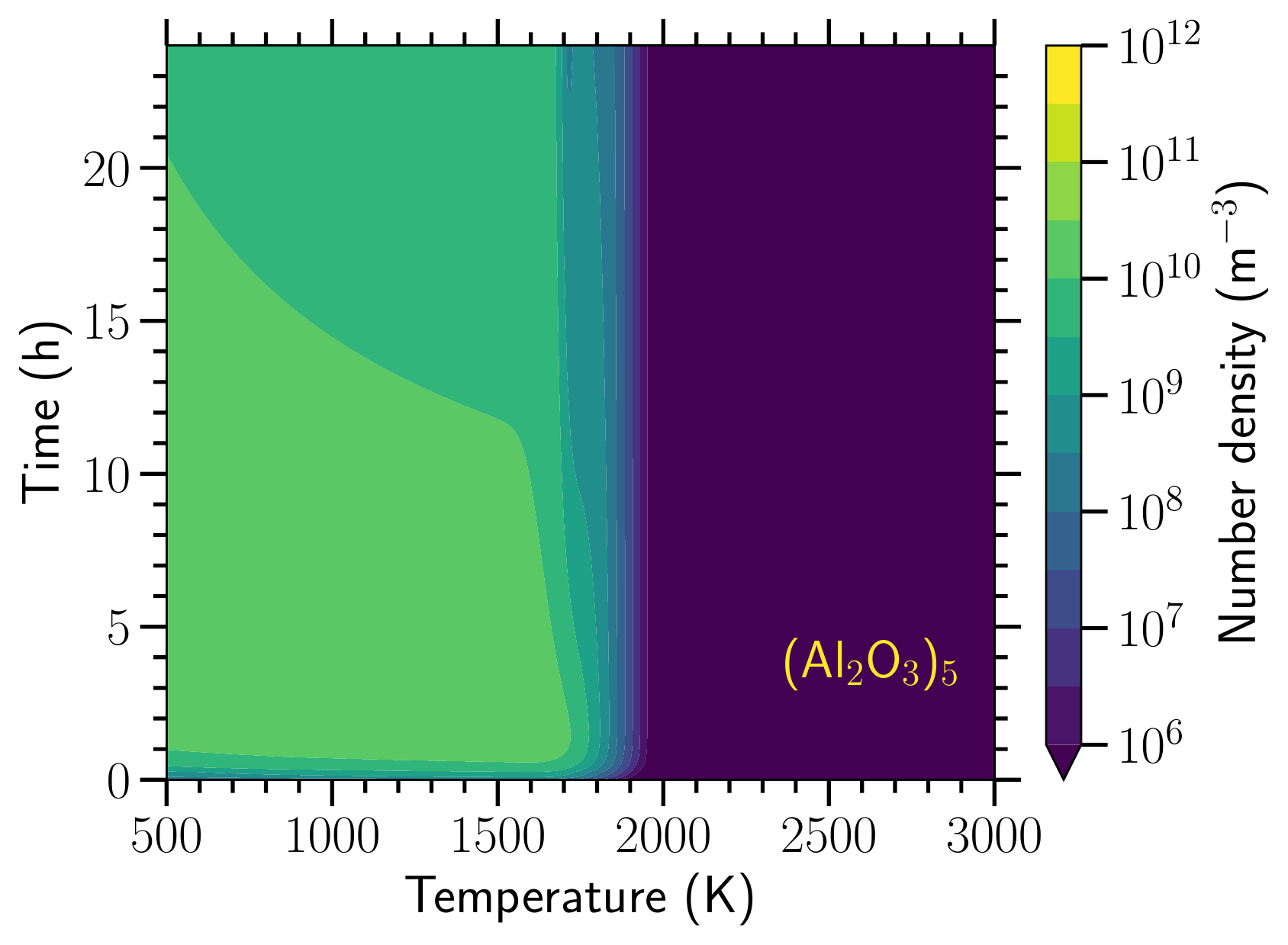}
        \includegraphics[width=0.32\textwidth]{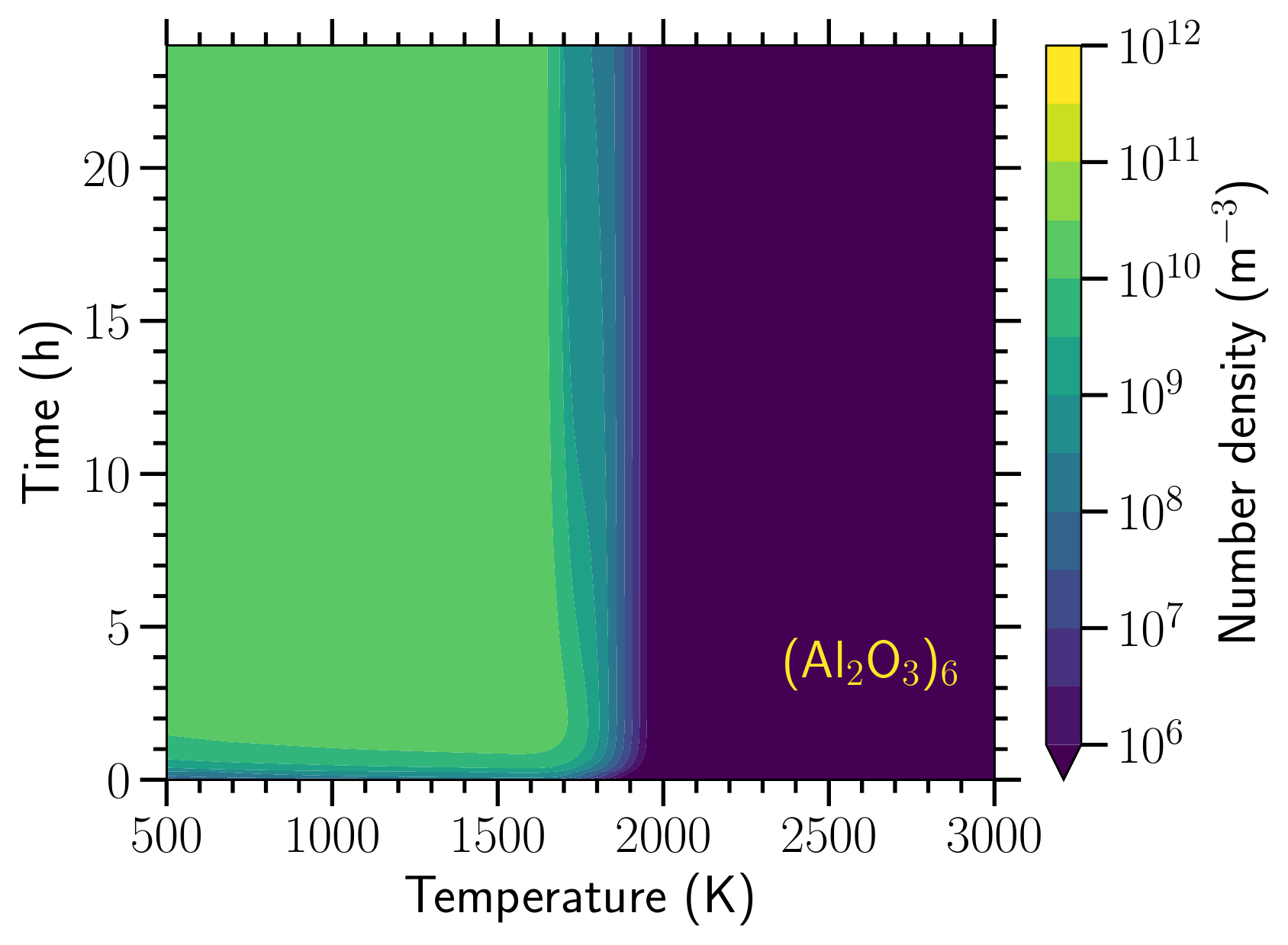}
        \includegraphics[width=0.32\textwidth]{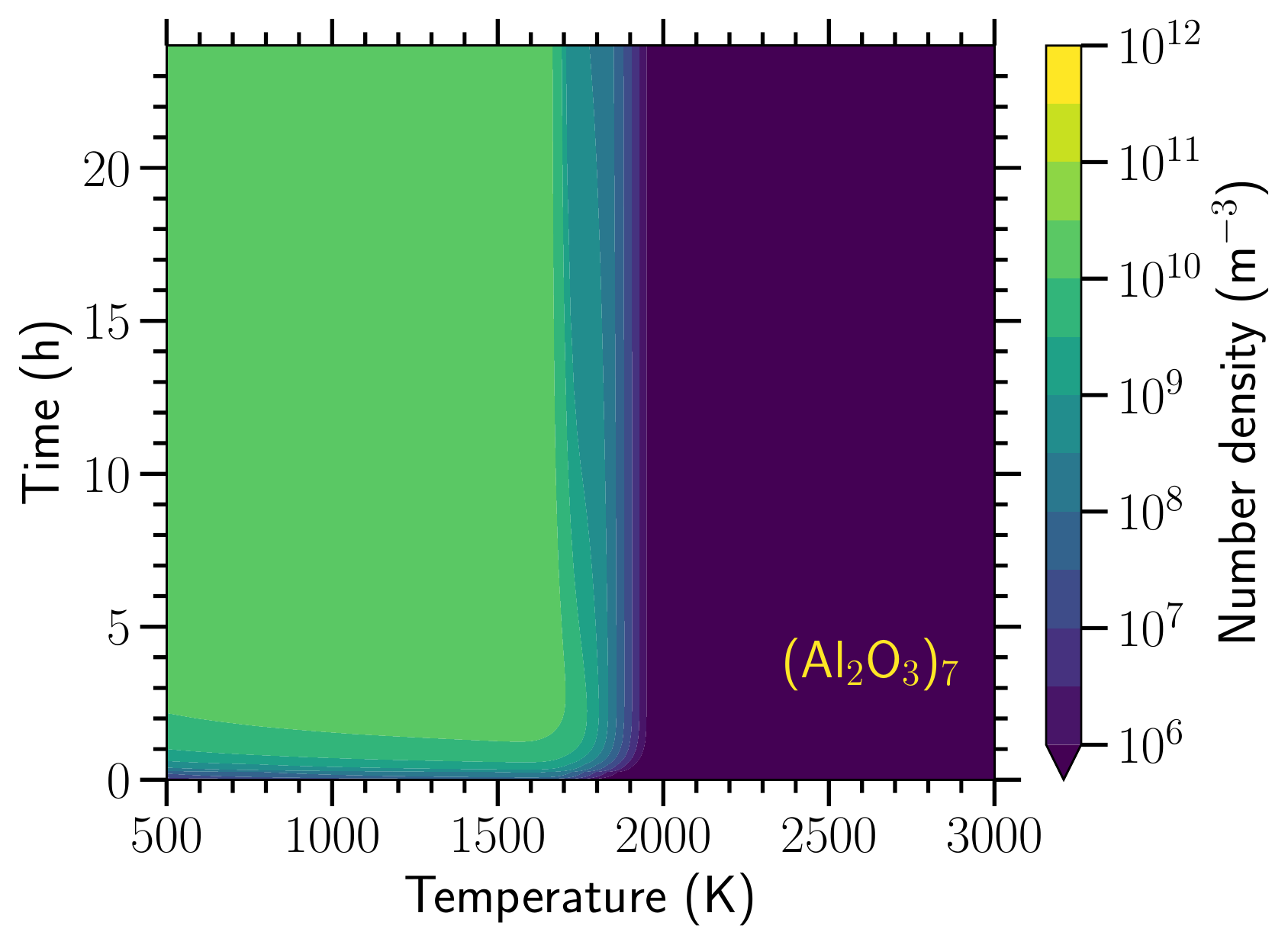}
        \includegraphics[width=0.32\textwidth]{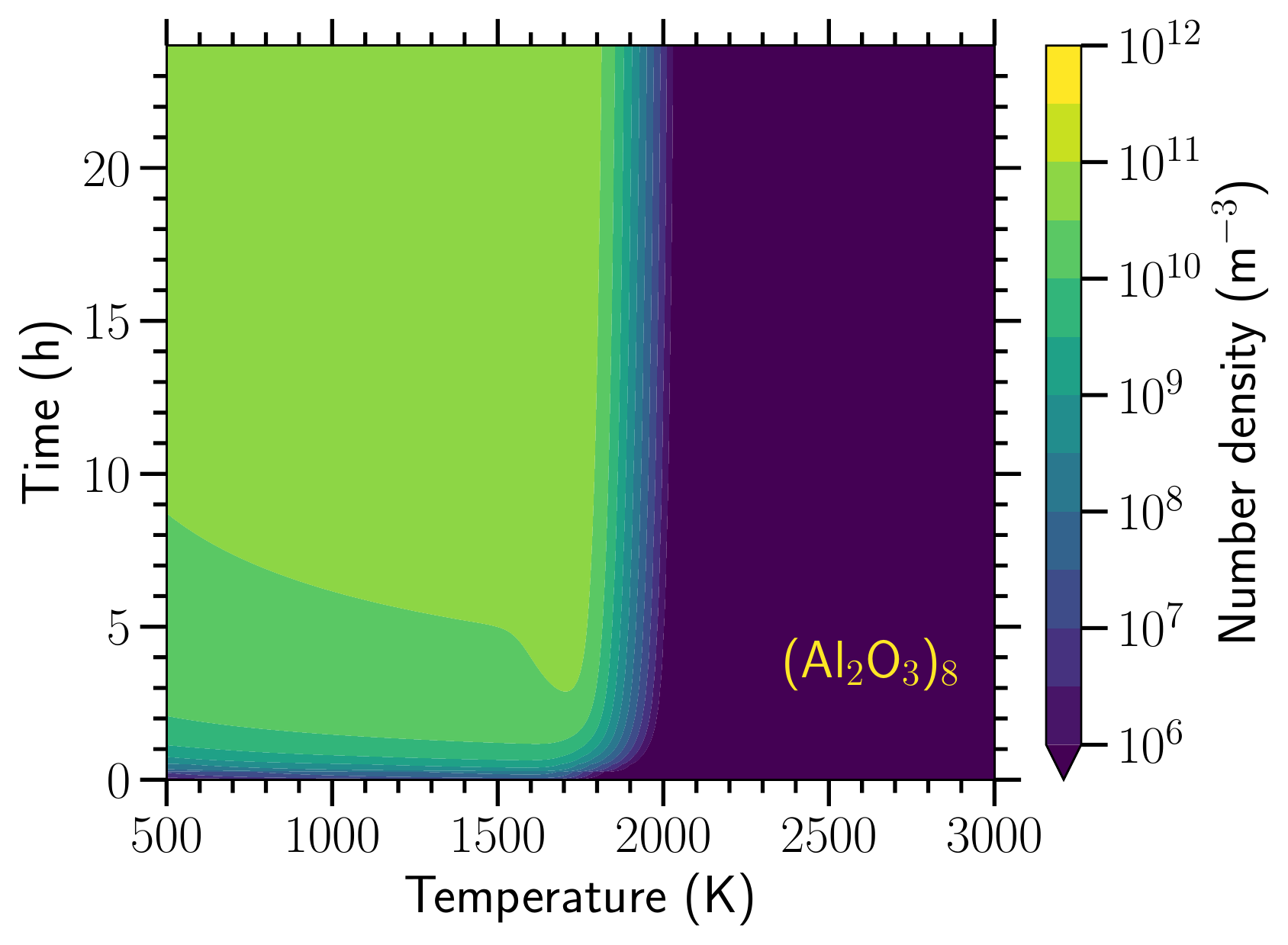}
        \end{flushleft}
        \caption{Refined temporal evolution of the absolute number density of all \protect\Al{1}-clusters at the benchmark total gas density $\rho=\SI{1e-9}{\kg\per\m\cubed}$ for a closed nucleation model using the polymer nucleation description.}
        \label{fig:Al2O3_clusters_general_time_evolution_short}
    \end{figure*}

        This section contains a more complete overview of the closed nucleation models using the polymer nucleation description of all nucleation clusters results.\\\\
        \Ti{1}-clusters: Figs.\,\ref{fig:TiO2_clusters_general_norm_same_scale} to \ref{fig:TiO2_clusters_general_time_evolution}\\
        \Mg{1}-clusters: Figs.\,\ref{fig:MgO_clusters_general_norm_same_scale} to \ref{fig:MgO_clusters_general_time_evolution_short}\\
        \SiO{1}-clusters: Fig.\,\ref{fig:SiO_clusters_general_norm_same_scale}\\
        \Al{1}-clusters: Figs.\,\ref{fig:Al2O3_clusters_general_norm_same_scale} to \ref{fig:Al2O3_clusters_general_time_evolution_short}
        
        \subsubsection{Polymer nucleation compared with equilibrium}
\begin{figure*}
        \centering
        \includegraphics[scale=0.65]{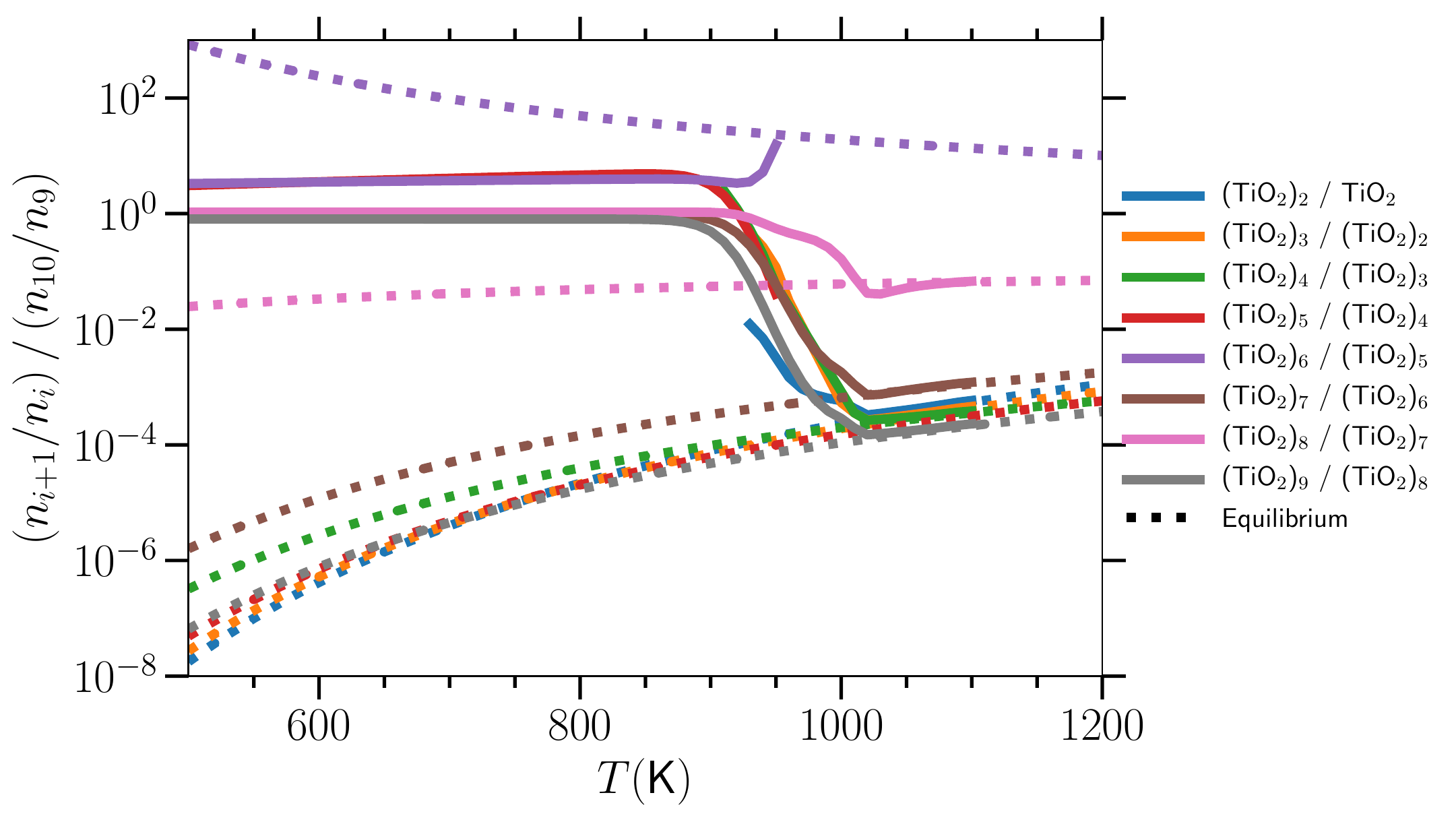}
        \caption{The relative ratios of \ch{TiO2}-clusters do not reach the equilibrium ratios in the entire temperature range. At the highest temperatures at which the nucleation is feasible, the model results (full lines) correspond to the equilibrium ratios (dotted line). At lower temperatures, the clusters need more time to reach the equilibrium ratios since the interaction probability is lower. This continuous evolution is also visible in Fig.~\ref{fig:TiO2_clusters_general_time_evolution}. The results are of the closed polymer nucleation model for the benchmark total gas density $\rho = \SI{1e-9}{\kg\per\m\cubed}$ at the final time step (one year). The figure shows the ratios of two clusters w.r.t. the ratio of both largest clusters. If the number density of any of the four clusters is below the numerical solver accuracy of \SI{1e-20}{\per\cm\cubed}, the ratios are not shown.}
        \label{fig:equ_ratios_TiO2}
    \end{figure*}

\begin{figure*}
        \centering
        \includegraphics[scale=0.65]{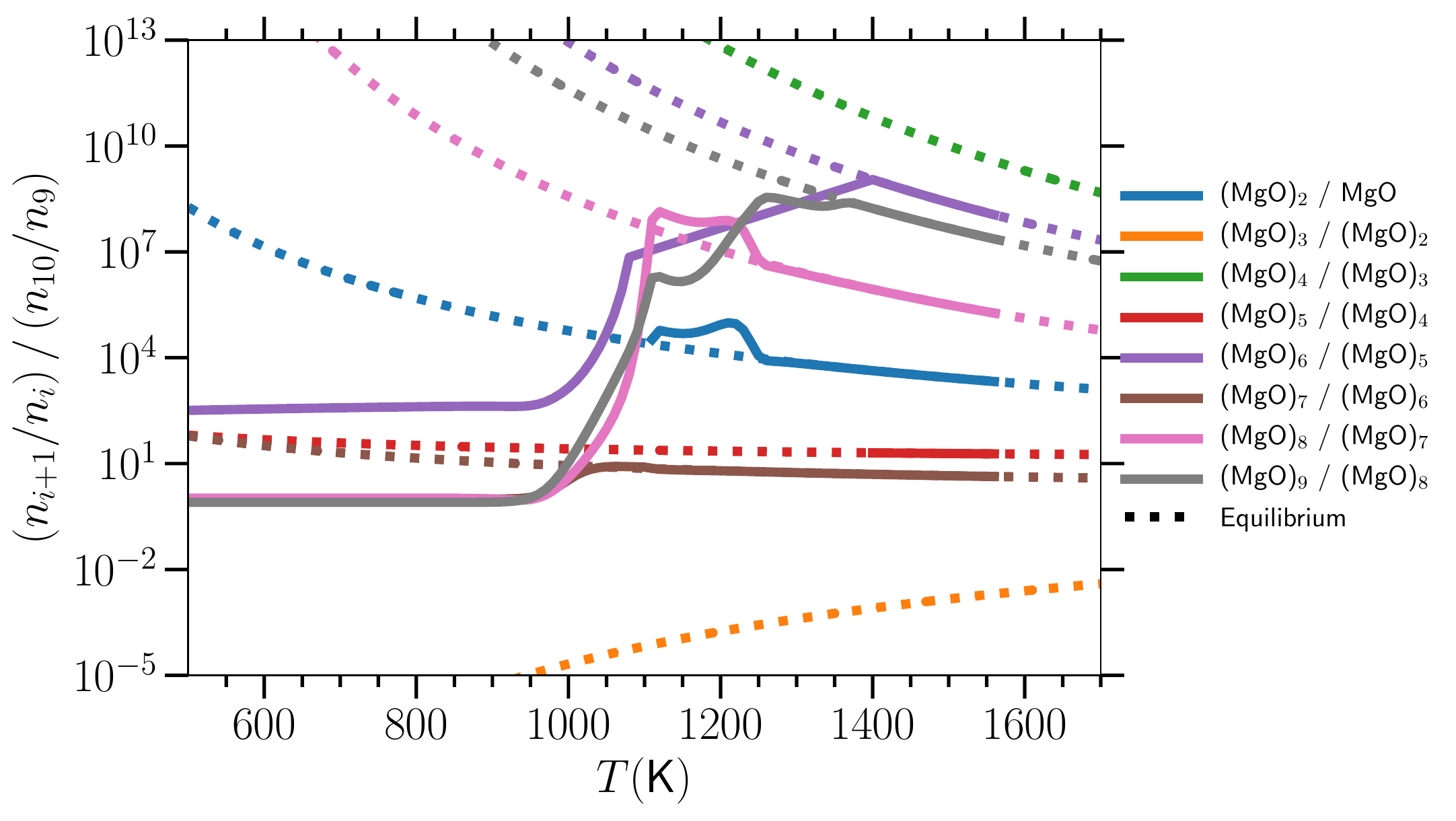}
        \caption{The relative ratios of \ch{MgO}-clusters do not reach the equilibrium ratios in the entire temperature range. At the highest temperatures at which the nucleation is feasible, the model results (full lines) correspond to the equilibrium ratios (dotted line). At lower temperatures, the clusters need more time to reach the equilibrium ratios since the interaction probability is lower. This transition is visible between \SIrange{1000}{1300}{\K}. The continuous evolution is also visible in Fig.~\ref{fig:MgO_clusters_general_time_evolution}. The results are of the closed polymer nucleation model for the benchmark total gas density $\rho = \SI{1e-9}{\kg\per\m\cubed}$ at the final time step (one year). The figure shows the ratios of two clusters w.r.t. the ratio of both largest clusters. If the number density of any of the four clusters is below the numerical solver accuracy of \SI{1e-20}{\per\cm\cubed}, the ratios are not shown. Note that no model results involving \ch{(MgO)_3} are visible since this cluster does not exists under the local conditions.}
        \label{fig:equ_ratios_MgO}
    \end{figure*}

\begin{figure*}
        \centering
        \includegraphics[scale=0.65]{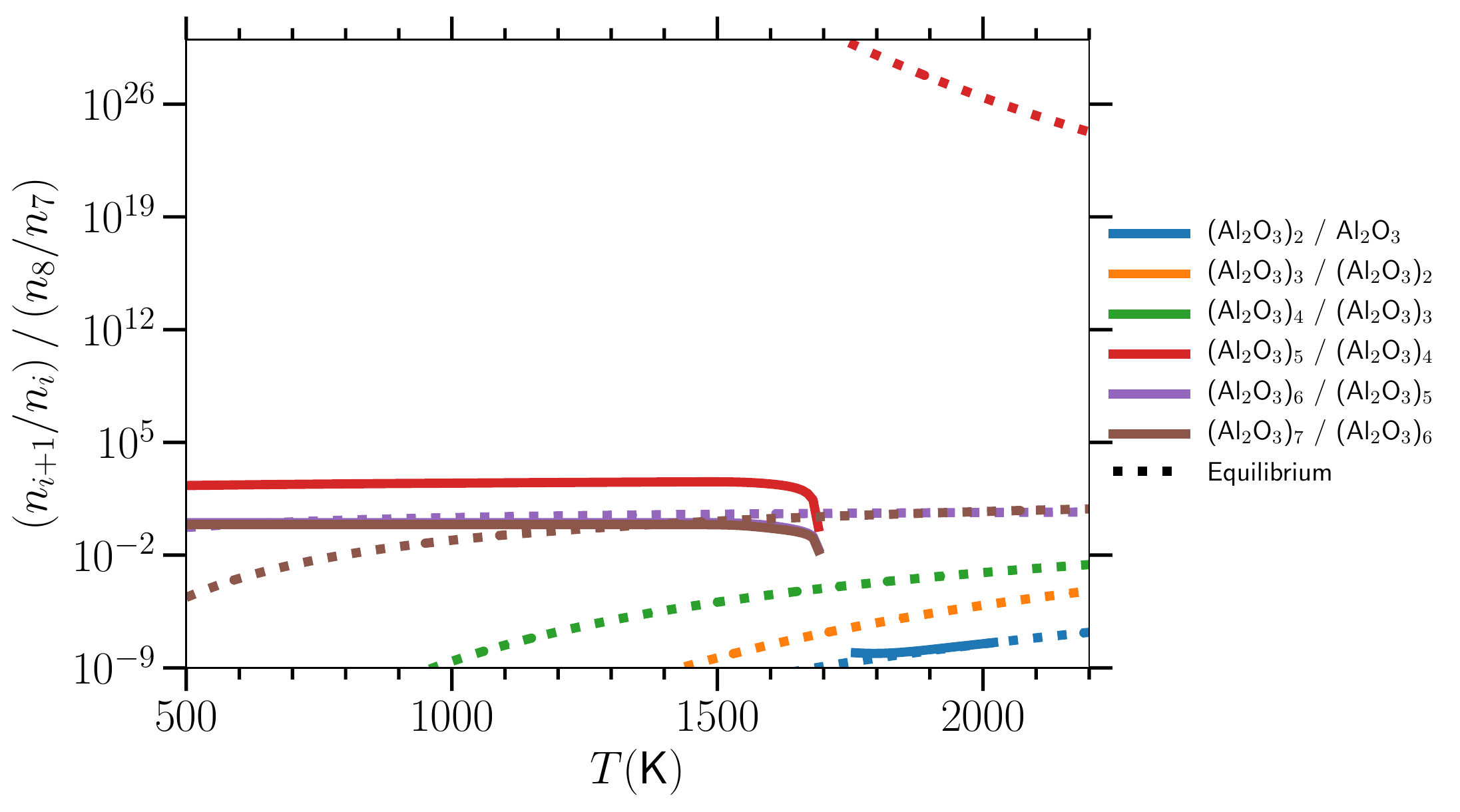}
        \caption{The relative ratios of \ch{Al2O3}-clusters (full lines) do not correspond to the equilibrium ratios (dotted line). The figure shows the ratios of two clusters w.r.t. the ratio of both largest clusters. If the number density of any of the four clusters is below the numerical solver accuracy of \SI{1e-20}{\per\cm\cubed}, the ratios are not shown. Due to the large variation in number densities of the clusters in different temperature regimes (order of magnitude), it is often impossible to compare ratios of the clusters. This variation is more clearly visible in Fig.~\ref{fig:Al2O3_clusters_general_time_evolution}. The results are of the closed polymer nucleation model for the benchmark total gas density $\rho = \SI{1e-9}{\kg\per\m\cubed}$ at the final time step (one year).}
        \label{fig:equ_ratios_Al2O3}
    \end{figure*}

        This section contains figures which compare the relative ratios of nucleation clusters of the closed nucleation models w.r.t. the equilibrium ratios (Figs.~\ref{fig:equ_ratios_TiO2} to \ref{fig:equ_ratios_Al2O3}).
    
    \subsection{Comprehensive chemical nucleation networks}\label{app:res_full_ntw}
    This section contains a more complete overview of all nucleation clusters results in the comprehensive chemical nucleation model using the polymer nucleation description. No \ch{Mg}-related figures are shown as it remains completely atomic.
    
        \subsubsection{\ch{Ti}-bearing species}
\begin{figure*}
        \begin{flushleft}
        \includegraphics[width=0.32\textwidth]{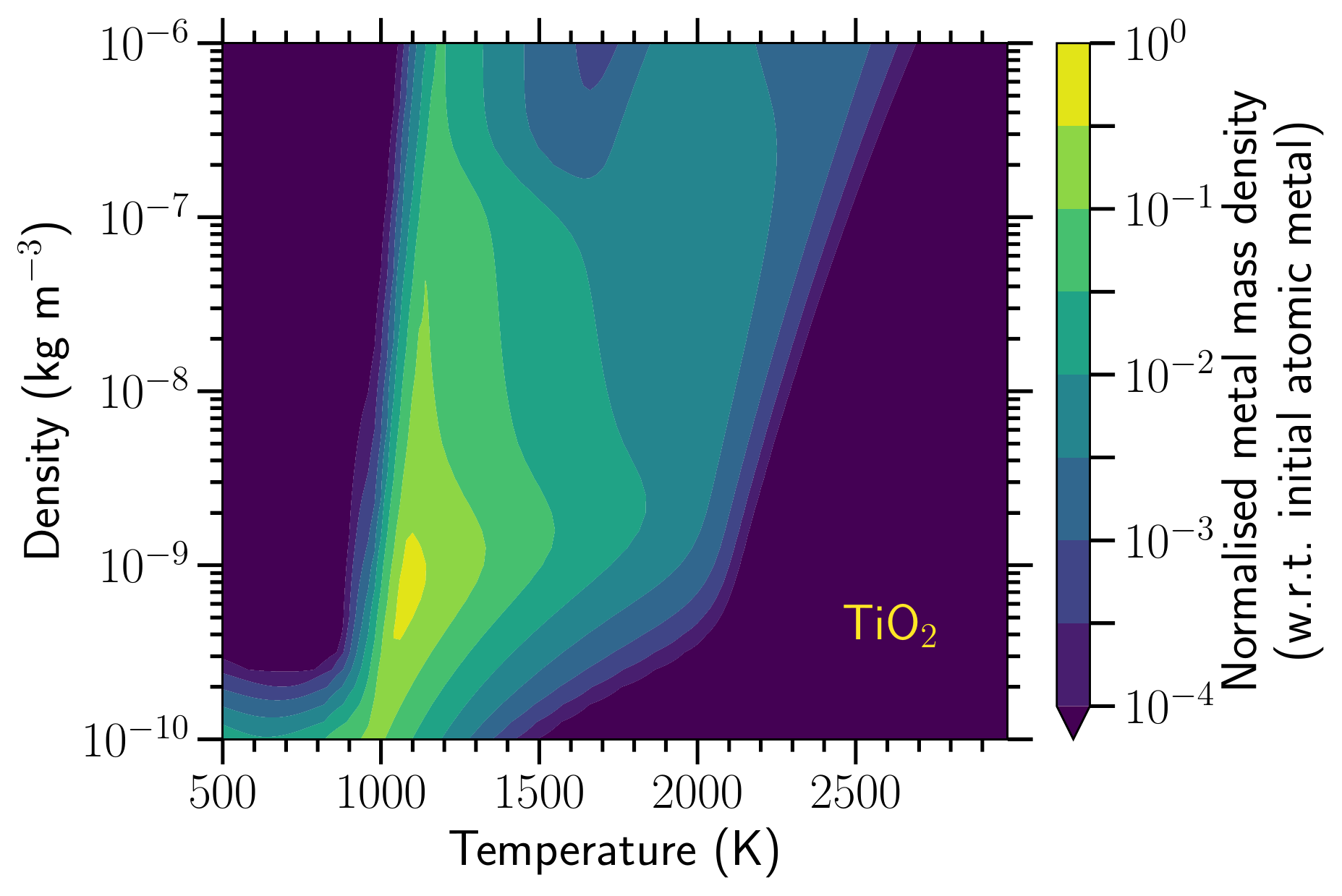}
        \includegraphics[width=0.32\textwidth]{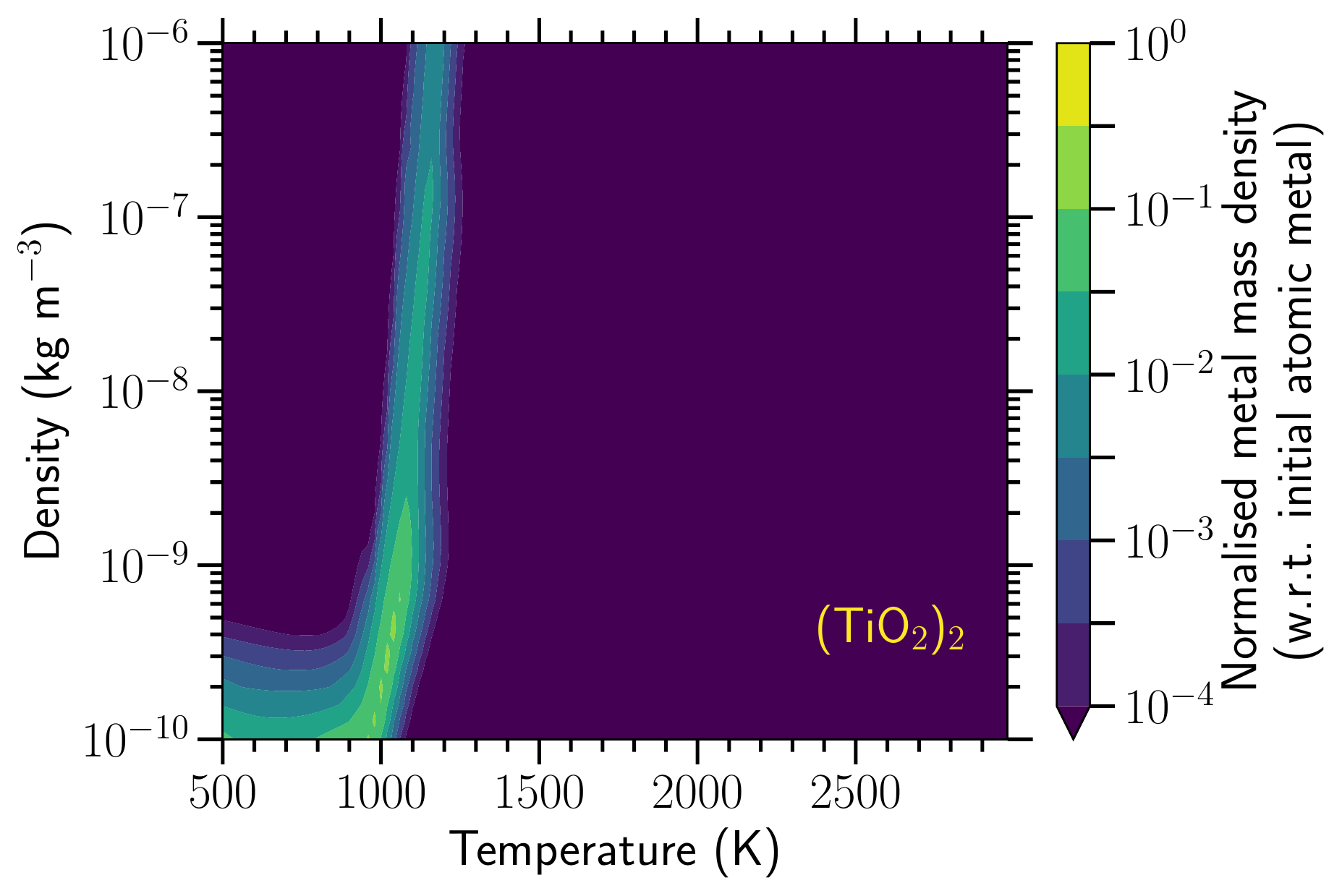}
        \includegraphics[width=0.32\textwidth]{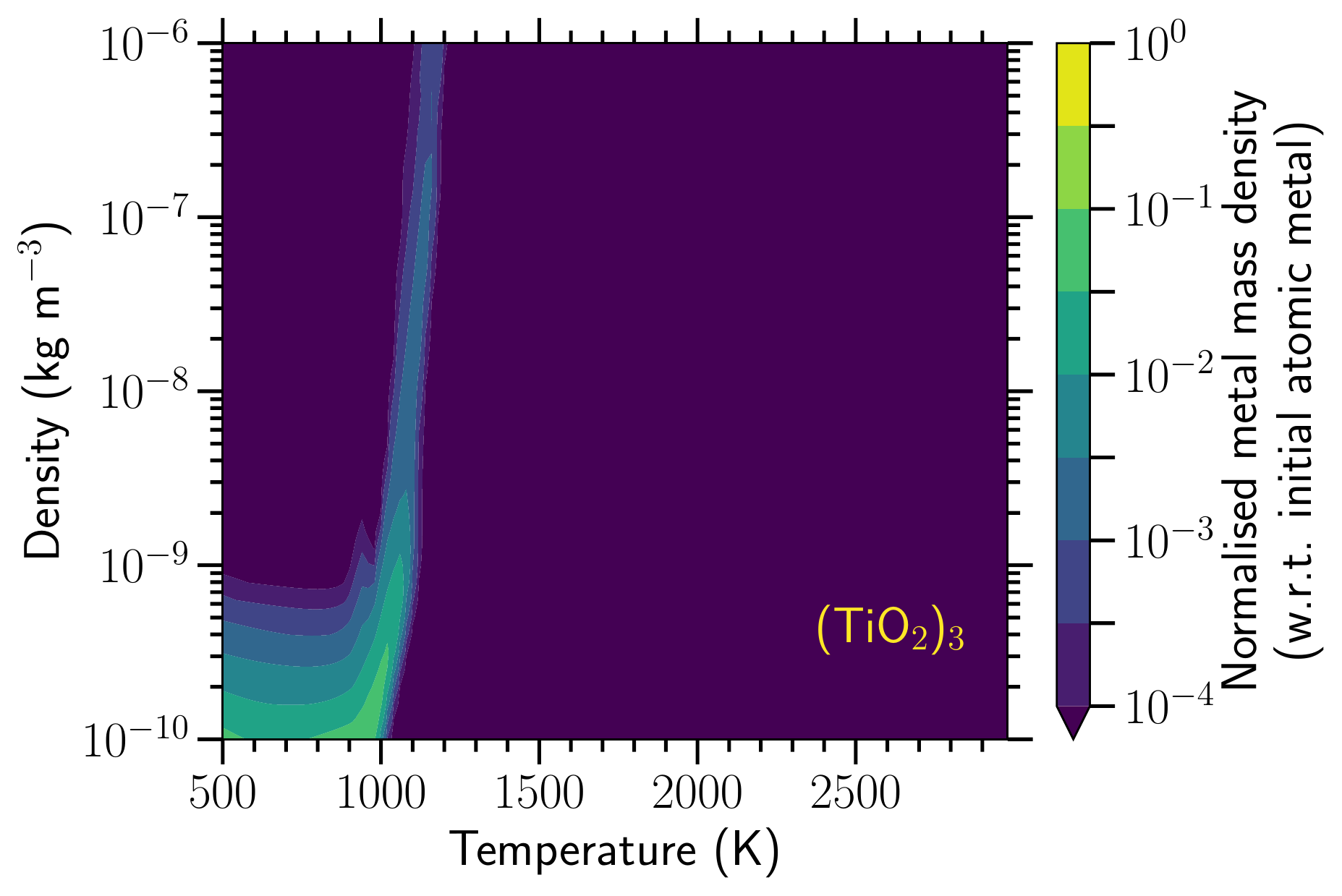}
        \includegraphics[width=0.32\textwidth]{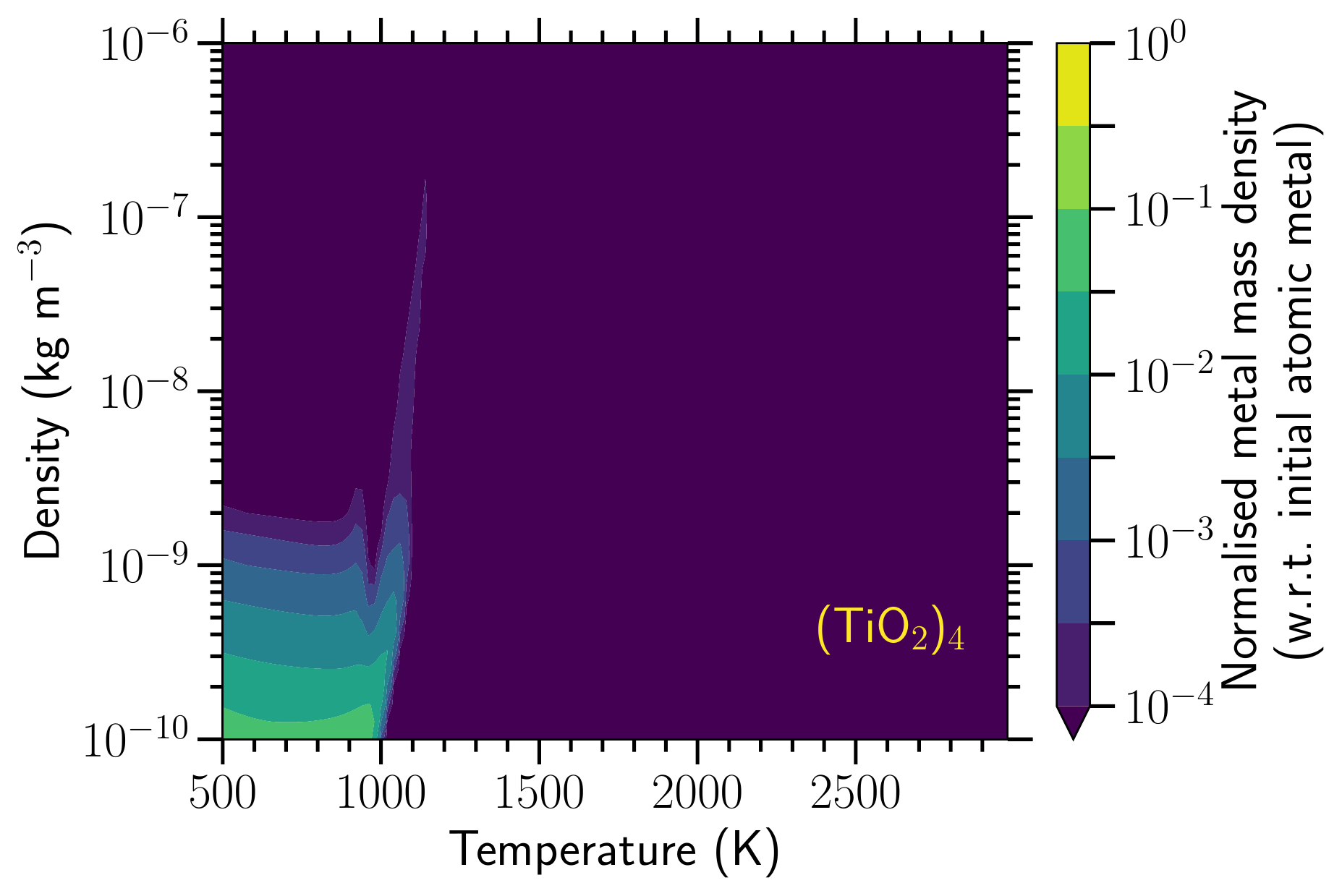}
        \includegraphics[width=0.32\textwidth]{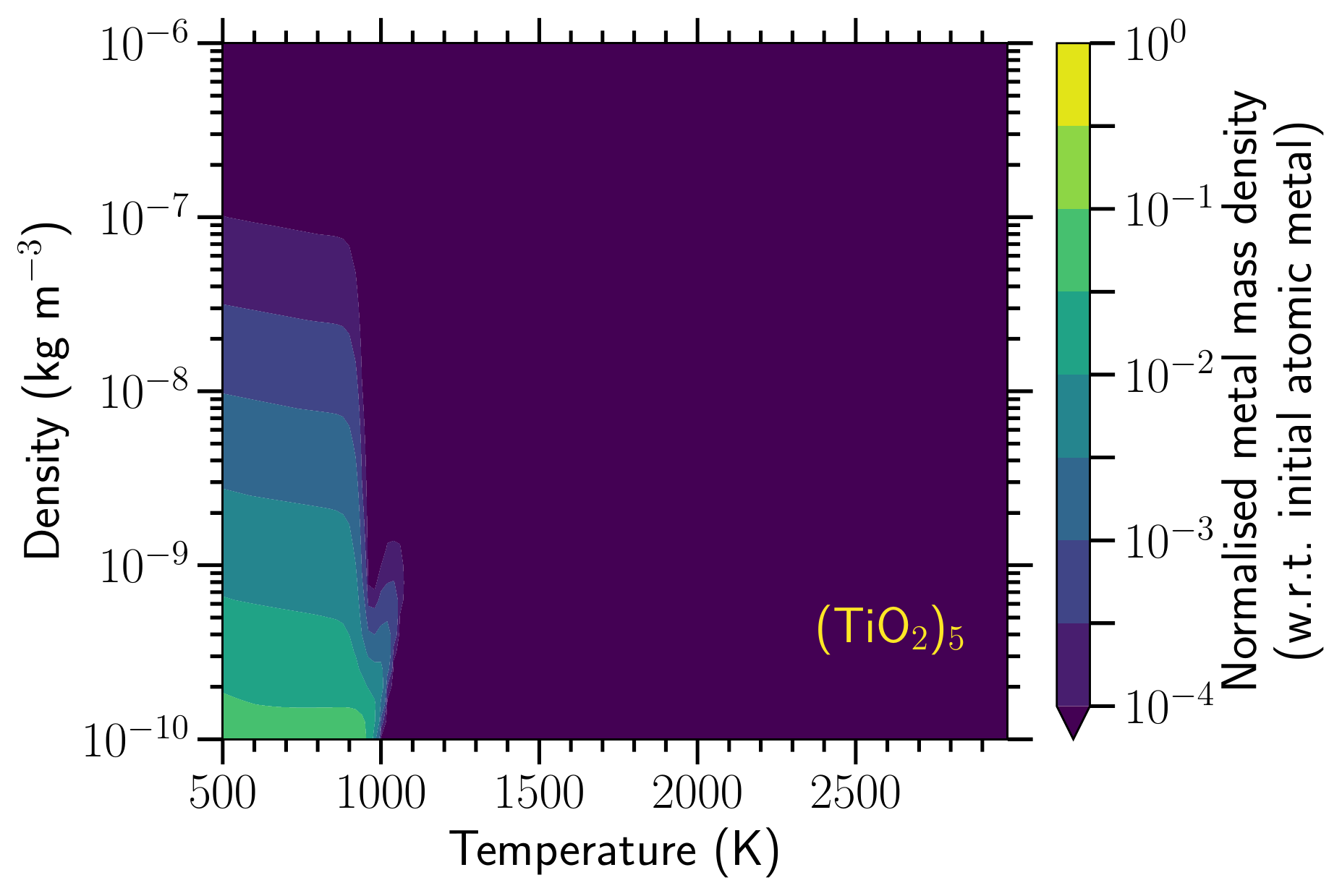}
        \includegraphics[width=0.32\textwidth]{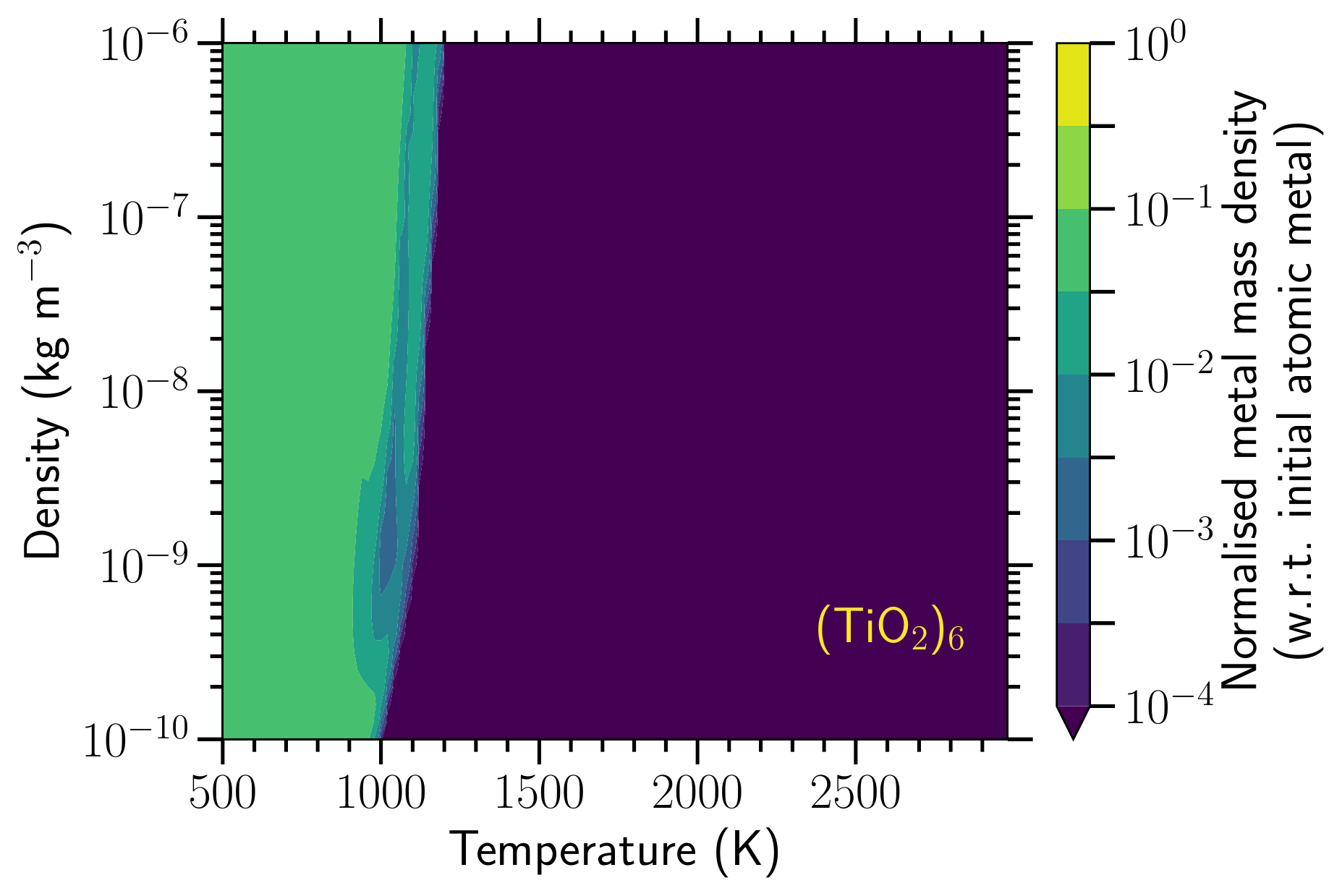}
        \includegraphics[width=0.32\textwidth]{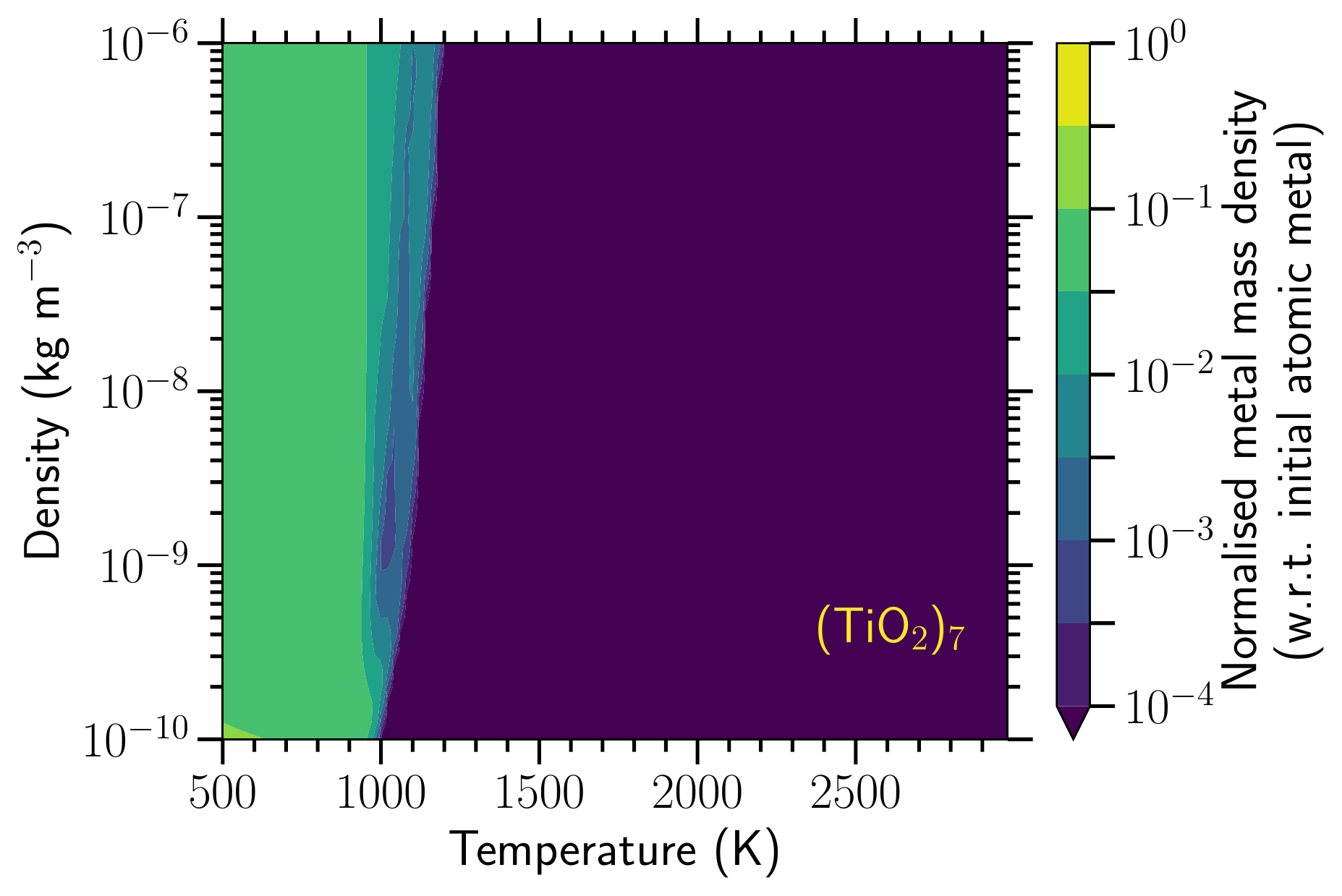}
        \includegraphics[width=0.32\textwidth]{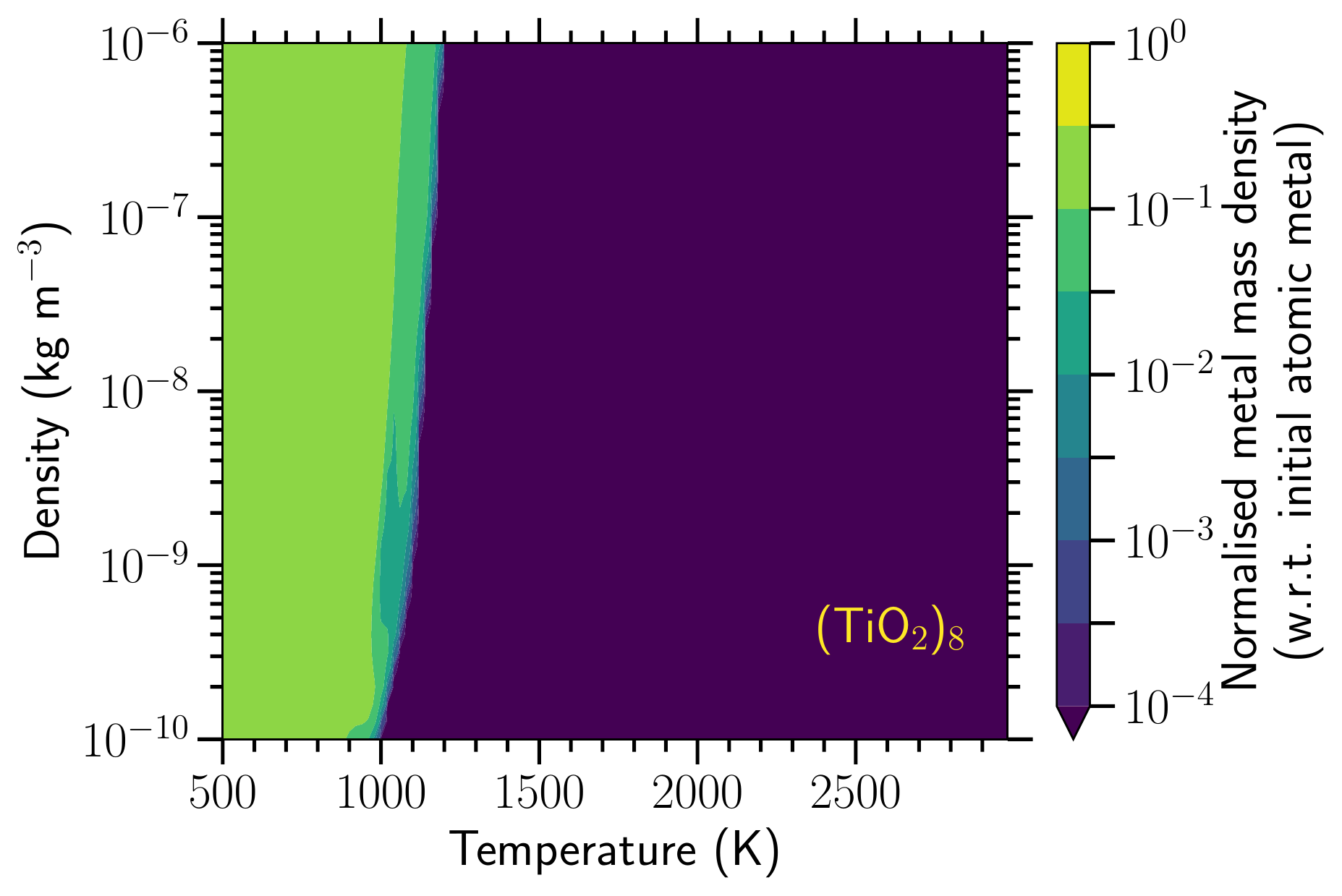}
        \includegraphics[width=0.32\textwidth]{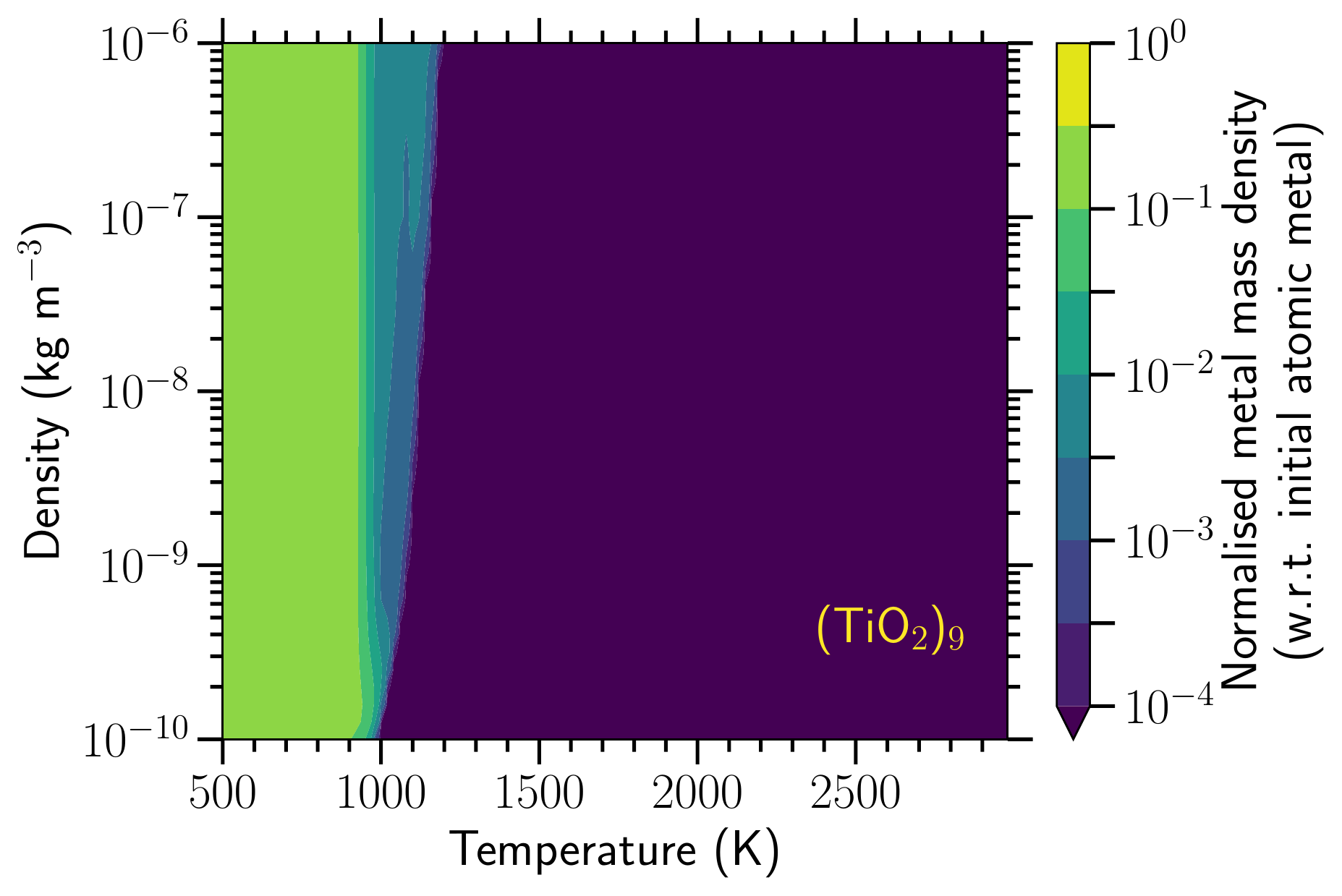}
        \includegraphics[width=0.32\textwidth]{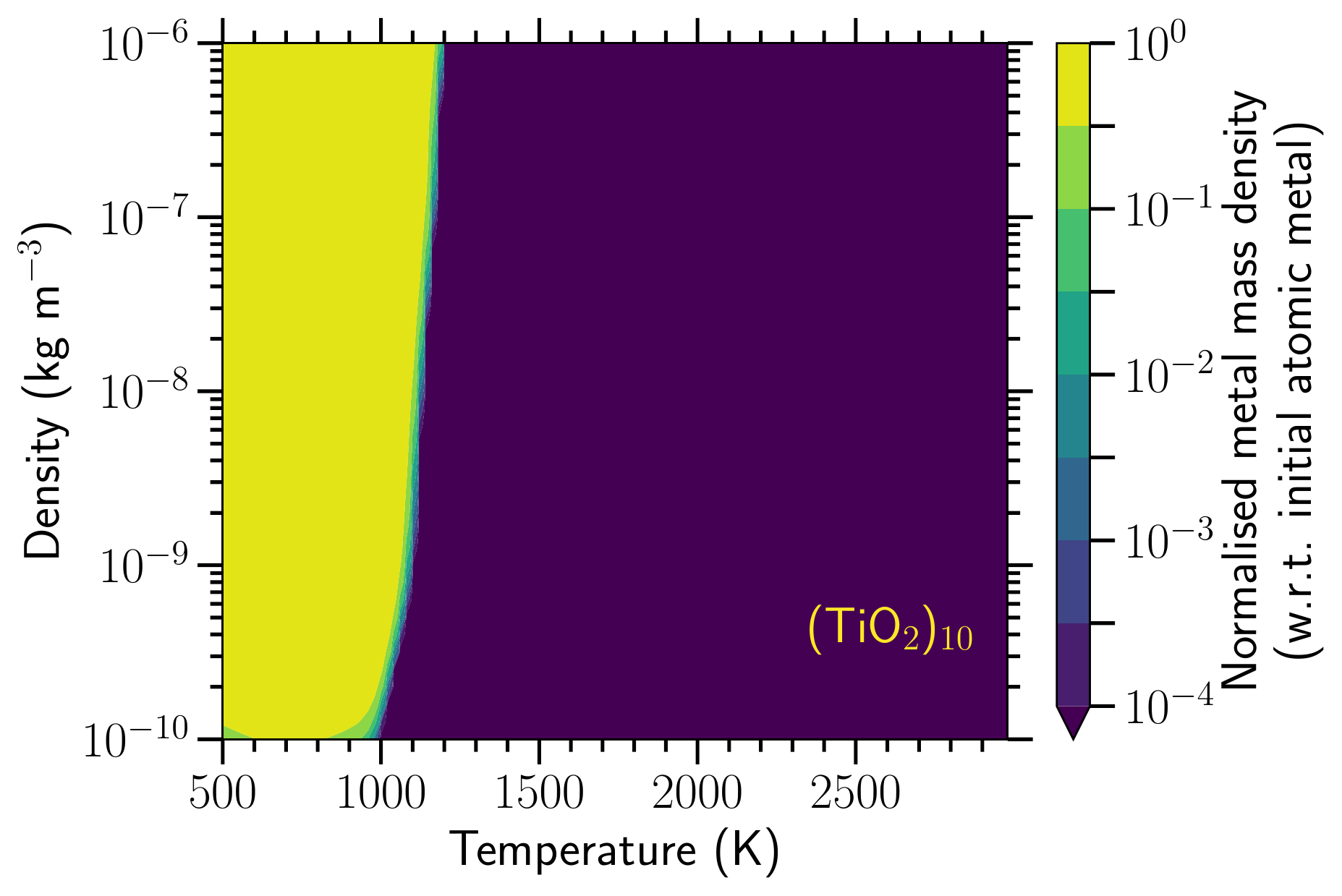}
        \includegraphics[width=0.32\textwidth]{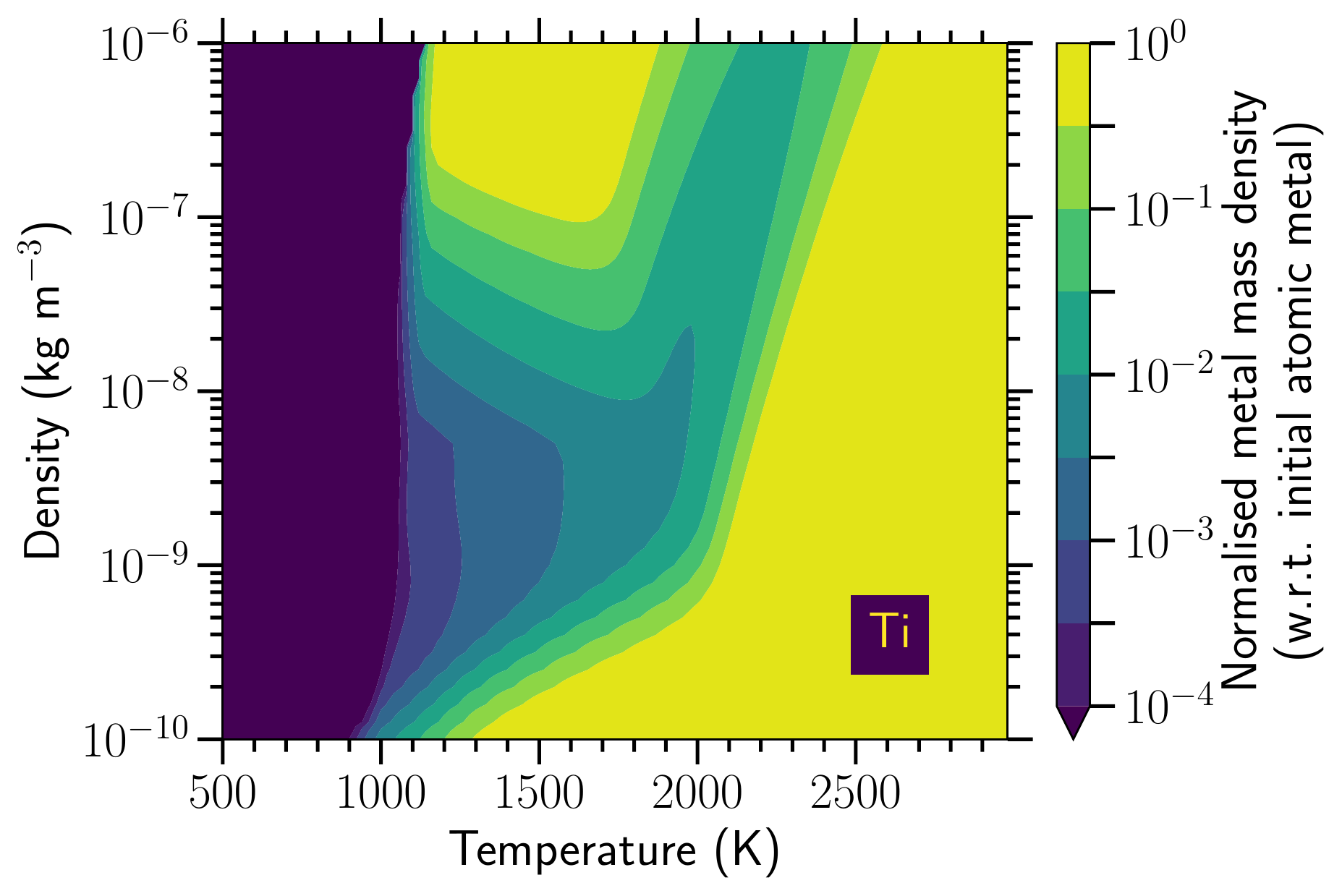}
        \includegraphics[width=0.32\textwidth]{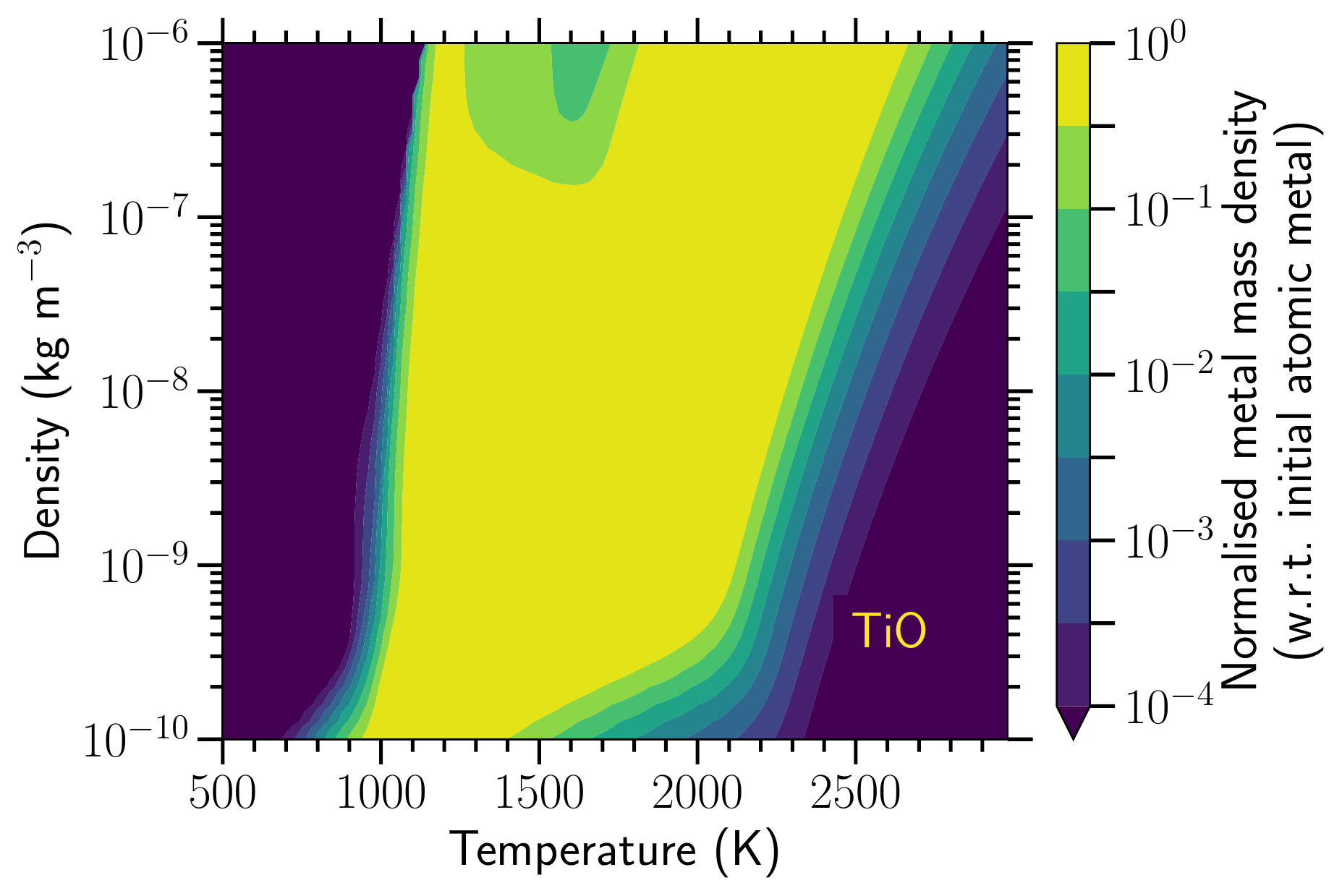}
        \end{flushleft}
        \caption{Overview of the normalised mass density after one year of all \ch{Ti}-bearing species for the comprehensive chemical nucleation model using the polymer nucleation description.}
        \label{fig:full_ntw_Ti-molecules_norm_same_scale}
    \end{figure*}

\begin{figure*}
        \begin{flushleft}
        \includegraphics[width=0.32\textwidth]{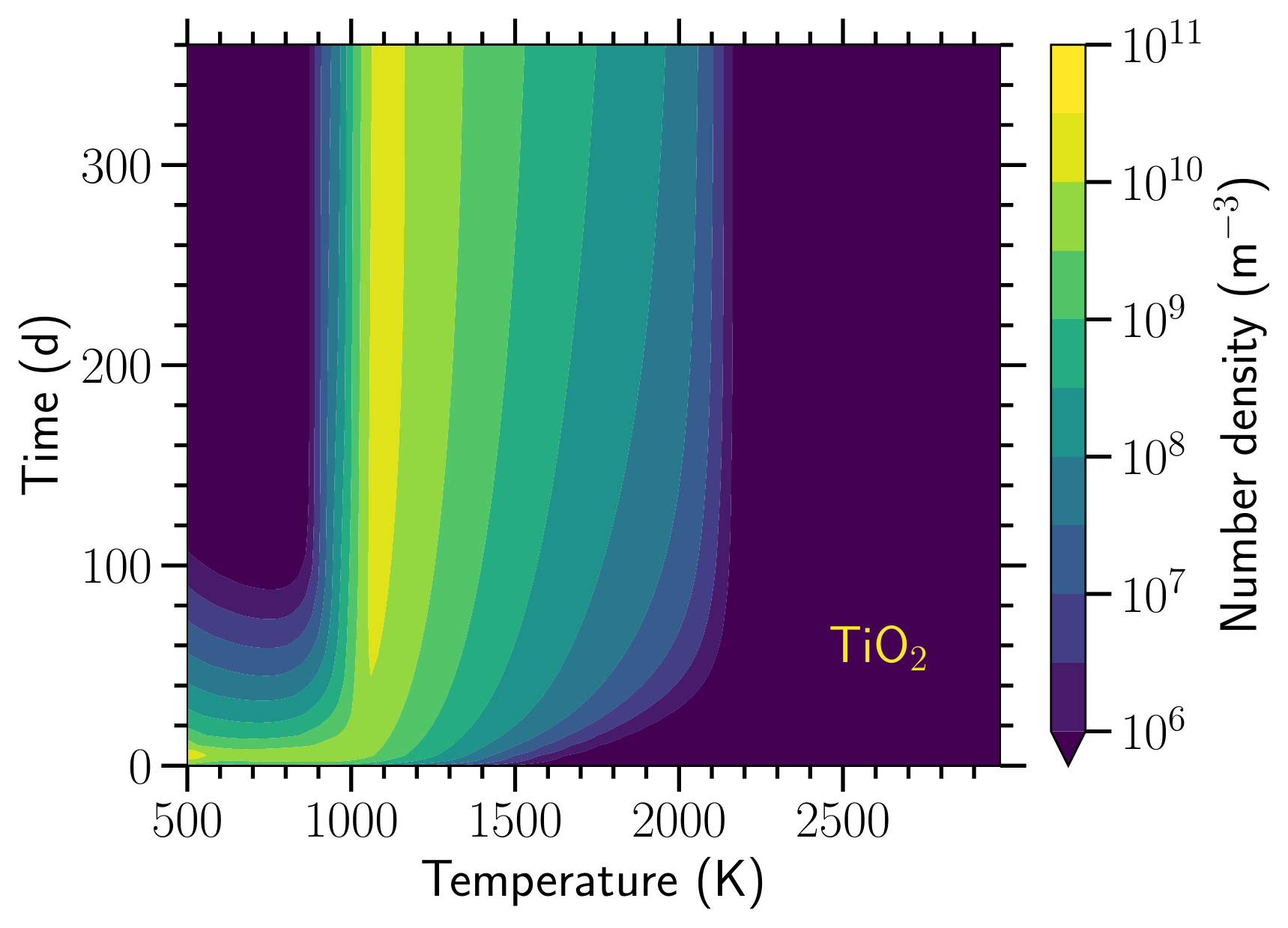}
        \includegraphics[width=0.32\textwidth]{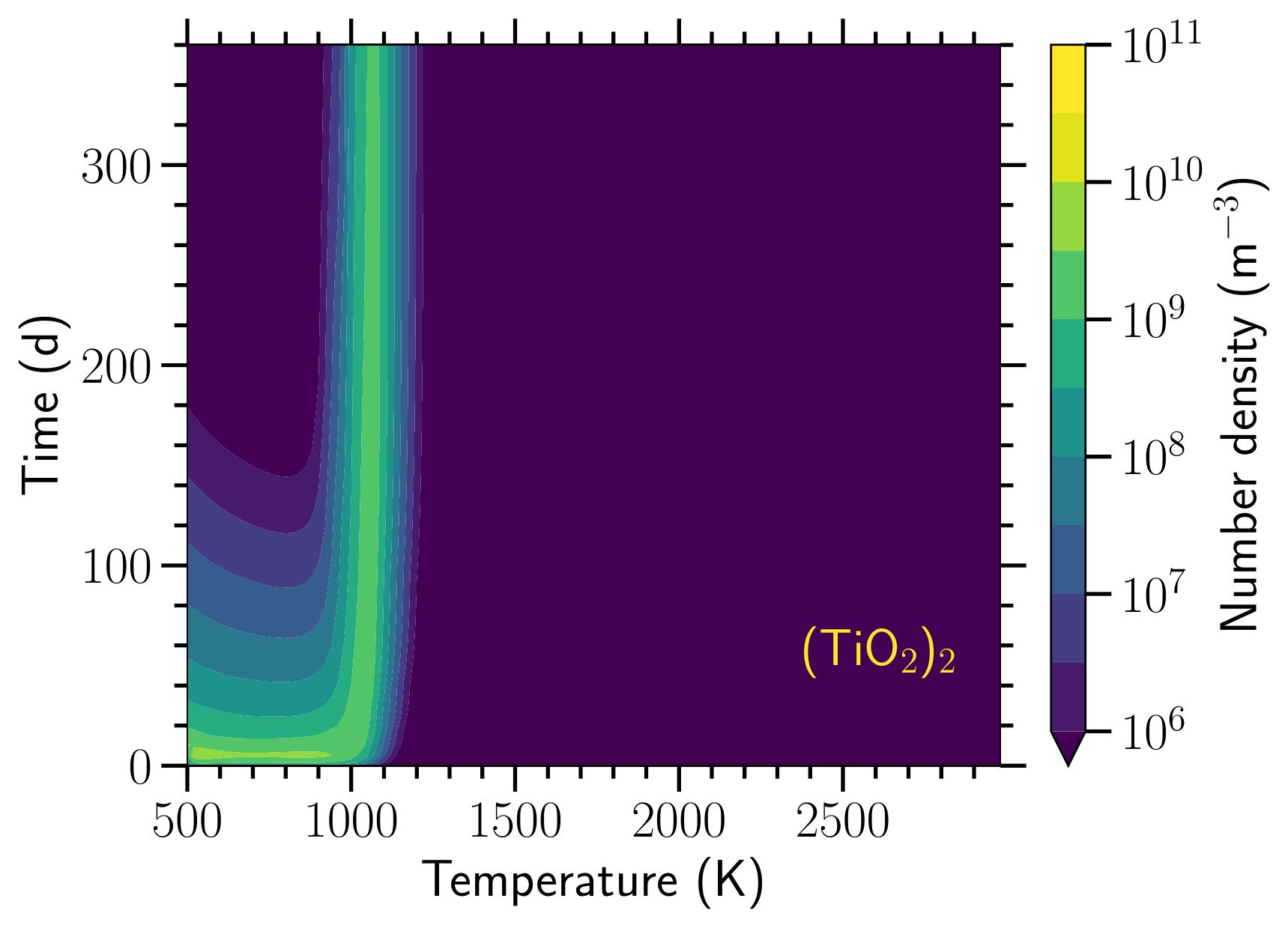}
        \includegraphics[width=0.32\textwidth]{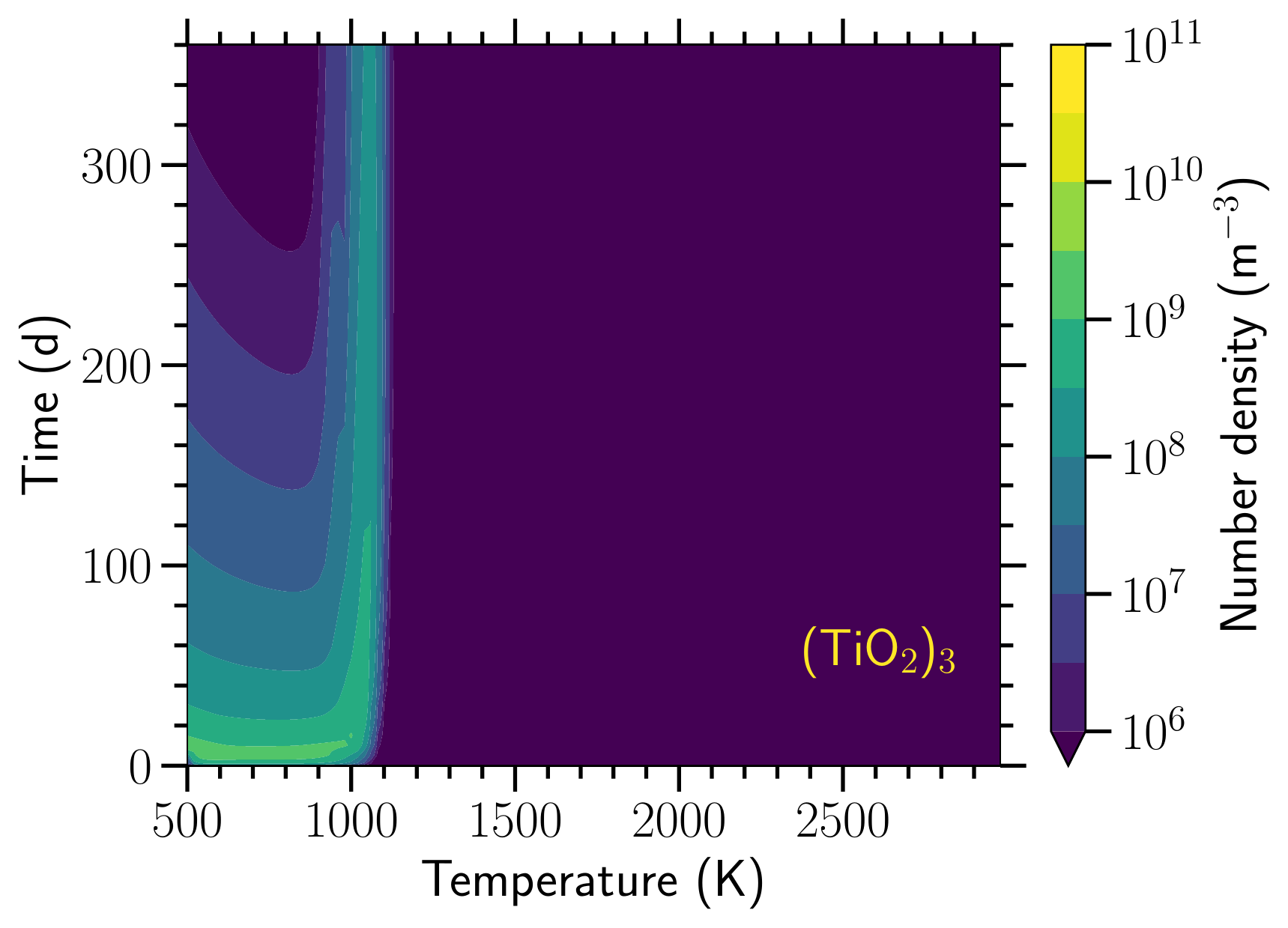}
        \includegraphics[width=0.32\textwidth]{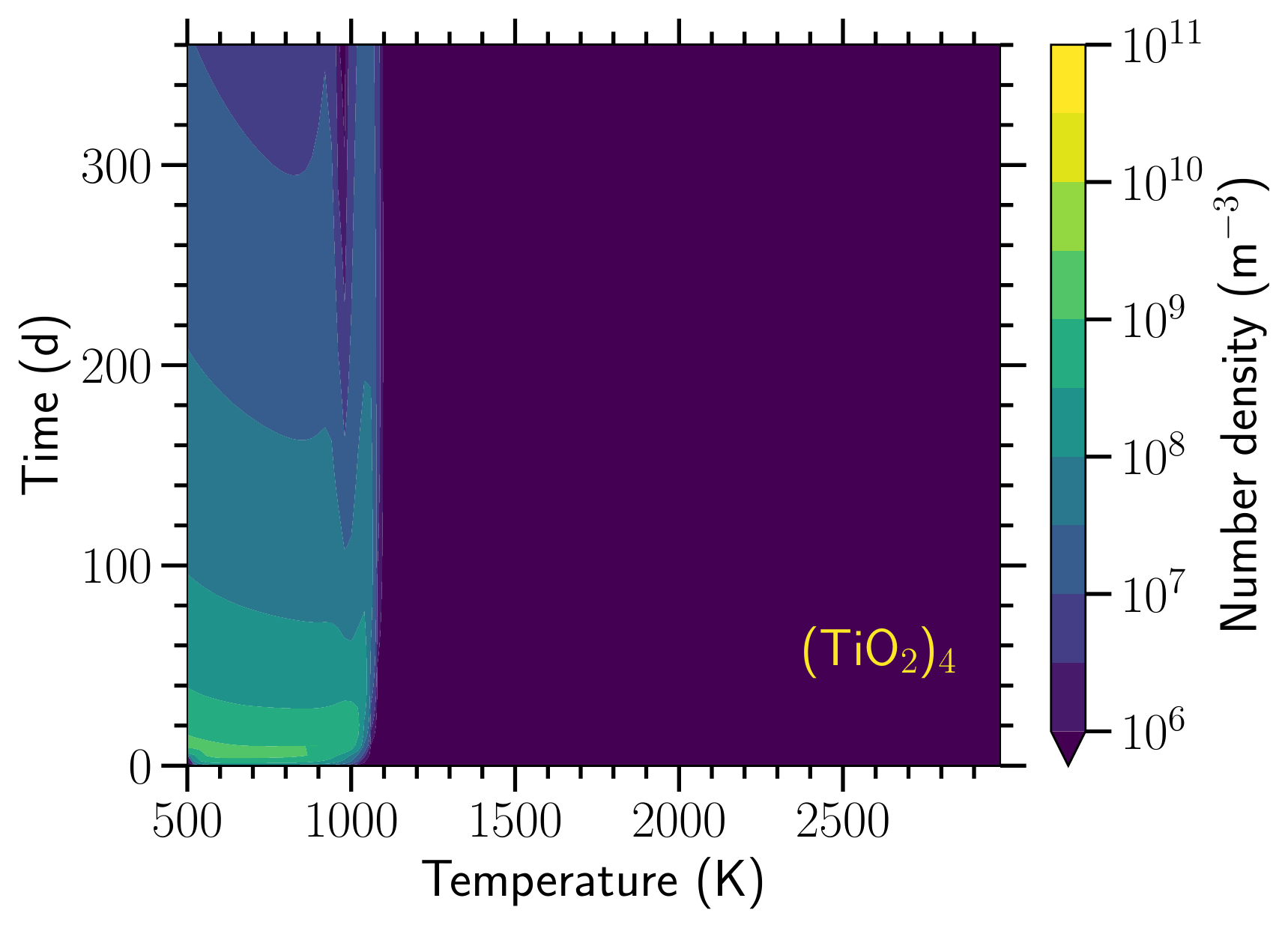}
        \includegraphics[width=0.32\textwidth]{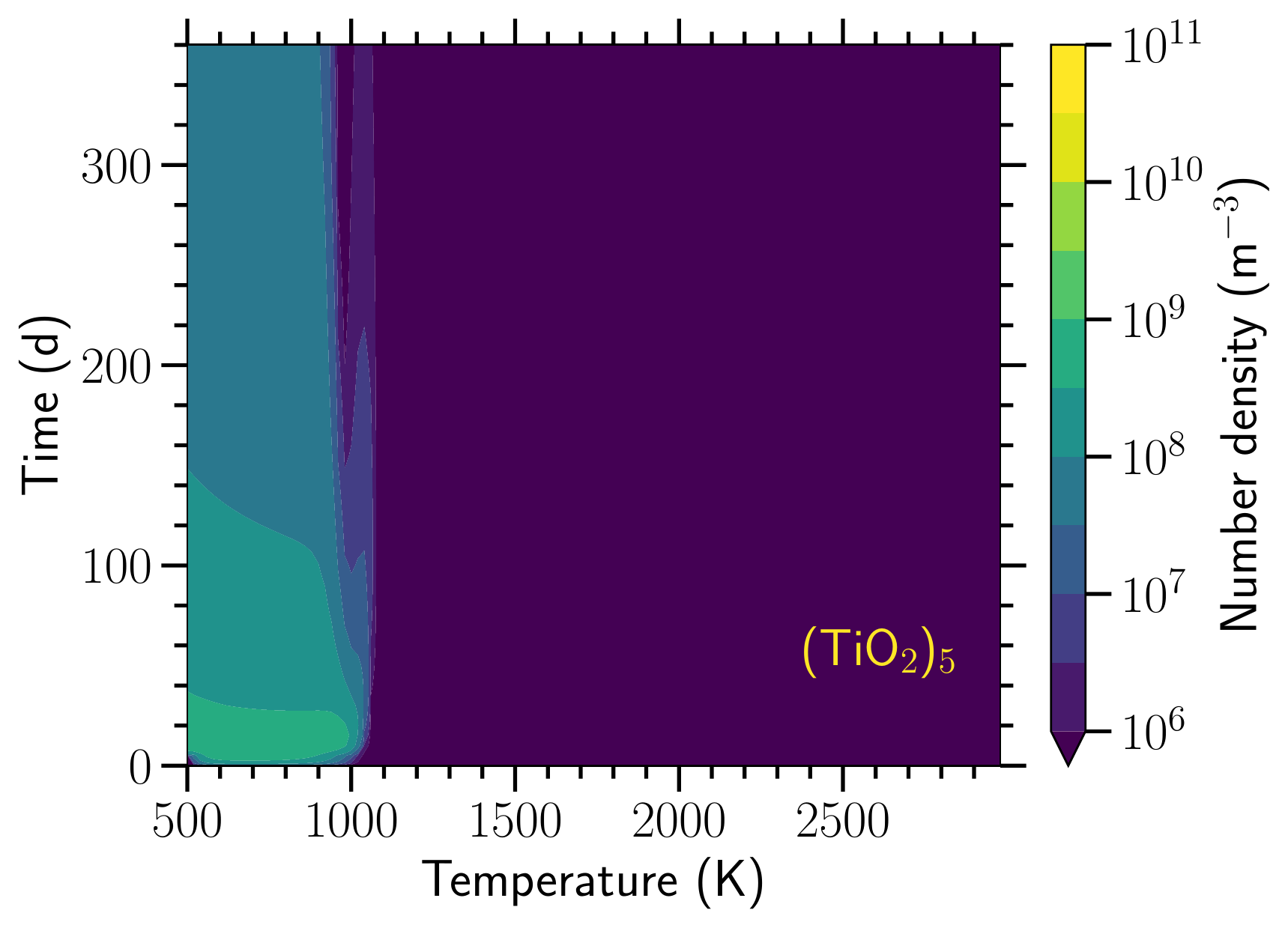}
        \includegraphics[width=0.32\textwidth]{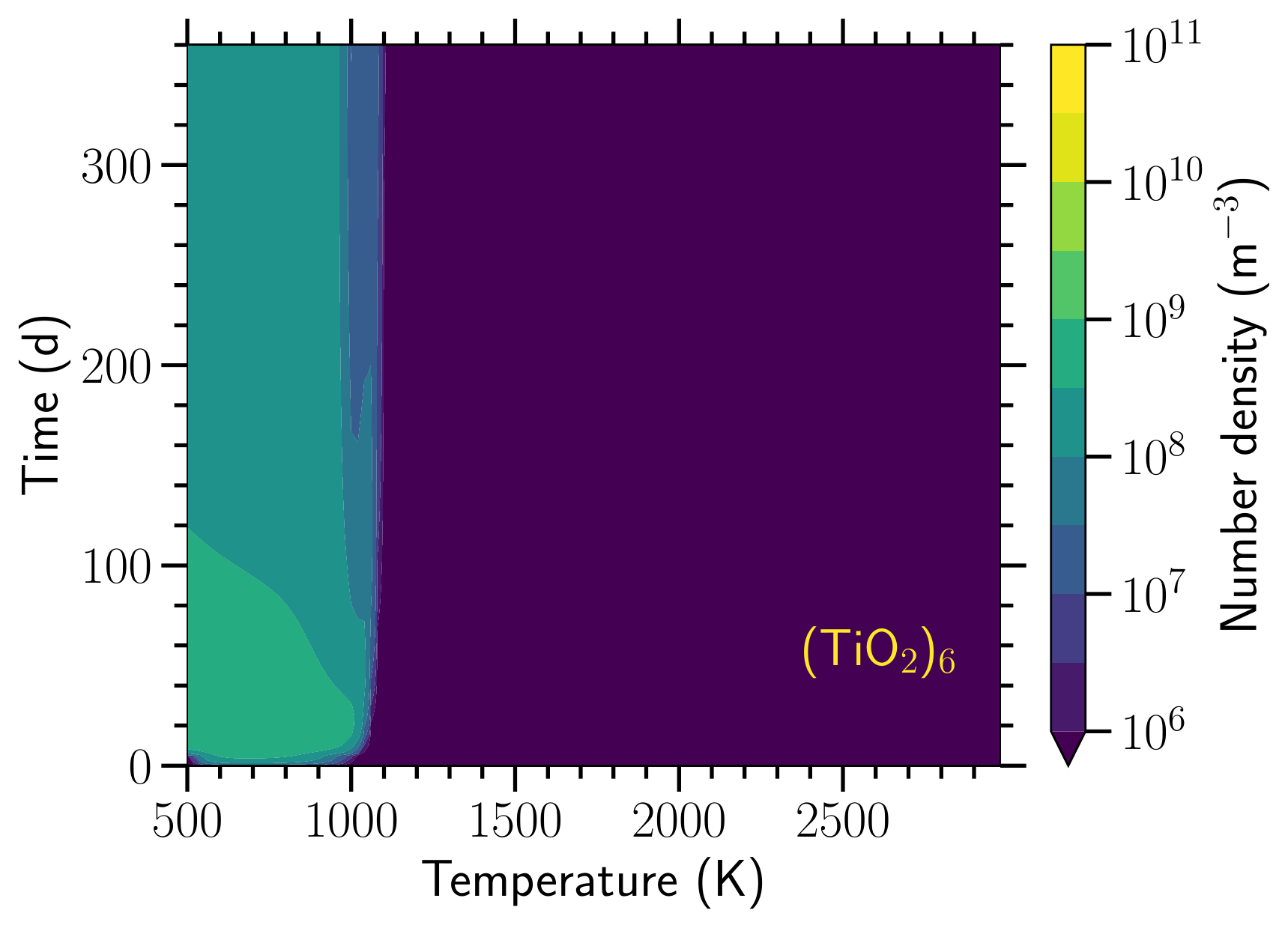}
        \includegraphics[width=0.32\textwidth]{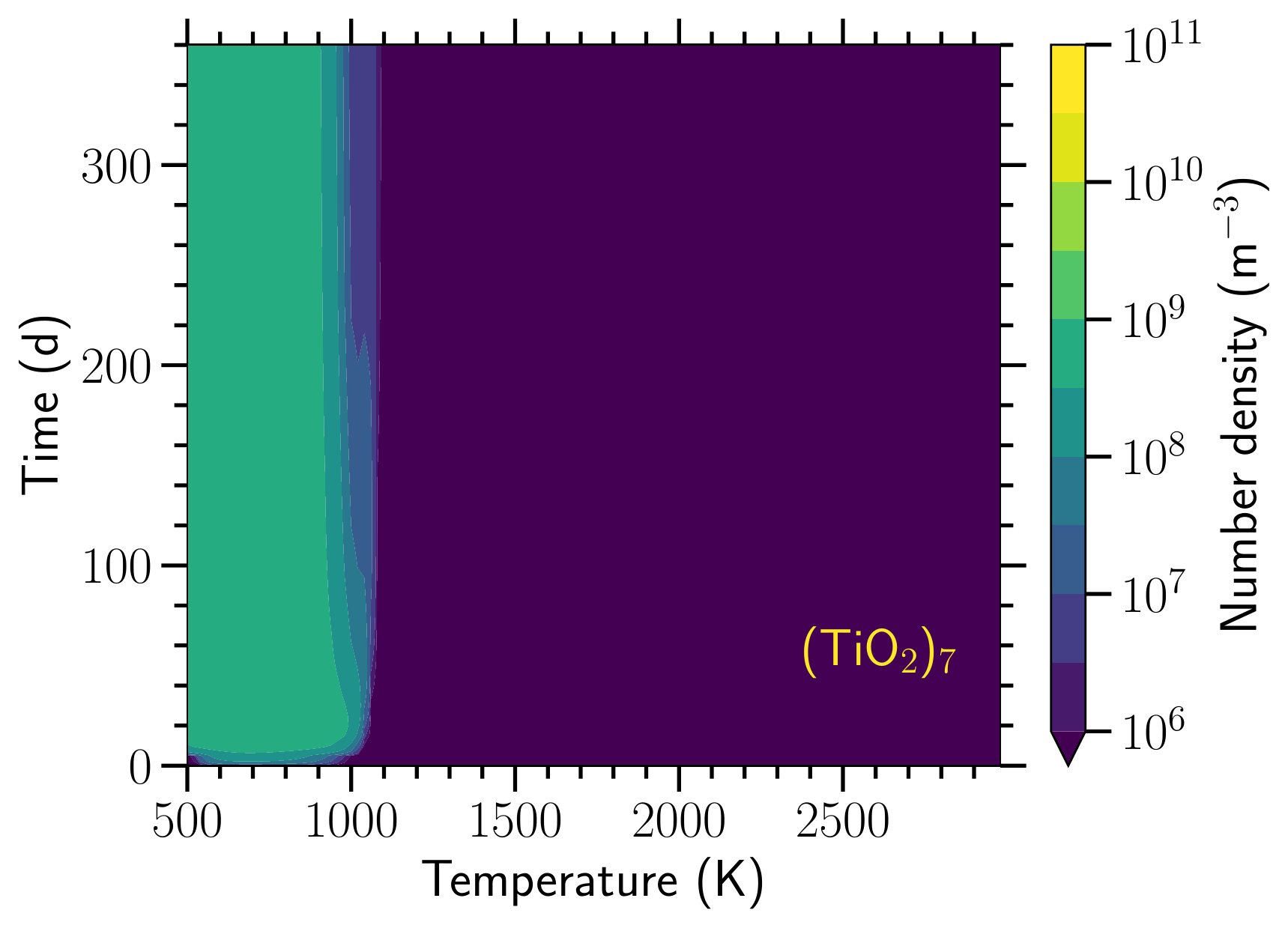}
        \includegraphics[width=0.32\textwidth]{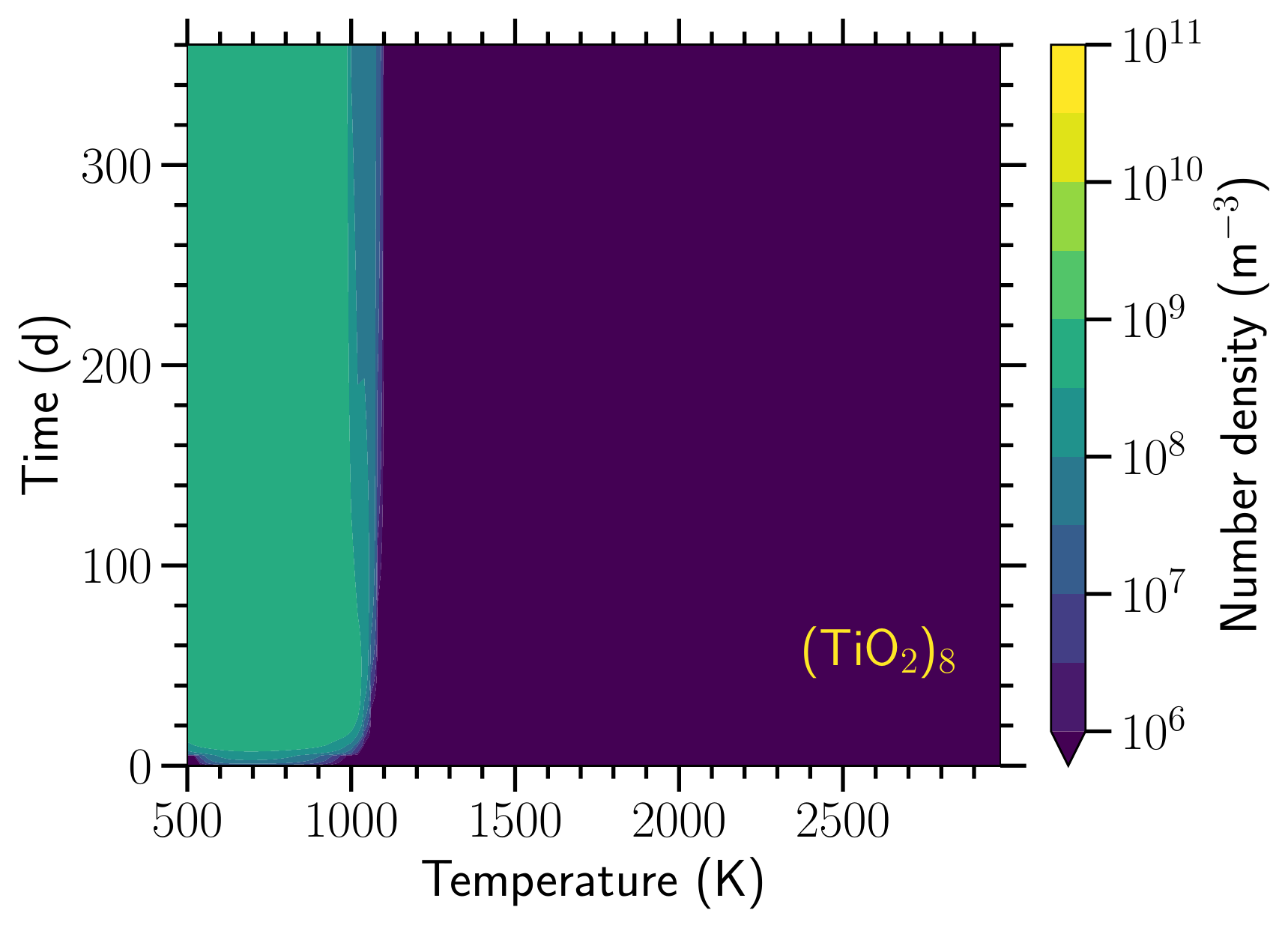}
        \includegraphics[width=0.32\textwidth]{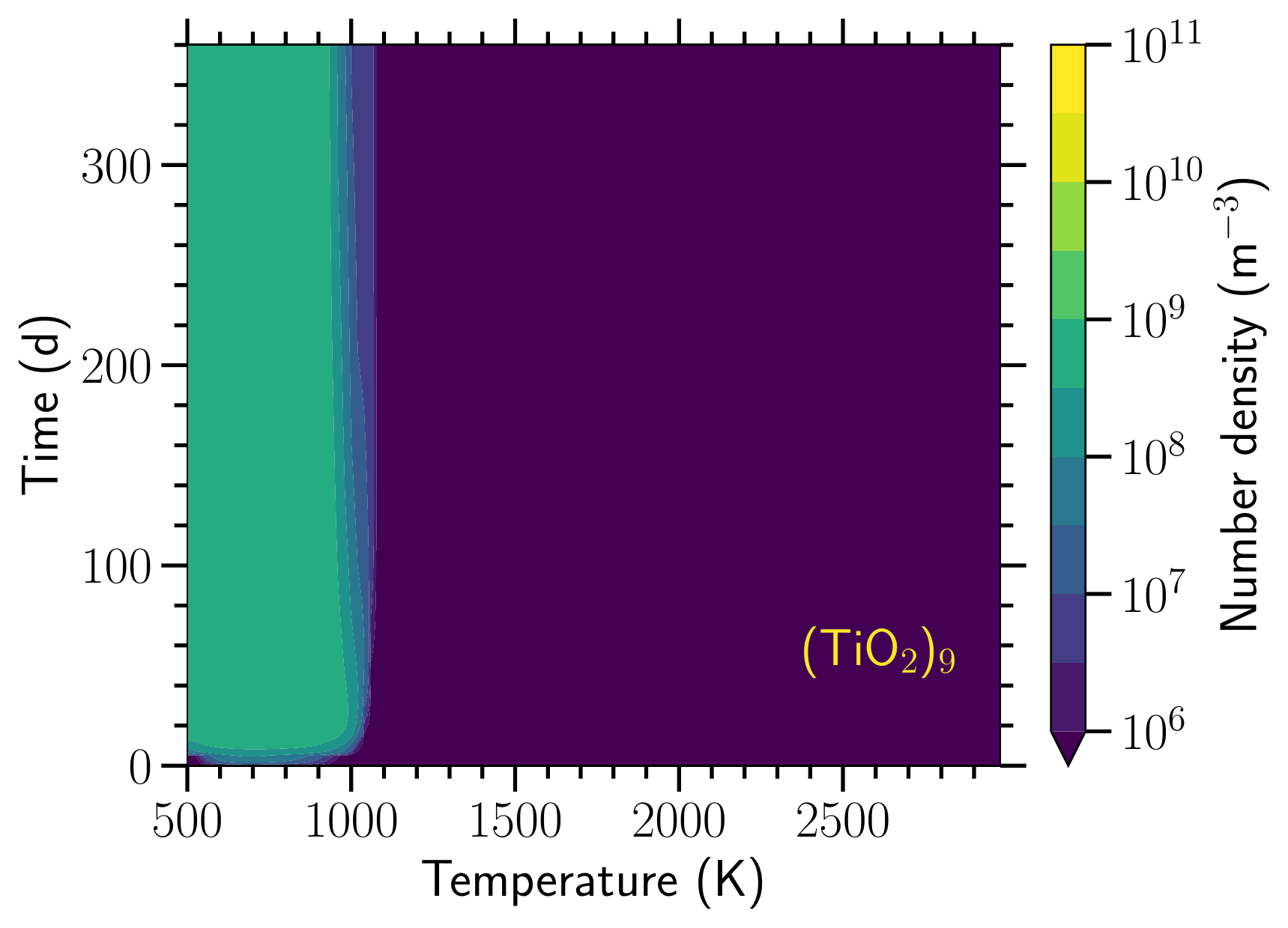}
        \includegraphics[width=0.32\textwidth]{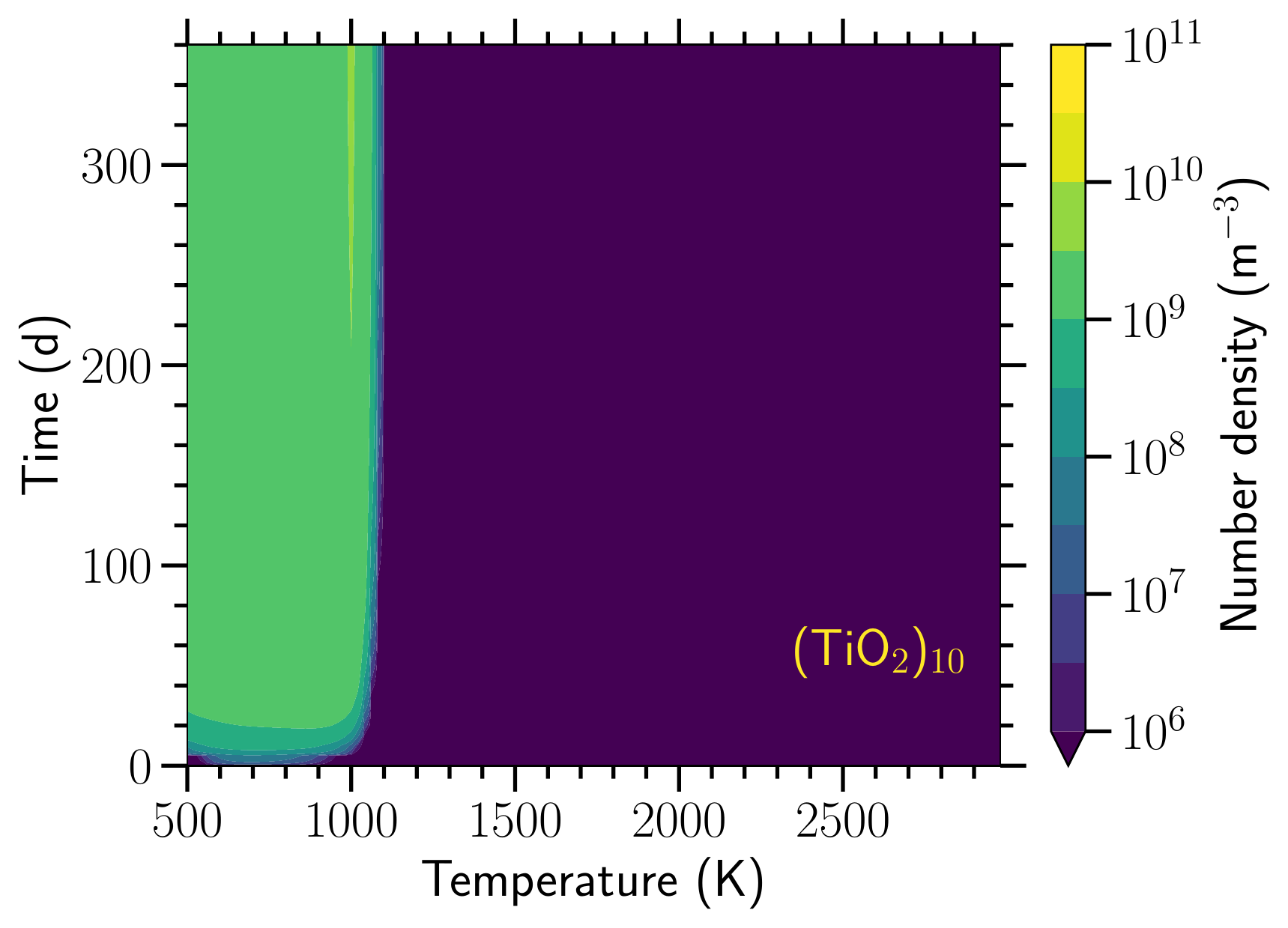}
        \includegraphics[width=0.32\textwidth]{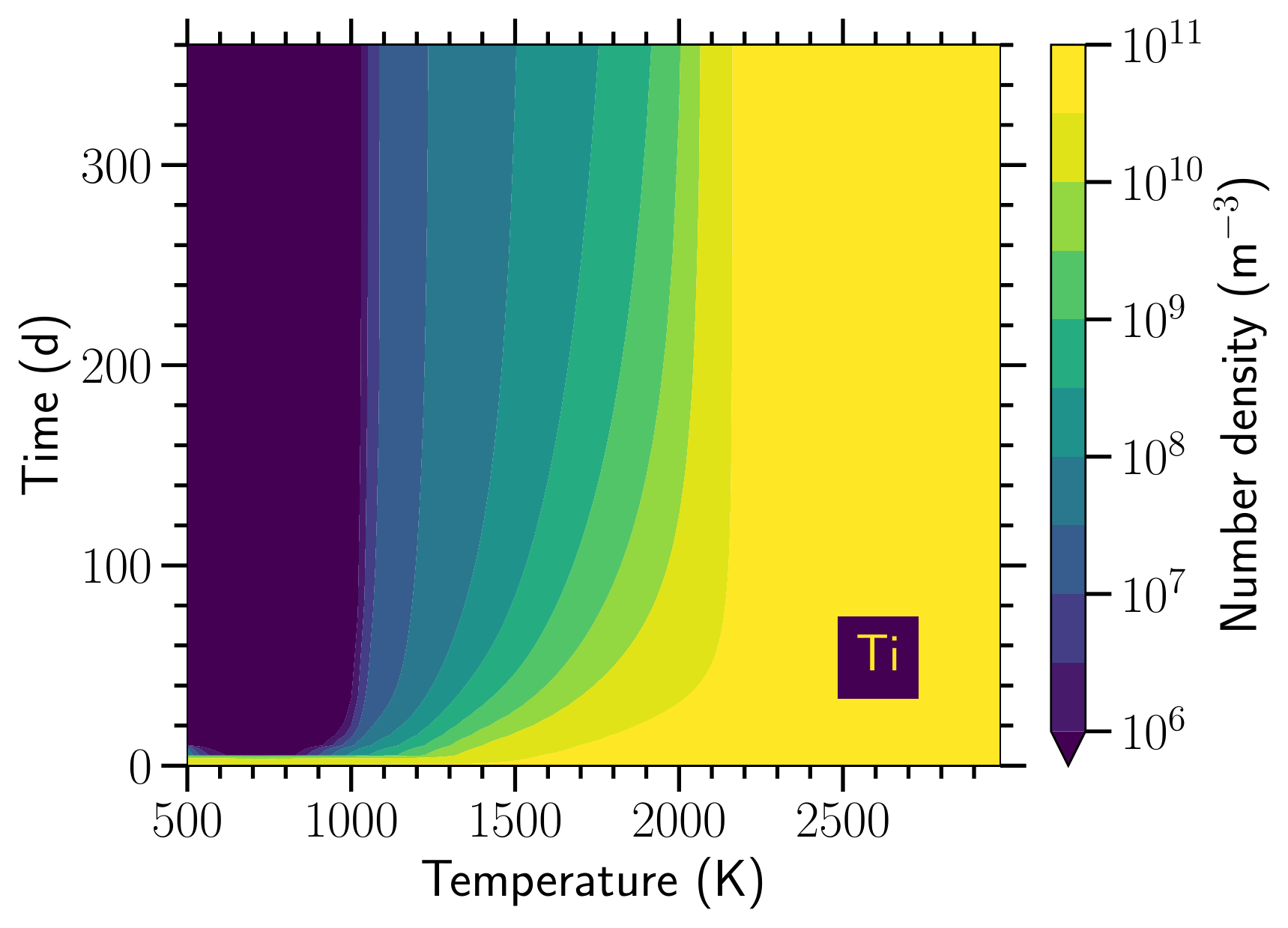}
        \includegraphics[width=0.32\textwidth]{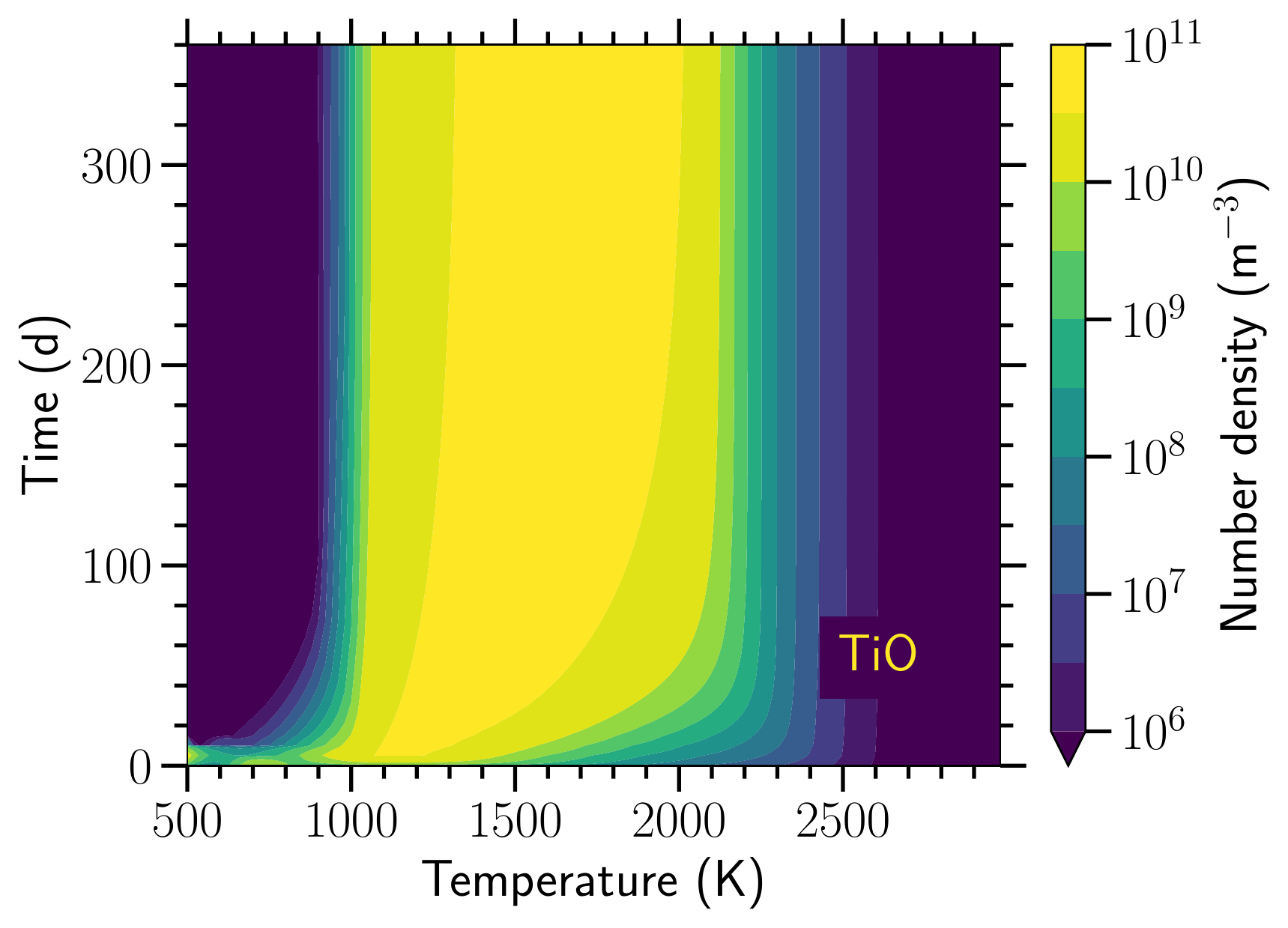}
        \end{flushleft}
        \caption{Temporal evolution of the absolute number density of all \ch{Ti}-bearing species at the benchmark total gas density $\rho=\SI{1e-9}{\kg\per\m\cubed}$ for the comprehensive chemical nucleation model using the polymer nucleation description.}
        \label{fig:full_ntw_Ti-molecules_time_evolution}
    \end{figure*}

        This section contains a more complete overview of all \ch{Ti}-bearing species results in the comprehensive chemical nucleation model using the polymer nucleation description \figs{\ref{fig:full_ntw_Ti-molecules_norm_same_scale}}{\ref{fig:full_ntw_Ti-molecules_time_evolution}}.
        
        \subsubsection{\ch{Si}-bearing species}
\begin{figure*}
        \begin{flushleft}
        \includegraphics[width=0.32\textwidth]{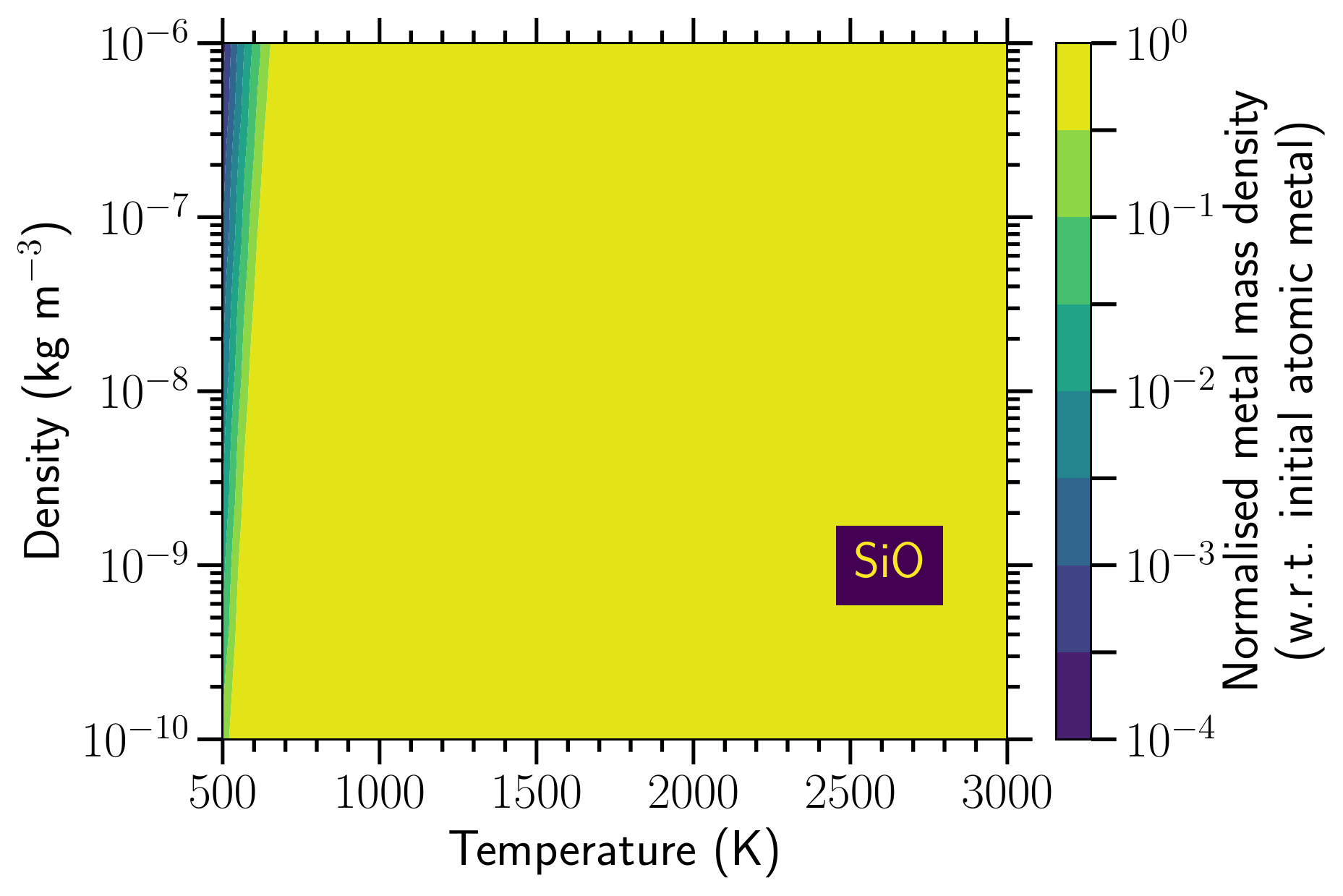}
        \includegraphics[width=0.32\textwidth]{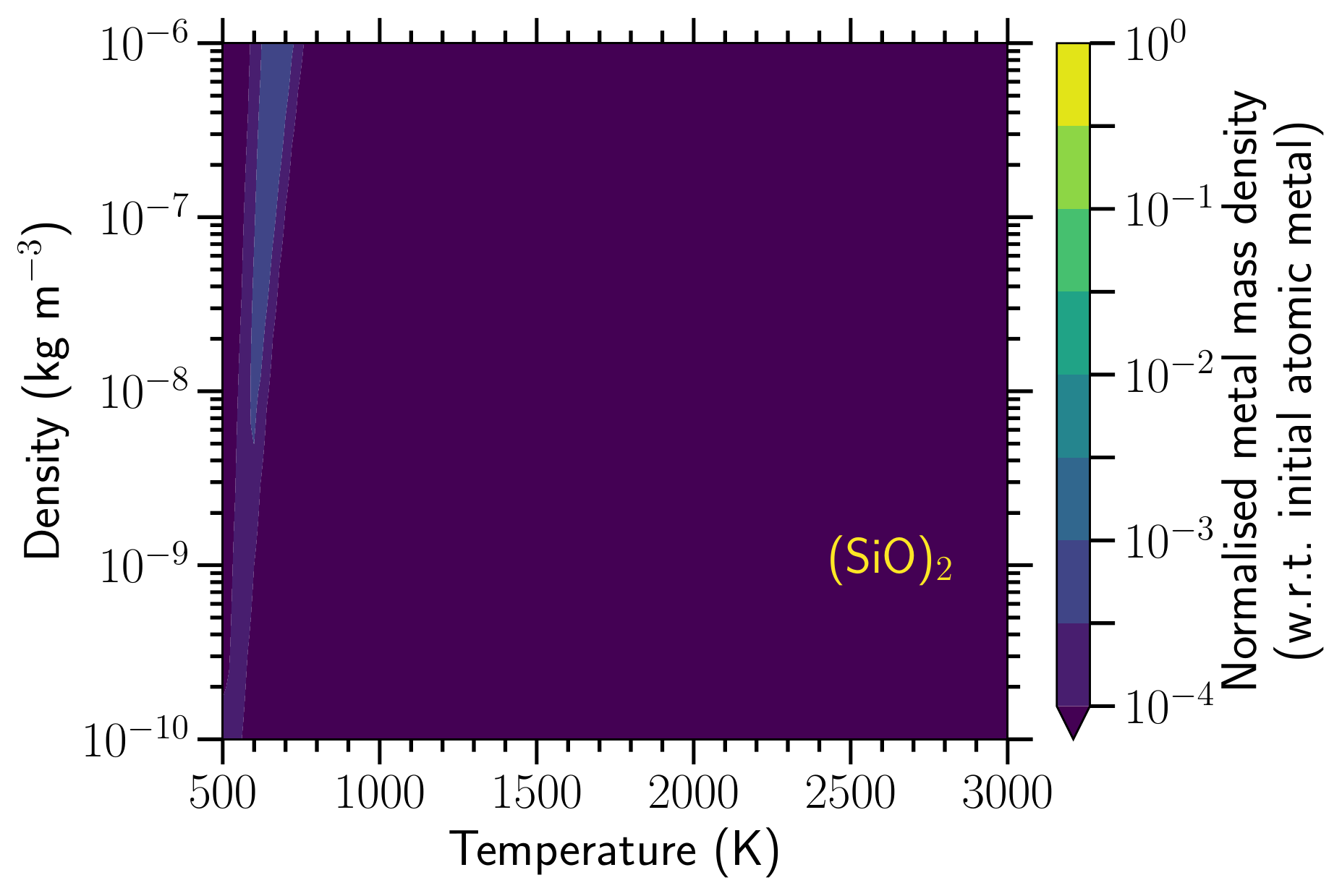}
        \includegraphics[width=0.32\textwidth]{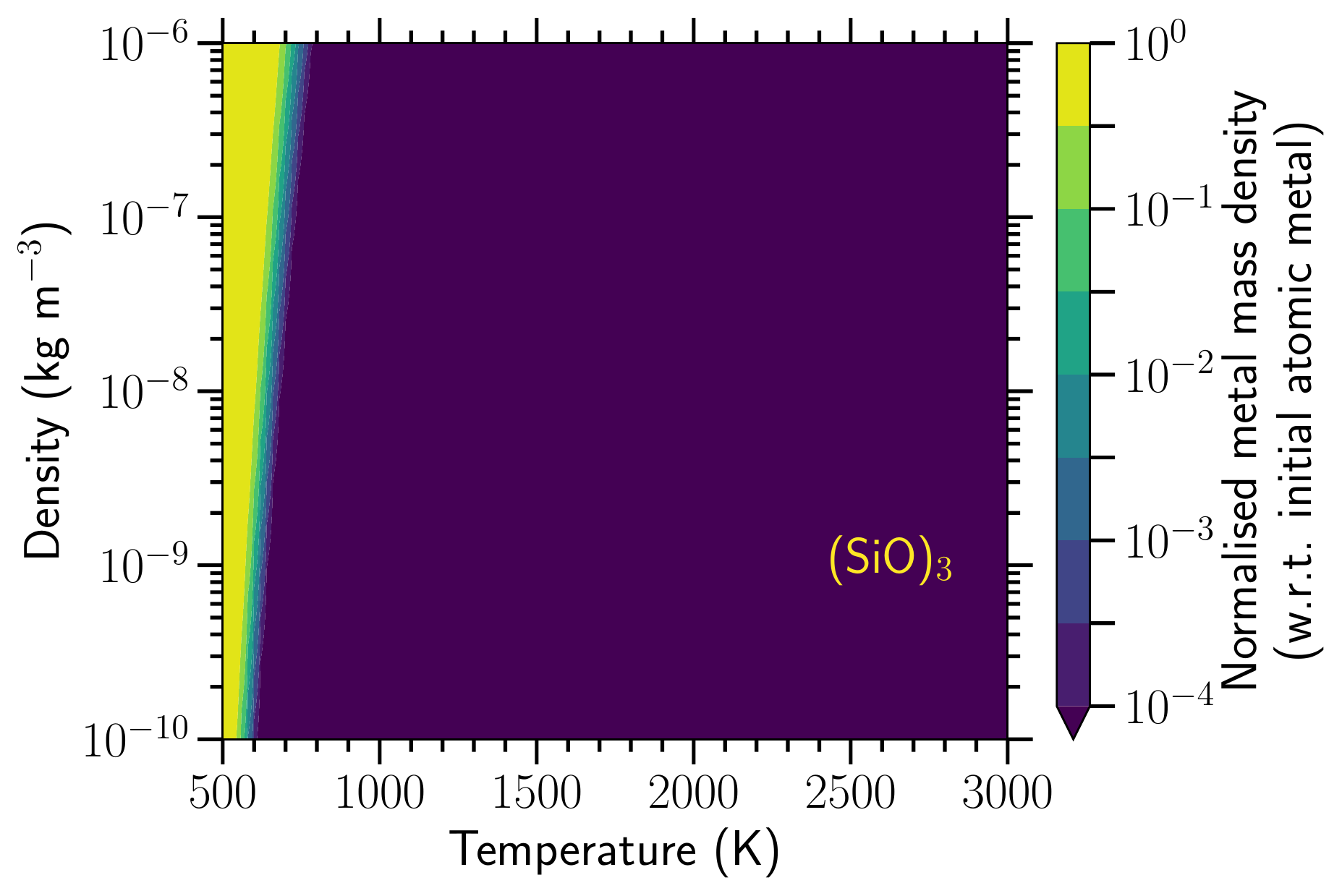}
        \includegraphics[width=0.32\textwidth]{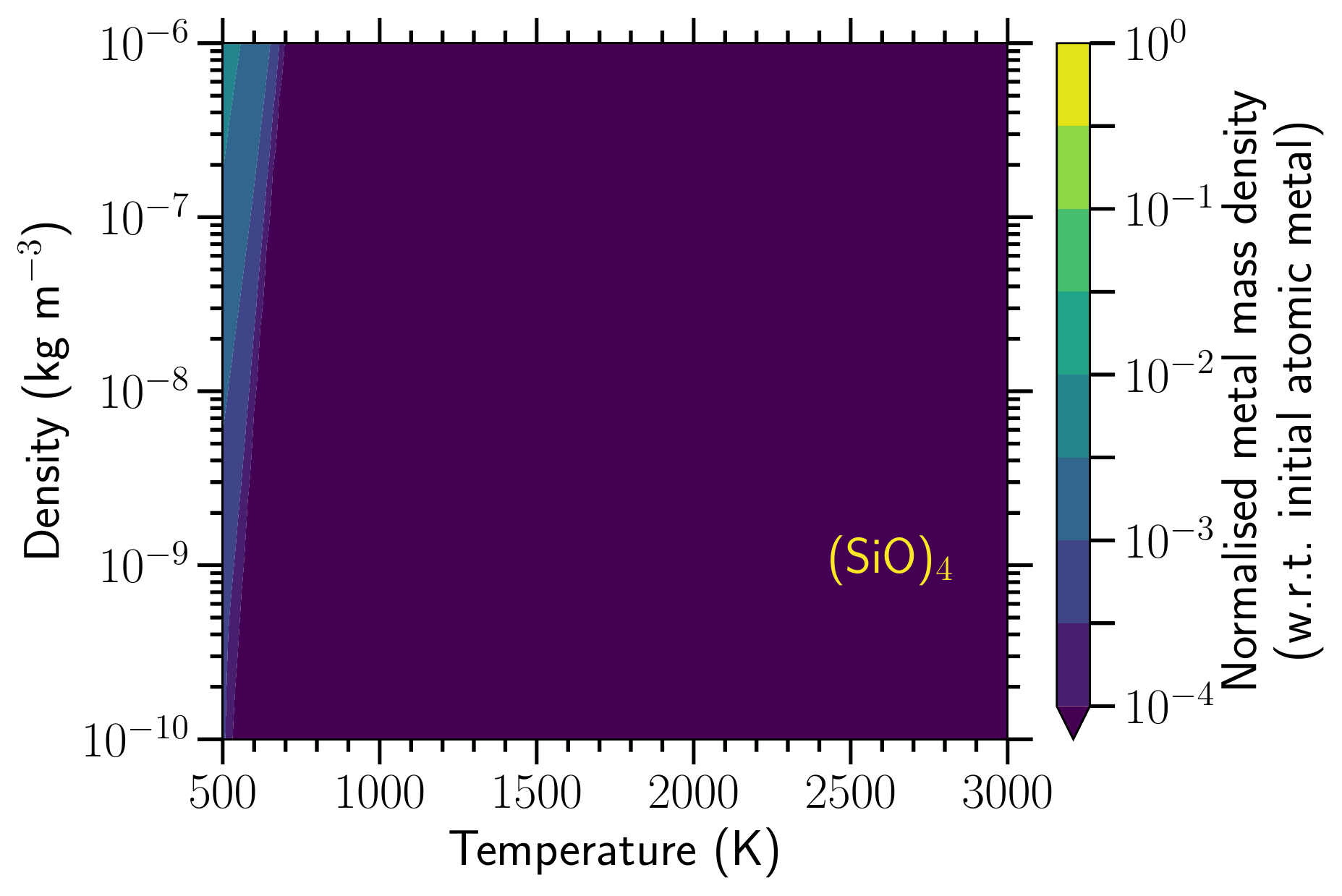}
        \includegraphics[width=0.32\textwidth]{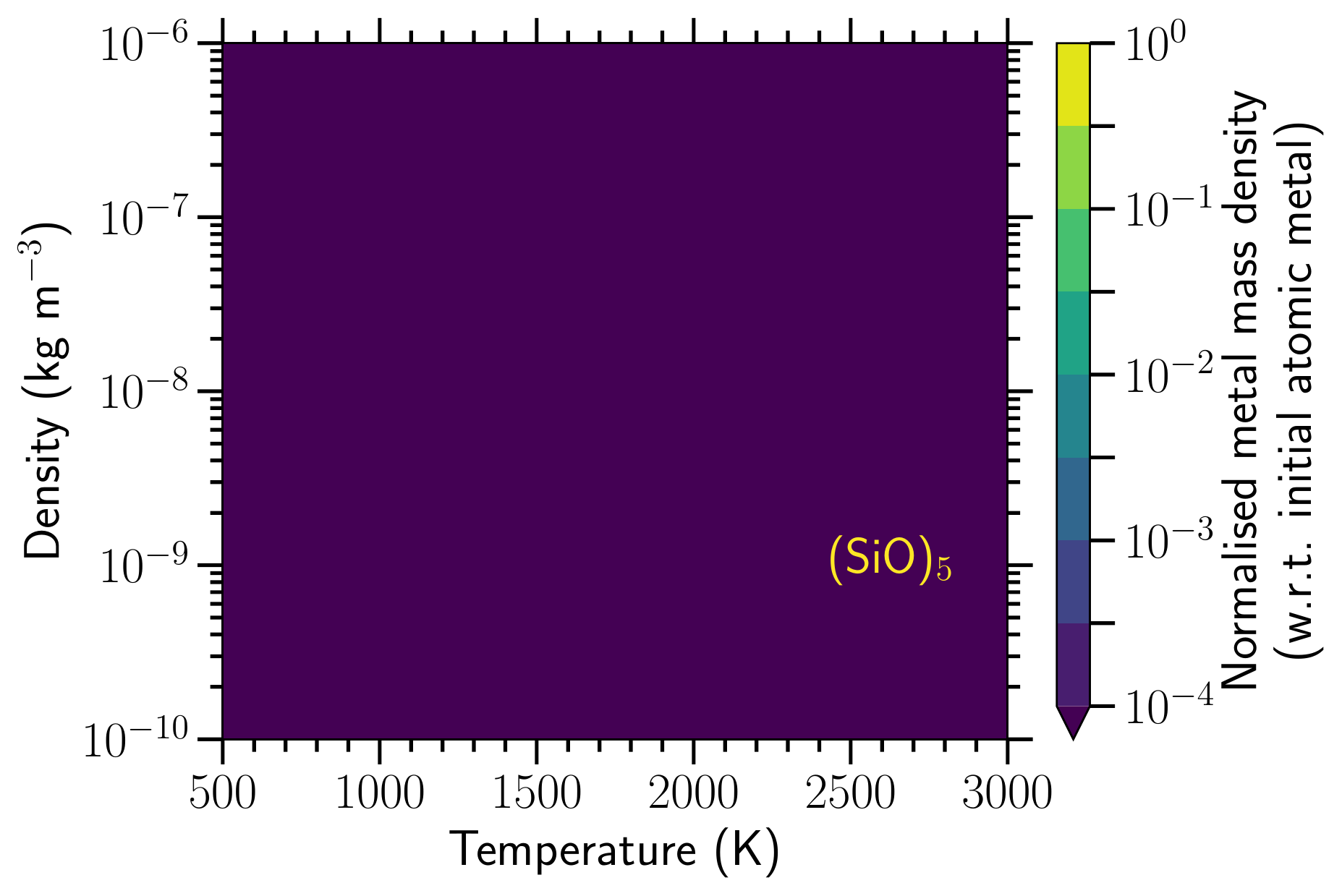}
        \includegraphics[width=0.32\textwidth]{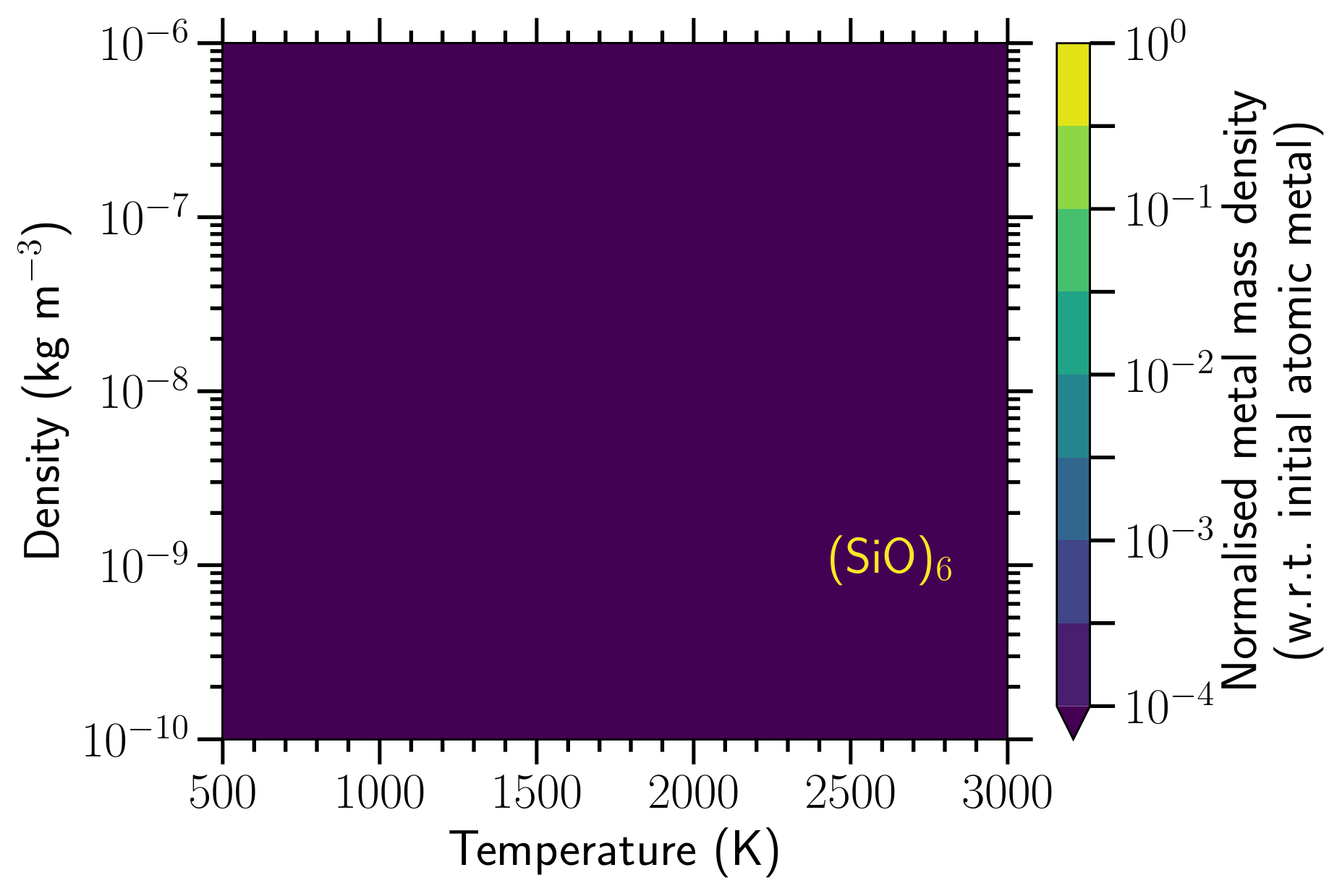}
        \includegraphics[width=0.32\textwidth]{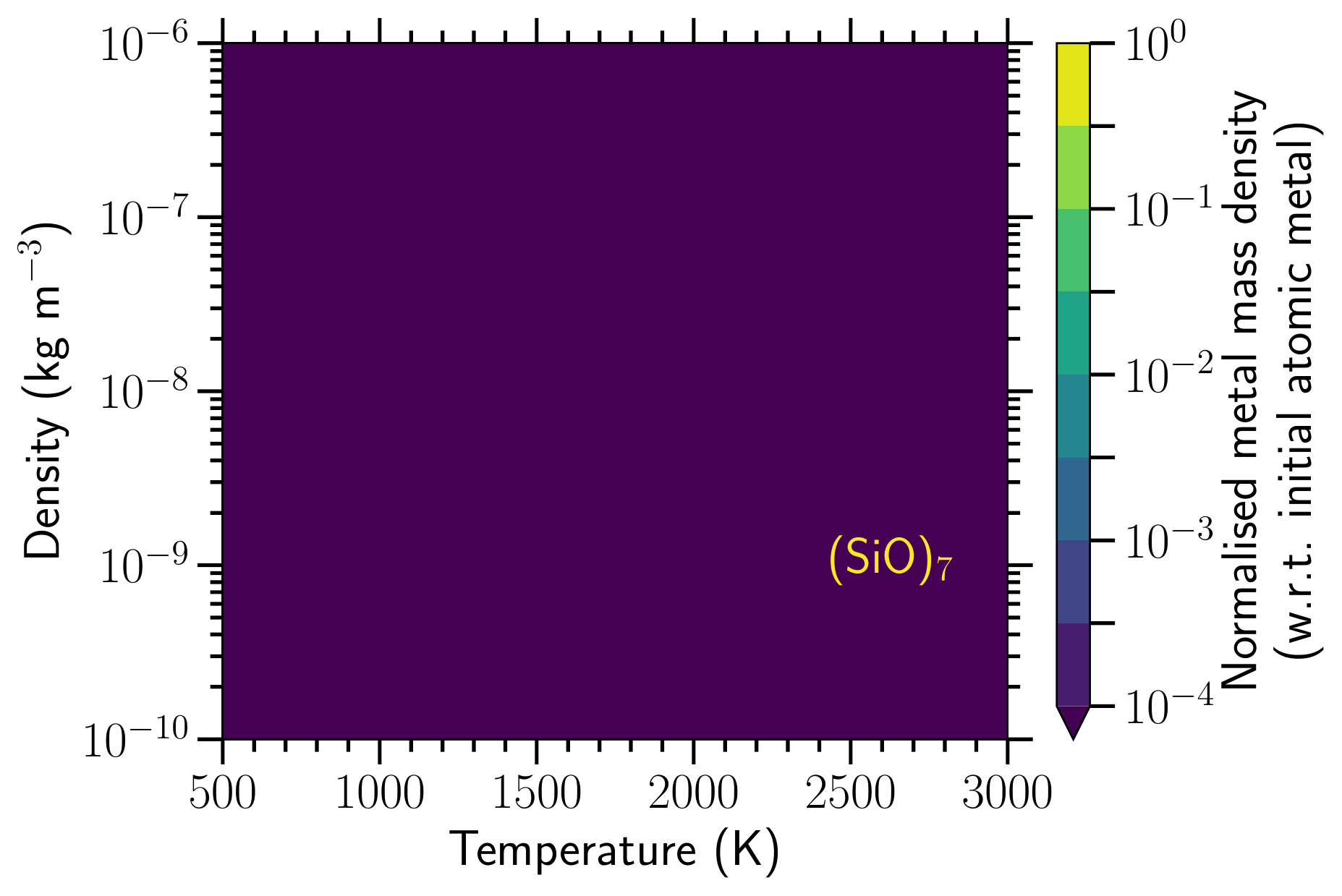}
        \includegraphics[width=0.32\textwidth]{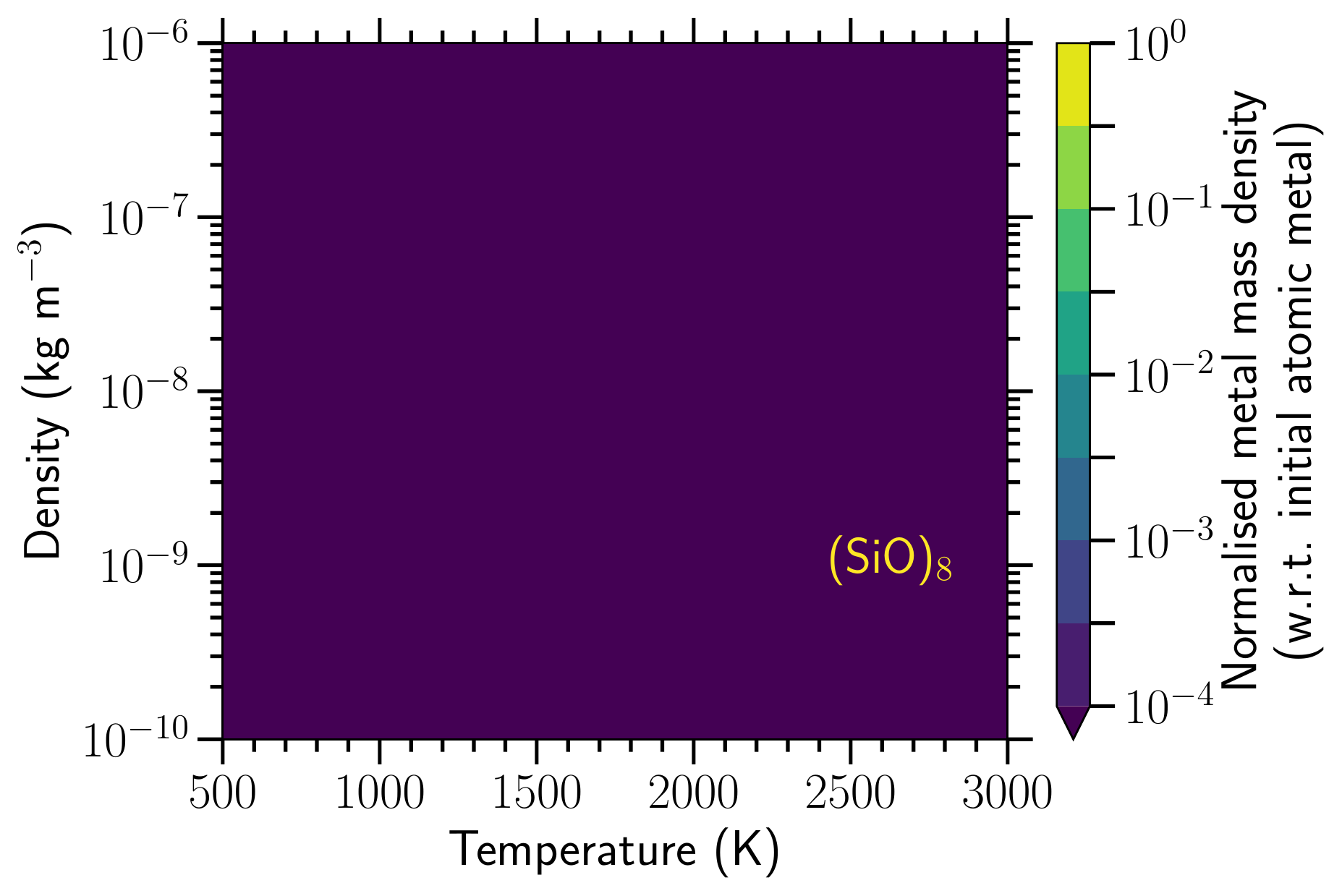}
        \includegraphics[width=0.32\textwidth]{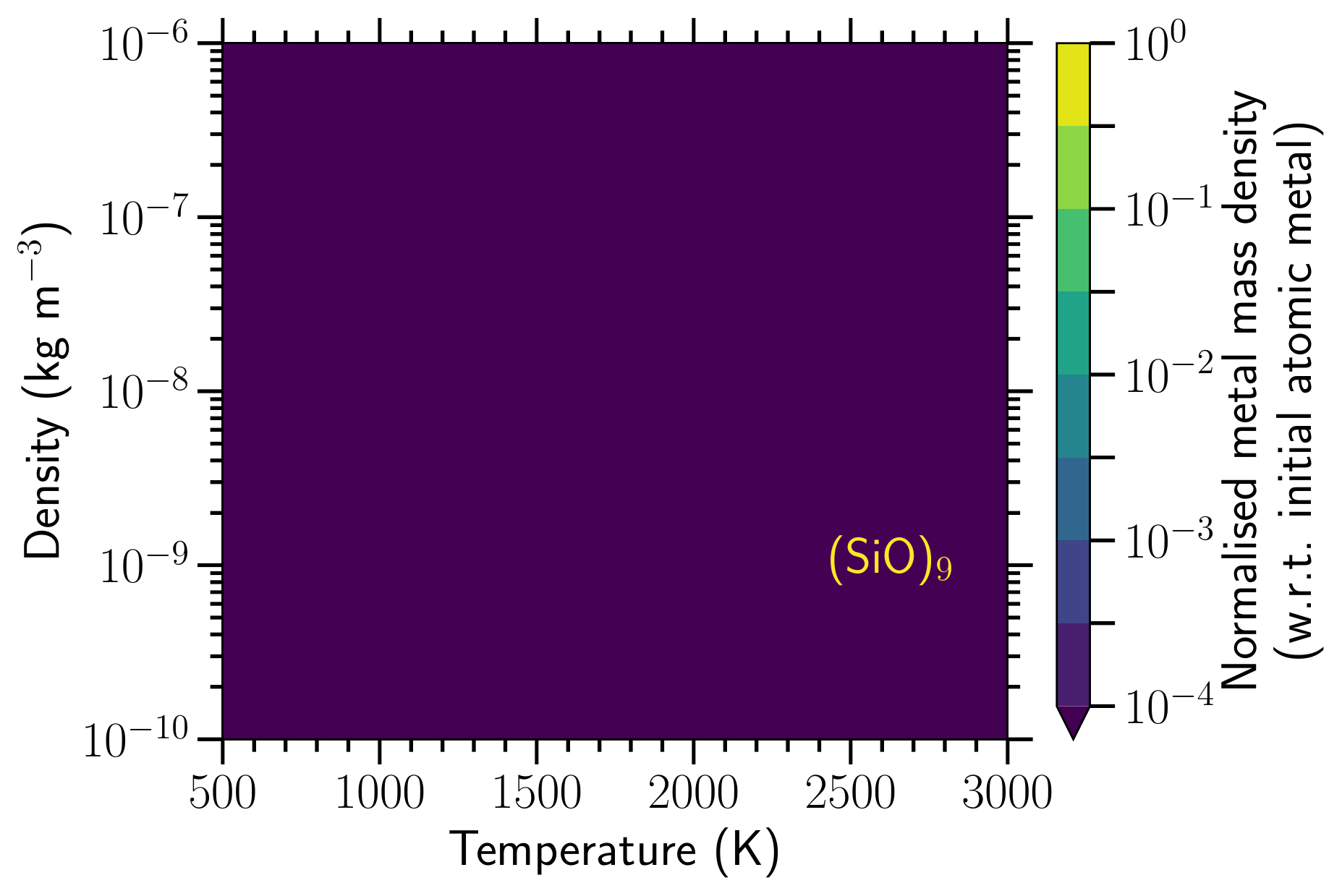}
        \includegraphics[width=0.32\textwidth]{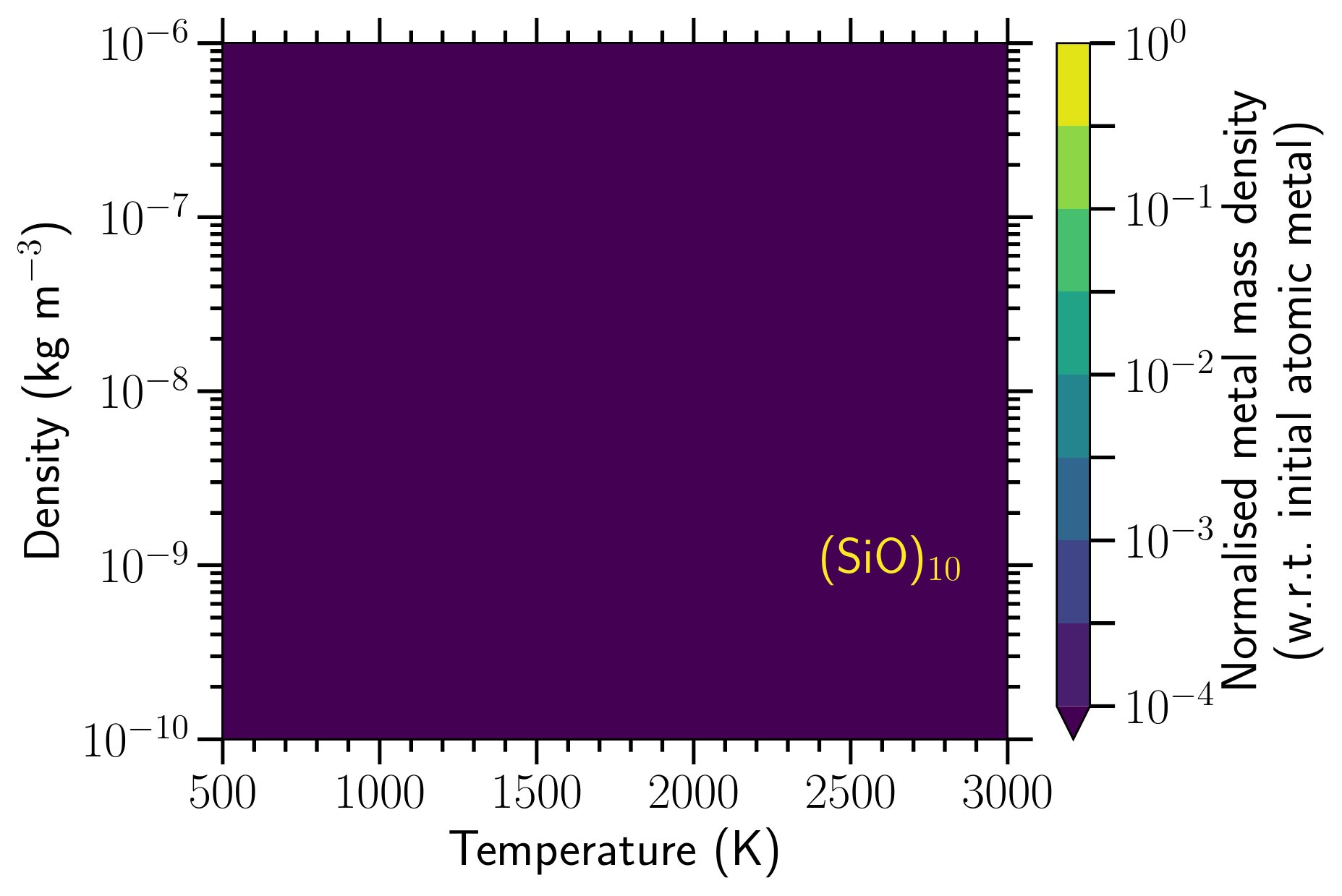}
        \includegraphics[width=0.32\textwidth]{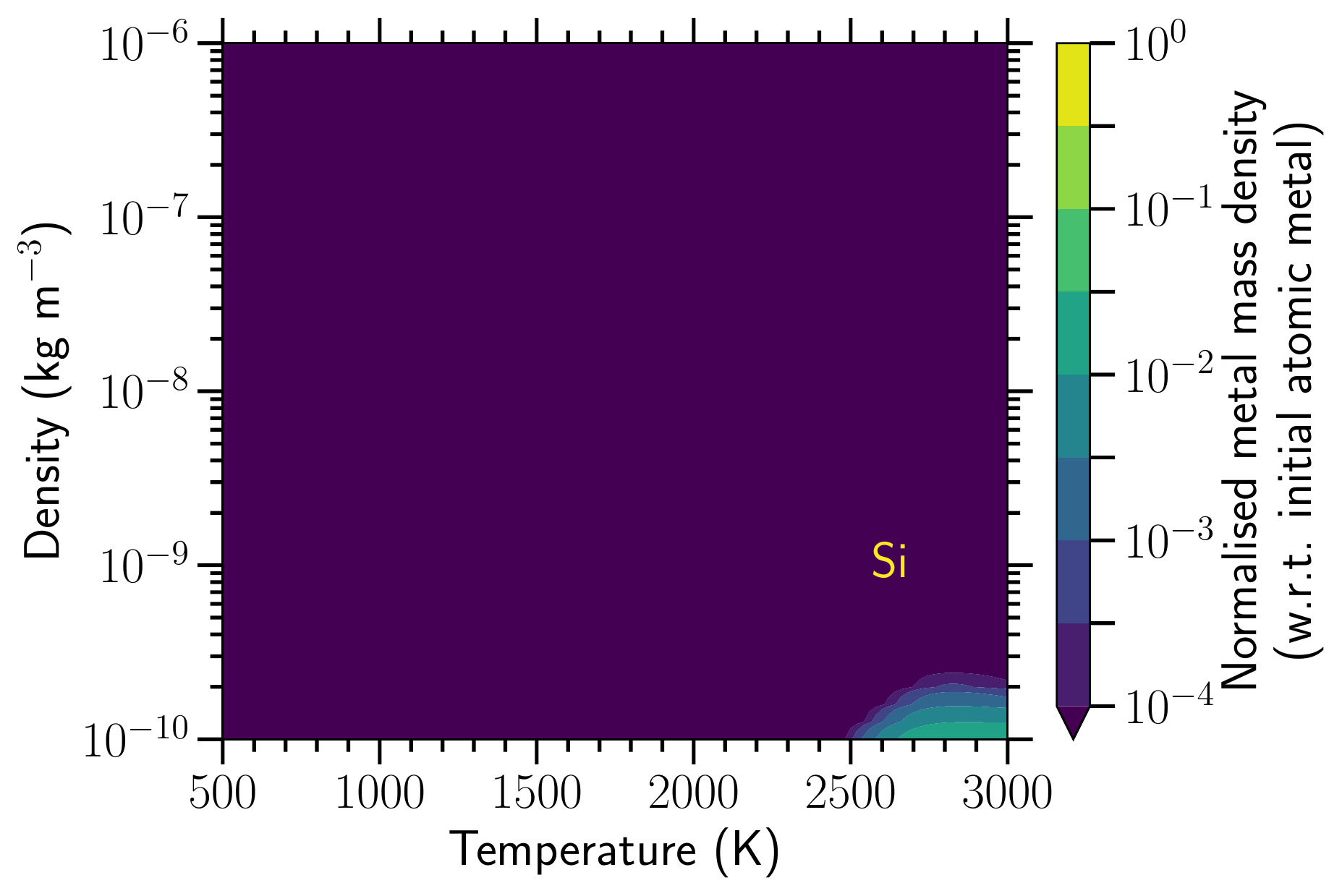}
        \end{flushleft}
        \caption{Overview of the normalised mass density of all \ch{Si}-bearing species for the comprehensive chemical nucleation model using the polymer nucleation description. Species with zero abundance are not shown.}
        \label{fig:full_ntw_Si-molecules_norm_same_scale}
    \end{figure*}

        This section contains a more complete overview of all \ch{Si}-bearing species results in the comprehensive chemical nucleation model using the polymer nucleation description \fig{\ref{fig:full_ntw_Si-molecules_norm_same_scale}}.
        
        \subsubsection{\ch{Al}-bearing species}
\begin{figure*}
        \begin{flushleft}
        \includegraphics[width=0.32\textwidth]{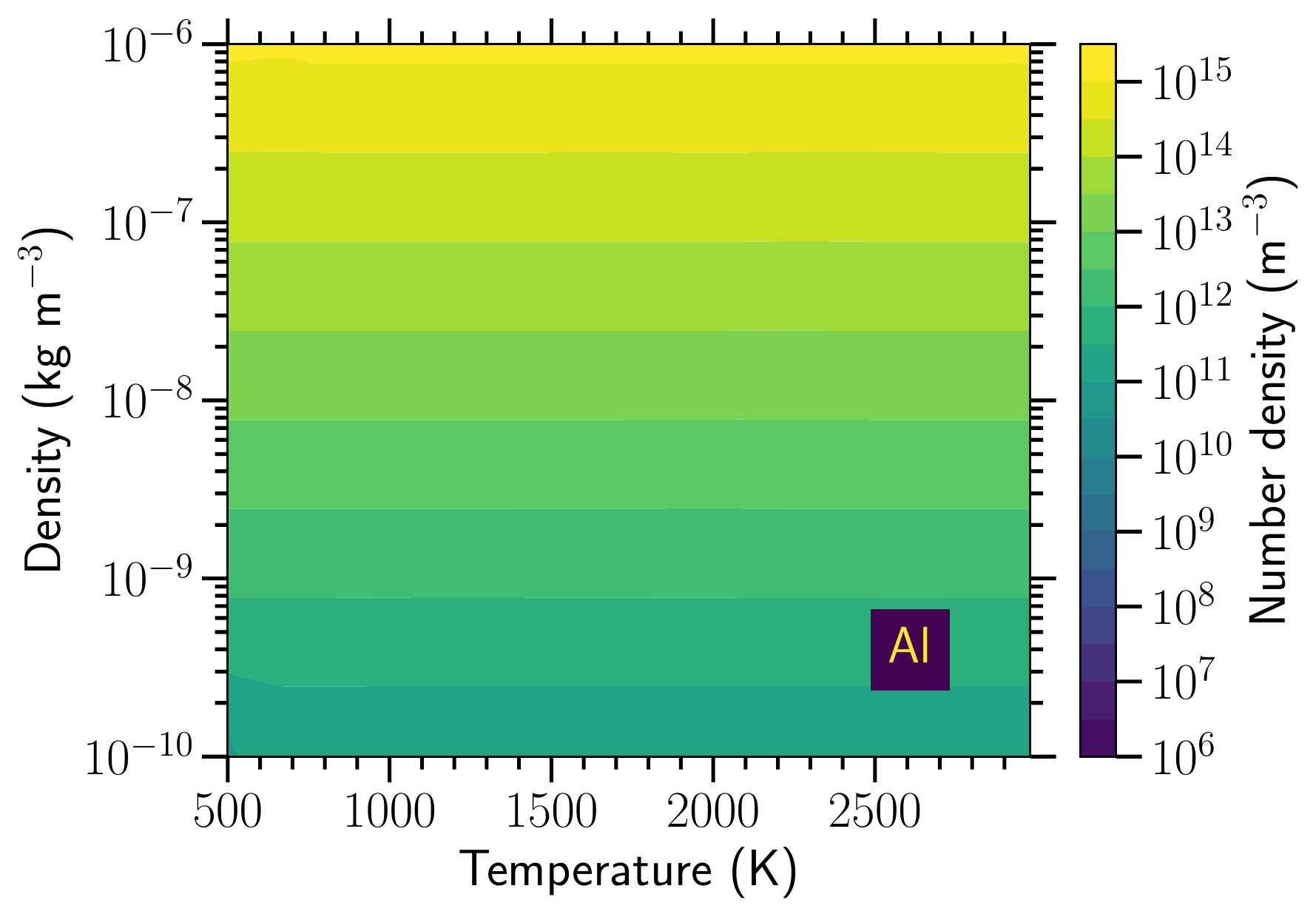}
        \includegraphics[width=0.32\textwidth]{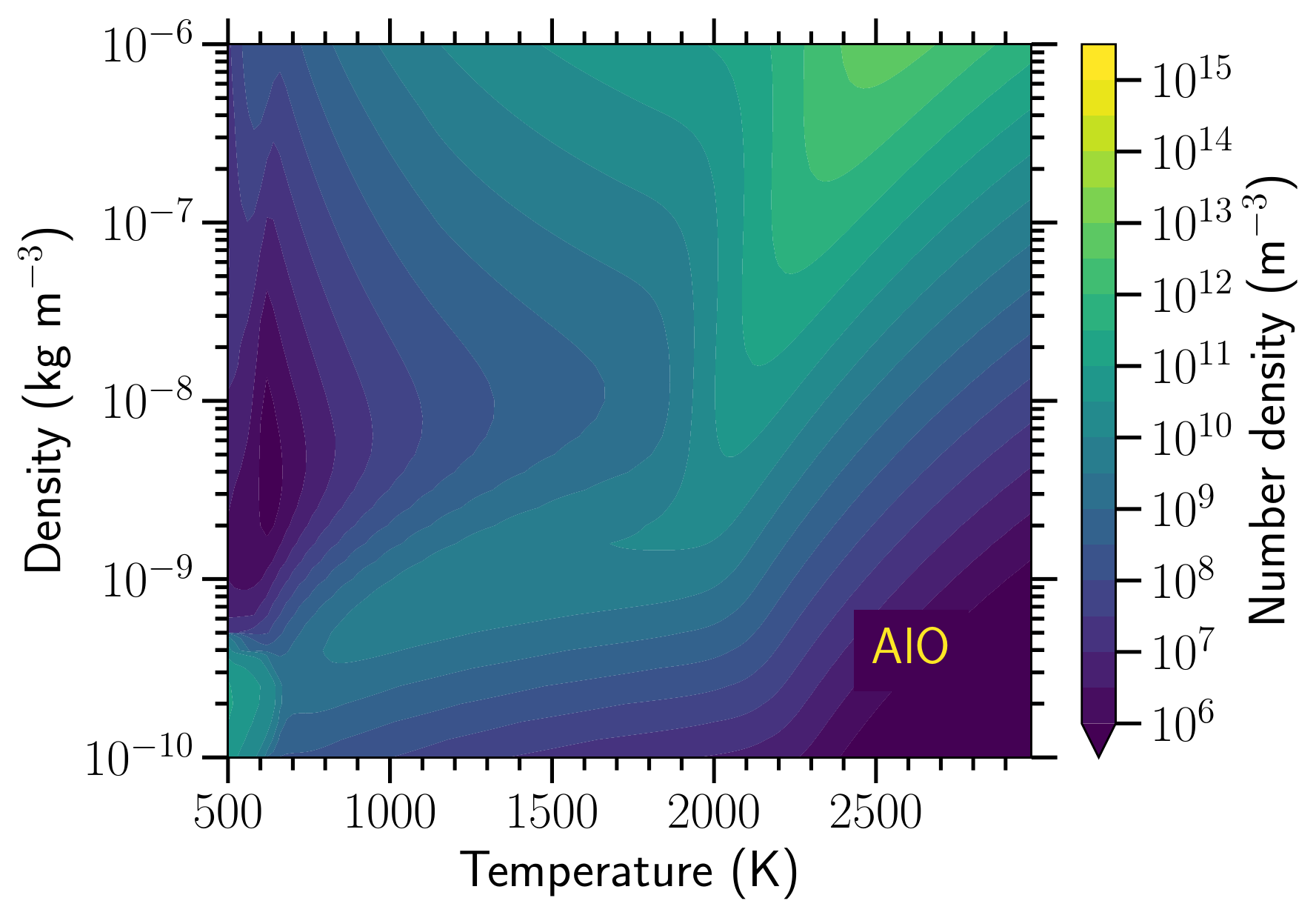}
        \includegraphics[width=0.32\textwidth]{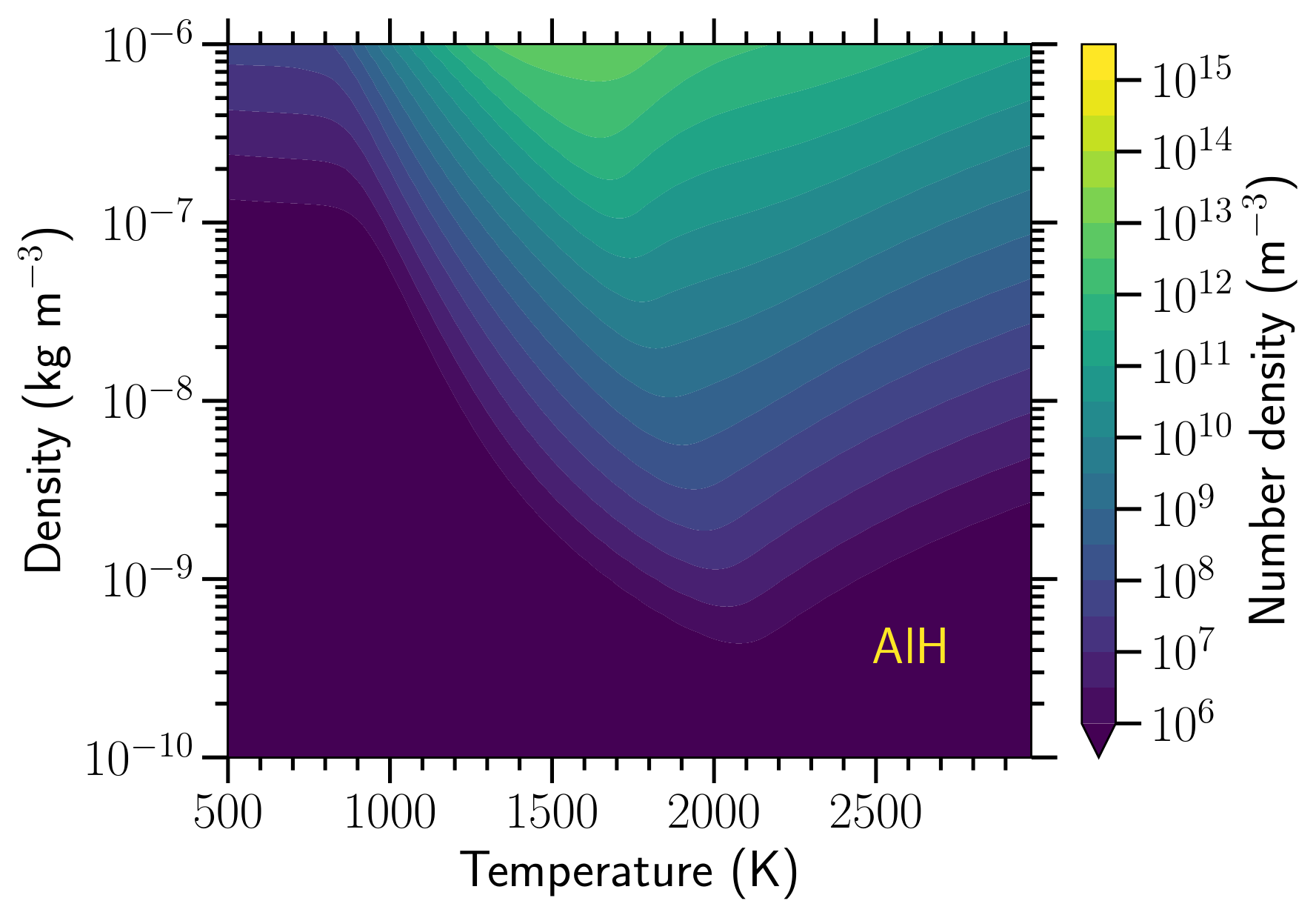}
        \includegraphics[width=0.32\textwidth]{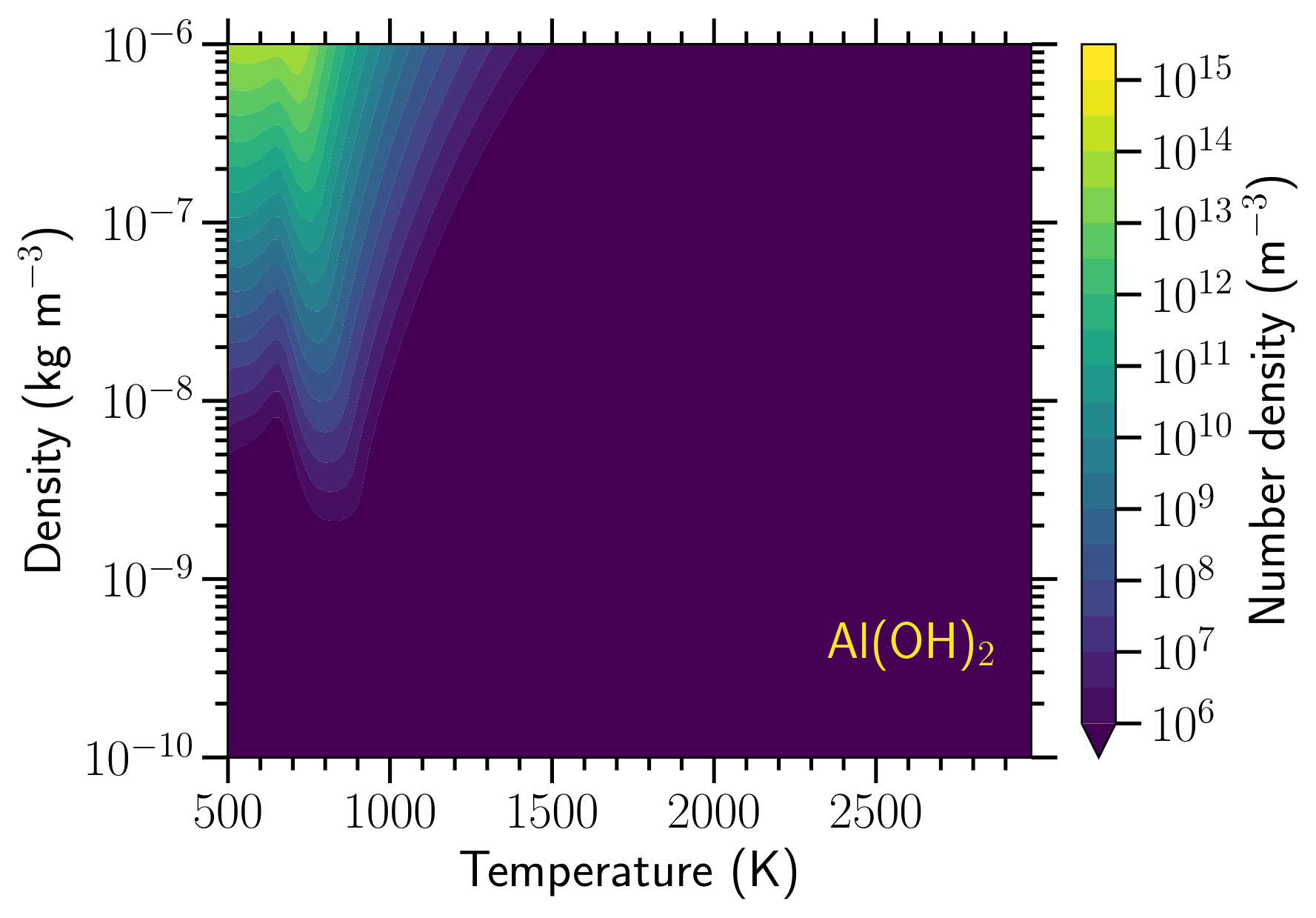}
        \includegraphics[width=0.32\textwidth]{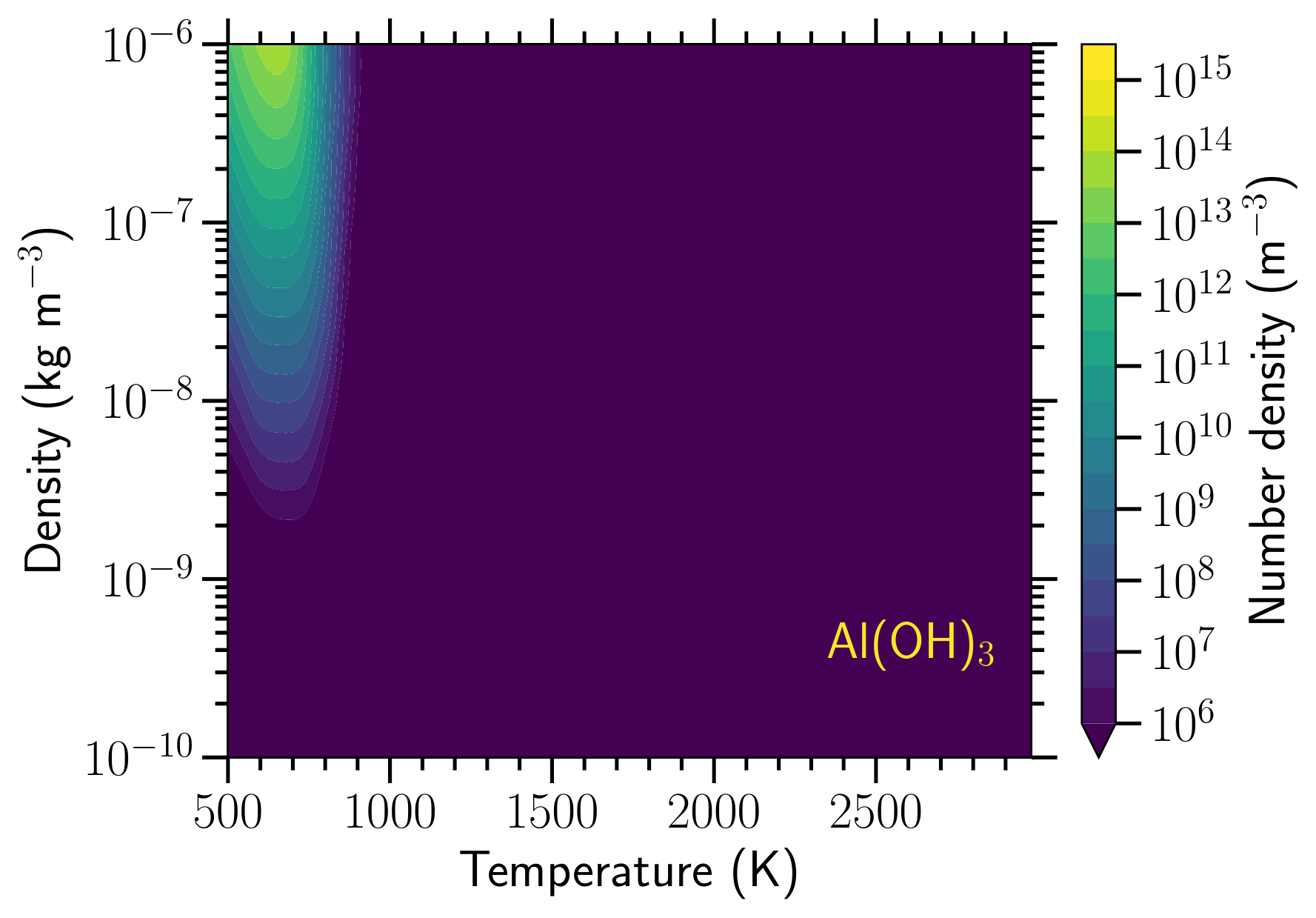}
        \includegraphics[width=0.32\textwidth]{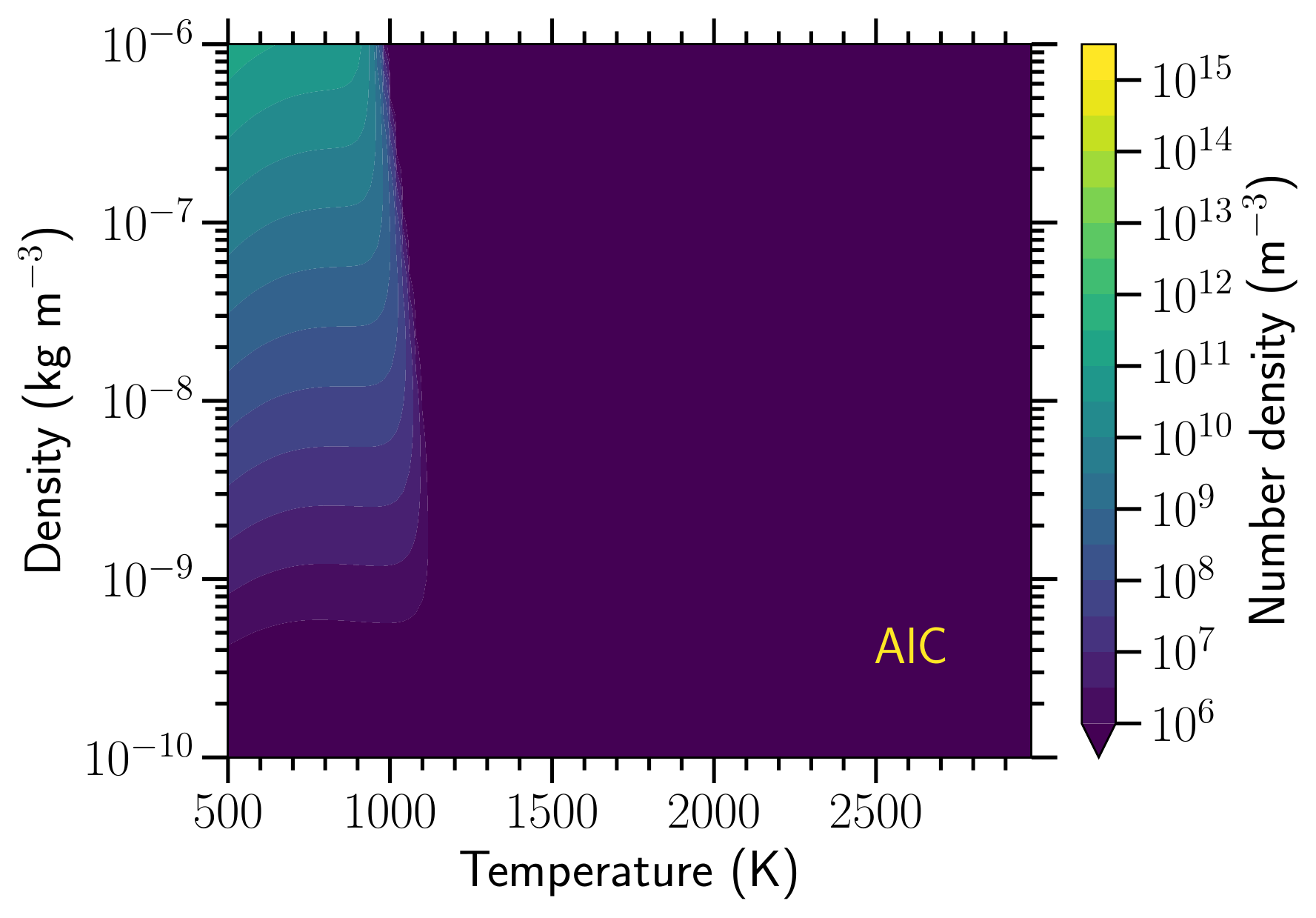}
        \includegraphics[width=0.32\textwidth]{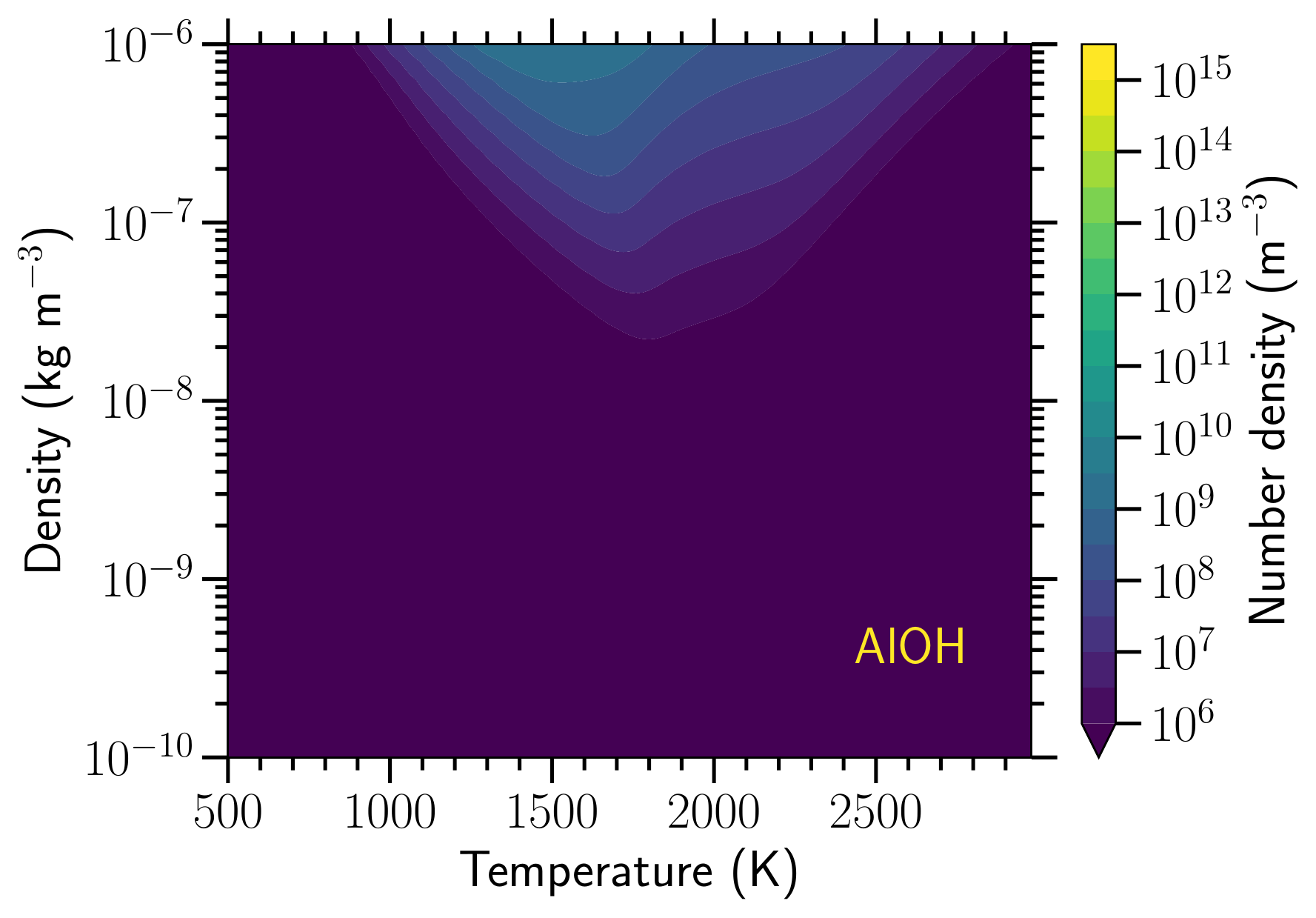}
        \includegraphics[width=0.32\textwidth]{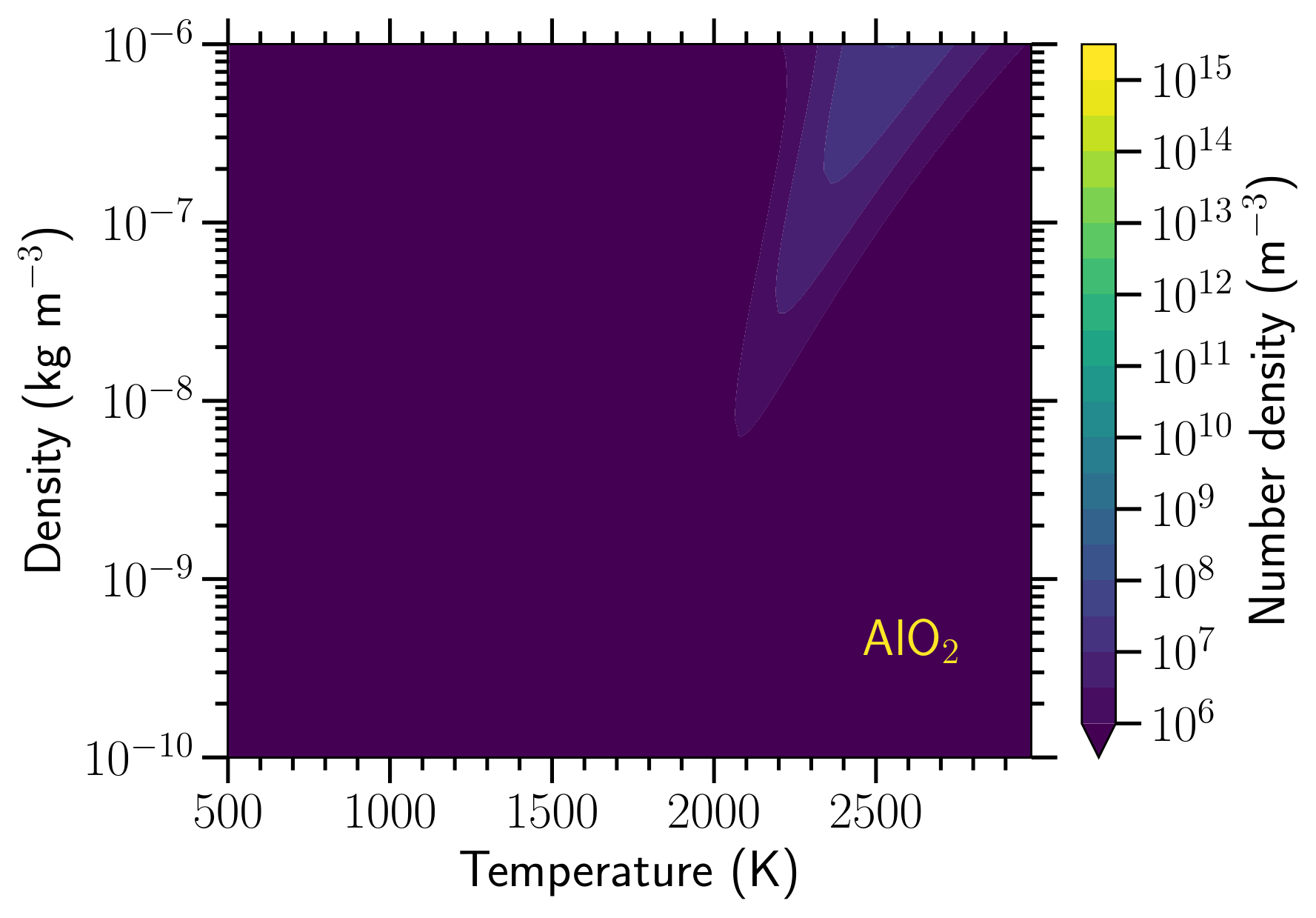}
        \end{flushleft}
        \caption{Overview of the absolute number density after one year of all \ch{Al}-bearing species for the comprehensive chemical nucleation model using the polymer nucleation description. Species with zero abundance are not shown.}
        \label{fig:full_ntw_Al-molecules}
    \end{figure*}

\begin{figure*}
        \begin{flushleft}
        \includegraphics[width=0.32\textwidth]{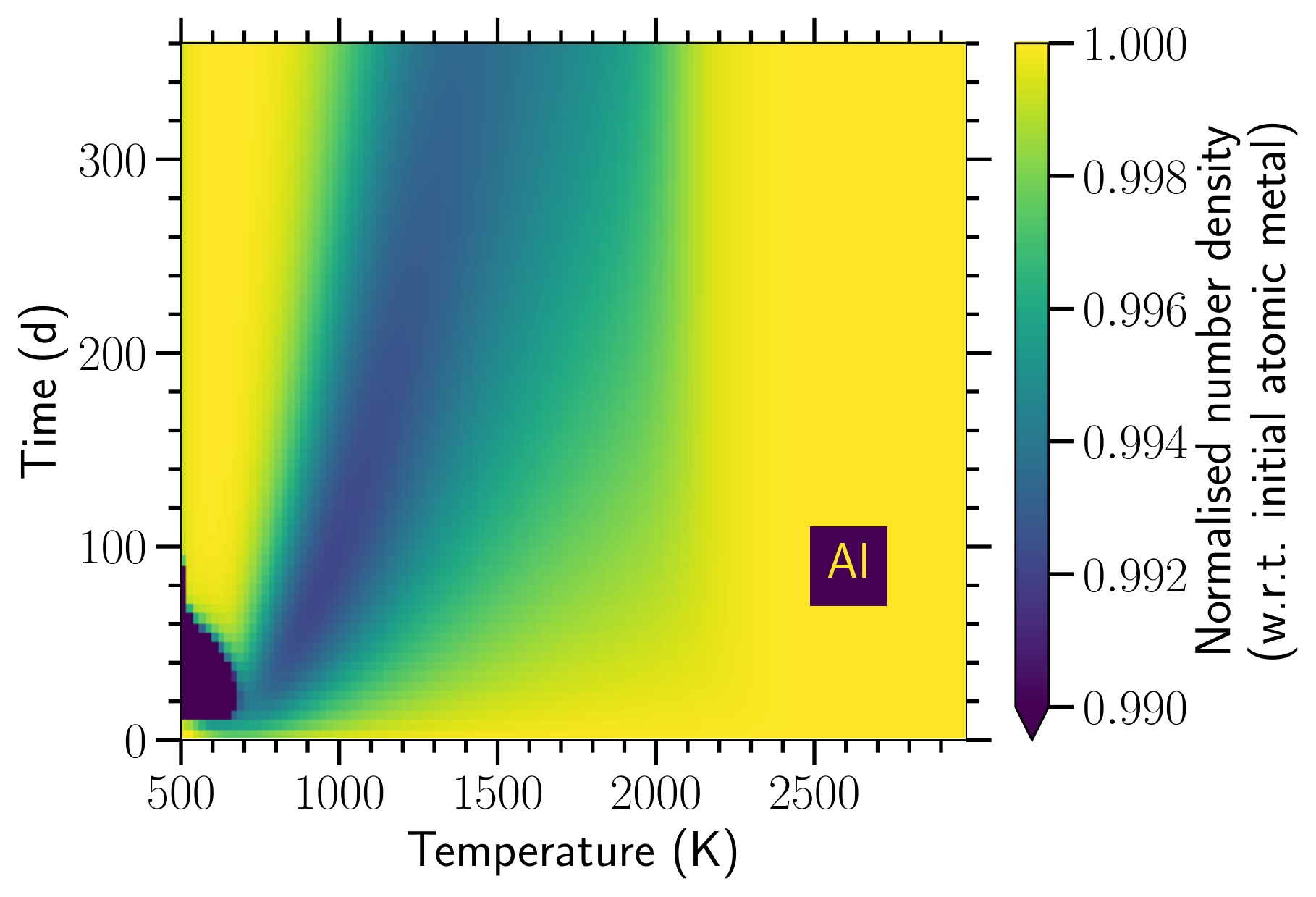}
        \includegraphics[width=0.32\textwidth]{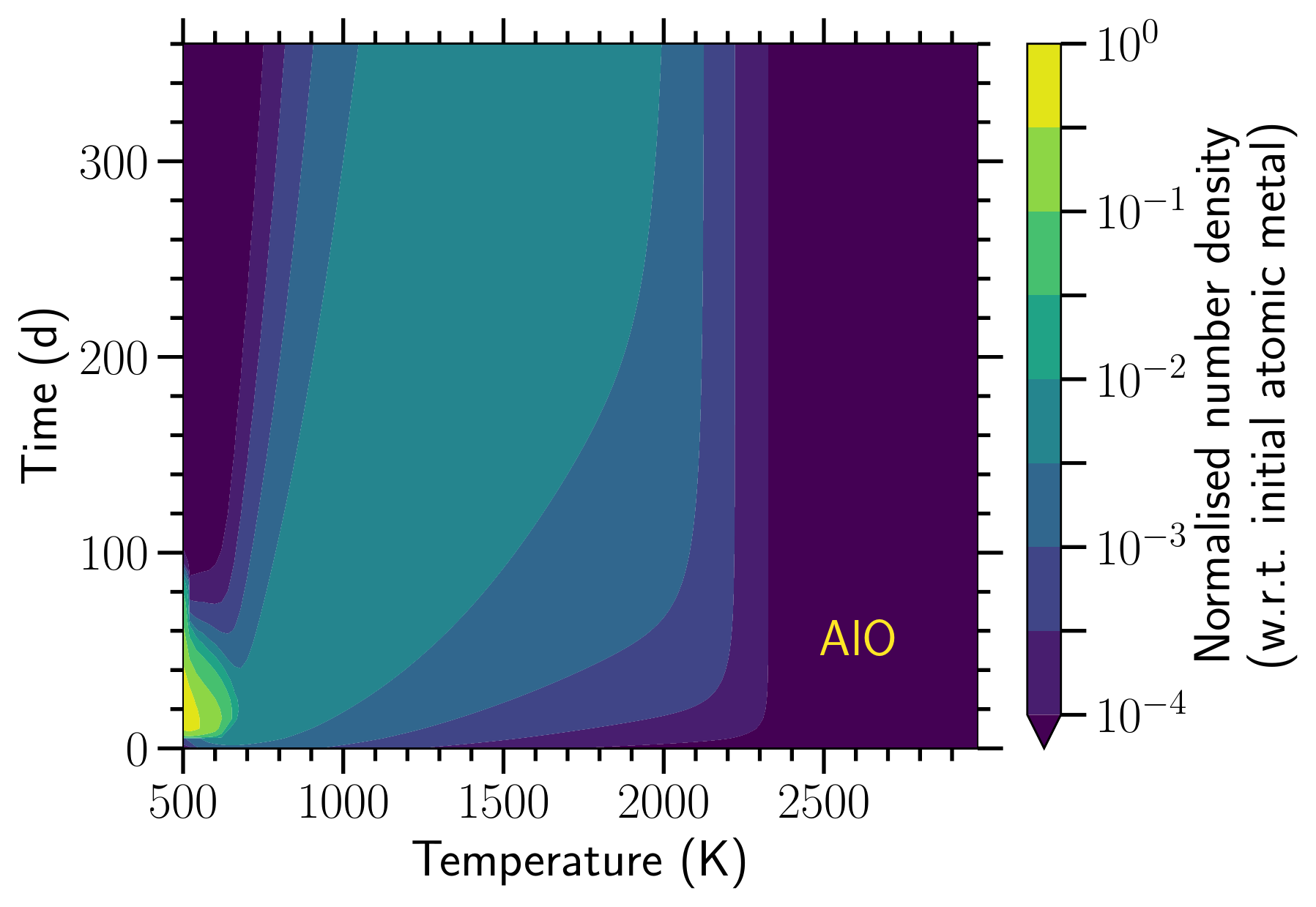}
        \end{flushleft}
        \caption{Temporal evolution of the absolute number density of all \ch{Al}-bearing species at the benchmark total gas density $\rho=\SI{1e-9}{\kg\per\m\cubed}$ for the comprehensive chemical nucleation model using the polymer nucleation description. Species with zero abundance are not shown.}
        \label{fig:full_ntw_Al-molecules_norm_same_scale_time_evolution}
    \end{figure*}

        This section contains a more complete overview of all \ch{Al}-bearing species results in the comprehensive chemical nucleation model using the polymer nucleation description         \figs{\ref{fig:full_ntw_Al-molecules}}{\ref{fig:full_ntw_Al-molecules_norm_same_scale_time_evolution}}.

\section{Chemical network}\label{app:chem_network}
This appendix lists all the used reactions with their reaction rate coefficient and the source of this data (Tab.~\ref{tab:full-chem-network}). This is the comprehensive chemical network used in this paper. Subsets of this network are not explicitly listed, i.e. the closed nucleation networks.\\\\
\onecolumn
\renewcommand{\arraystretch}{1.5}
\LTcapwidth=\linewidth
 } \\\\
\end{tabular}

\textbf{References:} (1) \cite{Washburn2008}, (2) \cite{Sharipov2012}, (3) \cite{Starik2014}, (4) \cite{Sharipov2011}, (5) \cite{Sharipov2016}, (6)  \NUCLEATION, (7) \cite{Swihart2003}, (8)  \UMIST, (9)  \KIDA, (10)  \cite{Verner1996}, (11)  \cite{Janev1987}, (12)  \cite{Forrey2013}, (13)  \cite{Glover2008}, (14) \cite{DeMore1997}, (15)  \cite{Abel1997}, (16)  \cite{Poulaert1978}, (17)  \cite{Capitelli2007}, (18)  \cite{Glover2010}, (19) \cite{Verner1996}, (20) \cite{Plane2015}, (21) \cite{Whalley2011}, (22) \cite{Plane2012}, (23) \cite{Plane1995}, (24) \cite{Rollason2001}, (25) \cite{Langowski2015}, (26) \cite{Atkinson2004}, (27) \cite{Campbell1993}, (28) \cite{Ritter1989}, (29) \cite{Higuchi2008} and (30) \cite{Plane2013}\\\\
\textbf{Notes:} \rev{The reactants \ch{M} act as catalyists and can be any species. Therefore the total number density of the gas $n_\text{tot}$ is used as its density.} References of reactions that contain the equilibrium ratio function $\EqRatio$ refer to the references of the reversed reaction. \rev{Not all reaction rate coefficients are valid in the considered temperature range. However, due to a lack of literature data, those coeffiecients are extrapolated in temperature when necessary.}

\label{lastpage}
\end{document}